\documentclass{aastex7}
\usepackage[utf8]{inputenc}
\usepackage{textcomp}
\usepackage{amsmath}
\usepackage{newunicodechar}
\newunicodechar{−}{-}

\graphicspath{{./}{figures/}}
\newif\ifapjfigset \apjfigsetfalse
\usepackage{tikz}
\usetikzlibrary{positioning,arrows.meta}

\begin{document}

\title{Discovery of over a thousand variable stars in Terzan 5 and Liller 1 with JWST}

\author[0000-0002-7226-836X]{Kevin B. Burdge}
\email[show]{kburdge@mit.edu}
\affiliation{Department of Physics, Massachusetts Institute of Technology, Cambridge, MA 02139, USA}
\affiliation{Kavli Institute for Astrophysics and Space Research, Massachusetts Institute of Technology, Cambridge, MA 02139, USA}

\author[0000-0002-4770-5388]{Ilaria Caiazzo}
\email{ilaria.caiazzo@ist.ac.at}
\affiliation{Institute of Science and Technology Austria (ISTA), Am Campus 1, 3400 Klosterneuburg, Austria}

\author[0009-0003-4448-3681]{Malina M. Desai}
\email{mmdesai@mit.edu}
\affiliation{Department of Physics, Massachusetts Institute of Technology, Cambridge, MA 02139, USA}
\affiliation{Kavli Institute for Astrophysics and Space Research, Massachusetts Institute of Technology, Cambridge, MA 02139, USA}

\author[0000-0002-7104-2107]{Cristina Pallanca}
\email{cristina.pallanca3@unibo.it}
\affiliation{Dipartimento di Fisica e Astronomia ``Augusto Righi'', Universit\`a di Bologna, Via Gobetti 93/2, 40129 Bologna, Italy}
\affiliation{INAF – Osservatorio di Astrofisica e Scienza dello Spazio di Bologna (OAS), 
  Via Piero Gobetti 93/3, I-40129 Bologna, Italy}

\author[0000-0002-7303-8144]{Vera L. Berger}
\email{vlberger@space.mit.edu}
\affiliation{Department of Physics, Massachusetts Institute of Technology, Cambridge, MA 02139, USA}
\affiliation{Kavli Institute for Astrophysics and Space Research, Massachusetts Institute of Technology, Cambridge, MA 02139, USA}

\author[0000-0002-8810-858X]{Kaley Brauer}
\email{kaley.brauer@cfa.harvard.edu}
\affiliation{Center for Astrophysics | Harvard \& Smithsonian, Cambridge, MA 02138, USA}

\author[0000-0001-6464-3257]{Matteo Correnti}
\email{matteo.correnti@inaf.it}
\affiliation{INAF Osservatorio Astronomico di Roma, Via Frascati 33, 00078, Monteporzio Catone, Rome, Italy}
\affiliation{ASI-Space Science Data Center, Via del Politecnico, I-00133, Rome, Italy}

\author[0000-0002-2218-2306]{Paul Draghis}
\email{pdraghis@mit.edu}
\affiliation{Kavli Institute for Astrophysics and Space Research, Massachusetts Institute of Technology, Cambridge, MA 02139, USA}

\author[0000-0002-6871-1752]{Kareem El-Badry}
\email{kelbadry@caltech.edu}
\affiliation{Division of Physics, Mathematics and Astronomy, California Institute of Technology, Pasadena, CA 91125, USA}

\author[0000-0002-2165-8528]{Francesco R. Ferraro}
\email{francesco.ferraro3@unibo.it}
\affiliation{Dipartimento di Fisica e Astronomia ``Augusto Righi'', Universit\`a di Bologna, Via Gobetti 93/2, 40129 Bologna, Italy}
\affiliation{INAF – Osservatorio di Astrofisica e Scienza dello Spazio di Bologna (OAS), 
  Via Piero Gobetti 93/3, I-40129 Bologna, Italy}
  
\author[0000-0001-5848-0180]{Denis Gonz\'alez-Caniulef}
\email{dgonzalez-ca@irap.omp.eu}
\affiliation{University of Toulouse, CNES, CNRS, IRAP, Toulouse, France}

\author[0000-0002-4086-3180]{Kyle Kremer}
\email{kykremer@ucsd.edu}
\affiliation{Department of Astronomy and Astrophysics, University of California, San Diego, La Jolla, CA 92093, USA}

\author[0000-0001-5613-4938]{Barbara Lanzoni}
\email{barbara.lanzoni3@unibo.it}
\affiliation{Dipartimento di Fisica e Astronomia ``Augusto Righi'', Universit\`a di Bologna, Via Gobetti 93/2, 40129 Bologna, Italy}
\affiliation{INAF – Osservatorio di Astrofisica e Scienza dello Spazio di Bologna (OAS), 
  Via Piero Gobetti 93/3, I-40129 Bologna, Italy}

\author[0000-0001-7931-0607]{Dongzi Li}
\email{dzli@tsinghua.edu.cn}
\affiliation{Department of Astronomy, Tsinghua University, Beijing 100084, China}

\author[0000-0001-9611-0009]{Jessica R. Lu}
\email{jlu.astro@berkeley.edu}
\affiliation{Department of Astronomy, University of California, Berkeley, CA 94720-3411, USA}

\author[0000-0001-6331-112X]{Geoffrey Mo}
\email{gmo@caltech.edu}
\affiliation{The Observatories of the Carnegie Institution for Science, 813 Santa Barbara Street, Pasadena, CA 91101, USA}
\affiliation{Division of Physics, Mathematics and Astronomy, California Institute of Technology, Pasadena, CA 91125, USA}

\author[0000-0002-0940-6563]{Mason Ng}
\email{masonng@mit.edu}
\affiliation{Department of Physics, McGill University, 3600 rue University, Montr\'eal, QC H3A 2T8, Canada}
\affiliation{Trottier Space Institute, McGill University, 3550 rue University, Montr\'eal, QC H3A 2A7, Canada}

\author[0000-0002-9556-1876]{Roger W. Romani}
\email{rwr@stanford.edu}
\affiliation{Department of Physics, Stanford University, Stanford, CA 94305, USA}
\affiliation{Kavli Institute for Particle Astrophysics and Cosmology, Stanford University, Stanford, CA 94305, USA}

\author[0000-0002-6823-2073]{Kaitlyn Shin}
\email{kaitshin@caltech.edu}
\affiliation{Division of Physics, Mathematics and Astronomy, California Institute of Technology, Pasadena, CA 91125, USA}

\author[0000-0002-6442-6030]{Daniel R. Weisz}
\email{dan.weisz@berkeley.edu}
\affiliation{Department of Astronomy, University of California, Berkeley, CA 94720-3411, USA}
\affiliation{Miller Institute for Basic Research, University of California Berkeley, Berkeley, CA, 94720, USA}

\author[0009-0001-1480-198X]{Giorgia Zullo}
\email{giorgia.zullo2@unibo.it}
\affiliation{Dipartimento di Fisica e Astronomia ``Augusto Righi'', Universit\`a di Bologna, Via Gobetti 93/2, 40129 Bologna, Italy}
\affiliation{INAF – Osservatorio di Astrofisica e Scienza dello Spazio di Bologna (OAS), 
  Via Piero Gobetti 93/3, I-40129 Bologna, Italy}


\begin{abstract}
The Bulge Fossil Fragments Terzan 5 and Liller 1 host some of the densest stellar environments in the Galaxy, yet their variable star populations have remained largely unexplored due to crowding and heavy obscuration. Here, we identify 1,315 variable sources in these two systems in what is to our knowledge the first JWST time-series variability census of a globular cluster-like object. By differencing consecutive non-destructive reads within each NIRCam integration, we transform sample-up-the-ramp imaging into an infrared movie sampled every 21 seconds. More than a century of observations has cataloged 5,604 variables, the majority of them pulsating stars, across 151 Galactic globular clusters. Our census alone amounts to 23 percent of that total and contains more photometrically variable binaries than had previously been cataloged in all globular clusters combined. The field of Liller 1, with zero previously known variables, now hosts 915, nearly double the count in the runner-up $\omega$~Centauri. We detect dramatic infrared variability from the Rapid Burster, revealing the counterpart sought since 1976 (see \citealt{Desai2026}), and blindly recover a signal coincident with the prototype redback pulsar PSR~J1748$-$2446A at its radio-timed orbital period, identifying the first optical or infrared counterpart to any of Terzan 5's pulsars. More broadly, we present a technique that recovers group-level time-series photometry from JWST observations, enabling time-domain studies with an observatory whose field of view is naturally matched to the Galaxy's densest stellar systems.
\end{abstract}



\section{Introduction}\label{sec:intro}

Globular clusters host some of the most dynamically active stellar populations in the Galaxy. These gravitationally bound systems span masses of $\sim 10^3$--$10^6\,M_{\odot}$ \citep{BaumgardtHilker2018}, with the most massive lying within an order of magnitude of the mass of the Milky Way's nuclear star cluster \citep{Neumayer2020}. Their central stellar densities can exceed $10^6\,M_{\odot}\,\mathrm{pc}^{-3}$ \citep[e.g.,][]{Lanzoni2010,BaumgardtHilker2018}, driving frequent dynamical encounters that are thought to continually reshape their binary populations.

The high stellar encounter rates in globular cluster cores drive the formation and hardening of compact binaries through three-body exchanges, tidal capture, and direct collisions \citep{Hut1992,Ivanova2008}. The number of X-ray sources per cluster correlates with the stellar encounter rate \citep{Pooley2003}, indicating that dynamics, rather than the primordial binary fraction, governs the exotic binary populations in dense clusters. These dynamically formed systems include low-mass X-ray binaries \citep{Clark1975}, millisecond pulsars \citep{Ransom2005}, and cataclysmic variables \citep{Grindlay2001}. The merging black hole binaries detected by LIGO/Virgo may also form preferentially through such encounters \citep{PortegiesZwart2000,Rodriguez2016,Antonini2019}.

Two globular cluster-like stellar systems stand out as the most extreme known environments for understanding the role of dynamical interactions in shaping binary populations: Liller~1 and Terzan~5. Residing in the Galactic bulge, they host stellar encounter rates higher than those of any Galactic globular cluster \citep{VerbuntHut1987,Saracino2015}, and are among the densest and most massive stellar systems in the Milky Way \citep{Lanzoni2010,Saracino2015,BaumgardtHilker2018}. Remarkably, both host multiple stellar populations spanning $\Delta t > 7$--8~Gyr in age and up to $\sim$1~dex in metallicity \citep{Ferraro2009,Ferraro2021,Origlia2011,Origlia2013,Crociati2023}. Age differences of this magnitude are known in no other Galactic globular cluster-like objects: $\omega$~Centauri exhibits an even larger metallicity spread \citep{JohnsonPilachowski2010} but predominantly ancient populations \citep{Clontz2024}, and the young populations projected on M54 belong to the surrounding Sagittarius nuclear star cluster \citep{Siegel2007,Bellazzini2008}. The age spreads in Liller~1 and Terzan~5 are thus evidence of recurrent star formation spanning most of cosmic time within these two systems. This has led to the hypothesis that they are ``Bulge Fossil Fragments'', surviving remnants of the massive primordial structures that coalesced to form the Galactic bulge \citep{Ferraro2021}. Some simulations suggest that Terzan~5 may have produced many merging black hole binaries over its lifetime, well in excess of the yield of a typical globular cluster, and that Liller~1, the other Bulge Fossil Fragment, may have been comparably efficient, which would make the two systems candidate factories of gravitational-wave sources \citep{Ferraro2026}.

\begin{figure*}
\centering
\includegraphics[width=\textwidth]{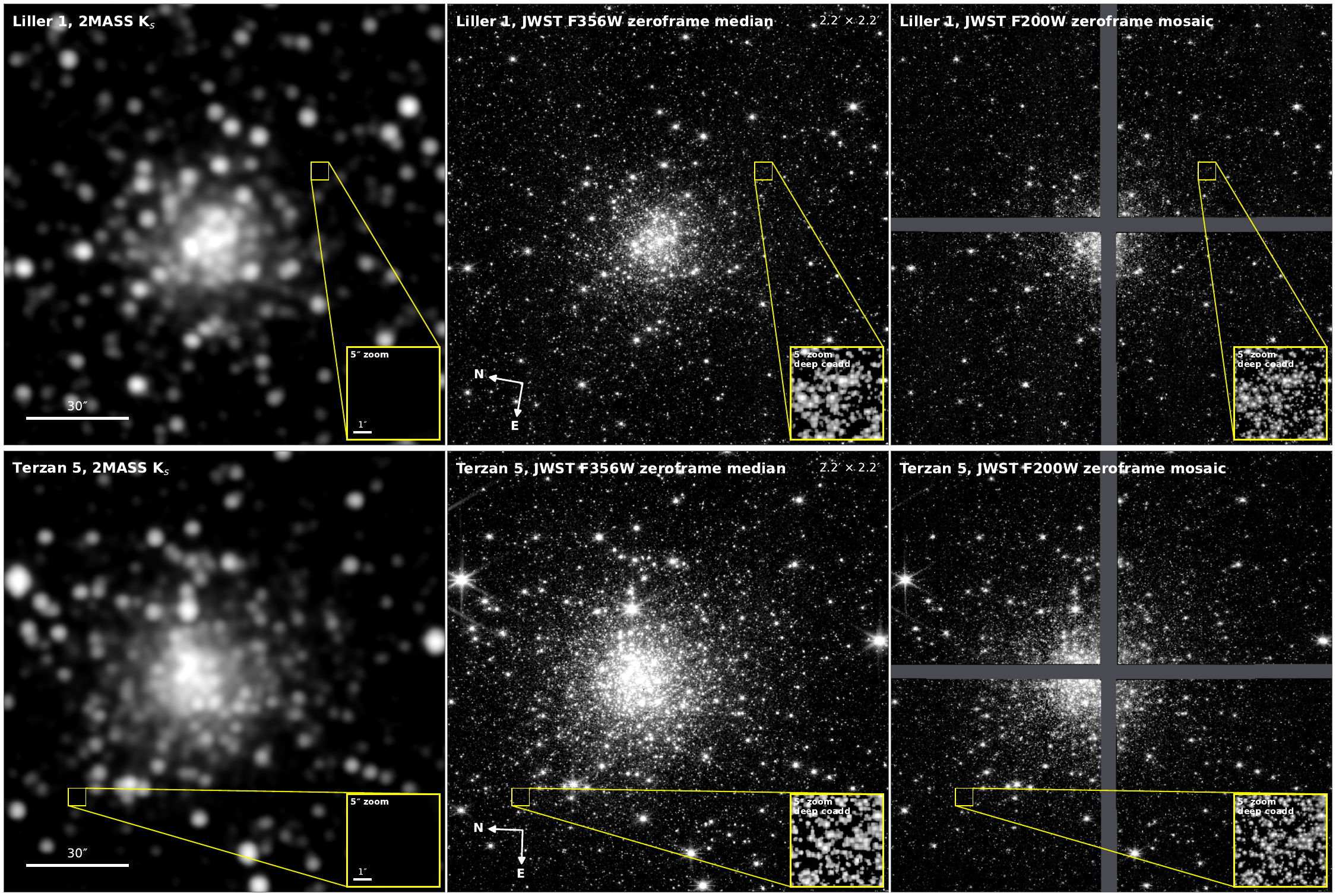}
\caption{Liller~1 (top) and Terzan~5 (bottom) as seen by 2MASS in $K_s$ \citep[left;][]{Skrutskie2006} and by JWST/NIRCam in F356W (center) and F200W (right), on the same $2\farcm2 \times 2\farcm2$ field, pixel grid, and orientation. The full-field JWST panels are medians of the zeroframes, the first read of each integration; the gray bands in the F200W mosaic are the gaps between the four short-wavelength detectors. The insets enlarge a $5\arcsec \times 5\arcsec$ region in each cluster's outskirts (yellow box), taken for JWST from the deep coadd of all 12 exposures: where 2MASS records smooth, unresolved light, NIRCam resolves more than a hundred stars.}
\label{fig:twomass_jwst}
\end{figure*}

Despite their extraordinary astrophysical importance, the variable-star populations of Liller~1 and Terzan~5 have remained almost entirely unexplored. Time-domain studies of globular clusters have historically concentrated on nearby, relatively unobscured systems, and these two bulge systems present nearly the worst case on both counts. Extreme foreground extinction ($E(B-V) \approx 4.5$ for Liller~1, \citealt{Pallanca2021}; $E(B-V) = 2.15$--$2.82$ across the face of Terzan~5, \citealt{Massari2012}) renders them effectively inaccessible to optical and ultraviolet surveys, while severe crowding challenges ground-based photometry even in the infrared, where the extinction is more forgiving. Figure~\ref{fig:twomass_jwst} illustrates the problem: at the $\sim$3\arcsec\ resolution of the Two Micron All Sky Survey (2MASS), the cores of both systems blend into unresolved light, and a 5\arcsec\ patch of their outskirts in which 2MASS registers no individual star holds more than a hundred stars resolved by NIRCam.

Previous attempts to identify infrared counterparts to their known X-ray populations have had limited success. For example, the Rapid Burster (MXB~1730$-$335) is one of only two known X-ray bursters exhibiting type~II bursts (driven by accretion instabilities rather than thermonuclear flashes; \citealt{Court2018}), yet its counterpart remained unidentified for nearly five decades despite extensive searches \citep{Homer2001}. A candidate was proposed only in 2025, but incomplete optical and infrared sampling and the lack of a secure orbital period prevented an unambiguous association \citep{Pallanca2025}. Our data reveal that candidate to be an unrelated eclipsing binary and identify the true counterpart (\S\ref{sec:rapid_burster_discussion}). A dedicated analysis of the counterpart, based primarily on the zeroframe imaging, is presented in a simultaneously submitted companion paper \citep{Desai2026}. Meanwhile, Terzan~5 hosts 49 confirmed millisecond pulsars, a population larger than that of any Galactic globular cluster and one that includes the fastest-spinning pulsar known, at 716~Hz \citep{Ransom2005,Hessels2006,Cadelano2018,Padmanabh2024}, as well as more than 200 cataloged \textit{Chandra} X-ray sources \citep{Heinke2006,Bahramian2020,Kumawat2025}. However, no companion to any of its pulsars has ever been detected at optical or infrared wavelengths, the wavelengths at which the companion star itself, rather than the pulsar, would be seen. Liller~1 poses an even starker puzzle. It is as gamma-ray luminous as the brightest globular clusters in the \textit{Fermi} sky, implying a substantial hidden millisecond pulsar population \citep{Tam2011}, yet no radio pulsar has ever been reported there, probably because of extreme interstellar scattering toward the inner bulge. For both clusters, infrared time-domain observations are the most powerful remaining channel for identifying the stellar companions in these X-ray and pulsar populations.

The existing census of variable stars in Galactic globular clusters \citep[][2017 update]{Clement2001} comprises 5,604 objects across 151 cluster fields, accumulated over more than a century of ground-based and space-based observations. The majority ($\sim$3,100) are RR~Lyrae pulsators on the horizontal branch, intrinsically interesting but unrelated to dynamical binary formation. Eclipsing binaries, the systems most directly relevant to the dynamical binary population, number only $\sim$400 across all clusters combined. A benchmark close-binary search in a single cluster is the 8.3-day Hubble Space Telescope (HST) campaign on 47~Tucanae by \citet{Albrow2001}, which monitored 46,000 main-sequence stars and yielded 26 eclipsing binaries. Among the largest reported single-survey yields of new variables of any kind in one cluster are $\sim$180 in M62 \citep{Contreras2010} and $\sim$117 in $\omega$~Centauri \citep{Kaluzny2004}. These surveys have been restricted to optically accessible, relatively nearby clusters, so that the most dynamically active clusters in the Galactic bulge, precisely those predicted to host the richest compact binary populations, have remained essentially unexplored. The \citet{Clement2001} catalog lists zero variables in Liller~1 and 13 in the field of Terzan~5 in its August 2019 update \citep{FigueraJaimes2024}, of which spectroscopy confirms $\sim$7 as members \citep{Origlia2019}. The current online listing has since grown to 53 entries, most of them \textit{Gaia}-flagged long-period variables (\S\ref{sec:detection}), and a recent dedicated ground-based lucky-imaging campaign on Terzan~5 identified four new variables \citep{FigueraJaimes2024}.

The James Webb Space Telescope \citep{Rigby2023} changes this landscape. Its infrared sensitivity and diffraction-limited angular resolution ($\sim$0\farcs07 at 2~\micron) resolve these extreme environments, and we show here that a third capability, its non-destructive detector readout, turns standard NIRCam imaging into a high-cadence time-domain survey of them. During each JWST integration, the accumulated charge is read non-destructively several times (in \emph{groups}) before the detector is reset. Our method uses these reads in three steps. First, we difference consecutive groups, producing an image of the photons collected between each pair of reads. Second, we perform photometry on each of these difference images, rather than on the single image obtained from the full integration. Third, we combine the resulting measurements into lightcurves for every star in the NIRCam field. In our observations this yields a time resolution of $\sim$21~s, increasing the number of measurements per integration by a factor of $N_{\rm groups}-1$. An additional advantage is that stars that saturate late in the ramp can still be measured from their earlier, unsaturated group differences.

This approach is fundamentally different from JWST's standard time-series observation (TSO) products. In the standard imaging reduction, the pipeline fits a slope to the groups within each integration, yielding one flux measurement per integration. The group-level temporal information therefore does not survive in the standard products, even though the individual reads remain available in the raw data, and our method recovers this otherwise unused information. The observing strategy also differs. TSO observations are generally designed around a single target, often on a small subarray, whereas we apply our method to ordinary full-array NIRCam imaging. The method can therefore be applied to archival observations that were never designed for time-domain science, and because it uses the full array, it produces lightcurves for every measurable star in the field at once.

Group-level differences have previously been used with JWST for individual targets. \citet{Schlawin2023} computed photometry on group-by-group difference images of defocused commissioning imaging to sense primary-mirror tilt events at sub-integration cadence, and \citet{Borowski2025} extracted a group-level MIRI lightcurve of the X-ray binary V404~Cygni to resolve a mid-infrared flare. The key advance here is to apply this approach systematically to every pixel of a full NIRCam field, turning standard JWST imaging into a blind, high-cadence variability survey.

\begin{figure*}
\centering
\includegraphics[width=\textwidth,height=0.86\textheight,keepaspectratio]{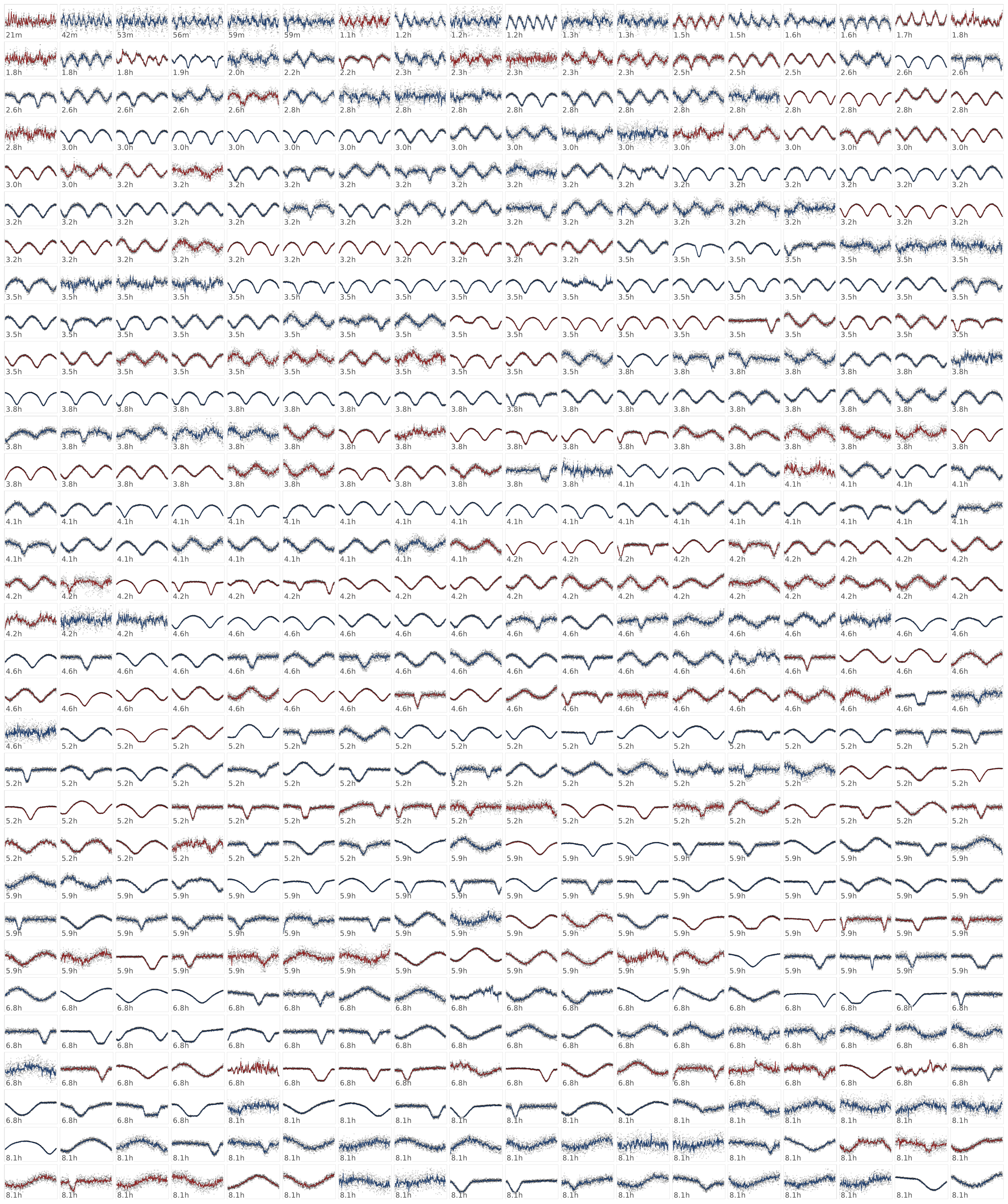}
\caption{Lightcurves of 576 of the 1,315 variable stars discovered in this work. Each tile shows the unfolded, pipeline-corrected 21-second-cadence lightcurve of one variable over a single continuous $\sim$7-hour JWST visit (black points), with the 3-minute median profile overplotted in dark red for Terzan~5 and dark blue for Liller~1. Each lightcurve is normalized to its own variability amplitude, and each tile shows the adopted lightcurve of the source, the version selected among the corrections of \S\ref{sec:corrections} (\S\ref{sec:strategy}). Tiles are sorted by the best-fit Lomb--Scargle period (bottom-left labels), which is not necessarily a binary orbital period, from 21~minutes to $\sim$8~hours, so eclipsing and contact binaries sweep from many cycles per visit (top) to single deep events (bottom). The full catalog is archived online (Appendix~\ref{app:repro}).}
\label{fig:variable_collage}
\end{figure*}

In this work, we present what is to our knowledge the first JWST time-series variability census of a globular cluster-like object. We identify 1,315 variable sources in $2\farcm2 \times 2\farcm2$ NIRCam fields centered on the cores of Liller~1 and Terzan~5, more than half the number of non-RR~Lyrae variables known across all Galactic globular clusters, and roughly seven times the largest single-survey yield of new variables previously reported for any cluster of which we are aware \citep{Contreras2010}. Figure~\ref{fig:variable_collage} illustrates the scale of this census, displaying 576 of the 1,315 variables, each as its unfolded, pipeline-corrected 21-second-cadence lightcurve over a single continuous $\sim$7-hour stare, sorted by variability timescale and color-coded according to whether they belong to the Terzan~5 or Liller~1 field. These variables are drawn from a different population than that of previous surveys. Rather than horizontal-branch pulsators, they represent the close binaries, accretion-powered systems, and dynamically formed compact objects that make these clusters unique astrophysical laboratories. We describe our methodology in detail below. We first explain how we extract group-level lightcurves from NIRCam's non-destructive reads, and then introduce an autocorrelation method to identify variable sources. This method takes advantage of the pixel-level stability provided by undithered staring observations. A multi-stage correction pipeline then recovers photometry from even heavily saturated ramps, while a separate cleaning-and-stitching pipeline extends the group-level photometry to dithered observations. In \S\ref{sec:discussion} we focus on some selected individual discoveries, as the infrared counterparts to the Rapid Burster and a Terzan~5 millisecond pulsar. The central result, however, is the census itself, and the capability it demonstrates. These observations target stars at distances of 6--8~kpc and through extreme foreground extinction, reaching $E(B-V) \approx 4.5$, or roughly 11 magnitudes of visual extinction, toward Liller~1 ($R_V=2.5$; \citealt{Pallanca2021}). Despite these challenging conditions, our pipeline delivers lightcurves at 21.5-second cadence with per-point precision reaching $\sim$0.5\%. Crucially, this capability is not specific to these observations: it leverages temporal information already present in JWST imaging data. We can apply group-level photometry to any sufficiently sampled JWST observation, of any field, at zero observing cost, since each already carries sub-integration temporal resolution in its group-level reads, though for dithered programs we must first run the correction-and-stitching treatment we demonstrate on our first visit. We unlock the densest fields by staring without dithering, which hands every pixel an uninterrupted stable time series, allowing us to detect variables on the lag-1 autocorrelation image built from those time series. Together, undithered staring and autocorrelation detection apply directly to the hardest targets, such as other globular clusters and the Galactic center. We present the resulting catalog of 1,315 variables as the foundation for a systematic study of binary populations and dynamical processes in these extraordinary stellar environments.

\section{Observations}\label{sec:observations}

We observed the dense stellar clusters Liller~1 and Terzan~5 using the Near Infrared Camera (NIRCam; \citealt{Rieke2023}) on the James Webb Space Telescope (JWST; \citealt{Rigby2023}) as part of Program GO-5381 (PI: Burdge). The observations comprised four visits using NIRCam Module~B in the F200W (short-wavelength, SW) and F356W (long-wavelength, LW) filters simultaneously, with the BRIGHT2 readout pattern, in which each stored group is the average of two detector frames, so that one group spans $t_\mathrm{group} = 21.47$~s. We number the four visits sequentially as Segments~1 to~4 (Table~\ref{tab:observations}). Terzan~5 was observed in Segments~1 and~2 and Liller~1 in Segments~3 and~4. Each segment is one complete visit, not one of the files into which the JWST archive splits a long exposure. We use the F200W and F356W filters to maximize our sensitivity to faint sources, combined with the BRIGHT2 readout pattern to reduce the number of saturated groups on bright sources and to maximize the temporal resolution of the observations while remaining within reasonable data-volume limits. Each visit consisted of a continuous $\sim$7-hour sequence of 12 exposures of 9 integrations each, with 10 groups per integration. All visits after the first were undithered stares. Table~\ref{tab:observations} summarizes the observing parameters.

JWST's data-volume limits, rather than the available observing time, shaped the setup, because retaining every group-level read at the BRIGHT2 cadence produces data at a rate far exceeding that of standard imaging programs. We thus observed with a single NIRCam module rather than both, sacrificing half the field of view to stay within the achievable data rate, and capped each visit at $\sim$7 hours, not by choice (a single uninterrupted 14-hour stare would have been scientifically preferable to two 7-hour visits for period recovery) but because the accumulated science data had to be downlinked before the next visit could begin.

The first visit (Terzan~5 Segment~1) employed a 12-point dither pattern, following standard practice for JWST imaging. Stepping the point spread function (PSF) across different pixels averages over flat-field and pixel-response systematics in the absolute photometry and improves sampling of the PSF \citep{FruchterHook2002, Anderson2009, Anderson2011}. For precision \emph{relative} photometry of a crowded field, however, we found that dithering is counterproductive. Each dither step re-registers every star's PSF onto a different set of pixels, and because each pixel has its own response, the raw lightcurves exhibit, at every dither position, discontinuous jumps that must be modelled before the lightcurves can be stitched back together (the effect is directly visible in the raw Segment~1 lightcurves of \S\ref{sec:seg1_extraction}). Dithering is designed to tame these pixel-to-pixel systematics, both by averaging them down and by quantifying, through the scatter among dither positions, their contribution to the error budget of calibrated apparent magnitudes. In the time domain, the very same systematics instead become a dominant noise source. An undithered stare, by contrast, lets each pixel sample the same portion of the same stellar PSFs continuously for the full $\sim$7-hour visit. This is the same strategy JWST adopts in its time-series observing modes, which forgo dithering to maximize photometric stability \citep[cf.][]{Beichman2014, Schlawin2023}. Consecutive frames can be cleanly differenced with no registration step, and per-pixel temporal statistics, such as the lag-1 autocorrelation we use for source detection (\S\ref{sec:detection}), measure real variability rather than pointing-induced pixel-response changes. The tradeoff is poorer sampling of the PSF, which matters little for our time-series goals. Thus, we removed dithering for all subsequent visits, vastly improving the lightcurve quality. In making this change, we followed the JWST monitoring program of Sgr~A*, which likewise began with dithered visits and switched to undithered staring after finding that dithering degraded the relative photometric precision \citep{YusefZadeh2025}. 

Terzan~5 Segment~1 is excluded from the blind variability search presented here. For the sources independently identified in the undithered Segment~2, we extract second-epoch lightcurves from the dithered Segment~1 data using a dedicated reduction pipeline (\S\ref{sec:seg1_extraction}). These provide an additional epoch, separated by $\sim$18.6~days from Segment~2, that is useful for verifying period stability, characterizing accretion variability, and identifying long-term flux changes, but is not used to discover new variables. We would like to note that this recovery is only partial. Even after the dedicated cleaning and stitching, the per-point precision of the reconstructed Segment~1 lightcurves remains significantly poorer than that of the undithered Segment~2 for the large majority of sources, because the dither-induced pixel-response systematics can be modelled and substantially suppressed, but not erased. 

For Liller~1, the two segments are separated by 1.45~days, providing a combined baseline of $\sim$1.75~days and enabling coherent period searches with substantially finer frequency resolution than either segment alone (\S\ref{sec:period}). For Terzan~5, only a single undithered segment is available; however, once we apply the dedicated cleaning-and-stitching reduction to the Segment~1 data, taken 18 days before Segment~2, the signals recovered there allow us to measure precise periods for many sources by leveraging that baseline.

\begin{deluxetable*}{lcccccccc}
\tablecaption{Summary of Observations\label{tab:observations}}
\tablewidth{0pt}
\tablehead{
\colhead{Target} & \colhead{Segment} & \colhead{Obs.} & \colhead{Date (UTC)} & \colhead{MJD Start} & \colhead{Duration} & \colhead{Dithered?} & \colhead{$N_\mathrm{int}$} & \colhead{$N_\mathrm{groupdiff}$} \\
 & & & & \colhead{(UTC)} & \colhead{(hr)} & & &
}
\startdata
Terzan~5 & 1\tablenotemark{a} & 001 & 2025 Apr 02 & 60767.904 & 7.0 & Yes (12-pt) & 108 & 972 \\
Terzan~5 & 2 & 002 & 2025 Apr 21 & 60786.488 & 7.1 & No & 108 & 972 \\
Liller~1 & 3 & 003 & 2025 Apr 22 & 60787.505 & 7.1 & No & 108 & 972 \\
Liller~1 & 4 & 004 & 2025 Apr 23 & 60788.956 & 7.1 & No & 108 & 972 \\
\enddata
\tablenotetext{a}{Excluded from the blind variability search due to dithering artifacts in relative photometry. After running source detection on Segment~2, we nonetheless extract second-epoch lightcurves from this visit for the detected sources (\S\ref{sec:seg1_extraction}).}
\tablecomments{All visits used NIRCam Module~B with F200W/CLEAR (SW, nrcb1--4, $0\farcs031$\,pixel$^{-1}$) and F356W/CLEAR (LW, nrcblong, $0\farcs063$\,pixel$^{-1}$) simultaneously. Readout pattern: BRIGHT2, 10~groups/integration, 9~integrations/exposure, 12~exposures/visit, $t_\mathrm{group} = 21.47354$\,s, $t_\mathrm{int} = 214.7$\,s. Each group-differenced cube contains $N_\mathrm{groupdiff} = 9 \times 12 \times 9 = 972$ frames at an effective time resolution of 21.47\,s. Program GO-5381 (PI: Burdge). The Obs.\ column gives the observation number within the program. This number identifies the corresponding MAST products (filenames \texttt{jw05381<obs>...}).}
\end{deluxetable*}

\section{Analysis}\label{sec:analysis}

We developed a custom pipeline to detect variable sources, extract lightcurves, and produce publication-quality photometry from the group-differenced image stacks. The pipeline reduces the data and builds the group-differenced cubes, constructs a variability reference image and detects sources on it, performs aperture photometry with outlier rejection, corrects the lightcurves for saturation artifacts and residual detector systematics through a multi-stage pipeline, searches for periods, and cross-matches sources between independent observations. We describe each step below.

\subsection{Data reduction and group-differenced cube construction}\label{sec:reduction}

NIRCam's HgCdTe H2RG detectors employ a non-destructive readout pattern, recording the accumulated charge in each pixel at multiple intervals during each integration (``sampling up the ramp''). The standard JWST pipeline fits a slope to each ramp, producing a single flux measurement per pixel per integration. We instead leverage the individual group reads to construct group-differenced images, differences between consecutive ramp samples, which provide $N_\mathrm{groups} - 1 = 9$ flux measurements per integration (adjacent differences share a group, so they are not independent; \S\ref{sec:photerr}) at a cadence of $t_\mathrm{group} = 21.47$\,s, yielding 972 time samples per 7.1-hour visit. All time stamps are barycentric and refer to the mid-exposure of each sample (\S\ref{sec:timing}).

We process the raw data through the JWST calibration pipeline \citep{Bushouse2023} ourselves rather than using the default products from MAST, because the standard archive reductions do not preserve the intermediate calibrated ramp files. We need these ramp files because they contain the individual group reads with detector-level corrections applied, including the nonlinearity correction from the \texttt{Detector1Pipeline} stage (\texttt{calwebb\_detector1}), while preserving the full ramp structure needed for group-differencing. We run \texttt{Detector1Pipeline} on the uncalibrated files with \texttt{save\_calibrated\_ramp=True}, then \texttt{Image2Pipeline} on the resulting rate images to produce calibrated integration-level products with WCS and SIP (simple imaging polynomial) distortion solutions. For sources processed with the saturation correction, individual group differences are excluded where the group in question is itself too near saturation for the saturation correction to be trusted, rather than being corrected and retained (\S\ref{sec:saturation}).

\begin{figure*}
\centering
\includegraphics[width=\textwidth]{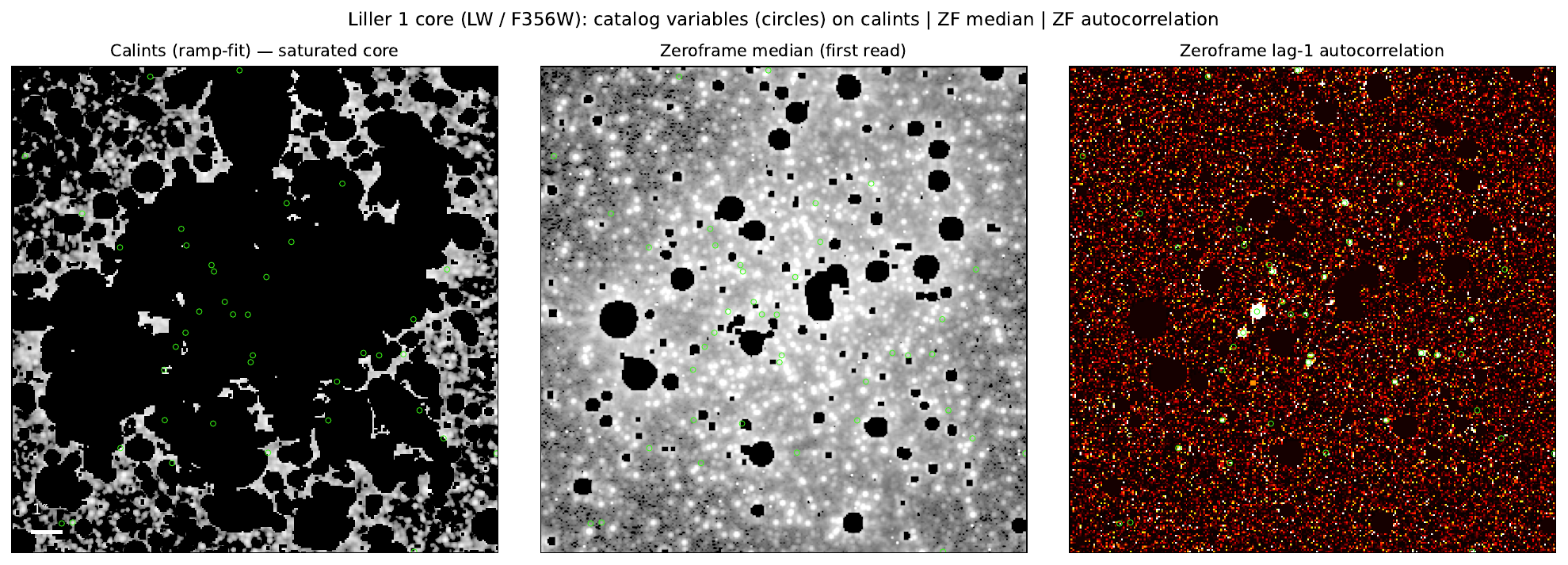}
\includegraphics[width=\textwidth]{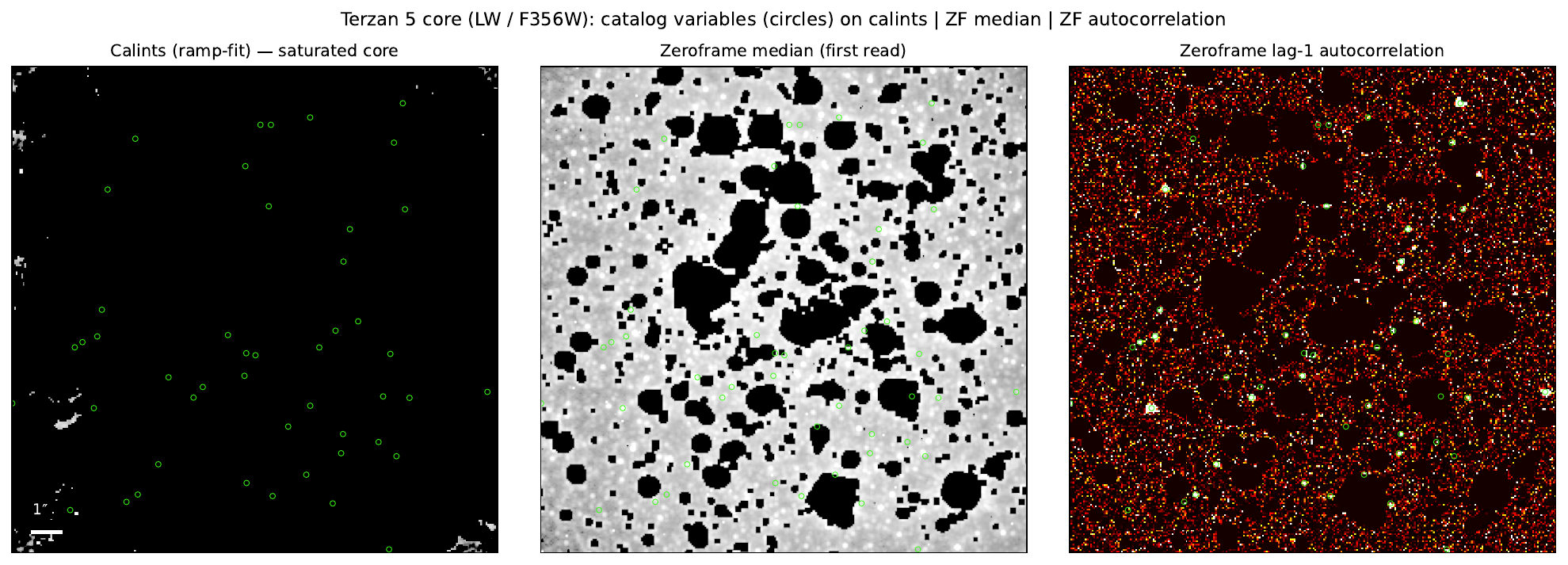}
\caption{The cores of Liller~1 (top row) and Terzan~5 (bottom row) in F356W; each cutout spans 17\farcs6, under 2\% of the area of the NIRCam field surveyed in this work. \textit{Left:} the conventional science image produced by the standard JWST pipeline (median of the calibrated integrations). The brightest giants saturate, and the pipeline masks saturated pixels to NaN (black), so most of each core is blank; in Terzan~5 almost no unmasked pixels remain. \textit{Middle:} median of the zeroframes, the first read of each integration, which recovers most of each core; only the very brightest giants saturate even in this 10.7~s read, and their cores appear black. \textit{Right:} the lag-1 temporal autocorrelation image (Eq.~\ref{eq:autocorr}) of the zeroframe series, the reference image on which we detect variables (for the wider field we apply the same statistic to the calibrated integrations). The color scale runs from black (zero or negative autocorrelation) through red and yellow to white (strong positive), so constant stars, however bright, average to zero and appear dark, and the variables stand out on their own. Green circles mark the variables in our catalog that fall within the cutouts.}
\label{fig:field_overview}
\end{figure*}

\subsection{Detection of variable sources}\label{sec:detection}

To detect variable sources, we construct a variability reference image based on the lag-1 temporal autocorrelation of the calibrated integration-level data products (calints files). For each pixel, we compute
\begin{equation}\label{eq:autocorr}
    \rho_1 = \frac{\sum_{i=1}^{N-1}(f_i - \bar{f})(f_{i+1} - \bar{f})}{\sum_{i=1}^{N}(f_i - \bar{f})^2} \;,
\end{equation}
where $f_i$ is the pixel value in the $i$-th integration and $\bar{f}$ is the temporal mean. For non-variable sources, the frame-to-frame fluctuations are dominated by uncorrelated noise (photon noise and read noise), yielding $\rho_1 \approx 0$ regardless of source brightness. Variable sources, however, exhibit temporally correlated flux changes between consecutive integrations, producing $\rho_1 \gg 0$. We note that this autocorrelation-based reference image offers a key advantage over the traditional variance-over-mean approach, in that it is brightness-independent, so bright but non-variable sources do not appear as false positives (Figure~\ref{fig:field_overview}). In a direct comparison on the same data, detection on the autocorrelation image produced roughly one-third as many spurious detections as detection on a variance-over-mean image.

We perform source detection on the autocorrelation reference image using PSF-matched filtering. The autocorrelation image is a per-pixel statistic rather than a flux image, so it is not itself a PSF-convolved quantity: each pixel carries an independent estimate of $\rho_1$, and a variable star imprints a PSF-shaped \emph{pattern} of correlated pixels on it rather than a PSF-shaped flux profile. Filtering with the PSF is therefore still appropriate, and it is what separates a genuine variable, which correlates a whole PSF footprint of pixels, from a single hot or noisy pixel, which does not. We convolve the reference image with a model of the NIRCam PSF generated using \texttt{WebbPSF} \citep{Perrin2014} (since renamed \texttt{STPSF}) with in-flight optical path difference maps and resampled to the detector pixel scale. Local maxima (within a $3\times3$ pixel neighborhood) of the convolved image are identified as candidate variable sources if they exceed $3\sigma$ on the image built from the full integrations, or $5\sigma$ on the zeroframe image introduced below, where $\sigma$ is the sigma-clipped standard deviation of the autocorrelation image scaled by the PSF kernel norm. 

We refer to everything built from the full integrations, namely the calints autocorrelation image and the group-differenced lightcurves, as the \emph{ramp} products. In parallel with them, we leverage the \emph{zeroframe} recorded with each integration: the initial detector read taken immediately after reset, before charge accumulates, which preserves all but the very brightest stars as unsaturated. Discarding the first integration of each exposure, whose zeroframe is affected by reset settling, yields a 96-frame time series per visit (108 integrations less the 12 discarded first integrations, one per exposure) at the $\sim$215~s integration cadence. We construct a lag-1 autocorrelation image from the resulting zeroframe cube in exactly the same way, detect sources on this image at $5\sigma$, and extract zeroframe lightcurves with the same aperture photometry. We call these the \emph{zeroframe} products. Their outlier rejection is applied in consecutive chunks of 4 points rather than the 18 used for the ramp lightcurves, since the zeroframe series is sampled ten times more sparsely (one point per integration), so a chunk of 18 points would span $\sim$1~hr and could remove real variability (\S\ref{sec:outlier}). The zeroframe products, immune to saturation for all but the very brightest stars, confirm bright-star variability throughout this work. Table~\ref{tab:detections} lists the raw detection counts from both products for each segment and detector. The $3\sigma$ ramp threshold is deliberately permissive, and these candidates are winnowed by the classification and deduplication steps below to the final catalog of 1,315 variables, whose deduplication groupings are archived with the data products (Appendix~\ref{app:repro}).

The previously cataloged variables offer a small but concrete recovery check, and the composition of that sample matters more than its size. The current \citet{Clement2001} listing for Terzan~5 carries 53 entries, of which 44 are Mira, semiregular, or long-period variables of hundreds of days' period, which no seven-hour visit can recover. Thirty-four of the 53 were flagged as variables by \textit{Gaia}~DR3 and confirmed photometrically from the ground by \citet{FigueraJaimes2024}, who also discovered four semiregular variables of their own. Four of the 53 are RR~Lyrae, with periods of 14.2--21.4~h that exceed the twelve-hour bound of our search grid, and two are eclipsing binaries, at 6.98 and 7.24~h. Of these six known short period variables, one RR~Lyrae (V12) falls outside our detectors and the 7.24-h eclipsing binary (G10) lies in a region masked by saturation in both channels, leaving four we could have detected. We recover three, at 0\farcs09, 0\farcs42, and 0\farcs48 from their catalog positions. The RR~Lyrae V3 and V11 are recovered as variables without a measured period, since their periods exceed our grid. The eclipsing binary V4 is recovered with a fitted orbital period of 13.8~h, twice the 6.98~h catalog value: the binary model resolves the half-period ambiguity in favor of the longer period. The fourth, the RR~Lyrae V13, has a 21.4-h period that produces only a slow drift within a seven-hour visit. It reaches the 3$\sigma$ detection level in the autocorrelation image but we judged it not to be a real variable on visual inspection (\S\ref{sec:classification}). Liller~1 had no previously cataloged variables of any kind.

\begin{deluxetable*}{lllcrr}
\tablecaption{Raw PSF-matched detection yield per segment and detector\label{tab:detections}}
\tablewidth{0pt}
\tablehead{
\colhead{Target} & \colhead{Segment} & \colhead{Detector} & \colhead{Filter} & \colhead{$N_\mathrm{det}$ (ramp, $3\sigma$)} & \colhead{$N_\mathrm{det}$ (zeroframe, $5\sigma$)}
}
\startdata
Terzan~5 & 2 & nrcb1    & F200W & 11,202 & 2,412 \\
         &   & nrcb2    & F200W & 11,145 & 1,052 \\
         &   & nrcb3    & F200W & 13,771 & 1,415 \\
         &   & nrcb4    & F200W & 16,562 & 1,649 \\
         &   & nrcblong & F356W & 25,513 &   697 \\
Liller~1 & 3 & nrcb1    & F200W & 13,765 & 3,094 \\
         &   & nrcb2    & F200W & 13,595 & 1,255 \\
         &   & nrcb3    & F200W & 15,048 & 1,202 \\
         &   & nrcb4    & F200W & 16,372 & 1,677 \\
         &   & nrcblong & F356W & 20,309 &   791 \\
Liller~1 & 4 & nrcb1    & F200W & 18,680 & 3,124 \\
         &   & nrcb2    & F200W & 17,339 & 1,777 \\
         &   & nrcb3    & F200W & 21,291 & 1,719 \\
         &   & nrcb4    & F200W & 21,476 & 2,556 \\
         &   & nrcblong & F356W & 21,421 &   888 \\
\hline
\multicolumn{4}{l}{Total} & 257,489 & 25,308 \\
\enddata
\tablecomments{Number of candidate sources returned by the PSF-matched filter (\S\ref{sec:detection}) on each autocorrelation reference image, prior to human REAL/FAKE vetting and morphological classification (\S\ref{sec:classification}) and spatial deduplication across detectors and segments (\S\ref{sec:dedup}). The $3\sigma$ ramp threshold is intentionally permissive. The overwhelming majority of these peaks are noise or artifacts rejected during vetting, and a source detected on multiple detectors or in both segments is counted here once per detection. Terzan~5 Segment~1 (dithered) is not searched. F200W = SW channel (nrcb1--4). F356W = LW channel (nrcblong).}
\end{deluxetable*}

\subsection{Extraction of lightcurves}\label{sec:extraction}

We extract lightcurves by performing aperture photometry at the position of each detected source on the group-differenced image cube. We use a circular aperture with a radius of 1.5~pixels, smaller than the $\sim$2.0~pixel FWHM of the NIRCam short-wavelength PSF at the 0\farcs031/pixel plate scale in F200W, to minimize flux contamination from neighboring sources in these extremely crowded fields. For the long-wavelength channel (F356W, 0\farcs063/pixel), the same aperture radius in pixels is used. This is not an approximation of convenience: the plate scale and the wavelength both roughly double between the channels, so the PSF is about twice as wide in arcseconds but covers the same number of pixels, and a fixed pixel radius thus encircles a similar fraction of the PSF in both. All pixels whose centers lie within the aperture radius are summed with unit weight (no fractional-pixel weighting). Any pixel that is NaN in \emph{any} frame is excluded from the aperture in \emph{all} frames, ensuring a consistent effective aperture across all time steps.

\subsubsection{Time stamps and absolute timing}\label{sec:timing}

Because these lightcurves are intended for correlation with pulsar radio ephemerides, X-ray timing, and other external clocks, we define the time stamps carefully. All times in this work and in the released catalog are barycentric MJD on the TDB scale ($\mathrm{BMJD_{TDB}} = \mathrm{BJD_{TDB}} - 2400000.5$), inherited from the group-level barycentric times computed by the level-1b JWST pipeline.

Every sample is stamped at its \emph{mid-exposure}, the temporal midpoint of the interval of light collection it measures, independent of how the source's flux varies within that interval. For group-differenced data, locating this midpoint involves a subtlety. In the BRIGHT2 readout, each stored group is the average of two consecutive frames ($t_\mathrm{frame} = 10.737$~s). For a ramp accumulating linearly in time, an averaged group is exactly equivalent to an instantaneous charge sample at the mean of its two frame read times, $t_\mathrm{reset} + (2k - 0.5)\,t_\mathrm{frame}$ for group $k$. The difference of groups $k{+}1$ and $k$ thus measures the charge collected between those two equivalent sample times, and we stamp it at their midpoint, $t_\mathrm{reset} + (2k + 0.5)\,t_\mathrm{frame}$. The zeroframe, the single first read of each integration, is stamped analogously at $t_\mathrm{reset} + t_\mathrm{frame}/2 \approx 5.4$~s. In the released catalog, lightcurves that have passed through the corrections of \S\ref{sec:corrections} store times in hours from a per-segment zero point, which the catalog tabulates (table \texttt{time\_reference}). We verified that the group-differenced, zeroframe, and corrected lightcurves agree on a common absolute time grid to better than 0.1~s.

One caveat attaches to the mid-exposure convention. The mid-exposure time coincides with the flux-weighted mean photon arrival time only for a source whose flux is constant, or varies linearly, across the sample. The sample itself is not a top-hat. Differencing two frame-averaged groups weights three consecutive frame intervals by $\tfrac12$, 1, $\tfrac12$, so each difference draws on $\approx$32~s of light with its symmetry center at the quoted mid-exposure, and neighboring differences share one frame. On the steep ingress or egress of a sharp eclipse, the flux-weighted mean arrival time of the collected photons is displaced from the temporal midpoint, toward the brighter end of the interval, by at most a fraction of the sample duration. The effect is negligible for the periods and eclipse durations in this catalog, where an ingress spans many samples, and we do not correct it. An analysis fitting eclipse contact times at the few-second level, however, should account for it.

Two absolute-timing terms remain uncorrected. The group times refer to the readout of the full array, whereas the detector is read out sequentially over one frame time, so the true sample time of a given source depends on its position on the detector, introducing an offset of up to one frame time ($\leq 10.7$~s). We do not remove it. It is constant for a given source, so it cancels in every relative and periodic analysis in this paper, and correcting it would require mapping each source's detector row through the readout order for a gain no result here depends on. A user needing absolute times at the sub-second level can recover it from the stored detector position. The second term is the observatory clock itself. \citet{Shaw2025} calibrated it against an eclipsing double white dwarf and found a clock accuracy of $0.12 \pm 0.06$~s, a factor of $\sim$5 better than the pre-launch requirement. Absolute event times in this catalog are thus reliable at the few-second level as stored (readout-phase dominated), and at the $\sim$0.1~s level (clock dominated) if the readout phase of a specific source is accounted for, sufficient for phase-connecting infrared eclipses and flares with radio and X-ray ephemerides.

\subsubsection{Outlier rejection}\label{sec:outlier}

Cosmic ray events and other transient detector artifacts can introduce into the extracted lightcurves outliers that mimic astrophysical variability. To remove these, we apply an interquartile range (IQR)-based clipping algorithm. The lightcurve is divided into contiguous chunks of 18~consecutive measurements (corresponding to two JWST integrations). Within each chunk, we compute the first and third quartiles ($Q_1$ and $Q_3$) and reject any measurement falling outside the range $[Q_1 - 2\,\mathrm{IQR},\; Q_3 + 2\,\mathrm{IQR}]$, where $\mathrm{IQR} = Q_3 - Q_1$. When the number of points is not evenly divisible by the chunk size, the pipeline handles the trailing partial chunk in one of two ways, both preserved as options in the released code. The initial extraction described here and the forced photometry used to build the catalog (\S\ref{sec:catalog}) leave it (at most 17 points) unclipped, whereas the corrections of \S\ref{sec:corrections} merge it into the preceding chunk. The distinction matters only for exact reproduction of the published lightcurves. We compute the IQR locally within each chunk rather than globally because the lightcurves may contain genuine astrophysical variability on timescales of minutes to hours, and a global rejection would risk masking real signals, whereas local clipping removes impulsive outliers (cosmic rays) while preserving smooth variations that evolve over many chunks.

During initial lightcurve extraction, a single pass of IQR clipping is applied. In the subsequent correction pipeline (\S\ref{sec:corrections}), we apply IQR clipping in two consecutive passes after each correction, as a single pass can leave residual outliers when multiple deviant points in the same chunk inflate the IQR.

\subsection{Lightcurve correction pipeline}\label{sec:corrections}

Bright and moderately bright sources exhibit systematic artifacts in their group-differenced lightcurves due to detector nonlinearity near saturation. These manifest as ``banding'', a repeating pattern within each integration in which later group differences are systematically suppressed relative to the first (Figure~\ref{fig:satcorr_sw}, panel~a). In this section, we describe the multi-stage correction pipeline that we have developed to address these artifacts while preserving genuine astrophysical variability. Appendix~\ref{app:underhood} shows the fitted models underlying each correction on a heavily saturated example source. For each source, segment, and wavelength channel, we evaluate four correction strategies, one of which is to apply no correction, and select the one that minimizes the standard deviation within each integration averaged over all integrations, which we call the \emph{integration scatter} (a direct measure of banding severity). A 10\% improvement gate applies, meaning a correction is adopted only if it reduces the integration scatter to below 90\% of that of the uncorrected lightcurve (\S\ref{sec:strategy}).

\subsubsection{Saturation correction}\label{sec:saturation}

Our saturation correction leverages the empirical observation that, for a given pixel, the ratio of each group-difference to the first group-difference ($g_i/g_0$) is a deterministic function of the per-read signal level of that pixel ($g_0$), which sets how far up the well the pixel climbs during the integration. We fit this relationship using a quadratic polynomial for each group index $i$ and each pixel in the photometric aperture, working from the uncalibrated (\texttt{uncal}) data rather than the JWST pipeline's linearized ramp files, as we find that the standard linearity correction can introduce additional artifacts near saturation.

For each pixel, we first apply temporal clipping to the $g_0$ time series using a running median filter (window = 7 integrations) to identify and mask cosmic ray showers that can affect multiple pixels simultaneously. We then fit the ratio model $g_i/g_0 = a_2 g_0^2 + a_1 g_0 + a_0$ for each subsequent group $i$. Groups are corrected only while their median ratio stays at or above 3\% and the residual scatter about the quadratic fit stays at or below 5\% of that median ratio, and the first group failing either test is excluded together with all later groups. This adaptive group selection thus automatically excludes groups that are too heavily saturated to correct. Each pixel's corrected values are normalized by the pixel median before aperture summation, and at each time step we reject any pixel whose normalized value deviates by more than $3\sigma$ from those of the other aperture pixels. A second pass of IQR clipping is applied to the final aperture-summed lightcurve. For Terzan~5~\#133, this step collapses the separate group-difference tracks of Figure~\ref{fig:satcorr_sw}a onto a single coherent lightcurve that reveals a deep eclipse (Figure~\ref{fig:satcorr_sw}b). The fitted ratio models and one corrected integration for the same source are shown in Figure~\ref{fig:underhood}c,d.

\subsubsection{Slope correction}\label{sec:slope}

\begin{figure}
\centering
\includegraphics[width=\columnwidth]{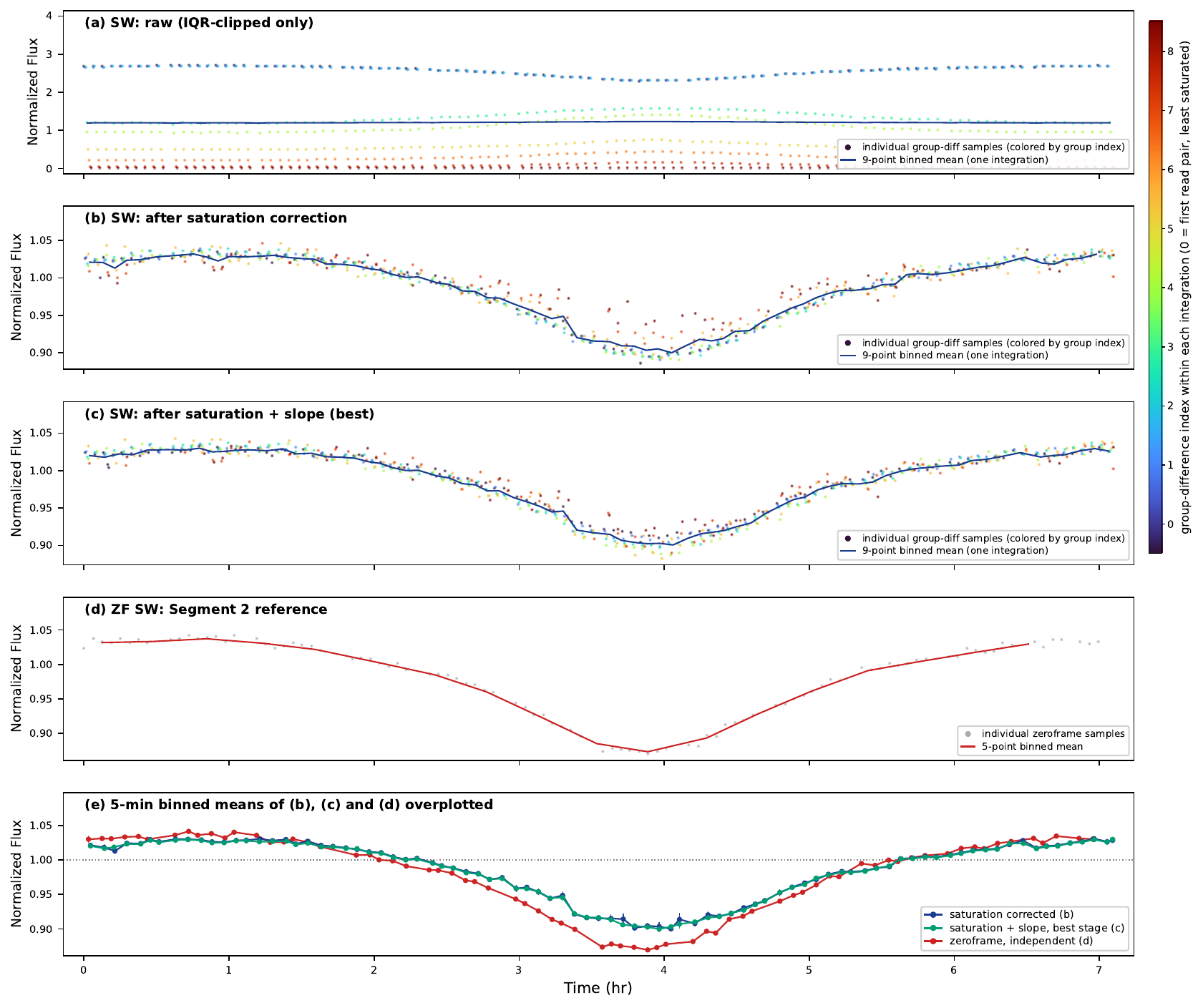}
\caption{Saturation correction of the SW (F200W) lightcurve of Terzan~5 \#133
(Segment~2), an eclipsing binary.  Points are color-coded by group-difference index within
each integration. Group~0 (blue) retains the most signal, while later groups
(green through red) are progressively more saturated. In (a)--(c) the solid line is the 9-point binned mean, one integration per point.
\textbf{(a)}~Uncorrected IQR-clipped lightcurve showing extreme banding, in which each group forms a distinct horizontal track, with the most saturated groups (red) driven
to near-zero or negative normalized flux.
\textbf{(b)}~After the per-pixel quadratic ratio-model saturation correction
(\S\ref{sec:saturation}), the groups collapse onto a coherent signal, revealing
a deep eclipse near $t \approx 4$~hr in which the flux drops to $\sim$88\% of the
out-of-eclipse median (5-minute binned means, panel~e).  Residual scatter is
largest for the most saturated groups (orange/red points).
\textbf{(c)}~The lightcurve after combined saturation + slope correction, the
best strategy for this source, in which the intra-integration slopes are further
flattened.
\textbf{(d)}~Independent zeroframe (ZF) SW lightcurve (96~points; the solid line is the 5-point binned mean), which samples only the first read of each integration and is
therefore immune to saturation.  The same eclipse is recovered at the same epoch
and duration, at slightly larger fractional depth than in the corrected ramp
photometry (15\% against 12\%, measured on the 5-minute binned means of panel~e as the minimum relative to the
out-of-eclipse median; the two apertures differ in crowding dilution).
\textbf{(e)}~The 5-minute binned means of (b), (c) and (d) overplotted on a
common axis, with whiskers giving the standard error of the mean within each bin. The saturation-corrected and saturation$+$slope reconstructions
are indistinguishable at this level, and the independent zeroframe photometry
tracks the same eclipse, confirming the astrophysical origin of the corrected
ramp signal. The colorbar at right keys the group-difference index in panels
(a)--(c), and each panel's legend identifies the individual samples and the
binned mean.}
\label{fig:satcorr_sw}
\end{figure}

Even after IQR clipping (or saturation correction), however, some lightcurves exhibit a residual flux-dependent linear slope within each integration, a subtler manifestation of detector nonlinearity. We model this by fitting a quadratic relationship between each integration's median flux and its within-integration linear slope. The predicted slope is then subtracted from each integration. Applied after the saturation correction, it further flattens the intra-integration slopes of \#133 (Figure~\ref{fig:satcorr_sw}c), and this combined correction is the best strategy for that source. As an independent check, the zeroframe photometry of \#133, which samples only the first read of each integration and so is unaffected by saturation, recovers the same eclipse at the same epoch (Figure~\ref{fig:satcorr_sw}d), and its binned lightcurve tracks the corrected ramp photometry (Figure~\ref{fig:satcorr_sw}e). Appendix~\ref{app:underhood} (Figure~\ref{fig:satcorr_lw}) illustrates the full correction sequence (slope only, saturation only, and combined) for a saturated long-wavelength source, with the result validated against the simultaneously observed short-wavelength lightcurve.

\subsubsection{Strategy selection}\label{sec:strategy}

For each source, segment, and channel, we compute the integration scatter of four strategies: (1)~the uncorrected lightcurve (IQR clipping only), (2)~the saturation-corrected lightcurve, (3)~the slope-corrected lightcurve, and (4)~the lightcurve with both saturation and slope corrections. The strategy with the lowest integration scatter is adopted, provided it improves on the uncorrected lightcurve by at least 10\% (Table~\ref{tab:correction_summary}). We call the selected version the \emph{adopted lightcurve} of that source, segment, and channel. It is the lightcurve released in the catalog and shown in the figures of this paper, except for the one source whose PSF-wing reduction replaces it (\S\ref{sec:special}).

\subsubsection{Background amplitude rescaling}\label{sec:background}

Because the lightcurves are normalized to unit median flux, the measured
fractional variability amplitude is diluted by the contribution of unresolved
background light within the photometric aperture.  We tested, but do not adopt, an estimate of that contamination based on the static background level measured from an annulus around each source
($r_{\rm in} = 10$, $r_{\rm out} = 16$~px for the short-wavelength channel;
$r_{\rm in} = 5$, $r_{\rm out} = 8$~px for the long-wavelength channel,
matching physical scales at the respective plate scales).  The median
background per pixel is multiplied by the aperture area to obtain a total
background flux $B$, and the normalized lightcurve $f_n$ is rescaled as
\begin{equation}
    f_{\rm corr} = \frac{f_n \, \tilde{F} - B}{\tilde{F} - B},
\end{equation}
where $\tilde{F}$ is the median unnormalized aperture flux.

We do not apply this rescaling to any published lightcurve, in any segment. In a core this crowded, the annulus does not measure the light blended into the aperture; it
samples a different patch of an unresolved and highly structured stellar
background, so it over- or under-corrects by an amount whose sign we cannot
predict, and for the most heavily diluted sources the implied amplification
reached a factor of $\sim$800. Rather than commit a correction of unknown sign
to the flux, we publish the aperture photometry without this rescaling and carry the dilution where it
can be constrained per source, as the dilution term fitted in the
binary models of \S\ref{sec:classification}. We release the annulus background level of
every source as a separate catalog quantity, so that a reader who wants the
rescaling above can apply it.

\subsubsection{Amplitude estimation}\label{sec:amplitudes}

In fields this crowded, much of the light inside a photometric aperture
belongs to blended neighboring stars rather than to the variable, so the fractional amplitudes of the normalized lightcurves are diluted by light that is not the star's own. As a result, we constrain some quantities far better than others. The differential signal (the flux \emph{changes} measured on the detector, in DN) is robust, since neighbors contribute constant light that cancels
in the differential, and the modulation shape is traced at $\sim$0.5\% per
21.5-s sample (\S\ref{sec:photerr}). Converting that differential into a
\emph{fractional} amplitude, however, requires knowing how bright the
underlying star itself is within the aperture, and that dilution factor is
far harder to measure than the modulation it scales.

The catalog lightcurves, and all figures in this paper, show aperture photometry normalized to unit median flux, with no dilution correction of any kind applied in any segment (\S\ref{sec:background}). We call amplitudes measured on this scale \emph{aperture amplitudes}, whatever detector corrections the lightcurve has received. That choice buys a
property worth more to us than a central value: on this scale every fractional amplitude is a \emph{lower bound} whenever the blended light is constant, because constant blended light can only dilute a fractional signal and never inflate it.

As an alternative dilution estimate, we compare each source's aperture flux with DOLPHOT~3.1 \citep{Dolphin2000,Dolphin2016} PSF photometry of the same data, run with the NIRCam module developed for the JWST Resolved Stellar Populations Early Release Science program \citep{Weisz2023,Weisz2024}, since PSF fitting deblends each star from its neighbors. We match 1,312 of 1,315 sources within 0\farcs15, with a median separation $<$10~mas. Expressing the
DOLPHOT count rate as the expected flux in our $r = 1.5$~px aperture (rate
$\times$ 21.47~s per group difference $\times$ the encircled-energy
fraction, 0.49 in F200W and 0.57 in F356W) implies that the median
variable contributes only $\sim$25--40\% of its aperture flux, i.e.,
intrinsic fractional amplitudes 2.5--4$\times$ larger than the
aperture amplitudes of the catalog. The DOLPHOT-based estimate is not uniformly trustworthy,
however. For $\sim$3\% of sources the implied amplitude is unphysical. These are classified eclipsing binaries whose implied eclipse depth exceeds 100\%, and they expose deblending and cross-match ambiguities in severe crowding (deblending errors, saturated DOLPHOT fluxes, or a match landing on the wrong star). We store the DOLPHOT-based dilution estimate per
source in the catalog, alongside the annulus background level and the
DOLPHOT F200W and F356W apparent magnitudes of each matched counterpart, rather than adopting it as the catalog flux scale.

We measure the \emph{detection} of variability and the \emph{shape} of the modulation at high signal-to-noise. The quoted fractional amplitudes are far less secure, inheriting the systematic uncertainty of the dilution factor, PSF encircled-energy and centering terms at the few-percent level, DOLPHOT deblending in severe crowding, and saturated DOLPHOT fluxes for the brightest stars. The most robust statement available for each source is a \emph{lower bound} on its amplitude, the aperture amplitude, with no dilution correction at all, since constant blended light can only dilute the fractional signal (a variable neighbor can violate this, as discussed in \S\ref{sec:photerr}). The DOLPHOT-based correction
attempts to correct for the blended light, but disentangling more than a
thousand sources in the two densest known globular cluster-like objects in the Galaxy is
hard. Depending on the local crowding and saturation, PSF photometry can recover a source's flux well or, in worse cases, over-subtract its neighbors or leave residual blends, so the DOLPHOT-based amplitude can fall above or below the truth for a given source. We treat individual
fractional amplitudes as uncertain, and in any case quote both values in the catalog.

In a few of the faintest or most saturation-affected lightcurves the noisiest points fall formally below zero flux. We do not clip these, but flag every affected lightcurve with a per-source bitmask in the catalog (\S\ref{sec:catalog}), evaluated on the catalog flux scale.

\subsubsection{Special reductions}\label{sec:special}

For extremely saturated sources where even the core aperture pixels are unusable, we extract lightcurves from the wings of the PSF using annular apertures that exclude the saturated core. We apply this special reduction to a single source, the Rapid Burster (MXB~1730$-$335) in Liller~1, using an annular aperture with inner radius 3~pixels and outer radius 5~pixels, followed by the same saturation correction pipeline applied to the wing pixels. For this source the wing lightcurves replace the standard short-wavelength lightcurves of Segments~3 and~4 in the catalog.

This wing extraction is complementary to the zeroframe analysis of the companion paper \citep{Desai2026}. Because the zeroframe is read before appreciable charge has accumulated, it leaves even the source core unsaturated, and \citet{Desai2026} leverage this to recover core photometry, at the price of a single sample per integration, i.e., the $\sim$215~s integration cadence. Our wing extraction instead sacrifices the saturated core in order to retain the full group-level $\sim$21.5~s temporal resolution of the ramp. The technique is general and could be applied to other heavily saturated sources. We reserve it here for the Rapid Burster alone, given the exceptional importance of establishing its long-sought infrared counterpart.

\subsection{Dithered Segment Extraction}\label{sec:seg1_extraction}

The Terzan~5 Segment~1 visit was dithered (12 exposures over $\sim$7~hr, each with 9 integrations of 9 group differences), which breaks the assumptions of the undithered pipeline: every exposure places each star on different pixels, so each exposure block carries its own flat-field level, its own blend environment, and its own saturation behavior. We therefore take the sources that we detect in Segment~2 and we extract their lightcurves in Segment~1 with a dedicated reduction. For each exposure, we first perform aperture photometry at the WCS-projected source positions and then we correct for saturation using the median flux ratio measured separately for each exposure and group. We then clean and correct the individual exposure lightcurves using double IQR clipping and an integration-slope correction, before normalizing each exposure block. After rejecting anomalous exposures with a $4\times$IQR criterion, we stitch the twelve blocks into a continuous lightcurve by fitting a quadratic to each block and offsetting each block so that it continues the trend of the preceding block across the gap (a curvature-aware stitching, Appendix~\ref{app:seg1_detail}), and we perform a final outlier rejection.

\begin{figure*}
    \centering
    \includegraphics[width=\textwidth,height=0.76\textheight,keepaspectratio]{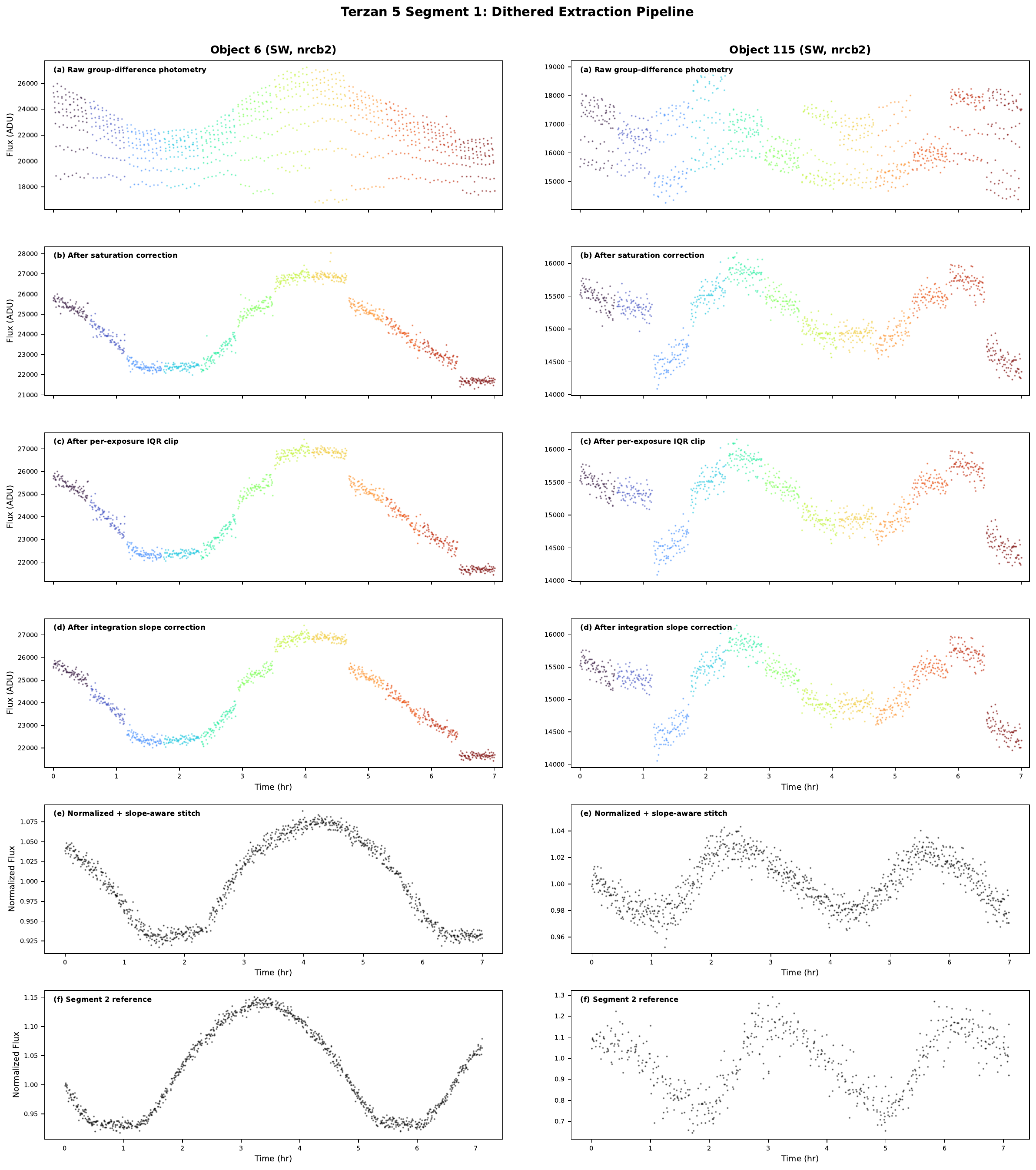}
    \caption{Processing stages for the dithered Segment~1 extraction, shown
    for two variable stars in Terzan~5.  Each column shows one source, and rows show successive pipeline stages.  Points are color-coded by exposure
    (12 dithered exposures spanning $\sim$7~hr).
    \textbf{(a)}~Raw group-difference aperture photometry, showing saturation
    structure (declining flux in later group differences) and flat-field offsets
    between dither positions.
    \textbf{(b)}~After per-group median-ratio saturation correction: the horizontal band structure collapses into coherent per-exposure tracks.
    \textbf{(c)}~After per-exposure IQR outlier rejection (two passes,
    $2\times\mathrm{IQR}$ threshold).
    \textbf{(d)}~After integration slope correction, removing residual
    intra-integration saturation ramps.
    \textbf{(e)}~Final stitched lightcurve after per-block normalization,
    curvature-aware stitching, and exposure rejection.
    \textbf{(f)}~Independent Segment~2 (undithered) lightcurve for comparison.
    Object~6 (\textit{left}) is a moderate-amplitude variable, while Object~115 (\textit{right}) shows stronger within-integration structure
    in the raw data (its group differences rise by $\sim$10\% through each ramp) but yields a clean multi-cycle lightcurve after
    processing.  Observing Terzan~5 first with dithers (Segment~1) and then
    undithered (Segment~2) provided the key dataset for developing this
    correction pipeline: the undithered Segment~2 lightcurves provided an
    excellent ground truth against which the dithered reconstruction could be
    checked, in morphology and period rather than in phase or amplitude, since
    rows~(e) and~(f) are different visits.}
    \label{fig:seg1_cleaning}
\end{figure*}

Figure~\ref{fig:seg1_cleaning} shows the full progression on two representative sources. The raw dithered photometry (panel~a) is dominated by tens-of-percent jumps between dither positions and coherent within-integration structure. The final stitched lightcurves (panel~e) recover clean variability, and their periods and morphologies agree with those of the independent, undithered Segment~2 lightcurves of the same stars (panel~f), obtained 18.6 days later, which validates this reduction for period recovery. The fractional amplitudes are not expected to agree between the two segments, since the two extractions sample different pixels and blend environments (Appendix~\ref{app:seg1_detail}). Each step, with all parameters and the stitching algorithm, is specified in Appendix~\ref{app:underhood}, which also shows the fitted models for one of these sources (Figure~\ref{fig:underhood_stitch}). These second-epoch lightcurves are used for period verification and long-term variability characterization, not for source discovery (\S\ref{sec:observations}).

\subsection{Photometric uncertainties}\label{sec:photerr}

Our uncertainty model is calibrated empirically from the scatter measured in apertures centered on real stars, rather than from an analytic propagation of the noise in individual pixels. Crowding enters the photometric uncertainties through the calibration itself rather than through a model of it. Any blended light from neighboring stars that falls within an aperture contributes both to its measured flux and to the scatter used to calibrate the noise model. The photon noise associated with this contaminating flux is therefore automatically included in the inferred uncertainties, without requiring an explicit model of crowding.  What the model does not capture is a systematic: a neighbor's own variability leaking into the aperture, which is not noise and is addressed instead by the cross-channel and cross-segment consistency checks of \S\ref{sec:classification}.

We assign per-point uncertainties to every catalog lightcurve using an empirical, aperture-level photon-transfer calibration rather than analytic per-pixel error propagation. For each detector and segment, we extract aperture photometry with the same estimator as the catalog lightcurves (exact-fraction $r=1.5$~px apertures, \S\ref{sec:catalog}) at $\sim$1,500 star centers, identified as local maxima of the median group-difference image, distributed across the full range of fluxes and excluding all positions within 1\arcsec\ of a catalog variable. Before measuring the scatter, we remove each aperture's banding (Figure~\ref{fig:satcorr_sw}a). For each group-difference index within an integration, we compute the median offset of the samples at that index from the aperture's overall median, and subtract this offset, which we call the group-index template, from every sample. The template is the repeating within-integration pattern averaged over the visit, so removing it leaves the white-noise component that the calibration is meant to isolate. We then measure each aperture's point-to-point scatter from the median absolute deviation (MAD) of successive differences, $\sigma_{\rm ptp} = 1.4826\,{\rm MAD}(\Delta f)/\sqrt{2}$. The factor 1.4826 converts a MAD into the standard deviation of a Gaussian, and dividing by $\sqrt{2}$ accounts for differencing two samples. This statistic is insensitive to astrophysical variability on timescales longer than the 21.5~s cadence. Successive BRIGHT2 differences are not strictly independent, since neighboring differences share one frame (\S\ref{sec:timing}), so this statistic estimates the point-to-point rather than the marginal variance. The test described below, which checks how often the normalized residuals of the catalog lightcurves fall within 1, 2, and 3$\sigma$, is what validates the attached uncertainties.  The resulting variance--flux relation is fit per detector with the two-parameter model
\begin{equation}
\sigma^2(F) \;=\; \alpha F + \beta \qquad [\mathrm{DN^2\ per\ group\ difference}],
\end{equation}
where the photon coefficient $\alpha$ and the detector-noise floor $\beta$ (read noise, $1/f$, dark and background shot noise) are determined from a fit to the median trend of the variance--flux relation over $200 < F < 2\times10^4$~DN, with two-sided $\sigma$ clipping. Above $\sim$$2\times10^4$~DN per difference, partial saturation compresses the response and the measured scatter falls \emph{below} the Poisson expectation for the reported flux (variance scales with the square of the local response derivative while the mean scales with its average). This regime is excluded from the fit, and we regard the precision of sources whose apertures reach it as limited by the saturation correction rather than by photon noise.

\begin{table}
\centering
\caption{Empirical aperture noise model, $\sigma^2(F) = \alpha F + \beta$, fitted independently per detector and field. The photon coefficient $\alpha$ is reproduced across the two clusters and three undithered visits to within 3\% of its per-detector mean, identifying it as a per-detector constant. For a successive-difference estimator applied to BRIGHT2 group differences, in which independent Poisson increments enter each difference with weights $\tfrac12$, 1, $\tfrac12$ and neighboring differences share one frame, the photon term is $(5/8)F/g$. The fitted $\alpha$ lies below the value this expression gives for the CRDS reference-file gains (1.6--2.9~e$^-$/DN across the Module~B detectors), consistent with the variance suppression that interpixel capacitance imprints on photon-transfer measurements of HgCdTe arrays \citep{Moore2006}.}
\label{tab:noise_model}
\begin{tabular}{lcccc}
\hline
Detector & $\alpha$ (Ter\,5) & $\alpha$ (L1 Seg\,3) & $\alpha$ (L1 Seg\,4) & $\beta$ (DN$^2$) \\
\hline
nrcb1    & 0.245 & 0.245 & 0.245 & 633--636 \\
nrcb2    & 0.244 & 0.242 & 0.244 & 657--671 \\
nrcb3    & 0.246 & 0.249 & 0.250 & 695--696 \\
nrcb4    & 0.274 & 0.274 & 0.274 & 670--684 \\
nrcblong & 0.296 & 0.298 & 0.304 & 538--552 \\
\hline
\end{tabular}
\end{table}

\begin{figure*}
\centering
\includegraphics[width=\textwidth]{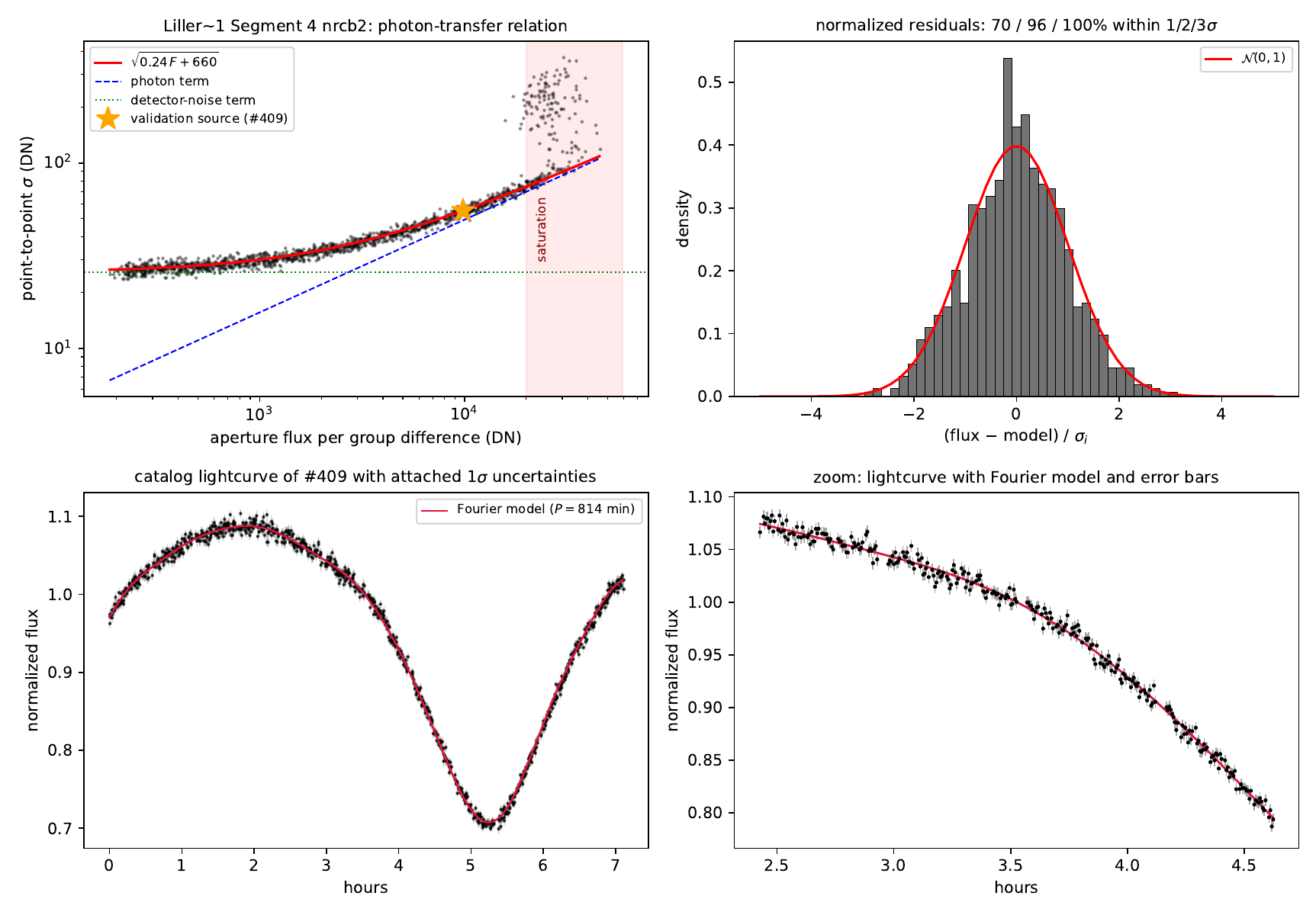}
\caption{Photometric error model calibration and validation (Liller~1, Segment~4, nrcb2 shown, though all 15 detector/segment combinations are treated identically). \textit{Top left:} aperture-level photon-transfer relation from star-centered calibration apertures, with the two-parameter model, its photon and detector-noise components, and the saturation-compressed regime (shaded) excluded from the fit. \textit{Top right:} normalized residuals, (flux $-$ model)/$\sigma$, of the validation source against an 8-term Fourier model at its adopted period, using the uncertainties attached in the catalog: 70/96/100\% of points fall within 1/2/3$\sigma$ ($\chi^2/\mathrm{dof}=0.93$). \textit{Bottom:} the validation source's catalog lightcurve (\#409, a 13.6-hr contact binary) with its attached per-point 1$\sigma$ uncertainties and the Fourier model overplotted, with a zoom on a $\sim$2-hr window at right.}
\label{fig:photerr_calibration}
\end{figure*}

\begin{figure}
\centering
\includegraphics[width=\columnwidth]{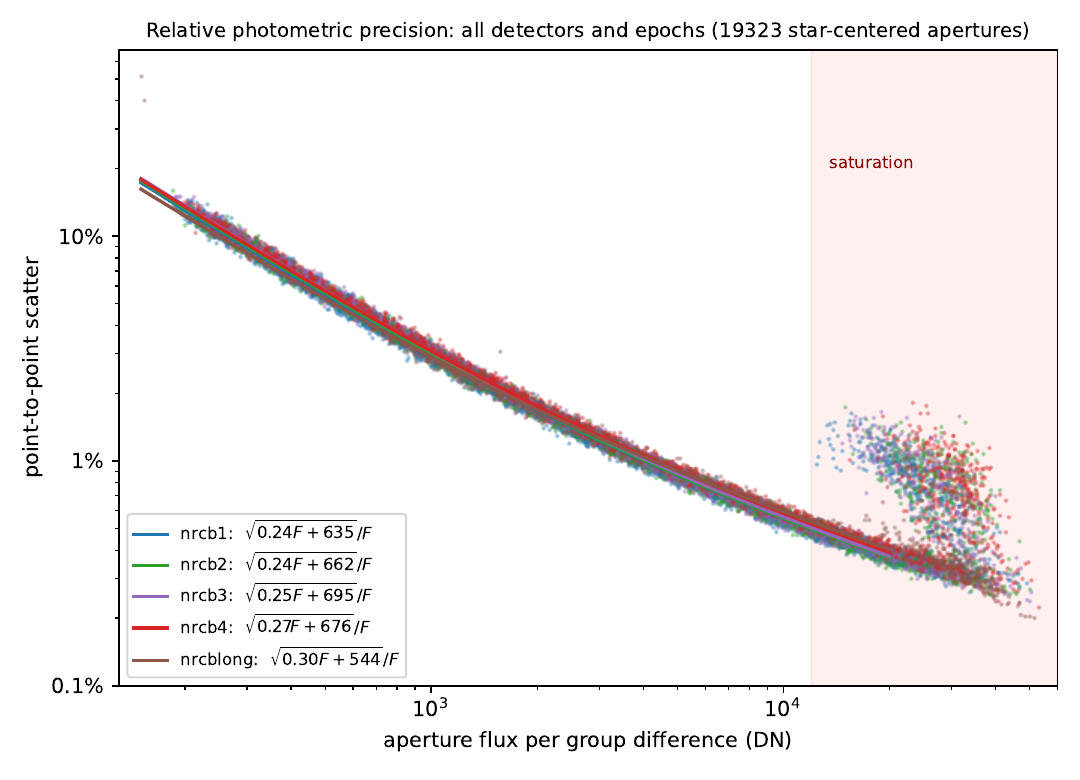}
\caption{Relative photometric precision of the survey. Fractional point-to-point scatter versus aperture flux for all 19,323 star-centered calibration apertures across the detector/segment combinations of both clusters, with the per-detector noise models overplotted. All detectors and visits collapse onto a single locus, from the detector-noise-dominated faint end to $\sim$0.5\% per 21.5~s point at the onset of saturation ($\sim$$1.2\times10^4$~DN, shaded). Beyond this point the later groups of each integration begin to saturate, and because these lightcurves are built from group differences, even a constant star then yields inconsistent differences within every integration, abruptly adding scatter that is not photometric. This is the regime the saturation correction of \S\ref{sec:saturation} addresses.}
\label{fig:photerr_fractional}
\end{figure}

The fitted $\alpha$ values (Table~\ref{tab:noise_model}) are stable, with each detector returning the same coefficient in both clusters and all three undithered visits to within 3\% of its mean, confirming that the model captures fixed detector physics rather than field-dependent systematics. Notably, no additional variance term proportional to $F^2$ (a fractional scatter proportional to flux, the signature of pointing jitter) is required once the group-index template is removed, indicating that in undithered staring mode, pointing-jitter noise at star centers is negligible. We can put a number on that, since the 96 zeroframes of each visit provide independent images: the common-mode centroid of several tens of bright stars per detector is stable to 0.007--0.016~px rms (0.2--0.5~mas), with the largest excursion over a full visit reaching only 0.02--0.07~px (0.5--2.2~mas), against a per-frame measurement floor of 0.002--0.006~px. A fixed aperture centered on a star sits at an extremum of enclosed flux, so an offset this small enters the photometry only at second order. A calibration that omits the template subtraction, or that places its apertures on bright-star wings rather than star centers, shows an apparent systematic floor of about 1\%, which arises entirely from the within-integration banding and from the wing placement.

Per-point uncertainties are attached to the catalog lightcurves by evaluating the model at each point's instantaneous flux and mapping the resulting uncertainty back onto the catalog normalization: $F_i = f_i\,F_{\rm med}$ and $\sigma_{f,i} = \sigma(F_i)/F_{\rm med}$, where $f_i$ is the normalized catalog flux and $F_{\rm med}$ the median unnormalized aperture flux. Because the catalog lightcurves carry no dilution correction (\S\ref{sec:background}), the flux and its uncertainty are normalized by the same quantity, and the fractional uncertainty is independent of any blend estimate. We then test these uncertainties on real variables. If the uncertainties are correct, the residuals of a lightcurve about a smooth model of its variability, divided by the attached uncertainties, should follow a unit Gaussian. We apply this test to bright variables that do not saturate, so that their catalog lightcurves are the uncorrected group-difference photometry, and model each one with a Fourier series of eight harmonics at its adopted period, enough to follow the variability without absorbing the point-to-point noise. The corrected lightcurves inherit these uncertainties through the same flux mapping and were not validated separately. For the validation exemplar (Terzan~5 \#60, a 7.5-hr contact binary at $\sim$9,500~DN), 68.4\%, 94.8\%, and 100.0\% of points fall within 1, 2, and 3$\sigma$ (Gaussian expectation 68.3/95.4/99.7) with $\chi^2/{\rm dof} = 1.02$. Figure~\ref{fig:photerr_calibration} shows the equivalent validation in Liller~1, and Figure~\ref{fig:photerr_fractional} the resulting survey-wide precision--flux relation. Across all 15 detector/segment combinations the single-star validations give $\chi^2/{\rm dof} = 0.92$--1.61, running systematically higher in the long-wavelength channel (with per-point uncertainties that are slightly more underestimated). We therefore apply pooled per-channel rescale factors (1.05 SW, 1.24 LW) to the attached uncertainties. At the survey's 21.5~s cadence the model implies a per-point fractional precision of $\sim$6~ppt for a $10^4$~DN source, a precision that reaches the millimagnitude regime in the binned lightcurves.

\subsection{Period search}\label{sec:period}

We search for periodic signals in each corrected lightcurve using two complementary algorithms. For sinusoidal variability (e.g., ellipsoidal modulations, pulsations), we compute the Lomb--Scargle periodogram \citep{Lomb1976,Scargle1982} over a frequency grid of 5000 points spanning periods from 20~min to 12~hr. We adopt a minimum period of 20~min because shorter periods approach the integration timescale ($\sim$3.2~min for 9 group differences at 21.5~s each) and are prone to aliasing artifacts. We quantify the significance of the highest peak as $(\mathrm{peak\;power} - \mathrm{median\;power}) / \mathrm{MAD}$, where MAD is the median absolute deviation of the periodogram power values. For transit- and eclipse-like signals, we apply the Box Least Squares (BLS) algorithm \citep{Kovacs2002} with 1000 linearly spaced trial periods from 20~min to 12~hr and 30 geometrically spaced trial durations from 1~min up to 0.95 times the minimum trial period (19~min). Deep eclipses of longer duration are instead recovered by the Lomb--Scargle search and the sliding-box single-event search (\S\ref{sec:singletx}). For BLS we quantify significance as the signal detection efficiency, $(\mathrm{peak\;power} - \mathrm{mean\;power})/\sigma$ of the power spectrum. We report the best-fit period and significance from both algorithms independently for each source, as the two methods are sensitive to different variability morphologies. Throughout this work, ``significant periodicity'' denotes a Lomb--Scargle significance above 50, a threshold chosen empirically to lie well above the noise tail of the significance distribution. (An earlier search that extended to periods as short as 1~min found essentially all peaks below 20~min to be aliases of the integration structure, which motivated the floor adopted here.)

\subsection{Sliding-box single-event search}\label{sec:singletx}

The Lomb--Scargle and BLS searches are sensitive to repeating signals, since their power is integrated over multiple cycles. A single transit-like dip occurring during a
$\sim$7-hour visit still injects power into the periodogram, but as broadband
structure spread across frequencies rather than a coherent peak at a
well-defined period. With only one event there is no repetition to define a
period, so any recovered ``best'' period is unconstrained (essentially any
value $\gtrsim$ the visit length) and carries no significance as a periodicity.
Such one-off events are thus not reliably recovered by either algorithm. To recover them, and as an independent
completeness check on the autocorrelation-based variable source
detection, we run a brightness-independent sliding-box matched filter on every
group-differenced cube (Terzan~5 Segment~2 and Liller~1 Segments~3 and~4).

For each pixel time series we apply an asymmetric isolated-outlier clip designed
to remove cosmic-ray spikes without erasing real eclipses. At each iteration we
compute a local $5$-frame running mean and the per-pixel median absolute
deviation, replace any sample exceeding $+3\sigma$ with the local mean, and clip negative excursions only beyond $-10\sigma$.  In practice, two iterations of this clipping suffice.
We then slide a top-hat box of $N_w$ frames across the cleaned lightcurve and
compute, for each center time $t_0$ and width $w$,
\begin{equation}
    \mathrm{depth}(t_0, w) = \tilde{f} - \langle f \rangle_{\rm box}, \qquad
    \mathrm{SNR}(t_0, w) = \mathrm{depth}\,\sqrt{N_w}\,/\,\mathrm{MAD},
\end{equation}
where $\tilde{f}$ is the global median, $\langle f \rangle_{\rm box}$ is the
within-box mean (computed in $\mathcal{O}(1)$ per center via a cumulative-sum
buffer), and MAD is the per-pixel median absolute deviation of the cleaned lightcurve. We sample 20 widths log-uniformly between 10~min and 3~hr, oversample the center grid by a factor of four within each width, and require the box to lie at least 30~min from either edge of the visit so detections correspond to fully contained events rather than drifts within a ramp. Per pixel we record the best (depth, width, $t_0$, SNR).

The clip is deliberately asymmetric, and that asymmetry bounds what this particular search can find: it is tuned for dips, and a positive impulsive event of a few minutes' duration would be suppressed along with the cosmic rays. The variability \emph{detection} of \S\ref{sec:detection} carries no such bias, since the lag-1 autocorrelation responds to correlated flux changes of either sign, and it is what recovers the flare sources in the catalog. The two searches also have different timescale floors: the autocorrelation detection is built from the integration-level calints, so an event must persist across consecutive $\sim$215~s integrations to register, whereas the sliding-box search runs on the 21.47~s group differences with a minimum box of 10~min. A matched-filter search for positive excursions, which would reach stellar flares, self-lensing, and microlensing peaks, is a straightforward extension we have not run here.

For Terzan~5 Segment~2 we performed an independent search using the \texttt{DOLPHOT}
photometric catalog as the source list (running the matched filter on the
catalog-extracted lightcurves rather than at every pixel).  After the cuts
described above, this search returns the cleaned lightcurves of all the
variables already identified by the autocorrelation method,
plus a small number of additional shallow-transit candidates that are flagged
for visual inspection. That the matched filter re-detects every Segment~2 variable previously classified as REAL indicates that the autocorrelation search (\S\ref{sec:detection}) does not systematically miss short-duration single events of a few-percent depth. This test is one-sided, and we do not read it as a completeness
measurement. Because the matched filter runs on a \texttt{DOLPHOT} source
list, it inherits that catalog's own incompleteness in the most crowded
central regions, so agreement between the two searches cannot distinguish a
complete input list from two methods failing on the same stars. What the test does establish is that, within the shared DOLPHOT source list, the two searches recover the same variables. A quantitative completeness assessment requires signal injection and recovery, which we defer to a dedicated transit-search paper. The matched-filter candidates
that pass visual inspection contribute to the shallow-transit candidates
discussed in \S\ref{sec:planets}.

\subsection{Source classification}\label{sec:classification}

We deliberately place no automated variability cut between detection and classification, because any threshold we could write down would be tuned to the variables we already expect. The $3\sigma$ autocorrelation detection is instead left permissive, and every candidate it returns is judged by eye. We generate a multi-panel diagnostic plot for each one, showing the detection lightcurve alongside forced photometry at the same sky position from independent observations (other segments and wavelength channels), and classify it as a genuine variable (``REAL'') or an artifact (``FAKE''). During the same inspection we gave almost every REAL variable a provisional class by eye from the shape of its lightcurve: one of the eclipsing-binary types of the General Catalogue of Variable Stars \citep{Samus2017}, EA (Algol type), EB ($\beta$~Lyrae type), or EW (W~UMa type), a transit-like event, a non-eclipsing class such as post-common-envelope binary (PCEB), accreting binary, flare source, or pulsator, or ``uncertain'' where no type was clear. The starting sample was large: $\sim$133{,}600 candidate detections across both clusters, both channels, and all visits (the per-image detections of Table~\ref{tab:detections} after spatial deduplication within each detector and the merging of coincident ramp and zeroframe detections into one diagnostic), of which 4{,}244 were labeled REAL, deduplicating to the 1{,}315 unique objects of the published catalog, whose deduplication groupings are archived with the data products (Appendix~\ref{app:repro}). While labor-intensive, this approach avoids the loss of unusual or low-amplitude variables that automated filters would reject, and leverages the astronomer's ability to recognize correlated signals across independent datasets.

Beyond the REAL/FAKE decision, we classify each confirmed variable by fitting it as a binary where we can. Using \texttt{PHOEBE}~2.5 \citep{Prsa2016,Conroy2020}, we model each source as an eclipsing or ellipsoidal binary in a detached, semi-detached, or contact (overcontact) Roche geometry. The fit varies the mass ratio, inclination, temperature ratio, the fill factors or fractional radii of the two components, and the orbital period and epoch, with a separate dilution term for each segment and wavelength channel, a scale and an offset solved by linear least squares at each step of the fit, to absorb light from blended neighbors. Where a lightcurve required it, the model also includes a starspot, an eccentric orbit, or a linear baseline trend in time. The stellar fluxes are computed using either blackbody or Castelli--Kurucz atmospheres. Each model is fitted jointly to both visits of its cluster, and the adopted period is that of the joint fit. For Liller~1 both visits, Segments~3 and~4, constrain the period and the lightcurve shape. For Terzan~5 the dithered Segment~1 enters the fit only to refine the period, since its reconstructed lightcurve (\S\ref{sec:seg1_extraction}) is not trusted to constrain the shape, which is set by the undithered Segment~2. We inspect every fit by eye and either accept it or set it aside, and the accepted models and all acceptance decisions are archived with the catalog (Appendix~\ref{app:repro}). Our purpose here is classification, not characterization: we use the fitted Roche geometry to assign each source a morphology, and defer careful modelling of individual systems, with the uncertainties such modelling requires, to future work.

Each of the 1,315 sources falls into one of five groups. For 721 sources we accepted a model and the two visits constrain the period, and we adopt the fitted morphology as the classification: 288 detached, 260 contact, and 173 semi-detached systems (for one of them, Terzan~5 \#304, the number of minima per orbit is ambiguous and both period readings are given in Appendix~\ref{app:atlas}). A further 194 sources have an accepted model but only a single eclipse within the two visits, so the period is unconstrained, and a single event cannot distinguish one Roche geometry from another. We class these as single-transit sources, together with 11 transit-like candidates for which no model was accepted, and quote no period for them (\S\ref{sec:planets}). For 309 sources no \texttt{PHOEBE} model was accepted, and we keep the class we assigned by eye, which is EA for 15 of them, EB for 4, and ``uncertain'' for the remaining 290. The 39 sources whose lightcurves we judged unfit for modelling carry no class. The remaining 41 belong to classes a binary model cannot express: the redback companion of PSR~J1748$-$2446A (\S\ref{sec:psrj1748}), the 20.9-minute candidate ultracompact X-ray binary or intermediate polar (UCXB/IP, \S\ref{sec:xray_cross}), 20 candidate PCEBs, and smaller numbers of candidate accreting binaries, X-ray binaries, flare sources, and pulsators, among them the RR~Lyrae variable Terzan~5 V3 of the Clement catalog. These 41 also keep their by-eye classes, and all save the redback's are candidate designations from lightcurve morphology alone. In total, the classifications of 915 sources (the 721 with a constrained period and the 194 single-transit sources with a model) rest on an accepted model (for two of the single-transit sources the accepted model is from an earlier, simpler fit rather than \texttt{PHOEBE}, and it is not drawn in Appendix~\ref{app:atlas}).

The fits rest on broadband photometry alone, without radial velocities, so they determine geometry far better than they determine physical parameters, and we release the morphologies and fitted ephemerides only. The lightcurve of every classified source, with its model overlaid where one was accepted, is shown in Appendix~\ref{app:atlas}. Because both the model acceptances and the by-eye classes rest on human judgment, we release both with the catalog (Appendix~\ref{app:repro}) so that any analysis built on them is reproducible.

\subsection{Astrometric calibration}\label{sec:astrometry}

Throughout this work, detector pixel positions are the ground truth: every lightcurve is extracted at a fixed detector position. Celestial coordinates enter only when detections must be related across detectors, visits, or external catalogs. For that we place all detections on a common Gaia-tied frame with a two-stage calibration. We first align the long-wavelength detector (nrcblong), which spans the full module field of view, directly to Gaia~DR3, matching sources detected on its zeroframe median image to Gaia positions proper-motion-propagated to the observation epoch ($\sim$9.3~yr baseline). This alignment yields frame ties good to 1--2~mas with post-fit residuals of 5--12~mas per star. We then register each short-wavelength detector onto this Gaia-tied LW frame using several hundred cross-matched stellar detections per detector, far more than the number of usable Gaia stars in these crowded, reddened fields, so the transfer uncertainty is negligible ($\sim$0.2~mas). Source positions are then computed by mapping each object's refined detector centroid ($\sim$3~mas centroid precision) through the resulting per-detector solutions, giving an absolute systematic floor of $\approx$3--4~mas for well-detected sources. The full procedure, per-detector solutions, residual distributions, and error budget are presented in Appendix~\ref{app:astrometry}. All celestial coordinates used below, for cross-segment matching, deduplication, the published catalog, and X-ray/radio cross-identification (\S\ref{sec:xray_cross}), are on this frame.

\subsection{Cross-segment matching and deduplication}\label{sec:crossmatch}

Each visit to a given target constitutes an independent observation, with source detection and lightcurve extraction performed separately. To build a unified source catalog and assess the reliability of our detections, we cross-match the REAL-labelled source lists of the two Liller~1 visits using their celestial coordinates (\S\ref{sec:astrometry}). We adopt a matching radius of 0\farcs3 ($\sim$10~NIRCam short-wavelength pixels). This is far larger than the $\approx$3--4~mas astrometric floor of \S\ref{sec:astrometry}, and deliberately so: the radius is not set by how well we know a position but by how far an independent \emph{detection} of the same star can land from it. Detections are peaks in an autocorrelation image rather than centroids of a flux profile, and around a bright or saturated star a detection can attach to a PSF wing several pixels from the star itself, so the scatter between two independent detections of one variable is a pixel-scale quantity, not a milliarcsecond one. We control the resulting false-match rate by enforcing one-to-one matching rather than by tightening the radius: when several sources from one segment match the same source in the other, only the closest pair is retained.

Sources matched in both Liller~1 segments are the most secure detections. Sources appearing in only one segment are retained in the catalog and are examined with particular care when classified by eye. A single-segment detection does not mean the source is spurious: a class of real, and often high signal-to-noise, variables shows a single, clean eclipse in one visit and is flat in the other, exactly as expected for an eclipsing binary whose orbital period is long compared to the $\sim$7-hour visit, so that an eclipse falls within one window but not the other. Discarding single-segment sources would systematically remove precisely these longer-period eclipsing systems. We therefore retain them and rely on the by-eye REAL/FAKE classification to separate genuine one-segment variables from artifacts.

\subsubsection{Deduplication}\label{sec:dedup}

In crowded fields, the wings of bright variable sources can produce secondary detections at nearby positions that track the variability of the primary source rather than representing independent variables. To identify and remove such duplicates, we search for source pairs within 0\farcs3. When duplicates are found, we retain the detection with the higher signal-to-noise ratio. The deduplication is applied non-transitively: after removing a duplicate, we re-check distances from the surviving source to avoid cascading removals that could erroneously merge distinct sources. The final merging of detections into the unique objects of the catalog is done by hand (\S\ref{sec:catalog}).

\subsection{Catalog construction}\label{sec:catalog}

We build a master catalog by cross-matching the manually classified REAL sources across all detectors (nrcb1--4 for SW, nrcblong for LW), observation segments, and ramp and zeroframe products. In these extremely crowded cores, automated positional matching alone is unreliable for establishing the final source list. The wings of bright variables, sources straddling detector boundaries, and the same star detected at slightly offset positions in different segments or channels can each generate multiple catalog entries for a single physical object, while truly distinct variables can lie within a fraction of an arcsecond of one another. We perform the final deduplication by hand. Candidate detections are first grouped by sky position (within a $1\arcsec$ linking radius) into per-object collections. We then inspect each collection visually, comparing positions and phased lightcurves across all detectors, segments, and channels, to decide whether it represents a single source or several blended variables, and split a collection into multiple objects only where distinct, independently varying signals are clearly present. This manual deduplication yields 1,315 unique objects: 915 in Liller~1 and 400 in Terzan~5. Because this step relies on human judgment rather than a fixed algorithm, the resulting deduplication groupings are released with the data products (Appendix~\ref{app:repro}) so that the catalog is exactly reproducible. We note that automated positional matching at a fixed radius yields a $\sim$4--8\% larger count, dominated by unmerged wing detections and blends in the cores.

For each source, we refine the centroid by two-dimensional Gaussian fitting on the autocorrelation image (Appendix~\ref{app:astrometry}) and perform forced aperture photometry at the refined position across all available data cubes, producing lightcurves from both segments and both wavelength channels where the source falls on the detector. Which autocorrelation image is used depends on the source's brightness. For sources with a strong zeroframe detection (${\rm SNR} > 10$), i.e., the bright stars whose saturating cores corrupt the ramp-based autocorrelation, we centroid on the \emph{zeroframe} autocorrelation image, built from the 10.7~s first reads, which are unsaturated for all but the very brightest stars (the Rapid Burster is the archetypal case). For all other sources we use the ramp autocorrelation image, preferring the short-wavelength channel where available. This forced-photometry step deliberately differs from the initial extraction (\S\ref{sec:extraction}) in one respect: it uses exact fractional-pixel-overlap apertures (same $r = 1.5$~px radius) at the refined sub-pixel position, rather than whole-pixel apertures at the integer detection position. Both conventions are preserved in the released code. The lightcurve correction pipeline (\S\ref{sec:corrections}) is then applied to all ramp lightcurves. Table~\ref{tab:correction_summary} summarizes the outcome of the correction strategy selection across all 4,246 source/segment/channel combinations with usable lightcurves (a count of lightcurves, unrelated to the 4,244 REAL labels of \S\ref{sec:classification}, which count detections).

\begin{deluxetable}{lcc}
\tablecaption{Lightcurve correction strategy selection across all sources.\label{tab:correction_summary}}
\tablehead{
\colhead{Adopted strategy} & \colhead{Count} & \colhead{Fraction}
}
\startdata
None (IQR clipping only) & 2917 & 68.7\% \\
Saturation + slope correction & 767 & 18.1\% \\
Slope correction only & 549 & 12.9\% \\
Saturation correction only & 11 & 0.3\% \\
PSF-wing reduction\tablenotemark{a} & 2 & $<$0.1\% \\
\hline
\textbf{Total} & \textbf{4246} & \\
\enddata
\tablenotetext{a}{Not one of the four strategies of \S\ref{sec:strategy}. The PSF-wing lightcurves of the Rapid Burster (\S\ref{sec:special}) replace its standard short-wavelength lightcurves in Segments~3 and~4.}
\end{deluxetable}

The majority of lightcurves (69\%) require no correction beyond IQR clipping. The combined saturation plus slope correction is the most common improvement (18\%) and is concentrated among the brightest sources, where later groups in each integration are suppressed by detector nonlinearity and the residual flux-dependent slopes benefit from the additional slope correction. Slope correction alone improves a further 13\%, mostly sources of moderate brightness in which residual nonlinearity produces flux-dependent within-integration slopes but not severe banding.

\subsection{Estimating apparent magnitudes}

We use version 3.1 of the \texttt{DOLPHOT} PSF-photometry package \citep{Dolphin2000,Dolphin2016}, with the NIRCam module of \citet{Weisz2024}, to establish a deep source catalog of each cluster and estimate apparent magnitudes (VEGAMAG) in both NIRCam filters for the variable sources with a DOLPHOT counterpart (\S\ref{sec:amplitudes}).

Table~\ref{tab:catalog} presents the master catalog. For each source it lists the Gaia-tied ICRS position with its per-source uncertainty, the detector and autocorrelation signal-to-noise ratio of the detection, the adopted classification (\S\ref{sec:classification}), and the adopted period (\S\ref{sec:period}). Per-source positional uncertainties combine the Gaussian centroid error with the Gaia frame-tie and LW$\to$SW transfer systematics in quadrature (Appendix~\ref{app:astrometry}), and are typically 3--4~mas for well-detected isolated sources, rising to $\sim$10~mas in the most crowded, blended regions. The full table for all 1,315 sources is available in machine-readable form, together with a per-source 8-bit data-quality flag marking which of the catalog lightcurves (each segment, channel, and ramp or zeroframe product) contain formally negative normalized-flux points, found in the faintest and most saturation-affected apertures (\S\ref{sec:amplitudes}).

\begin{deluxetable*}{lccccccccc}
\tabletypesize{\scriptsize}
\tablecaption{Variable star catalog (first 15 of 1,315 sources; the full table is available in machine-readable form).\label{tab:catalog}}
\tablewidth{0pt}
\tablehead{\colhead{ID} & \colhead{R.A.} & \colhead{Decl.} & \colhead{$\sigma_\alpha$} & \colhead{$\sigma_\delta$} & \colhead{Det.} & \colhead{S/N} & \colhead{Type} & \colhead{$P$} & \colhead{$P$ source} \\
 & \colhead{(deg)} & \colhead{(deg)} & \colhead{(mas)} & \colhead{(mas)} & & & & \colhead{(min)} & }
\startdata
T5~0 & 267.029082 & -24.782450 & 5.6 & 5.6 & nrcb2 & 39.5 & contact & 895.5 & PHOEBE \\
T5~1 & 267.005265 & -24.790062 & 3.9 & 3.9 & nrcb1 & 38.9 & contact & 703.9 & PHOEBE \\
T5~2 & 267.020997 & -24.768474 & 4.0 & 4.0 & nrcb4 & 38.7 & semi-detached & 1829.8 & PHOEBE \\
T5~3 & 267.010799 & -24.769378 & 3.7 & 3.7 & nrcb3 & 38.6 & contact & 649.9 & PHOEBE \\
T5~4 & 267.000934 & -24.784767 & 4.9 & 5.3 & nrcb1 & 38.6 & single-transit & \nodata & \nodata \\
T5~5 & 267.011785 & -24.796570 & 5.9 & 6.0 & nrcb1 & 38.6 & \nodata & 453.4 & LS \\
T5~6 & 267.022304 & -24.796452 & 3.9 & 3.9 & nrcb2 & 38.4 & contact & 580.6 & PHOEBE \\
T5~7 & 267.032327 & -24.783095 & 3.6 & 3.6 & nrcb2 & 38.2 & contact & 633.7 & PHOEBE \\
T5~8 & 267.018260 & -24.774920 & 5.8 & 5.8 & nrcb3 & 38.0 & detached & 1760.9 & PHOEBE \\
T5~9 & 267.003712 & -24.779987 & 3.6 & 3.6 & nrcb3 & 37.7 & contact & 453.0 & PHOEBE \\
T5~10 & 267.030231 & -24.766513 & 4.0 & 4.0 & nrcb4 & 37.4 & contact & 576.7 & PHOEBE \\
T5~11 & 267.019758 & -24.799162 & 3.3 & 3.3 & nrcb2 & 37.3 & single-transit & \nodata & \nodata \\
T5~12 & 267.016461 & -24.779536 & 3.6 & 3.6 & nrcb3 & 37.2 & detached & 932.1 & PHOEBE \\
T5~13 & 267.017973 & -24.783817 & 3.8 & 3.8 & nrcb1 & 37.2 & semi-detached & 750.9 & PHOEBE \\
T5~14 & 267.009268 & -24.773953 & 3.4 & 3.4 & nrcb3 & 37.2 & contact & 617.3 & PHOEBE \\
\enddata
\tablecomments{Positions are barycentric ICRS from the Gaia-tied astrometric solution (Appendix~\ref{app:astrometry}). Per-source uncertainties $\sigma_\alpha,\sigma_\delta$ combine the Gaussian centroid error with the Gaia frame-tie and LW$\to$SW transfer systematics in quadrature. ``Type'' is the adopted classification (\S\ref{sec:classification}): the fitted Roche morphology where a binary model was accepted and the period constrained, ``single-transit'' where an accepted fit covers a single event and so constrains no period, and otherwise the class we assigned by eye (EA, Algol type, EB, $\beta$~Lyrae type, or ``uncertain''). The non-eclipsing classes (PCEB, accreting binary, X-ray binary, UCXB/IP, flare, pulsator, irregular) are candidate designations from lightcurve morphology alone, with the exception of ``redback'', the ephemeris-confirmed companion of PSR~J1748$-$2446A. ``\nodata'' in this column denotes an unclassified variable. $P$ is the adopted period (\S\ref{sec:period}), and its source is ``PHOEBE'' (fitted orbital period), ``2x'' (twice the Lomb--Scargle period, the rule we applied to double-minimum lightcurves before the binary fits, kept for sources without an accepted model), or ``LS'' (best Lomb--Scargle peak), with ``\nodata'' in both columns where there is no period. A single-transit source carries no period, since one event constrains none, and for other classes without a measured orbital period an ``LS'' value is the periodogram peak, not an orbital period. The full machine-readable table includes all 1,315 sources and the Lomb--Scargle significance, and gives the classes as the catalog codes (e.g., single\_transit, accreting\_binary).}
\end{deluxetable*}

\section{Discussion}\label{sec:discussion}

\begin{figure*}
\centering
\includegraphics[width=0.95\textwidth,height=0.80\textheight,keepaspectratio]{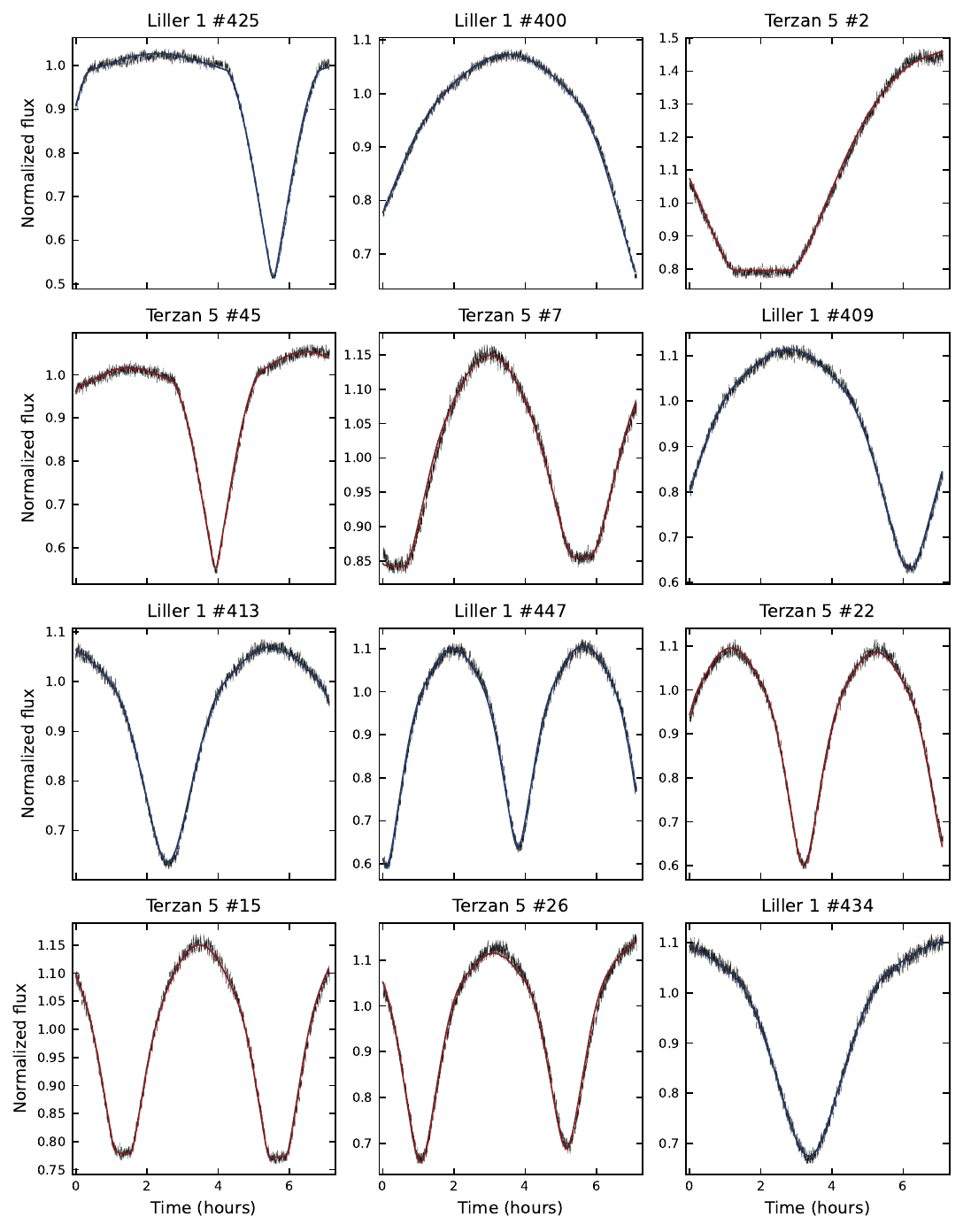}
\caption{Twelve high-signal-to-noise lightcurves from the catalog, shown as the unfolded, pipeline-corrected 21.5-second-cadence photometry of a single
$\sim$7-hour visit with $\pm 1\sigma$ error bars from the photometric error
model (\S\ref{sec:photerr}) and fitted \texttt{PHOEBE}~2.5
\citep{Prsa2016,Conroy2020} binary models (\S\ref{sec:classification}; dark
red = Terzan~5, dark blue = Liller~1). The
per-point precision (0.4--0.7\% per 21.5-s sample) is high enough that
smooth overlays require physical binary models.}
\label{fig:showcase_lcs}
\end{figure*}

\begin{figure}
\centering
\includegraphics[width=\columnwidth,height=0.78\textheight,keepaspectratio]{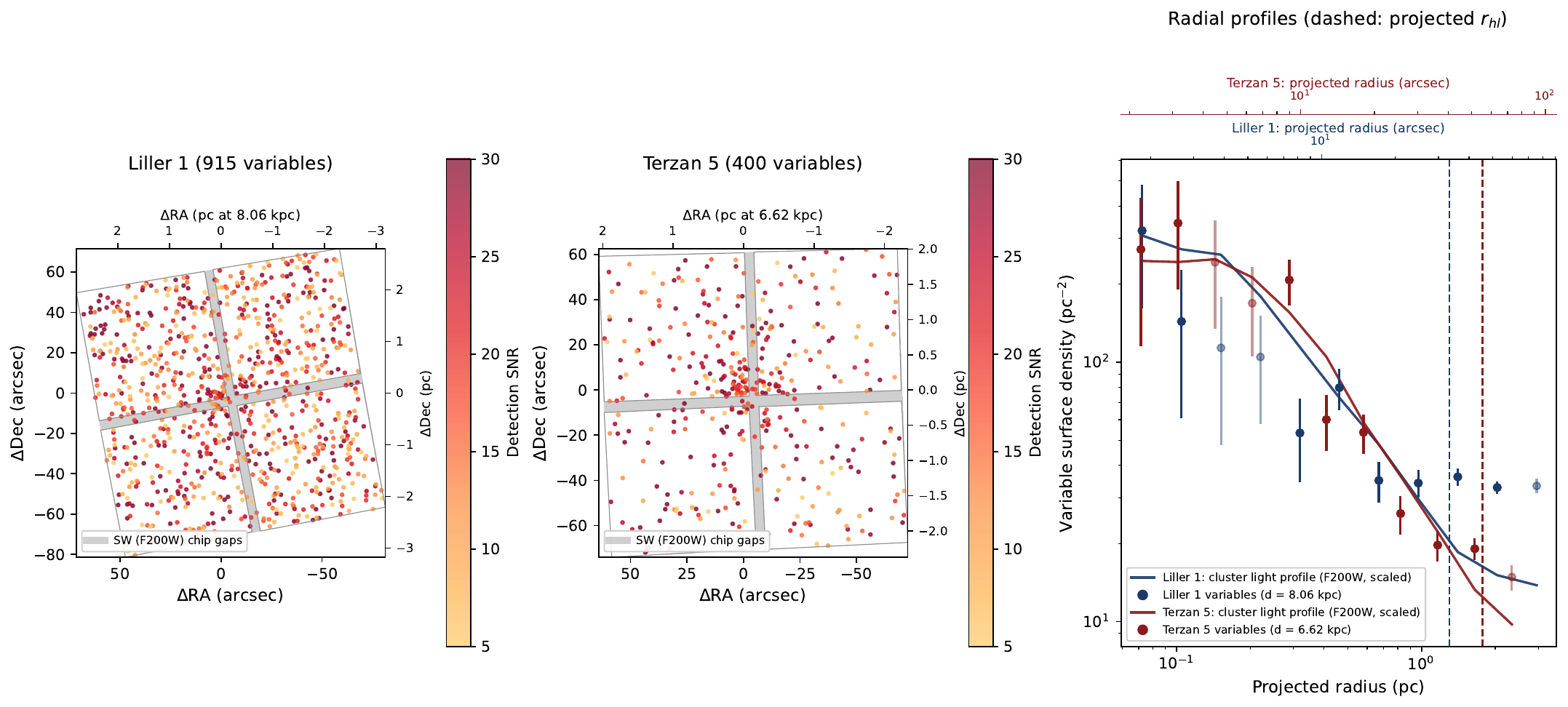}
\caption{Spatial distribution of the detected variables. \textit{Left,
center:} positions relative to the cluster centers
\citep{BaumgardtHilker2018}, colored by detection signal-to-noise ratio, in
arcseconds (lower and left axes) and parsecs (upper and right axes). Light
gray bands mark the gaps between the four short-wavelength detectors, which
the gap-free long-wavelength channel fills: variables inside the bands are
F356W-only detections. \textit{Right:} radial surface-density profiles of
the variables (error bars are 1$\sigma$ Poisson uncertainties on the counts in each annulus; faded points are annuli with under 60\% of their area inside the footprint), with dashed lines at each cluster's projected half-light
radius \citep{BaumgardtHilker2018,BaumgardtVasiliev2021}, compared with the
azimuthally averaged F200W surface brightness of the cluster (solid curves)
measured in the same annuli and rescaled so that only the shape is compared.
The Terzan~5 profile declines roughly tenfold from the core to beyond
$r_{hl}$, while Liller~1 shows a comparably dense core atop an extended
plateau that persists across the field, consistent with a broader intrinsic
distribution and/or blended field variables along this denser bulge sight
line.}
\label{fig:spatial_distribution}
\end{figure}

\subsection{The variable star population}\label{sec:population}

\begin{figure*}
\centering
\includegraphics[width=0.85\textwidth]{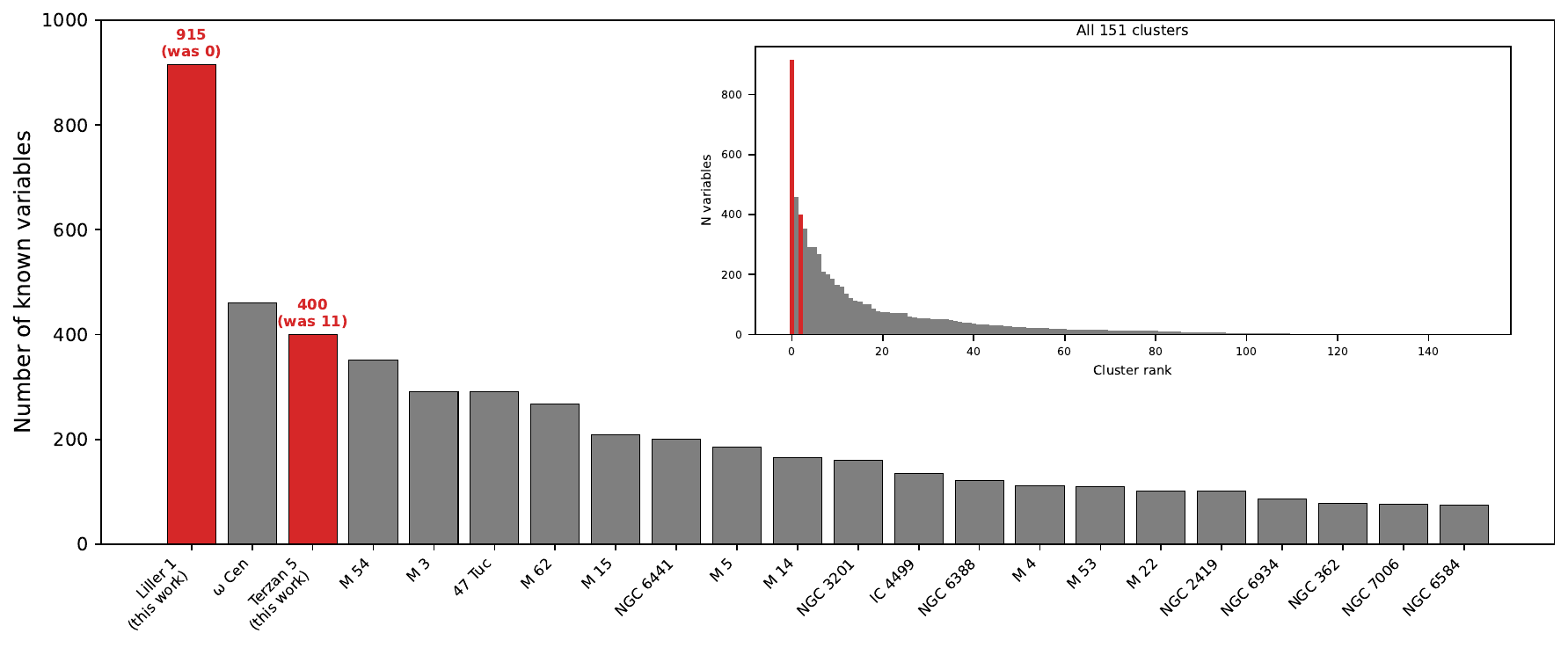}
\caption{Number of known variable stars in the fields of Galactic globular clusters. Gray bars show the top 20 clusters in the \citet{Clement2001} catalog (2017 update, 5,604 variables across 151 clusters). Red bars show our JWST detections, with each cluster's count in that same catalog release in parentheses. Liller~1 had \emph{zero} confirmed variables prior to this work. By this count it now has more variables in its field than any Galactic globular cluster, nearly twice as many as $\omega$~Centauri, the previous record holder. The inset ranks all 151 clusters by variable count, with Terzan~5 and Liller~1 updated to our counts. We have not established cluster membership for our detections, so these are counts of variables in each cluster's field rather than of confirmed members.}
\label{fig:clement_bar}
\end{figure*}

Figure~\ref{fig:showcase_lcs} illustrates the photometric quality of the catalog through twelve of its highest-signal-to-noise lightcurves, all eclipsing or contact binaries, tracked at the few-mmag level by the fitted eclipsing-binary models of \S\ref{sec:classification}, and Figure~\ref{fig:spatial_distribution} shows the spatial distribution of the detected variables within each cluster. Our census transforms the landscape of known globular-cluster variables (Figure~\ref{fig:clement_bar}). Against the 53 variables previously cataloged in the fields of these two clusters, all of them in Terzan~5 and only two of them eclipsing (\S\ref{sec:detection}), we detect 1,315. With 915 new detections, Liller~1, which had \emph{zero} previously confirmed variables, becomes, by detection count against the 2017 Clement release, the globular cluster-like object with the most known variables in its field, with nearly twice as many as the previous record holder $\omega$~Centauri (460 cataloged variables). Terzan~5 makes the same jump on a smaller scale, going from 53 cataloged variables to 400 and entering the all-cluster ranking in third place. The combined yield represents 23\% of the entire 5,604-object Galactic census \citep{Clement2001}. These counts refer to variables in the JWST field of each cluster. The central concentration of the detections (Figure~\ref{fig:spatial_distribution}), pronounced in Terzan~5, and present as a dense core atop a flatter extended component in Liller~1, indicates that most Terzan~5 detections and at least the concentrated Liller~1 component are cluster members, but a proper-motion membership analysis is needed to quantify the field contribution, particularly along the denser Liller~1 sight line. While JWST has been used for multi-epoch imaging of globular clusters, including both of these targets, for population and astrometric science \citep{Milone2026}, this is the first JWST program to leverage the time axis for a variability census of a globular cluster-like object.

The character of our detections also differs from that of previous surveys, since the Clement catalog is $\approx$55\% RR~Lyrae, whose pulsations trace the horizontal branch rather than the dynamical binary population. Our survey, operating at 2~\micron, where RR~Lyrae amplitudes are small \citep{Braga2018}, and optimized for correlated variability rather than pulsation amplitude, instead preferentially detects eclipsing binaries, ellipsoidal variables, and accretion-powered systems, precisely the populations most directly shaped by the dynamical encounters that make dense cluster cores unique. Our classification (\S\ref{sec:classification}) demonstrates this: of the sources for which we accepted a binary model, 721 have a constrained orbital period and a Roche morphology (288 detached, 260 contact and 173 semi-detached), against the 26 eclipsing binaries from the most comparable previous survey, the 8.3-day HST campaign on 47~Tucanae \citep{Albrow2001}. The predominance of binaries in our sample is therefore not unexpected. Terzan~5 and Liller~1 have among the highest known stellar encounter rates in the Galaxy \citep{VerbuntHut1987,Saracino2015}, with the rate in Terzan~5 alone $\sim$7 times that of 47~Tucanae \citep{Bahramian2013}. Such encounters can form new binaries, harden existing ones, and drive them toward interaction, naturally producing the compact binary populations to which our survey is particularly sensitive.

\begin{figure*}
\centering
\includegraphics[width=\textwidth]{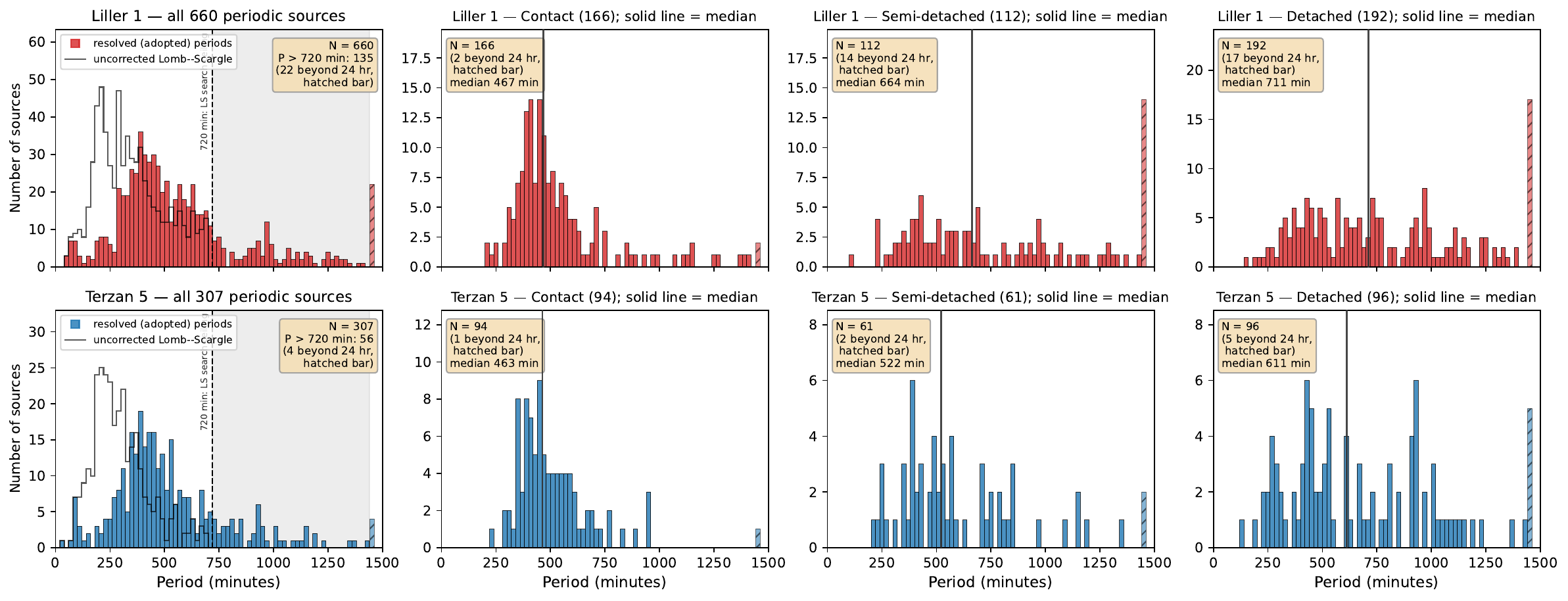}
\caption{Period distribution of the periodic variables in Liller~1 (top row) and Terzan~5 (bottom row). \textit{Left column:} every source with a significant Lomb--Scargle detection (660 in Liller~1, 307 in Terzan~5). Filled histograms show the adopted period: the fitted orbital period where a \texttt{PHOEBE} model was accepted, and otherwise the best Lomb--Scargle period or twice that value, as appropriate for double-minimum lightcurves. Unfilled outlines show the raw Lomb--Scargle periods. The search extends only to 12~hr (dashed line), so periods beyond it (shaded) come only from fitted or doubled values. \textit{Remaining columns:} the fitted orbital periods of the sources with an accepted binary model, split by Roche morphology: contact (260), semi-detached (173) and detached (288), with the class median drawn as a solid vertical line. The contact systems peak between 0.25 and 0.33~d (median 0.32~d), bracketing the known W~UMa peak \citep{Rucinski2007} but without its sharp 0.22~d cutoff (\S\ref{sec:population}). Periods approaching or exceeding the 7-hr visit duration are increasingly uncertain, and the longest, up to $\sim$44~hr for contact systems, rest on models fitted to less than one cycle per visit. In all panels, hatched bars at the right edge collect periods beyond 24~hr; the 205 single-transit sources carry no measured period and are not shown.}
\label{fig:period_histogram}
\end{figure*}

\begin{figure*}
\centering
\includegraphics[width=\textwidth,height=0.82\textheight,keepaspectratio]{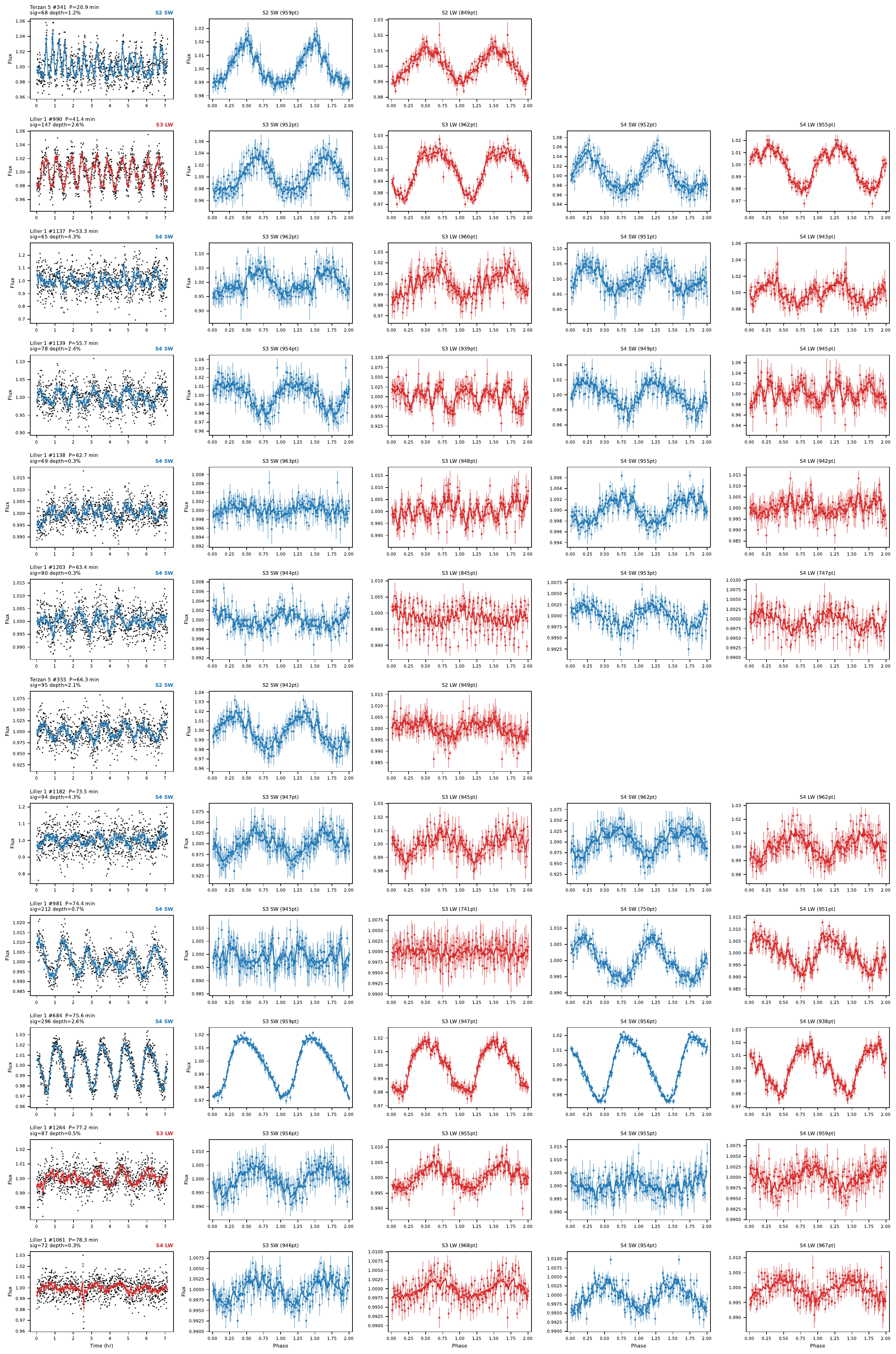}
\caption{Short-period variable candidates ($P < 80$~min), ordered by period. Left panels show the unfolded lightcurve from the segment and channel with the strongest Lomb--Scargle detection, named in the upper-right label of each panel. Black points are the individual 21-second samples, and the colored points are means of consecutive blocks of 7 to 9 samples with standard-error-of-the-mean bars, joined by a line in the color of that channel. Right panels show the lightcurve of each segment and channel folded on the same period and plotted over two cycles, as means in 100 phase bins with standard-error-of-the-mean bars, with a five-bin running mean drawn through them as the solid line. Short-wavelength (F200W) data are shown in blue and long-wavelength (F356W) data in red throughout. The two shortest-period sources are \#341 in Terzan~5 ($P = 20.9$~min, recovered independently in both wavelength channels) and \#990 in Liller~1 ($P = 41.4$~min, detected independently in both segments and both wavelength channels).}
\label{fig:short_period_collage}
\end{figure*}

The detected variables span a wide range of variability timescales. We identify 660 sources with significant periodicity in Liller~1 and 307 in Terzan~5. A periodogram cannot be taken at face value for a binary with two minima per orbit, because it locks onto the half period. Where we accepted a binary model, we thus adopt its fitted orbital period directly (\S\ref{sec:classification}), which resolves the ambiguity in all but one case (Terzan~5 \#304) by fitting the two minima rather than by doubling a peak. The earlier heuristic of doubling the best Lomb--Scargle period survives only for a handful of sources with no adopted model. Restricted to the systems the models call contact, the distribution (Figure~\ref{fig:period_histogram}) peaks between 0.25 and 0.33~d with a median of 0.32~d, bracketing the $\approx$0.27~d peak of the well-established W~UMa period distribution \citep{Rucinski2007}. We do not, however, reproduce the sharp cutoff near 0.22~d that accompanies it in that literature: 16 of the 228 contact systems whose period falls inside the search window, and 59 of the 721 sources with a fitted orbital period, lie below 0.22~d, and the distribution declines smoothly through the cutoff rather than terminating at it. \citet{ElBadry2022} find the same thing in a ZTF sample of low-mass detached eclipsing binaries, where the lowest-mass bins peak below 0.22~d and only the highest-mass bin shows the classical cutoff, and attribute the difference to sensitivity to lower-mass systems. Our survey observes at 2~\micron{} in clusters whose main sequences we resolve well below the turnoff, so the same explanation applies naturally here. The long-period end of the distribution is limited by the search rather than by the population. Our Lomb--Scargle grid extends only to 12~hours, so no periodogram peak beyond 720~minutes can be recovered, and longer periods enter the catalog only through the fitted binary models. The decline approaching that bound is thus a property of the search window. Among these periodic sources, 12 exhibit significant periods shorter than 80~minutes in at least one segment and channel (Figure~\ref{fig:short_period_collage}), including candidates with periods as short as $\sim$21~minutes that are recovered independently in both wavelength channels. These may represent ultracompact binaries or AM~CVn systems formed through dynamical interactions, or, in the case of the X-ray-matched 20.9-minute candidate (\S\ref{sec:xray_cross}), an intermediate polar detected at its spin period, which we judge the more likely interpretation for that source. Quantitative modeling of individual systems will be presented in later work.

\subsection{Location of the variables in the color--magnitude diagram}\label{sec:cmd}

\begin{figure*}
\centering
\includegraphics[width=\textwidth]{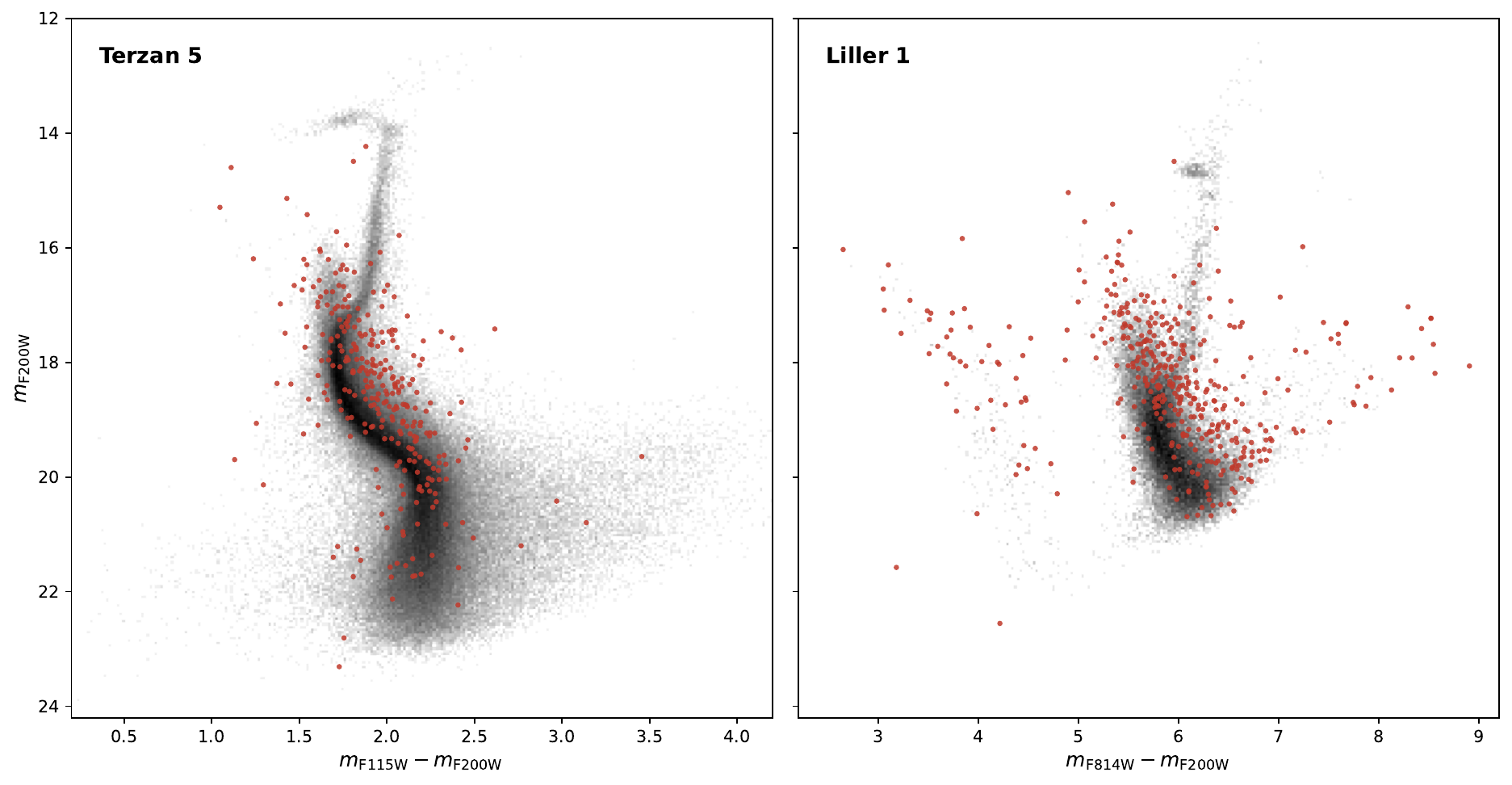}
\caption{Location of the variables in the color--magnitude diagrams of Terzan~5 (left) and Liller~1 (right). Gray: reference stars from the HST$+$JWST catalogs of \citet{Zullo2026} for Terzan~5 (275,437 stars) and \citet{ZulloInPrep} for Liller~1 (45,591 stars). Red: the 299 and 408 variables with a secure or likely catalog counterpart. Catalog magnitudes are means over the comparison observations, which sample each source's variability unevenly. In both clusters the locus of the variables sits slightly above the main sequence, as expected for binaries, whose combined light exceeds that of a single star. In Liller~1 the CMD alone shows that at least some of the variables are foreground interlopers, belonging to the blue population of sources to the left of the cluster sequence.}
\label{fig:cmd_variables}
\end{figure*}

Figure~\ref{fig:cmd_variables} places the variables on the color--magnitude diagrams of their host clusters. To do so, we cross-matched our DOLPHOT source lists with the joint HST$+$JWST photometric catalogs of Terzan~5 \citep{Zullo2026} and Liller~1 \citep{ZulloInPrep}. Counterparts were identified individually through visual inspection, using positional coincidence, consistency of measured apparent magnitudes, and evidence of variability in the comparison data. This approach accommodated small, chip-dependent positional offsets between the catalogs that complicated direct positional matching. Sources satisfying all three criteria were flagged as secure identifications, while less certain associations were assessed on a case-by-case basis, and we plot the 299 Terzan~5 and 408 Liller~1 variables whose counterpart is secure or likely. Because the two clusters were observed in different filter sets, we show each in the color that best resolves its sequence, F115W$-$F200W for Terzan~5 and F814W$-$F200W for Liller~1, so the two abscissae are not directly comparable. The resulting CMD positions may be affected by uneven sampling of the sources' variability in the comparison observations.

In both clusters the variables trace the main sequence from the turnoff downward, displaced above the ridge line, as expected for unresolved binaries whose combined light exceeds that of either component. We go no further than this here. Future work will refine these matches, investigate additional structure in the CMD, and use proper motions to determine what fraction of the population consists of field interlopers.

\subsection{Radial distribution of the variables}\label{sec:radial}

To quantify the radial distributions (right panel of Figure~\ref{fig:spatial_distribution}), we compute variable surface-density profiles by counting sources in logarithmically spaced annuli around each cluster center. An annulus drawn around the cluster center does not lie entirely on the detector, because the NIRCam module footprint is square and is split by gaps between its four detectors. We therefore compute each annulus area as only the sky we actually observed, taking the footprint to be the union of the four short-wavelength detector quadrants mapped through their individual astrometric solutions (Appendix~\ref{app:astrometry}). This excludes both the sky beyond the field of view and the inter-chip gaps. The footprint fraction of each annulus is evaluated by Monte Carlo, annuli with $<$15\% coverage are discarded, and those below 60\% are flagged. Angular radii are converted to parsecs using the Gaia-era distances of \citet{BaumgardtVasiliev2021}, and uncertainties are Poissonian. Because the footprint is the short-wavelength one, the 63 variables detected only in the gap-free long-wavelength channel (45 in Liller~1 and 18 in Terzan~5) are excluded from the profiles. The inter-chip gaps cover 7\% of the module footprint. For comparison with the cluster as a whole, we measure the azimuthally averaged F200W surface brightness of the NIRCam mosaic in the same annuli and footprint, rescaled by a single multiplicative factor. We use integrated light rather than resolved star counts because star counts are crowding-suppressed in these cores. The outermost points of the light profiles include a foreground and background bulge contribution. Within $\sim$1~pc of the center, where the uncertainties are largest, the variable surface density follows the cluster light in both clusters. The two profiles differ mainly in their outer parts. The Terzan~5 variables continue to track the cluster light to the edge of the field, declining roughly tenfold from the core to beyond the projected half-light radius. Beyond $\sim$1~pc, the Liller~1 variables instead level off onto a plateau above the light profile that persists across the field of view, consistent with a broader intrinsic distribution, a larger contribution of blended field variables along its denser bulge sight line, or both.

\subsection{Implications for dynamical binary formation}

As the two clusters with the highest known stellar encounter rates in the Galaxy (\S\ref{sec:intro}), Liller~1 and Terzan~5 are predicted by the empirical scaling between encounter rate and X-ray source population \citep{Pooley2003} to host the richest populations of compact binaries formed dynamically, through tidal capture, three-body exchange, and direct collisions \citep{Hut1992,Ivanova2008}. Our variable star census provides the first opportunity to test this prediction in the infrared, where extinction no longer prohibits detection (Figure~\ref{fig:spatial_distribution}).

The recent proposal that Bulge Fossil Fragments act as efficient ``factories'' of gravitational-wave progenitors (\S\ref{sec:intro}; \citealt{Ferraro2026}) further motivates characterizing their binary populations. The eclipsing and contact systems we detect are predominantly stellar binaries rather than compact-object binaries themselves, and their abundance should not be read as a simple tracer of the dynamical processing that manufactures compact binaries \citep{Ivanova2008,Ye2020}. The two populations arise through different channels. Close main-sequence pairs form readily in the field through ordinary binary evolution, and the same dynamical encounters that assemble compact binaries can act to \emph{suppress} the stellar-binary population by ionizing soft pairs and disrupting or exchanging away primordial binaries. Thus, a cluster with a dynamically enhanced X-ray-binary population need not host an enhanced close stellar-binary population, and may even host a depleted one. Whether the close-binary populations of the most dynamically active clusters are enhanced, suppressed, or merely reshaped relative to the field is, however, an open empirical question, and this census provides hundreds of close binaries with periods and amplitudes in the two most extreme such environments, the sample against which dynamical models can now be tested. The shortest-period systems are of additional interest in the gravitational-wave context. At the $\sim$8~kpc distances of these clusters, our sub-80-minute candidates would not be expected to be detectable by LISA. In the calculation of \citet{Kremer2018}, which depends on the component masses, eccentricities, and mission duration, only the most compact double white dwarfs, with periods $\lesssim$10~min, are detectable, and several such systems are predicted to reside in Milky Way globular clusters. The masses and even the orbital nature of our candidates are unknown, so their LISA detectability is not established. Our survey nonetheless demonstrates photometric sensitivity in the ultracompact regime. The census of confirmed ultracompact X-ray binaries in Galactic globular clusters numbers only eight systems \citep{Wang2026}, so even one confirmation from our sub-80-minute candidates would be a notable addition. The same observing mode could eventually identify infrared counterparts to LISA gravitational-wave sources in clusters, once such sources are found.

\subsection{Cross-identification with X-ray sources}\label{sec:xray_cross}

Both clusters host rich populations of \textit{Chandra} X-ray sources. In Terzan~5, 212 are cataloged within 1.2 times the half-light radius \citep{Bahramian2020}, including 22 candidate quiescent low-mass X-ray binaries (LMXBs), over twice as many as in any Galactic globular cluster \citep{Kumawat2025}. In Liller~1, \citet{Homer2001} detected the Rapid Burster plus three additional low-luminosity sources within two core radii. The overwhelming majority of these X-ray sources have no secure optical or infrared counterpart.

For a first census of X-ray--infrared associations, we cross-match our catalog against two \textit{Chandra} catalogs: the \textit{Chandra} Source Catalog 2.1 \citep{Evans2024}, whose astrometry is tied to \textit{Gaia}, for both clusters, and the deeper MAVERIC ACIS catalog \citep{Bahramian2020} for Terzan~5. For CSC~2.1, we measure and remove the residual frame offset, the median offset of the variable--X-ray source pairs closer than 0\farcs6 in each cluster ($\sim$30~mas for Terzan~5 and $\sim$210~mas for Liller~1), and count a match when the nearest catalog variable lies inside the frame-aligned 2$\sigma$ positional error ellipse. CSC~2.1 tabulates 95\%-confidence semi-axes and position angles, which we scale to 1$\sigma$ by the Rayleigh factor 2.45, and our own $\sim$3--10~mas uncertainties are negligible. The MAVERIC positions are tied to the radio frame, requiring a larger correction. A weighted-centroid fit to the 17 MAVERIC sources that fall within 50~mas of one of our variables once the correction is applied (the choice of tie sources and the offset are iterated to convergence) yields a frame tie of $(-198, -294)$~mas ($\pm$2.4~mas formal, and these tie sources register to 25~mas rms). The tabulated MAVERIC positional error, however, is a photon-statistics centroid uncertainty (as small as 4~mas) that omits registration systematics, so we adopt an effective per-source uncertainty $\sigma_{\rm eff} = (1.25\,\sigma_{\rm cat}) \oplus 74$~mas, with the systematic floor fitted by maximum likelihood from the secure matches, and again require the nearest variable to lie within 2$\sigma_{\rm eff}$.

\begin{figure*}
\centering
\includegraphics[width=\textwidth,height=0.82\textheight,keepaspectratio]{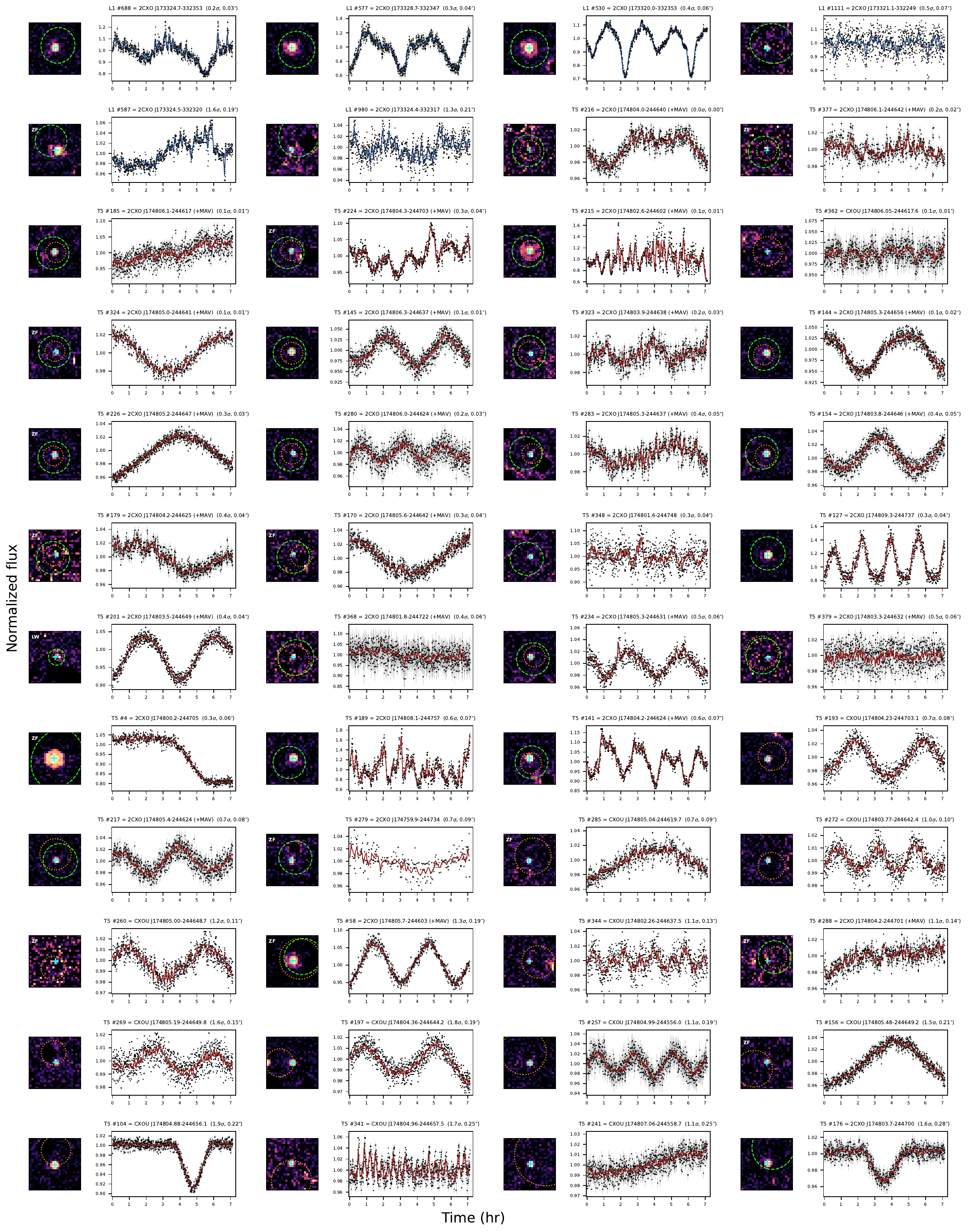}
\caption{Infrared variables inside the 2$\sigma$ positional error regions of \textit{Chandra} X-ray sources, from CSC~2.1 (both clusters) and the deeper MAVERIC catalog (Terzan~5). Each pair shows the autocorrelation-image cutout, with the JWST catalog position (cyan cross) and the X-ray error regions (dashed green for CSC~2.1, dotted orange for MAVERIC), beside the source's adopted lightcurve (\S\ref{sec:strategy}), plotted as normalized flux against time in hours through the visit; the colored line is a 1.5-minute median profile (3-bin running mean), and the gray error bars, where present, are the 1$\sigma$ per-point uncertainties of \S\ref{sec:photerr}. Cutouts marked ``ZF'' use the zeroframe autocorrelation image for stars that saturate during the ramp; the one marked ``LW'' falls in a short-wavelength chip gap and is shown in F356W. The 48 variables (blue: Liller~1; red: Terzan~5, sorted by separation) are all the matches within our footprint, and 23 of them are matched independently in both catalogs.}
\label{fig:xray_collage}
\end{figure*}

In Terzan~5, 64 CSC~2.1 sources fall within our short-wavelength footprint, and 29 of them (45\%) contain a catalog variable within their 2$\sigma$ error ellipse (26 within 1$\sigma$). The deeper MAVERIC catalog contributes 168 sources within the footprint, of which 36 match a variable at 2$\sigma_{\rm eff}$ (24 at 1$\sigma_{\rm eff}$). In Liller~1, 6 of 12 CSC~2.1 sources match, including the Rapid Burster itself (1.6$\sigma$). Together, these give 48 distinct X-ray-matched variables, 23 of them matched independently in both catalogs. Monte Carlo chance-alignment tests, randomly re-placing each 2$\sigma$ region at its own cluster-centric radius within the footprint, thereby sampling the same local density of variables, predict $\approx$4.7 chance matches across all 244 searched regions ($\approx$1.3 among the Terzan~5 CSC~2.1 regions, $\approx$3.0 among MAVERIC, and $\approx$0.4 among the Liller~1 CSC~2.1 regions). Even attributing every one of these to the matched sample implies an expected purity of $\approx$93\% for the 71 catalog-level associations, or $\approx$90\% for the 48 distinct variables. We adopt 2$\sigma$ as the matching threshold because it balances completeness against purity. A 2$\sigma$ Rayleigh region contains 86\% of true counterparts at high sample purity, whereas extending to 3$\sigma$ would add only $\approx$5 candidates, of which $\approx$3.5 are expected to be chance alignments. Figure~\ref{fig:xray_collage} shows all 48 matches, each as the variable's autocorrelation-image cutout, with the frame-aligned 2$\sigma$ regions overplotted, beside its lightcurve. The sample is dominated by clean eclipsing and ellipsoidal binaries, precisely the counterpart classes expected for quiescent LMXBs, cataclysmic variables, and active binaries. Among the individually notable members, the counterpart of Terzan~5~A enters at 1.1$\sigma_{\rm eff}$ (its association is independently secured by the radio timing ephemeris, \S\ref{sec:psrj1748}), and our 20.9-minute UCXB/IP candidate \#341 matches CXOU~J174804.96$-$244657.5 at 1.65$\sigma_{\rm eff}$ (individual chance probability 6\% in the crowded core), an X-ray association consistent with either an ultracompact X-ray binary at its orbital period or an intermediate polar at its spin period. Of the two, the intermediate polar is the more likely a priori, since intermediate polars are the more common class in the field, and in that case the 20.9-minute photometric signal would be the spin period of the accreting white dwarf rather than an orbital period. However, the abundance of ultracompact X-ray binaries in globular clusters keeps the ultracompact interpretation a strong possibility here, and only spectroscopy can separate the two. 

The two most valuable identifications rest on evidence beyond X-ray positional coincidence alone, namely those of the millisecond pulsar Terzan~5~A, whose counterpart (\#344) is phase-locked to the radio timing ephemeris (\S\ref{sec:psrj1748}), and the Rapid Burster, whose counterpart (\#587) is supported jointly by the X-ray and radio positions, 0\farcs19 from the CSC~2.1 position (1.6$\sigma$) and 0\farcs32 from the VLA position (\S\ref{sec:rapid_burster_discussion}). A systematic treatment incorporating the MAVERIC radio-continuum catalog, X-ray-to-infrared flux ratios, and spectral classification of the counterparts will be presented in a forthcoming paper.

\begin{deluxetable*}{lcccccccc}
\tabletypesize{\scriptsize}
\tablecaption{Infrared variables with \textit{Chandra} X-ray counterparts (2$\sigma$ positional match; \S\ref{sec:xray_cross}).\label{tab:xray}}
\tablewidth{0pt}
\tablehead{\colhead{ID} & \colhead{R.A.} & \colhead{Decl.} & \colhead{$\sigma_\mathrm{pos}$} & \colhead{Type} & \colhead{$P$} & \colhead{\textit{Chandra} source} & \colhead{Catalog} & \colhead{Sep., $n\sigma$} \\
 & \colhead{(deg)} & \colhead{(deg)} & \colhead{(mas)} & & \colhead{(min)} & & & \colhead{(\arcsec)}}
\startdata
L1~688 & 263.353016 & -33.398265 & 3.9 & accreting binary & 204.1$^{\rm LS}$ & 2CXO~J173324.7-332353 & CSC~2.1 & 0.03, 0.2 \\
L1~577 & 263.369542 & -33.396477 & 3.7 & X-ray binary & 176.7$^{\rm LS}$ & 2CXO~J173328.7-332347 & CSC~2.1 & 0.04, 0.3 \\
L1~530 & 263.333482 & -33.398162 & 3.4 & EB & 229.6$^{\rm LS}$ & 2CXO~J173320.0-332353 & CSC~2.1 & 0.06, 0.4 \\
L1~1111 & 263.338071 & -33.380412 & 3.7 & PCEB & 103.2$^{\rm LS}$ & 2CXO~J173321.1-332249 & CSC~2.1 & 0.07, 0.5 \\
L1~587 & 263.352446 & -33.388954 & 7.2 & X-ray binary & 705.2$^{\rm LS}$ & 2CXO~J173324.5-332320 & CSC~2.1 & 0.19, 1.6 \\
L1~980 & 263.351850 & -33.388055 & 6.0 & uncertain & 223.6$^{\rm LS}$ & 2CXO~J173324.4-332317 & CSC~2.1 & 0.21, 1.3 \\
T5~216 & 267.017073 & -24.777974 & 3.2 & detached & 732.3 & 2CXO~J174804.0-244640 & CSC~2.1 (+M) & 0.00, 0.0 \\
T5~185 & 267.025713 & -24.771637 & 3.5 & uncertain & \nodata & 2CXO~J174806.1-244617 & CSC~2.1 (+M) & 0.01, 0.1 \\
T5~215 & 267.011054 & -24.767438 & 6.8 & accreting binary & 494.8$^{\rm LS}$ & 2CXO~J174802.6-244602 & CSC~2.1 (+M) & 0.01, 0.1 \\
T5~362 & 267.025127 & -24.771662 & 4.3 & PCEB & 64.2$^{\rm LS}$ & CXOU~J174806.05-244617.6 & MAVERIC & 0.01, 0.1 \\
T5~324 & 267.020983 & -24.778151 & 6.5 & single-transit & \nodata & 2CXO~J174805.0-244641 & CSC~2.1 (+M) & 0.01, 0.1 \\
T5~145 & 267.026311 & -24.777182 & 3.5 & detached & 421.4 & 2CXO~J174806.3-244637 & CSC~2.1 (+M) & 0.01, 0.1 \\
T5~144 & 267.022401 & -24.782393 & 3.5 & contact & 588.8 & 2CXO~J174805.3-244656 & CSC~2.1 (+M) & 0.01, 0.1 \\
T5~377 & 267.025803 & -24.778594 & 8.2 & EB & 479.6$^{\rm LS}$ & 2CXO~J174806.1-244642 & CSC~2.1 (+M) & 0.02, 0.2 \\
T5~323 & 267.016313 & -24.777362 & 4.0 & uncertain & 178.6$^{\rm LS}$ & 2CXO~J174803.9-244638 & CSC~2.1 (+M) & 0.03, 0.2 \\
T5~280 & 267.025360 & -24.773451 & 3.7 & detached & 288.2 & 2CXO~J174806.0-244624 & CSC~2.1 (+M) & 0.03, 0.2 \\
T5~226 & 267.021756 & -24.779917 & 3.4 & detached & 1130.7 & 2CXO~J174805.2-244647 & CSC~2.1 (+M) & 0.03, 0.3 \\
T5~170 & 267.023545 & -24.778429 & 3.3 & detached & 908.0 & 2CXO~J174805.6-244642 & CSC~2.1 (+M) & 0.04, 0.3 \\
T5~224 & 267.018269 & -24.784436 & 4.6 & accreting binary & 415.0$^{\rm LS}$ & 2CXO~J174804.3-244703 & CSC~2.1 (+M) & 0.04, 0.3 \\
T5~348 & 267.006886 & -24.796721 & 4.1 & uncertain & 57.8$^{\rm LS}$ & 2CXO~J174801.6-244748 & CSC~2.1 & 0.04, 0.3 \\
T5~127 & 267.038787 & -24.793756 & 3.7 & accreting binary & 100.3$^{\rm LS}$ & 2CXO~J174809.3-244737 & CSC~2.1 & 0.04, 0.3 \\
T5~201 & 267.014877 & -24.780487 & 4.9 & semi-detached & 579.1 & 2CXO~J174803.5-244649 & CSC~2.1 (+M) & 0.04, 0.4 \\
T5~179 & 267.017768 & -24.773821 & 3.4 & detached & 439.2 & 2CXO~J174804.2-244625 & CSC~2.1 (+M) & 0.04, 0.4 \\
T5~154 & 267.016050 & -24.779536 & 3.1 & detached & 526.7 & 2CXO~J174803.8-244646 & CSC~2.1 (+M) & 0.05, 0.4 \\
T5~283 & 267.022503 & -24.777207 & 3.9 & uncertain & 350.0$^{\rm LS}$ & 2CXO~J174805.3-244637 & CSC~2.1 (+M) & 0.06, 0.5 \\
T5~368 & 267.007677 & -24.789577 & 4.5 & uncertain & 664.2$^{\rm LS}$ & 2CXO~J174801.8-244722 & CSC~2.1 (+M) & 0.06, 0.5 \\
T5~234 & 267.022282 & -24.775520 & 3.4 & contact & 350.4 & 2CXO~J174805.3-244631 & CSC~2.1 (+M) & 0.06, 0.5 \\
T5~379 & 267.013776 & -24.775699 & 8.2 & \nodata & 142.9$^{\rm LS}$ & 2CXO~J174803.3-244632 & CSC~2.1 (+M) & 0.06, 0.5 \\
T5~4 & 267.000934 & -24.784767 & 5.1 & single-transit & \nodata & 2CXO~J174800.2-244705 & CSC~2.1 & 0.06, 0.3 \\
T5~189 & 267.033927 & -24.799388 & 4.0 & accreting binary & 103.8$^{\rm LS}$ & 2CXO~J174808.1-244757 & CSC~2.1 & 0.07, 0.6 \\
T5~141 & 267.017546 & -24.773491 & 3.6 & accreting binary & 111.1$^{\rm LS}$ & 2CXO~J174804.2-244624 & CSC~2.1 (+M) & 0.07, 0.6 \\
T5~193 & 267.017588 & -24.784288 & 3.1 & semi-detached & 473.6 & CXOU~J174804.23-244703.1 & MAVERIC & 0.08, 0.8 \\
T5~217 & 267.022603 & -24.773441 & 3.2 & contact & 455.3 & 2CXO~J174805.4-244624 & CSC~2.1 (+M) & 0.09, 0.7 \\
T5~279 & 266.999998 & -24.792866 & 5.6 & \nodata & 467.8$^{\rm LS}$ & 2CXO~J174759.9-244734 & CSC~2.1 & 0.09, 0.7 \\
T5~285 & 267.020932 & -24.772265 & 3.6 & uncertain & 588.2$^{\rm LS}$ & CXOU~J174805.04-244619.7 & MAVERIC & 0.10, 0.7 \\
T5~272 & 267.015640 & -24.778525 & 3.3 & detached & 276.1 & CXOU~J174803.77-244642.4 & MAVERIC & 0.10, 1.0 \\
T5~260 & 267.020629 & -24.780375 & 3.3 & semi-detached & 530.0 & CXOU~J174805.00-244648.7 & MAVERIC & 0.11, 1.2 \\
T5~344 & 267.009365 & -24.777155 & 3.9 & redback & 108.0$^{\rm LS}$ & CXOU~J174802.26-244637.5 & MAVERIC & 0.13, 1.1 \\
T5~288 & 267.017633 & -24.783595 & 9.6 & uncertain & \nodata & 2CXO~J174804.2-244701 & CSC~2.1 (+M) & 0.14, 1.1 \\
T5~269 & 267.021766 & -24.780474 & 3.3 & detached & 413.5 & CXOU~J174805.19-244649.8 & MAVERIC & 0.15, 1.6 \\
T5~58 & 267.024168 & -24.767627 & 4.3 & contact & 383.2 & 2CXO~J174805.7-244603 & CSC~2.1 (+M) & 0.19, 1.3 \\
T5~197 & 267.018130 & -24.779090 & 3.3 & detached & 529.0 & CXOU~J174804.36-244644.2 & MAVERIC & 0.19, 1.8 \\
T5~257 & 267.020796 & -24.765647 & 3.5 & semi-detached & 274.6 & CXOU~J174804.99-244556.0 & MAVERIC & 0.19, 1.1 \\
T5~156 & 267.022808 & -24.780435 & 3.0 & contact & 946.3 & CXOU~J174805.48-244649.2 & MAVERIC & 0.21, 1.5 \\
T5~104 & 267.020339 & -24.782364 & 4.2 & single-transit & \nodata & CXOU~J174804.88-244656.1 & MAVERIC & 0.22, 1.9 \\
T5~341 & 267.020542 & -24.782739 & 4.4 & UCXB/IP & 20.9$^{\rm LS}$ & CXOU~J174804.96-244657.5 & MAVERIC & 0.25, 1.7 \\
T5~241 & 267.029616 & -24.766181 & 3.3 & uncertain & \nodata & CXOU~J174807.06-244558.7 & MAVERIC & 0.25, 1.1 \\
T5~176 & 267.015886 & -24.783359 & 3.5 & single-transit & \nodata & 2CXO~J174803.7-244700 & CSC~2.1 & 0.28, 1.6 \\
\enddata
\tablecomments{The nearest catalog variable inside each X-ray source's frame-aligned 2$\sigma$ positional region (\S\ref{sec:xray_cross}), for CSC~2.1 (both clusters) and the deeper MAVERIC ACIS catalog (Terzan~5). ``Catalog'' gives the matched catalog; ``(+M)'' marks sources matched independently in MAVERIC as well as CSC~2.1. ``Sep.'' is the frame-corrected separation to the quoted \textit{Chandra} source and ``$n\sigma$'' its position within that source's error region. Positional uncertainty $\sigma_\mathrm{pos}$ is the mean of $\sigma_\alpha,\sigma_\delta$ from Table~\ref{tab:catalog}. ``Type is as in Table~\ref{tab:catalog}, with ``\nodata for an unclassified variable. A period marked LS is the Lomb--Scargle peak of a source without an adopted orbital period (Table~\ref{tab:catalog}) and is not a claimed periodicity; for the Rapid Burster in particular the variability is aperiodic (\S\ref{sec:rapid_burster_discussion}). Identifications secured by radio timing or radio positions (Terzan~5~A, the Rapid Burster) are discussed in \S\ref{sec:psrj1748} and \S\ref{sec:rapid_burster_discussion}.}
\end{deluxetable*}

\subsection{The Rapid Burster}\label{sec:rapid_burster_discussion}

The Rapid Burster is a unique neutron-star low-mass X-ray binary, as only one other known source in the Galaxy shares its type~II (accretion-instability) X-ray bursts \citep{Court2018}. Its counterpart has been an open problem since the source's 1976 discovery \citep{Lewin1976}. Five decades of X-ray study and repeated optical/infrared searches \citep{Homer2001} were defeated by the crowding and extinction toward the core of Liller~1. Recently, \citet{Pallanca2025} proposed a candidate counterpart from HST and Gemini adaptive-optics imaging, a star showing $\sim$0.4--0.5~mag optical variability, while noting that a firm association would require orbital-period or time-resolved infrared information. Our time-series photometry provides exactly that test, and it resolves the question.

\begin{figure*}
\centering
\includegraphics[width=\textwidth]{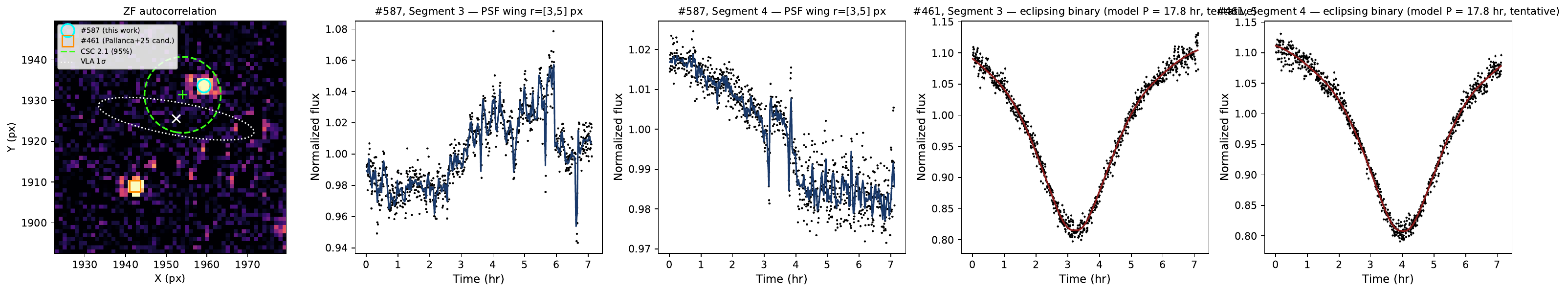}
\caption{Identification of the infrared counterpart of the Rapid Burster (MXB~1730$-$335) in Liller~1. \textit{Left:} zeroframe autocorrelation image (F200W, Segment~3) around the X-ray position, showing our counterpart \#587 (cyan circle), the \citet{Pallanca2025} candidate, source \#461 of our catalog (orange square), the \textit{Chandra} Source Catalog 2.1 95\%-confidence localization \citep[green dashed circle;][]{Evans2024}, and the VLA radio position with its 1$\sigma$ uncertainty \citep[white dotted ellipse;][]{vandenEijnden2024}. Both sources are strong variables, but only \#587 lies within the X-ray error region. \textit{Center:} F200W lightcurves of \#587 from Segments~3 and~4, extracted from the PSF wings to avoid the saturated core; the blue curve is a 2-minute median profile. The variability is aperiodic, with hour-timescale flux changes and an apparent state change in Segment~4, consistent with the source's known X-ray behavior. \textit{Right:} lightcurves of \#461, a likely contact binary and the origin of the optical variability behind the earlier candidacy, with our joint two-segment \texttt{PHOEBE} model overplotted (red). A dedicated analysis of the counterpart is presented in a companion paper \citep{Desai2026}.}
\label{fig:rapid_burster}
\end{figure*}

The \citet{Pallanca2025} candidate is source \#461 of our catalog (coincident to 0\farcs08): an eclipsing binary of two main-sequence stars with a 17.8-hour photometric orbital period, likely a contact binary \citep{Desai2026}, which naturally explains its strong optical variability (Figure~\ref{fig:rapid_burster}). The true counterpart is \#587, a distinct variable 0\farcs92 to its southwest. The astrometry favors \#587. The CSC~2.1 position of the Rapid Burster \citep[2CXO~J173324.5$-$332320, by far the brightest X-ray source in the field;][]{Evans2024} lies 0\farcs19 from \#587 but 0\farcs98 from \#461, which places \#461 outside the 0\farcs29 95\%-confidence ellipse, and the VLA radio position \citep{vandenEijnden2024} likewise prefers \#587 (0\farcs32 versus 0\farcs60 for \#461). Source \#587 shows just the behavior expected of the Rapid Burster, namely aperiodic, hour-timescale, accretion-driven flux changes, and it does so more violently than any other source in the Liller~1 field: its pixels show the largest flux changes of any on the detector, and among the 915 detected variables it is the single brightest source in the F356W zeroframe variability (autocorrelation) image of the field. We thus identify \#587 as the counterpart. The companion paper \citep{Desai2026} reaches the same conclusion independently, providing a Gaia-tied localization, resolving the counterpart from a blended red-giant neighbor with two-source PSF photometry, and reporting a fade of more than a magnitude between segments with flaring on minute timescales.

Our JWST observations detect clear infrared variability at the \textit{Chandra} position of the Rapid Burster (Figure~\ref{fig:rapid_burster}). Because the source's core pixels are heavily saturated in NIRCam, we extract its lightcurve from the wings of the point spread function using the annular-aperture special reduction of \S\ref{sec:special}. The resulting lightcurves show variability on timescales of hours, including an apparent state change in Segment~4, consistent with the known X-ray behavior. The source is clearly detected in the zeroframe autocorrelation image (Figure~\ref{fig:rapid_burster}, left panel), confirming its intrinsic variability. These observations (this work and \citealt{Desai2026}) represent the first detection of infrared variability from the Rapid Burster, and the direct test that \citet{Pallanca2025} identify as required to confirm or refute their candidate. 

\subsection{The infrared counterpart of \texorpdfstring{PSR~J1748$-$2446A}{PSR J1748-2446A}}\label{sec:psrj1748}

Terzan~5 hosts a population of millisecond pulsars (MSPs) richer than that of any Galactic globular cluster at the time of writing, with 49 confirmed through radio timing \citep{Ransom2005,Hessels2006,Padmanabh2024}. Among these is PSR~J1748$-$2446A (also known as Terzan~5~A or PSR~B1744$-$24A), only the second eclipsing binary pulsar known at the time of its discovery in 1990 and the first detected member of the ``redback'' class \citep{Lyne1990}, in which the pulsar wind irradiates and ablates a low-mass companion. The class was later defined and reviewed by \citet{Roberts2013}. Terzan~5~A is the brightest of the cluster's pulsars, with an 11.56~ms spin period and the shortest orbital period of any known redback \citep[$P_\mathrm{orb} = 108.93$~min;][]{Rosenthal2025}. Its radio eclipses are variable and longer than expected for a Roche-lobe-confined companion, and at times its radio pulsations become undetectable for hours \citep{Rosenthal2025}.

Despite precise radio timing positions and 35 years of effort since Terzan~5~A's discovery, no optical or infrared counterpart to any of Terzan~5's 49 MSPs has been unambiguously identified, owing to the extreme stellar crowding ($\rho_0 \approx 4\times10^6\,M_\odot\,\mathrm{pc}^{-3}$; \citealt{Lanzoni2010}) and heavy reddening. Even in X-rays, the soft thermal emission of MSPs is strongly attenuated by the high absorbing column toward the cluster \citep{Bogdanov2021}. Every MSP companion identified in any globular cluster to date, in 47~Tucanae, NGC~6397, NGC~6752, M71, and others, has been found at optical or near-ultraviolet wavelengths in clusters of low to moderate extinction \citep{Edmonds2001,Edmonds2002,Cadelano2015}. 

\begin{figure*}
\centering
\includegraphics[width=\textwidth]{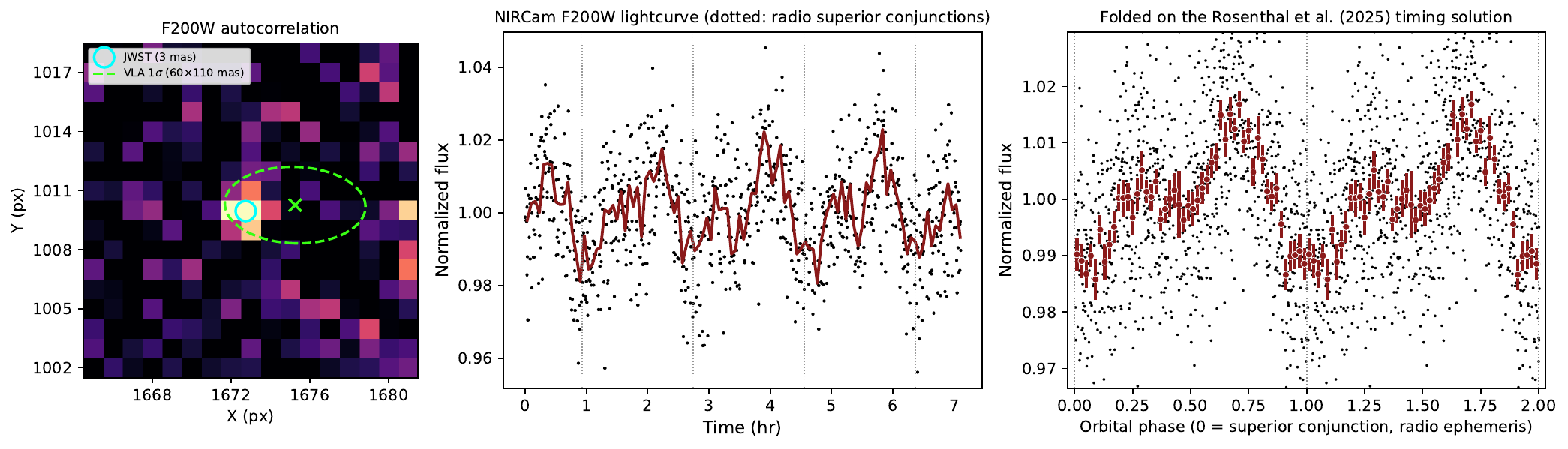}
\caption{First unambiguous optical or infrared counterpart of PSR~J1748$-$2446A (Terzan~5~A, the first redback ever detected; \citealt{Lyne1990}), identified 35 years after the pulsar's discovery. \textit{Left:} F200W autocorrelation image around the pulsar, with our JWST centroid (cyan circle) and the $1\sigma$ VLA radio position \citep[dashed green ellipse;][]{Urquhart2026}. \textit{Center:} unfolded NIRCam F200W lightcurve, with the orbital modulation visible directly; the dark red line is a 4-minute median profile. Dotted lines mark the radio-predicted superior conjunctions, which coincide with the flux minima. \textit{Right:} lightcurve folded on the \citet{Rosenthal2025} radio timing solution ($P_\mathrm{orb} = 108.93$~min), with phase 0 at the pulsar's superior conjunction (red points: means in 50 phase bins, with error bars giving the standard error of the mean); our blind search independently recovered $P = 108.0$~min. The flux minimum falls near the pulsar's superior conjunction, and the phase connection to the radio ephemeris is good to $\Delta\phi \approx 0.003$ (\S\ref{sec:timing}). None of Terzan~5's 49 pulsars had a known optical or infrared counterpart before this identification, which is also the first infrared detection of a millisecond pulsar companion in any globular cluster.}
\label{fig:psrj1748}
\end{figure*}

Our period search identifies a variable source (\#344) at the radio timing position of PSR~J1748$-$2446A with a Lomb--Scargle period of $108.0$~min, within $\sim$1\% of the precisely known orbital period of $108.9$~min from pulsar timing. Figure~\ref{fig:psrj1748} shows the lightcurve phase-folded on the full radio timing solution of \citet{Rosenthal2025} (orbital period, period derivative, and epoch of ascending node), propagated 89,507 orbits from their reference epoch to ours, with orbital phase zero defined at the pulsar's superior conjunction (the center of the radio eclipses). Because our timestamps are barycentric mid-exposures (\S\ref{sec:timing}) and the ephemeris wanders by $\lesssim$10~s \citep{Rosenthal2025}, the phase connection is accurate to $\Delta\phi \approx 0.003$. The infrared modulation matches the radio ephemeris in phase as well as in period: the flux minimum falls near the pulsar's superior conjunction ($\phi \approx 0.09$), and the maximum near $\phi \approx 0.7$. The modulation is detected only in F200W. The simultaneous F356W lightcurve shows no significant signal at the orbital period: a sinusoid at the radio period has a best-fit amplitude of only 0.3\% there, against 0.8\% in F200W at a similar per-point scatter, and the F356W periodogram shows no peak at that period. The long-wavelength aperture covers four times the sky area of the short-wavelength one, so it collects correspondingly more light from blended neighbors, and this dilution can suppress the signal below detection. Even with a detection in both bands, the uncertain and band-dependent dilution would prevent an amplitude comparison precise enough to separate irradiation, intrabinary-shock emission and ellipsoidal distortion, so we do not attribute the modulation to physical components, and we treat its amplitude as a lower limit. At our epoch the modulation is phase-locked to the pulsar clock predicted from three decades of radio timing, and this agreement alone, independent of any physical interpretation, is definitive evidence that the infrared source is the pulsar's companion. This blind period recovery constitutes the same gold-standard identification, periodicity matching the radio ephemeris, used for the first MSP companion detections in 47~Tucanae \citep{Edmonds2001,Edmonds2002}, and is far stronger evidence than positional coincidence alone. This is the first unambiguous optical or infrared counterpart identified for any of Terzan~5's pulsars (X-ray counterparts to several, including Ter5~A itself, were identified by \citealt{Bogdanov2021}), and the first infrared identification of an MSP companion in any globular cluster. It is the companion of the prototype redback, identified 35 years after the pulsar's discovery.

This companion stood out in our blind variability search. The companions of Terzan~5's other pulsars did not, most likely because they are considerably fainter. Rather than blind detection, a targeted search that folds each source at its precisely known radio ephemeris is the natural route to those fainter companions, and will be pursued in a dedicated accompanying paper.

This identification also inverts the usual discovery sequence for spider binaries, in which a radio pulsation search normally comes first and the counterpart second. Our blind period search recovered the orbit of PSR~J1748$-$2446A with no radio input whatsoever. Had the pulsar been unknown, source \#344 would still have stood out as a $\sim$1.8-hour periodic variable, a natural target for a pulsation search. In the Galactic center, no millisecond pulsar has yet been detected despite decades of searching, a ``missing pulsar problem'' whose proposed explanations, from extreme interstellar scattering to an intrinsic deficit of ordinary pulsars, remain debated \citep{DexterOLeary2014}. The same absence leaves open whether the Galactic center's diffuse GeV gamma-ray excess arises from dark matter annihilation \citep[e.g.,][]{Daylan2016} or from an unresolved MSP population \citep[e.g.,][]{Bartels2016}. Infrared time-domain surveys of the sort demonstrated here can identify candidate spider companions directly through their orbital variability, converting blind radio pulsation searches into targeted ones at known positions and orbital periods, including at frequencies and epochs chosen to defeat scattering. The same strategy applies to Liller~1 itself, whose gamma-ray luminosity implies a hidden MSP population \citep{Tam2011} despite zero radio detections, and to the nuclear star cluster.

\subsection{Prospects for planetary transits}\label{sec:planets}

The question of whether planets exist in globular clusters remains an open problem at the intersection of planetary science and stellar dynamics. The landmark HST transit search by \citet{Gilliland2000} monitored $\sim$34,000 stars in 47~Tucanae for 8.3~days and found zero transiting hot Jupiters, whereas $\sim$17 were expected based on field-star occurrence rates. A ground-based campaign by \citet{Weldrake2005} confirmed this null result in 47~Tuc's outer regions. These results have been widely interpreted as evidence that the globular cluster environment suppresses planet formation or survival.

However, \citet{Masuda2017} reassessed the 47~Tuc null result using the hot-Jupiter occurrence rates and planet radius distributions measured by \textit{Kepler}, showing that the expected yield drops from 17 to only $\sim$2--4 detections, rendering the null result far less constraining than originally believed, while noting that the available data did not permit controlling for 47~Tuc's low metallicity ([Fe/H]~$\approx -0.7$).

Liller~1 and Terzan~5 offer a fundamentally different test of this question. Both clusters harbor metal-rich stellar populations with [Fe/H]~$\approx +0.3$ \citep{Ferraro2009,Ferraro2021,Origlia2011,Crociati2023}, super-solar metallicities at which the giant-planet occurrence rate in the field is 3--5$\times$ higher than at solar \citep{FischerValenti2005}, with a steeper dependence still for hot Jupiters \citep{Petigura2018}. By the $\sim$$10^{2[\mathrm{Fe/H}]}$ giant-planet occurrence scaling \citep{FischerValenti2005}, these are thus the most favorable globular cluster environments for hot Jupiter occurrence ever surveyed for transits, and the first in which the metallicity-dependent occurrence rate favors rather than disfavors a detectable population. The expected yield, however, requires a quantitative calculation with injection--recovery tests, which we defer to a dedicated transit-search paper. The community has begun to pursue exactly this question from the ground. The MISHAPS survey is conducting a dedicated wide-field transit search for hot Jupiters in globular clusters, with first results for 47~Tucanae \citep{Crisp2025}. Recent dynamical modelling of 47~Tucanae finds high-eccentricity migration to be \emph{suppressed} at high stellar density, with ionization and tidal disruption outpacing hot Jupiter formation, a result \citet{Wirth2025} invoke to explain the absence of hot Jupiters in that cluster's core. No clusters have higher encounter rates than Liller~1 and Terzan~5 \citep{VerbuntHut1987,Saracino2015}.

\begin{figure*}
\centering
\includegraphics[width=\textwidth]{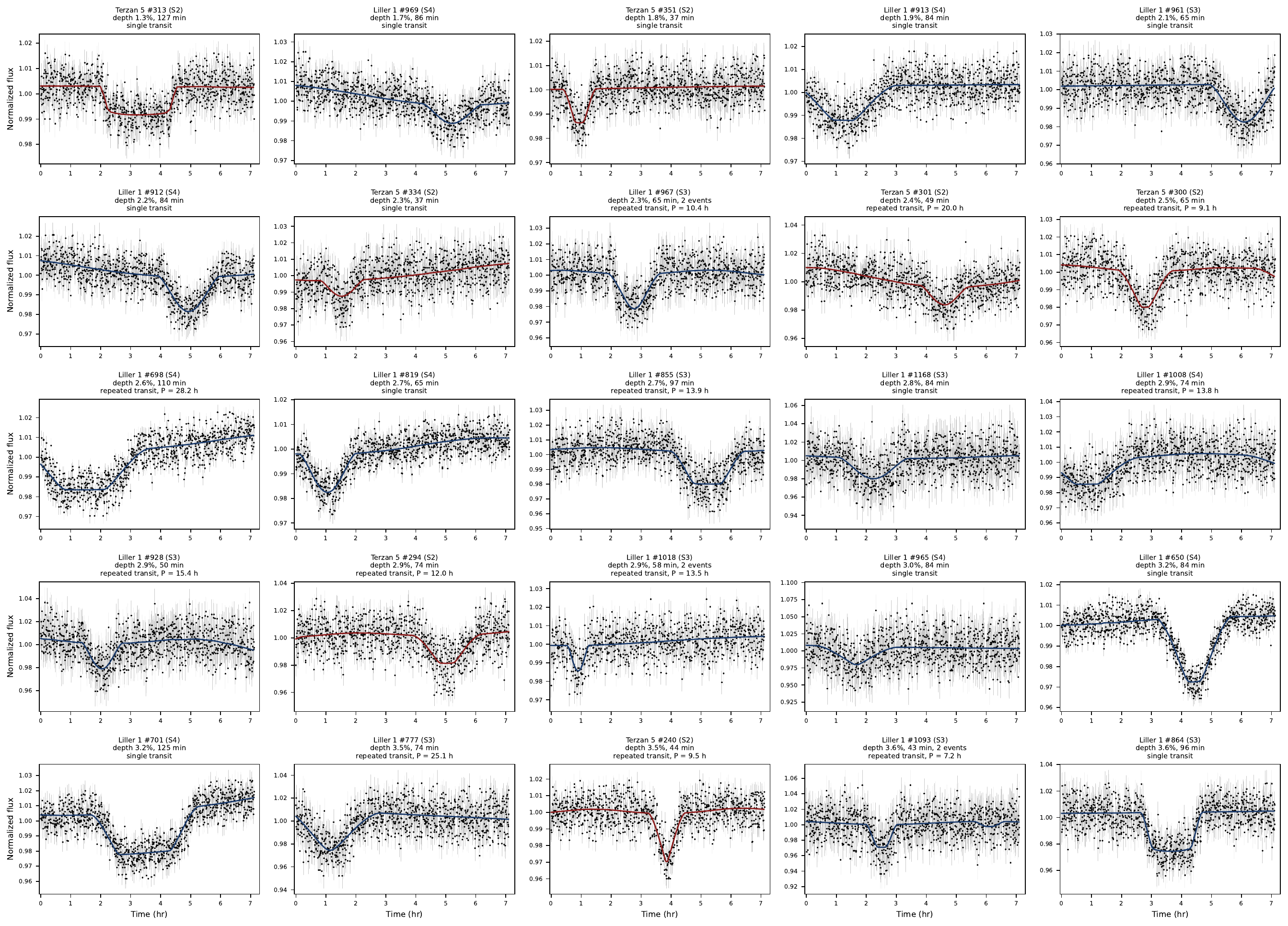}
\caption{The 25 shallowest transit-like candidates from our sliding-box search that passed visual inspection (\S\ref{sec:singletx}) and carry an accepted \texttt{PHOEBE} model, ordered by event depth (1.3--3.6\%, measured on the catalog lightcurves and therefore lower bounds with respect to crowding dilution, \S\ref{sec:amplitudes}). Each panel shows the adopted lightcurve of the segment in which the event was detected (black points, 21-s cadence, with gray 1$\sigma$ error bars from the photometric error model of \S\ref{sec:photerr}) with the model overplotted (dark red for Terzan~5, dark blue for Liller~1). Labels give the catalog identifier, the event depth and duration, and whether the transit is single or repeats (13 single, 12 repeated, the latter with the period of the repetition). A transit is labelled repeated when the source's accepted model carries a period the data constrain, i.e., when the source is not classed as single-transit (\S\ref{sec:classification}), and the quoted period is that of the model. For Liller~1 the period rests on Segments~3 and~4 together, so the other event may fall in the visit not shown and a panel with a single dip can still be labelled repeated. For Terzan~5 the shape is set by Segment~2, and the dithered Segment~1 enters the model fit only to refine the period. An event count, where given, is the number of events for that source that passed visual inspection. We do not name the occulting body, because a transit that recurs is equally consistent with a planet and with a grazing or diluted eclipsing binary. The adopted classes, from the model fits, are given in the atlas of Appendix~\ref{app:atlas}. As discussed in \S\ref{sec:planets}, blends and grazing eclipsing binaries are expected to dominate a sample of this kind.}
\label{fig:transit_candidates}
\end{figure*}

Our sliding-box single-event search identified dozens of shallow transit-like signals in both clusters (Figure~\ref{fig:transit_candidates}). The binned per-point precision of these lightcurves lies well below typical hot-Jupiter transit depths ($\sim$1\%), and the reduced limb darkening at 2~\micron\ produces sharper transit ingress and egress. JWST has the precision to detect planetary transits in these environments, should they occur. Follow-up spectroscopy to rule out large radial velocity variations is needed to distinguish planetary transits from grazing eclipsing binaries.

In fields this crowded, blends and grazing eclipsing binaries are expected to dominate any sample of shallow transit-like signals, and we treat our detections as candidate signals only. The stakes of confirmation are high, however, because the only confirmed planet in any globular cluster remains PSR~B1620$-$26b in M4 \citep{Sigurdsson2003}, a $\sim$2.5~$M_\mathrm{Jup}$ planet in a wide orbit around a millisecond pulsar triple system. A second candidate, the low-mass companion to PSR~J1701$-$3006H in M62, has a minimum mass in the planetary range but is not confirmed as a planet \citep{Vleeschower2024}. No planet has yet been found orbiting a main-sequence star in a globular cluster, transiting or otherwise. A confirmed detection in the metal-rich populations of Liller~1 or Terzan~5 would thus be a first, and would reshape our understanding of planet formation and survival in dense stellar environments.

\subsection{Methodological advances}\label{sec:method_advances}

The techniques developed in this work have applicability well beyond Liller~1 and Terzan~5, and several have no published precedent.

\begin{figure*}
\centering
\includegraphics[width=\textwidth]{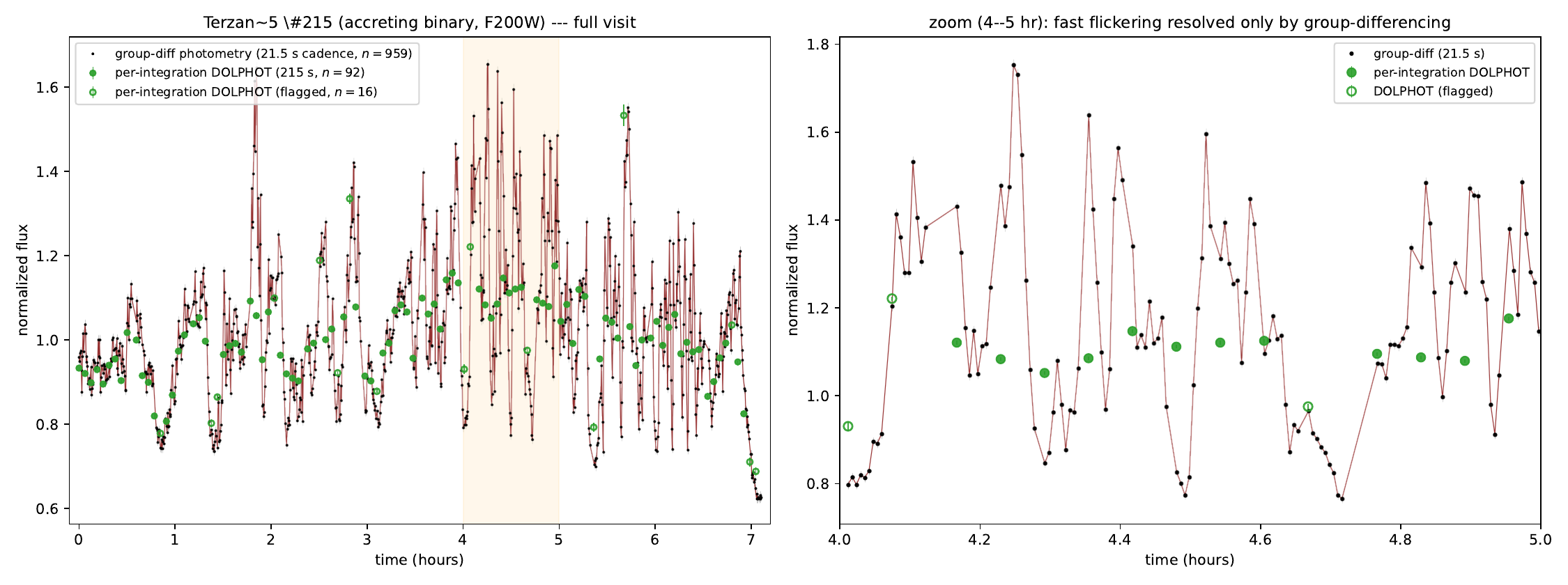}
\caption{Temporal information that group-level photometry recovers but per-integration products erase, for the candidate accreting binary Terzan~5~\#215 (F200W, Segment~2). \textit{Left:} the full $\sim$7-hour visit. \textit{Right:} a zoom on hours 4--5. Black points are our group-differenced lightcurve at the native 21.5~s cadence, connected by a line in the cluster color, with gray 1$\sigma$ error bars from the photometric error model (\S\ref{sec:photerr}). Green points are per-integration PSF photometry of the same source from \texttt{DOLPHOT} (215~s cadence, one point per integration), with $1\sigma$ uncertainties, and flagged integrations shown as open circles. The group cadence resolves a dense train of $\sim$2--3-minute accretion flares reaching nearly twice the median flux, whereas the per-integration photometry, averaging over $\sim$10 groups per point, registers only a smooth envelope and erases the flickering that is the physical signature of the accretion.}
\label{fig:group_vs_integration}
\end{figure*}

Differencing consecutive non-destructive reads has an established lineage in HST/WFC3 spatial-scan spectroscopy \citep{Deming2013}, and sub-exposure time-domain science from non-destructive reads has been proposed for the Roman Space Telescope \citep{TovarMendoza2023}. On JWST itself, group-by-group difference images have been used to sense mirror tilt events in defocused NIRCam commissioning imaging \citep{Schlawin2023}, and a group-level MIRI lightcurve resolved a flare of the X-ray binary V404~Cygni \citep{Borowski2025}, in both cases for a single target. However, converting standard JWST imaging into a uniform, sub-integration-cadence photometric survey of every source in the field, 972 frames at 21.5~s over a full NIRCam module, has not previously been demonstrated. Read-level differencing is also what separates our method from JWST's own time-series observation mode, whose standard products fit the ramp and retain no group-level information (\S\ref{sec:intro}). This is the fastest-cadence wide-field photometry yet published with JWST (faster sampling has been achieved only in small subarrays on single bright targets, e.g., the Sgr~A* monitoring of \citealt{YusefZadeh2025}), and it comes at zero cost in observing efficiency, as any sufficiently sampled NIRCam imaging program produces these data automatically. Retaining all group-level reads stressed JWST's downlink capacity rather than its observing time, and dictated our single-module field and 7-hour visit lengths (\S\ref{sec:observations}), a constraint that future programs in this mode, and future missions contemplating read-level downlink, should weigh explicitly. Slow variability, such as eclipses lasting hours, is also recovered at the 215-s integration cadence, and for it the group cadence mainly provides finer sampling. Fast, aperiodic variability is different: the group cadence resolves structure on timescales shorter than an integration that the per-integration products average away. Figure~\ref{fig:group_vs_integration} illustrates this with the candidate accreting binary Terzan~5~\#215, whose accretion flickering consists of tens-of-percent flux swings on a characteristic $\sim$2--3~minute timescale (autocorrelation $1/e$ time 2.5~min). The flickering is fully resolved in our group-differenced lightcurve, whereas in standard per-integration DOLPHOT PSF photometry of the same source each 215~s integration spans more than a full flare cycle, reducing it to a slow, low-amplitude envelope. The same 214.7~s ramps that yield a single calibrated image per integration contain, in their intermediate reads, an order of magnitude more temporal information.

The lag-1 autocorrelation is a well-known variability statistic for individual lightcurves \citep{Sokolovsky2017}, and per-pixel \emph{variance} images have been used to detect pulsars in radio imaging \citep{Dai2016}, but we are not aware of a per-pixel lag-1 temporal autocorrelation image having been used before as the primary variability \emph{detection} image in an astronomical survey. We adopt it because it is brightness independent (\S\ref{sec:detection}), so that faint variables are not buried by the bright constant stars that defeat variance-based detection in fields spanning a $10^5$-fold range in brightness. We can only build this image because we stare without dithering, since a pixel accumulates the uninterrupted time series the autocorrelation requires only if it watches the same piece of sky for the entire visit.

The saturation treatment is also worth placing in context. Our per-pixel ratio-model saturation correction extends the long lineage of saturated-star photometry (bleed-trail photometry on HST CCDs \citep{Gilliland2004}, halo photometry on Kepler/K2 \citep{White2017}, and PSF-wing photometry on Spitzer \citep{Su2022}) to the JWST ramp domain, leveraging the unsaturated early reads of each integration. Combined with PSF-wing extraction for the most extreme cases (\S\ref{sec:special}), it extends time-domain sensitivity across the full dynamic range of the detector.

These techniques also carry lessons for the future, as the Nancy Grace Roman Space Telescope's HgCdTe H4RG-10 detectors are likewise read non-destructively up the ramp \citep{Mosby2020}, and its Galactic Bulge Time Domain Survey will monitor crowded, reddened bulge fields much like ours \citep{Penny2019}. Our pipeline does not transfer wholesale, since Roman's survey will dither between pointings rather than stare as we do, but we think the lessons carry over directly. Read-level differencing can multiply the effective within-exposure cadence of Roman photometry for bright-source science at zero observing-time cost, as we demonstrate here for JWST \citep[cf.][]{TovarMendoza2023}. Because we dithered one visit (Segment~1) and left the other three undithered, we can see both sides of the dithering trade-off in crowded fields: dithering degrades the raw relative photometry substantially, but we recover most of the lost precision with the per-exposure ratio corrections and curvature-aware stitching we develop (\S\ref{sec:seg1_extraction}), so our Segment~1 pipeline is already a working precedent for what Roman's crowded-field time-series photometry will require. Our ramp-based saturation recovery, in turn, preserves the brightest giants, often the most scientifically valuable variables, that standard pipelines discard. Beyond these lessons, our results motivate a complementary observing mode for Roman: dedicated deep, \emph{undithered} stares at fields within the bulge survey footprint and at the Galactic center. Such stares, directly analogous to our visits, would deliver the highest relative-precision photometry the observatory can achieve in crowded fields and open a discovery window on timescales of minutes, shorter than the $\sim$15-minute cadence of the dithered, long-baseline survey \citep{Penny2019}, at modest observing cost and with the analysis techniques demonstrated here. The autocorrelation detection image does not depend on the instrument that produced the pixel time series, so these techniques should carry over to other infrared time-domain surveys.

\section{Conclusion}\label{sec:conclusion}

In this work, we present a framework for JWST time-series photometry in crowded fields. By leveraging uninterrupted, undithered stares with JWST/NIRCam, we extract precision photometry of variable sources in the most dynamically active known globular cluster-like objects in the Galaxy. We develop a pipeline that detects variable stars, extracts their lightcurves from the group-differenced ramp data, and corrects the systematic artifacts introduced by detector nonlinearity and saturation. The pipeline employs lag-1 temporal autocorrelation for brightness-independent variable source detection, and a multi-stage correction framework that adaptively selects among saturation correction, slope correction, and their combination based on empirical quality metrics. Group-level photometry applies to any sufficiently sampled JWST NIRCam imaging observation of any field, however sparse or crowded, because each such observation already carries sub-integration temporal resolution that we extract at zero observing cost. For dithered programs, however, one must first run the correction-and-stitching pipeline we demonstrate on our dithered first visit (\S\ref{sec:seg1_extraction}), which recovers clean variability at reduced but useful precision. In crowded fields, an undithered stare keeps every pixel on the same stars for the entire visit, and the lag-1 autocorrelation image built from those pixel series then picks out the variables regardless of their brightness.

Applying this pipeline to deep JWST observations of the bulge globular cluster-like objects Terzan~5 and Liller~1, we identify 1,315 unique variable sources: 915 in Liller~1 and 400 in Terzan~5. Liller~1, which had zero previously known variables, thereby becomes the globular cluster-like object with the most known variable stars in its field, surpassing the cataloged count of $\omega$~Centauri (460 sources) by nearly a factor of two. The saturation correction, which models per-pixel nonlinearity from the raw uncalibrated ramp data, recovers clean lightcurves for hundreds of bright sources that would otherwise be dominated by banding artifacts. For extremely saturated sources such as the Rapid Burster (MXB~1730$-$335) in Liller~1, we demonstrate that PSF-wing photometry using annular apertures can extract astrophysical variability that is completely inaccessible in the saturated core pixels.

Our observations pave the way for a renewed campaign to study variable star populations in dense environments such as globular clusters and the nuclear star cluster. Of particular interest are the close binary star populations in these regions, which should be heavily influenced by the high rate of dynamical interactions expected in these environments. Our analysis also reveals infrared counterparts to known compact objects in both clusters. In Liller~1, photometry of the PSF wings around a heavily saturated stellar core recovers the long-sought counterpart to the Rapid Burster, which we characterize in detail in a companion paper \citep{Desai2026}. In Terzan~5, we identify the companion of the eclipsing redback PSR~J1748$-$2446A, the first of the cluster's millisecond pulsars to have a counterpart identified at optical or infrared wavelengths, and our blind period search independently recovers its 1.8-hour orbital period, which radio timing had already established precisely \citep{Lyne1990,Rosenthal2025}.

The redback recovery demonstrates the possibility of an inversion of the usual discovery sequence, one in which the orbit is found photometrically first and the pulsar only afterwards, and that inversion already has a track record. \citet{Romani2011,Romani2012,Romani2014} and \citet{Romani2015} identified the orbital periods of several \textit{Fermi} black widow and redback candidates photometrically first, which motivated and constrained the pulsation searches that later confirmed them, in some cases only after years of effort, and \citet{Nieder2020} used optical orbital constraints to make an otherwise intractable gamma-ray pulsation search succeed. What is new in the cluster context is the waveband and the crowding. Liller~1 is the clearest case. Its gamma-ray luminosity implies a population of millisecond pulsars \citep{Tam2011}, none of which has been detected in the radio, probably because of extreme interstellar scattering along the sight line. Optical searches for their companions cannot penetrate the extinction, and blind radio searches are defeated by the scattering, but an infrared orbital ephemeris for a candidate spider companion would allow the same kind of tightly targeted radio or gamma-ray search that has succeeded in the field. The same argument applies to any sight line where scattering defeats a blind radio search, above all the Galactic center. In Terzan~5, where the millisecond pulsars are already known, folding our lightcurves at their radio ephemerides offers a route to the fainter companions that our blind search did not recover.

This work lays the groundwork for time-domain astronomy in the Galaxy's most crowded stellar fields. Every NIRCam integration is built from non-destructive reads, so every NIRCam image already contains a movie, and reading it costs nothing in observing time. The chief lesson of this first attempt is simple. There is enormous value in uninterrupted, undithered stares in crowded fields. Keeping every pixel on the same stars for the whole visit removes the flat-field and sub-pixel systematics that dithering introduces, which is the key ingredient of high-precision relative photometry, and it lets the pixel time series themselves reveal the variables through their temporal autocorrelation, regardless of brightness; dithered data demand a far more involved reconstruction and return less. Applied to under ten square arcminutes of sky, these methods yielded more than a thousand variable stars with lightcurves precise enough to demand physical binary models, in two systems that between them had eleven known variables before. The consequence is that the close binary populations of the most dynamically active stellar systems in the Galaxy are now accessible. Contact and detached binaries, post-common-envelope pairs, the companions of pulsars and accreting compact objects can be found, timed and modelled by the hundreds, and their orbital period distributions, their radial profiles and their abundances relative to the field will show directly how these environments make, harden and destroy compact binaries. Globular clusters, the nuclear star cluster and the Galactic bulge hold the Galaxy's oldest, densest and most dynamically processed stars, and this unprecedented view into the densest regions of the Galaxy may well transform our understanding of how dynamical interactions shape close binary populations.

\begin{acknowledgments}
This work is based on observations made with the NASA/ESA/CSA James Webb Space Telescope. The data were obtained from the Mikulski Archive for Space Telescopes at the Space Telescope Science Institute, which is operated by the Association of Universities for Research in Astronomy, Inc., under NASA contract NAS5-03127 for JWST. These observations are associated with program \#5381. K.B.B. and M.D. acknowledge support from NASA through grant JWST-GO-05381.005-A from the Space Telescope Science Institute, which is operated by the Association of Universities for Research in Astronomy, Inc., under NASA contract NAS5-03127.
F.R.F., B.L., C.P. and G.Z. acknowledge financial support from the project GENESIS – Searching for the primordial structures of the Universe in the heart of the Galaxy (Advanced Grant FIS-2024-02056, PI: Ferraro), funded by the Italian MUR through the Fondo Italiano per la Scienza (FIS3) call.
\end{acknowledgments}

\begin{contribution}

K.B.B. was PI of the JWST program on which this work is based (I.C. was the co-PI), and, making extensive use of Claude Code (Anthropic), developed all custom software, performed all analysis presented here, and wrote the manuscript. The agent worked under his direction throughout, implementing and running pipeline and analysis code and re-deriving quoted quantities directly from the data products, so that statements in the text could be checked against the catalog they describe. Results were sanity-checked in both directions, with K.B.B. reviewing the agent's output and the agent re-testing manuscript claims against the data. The agent was Anthropic's Claude Fable 5.1. In the final revision, OpenAI's GPT-6 Astra, run through Codex as an independent referee, read the manuscript in sections against its tables and figures and returned ranked findings on internal consistency, scope of claims, and clarity, which Fable 5.1 verified against the data products and implemented under K.B.B.'s review. Because the implementation relied so heavily on an agentic tool, we have carefully archived all critical elements underlying this work in public repositories. The complete analysis pipeline is publicly available in a GitHub repository, and the catalog, the intermediate data products, and the human-judgment records the catalog depends on (the REAL/FAKE vetting labels, the manual deduplication groupings, and the per-source record of which \texttt{PHOEBE} fits we accepted) are archived at Zenodo, so that any step can be re-run or independently audited (Appendix~\ref{app:repro}). K.B.B. is responsible for all results presented here. C.P. performed the cross-match of the DOLPHOT catalog to the Zullo et~al.\ photometric catalog to construct the CMDs with variables labeled. M.D. conducted a detailed analysis of the Rapid Burster, and much of that analysis informed the direction of this work. All authors contributed edits/comments on the manuscript.

\end{contribution}

\facilities{JWST(NIRCam), CXO(ACIS)}

\software{astropy \citep{2013A&A...558A..33A,2018AJ....156..123A,2022ApJ...935..167A},
          photutils \citep{photutils},
          scipy \citep{scipy},
          matplotlib \citep{matplotlib},
          WebbPSF \citep{Perrin2014},
          h5py,
          CuPy
          }

\appendix

\section{Correction pipeline internals}\label{app:underhood}

\subsection{The fitted models of the correction pipeline}

\begin{figure*}
\centering
\includegraphics[width=\textwidth]{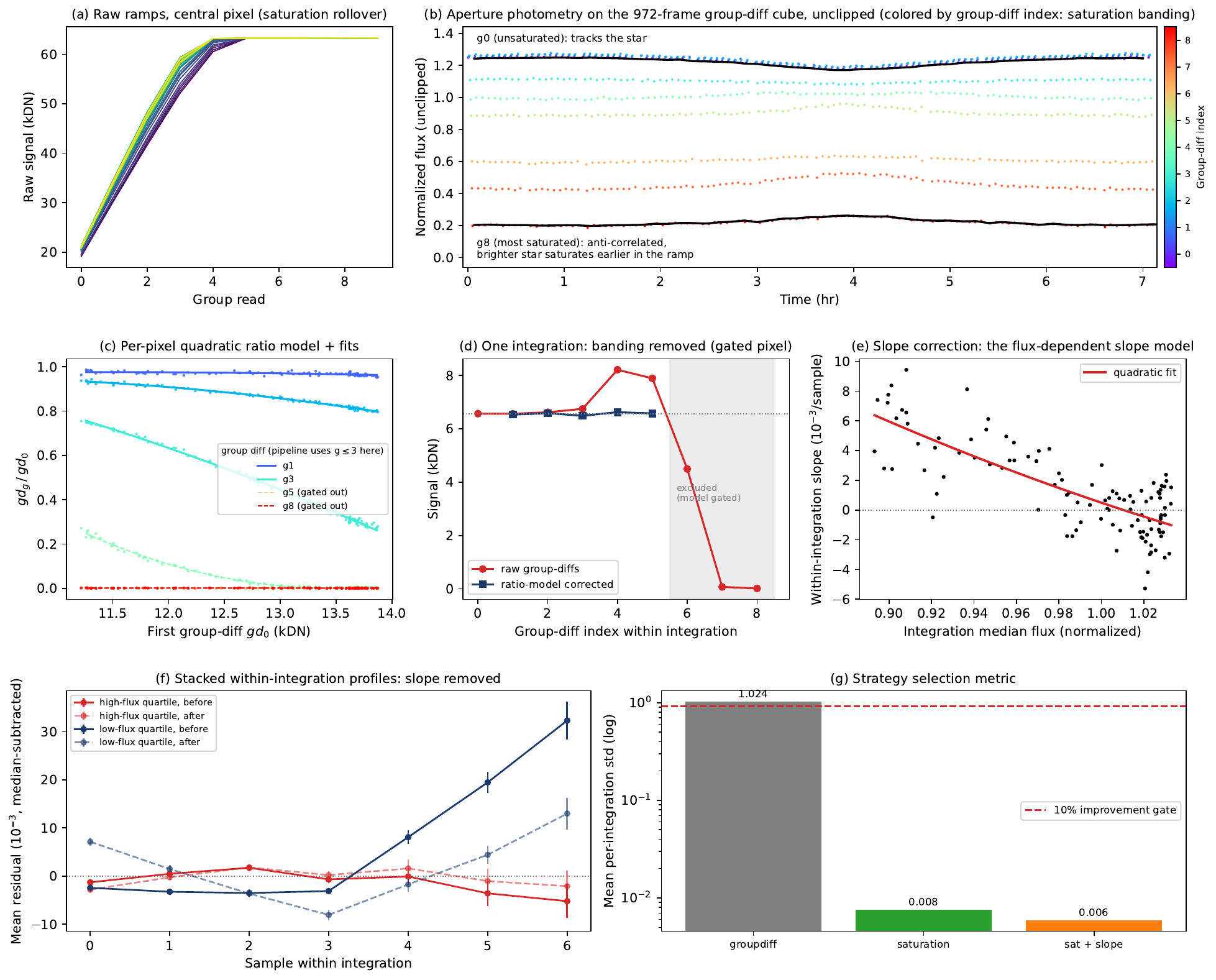}
\caption{The correction pipeline under the hood, on Terzan~5 \#133 (the Figure~\ref{fig:satcorr_sw} example source; Segment~2, nrcb4). \textbf{(a)}~Raw ramps of the central aperture pixel across 25 integrations (colored by flux). Saturation flattens the ramps by the fourth group read. \textbf{(b)}~Unclipped aperture photometry on the group-differenced cube, colored by group-difference index (the colorbar also keys the curves in panel~c). Black curves show the median tracks of the first (top) and last (bottom) differences. The former follows the star and the latter is anti-correlated with it, so the mixture of tracks carries no net signal. \textbf{(c)}~The per-pixel saturation model: group-difference ratios $gd_g/gd_0$ versus $gd_0$ with fitted quadratics, where dashed curves are groups the pipeline gates out rather than corrects (from the first group whose median ratio falls below 3\% or whose fit residual scatter exceeds 5\% of that ratio onward). \textbf{(d)}~One integration of a different, partially saturated aperture pixel (not the central pixel of panels a and c), with its own ratio model and gate (groups up to 5 retained, against up to 3 for the central pixel), before (red) and after (blue) division by the model ratios: the corrected differences return to the level of the unsaturated first difference (dotted line). The shaded region marks gated groups, which the pipeline excludes rather than corrects. \textbf{(e)}~The slope model: within-integration slope versus integration median flux with the fitted quadratic. \textbf{(f)}~Stacked within-integration residual profiles for the brightest and faintest flux quartiles, before (solid) and after (faded dashed) slope subtraction, with error bars giving the standard error of the mean over the stacked integrations. \textbf{(g)}~The integration scatter (log scale) of each correction strategy, with the 10\% improvement gate (dashed). A strategy is adopted only if its integration scatter falls below 90\% of that of the uncorrected lightcurve, and the lowest passing strategy wins.}
\label{fig:underhood}
\end{figure*}

Figure~\ref{fig:underhood} opens up the machinery of the lightcurve correction pipeline (\S\ref{sec:corrections}), using the paper's saturation-correction example source (Terzan~5 \#133, Segment~2, nrcb4; Figure~\ref{fig:satcorr_sw}). Every panel is produced by the same code that produced the catalog, so what is shown is exactly what the pipeline fits.

Panels~(a) through (d) dissect the saturation correction (\S\ref{sec:saturation}). Panel~(a) shows the raw non-destructive ramps of the central aperture pixel. The ramps roll over and saturate by the fourth group read, which is what renders standard slope-fitting photometry unusable for bright sources. Panel~(b) shows what this saturation does to the unclipped aperture photometry on the group-differenced cube: the characteristic banding. Each group-difference index forms its own track, and the tracks are not simply offset. The unsaturated first difference follows the star, while the most saturated difference is \emph{anti-correlated} with it (a brighter star saturates earlier in the ramp, removing flux from late differences). Median curves for both are drawn in black. Consequently, the uncorrected IQR-clipped lightcurve, which mixes all tracks, carries essentially no net signal for a bright source, and the saturation correction must rescale each pixel's group differences onto its unsaturated first difference to recover the coherent variability. Panel~(c) shows the quantity the correction models: for each pixel and each group-difference index $g$, the ratio $gd_g/gd_0$ (the $g_i/g_0$ of \S\ref{sec:saturation}) to the (least saturated) first group difference as a function of $gd_0$, together with the fitted quadratic ratio models. Early groups (blue) retain most of their signal. Later groups collapse toward zero as the ramp saturates. Groups are corrected only while their median ratio stays at or above 3\% and the residual scatter about the quadratic fit stays at or below 5\% of that ratio. The first group failing either test and all later groups are \emph{gated out}, meaning the pipeline discards those group differences entirely rather than attempting to correct them (dashed). For this heavily saturated pixel, only $g \leq 3$ survive, while a partially saturated neighboring pixel retains $g \leq 5$. Panel~(d) applies the fitted models to a single integration of that neighboring pixel. Dividing each group difference by its model ratio collapses the banded staircase onto the unsaturated level, with the gated groups shaded.

Panels~(e) through (g) dissect the slope correction (\S\ref{sec:slope}), which operates on the assembled lightcurve, and the choice between strategies. Panel~(e) shows the measured within-integration linear slope of each integration against its median flux, with the fitted quadratic relation. For this source, faint (in-eclipse) integrations ramp upward strongly, while bright ones drift slightly downward. Panel~(f) stacks the within-integration residual profiles of the brightest and faintest flux quartiles before and after subtracting the model-predicted slope. The strong flux-dependent ramp (reaching $\sim$0.03 in normalized flux at eclipse bottom, as predicted by the panel~(e) model) is removed, while only the slope, not the integration median, is touched, preserving the astrophysical signal. Panel~(g) shows the integration scatter (\S\ref{sec:strategy}) of each correction strategy, with the 10\% improvement gate. For this source the combined saturation + slope correction wins, reducing the integration scatter from 1.024 to 0.006 (0.008 for the saturation correction alone).

\begin{figure}
\centering
\includegraphics[width=0.88\columnwidth]{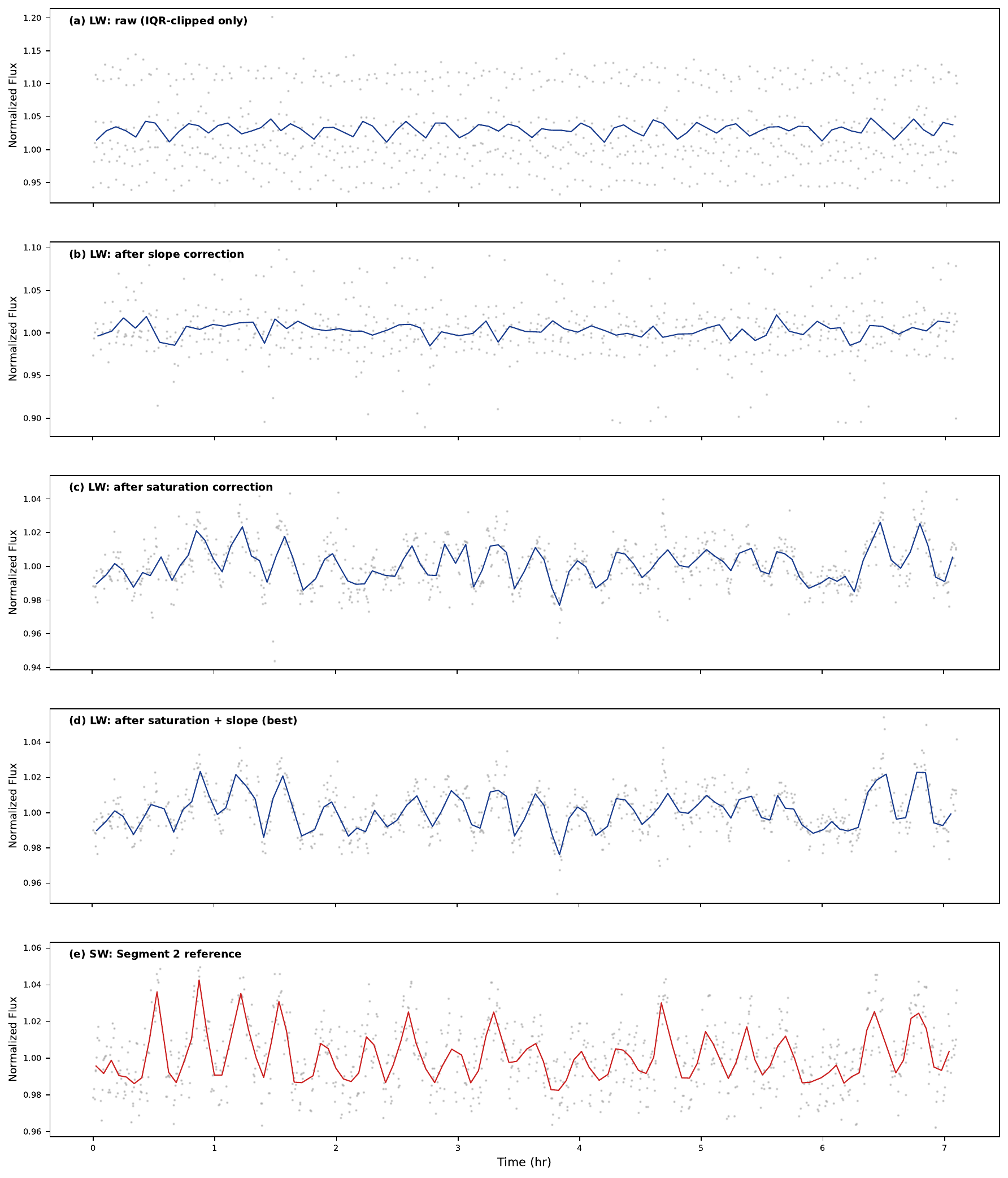}
\caption{Correction pipeline applied to the LW (F356W) lightcurve of
Terzan~5 \#341 (Segment~2), the 20.9-minute UCXB/IP candidate of
Figure~\ref{fig:short_period_collage}, with the simultaneously observed SW
lightcurve as an independent reference.  Blue lines in panels~(a)--(d) are the
binned mean.
\textbf{(a)}~Uncorrected IQR-clipped LW lightcurve, in which saturation banding
suppresses the true variability amplitude.
\textbf{(b)}~After slope correction alone, the banding is reduced but not
eliminated.
\textbf{(c)}~After saturation correction, the banding is largely removed and
coherent variability emerges.
\textbf{(d)}~After combined saturation and slope correction, the best strategy
for this source.
\textbf{(e)}~The SW (F200W) lightcurve of the same source (red binned line).
The source is background-dominated in SW, hence the large excursions, but the
binned trend shows the same 20.9-minute periodicity seen in the corrected LW
data.}
\label{fig:satcorr_lw}
\end{figure}

Figure~\ref{fig:satcorr_lw} shows the same machinery applied stage by stage to a saturated long-wavelength source, the 20.9-minute UCXB/IP candidate Terzan~5 \#341, illustrating each strategy of \S\ref{sec:strategy} in turn (slope only, saturation only, combined), with the recovered periodicity validated against the simultaneously observed SW lightcurve.

\subsection{The Segment-1 reduction in detail}\label{app:seg1_detail}

\paragraph{Aperture photometry.}
For each exposure, we project every cataloged source position onto the detector
using the per-exposure WCS from the \texttt{calints} file and perform circular
aperture photometry ($r = 1.5$~px) on each group-difference frame.  This
produces a raw lightcurve of 81 points per exposure in units of DN per
group-difference interval ($\Delta t_{\rm group} = 21.47$~s).

\paragraph{Saturation correction.}
Bright sources exhibit declining flux in successive group differences as the
accumulated signal approaches the detector's full-well capacity.  We correct
this by computing the median flux ratio of each group index $g$ relative to
$g = 0$ across all integrations within an exposure:
\begin{equation}
    R_g = \frac{\mathrm{median}(f_{g})}{\mathrm{median}(f_{g=0})},
\end{equation}
and dividing each measurement by the corresponding $R_g$.  Group differences
with $R_g < 0.05$ are discarded as too severely saturated.  The correction flattens coherent
within-integration structure of either sign. For Terzan~5 \#115 in
Figure~\ref{fig:seg1_cleaning}, whose group-difference medians \emph{rise} by
$\sim$10\% through the ramp ($R_g$ up to 1.12), it recovers the astrophysical
signal from data otherwise dominated by this structure and the
dither-dependent flat-field levels (compare panels~a and~b).

\paragraph{Outlier rejection.}
We apply a chunked interquartile range (IQR) clip independently within each
exposure, using chunks of 18 consecutive points and rejecting measurements
beyond $2 \times \mathrm{IQR}$ from the local quartiles.  The final chunk is
merged with the preceding one to avoid leaving a short unclipped tail.  Two
passes are applied to catch outliers initially masked by their neighbors
(panel~c).

\paragraph{Integration slope correction.}
Even after the group-ratio saturation correction, residual intra-integration
slopes remain, particularly for bright sources where higher-order saturation
effects cause flux to decrease systematically within each 9-point integration.
We model this by fitting a linear slope to each integration, then fitting a quadratic relation between integration median flux and signed slope.  The
predicted slopes are subtracted from each integration, flattening the
intra-integration structure while preserving inter-integration variability
(panel~d).

\paragraph{Normalization and stitching.}
Each exposure's corrected lightcurve is normalized to its median value.
The 12 normalized blocks are then stitched into a continuous lightcurve using
a curvature-aware continuity algorithm. We fit a quadratic model to each
block, predict the expected flux at the start of the next block from the
preceding block's fitted end value and the average of the two blocks'
endpoint derivatives across the inter-exposure gap, and apply an additive
offset to enforce continuity.  The quadratic (rather than linear) block model
matters wherever the lightcurve passes through an extremum within or near a block. There, a whole-block linear fit biases both the endpoint estimates
and the extrapolated slope, imprinting spurious jumps at precisely the joins
nearest the minima. Following the endpoint derivatives of a quadratic reduces
this failure mode.  This preserves real astrophysical variation
(including curvature and long-period variability spanning the full
observation window) while correcting for flat-field offsets between dither
positions.

Before stitching, we reject anomalous exposures where the source falls near a
detector edge or on a bad pixel region. Exposures whose raw median flux
deviates by more than $4 \times \mathrm{IQR}$ from the ensemble are discarded.
A final double-pass IQR clip is applied to the stitched lightcurve. As for the undithered segments, we apply no background rescaling (\S\ref{sec:background}), so the Segment~1 stitched lightcurves are on the same catalog flux scale. Their fractional amplitudes can still differ from those of Segment~2, since each dither position places the star in a different blend environment. Because each dither block is normalized to its own
median before stitching, the released annulus background levels for these sources
refer to the flux of an individual block rather than to the stitched lightcurve.

The result is a normalized lightcurve of typically $\sim$940 points spanning
$\sim$7~hr (panel~e of Figure~\ref{fig:seg1_cleaning}).  For sources also
observed in the undithered Segment~2 ($\sim$18.6~days later), the independent
Segment~2 lightcurve (panel~f) provides immediate validation. The periodic signals recovered from the dithered data are consistent in period with those measured from the undithered observations.

\begin{figure*}
\centering
\includegraphics[width=\textwidth]{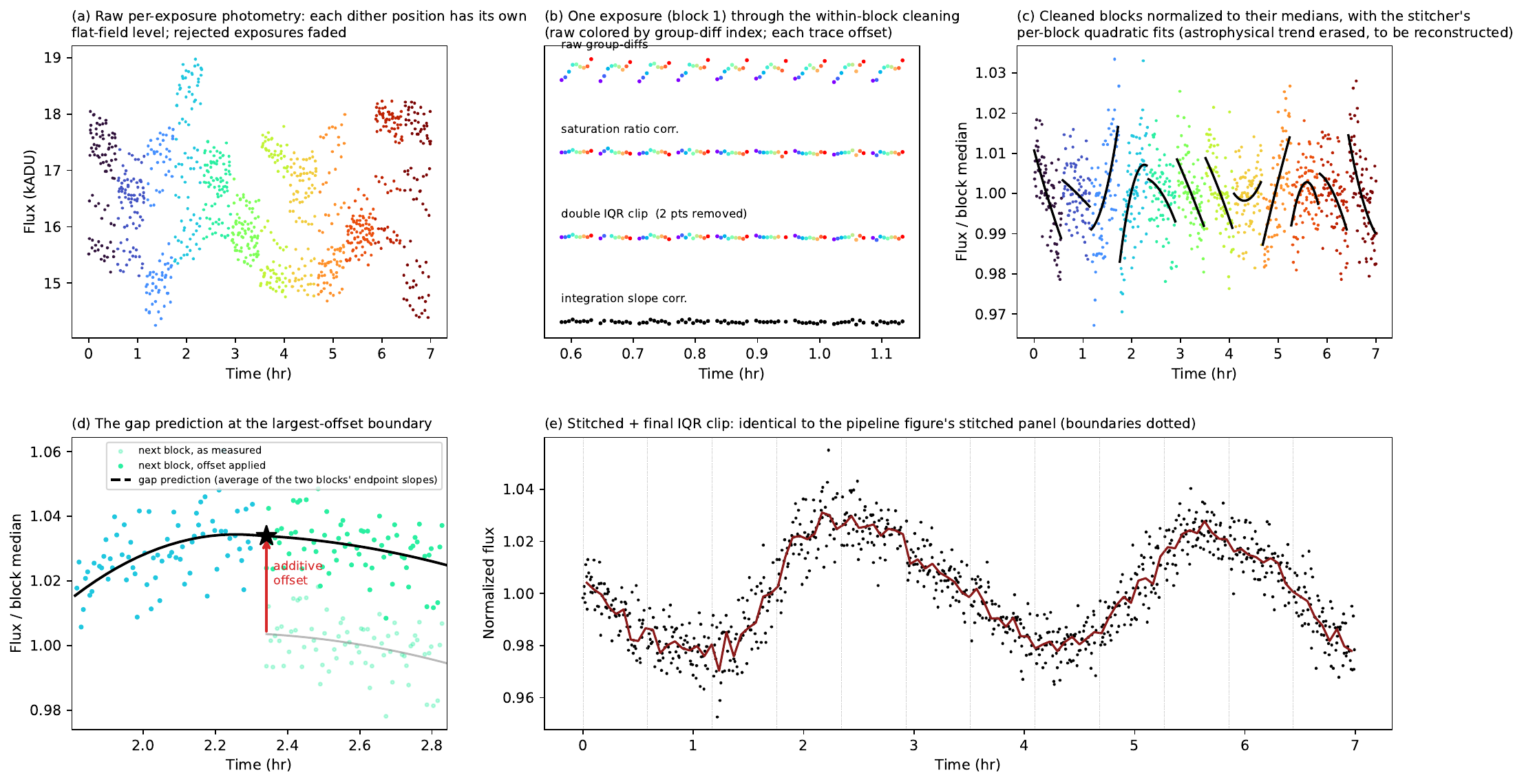}
\caption{The Segment~1 curvature-aware stitching under the hood, on Terzan~5 \#115, the source of the right column of Figure~\ref{fig:seg1_cleaning}. \textbf{(a)}~Raw per-exposure photometry (colored by exposure). Each dither position carries its own flat-field level. \textbf{(b)}~A single exposure through the within-block cleaning (traces offset). The raw group differences are colored by group index; their sawtooth is coherent within-integration structure. The subsequent traces show the per-group ratio correction, the double IQR clip, and the integration slope correction in turn. \textbf{(c)}~Cleaned blocks normalized to their medians, with the per-block quadratic fits of the stitching algorithm. \textbf{(d)}~The gap prediction at the largest-offset boundary. The fits of \emph{both} adjoining blocks (solid black) enter the prediction. The previous block's fit is extrapolated across the gap at the average of the two blocks' endpoint derivatives (dashed) to the predicted start (star). The additive offset (arrow) lifts the next block (as-measured fit in gray) so its fitted start lands on the prediction. \textbf{(e)}~The finished product after stitching and the final clip, on the catalog flux scale (no background rescaling, \S\ref{sec:background}), with a 4-minute median profile in dark red. The 3.1-hour modulation is recovered across all exposure boundaries (dotted), as in the stitched panel of Figure~\ref{fig:seg1_cleaning}.}
\label{fig:underhood_stitch}
\end{figure*}

Figure~\ref{fig:underhood_stitch} dissects the curvature-aware stitching at the heart of the dithered Segment~1 pipeline, on Terzan~5 \#115, the same source shown stage by stage in the right column of Figure~\ref{fig:seg1_cleaning}, and is generated by the same code, so the two figures are directly comparable. Panel~(a) shows the problem. In raw DN the twelve dither positions sit at flat-field levels differing by tens of percent. Panel~(b) puts a single exposure under the microscope, showing the within-block cleaning that precedes any fitting. The raw group differences form a sawtooth, as each integration shows a coherent pattern across its nine differences (for this source \emph{rising} $\sim$10\% from first to last, $R_g$ up to 1.12), which the per-group median-ratio correction flattens regardless of the pattern's sign. The double IQR clip then removes outliers, and the integration slope correction levels the residual within-integration tilts. Panel~(c) shows the cleaned blocks after per-block normalization, overlaid with the quadratic fit that the stitching algorithm makes to each block. The normalization erases the astrophysical trend along with the flat-field levels. Panel~(d) shows the reconstruction mechanism at the largest-offset boundary. The previous block's fit is extrapolated across the gap at the average of the two adjoining blocks' endpoint derivatives, and the next block receives the additive offset that places its fitted start on that prediction. Only the flat-field level is corrected, while real astrophysical slopes propagate through. Panel~(e) is the finished product (stitching plus the final outlier clip). The source's 3.1-hour periodic modulation, invisible in panel~(a), is recovered continuously across all twelve exposure boundaries, as in panel~(e) of Figure~\ref{fig:seg1_cleaning}.

\section{Astrometric Calibration}\label{app:astrometry}

No geometric-distortion resampling is applied to the images we measure. Every lightcurve is extracted in raw detector pixels, and the distortion solution enters only when a detector position is converted into a celestial one, as an a~posteriori mapping of the catalog rather than a rectification of the pixels. This preserves the native pixel-response behavior that the saturation correction depends on, and it costs nothing here because three of the four visits are undithered stares and the analysis is pixel-based throughout.

We refine the absolute astrometry of all NIRCam detectors using a two-stage procedure: (1)~alignment of the long-wavelength (LW; F356W, nrcblong) channel to Gaia~DR3, and (2)~alignment of each short-wavelength (SW; F200W, nrcb1--4) detector to the LW frame via cross-matched stellar detections. The LW detector spans the full module~B field of view, providing a single unified astrometric reference frame to which all four SW detectors can be tied. The SW-to-LW cross-match yields 289--631 reference sources per detector, more than ten times the 8--27 Gaia reference stars of the LW alignment that fall on any single SW detector, which reduces the uncertainty on the astrometric solution accordingly. We additionally refine individual source centroids using two-dimensional Gaussian fitting on autocorrelation reference images, achieving sub-pixel positional precision. All corrections are applied as rigid shifts to the WCS reference position (CRVAL), preserving the JWST calints SIP distortion polynomial.

\subsection{LW-to-Gaia Alignment}\label{sec:lw_gaia}

For each target and observation segment, we detect sources in the LW zeroframe median image constructed from uncal ZEROFRAME extensions, which preserve the saturated-pixel flux profile that the pipeline zeroframes discard by setting saturated pixels to zero. Source detection uses \texttt{DAOStarFinder} with a grid of FWHM values (4, 5, 6, 8, 10, 12, 15, 20, 25, 30~pixels). For each Gaia source, we take the nearest detection within the match radius in each FWHM catalog and select the one with the best (smallest $|r_1|$) DAOStarFinder roundness parameter. This ``best-roundness'' selection identifies the FWHM where the Gaussian model best matches the actual source profile, which varies with brightness due to saturation broadening of the PSF wings. We then filter to $|r_1| < 0.1$ to reject blends and artifacts, apply $3\times$IQR clipping to the positional residuals, and compute the median shift.

\begin{figure*}
\centering
\includegraphics[width=\textwidth]{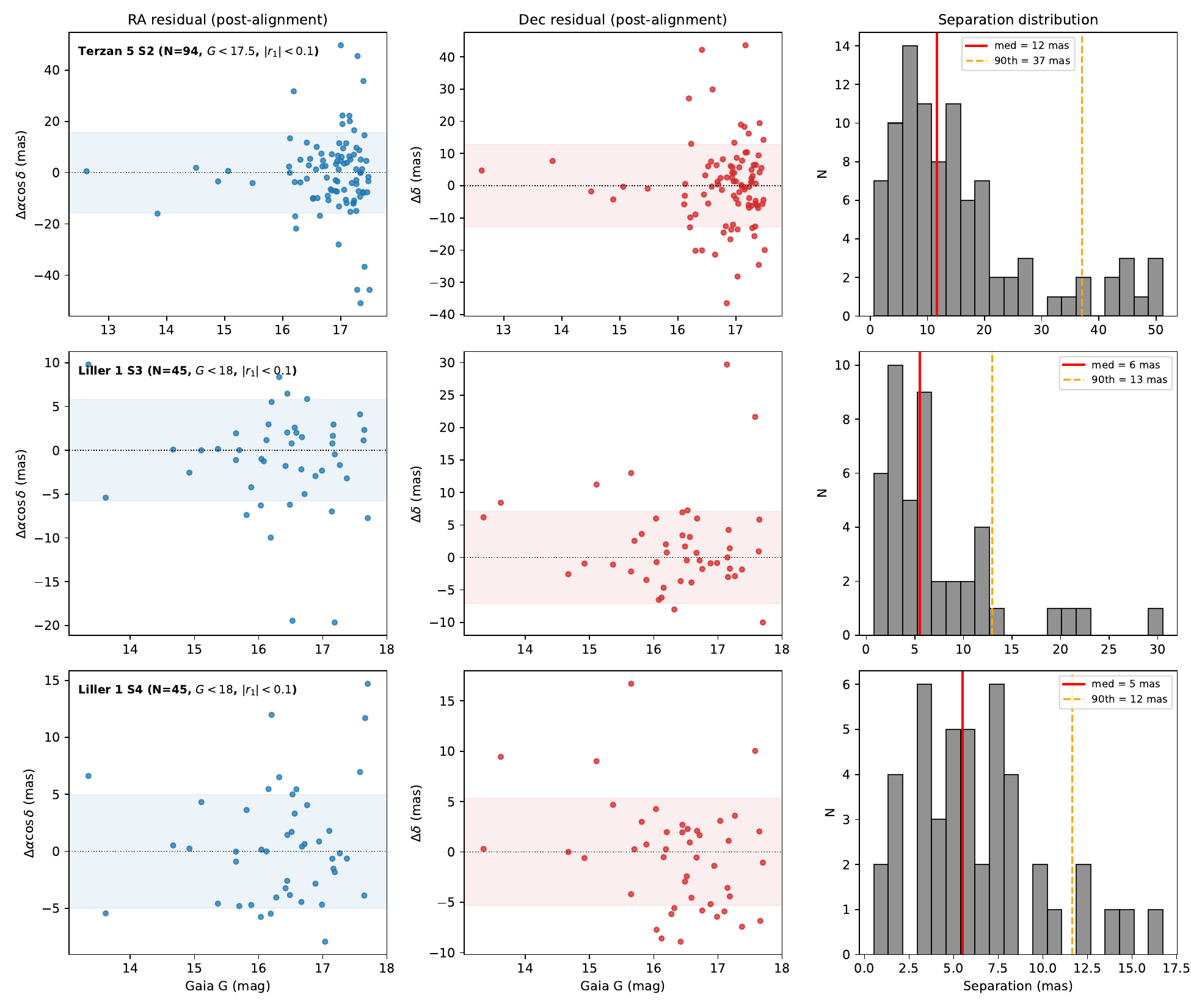}
\caption{Post-alignment astrometric residuals for the LW (F356W) channel matched to Gaia~DR3. Each row shows one target/segment combination: Terzan~5 Segment~2 (top), Liller~1 Segment~3 (middle), and Liller~1 Segment~4 (bottom), labelled with the number of reference stars, the Gaia magnitude cut, and the roundness cut $|r_1| < 0.1$, where $r_1$ is the \texttt{DAOStarFinder} roundness parameter used to reject blends and artifacts (\S\ref{sec:lw_gaia}). \textit{Columns:} RA offset versus Gaia $G$ magnitude, Dec offset versus $G$, and the histogram of total separations, with shaded bands showing the $\pm 1\sigma$ scatter. The median radial separation of 5--12~mas reflects the combined Gaia positional uncertainty and crowded-field centroiding floor.}
\label{fig:lw_gaia_resid}
\end{figure*}

\begin{figure*}
\centering
\includegraphics[width=\textwidth]{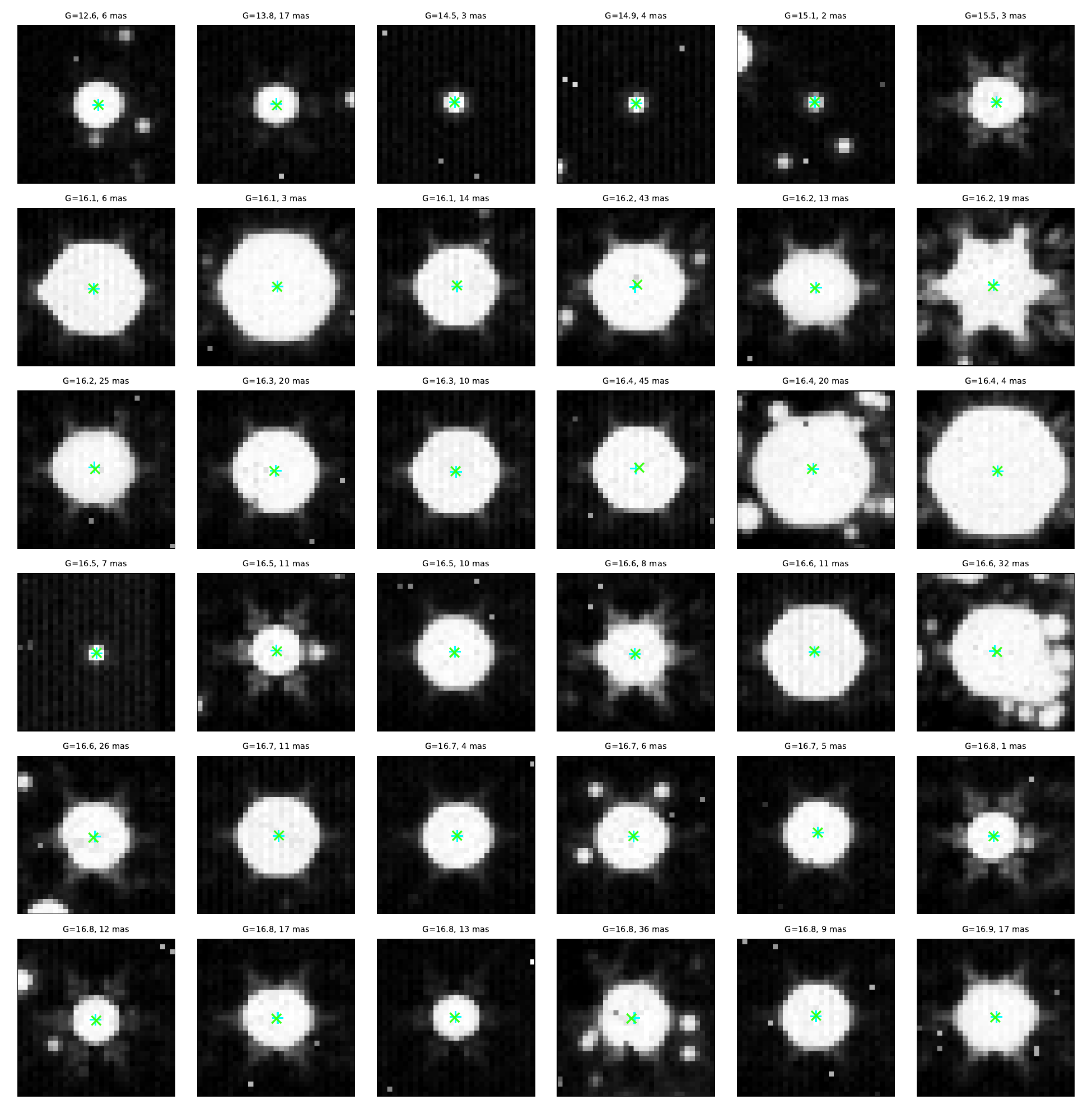}
\caption{Terzan~5 Segment~2: 36 representative Gaia-matched reference stars in the LW (F356W) zeroframe image after astrometric alignment. Each $31 \times 31$~pixel ($\sim 2\arcsec \times 2\arcsec$) cutout shows the JWST centroid (cyan cross) and proper-motion-propagated Gaia~DR3 position (green cross). The residual separation is labeled above each panel. The sources shown span $G = 12.6$--$16.9$~mag.}
\label{fig:cutouts_ter5}
\end{figure*}

Gaia~DR3 positions are propagated to the JWST observation epoch using catalog proper motions (propagation baselines of $\sim$9.3~yr). For Terzan~5, we use $G < 17.5$ and a $0\farcs5$ match radius. For Liller~1, $G < 18$ and $0\farcs3$. Table~\ref{tab:lw_gaia} summarizes the results. The correction is a pure translation, with no rotation or scale term: the rotation, plate scale and distortion all come from the JWST WCS (the SIP polynomial of the calints header), and only its reference position is shifted. Both clusters require shifts of $\sim$20--28~mas, with post-correction median residuals of 5--12~mas and uncertainties on the mean shift of 1--2~mas. Figure~\ref{fig:lw_gaia_resid} shows the post-alignment residuals, and Figure~\ref{fig:cutouts_ter5} shows representative cutouts of matched sources in the most crowded field, Terzan~5. The Liller~1 segments show equivalent agreement, with median residual separations of 5--6~mas (Figure~\ref{fig:lw_gaia_resid}, middle and bottom rows).

\begin{deluxetable}{lccccc}
\tablecaption{LW (F356W) to Gaia~DR3 Alignment\label{tab:lw_gaia}}
\tablewidth{0pt}
\tablehead{
\colhead{Target/Segment} & \colhead{$N$} & \colhead{$\Delta\alpha\cos\delta$} & \colhead{$\Delta\delta$} & \colhead{$\sigma_\mathrm{mean}$} & \colhead{Med.\ resid.} \\
 & & \colhead{(mas)} & \colhead{(mas)} & \colhead{(mas)} & \colhead{(mas)}
}
\startdata
Terzan~5 / Seg.~2  & 97\tablenotemark{a}  & $+27.2$ & $+7.1$  & 2.2 & 12 \\
Liller~1 / Seg.~3  & 46  & $+19.4$ & $+9.6$  & 1.3 & 6 \\
Liller~1 / Seg.~4  & 45  & $+18.9$ & $+12.3$ & 1.1 & 5 \\
\enddata
\tablenotetext{a}{Terzan~5 uses $G < 17.5$ and $0\farcs5$ match radius. Liller~1 uses $G < 18$ and $0\farcs3$. All use best-roundness FWHM selection, $|r_1| < 0.1$, $3\times$IQR clip, and median shift. $N$ counts the matched reference stars before the final clip. Figure~\ref{fig:lw_gaia_resid} shows the clipped samples (94, 45 and 45 stars).}
\end{deluxetable}

\subsection{SW-to-LW Alignment}\label{sec:sw_lw}

With the LW WCS tied to Gaia, we align each SW detector by cross-matching stellar detections between the SW and LW zeroframe median images. Sources are detected in the SW images using \texttt{DAOStarFinder} (FWHM\,$=$\,2~pixels, $10\sigma$ threshold) and in the LW images (FWHM\,$=$\,4~pixels, $10\sigma$). Each SW detection is paired with its nearest LW detection, and pairs separated by more than $0\farcs2$ are discarded. The pairing is not forced to be one-to-one, so when the SW channel resolves two stars that the LW channel blends, both can be paired with the same LW detection. The correction is the median positional offset, which such occasional mismatches barely affect, and like the LW correction it is a pure translation.

This approach yields 289--631 cross-matched reference sources per detector (Table~\ref{tab:sw_lw}), 13--78 times the 8--27 LW-alignment Gaia reference stars (Table~\ref{tab:lw_gaia}) that fall on each SW detector, and eliminates proper-motion propagation uncertainties since both channels observe simultaneously. The median per-source residual after correction is 2.5--6~mas across all detectors. The Liller~1 segments are mutually consistent, with shifts agreeing to $\sim$1~mas in right ascension and $\sim$3~mas in declination between Segments~3 and~4 for each detector.

\begin{deluxetable*}{llccccc}
\tablecaption{SW (F200W) to LW (F356W) Alignment\label{tab:sw_lw}}
\tablewidth{0pt}
\tablehead{
\colhead{Target/Segment} & \colhead{Detector} & \colhead{$N$} & \colhead{$\Delta\alpha\cos\delta$} & \colhead{$\Delta\delta$} & \colhead{Med.\ resid.} & \colhead{90th pct.} \\
 & & & \colhead{(mas)} & \colhead{(mas)} & \colhead{(mas)} & \colhead{(mas)}
}
\startdata
Terzan~5 / Seg.~2 & nrcb1 & 289 & $+29.5$ & $+9.1$ & 2.5 & 6.2 \\
                   & nrcb2 & 306 & $+22.5$ & $+8.3$ & 3.2 & 16.6 \\
                   & nrcb3 & 456 & $+29.6$ & $+1.8$ & 3.1 & 22.2 \\
                   & nrcb4 & 629 & $+22.8$ & $+1.9$ & 4.0 & 24.1 \\
\hline
Liller~1 / Seg.~3 & nrcb1 & 373 & $+21.0$ & $+11.7$ & 3.5 & 88.4 \\
                   & nrcb2 & 437 & $+15.3$ & $+12.1$ & 3.9 & 48.4 \\
                   & nrcb3 & 382 & $+20.5$ & $+4.9$ & 3.9 & 70.9 \\
                   & nrcb4 & 624 & $+14.9$ & $+5.6$ & 6.0 & 68.7 \\
\hline
Liller~1 / Seg.~4 & nrcb1 & 373 & $+20.6$ & $+14.6$ & 3.5 & 75.1 \\
                   & nrcb2 & 436 & $+14.9$ & $+14.2$ & 3.6 & 56.8 \\
                   & nrcb3 & 371 & $+19.9$ & $+7.7$ & 3.6 & 65.3 \\
                   & nrcb4 & 631 & $+14.1$ & $+8.7$ & 6.2 & 76.3 \\
\enddata
\end{deluxetable*}

\subsection{Centroid Refinement}\label{sec:centroid}

Source detection (\S\ref{sec:detection}) returns integer pixel positions from local maxima of the PSF-convolved autocorrelation image. To obtain sub-pixel precision, we refine each source centroid by fitting a two-dimensional circular Gaussian to a $9 \times 9$~pixel cutout of the autocorrelation image centered on the detection pixel. The cutout is taken from the zeroframe autocorrelation image for sources with a strong zeroframe detection (${\rm SNR} > 10$, bright stars whose saturating cores corrupt the ramp-based autocorrelation) and from the ramp autocorrelation image otherwise, choosing among the detections of a source in the order of preference: a bright SW zeroframe detection, a bright LW zeroframe detection, an SW ramp detection, and then any remaining SW zeroframe or LW detection. The fitted model is:
\begin{equation}
    I(x,y) = A \exp\!\left[-\frac{(x - x_0)^2 + (y - y_0)^2}{2\sigma^2}\right] + B,
\end{equation}
where the free parameters are the amplitude $A$, centroid $(x_0, y_0)$, width $\sigma$, and background $B$. The fit uses Levenberg--Marquardt optimization with initial guesses from the local maximum and edge pixels, and we reject fits where the centroid moves more than 4~pixels or the formal uncertainty exceeds 2~pixels.

This refinement achieves typical centroid precision of $\sim$0.1~pixels ($\sim$3~mas). For source \#344 (the infrared counterpart of PSR~J1748$-$2446A, Section~\ref{sec:psrj1748}), the sub-pixel correction more than doubled the Lomb--Scargle significance of the 108-minute orbital period.

\subsection{Systematic Error Budget}\label{sec:astrom_errors}

The total systematic uncertainty on any source position combines three contributions added in quadrature (Table~\ref{tab:astrom_errors}):
\begin{enumerate}
    \item \textbf{Gaia frame tie} ($\sigma_\mathrm{Gaia}$). The uncertainty on the mean LW-to-Gaia shift, ranging from 1.1~mas (Liller~1, $N \sim 45$) to 2.2~mas (Terzan~5, $N = 97$), which is the dominant systematic for the absolute reference frame.
    \item \textbf{LW-to-SW transfer} ($\sigma_\mathrm{LW \to SW}$). The uncertainty on the median SW-to-LW shift, $\sim$0.2~mas with hundreds of cross-matched sources, which is negligible.
    \item \textbf{Centroid precision} ($\sigma_\mathrm{centroid}$). The formal Gaussian fit uncertainty for individual sources, $\sim$3~mas at the SW plate scale, which dominates the per-source error.
\end{enumerate}

The total systematic floor is $\sigma_\mathrm{total} = \sqrt{\sigma_\mathrm{Gaia}^2 + \sigma_\mathrm{LW \to SW}^2 + \sigma_\mathrm{centroid}^2} \approx 3.2$--$3.7$~mas for well-detected, isolated sources, rising to $\sim$10~mas in the most crowded regions where blending degrades centroid accuracy.

\begin{deluxetable*}{llcccc}
\tablecaption{Absolute Astrometric Systematic Errors by Detector\label{tab:astrom_errors}}
\tablewidth{0pt}
\tablehead{
\colhead{Target/Segment} & \colhead{Detector} & \colhead{$\sigma_\mathrm{Gaia}$} & \colhead{$\sigma_\mathrm{LW \to SW}$} & \colhead{$\sigma_\mathrm{centroid}$} & \colhead{$\sigma_\mathrm{total}$} \\
 & & \colhead{(mas)} & \colhead{(mas)} & \colhead{(mas)} & \colhead{(mas)}
}
\startdata
Terzan~5 / Seg.~2 & nrcb1 & 2.2 & 0.15 & 3.0 & 3.7 \\
                   & nrcb2 & 2.2 & 0.18 & 3.0 & 3.7 \\
                   & nrcb3 & 2.2 & 0.15 & 3.0 & 3.7 \\
                   & nrcb4 & 2.2 & 0.16 & 3.0 & 3.7 \\
\hline
Liller~1 / Seg.~3 & nrcb1 & 1.3 & 0.18 & 3.0 & 3.3 \\
                   & nrcb2 & 1.3 & 0.18 & 3.0 & 3.3 \\
                   & nrcb3 & 1.3 & 0.20 & 3.0 & 3.3 \\
                   & nrcb4 & 1.3 & 0.24 & 3.0 & 3.3 \\
\hline
Liller~1 / Seg.~4 & nrcb1 & 1.1 & 0.18 & 3.0 & 3.2 \\
                   & nrcb2 & 1.1 & 0.17 & 3.0 & 3.2 \\
                   & nrcb3 & 1.1 & 0.19 & 3.0 & 3.2 \\
                   & nrcb4 & 1.1 & 0.25 & 3.0 & 3.2 \\
\enddata
\end{deluxetable*}

As an external consistency check, which cannot validate the few-mas figures above because the radio errors are far larger than the JWST error budget, we note that our JWST position for the infrared counterpart of PSR~J1748$-$2446A agrees with its VLA radio position \citep{Urquhart2026}: the total offset is 80~mas, well within the $60 \times 110$~mas ($1\sigma$) VLA error ellipse ($0.2\sigma$ in RA, $0.7\sigma$ in Dec). The right ascension of our JWST position agrees with that of the radio timing position from \citet{Rosenthal2025} to 0.3~mas.

\section{Data Availability and Reproduction}\label{app:repro}

The complete analysis pipeline used in this work is publicly available as the
\texttt{gc-variables-pipeline} package\footnote{\url{https://github.com/kburdge/gc-variables-pipeline}},
archived at Zenodo\footnote{\url{https://doi.org/10.5281/zenodo.21419760}}. The
raw data are public in the Mikulski Archive for Space Telescopes (MAST) under
JWST program GO-5381. To support exact reproduction, the Zenodo record also
hosts the final variable-star catalog, the group-differenced data cubes for a
demonstration subset, the human source-classification (``REAL''/``FAKE'')
labels described in \S\ref{sec:classification}, the manual deduplication groupings (\S\ref{sec:catalog}) that map detections to unique objects, and the binary-fit decisions behind the classifications of \S\ref{sec:classification} (for each source, the fitted \texttt{PHOEBE} model, whether the data constrain its period, and whether we accepted the fit). The last three encode the human-judgment steps in our analysis, and shipping them is what makes the final catalog and its classifications reproducible without re-doing the visual vetting, deduplication, and model acceptance.

\subsection{Pipeline overview}

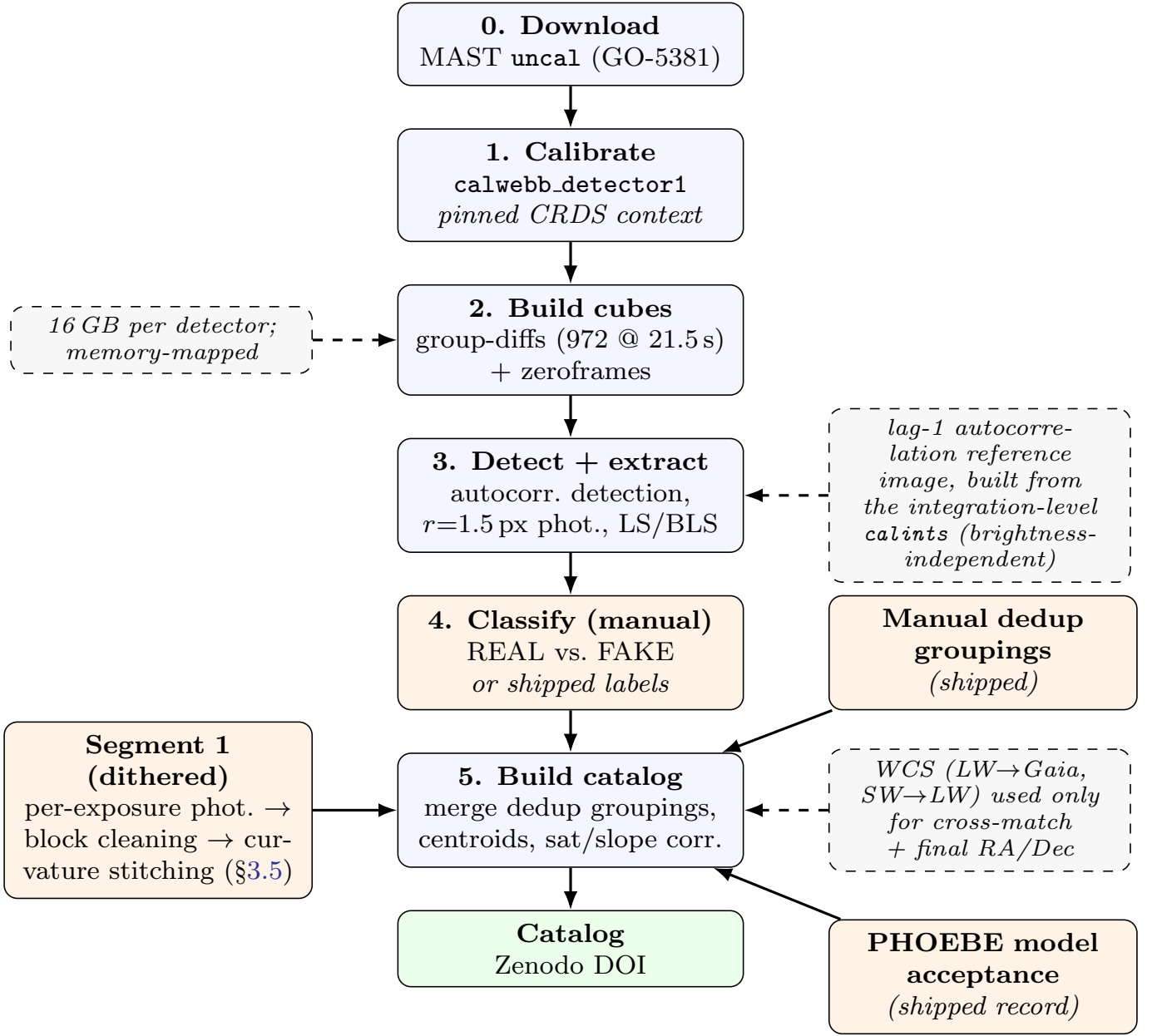
\begin{figure*}
\centering
\resizebox{\textwidth}{!}{%
\begin{tikzpicture}[
    node distance=4.5mm and 9mm,
    box/.style={draw, rounded corners, align=center, inner sep=4pt,
                font=\footnotesize, minimum height=7mm, text width=34mm, fill=blue!4},
    man/.style={box, fill=orange!10},
    pub/.style={box, fill=green!8},
    side/.style={draw, dashed, rounded corners, align=center, inner sep=3pt,
                 font=\scriptsize\itshape, text width=30mm, fill=gray!6},
    arr/.style={-{Latex[length=2mm]}, thick},
]
\node[box] (dl)  {\textbf{0. Download}\\MAST \texttt{uncal} (GO-5381)};
\node[box, below=of dl] (cal) {\textbf{1. Calibrate}\\\texttt{calwebb\_detector1}\\\textit{pinned CRDS context}};
\node[box, below=of cal] (cube) {\textbf{2. Build cubes}\\group-diffs (972 @ 21.5\,s)\\+ zeroframes};
\node[box, below=of cube] (det) {\textbf{3. Detect + extract}\\autocorr.\ detection,\\$r{=}1.5$\,px phot., LS/BLS};
\node[man, below=of det] (vet) {\textbf{4. Classify (manual)}\\REAL vs.\ FAKE\\\textit{or shipped labels}};
\node[box, below=of vet] (cat) {\textbf{5. Build catalog}\\merge dedup groupings,\\centroids, sat/slope corr.};
\node[pub, below=of cat] (out) {\textbf{Catalog}\\Zenodo DOI};

\draw[arr] (dl) -- (cal);
\draw[arr] (cal) -- (cube);
\draw[arr] (cube) -- (det);
\draw[arr] (det) -- (vet);
\draw[arr] (vet) -- (cat);
\draw[arr] (cat) -- (out);
\node[man, right=of vet, text width=30mm] (dedup) {\textbf{Manual dedup groupings}\\\textit{(shipped)}};
\draw[arr] (dedup) -- (cat);

\node[side, right=of det] (ac) {lag-1 autocorrelation reference image, built from the integration-level \texttt{calints} (brightness-independent)};
\draw[arr, dashed] (ac) -- (det);
\node[man, left=of cat, text width=30mm] (seg1) {\textbf{Segment 1 (dithered)}\\per-exposure phot.\ $\to$ block cleaning $\to$ curvature stitching (\S\ref{sec:seg1_extraction})};
\draw[arr] (seg1) -- (cat);
\node[man, right=of out, text width=30mm, yshift=-3mm] (phoebe) {\textbf{PHOEBE model acceptance}\\\textit{(shipped record)}};
\draw[arr] (phoebe) -- (cat);
\node[side, right=of cat] (wcs) {WCS (LW$\to$Gaia, SW$\to$LW) used only for cross-match + final RA/Dec};
\draw[arr, dashed] (wcs) -- (cat);
\node[side, left=of cube] (raw) {16\,GB per detector; memory-mapped};
\draw[arr, dashed] (raw) -- (cube);
\end{tikzpicture}%
}
\caption{End-to-end reproduction flow for the \texttt{gc-variables-pipeline}
package. Blue boxes are automated stages. Orange boxes mark inputs shipped as data products rather than re-derived, namely the human-judgment steps (REAL/FAKE classification, manual deduplication, binary-model acceptance) and the Segment~1 dithered-extraction products. Green boxes are the
published data products. Dashed nodes mark key methodological choices described
in the main text.}
\label{fig:repro_flow}
\end{figure*}

Figure~\ref{fig:repro_flow} summarizes the end-to-end flow from raw exposures to
the published catalog. The repository organizes the pipeline into six stage scripts (numbered 0--5), each documented there and cross-referenced to the relevant section of this paper.

The pipeline begins by retrieving the raw \texttt{uncal} exposures for program GO-5381 from MAST (\S\ref{sec:observations}) and calibrating them with \texttt{calwebb\_detector1} run with \texttt{save\_calibrated\_ramp=True}, followed by \texttt{Image2Pipeline} (\S\ref{sec:reduction}). Reproducing this step requires a \emph{pinned} CRDS reference context, and the published reduction used JWST calibration software version 1.17.1 with CRDS context \texttt{jwst\_1322.pmap}, which is set in the repository configuration. The calibrated ramps are then differenced group-to-group to synthesize the 972-frame, $21.47$\,s cadence cubes, and the zeroframes are stacked into 96-frame cubes (\S\ref{sec:reduction}). From the integration-level \texttt{calints} (ramp products) and the zeroframe cube (zeroframe products) the pipeline builds the lag-1 autocorrelation reference images, detects sources by PSF-matched filtering, performs $r=1.5$\,px aperture photometry, IQR-clips, and runs the Lomb--Scargle and BLS period searches (\S\ref{sec:detection}--\ref{sec:period}). Sources are then classified as real variables or artifacts by visual inspection of diagnostic plots (\S\ref{sec:classification}), though to reproduce the published catalog deterministically, the pipeline instead ingests the shipped classification labels. The final stage cross-matches and deduplicates the detections, refines centroids, extracts lightcurves, applies the multi-stage saturation/slope correction, and selects the adopted lightcurve for each source, segment, and channel (\S\ref{sec:corrections}--\ref{sec:catalog}).

Throughout the post-calibration stages the detector pixel position of each source is treated as the invariant quantity. The astrometric WCS (Appendix~\ref{app:astrometry}) is
used only to cross-match the short- and long-wavelength channels and to assign
final catalog coordinates.

\subsection{Levels of reproduction}

We provide three entry points of increasing cost. A \emph{demonstration} run
processes, on a single workstation, a single detector and segment
downloaded from MAST, exercising the full toolchain end-to-end (the
\texttt{calwebb\_detector1} calibration dominates the runtime). A \emph{reproduce-from-products}
run downloads the vetting labels, deduplication groupings and binary-fit decisions from Zenodo, builds the data cubes from the raw MAST data (stages 0--2), and re-derives the catalog without repeating any human-judgment step. A \emph{full} run also regenerates the automated products that the Zenodo archive otherwise supplies ready-made (the astrometric solutions and the Segment~1 dithered-extraction products), importing only the three human-judgment records. The LW-to-Gaia
astrometric solution of Appendix~\ref{app:astrometry} is itself reproducible
from the released code and the shipped Gaia DR3 catalog caches (we verified
that rerunning it reproduces the released WCS reference files exactly). The SW-to-LW transfer solutions are shipped as data products, as are the
Terzan~5 Segment~1 second-epoch lightcurves (their dedicated
dithered-extraction procedure, \S\ref{sec:seg1_extraction} and
Appendix~\ref{app:underhood}, is not yet part of the installable package, although the script we used is included in the repository archive). The full
run, however, entails a terabyte-scale download and multi-day computation, as the
group-differenced cubes are $\sim$16\,GB per detector. Command-by-command instructions for all three are
given in the repository documentation.

\section{Lightcurve atlas}\label{app:atlas}

This appendix presents the lightcurve of every source in the catalog, grouped by its adopted class and sorted within each group by adopted period (for single-transit sources, by the fitted period used only for ordering): the three fitted binary morphologies (contact, semi-detached, detached), the single-transit sources, the rarer classified variables, and those for which no model was accepted. The rarer classes (accreting binaries, pulsators, flare sources, and compact-object candidates) are also shown individually in the main text. Each source is shown as a four-panel block, both observing segments in both channels, with the catalog lightcurve drawn as black vertical bars spanning the 1$\sigma$ uncertainty of each plotted sample (\S\ref{sec:photerr}). Where a model was accepted, the colored curve is the accepted \texttt{PHOEBE} model and a narrow strip below each panel shows the residuals. The header names the fitted Roche geometry (detached, semi-detached, or contact), followed by ``spotted'', ``eccentric orbit'', or ``with linear trend'' where the model includes a starspot, an eccentric orbit, or a linear baseline trend in time (\S\ref{sec:classification}). For single-transit sources the header names only the Roche geometry of the fitted model, which is not a classification. Sources without an accepted model carry their by-eye class (EA for the Algol type, EB for the $\beta$~Lyrae type, or ``uncertain'') or ``no class''. A panel reading ``no SW'' or ``no LW'' has no lightcurve in that band and segment, and one reading ``SW unusable'' or ``LW unusable'' holds a lightcurve too faint against the background, or otherwise judged unreliable, to be shown. The machine-readable lightcurves, uncertainties, classifications, and adopted periods for all sources are included in the data release.

\ifapjfigset
\else

\subsection{Contact binaries}\label{app:atlas:contact}

\begin{figure*}
\centering
\includegraphics[width=0.98\textwidth,height=0.79\textheight,keepaspectratio]{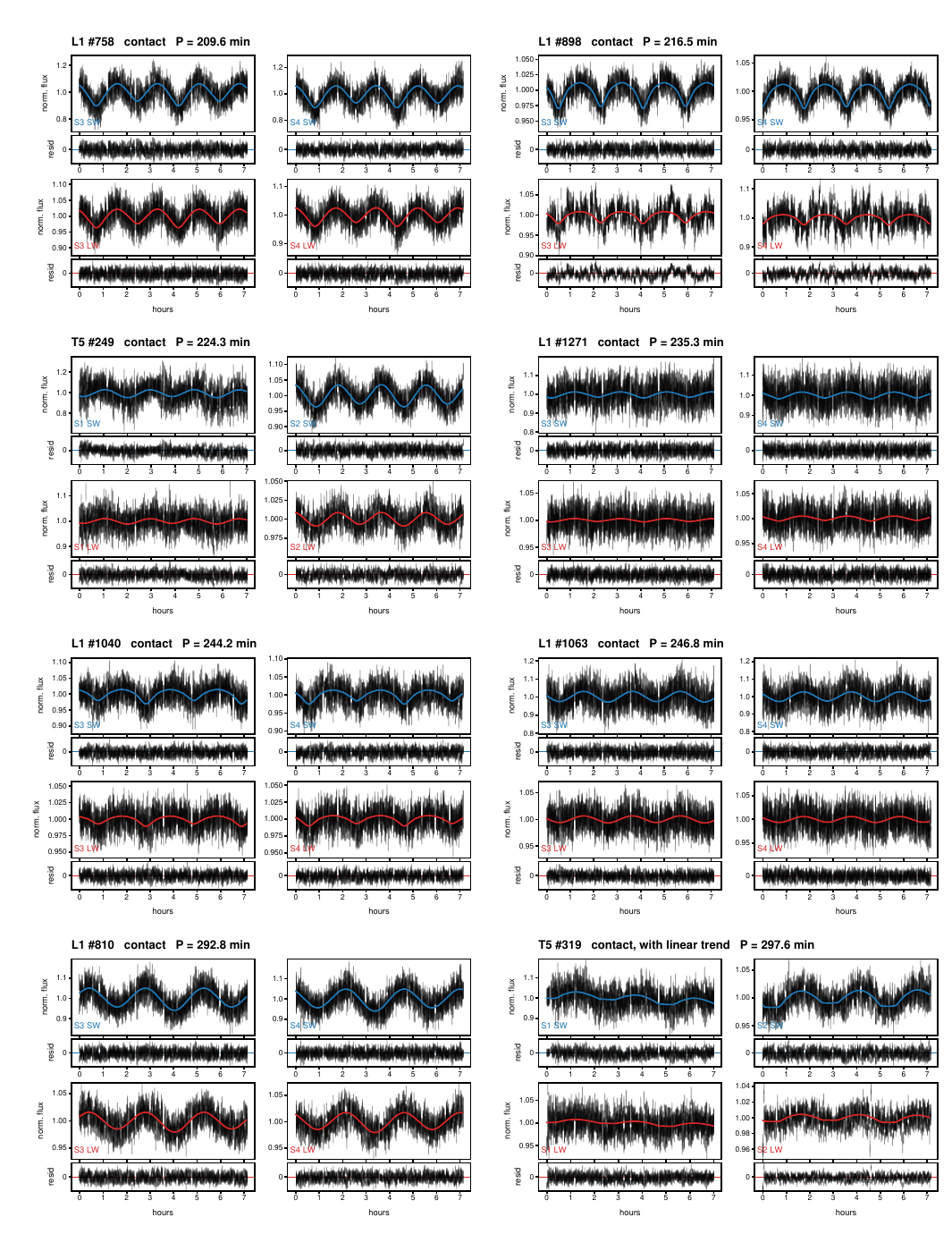}
\caption{Contact binaries: unfolded JWST NIRCam lightcurves of all 260 sources in this class (page 1 of 33), sorted by adopted period (sources without a period last). Each source is shown as a four-panel block. The two observing segments run left to right, with the short-wavelength F200W lightcurve (SW, \textbf{blue}) on top and the long-wavelength F356W lightcurve (LW, \textbf{red}) below, and the segment and band are labelled inside every panel. Time is hours from the start of that segment, and lightcurves are never phase-folded. Black vertical bars are the adopted lightcurve (\S\ref{sec:strategy}), each spanning the $1\sigma$ uncertainty of one plotted sample. The coloured curve is the accepted \texttt{PHOEBE} model, drawn in each panel as $A\,m(t)+B$ with the dilution terms $A$ and $B$ (\S\ref{sec:classification}) re-solved per panel, since blending differs between bands and segments. A depth difference between SW and LW is therefore not evidence of chromaticity. The narrow strip under each panel shows the residuals. The header gives the source, the fitted model (see the start of this appendix), and its orbital period, the adopted period.}
\end{figure*}
\clearpage

\begin{figure*}
\centering
\includegraphics[width=0.98\textwidth,height=0.94\textheight,keepaspectratio]{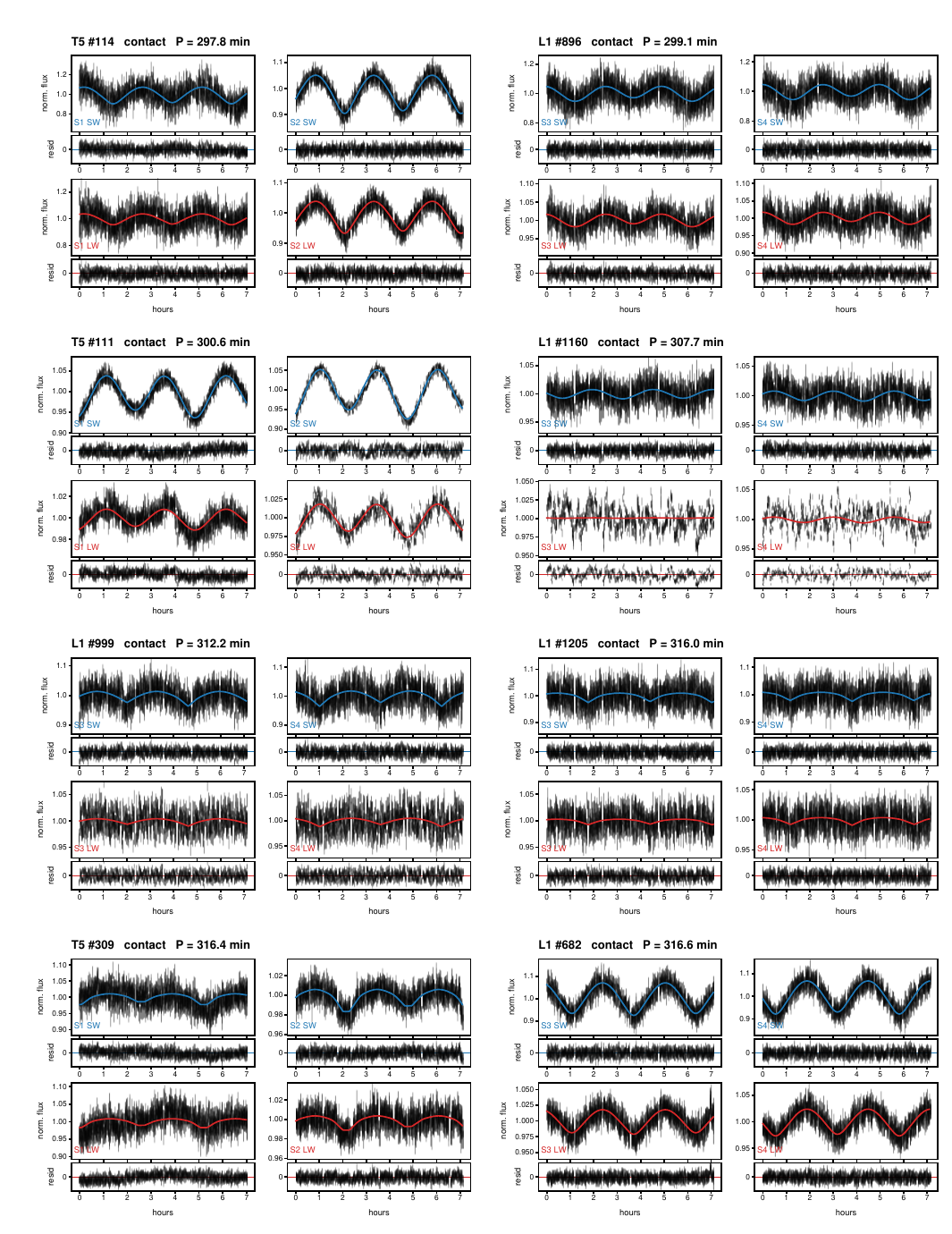}
\caption{Contact binaries, continued (page 2 of 33).}
\end{figure*}
\clearpage

\begin{figure*}
\centering
\includegraphics[width=0.98\textwidth,height=0.94\textheight,keepaspectratio]{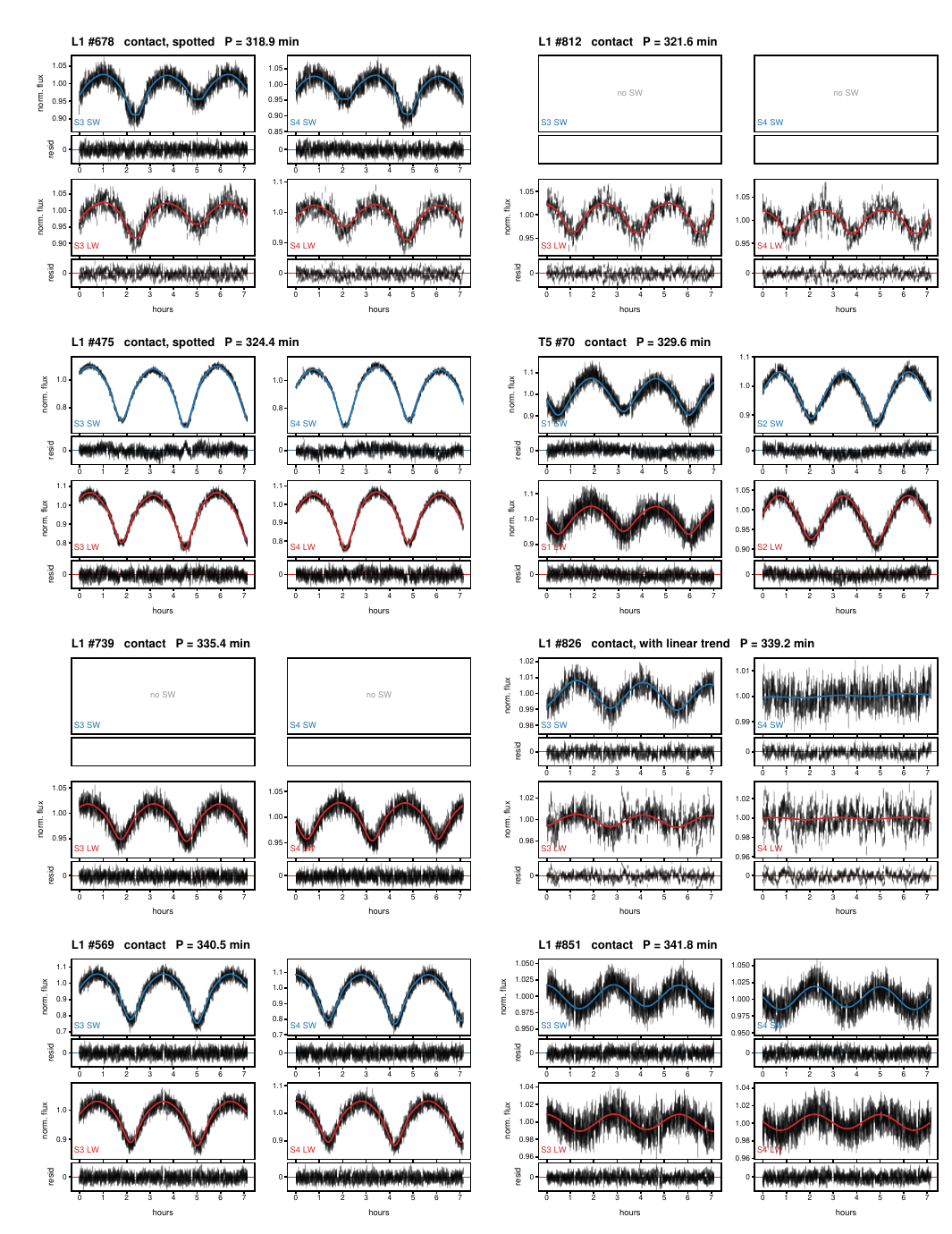}
\caption{Contact binaries, continued (page 3 of 33).}
\end{figure*}
\clearpage

\begin{figure*}
\centering
\includegraphics[width=0.98\textwidth,height=0.94\textheight,keepaspectratio]{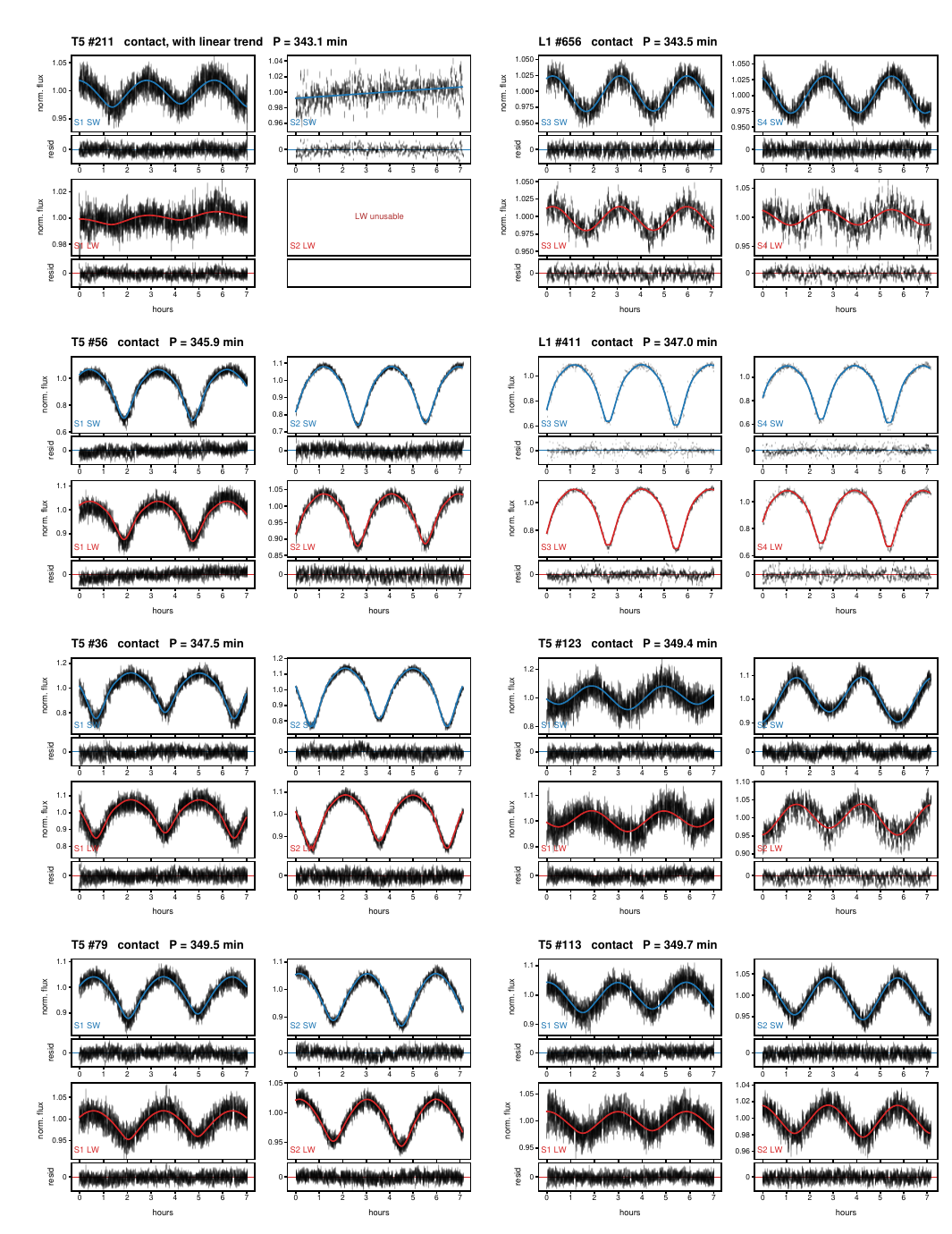}
\caption{Contact binaries, continued (page 4 of 33).}
\end{figure*}
\clearpage

\begin{figure*}
\centering
\includegraphics[width=0.98\textwidth,height=0.94\textheight,keepaspectratio]{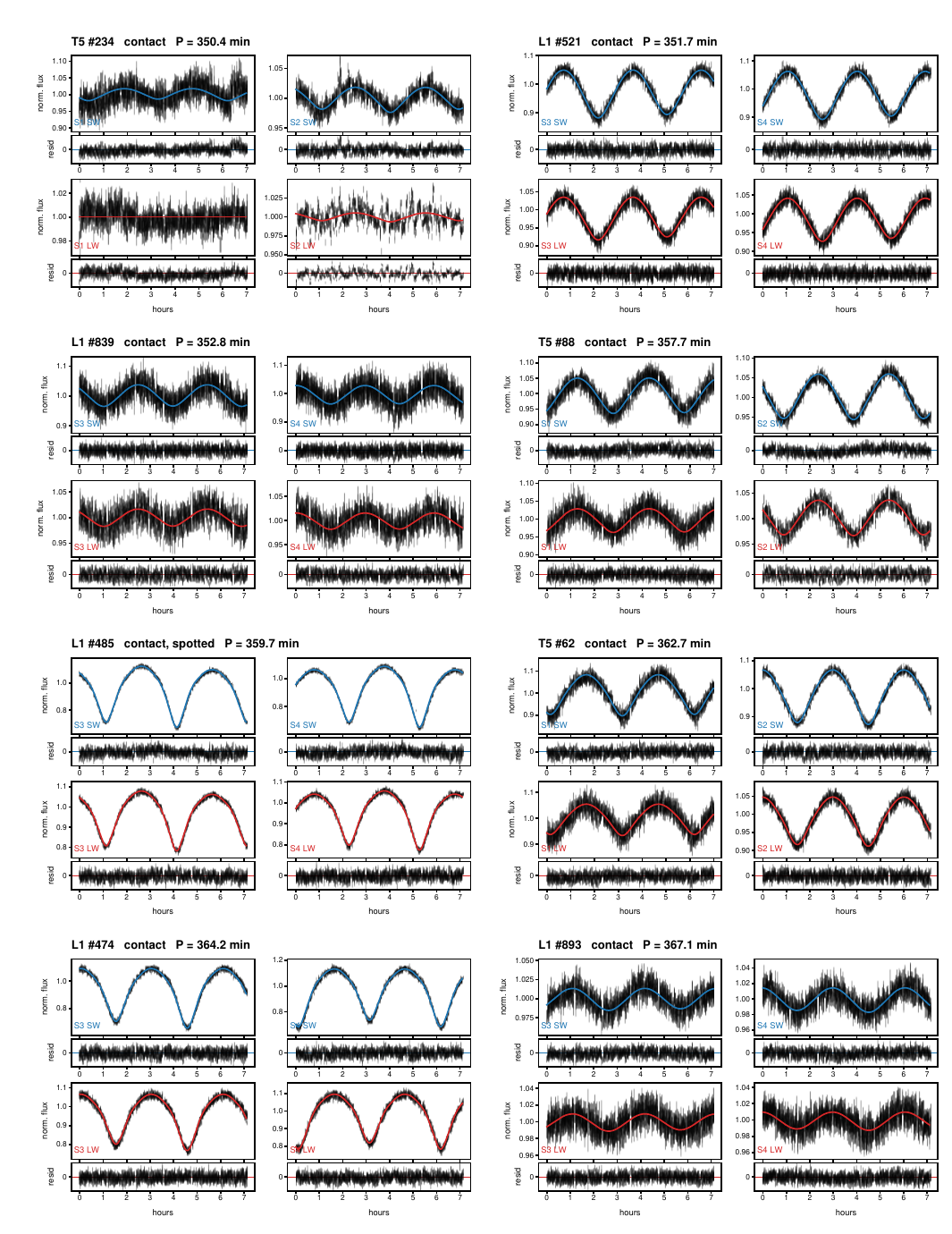}
\caption{Contact binaries, continued (page 5 of 33).}
\end{figure*}
\clearpage

\begin{figure*}
\centering
\includegraphics[width=0.98\textwidth,height=0.94\textheight,keepaspectratio]{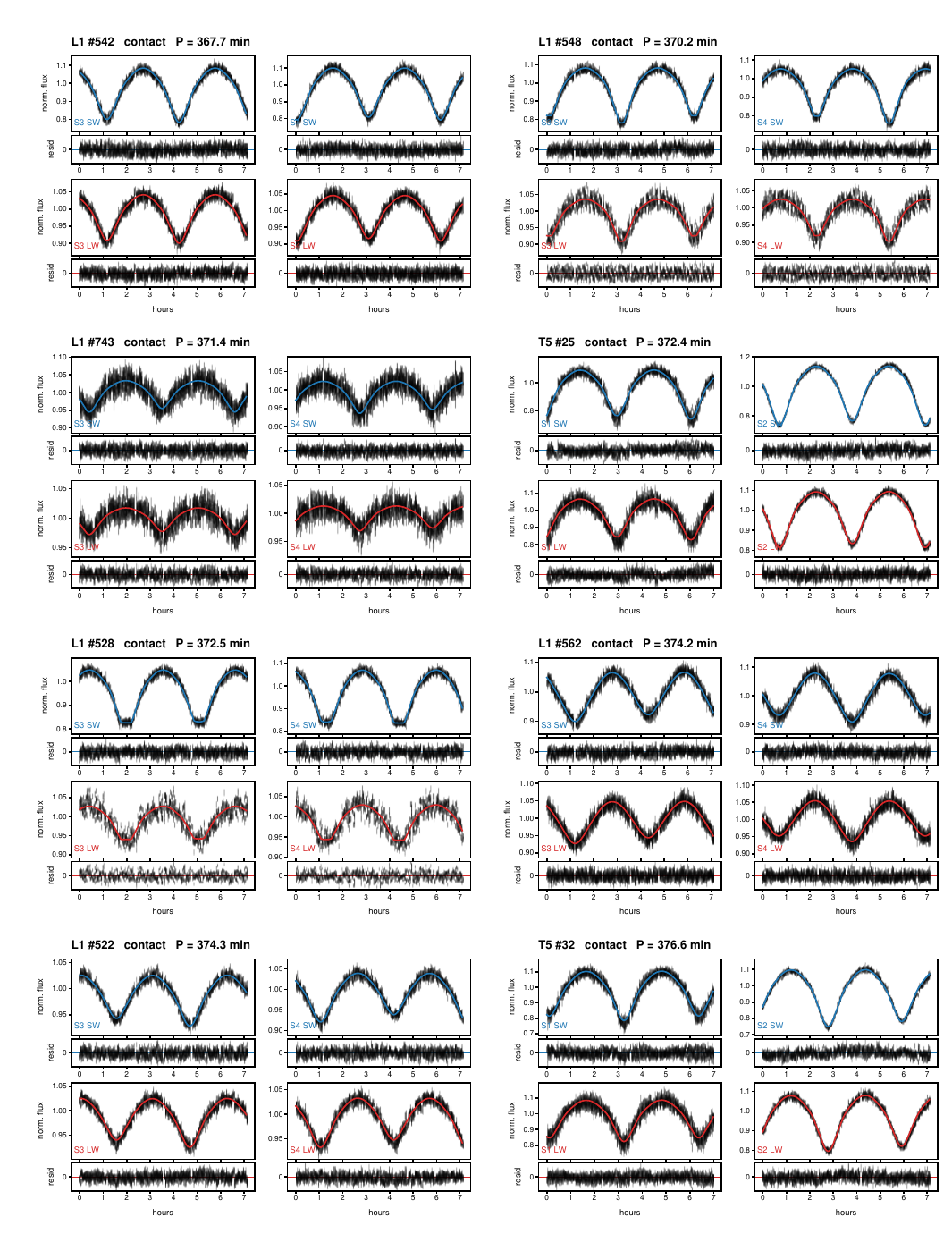}
\caption{Contact binaries, continued (page 6 of 33).}
\end{figure*}
\clearpage

\begin{figure*}
\centering
\includegraphics[width=0.98\textwidth,height=0.94\textheight,keepaspectratio]{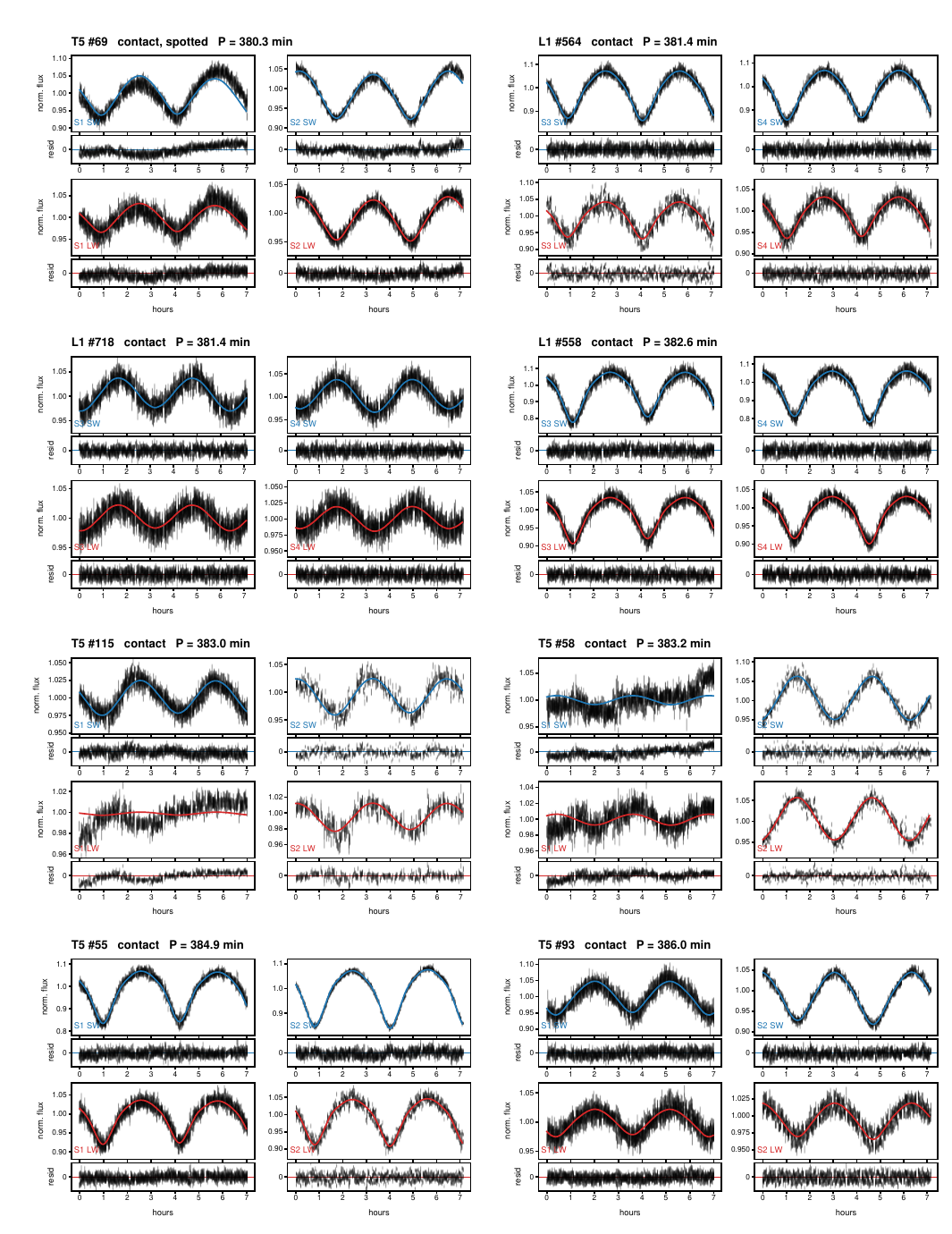}
\caption{Contact binaries, continued (page 7 of 33).}
\end{figure*}
\clearpage

\begin{figure*}
\centering
\includegraphics[width=0.98\textwidth,height=0.94\textheight,keepaspectratio]{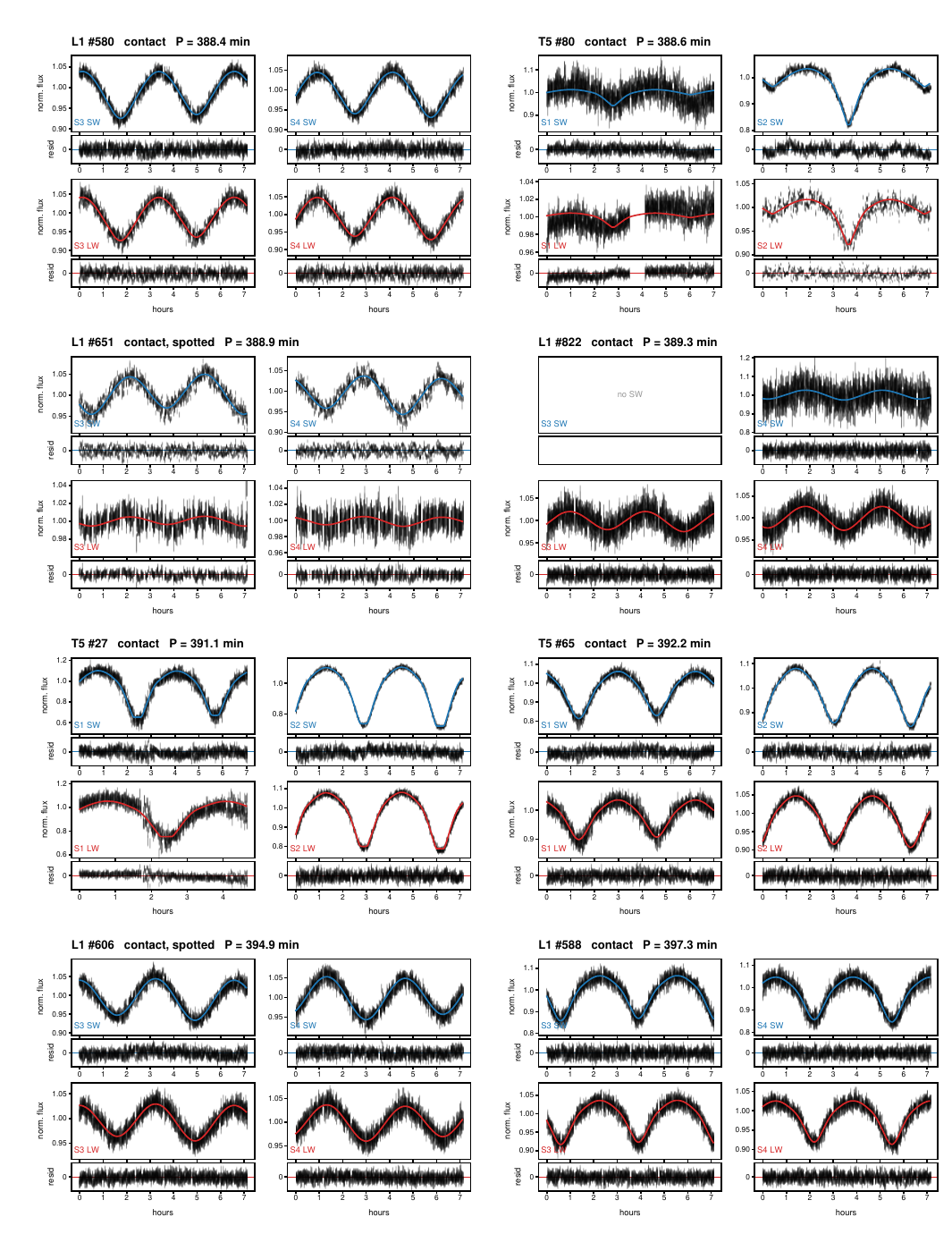}
\caption{Contact binaries, continued (page 8 of 33).}
\end{figure*}
\clearpage

\begin{figure*}
\centering
\includegraphics[width=0.98\textwidth,height=0.94\textheight,keepaspectratio]{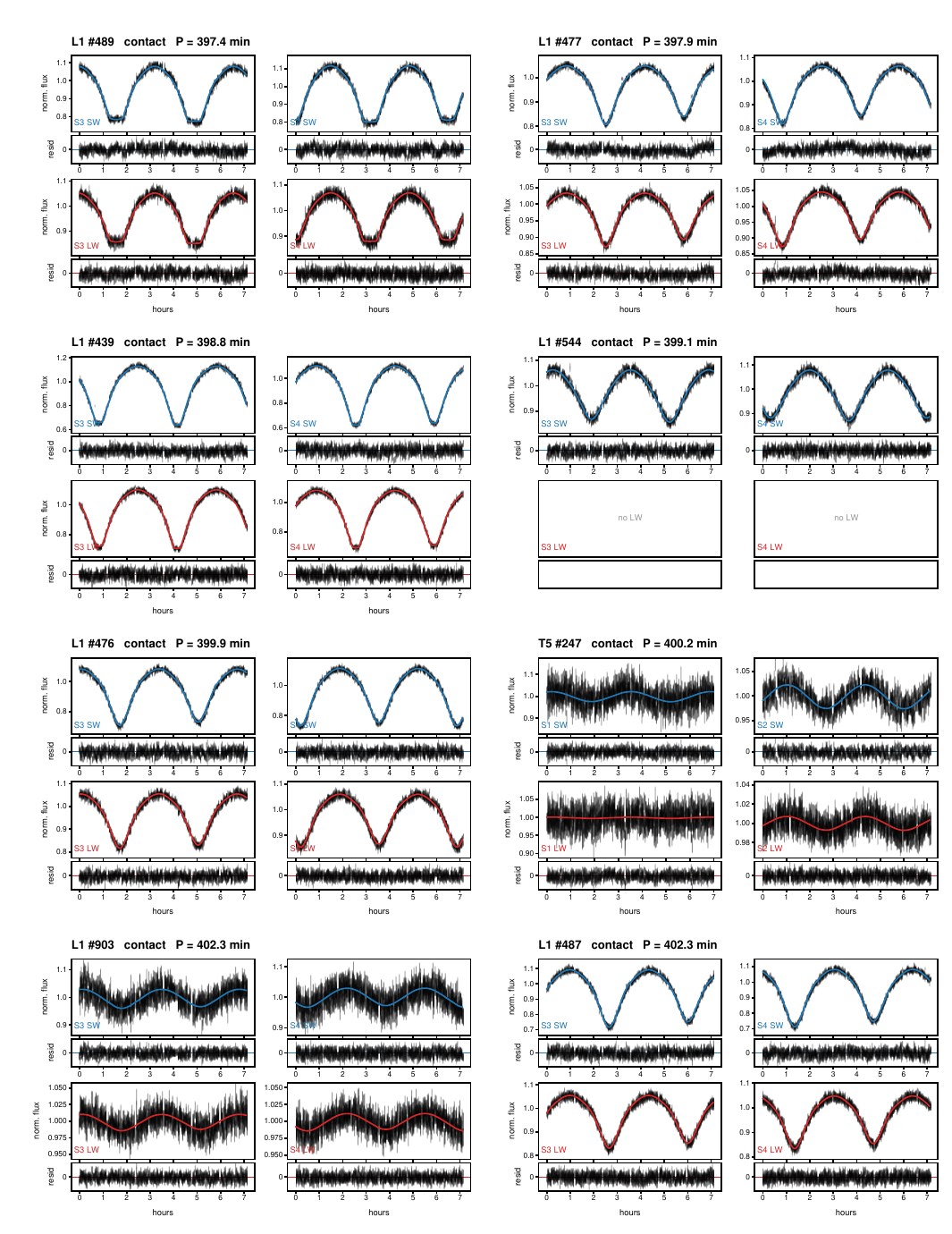}
\caption{Contact binaries, continued (page 9 of 33).}
\end{figure*}
\clearpage

\begin{figure*}
\centering
\includegraphics[width=0.98\textwidth,height=0.94\textheight,keepaspectratio]{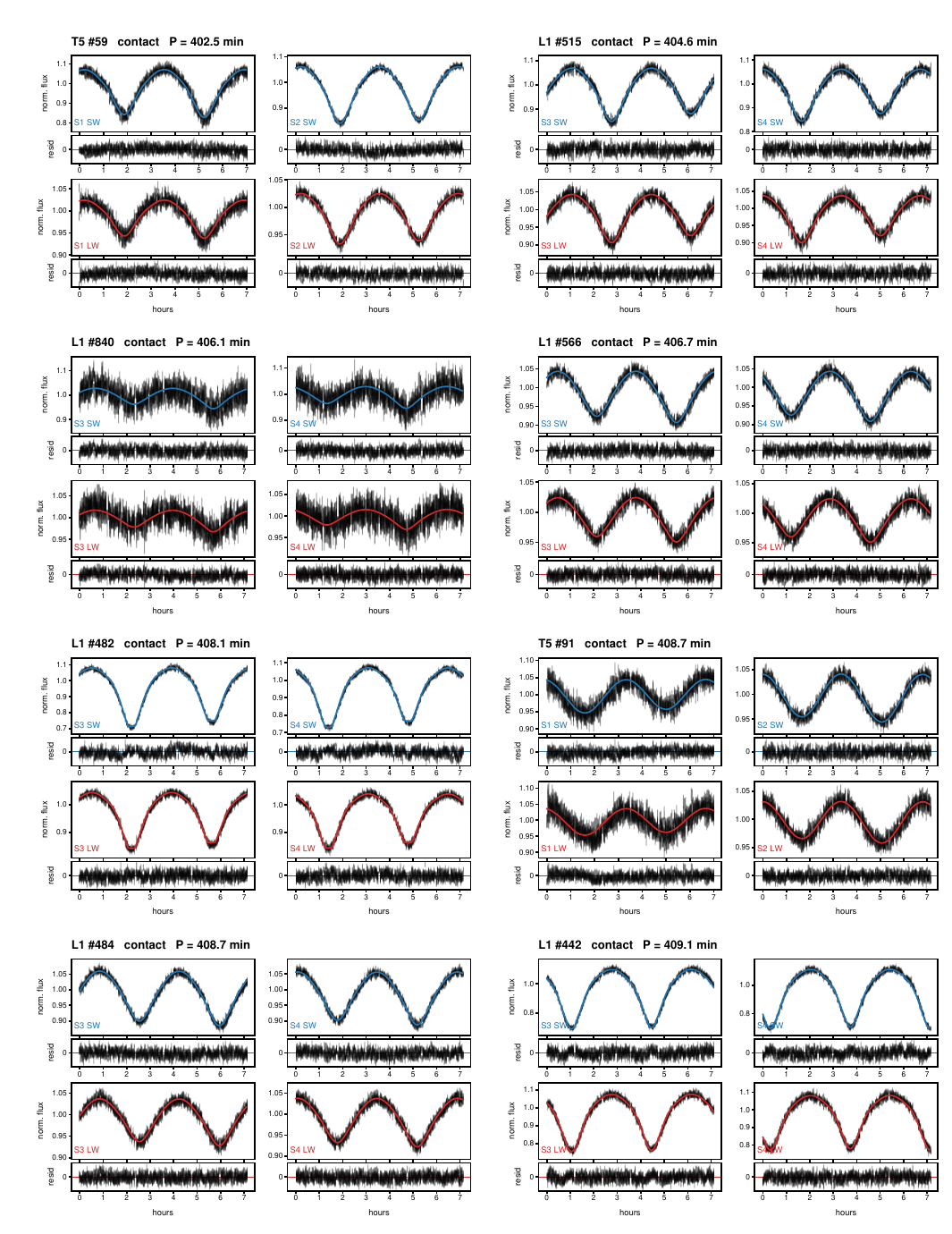}
\caption{Contact binaries, continued (page 10 of 33).}
\end{figure*}
\clearpage

\begin{figure*}
\centering
\includegraphics[width=0.98\textwidth,height=0.94\textheight,keepaspectratio]{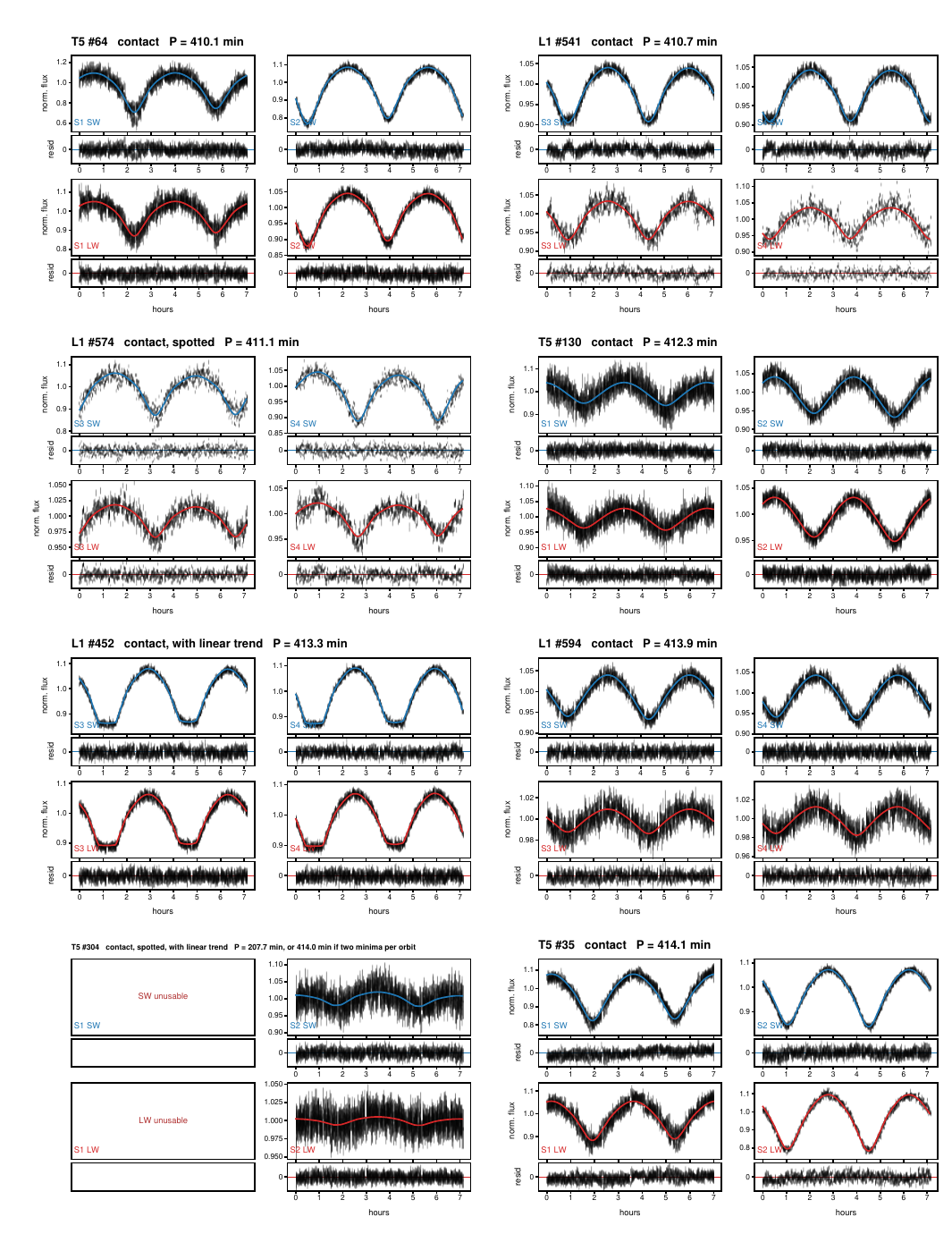}
\caption{Contact binaries, continued (page 11 of 33).}
\end{figure*}
\clearpage

\begin{figure*}
\centering
\includegraphics[width=0.98\textwidth,height=0.94\textheight,keepaspectratio]{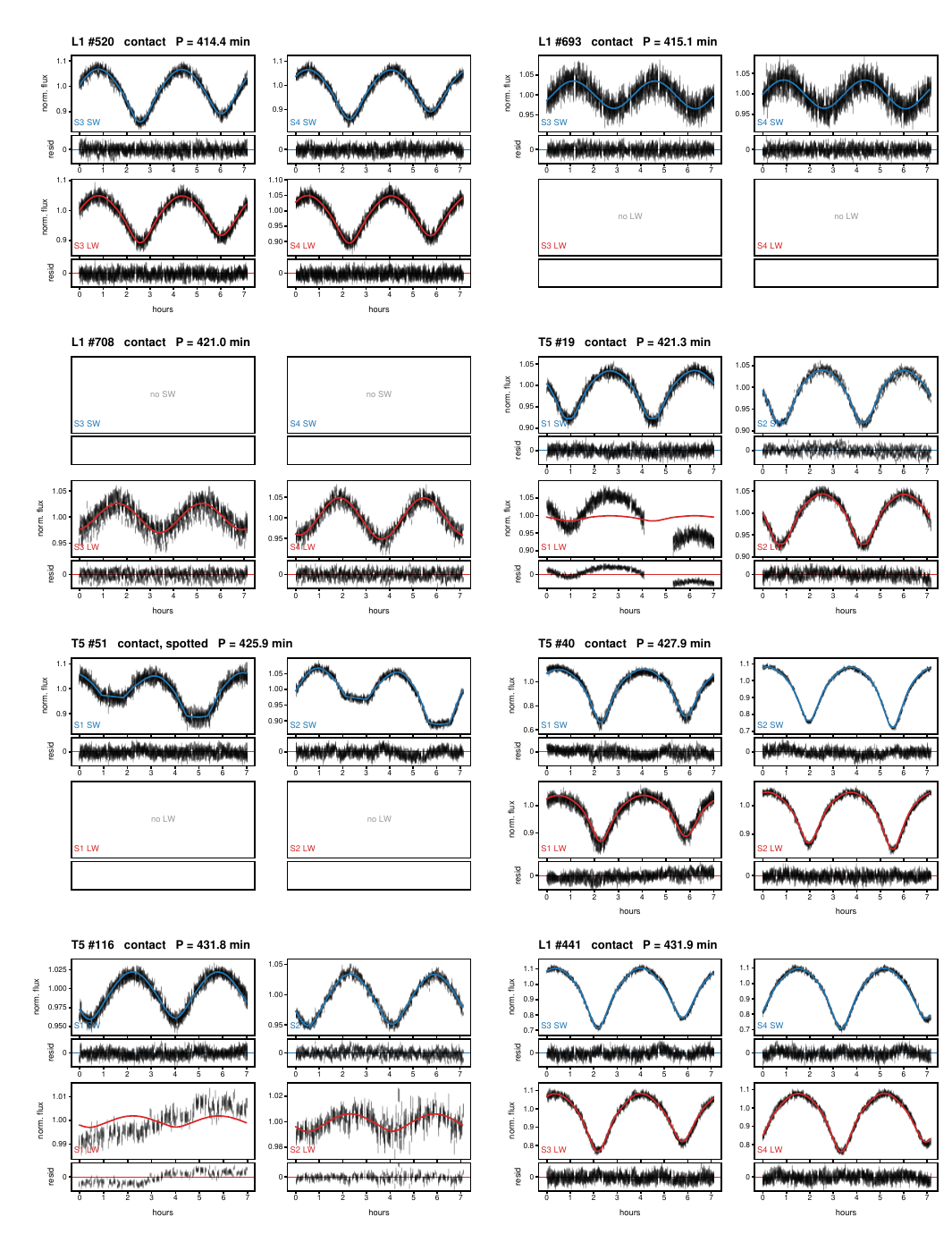}
\caption{Contact binaries, continued (page 12 of 33).}
\end{figure*}
\clearpage

\begin{figure*}
\centering
\includegraphics[width=0.98\textwidth,height=0.94\textheight,keepaspectratio]{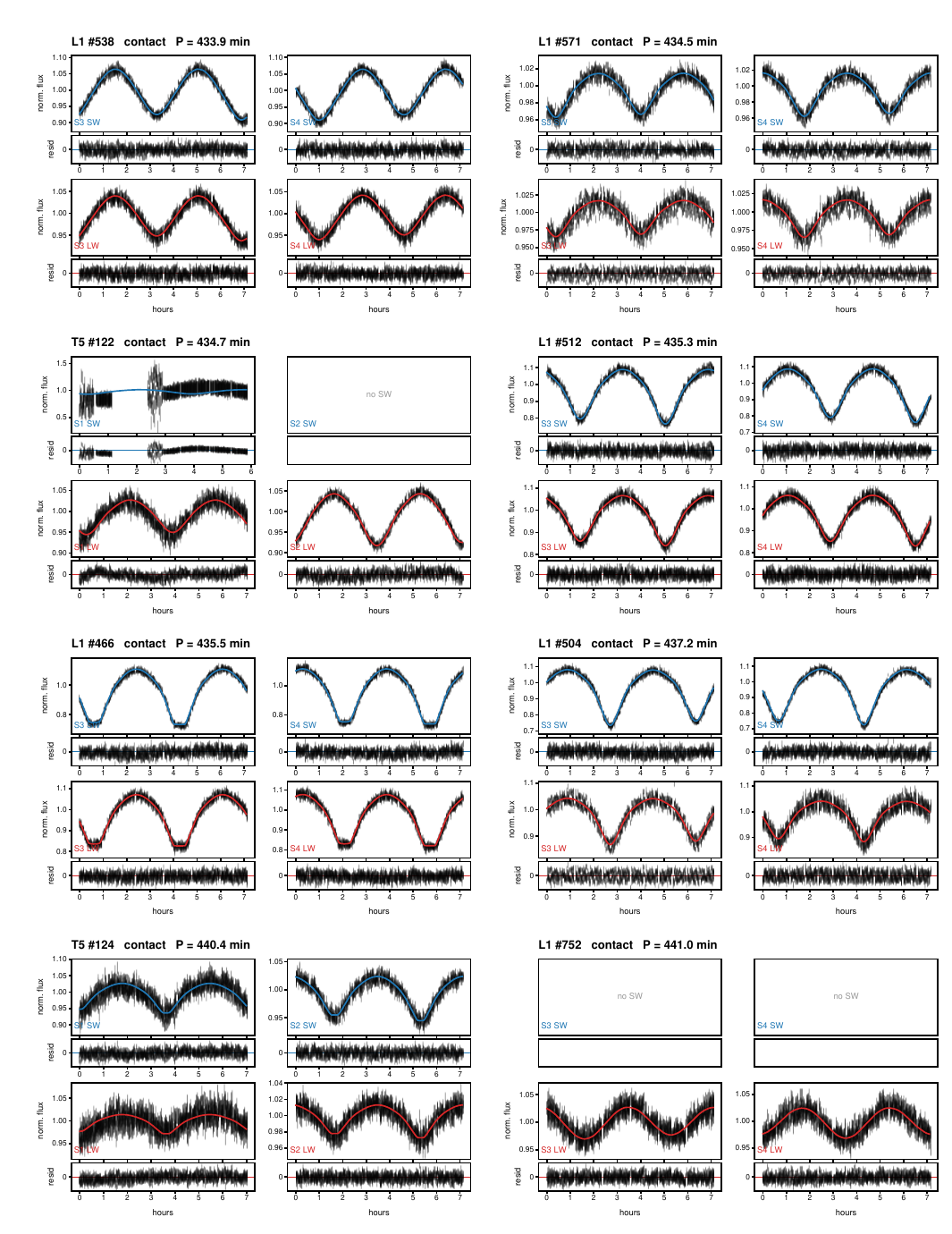}
\caption{Contact binaries, continued (page 13 of 33).}
\end{figure*}
\clearpage

\begin{figure*}
\centering
\includegraphics[width=0.98\textwidth,height=0.94\textheight,keepaspectratio]{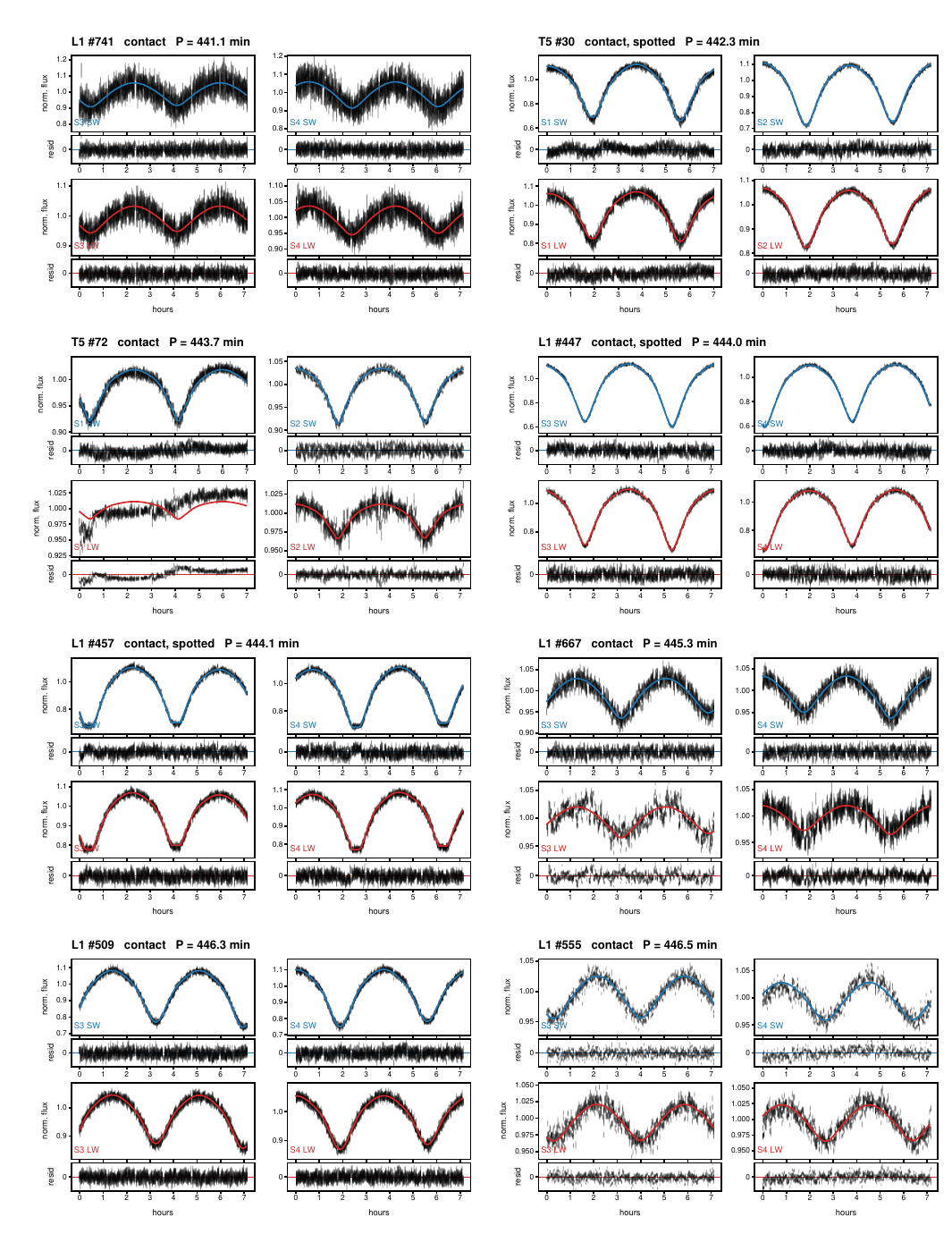}
\caption{Contact binaries, continued (page 14 of 33).}
\end{figure*}
\clearpage

\begin{figure*}
\centering
\includegraphics[width=0.98\textwidth,height=0.94\textheight,keepaspectratio]{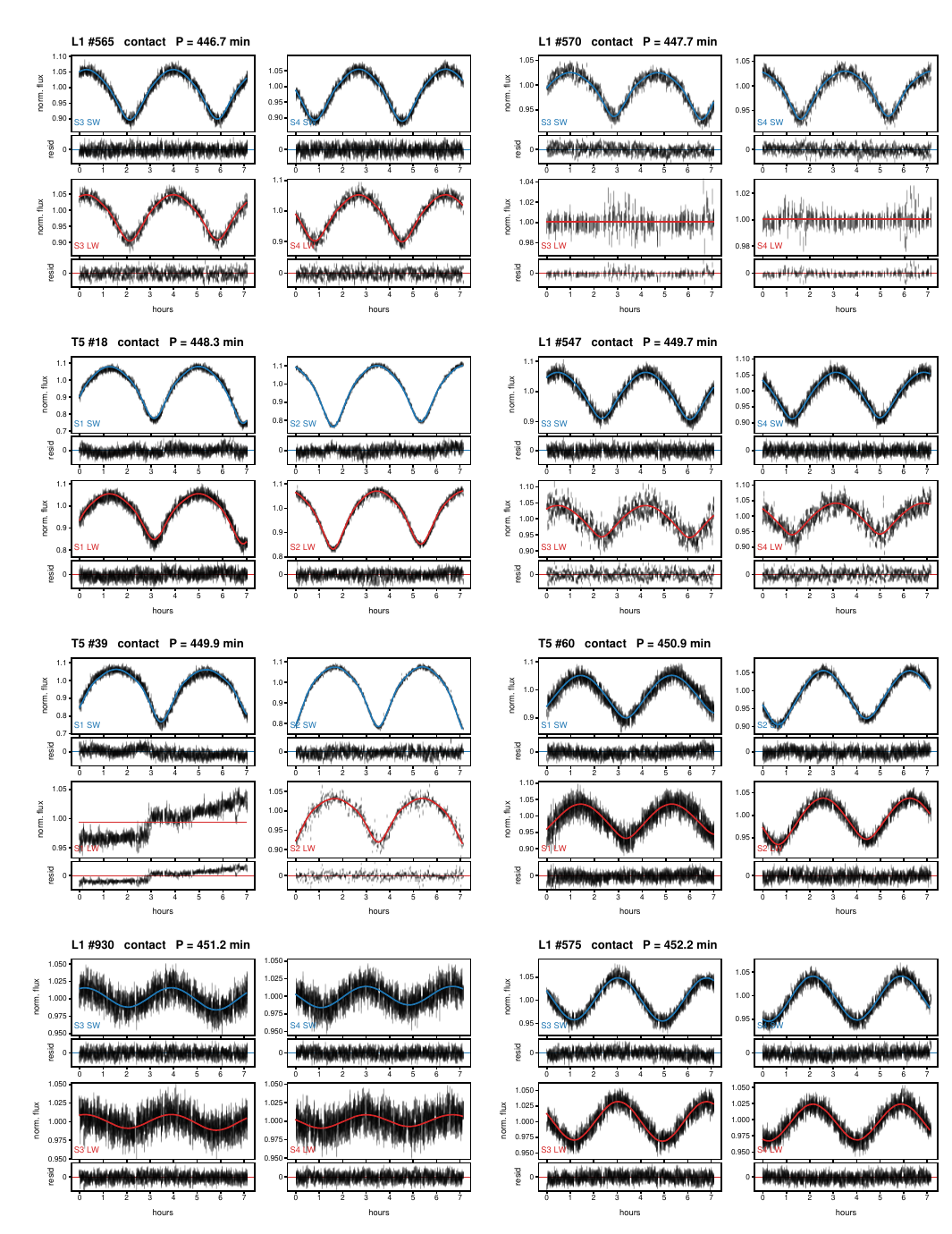}
\caption{Contact binaries, continued (page 15 of 33).}
\end{figure*}
\clearpage

\begin{figure*}
\centering
\includegraphics[width=0.98\textwidth,height=0.94\textheight,keepaspectratio]{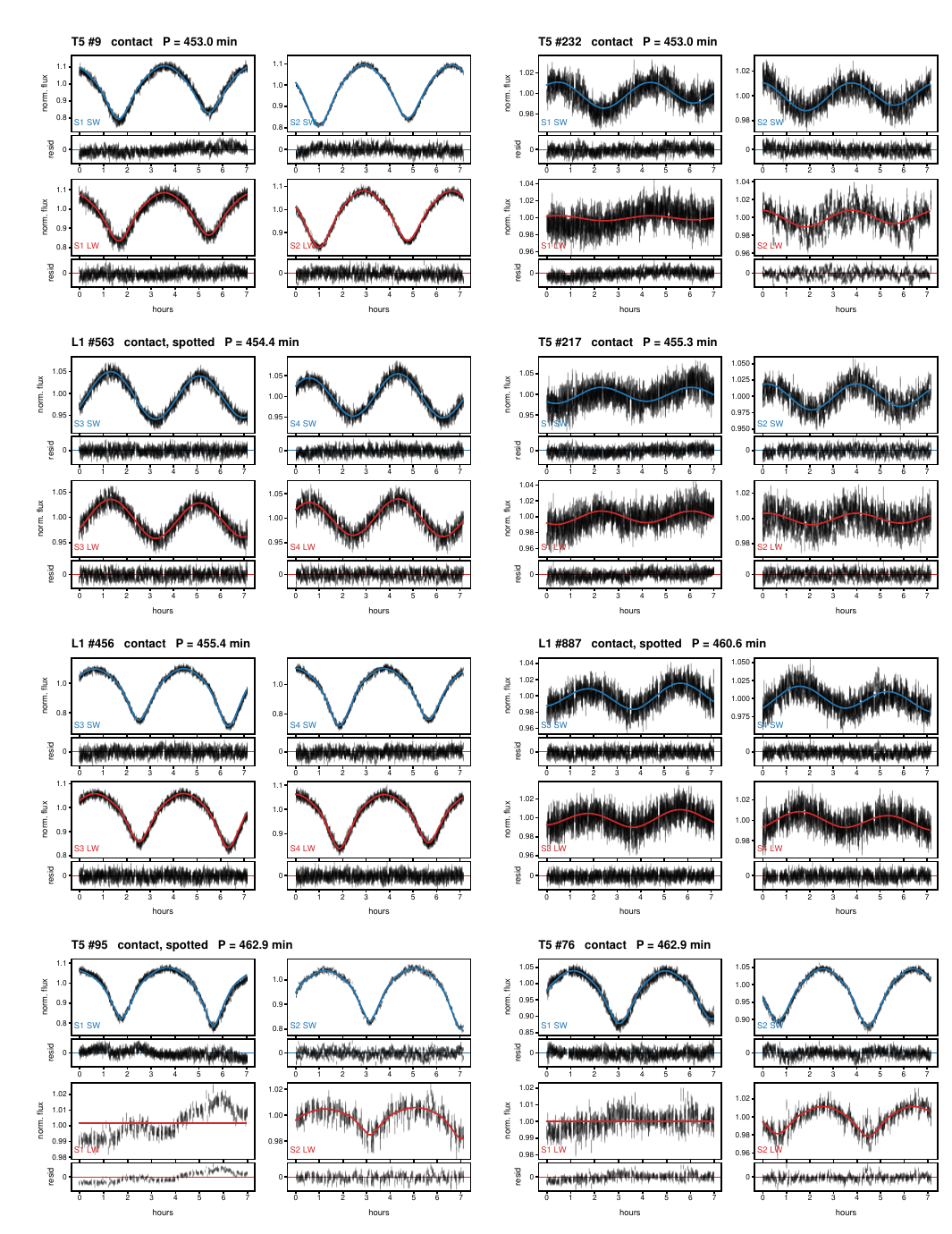}
\caption{Contact binaries, continued (page 16 of 33).}
\end{figure*}
\clearpage

\begin{figure*}
\centering
\includegraphics[width=0.98\textwidth,height=0.94\textheight,keepaspectratio]{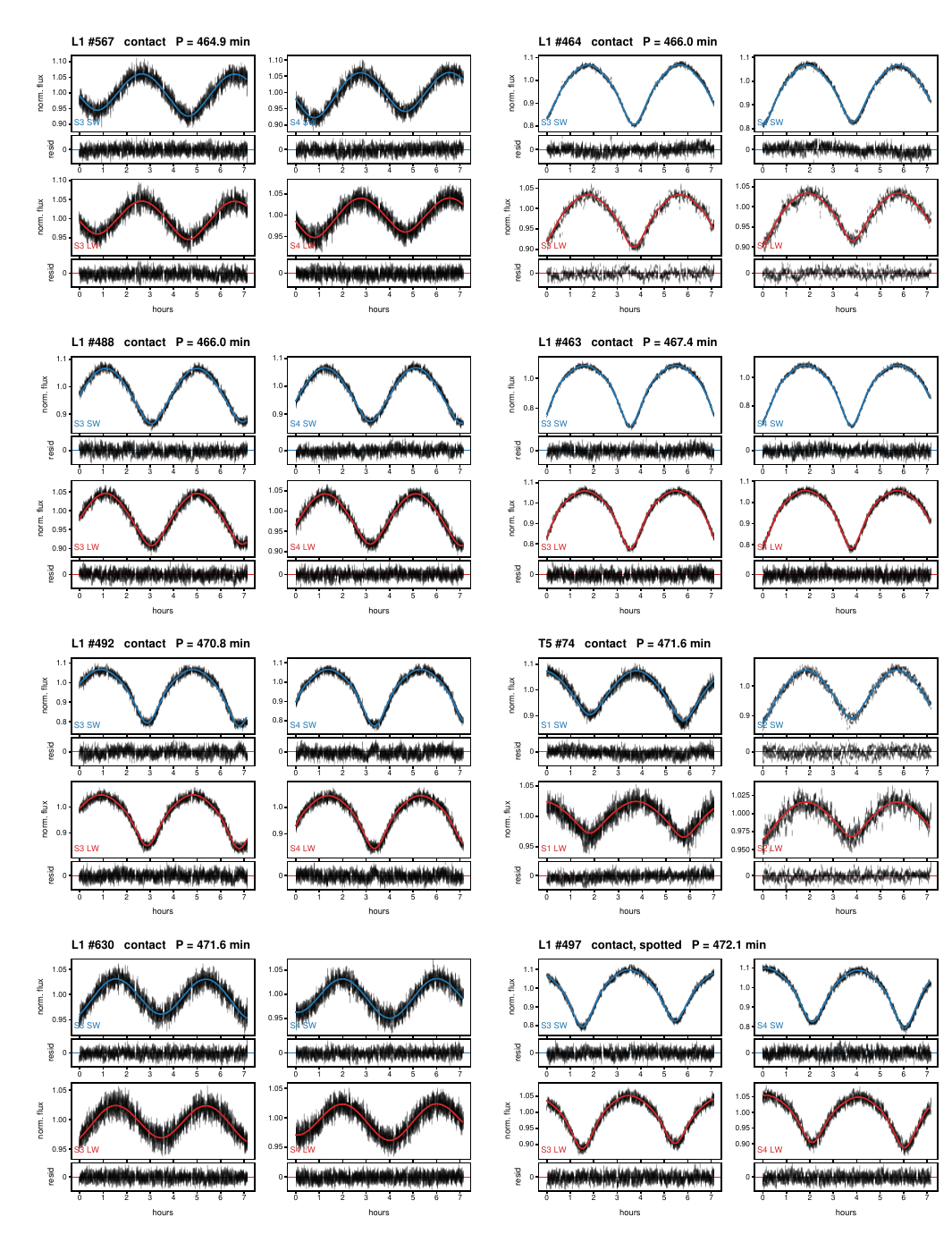}
\caption{Contact binaries, continued (page 17 of 33).}
\end{figure*}
\clearpage

\begin{figure*}
\centering
\includegraphics[width=0.98\textwidth,height=0.94\textheight,keepaspectratio]{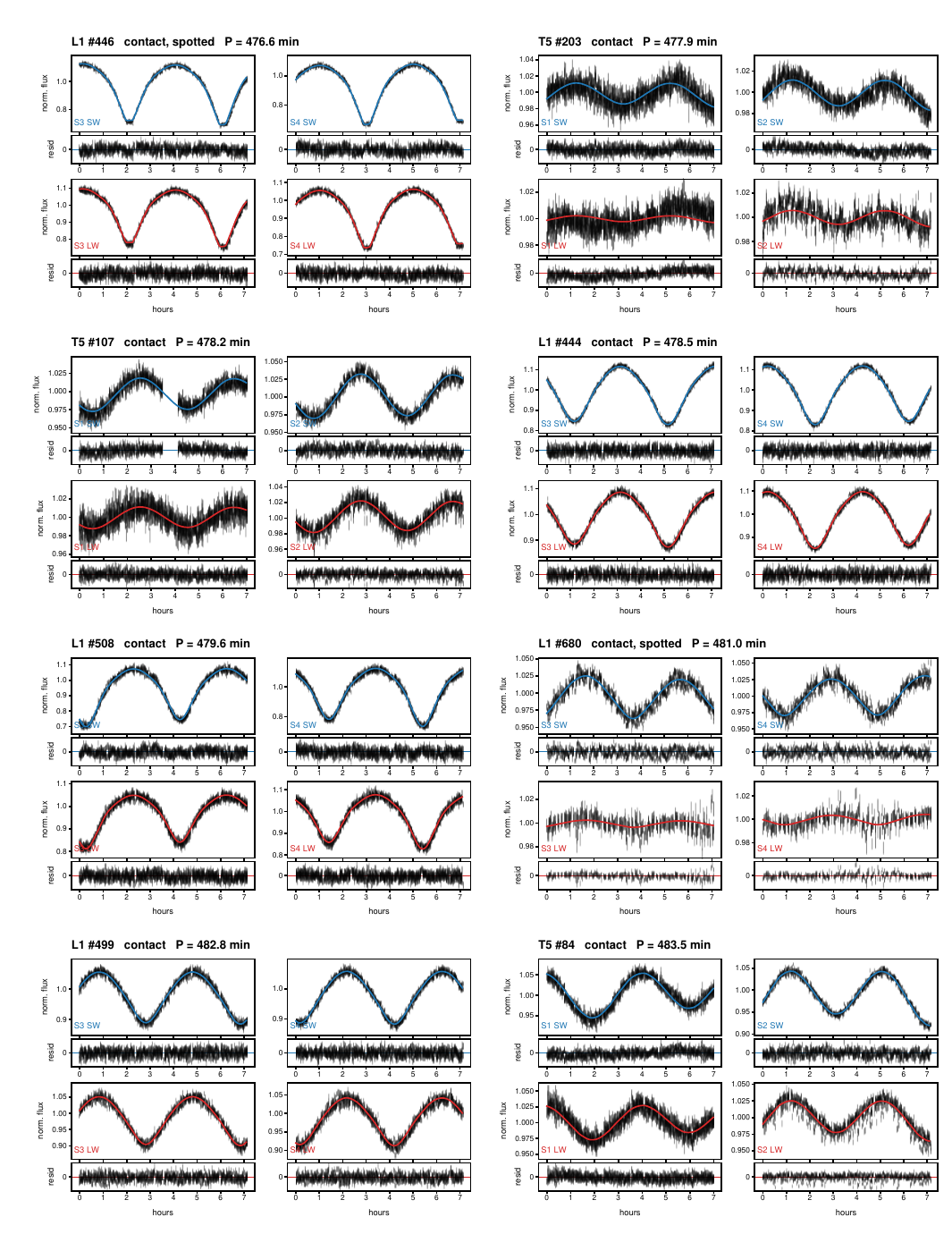}
\caption{Contact binaries, continued (page 18 of 33).}
\end{figure*}
\clearpage

\begin{figure*}
\centering
\includegraphics[width=0.98\textwidth,height=0.94\textheight,keepaspectratio]{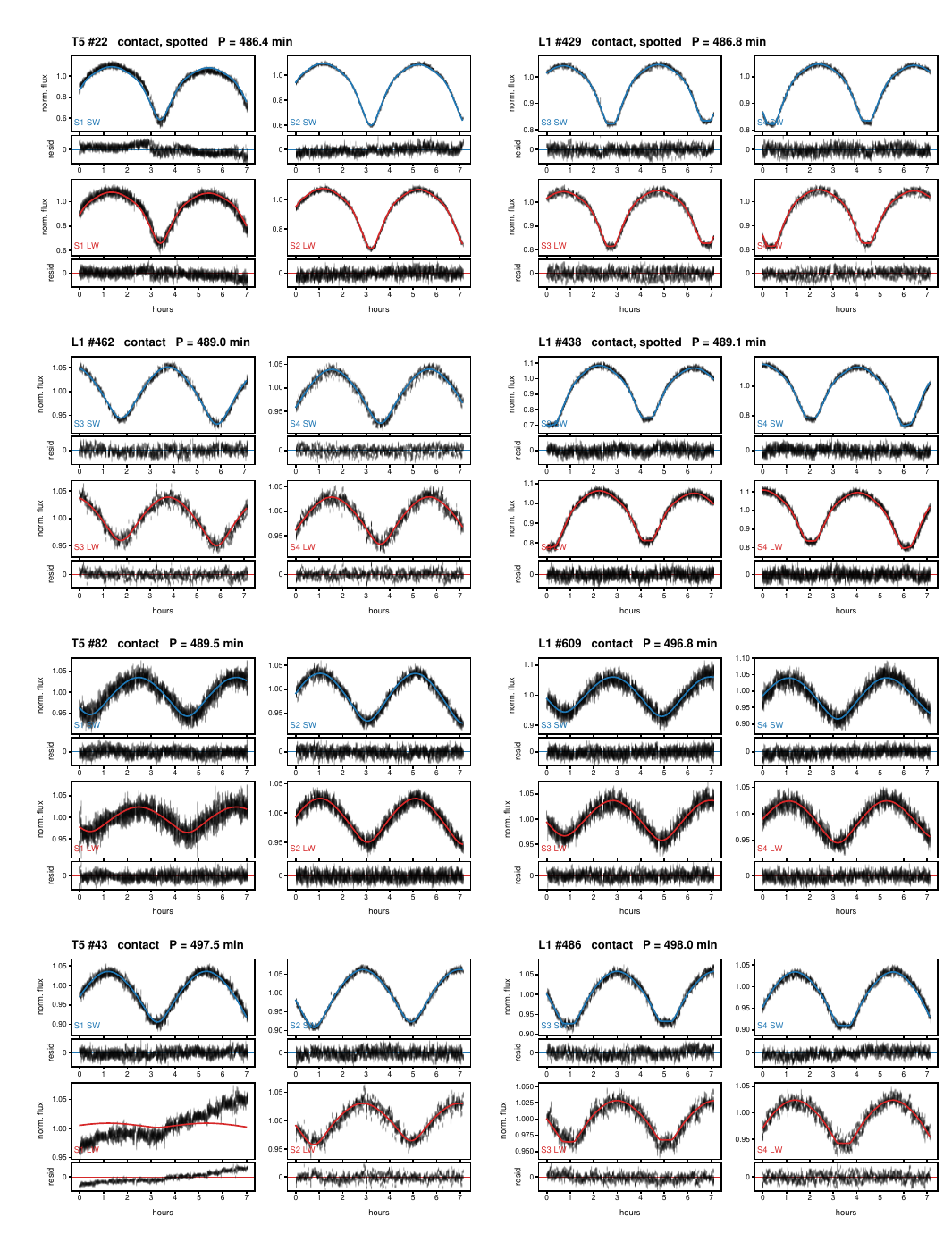}
\caption{Contact binaries, continued (page 19 of 33).}
\end{figure*}
\clearpage

\begin{figure*}
\centering
\includegraphics[width=0.98\textwidth,height=0.94\textheight,keepaspectratio]{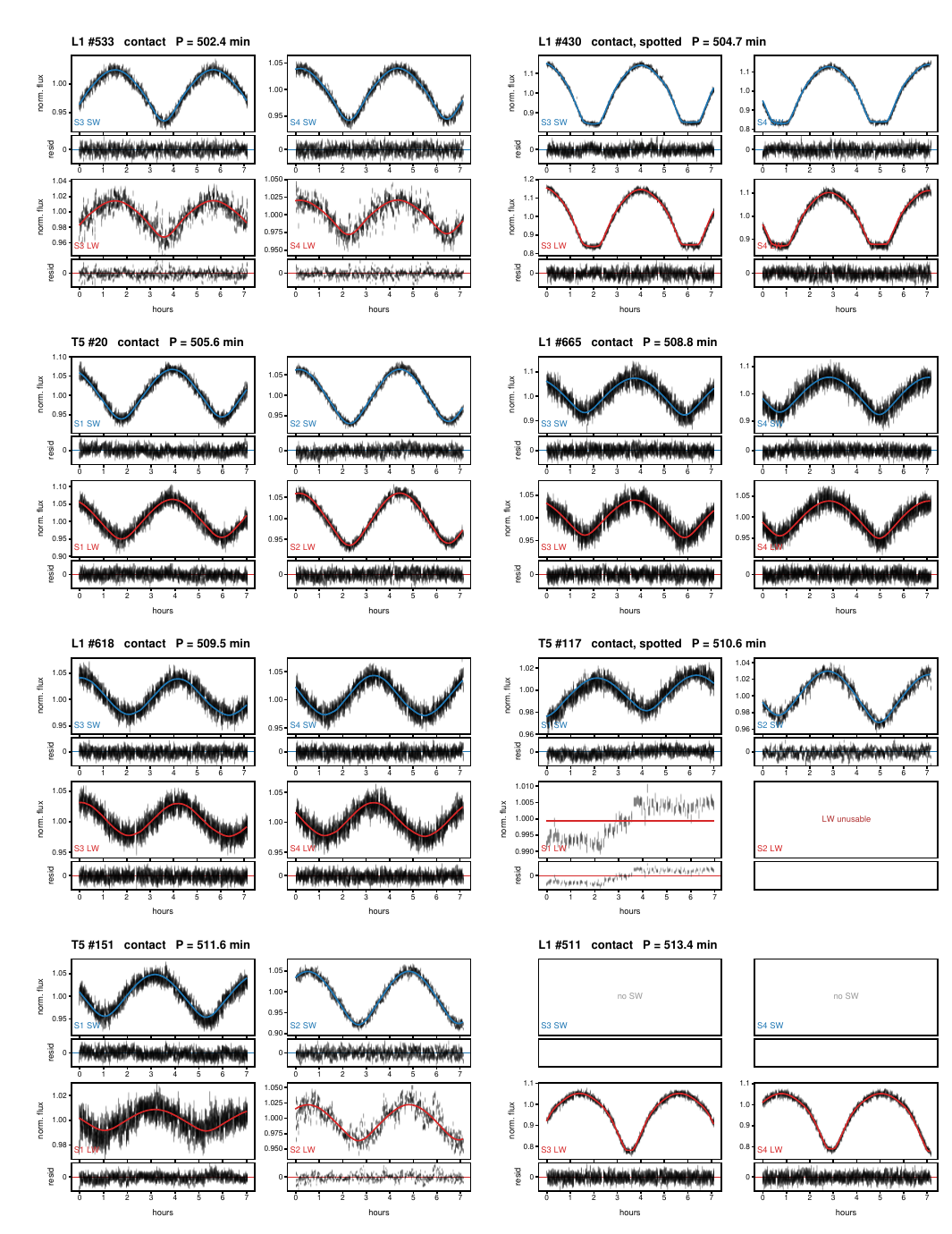}
\caption{Contact binaries, continued (page 20 of 33).}
\end{figure*}
\clearpage

\begin{figure*}
\centering
\includegraphics[width=0.98\textwidth,height=0.94\textheight,keepaspectratio]{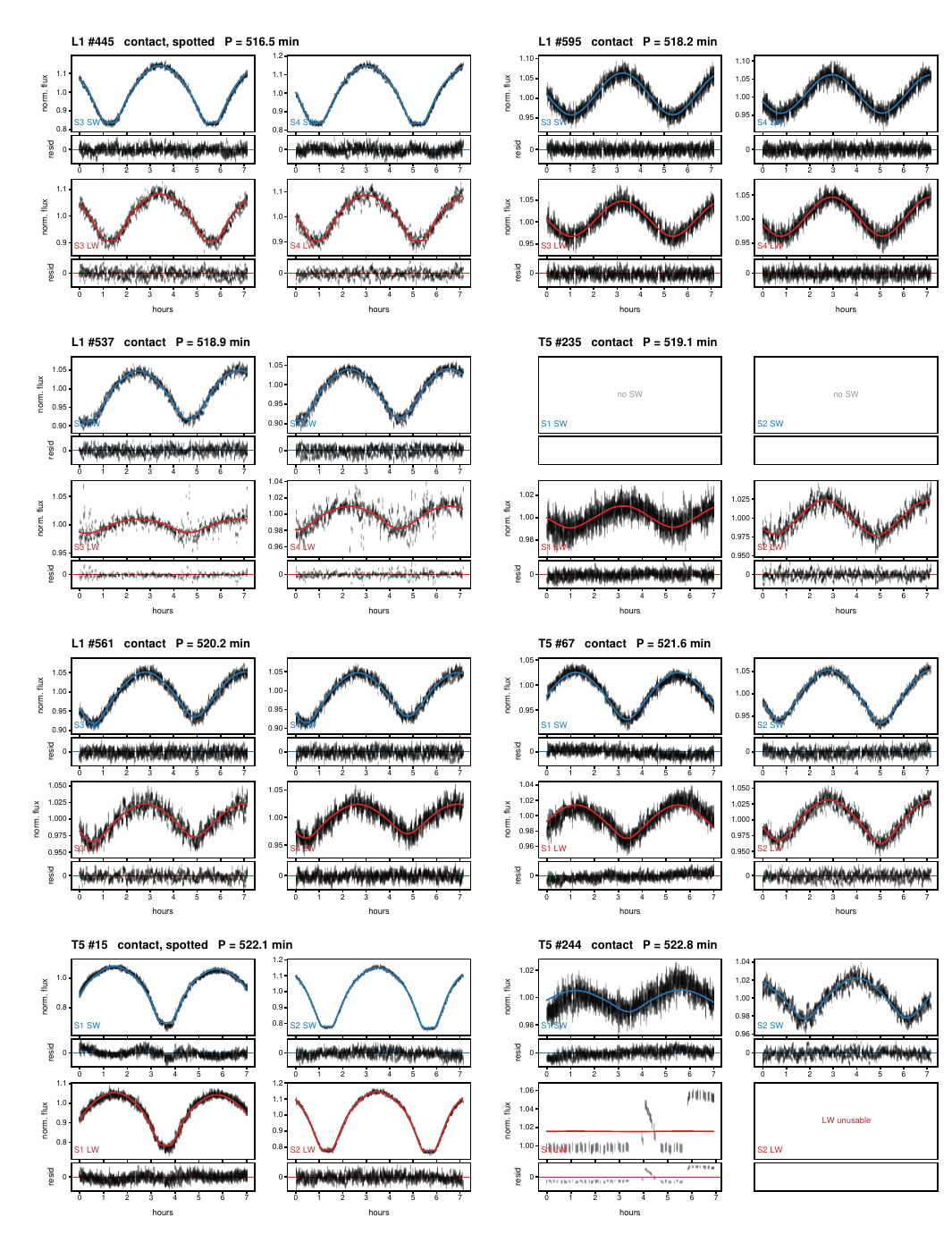}
\caption{Contact binaries, continued (page 21 of 33).}
\end{figure*}
\clearpage

\begin{figure*}
\centering
\includegraphics[width=0.98\textwidth,height=0.94\textheight,keepaspectratio]{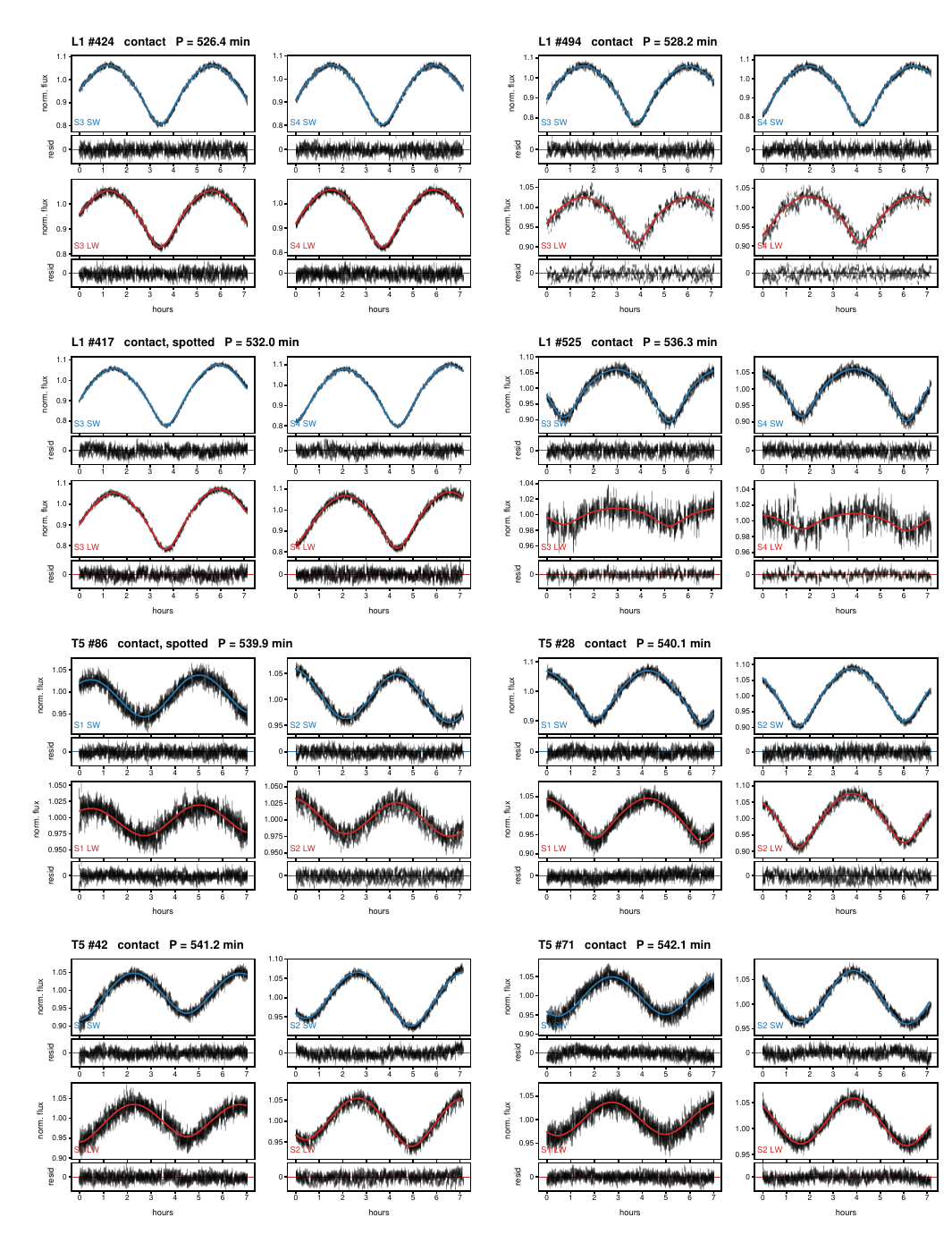}
\caption{Contact binaries, continued (page 22 of 33).}
\end{figure*}
\clearpage

\begin{figure*}
\centering
\includegraphics[width=0.98\textwidth,height=0.94\textheight,keepaspectratio]{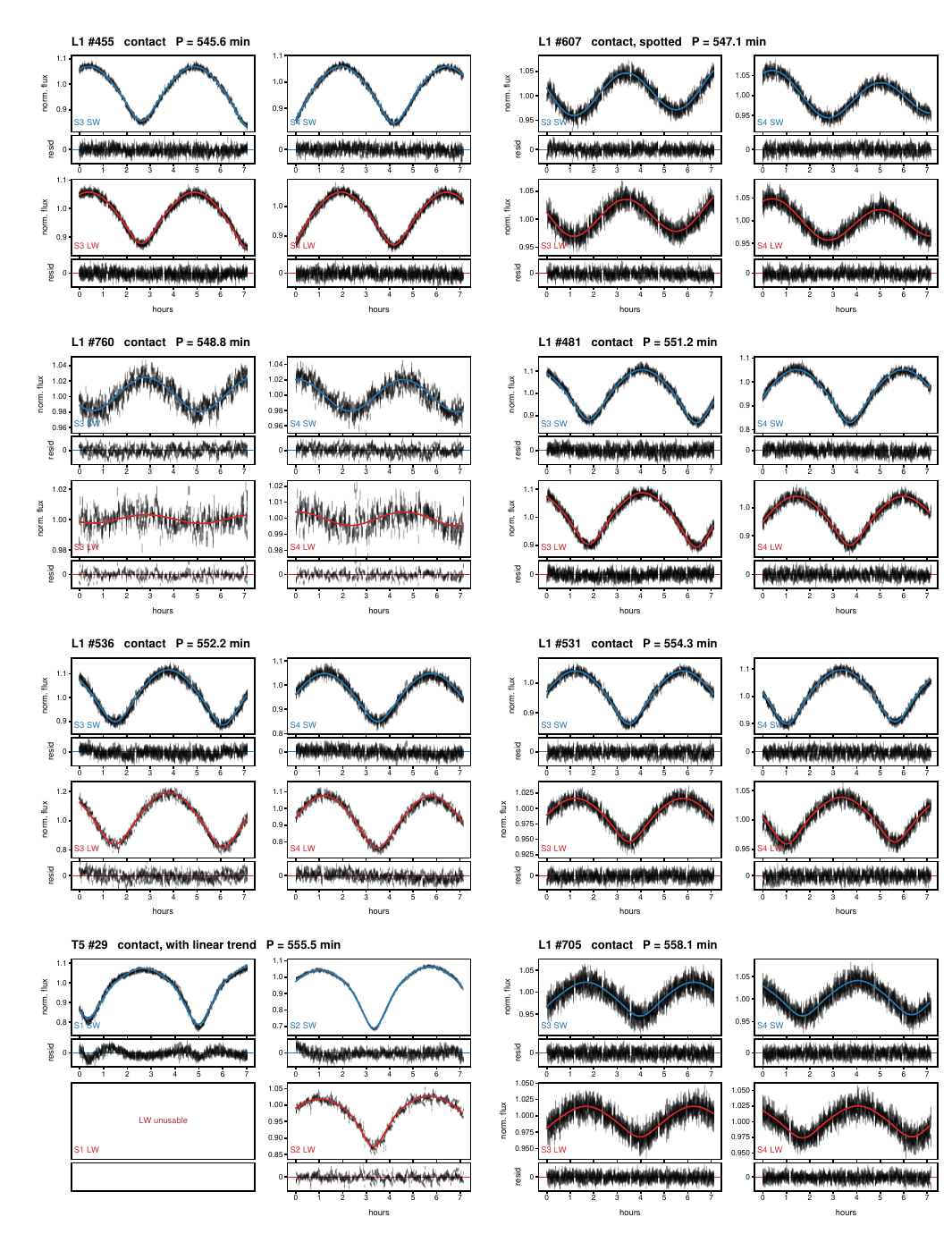}
\caption{Contact binaries, continued (page 23 of 33).}
\end{figure*}
\clearpage

\begin{figure*}
\centering
\includegraphics[width=0.98\textwidth,height=0.94\textheight,keepaspectratio]{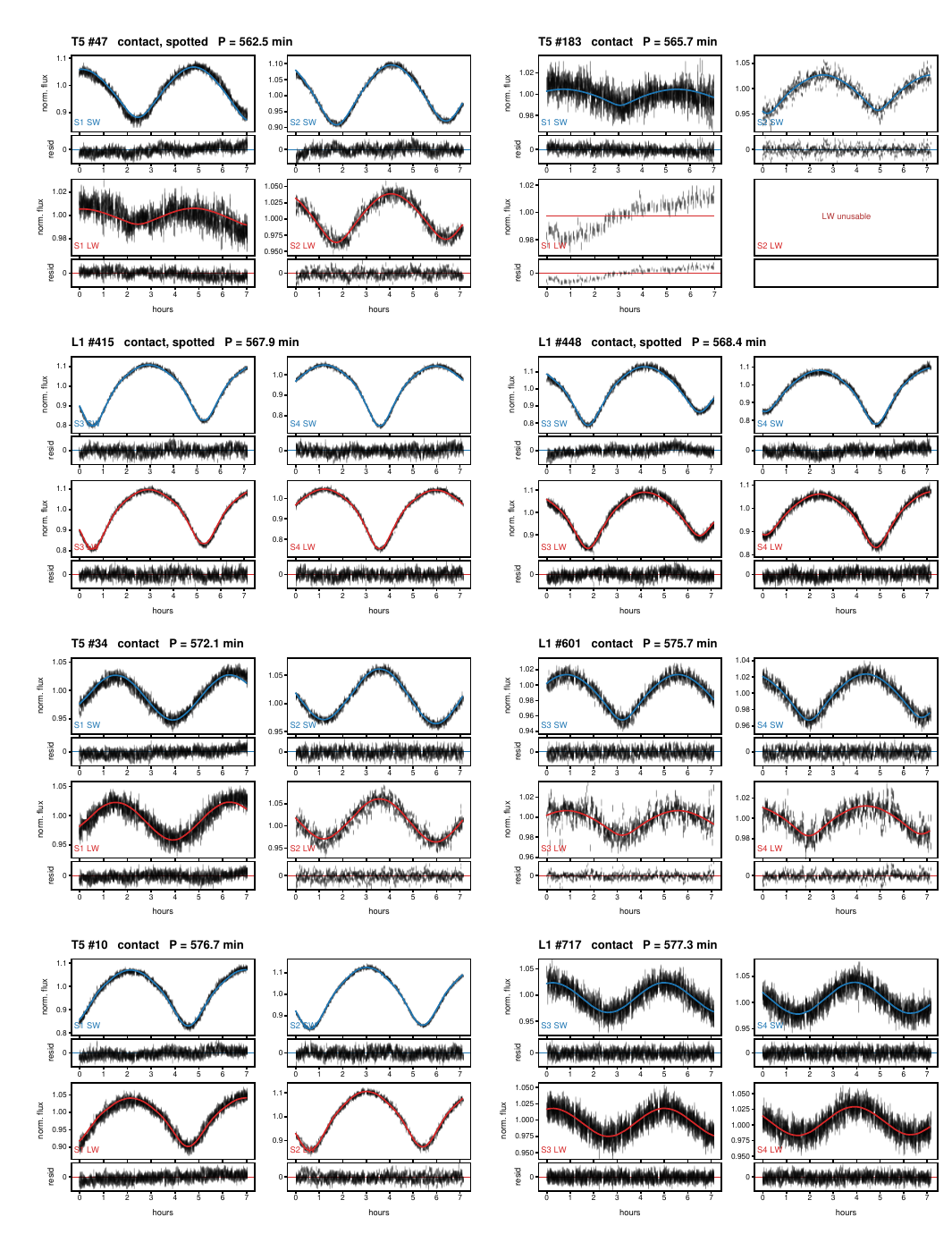}
\caption{Contact binaries, continued (page 24 of 33).}
\end{figure*}
\clearpage

\begin{figure*}
\centering
\includegraphics[width=0.98\textwidth,height=0.94\textheight,keepaspectratio]{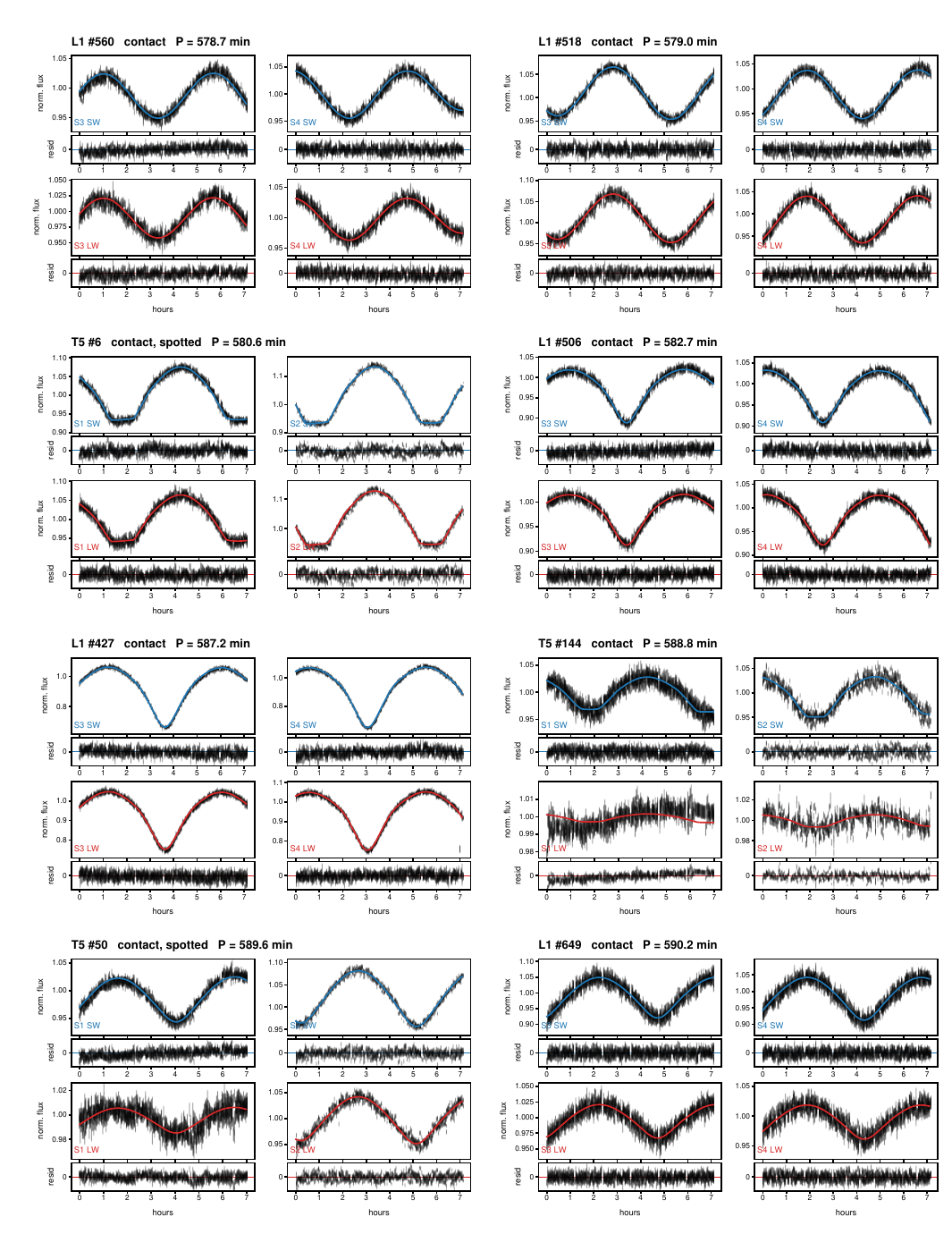}
\caption{Contact binaries, continued (page 25 of 33).}
\end{figure*}
\clearpage

\begin{figure*}
\centering
\includegraphics[width=0.98\textwidth,height=0.94\textheight,keepaspectratio]{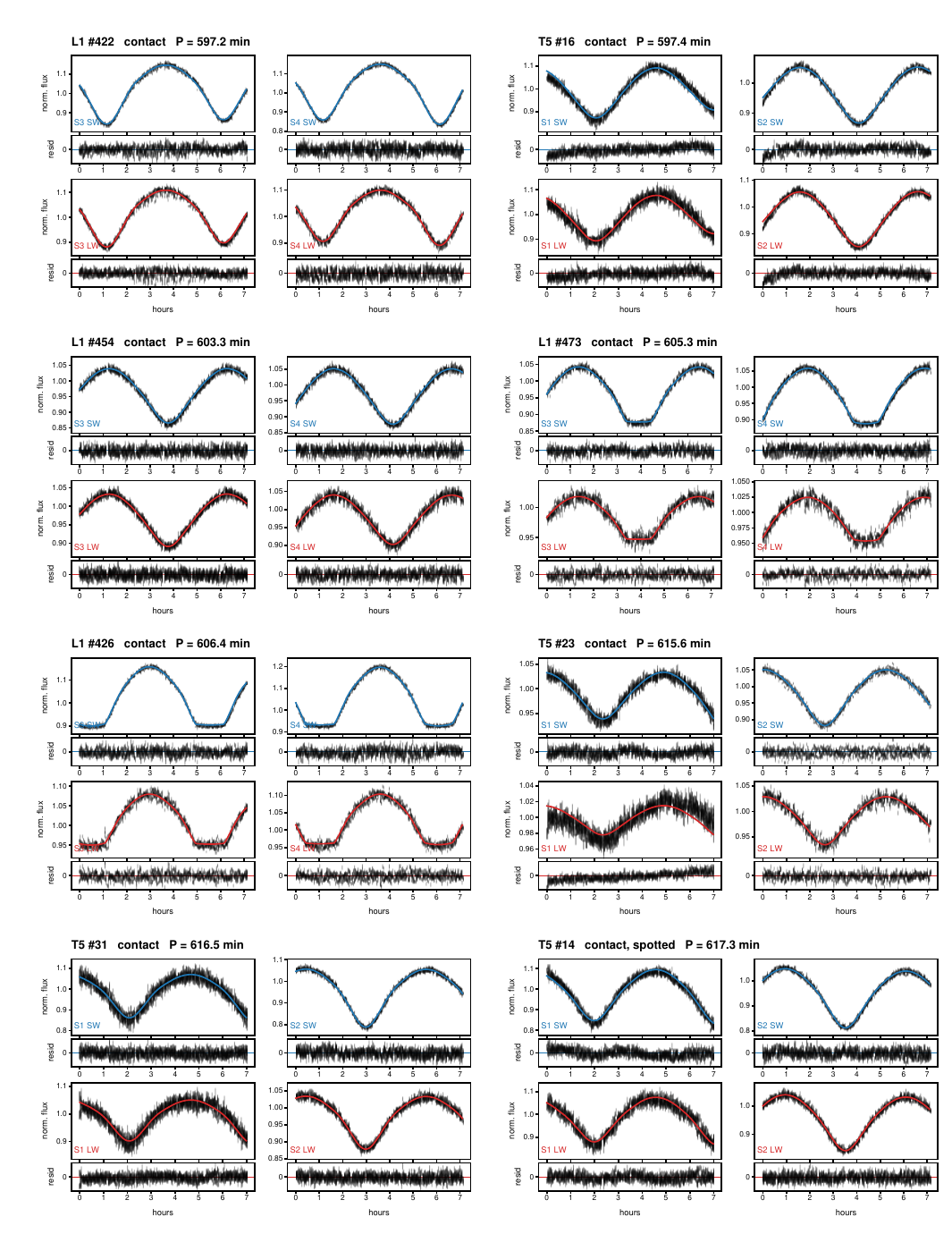}
\caption{Contact binaries, continued (page 26 of 33).}
\end{figure*}
\clearpage

\begin{figure*}
\centering
\includegraphics[width=0.98\textwidth,height=0.94\textheight,keepaspectratio]{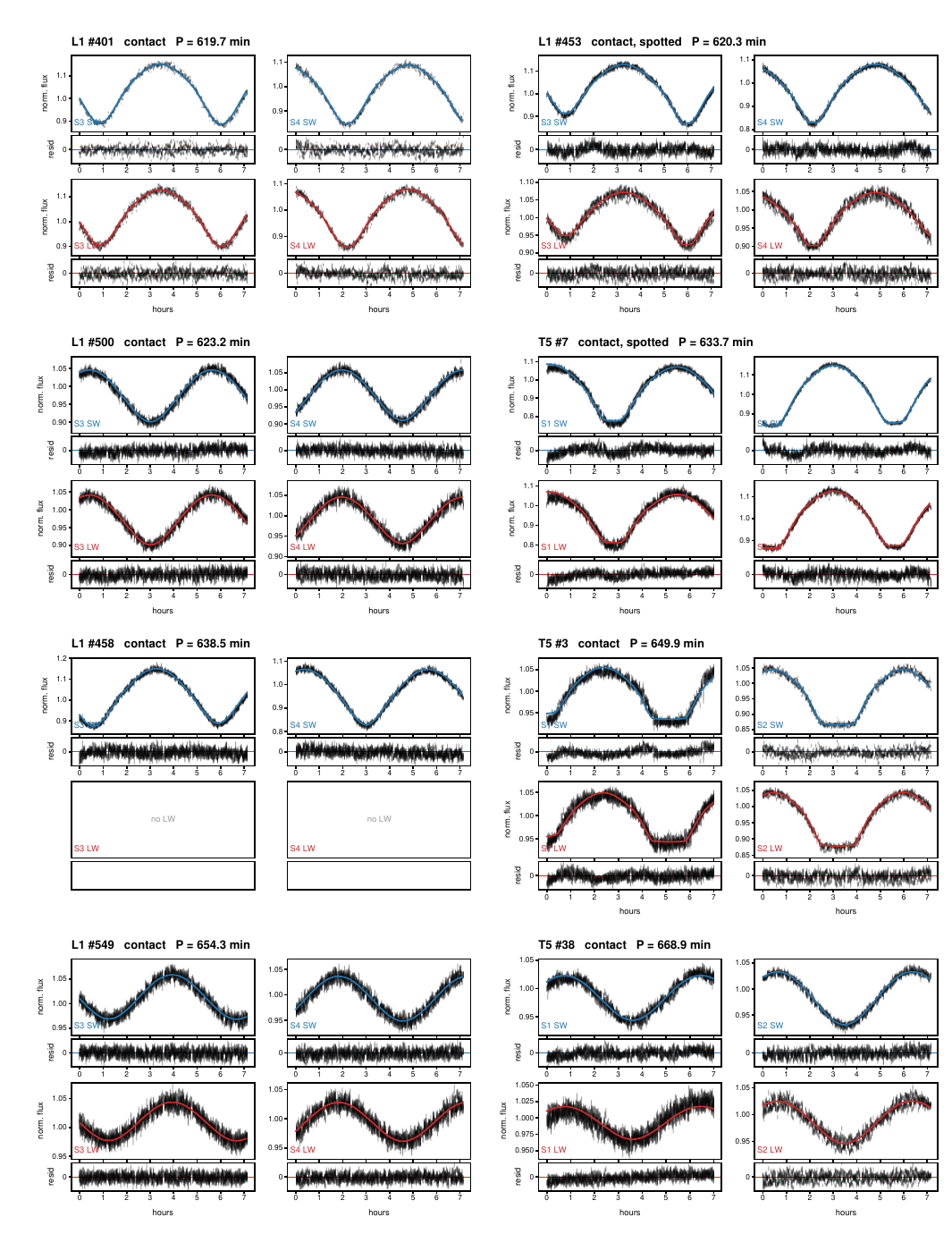}
\caption{Contact binaries, continued (page 27 of 33).}
\end{figure*}
\clearpage

\begin{figure*}
\centering
\includegraphics[width=0.98\textwidth,height=0.94\textheight,keepaspectratio]{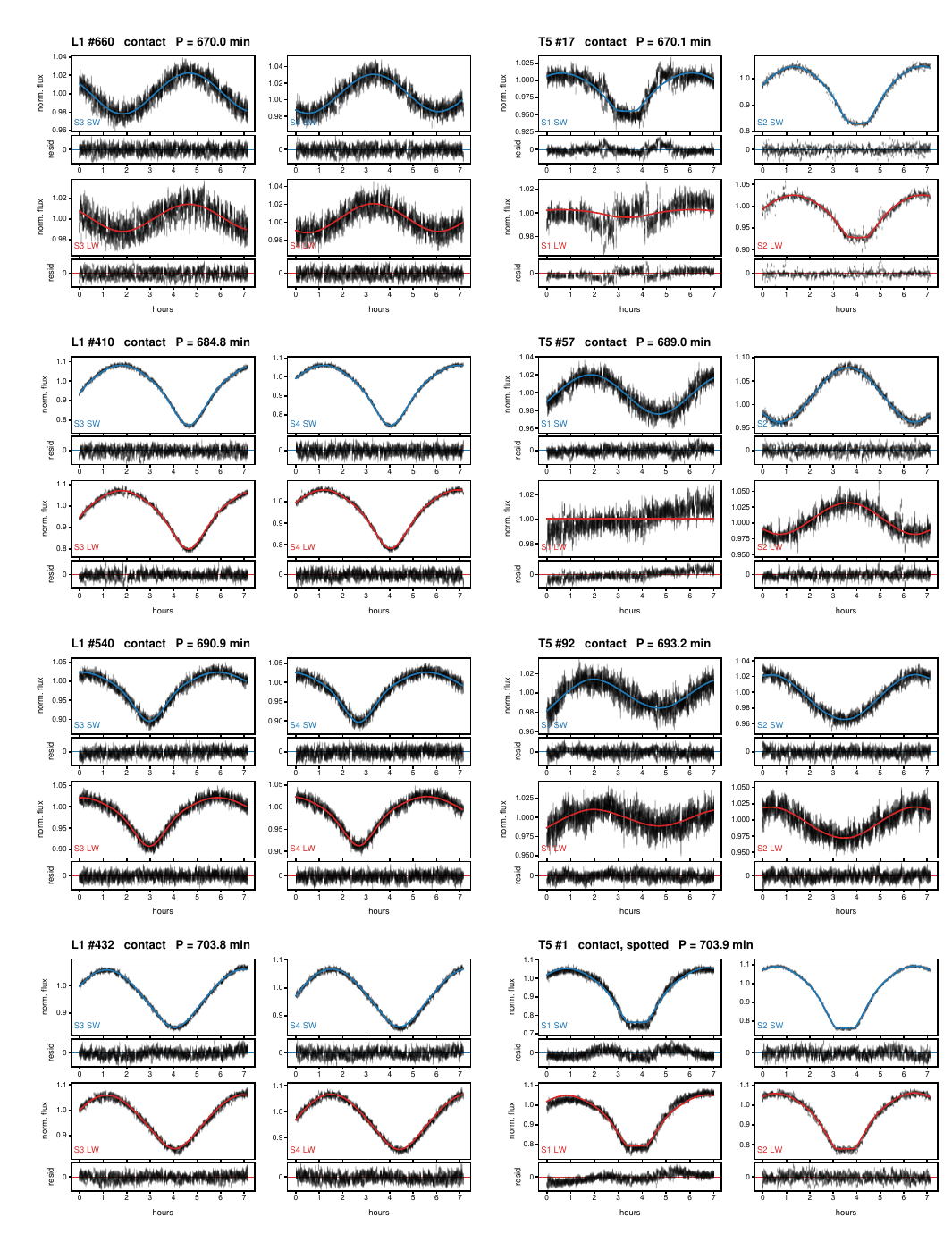}
\caption{Contact binaries, continued (page 28 of 33).}
\end{figure*}
\clearpage

\begin{figure*}
\centering
\includegraphics[width=0.98\textwidth,height=0.94\textheight,keepaspectratio]{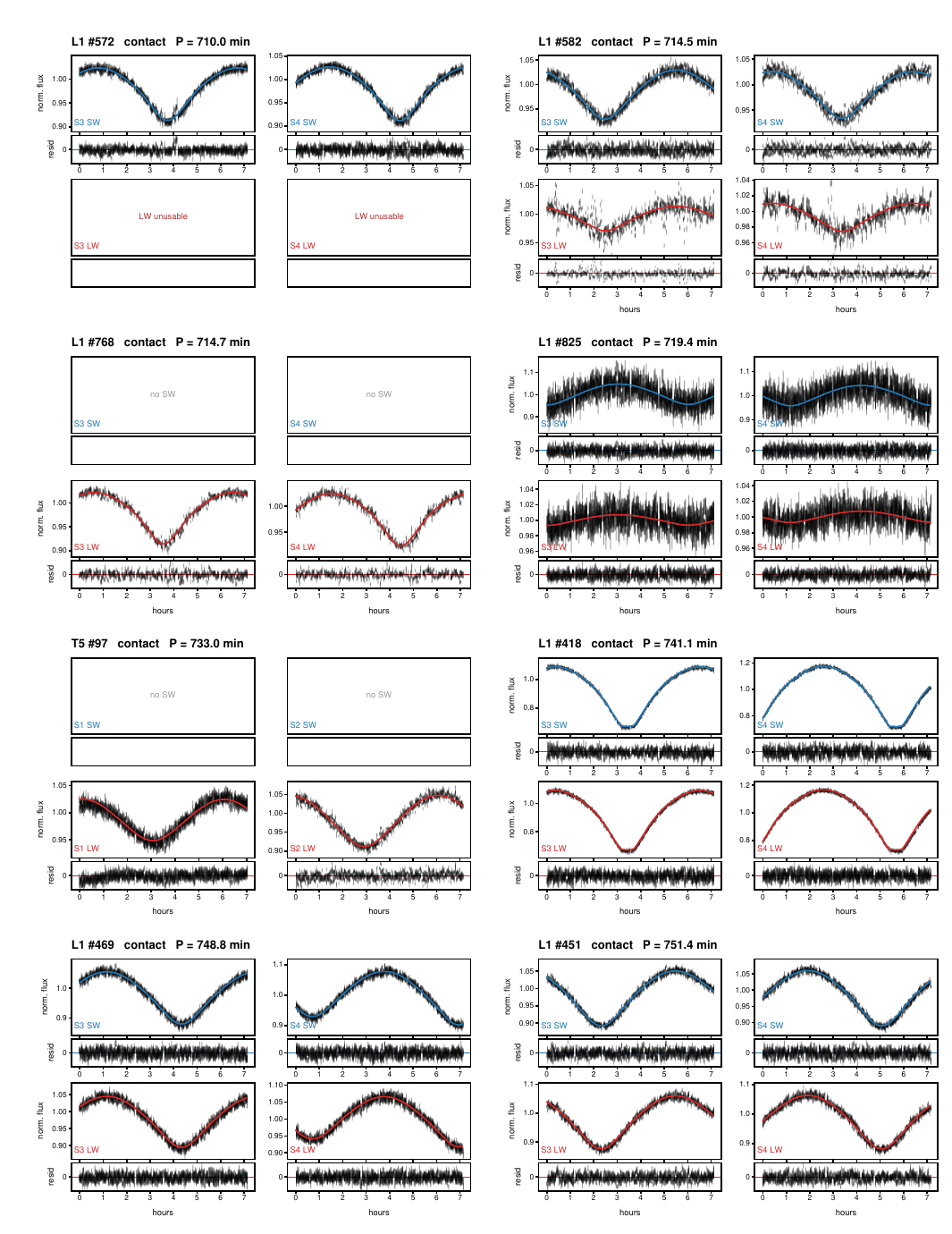}
\caption{Contact binaries, continued (page 29 of 33).}
\end{figure*}
\clearpage

\begin{figure*}
\centering
\includegraphics[width=0.98\textwidth,height=0.94\textheight,keepaspectratio]{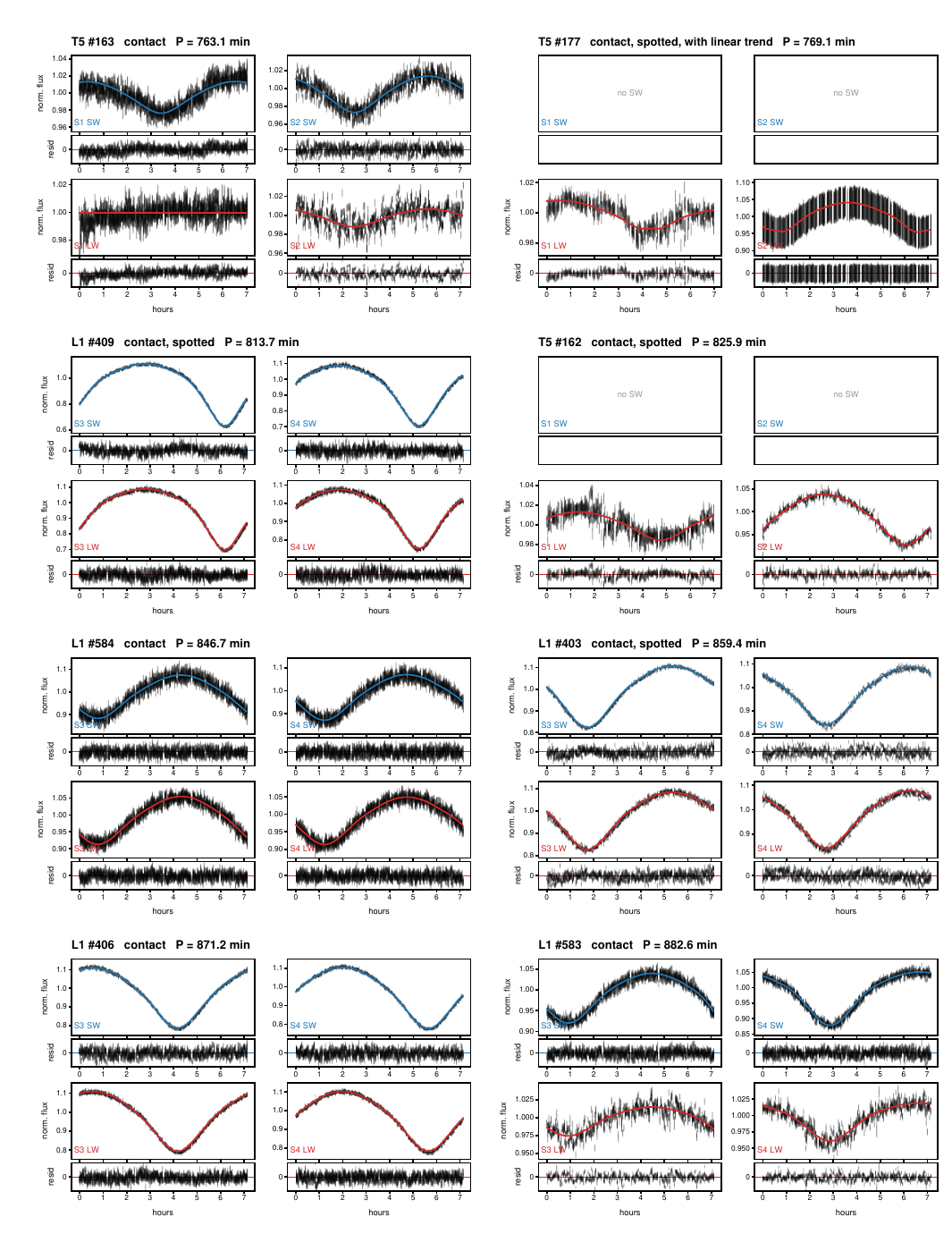}
\caption{Contact binaries, continued (page 30 of 33).}
\end{figure*}
\clearpage

\begin{figure*}
\centering
\includegraphics[width=0.98\textwidth,height=0.94\textheight,keepaspectratio]{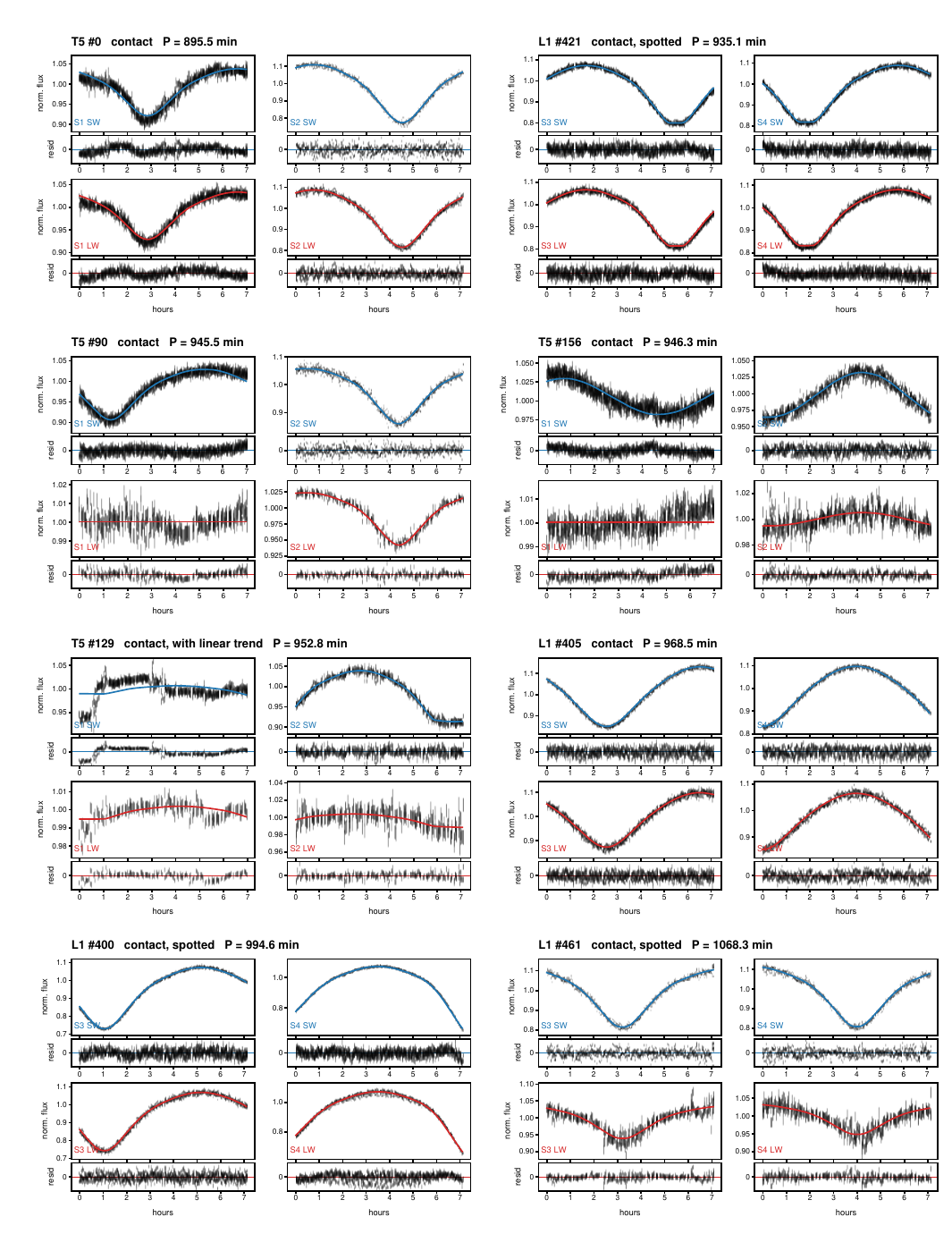}
\caption{Contact binaries, continued (page 31 of 33).}
\end{figure*}
\clearpage

\begin{figure*}
\centering
\includegraphics[width=0.98\textwidth,height=0.94\textheight,keepaspectratio]{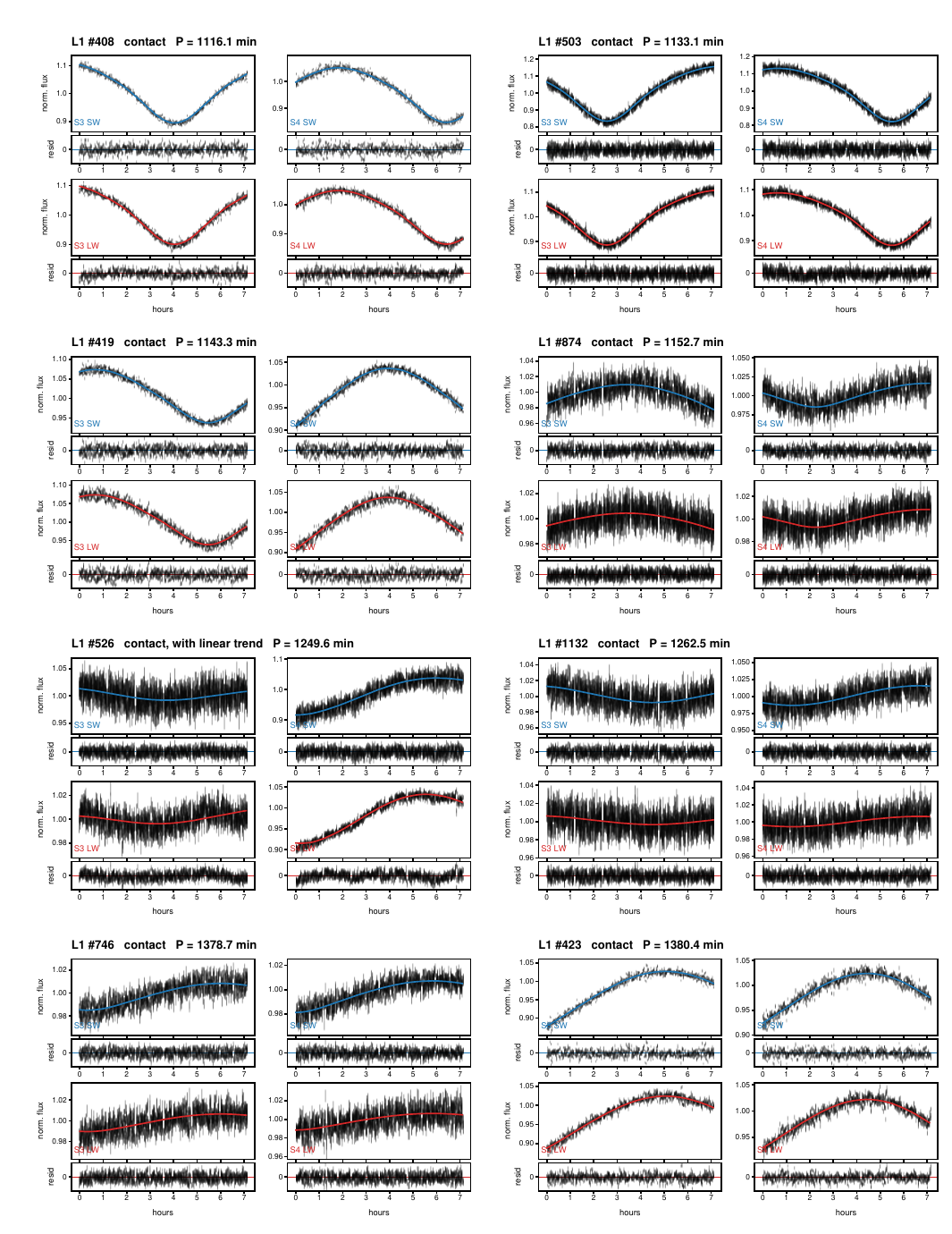}
\caption{Contact binaries, continued (page 32 of 33).}
\end{figure*}
\clearpage

\begin{figure*}
\centering
\includegraphics[width=0.98\textwidth,height=0.94\textheight,keepaspectratio]{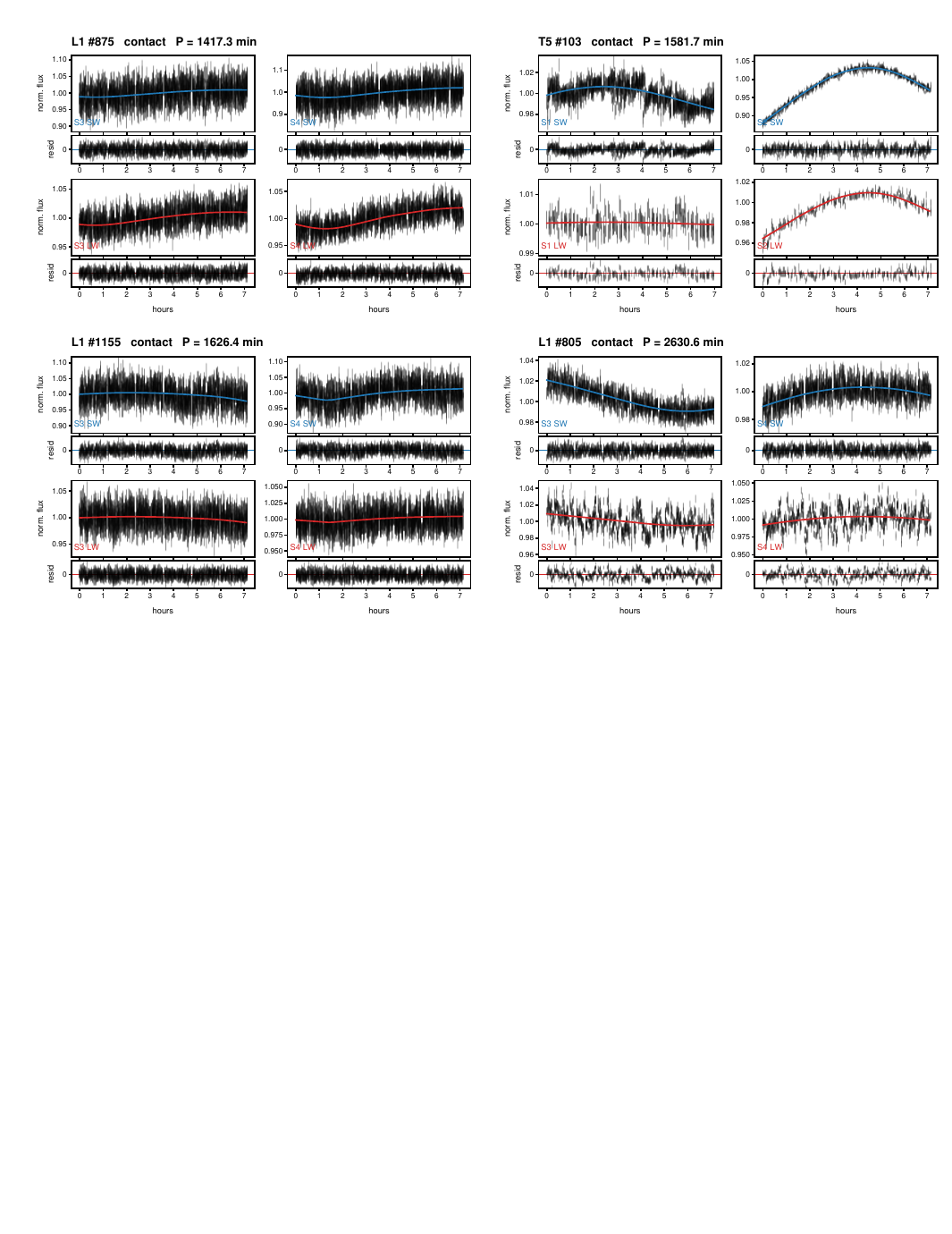}
\caption{Contact binaries, continued (page 33 of 33).}
\end{figure*}
\clearpage

\subsection{Semi-detached eclipsing binaries}\label{app:atlas:semidet}

\begin{figure*}
\centering
\includegraphics[width=0.98\textwidth,height=0.79\textheight,keepaspectratio]{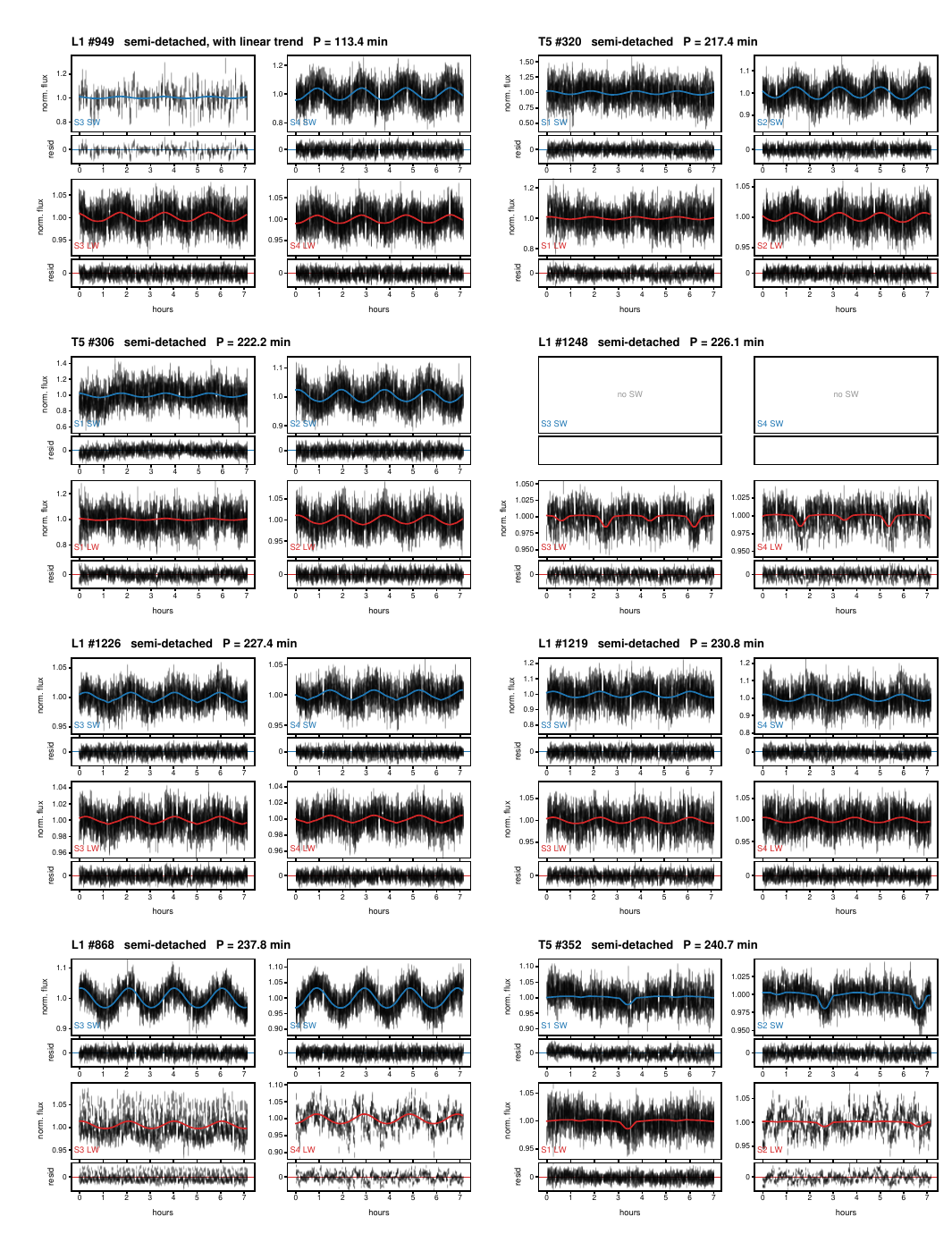}
\caption{Semi-detached eclipsing binaries: unfolded JWST NIRCam lightcurves of all 173 sources in this class (page 1 of 22), sorted by adopted period (sources without a period last). Each source is shown as a four-panel block. The two observing segments run left to right, with the short-wavelength F200W lightcurve (SW, \textbf{blue}) on top and the long-wavelength F356W lightcurve (LW, \textbf{red}) below, and the segment and band are labelled inside every panel. Time is hours from the start of that segment, and lightcurves are never phase-folded. Black vertical bars are the adopted lightcurve (\S\ref{sec:strategy}), each spanning the $1\sigma$ uncertainty of one plotted sample. The coloured curve is the accepted \texttt{PHOEBE} model, drawn in each panel as $A\,m(t)+B$ with the dilution terms $A$ and $B$ (\S\ref{sec:classification}) re-solved per panel, since blending differs between bands and segments. A depth difference between SW and LW is therefore not evidence of chromaticity. The narrow strip under each panel shows the residuals. The header gives the source, the fitted model (see the start of this appendix), and its orbital period, the adopted period.}
\end{figure*}
\clearpage

\begin{figure*}
\centering
\includegraphics[width=0.98\textwidth,height=0.94\textheight,keepaspectratio]{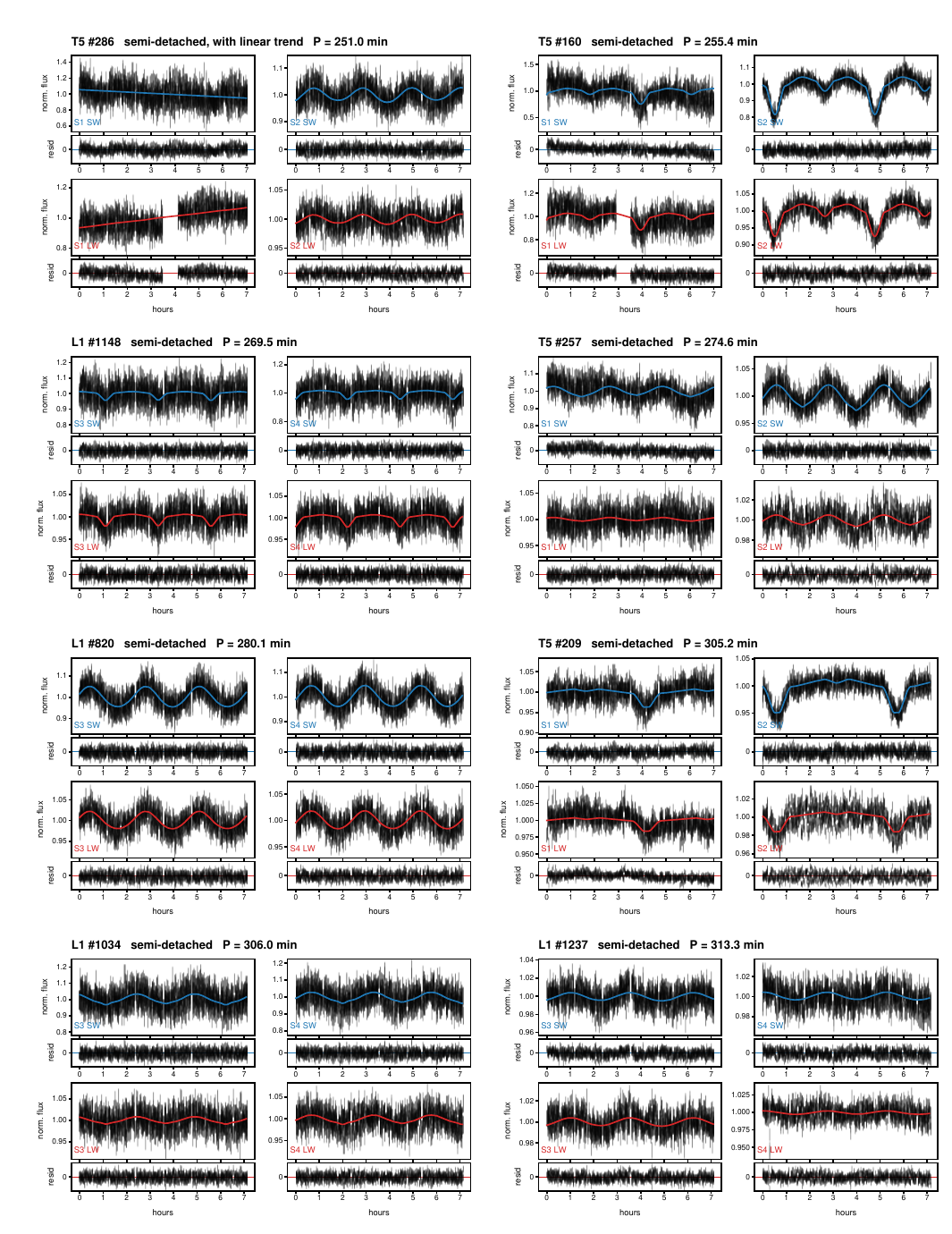}
\caption{Semi-detached eclipsing binaries, continued (page 2 of 22).}
\end{figure*}
\clearpage

\begin{figure*}
\centering
\includegraphics[width=0.98\textwidth,height=0.94\textheight,keepaspectratio]{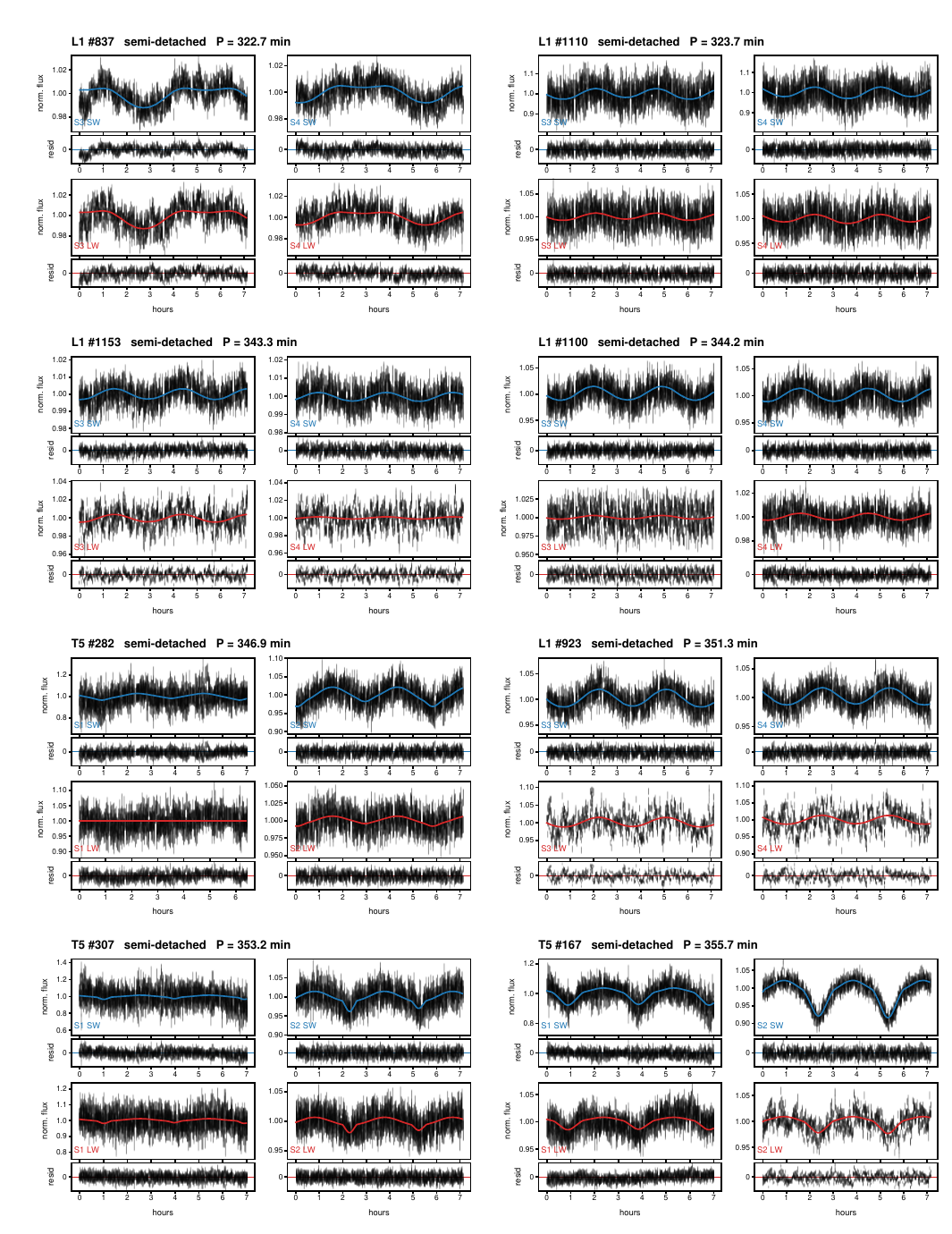}
\caption{Semi-detached eclipsing binaries, continued (page 3 of 22).}
\end{figure*}
\clearpage

\begin{figure*}
\centering
\includegraphics[width=0.98\textwidth,height=0.94\textheight,keepaspectratio]{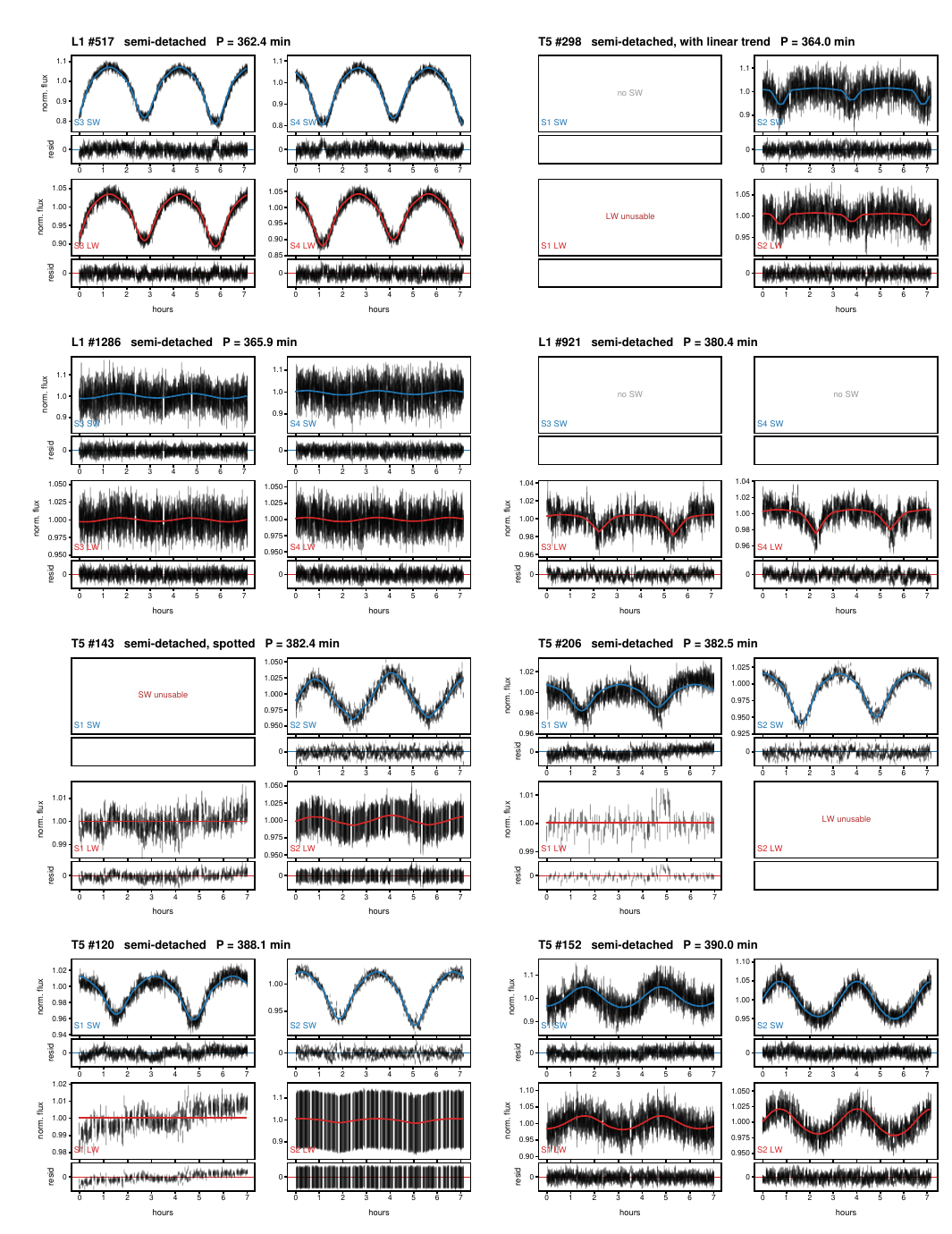}
\caption{Semi-detached eclipsing binaries, continued (page 4 of 22).}
\end{figure*}
\clearpage

\begin{figure*}
\centering
\includegraphics[width=0.98\textwidth,height=0.94\textheight,keepaspectratio]{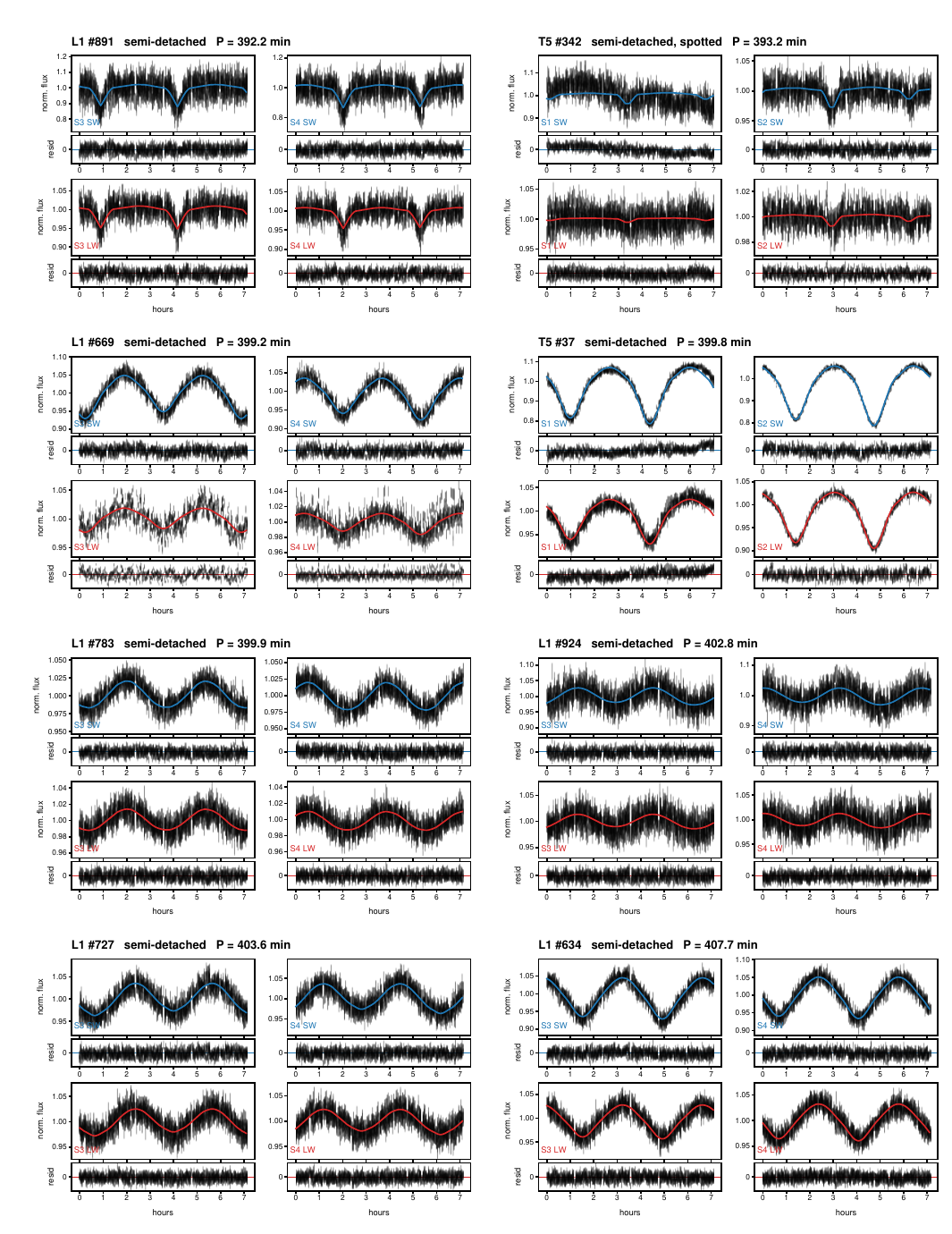}
\caption{Semi-detached eclipsing binaries, continued (page 5 of 22).}
\end{figure*}
\clearpage

\begin{figure*}
\centering
\includegraphics[width=0.98\textwidth,height=0.94\textheight,keepaspectratio]{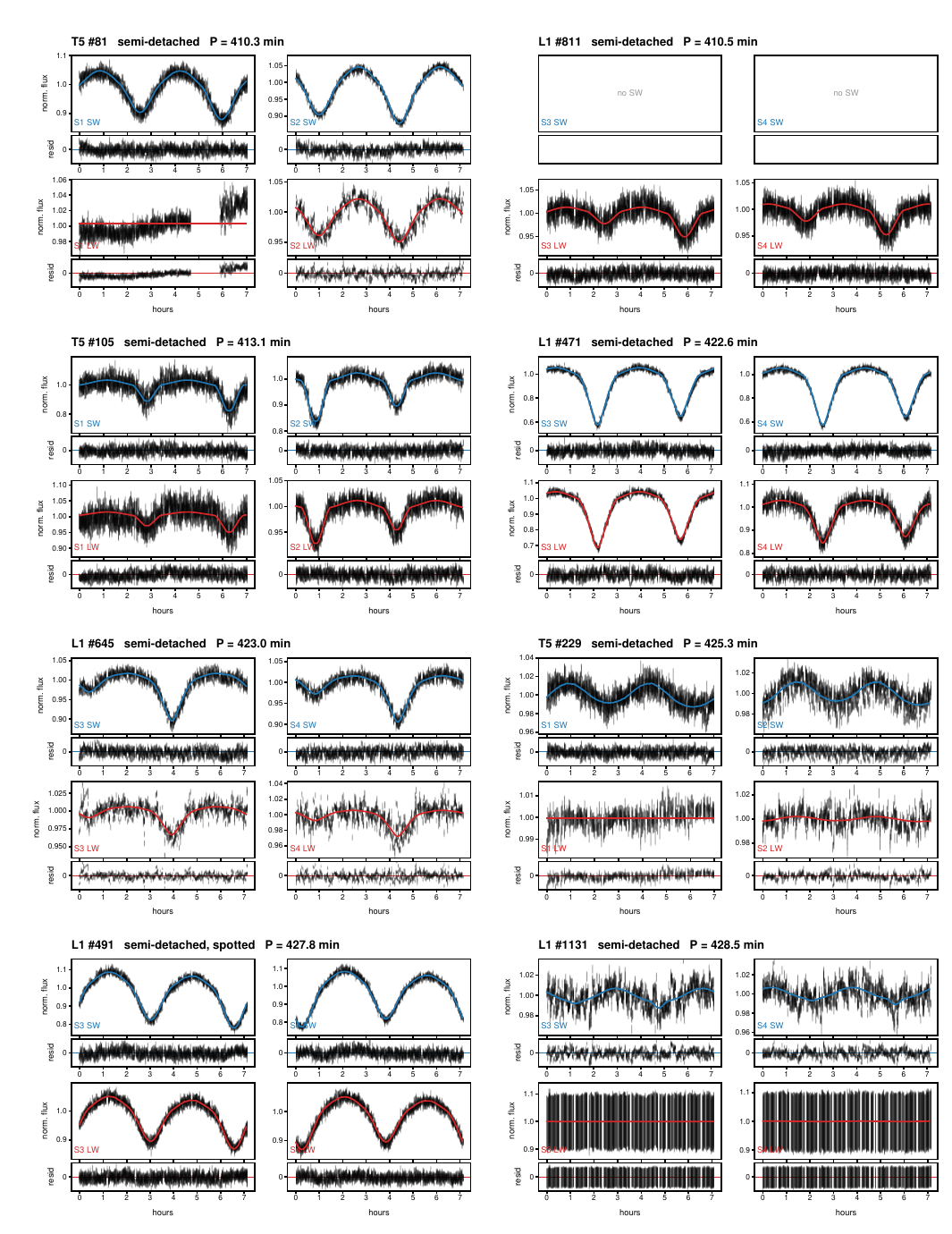}
\caption{Semi-detached eclipsing binaries, continued (page 6 of 22).}
\end{figure*}
\clearpage

\begin{figure*}
\centering
\includegraphics[width=0.98\textwidth,height=0.94\textheight,keepaspectratio]{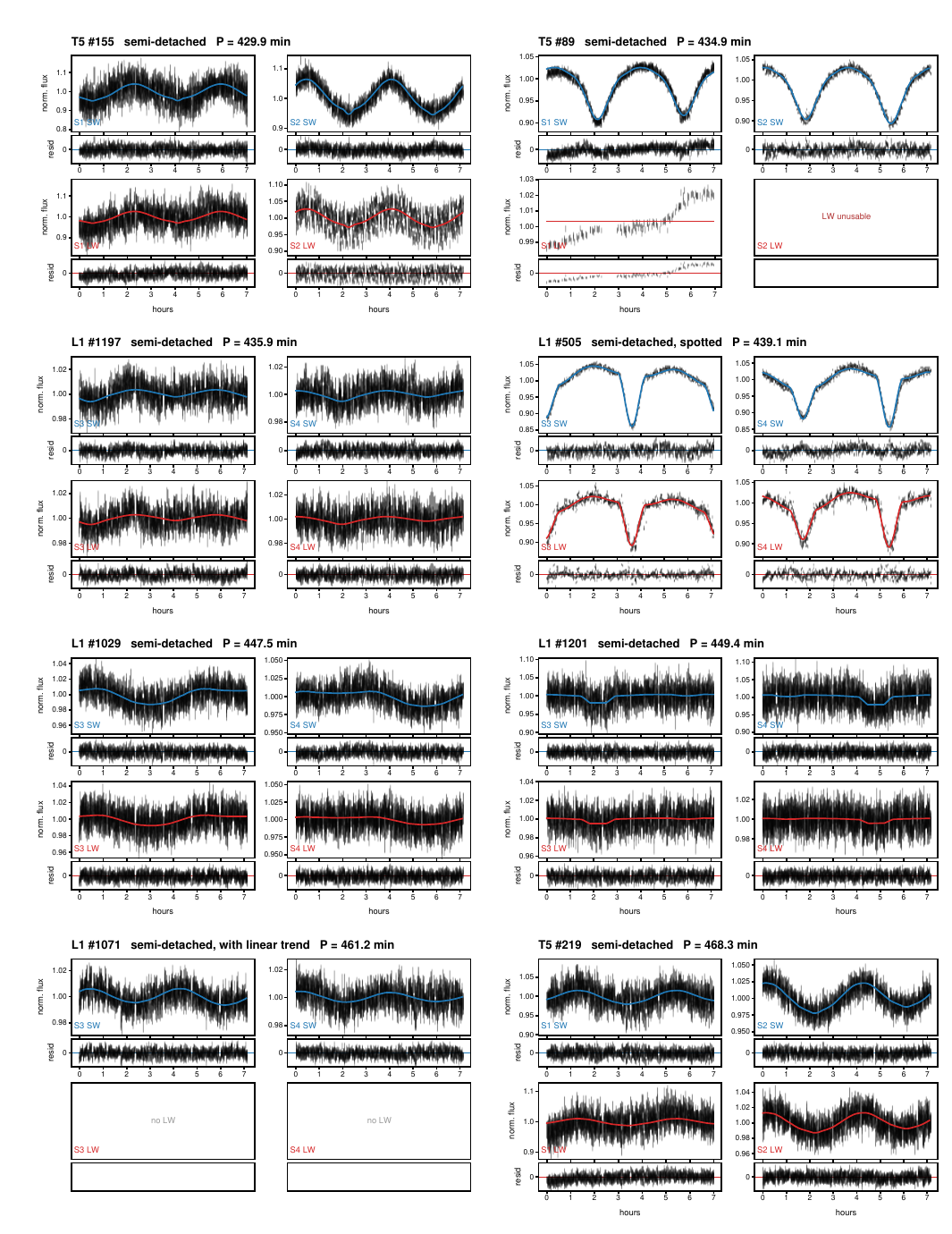}
\caption{Semi-detached eclipsing binaries, continued (page 7 of 22).}
\end{figure*}
\clearpage

\begin{figure*}
\centering
\includegraphics[width=0.98\textwidth,height=0.94\textheight,keepaspectratio]{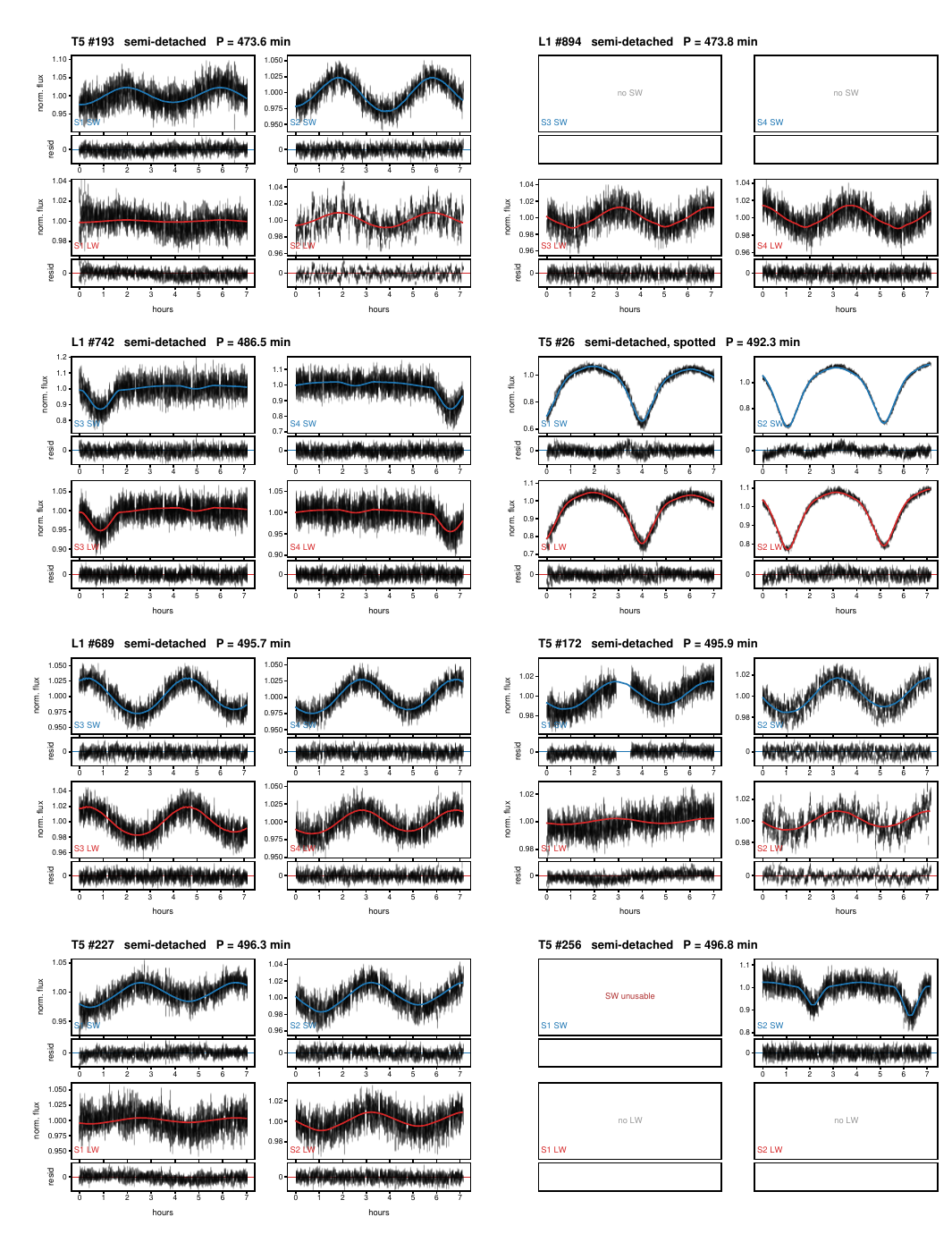}
\caption{Semi-detached eclipsing binaries, continued (page 8 of 22).}
\end{figure*}
\clearpage

\begin{figure*}
\centering
\includegraphics[width=0.98\textwidth,height=0.94\textheight,keepaspectratio]{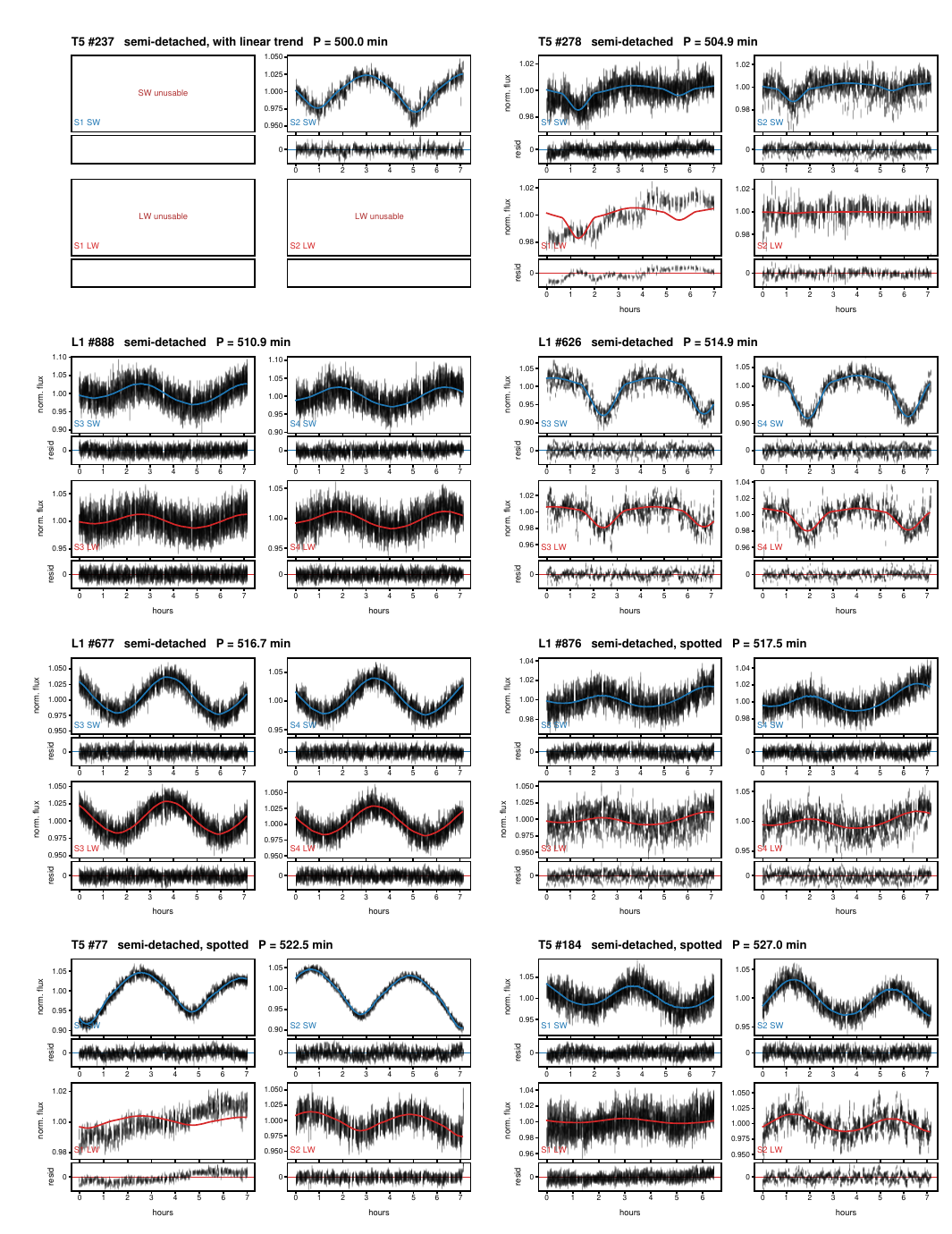}
\caption{Semi-detached eclipsing binaries, continued (page 9 of 22).}
\end{figure*}
\clearpage

\begin{figure*}
\centering
\includegraphics[width=0.98\textwidth,height=0.94\textheight,keepaspectratio]{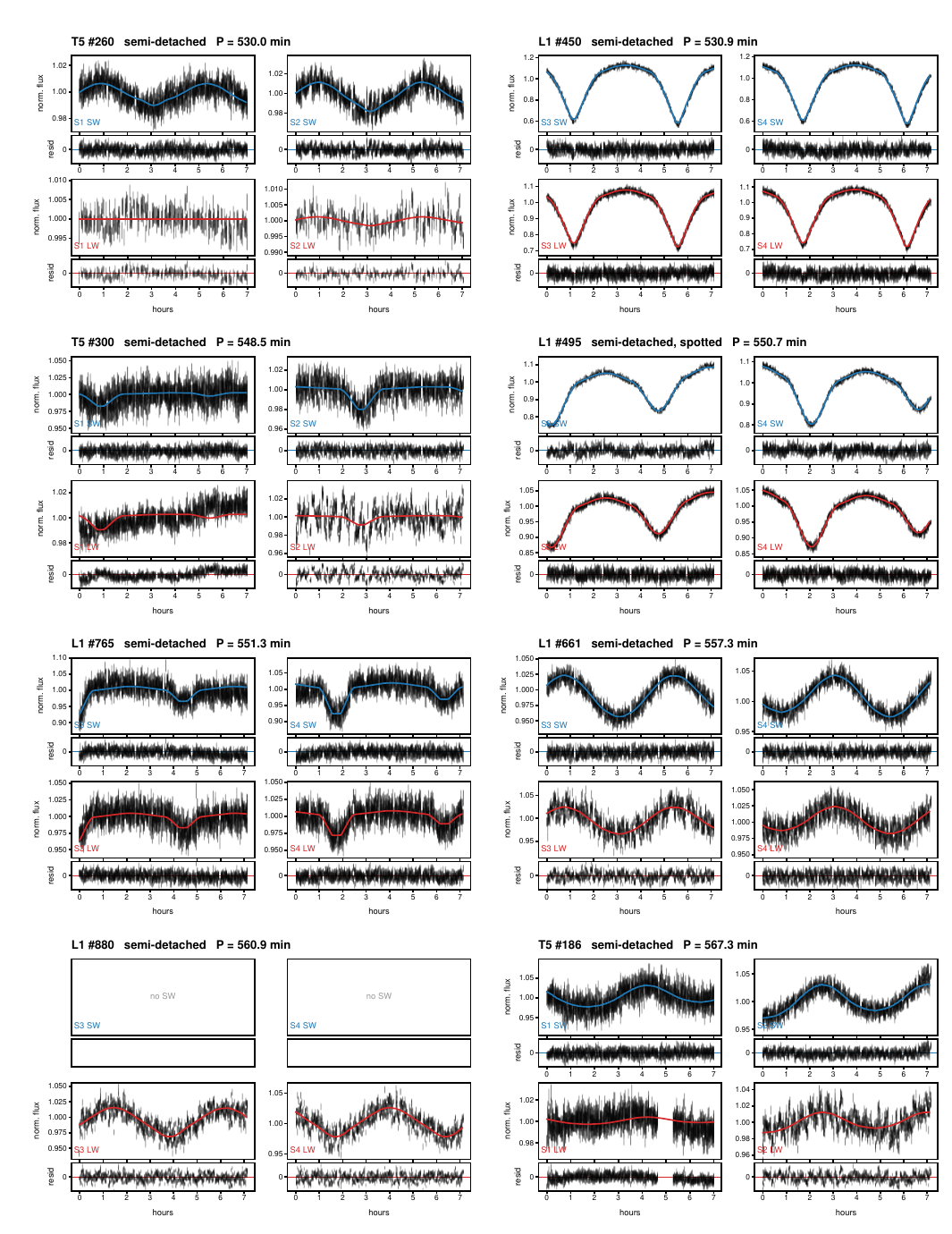}
\caption{Semi-detached eclipsing binaries, continued (page 10 of 22).}
\end{figure*}
\clearpage

\begin{figure*}
\centering
\includegraphics[width=0.98\textwidth,height=0.94\textheight,keepaspectratio]{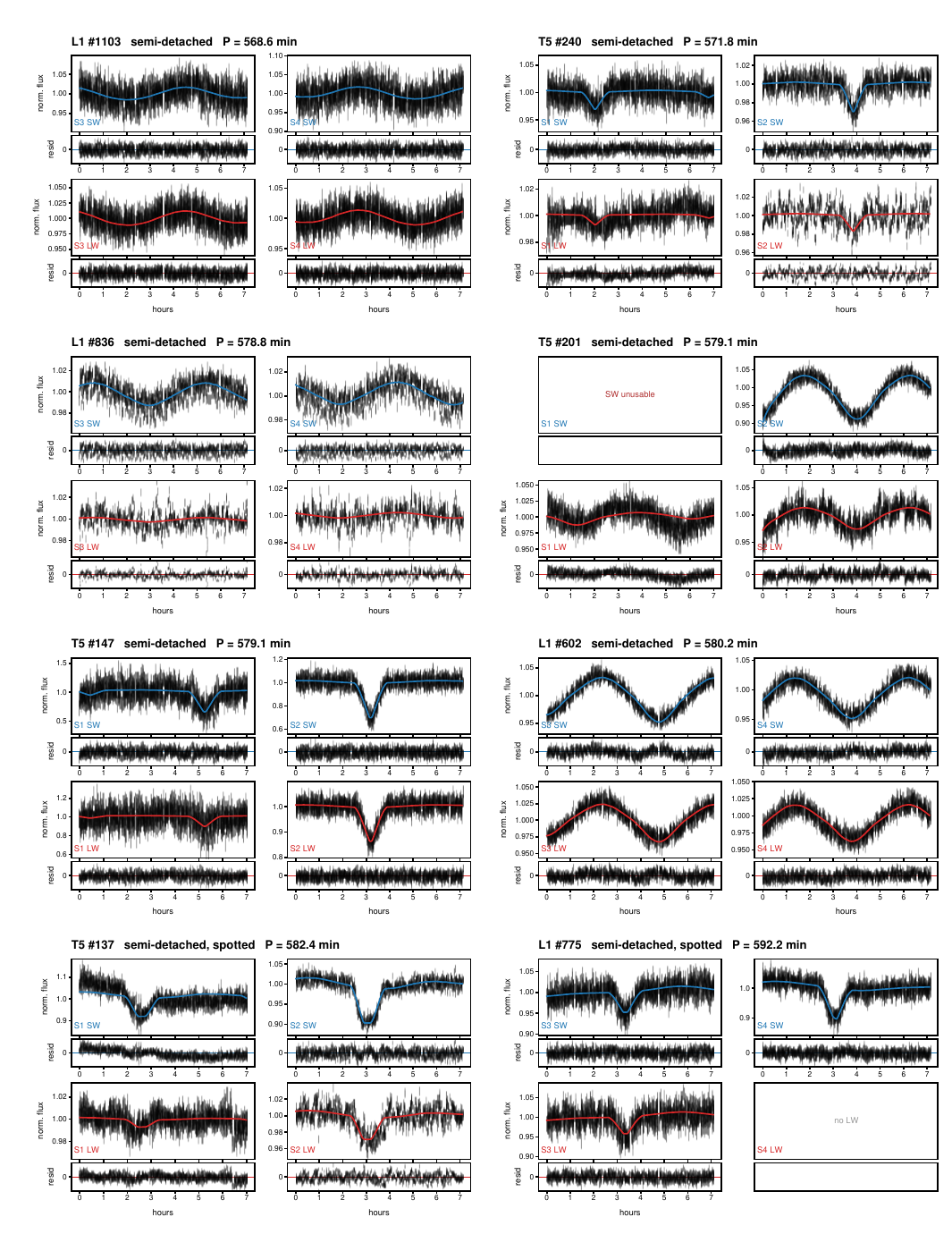}
\caption{Semi-detached eclipsing binaries, continued (page 11 of 22).}
\end{figure*}
\clearpage

\begin{figure*}
\centering
\includegraphics[width=0.98\textwidth,height=0.94\textheight,keepaspectratio]{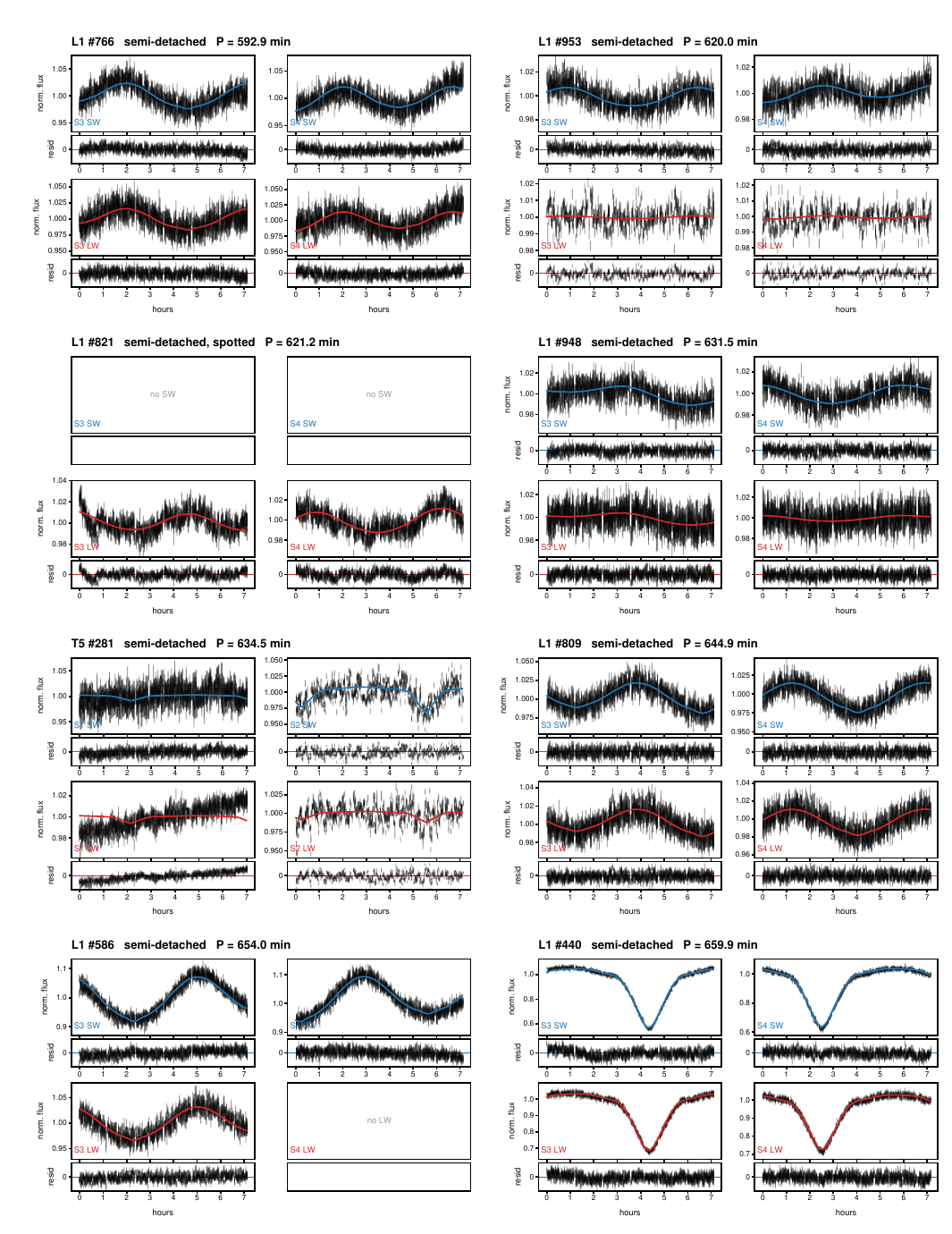}
\caption{Semi-detached eclipsing binaries, continued (page 12 of 22).}
\end{figure*}
\clearpage

\begin{figure*}
\centering
\includegraphics[width=0.98\textwidth,height=0.94\textheight,keepaspectratio]{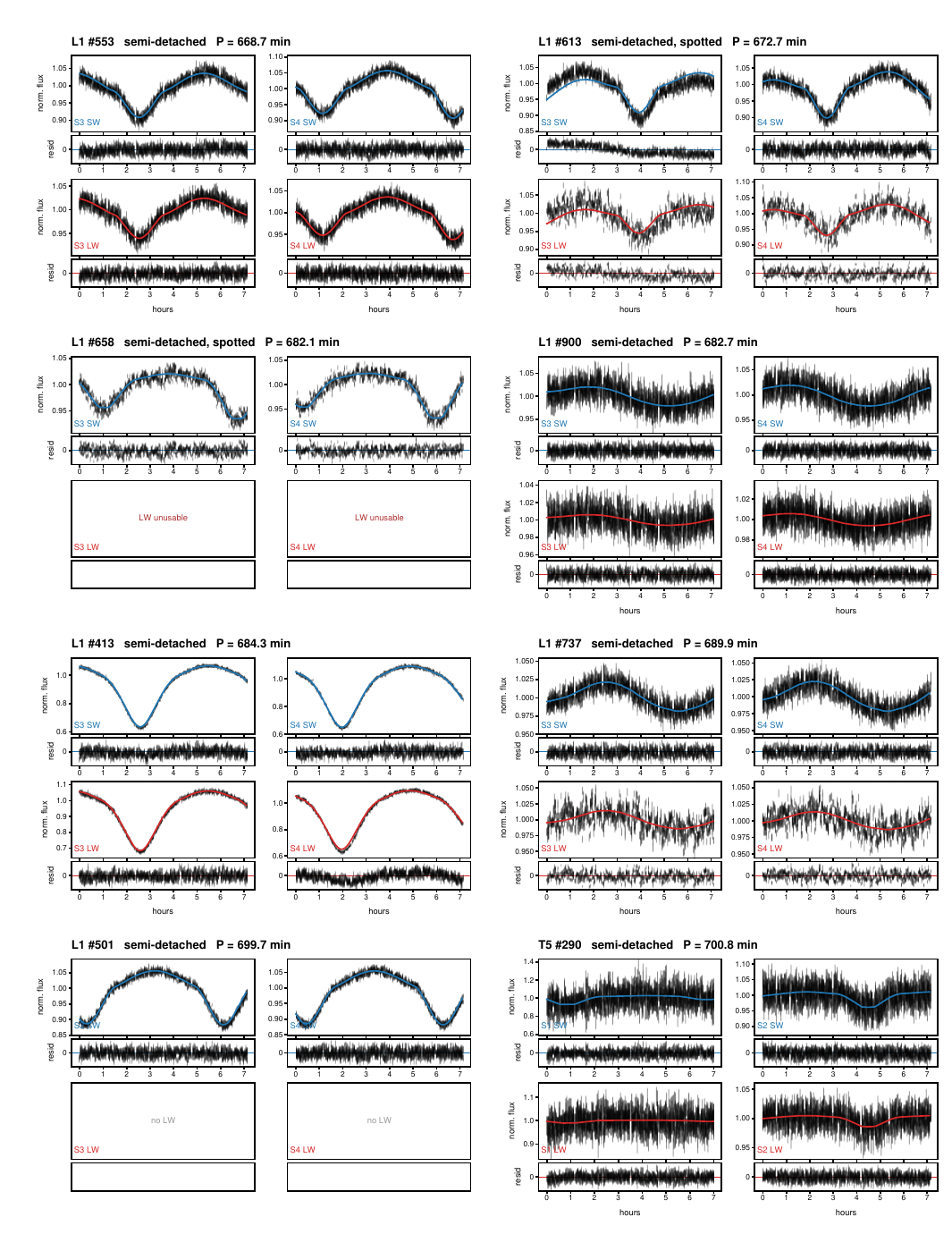}
\caption{Semi-detached eclipsing binaries, continued (page 13 of 22).}
\end{figure*}
\clearpage

\begin{figure*}
\centering
\includegraphics[width=0.98\textwidth,height=0.94\textheight,keepaspectratio]{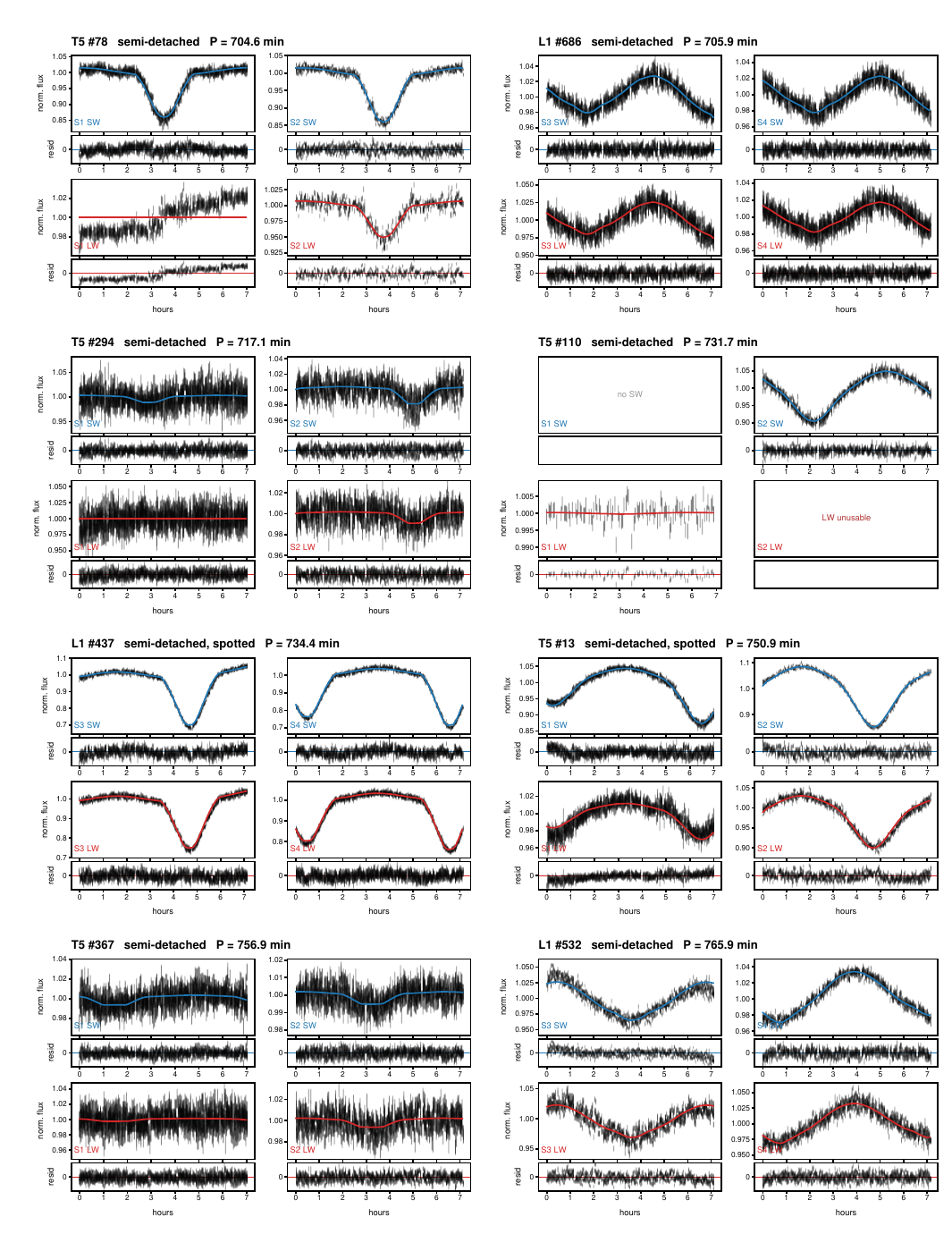}
\caption{Semi-detached eclipsing binaries, continued (page 14 of 22).}
\end{figure*}
\clearpage

\begin{figure*}
\centering
\includegraphics[width=0.98\textwidth,height=0.94\textheight,keepaspectratio]{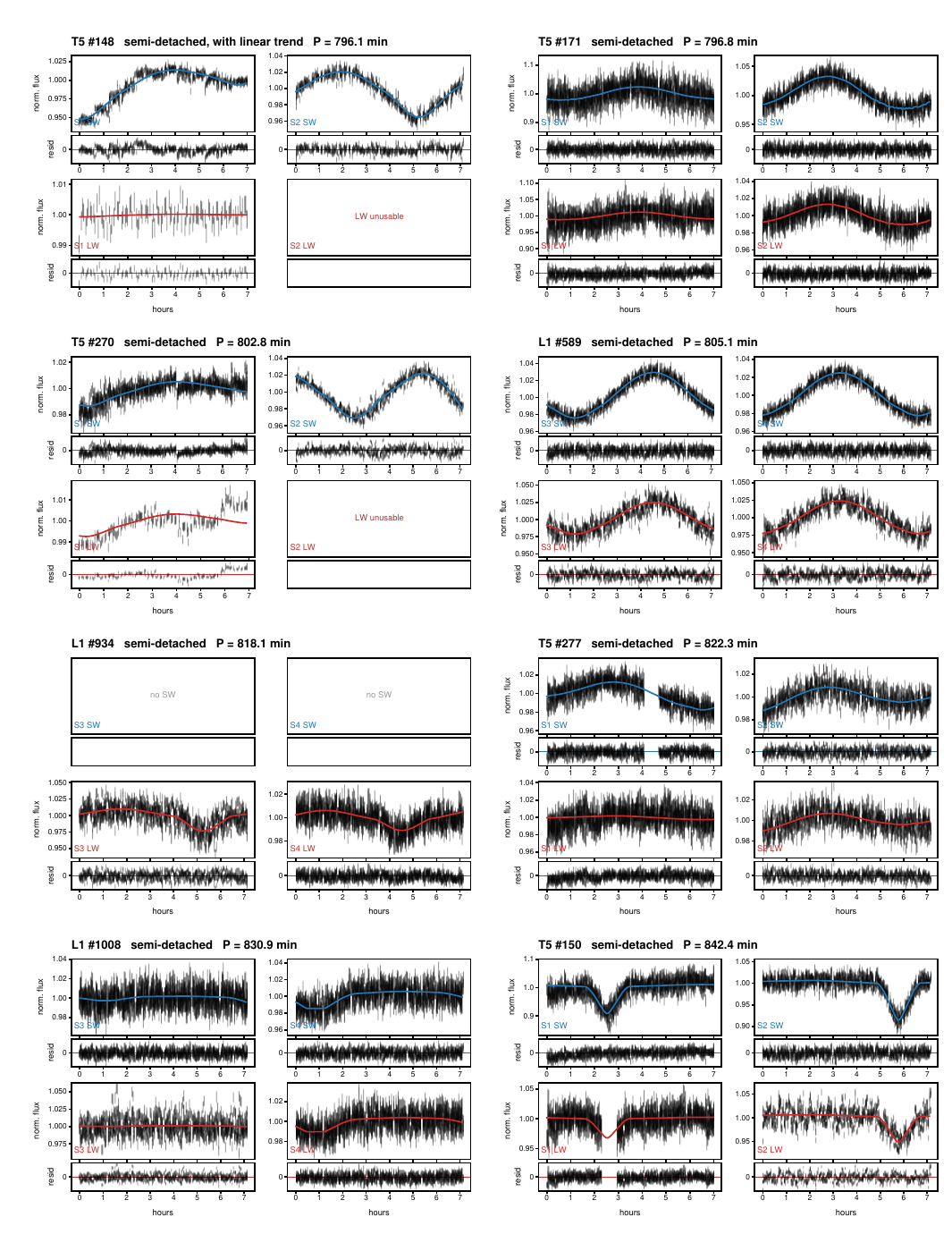}
\caption{Semi-detached eclipsing binaries, continued (page 15 of 22).}
\end{figure*}
\clearpage

\begin{figure*}
\centering
\includegraphics[width=0.98\textwidth,height=0.94\textheight,keepaspectratio]{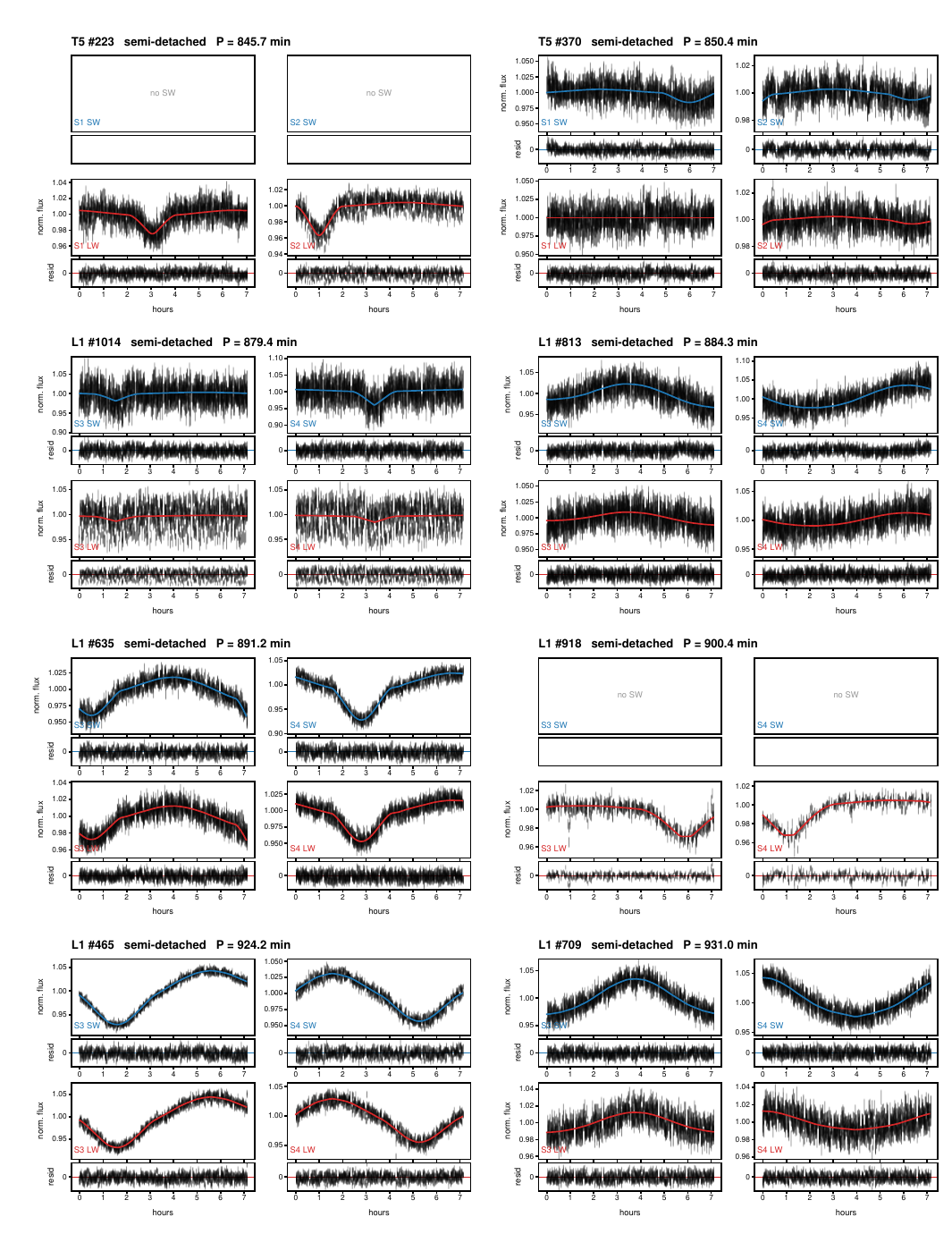}
\caption{Semi-detached eclipsing binaries, continued (page 16 of 22).}
\end{figure*}
\clearpage

\begin{figure*}
\centering
\includegraphics[width=0.98\textwidth,height=0.94\textheight,keepaspectratio]{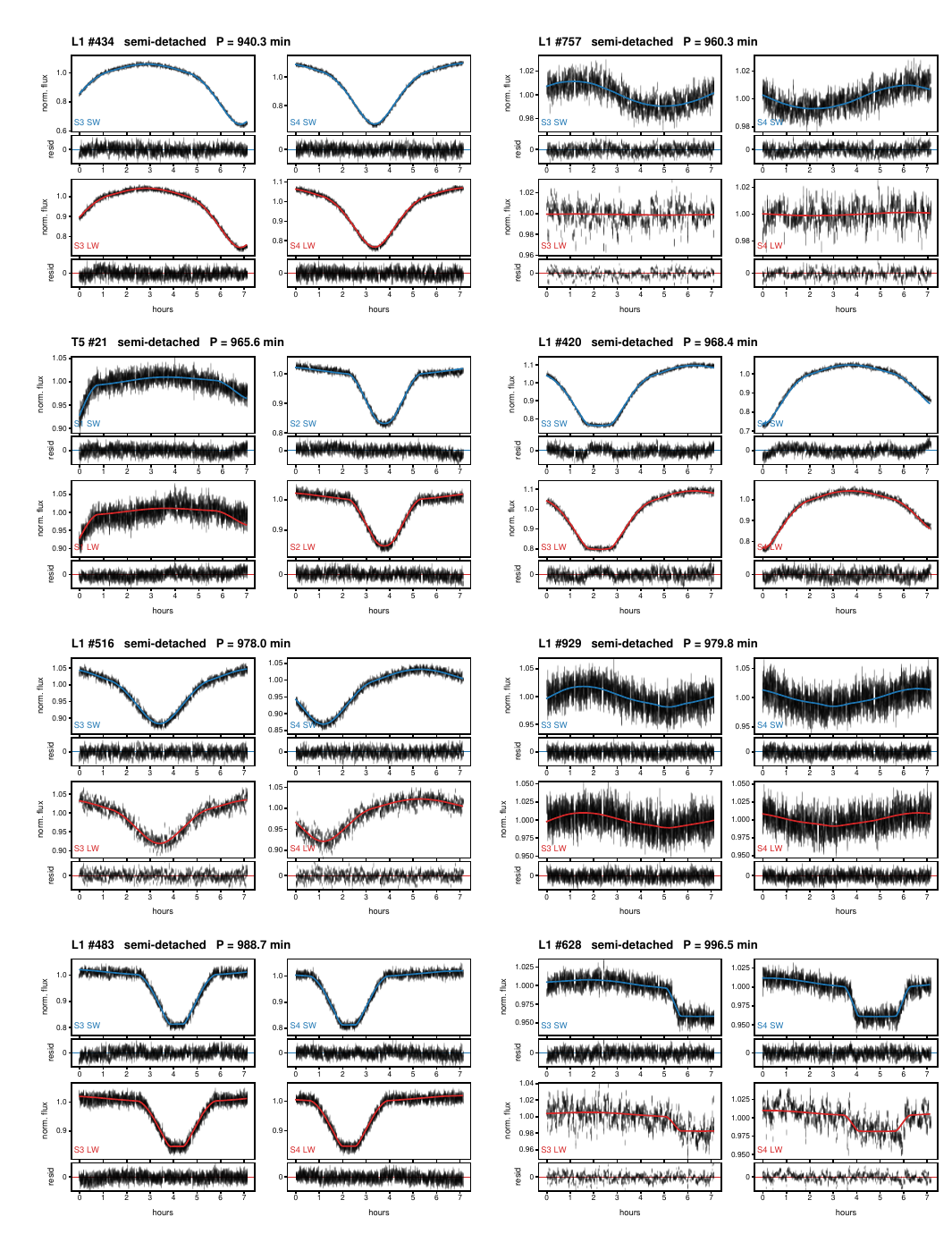}
\caption{Semi-detached eclipsing binaries, continued (page 17 of 22).}
\end{figure*}
\clearpage

\begin{figure*}
\centering
\includegraphics[width=0.98\textwidth,height=0.94\textheight,keepaspectratio]{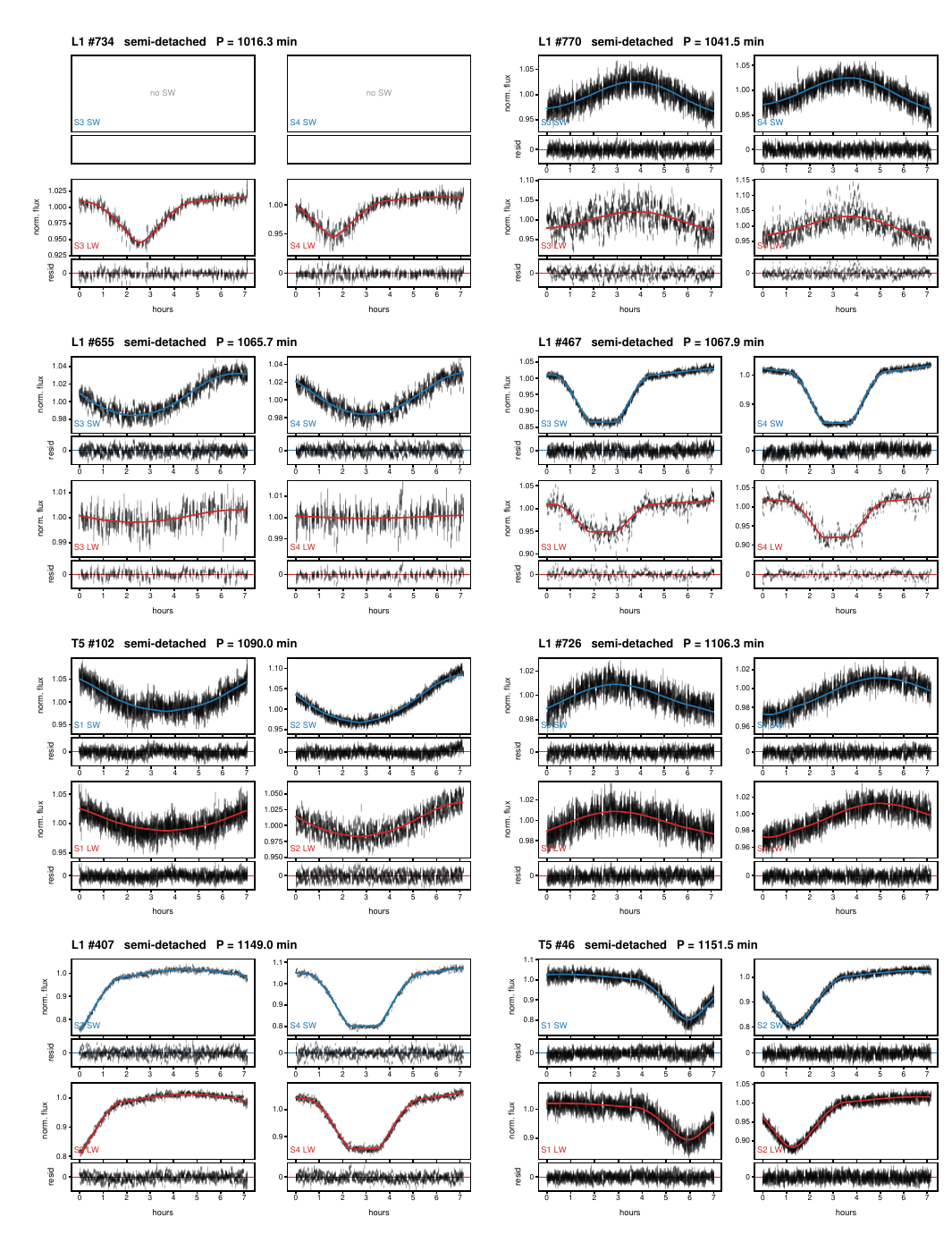}
\caption{Semi-detached eclipsing binaries, continued (page 18 of 22).}
\end{figure*}
\clearpage

\begin{figure*}
\centering
\includegraphics[width=0.98\textwidth,height=0.94\textheight,keepaspectratio]{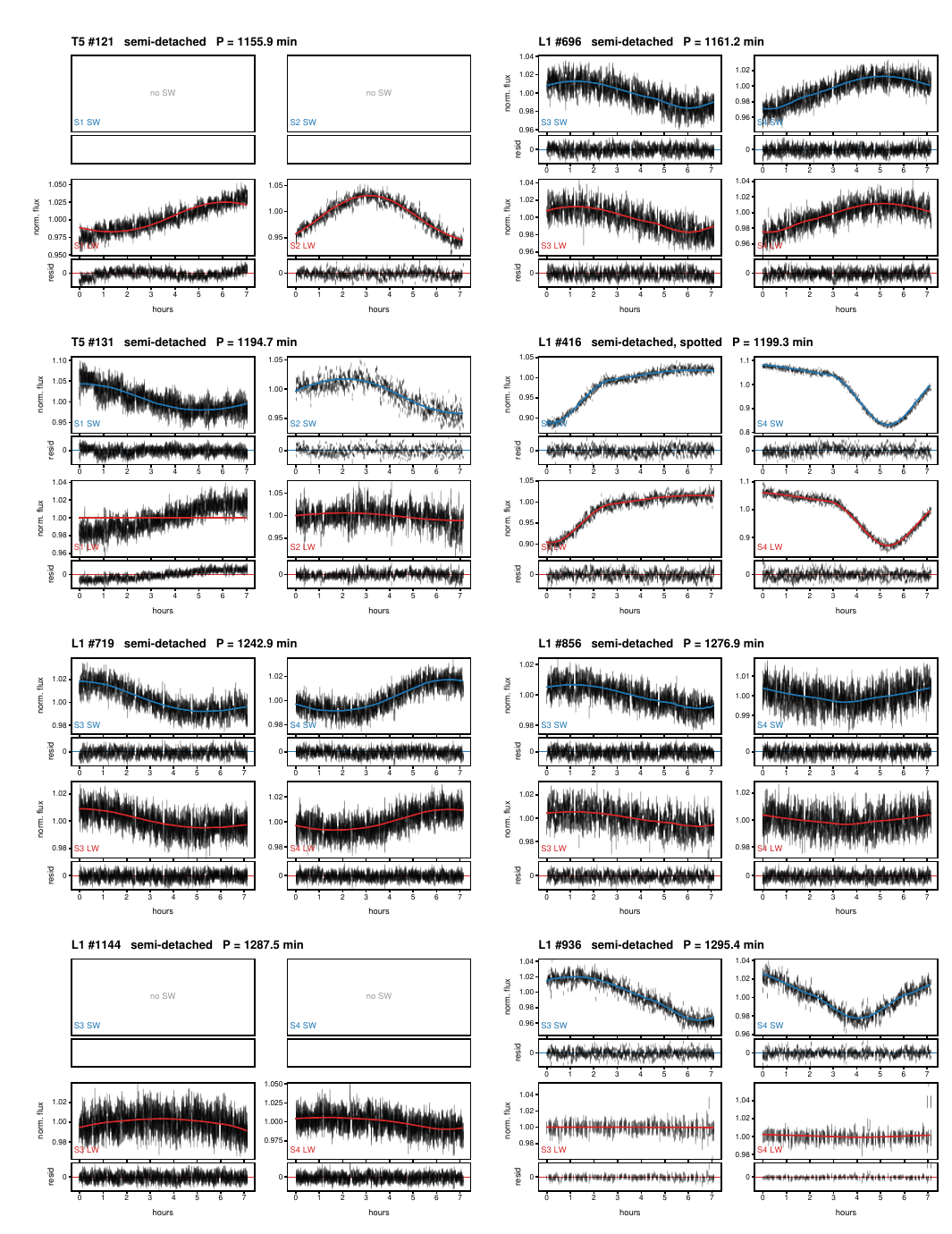}
\caption{Semi-detached eclipsing binaries, continued (page 19 of 22).}
\end{figure*}
\clearpage

\begin{figure*}
\centering
\includegraphics[width=0.98\textwidth,height=0.94\textheight,keepaspectratio]{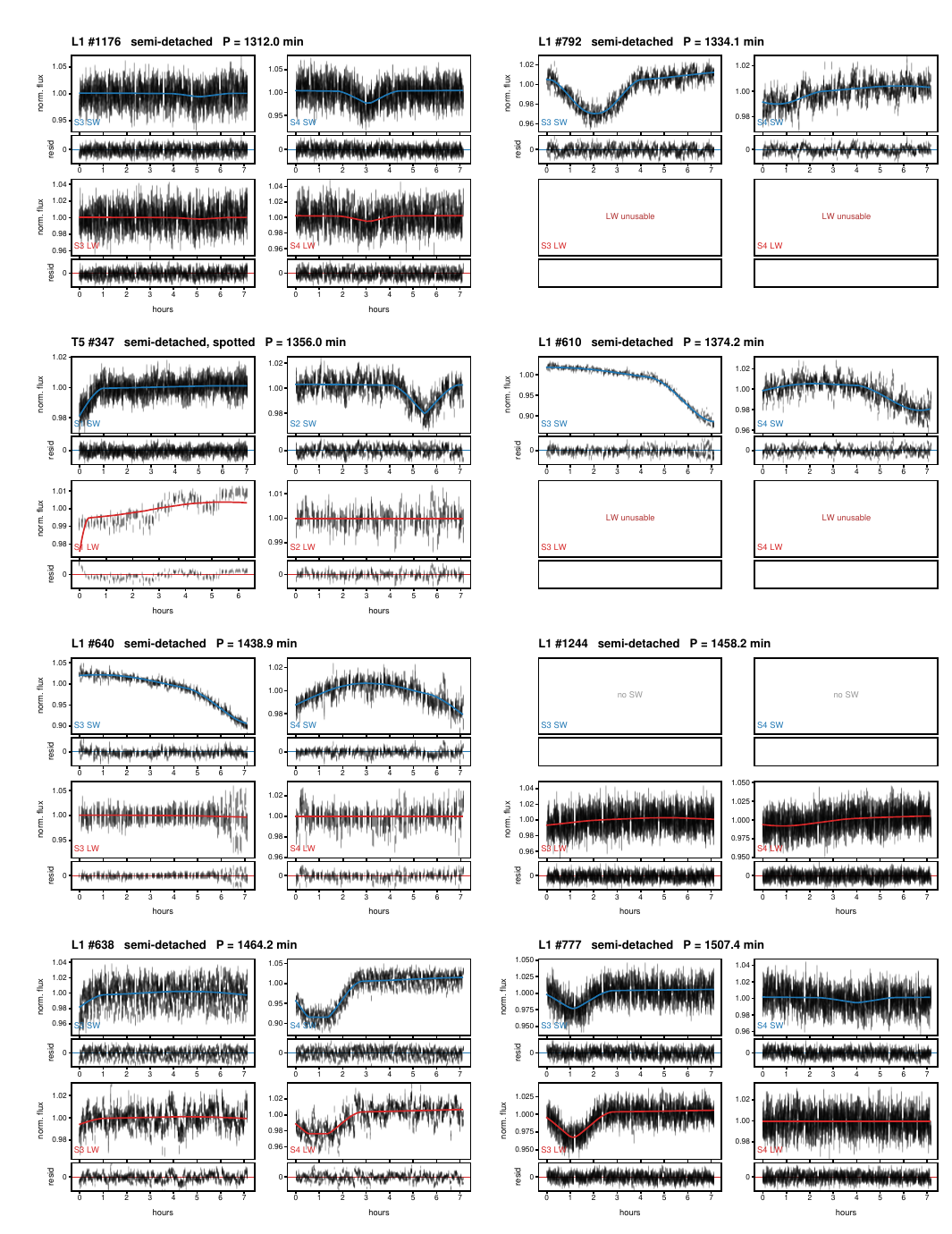}
\caption{Semi-detached eclipsing binaries, continued (page 20 of 22).}
\end{figure*}
\clearpage

\begin{figure*}
\centering
\includegraphics[width=0.98\textwidth,height=0.94\textheight,keepaspectratio]{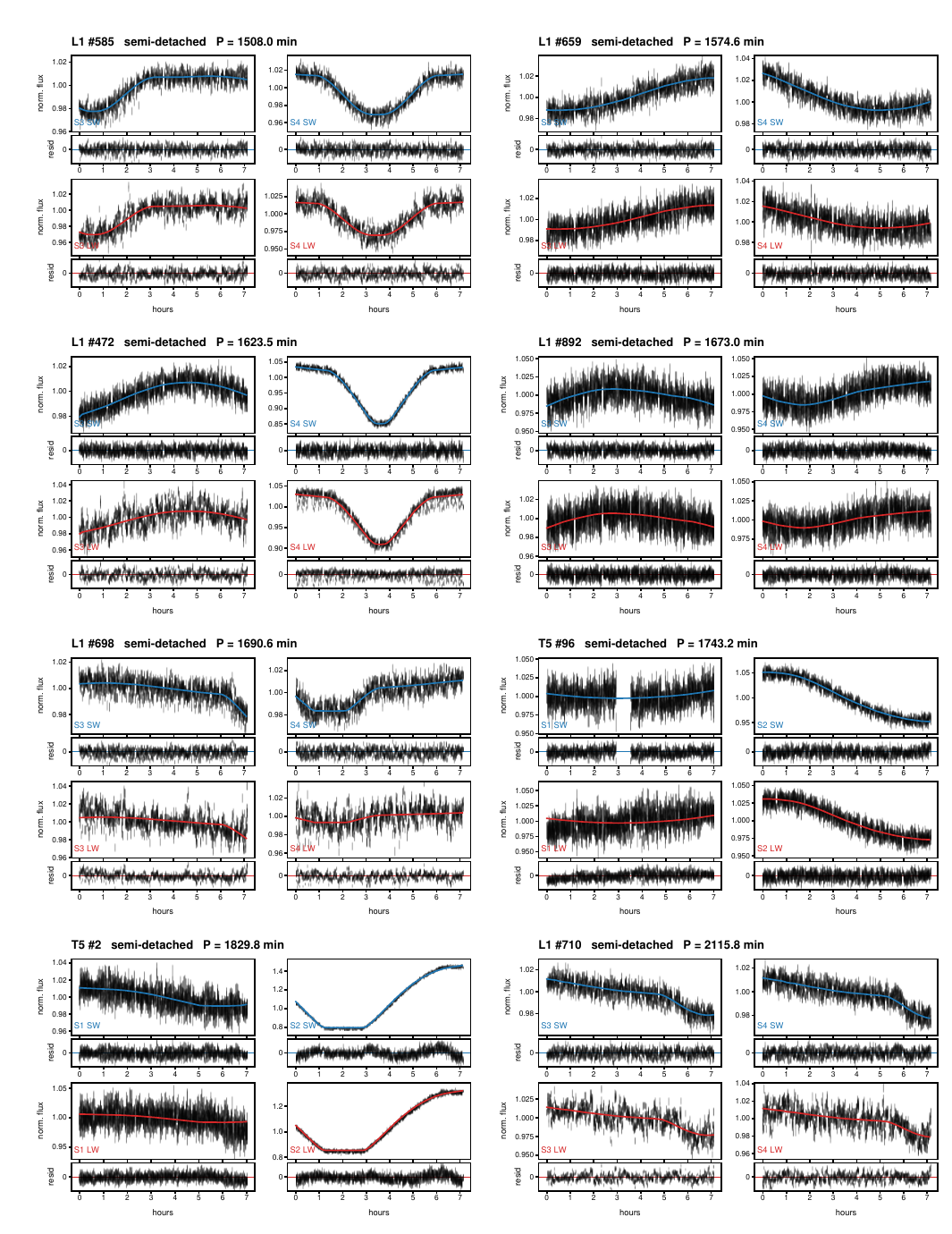}
\caption{Semi-detached eclipsing binaries, continued (page 21 of 22).}
\end{figure*}
\clearpage

\begin{figure*}
\centering
\includegraphics[width=0.98\textwidth,height=0.94\textheight,keepaspectratio]{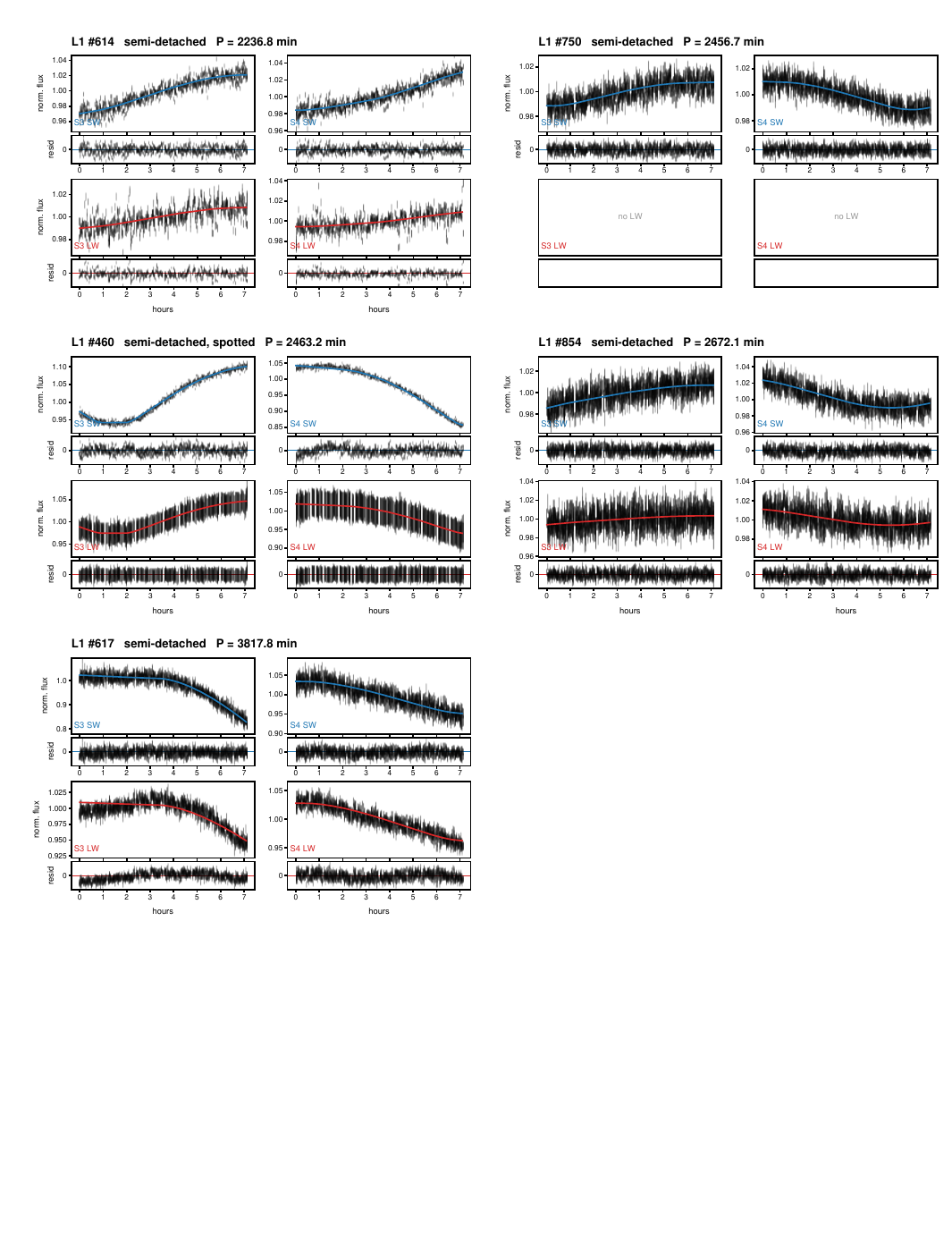}
\caption{Semi-detached eclipsing binaries, continued (page 22 of 22).}
\end{figure*}
\clearpage

\subsection{Detached eclipsing binaries}\label{app:atlas:detached}

\begin{figure*}
\centering
\includegraphics[width=0.98\textwidth,height=0.79\textheight,keepaspectratio]{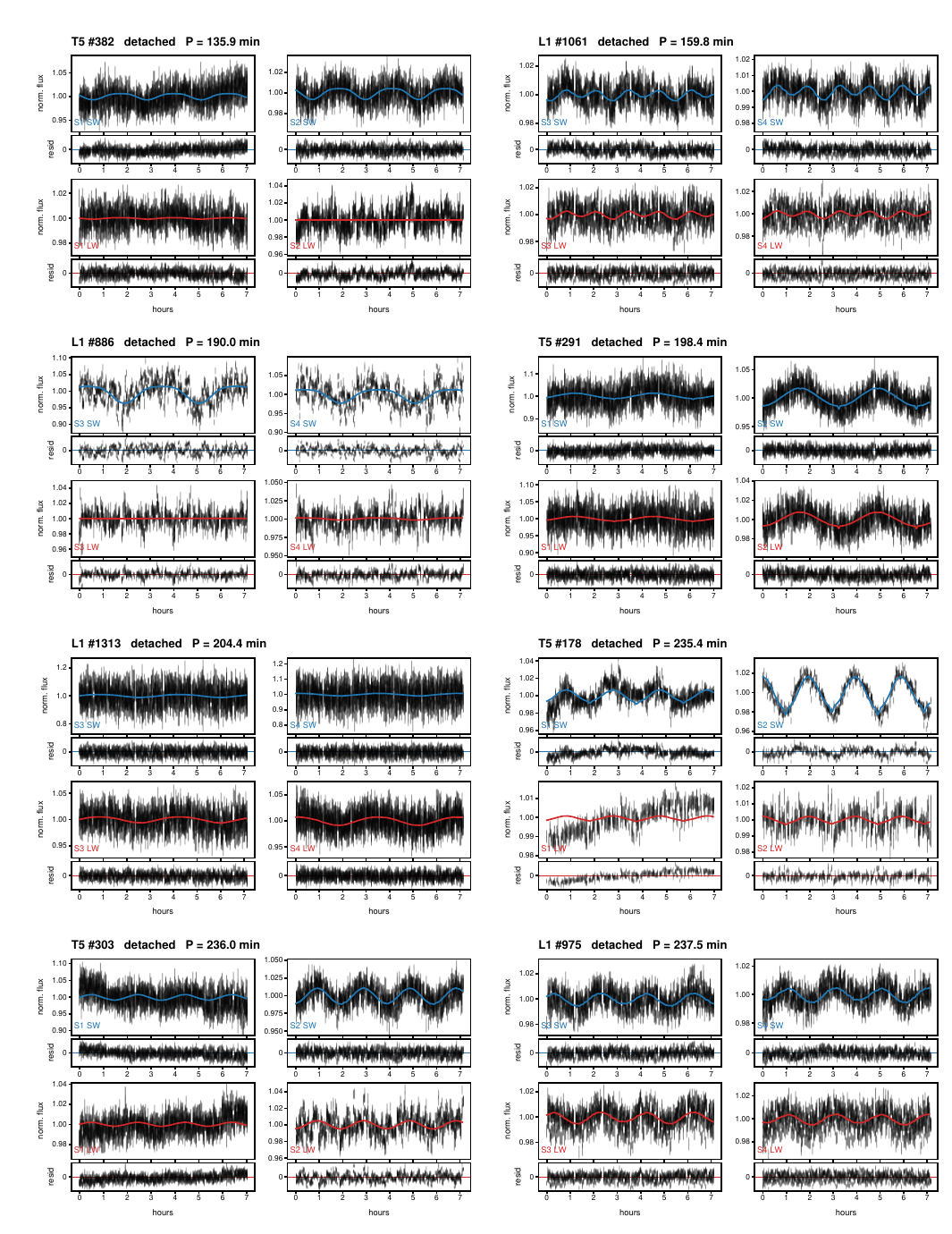}
\caption{Detached eclipsing binaries: unfolded JWST NIRCam lightcurves of all 288 sources in this class (page 1 of 36), sorted by adopted period (sources without a period last). Each source is shown as a four-panel block. The two observing segments run left to right, with the short-wavelength F200W lightcurve (SW, \textbf{blue}) on top and the long-wavelength F356W lightcurve (LW, \textbf{red}) below, and the segment and band are labelled inside every panel. Time is hours from the start of that segment, and lightcurves are never phase-folded. Black vertical bars are the adopted lightcurve (\S\ref{sec:strategy}), each spanning the $1\sigma$ uncertainty of one plotted sample. The coloured curve is the accepted \texttt{PHOEBE} model, drawn in each panel as $A\,m(t)+B$ with the dilution terms $A$ and $B$ (\S\ref{sec:classification}) re-solved per panel, since blending differs between bands and segments. A depth difference between SW and LW is therefore not evidence of chromaticity. The narrow strip under each panel shows the residuals. The header gives the source, the fitted model (see the start of this appendix), and its orbital period, the adopted period.}
\end{figure*}
\clearpage

\begin{figure*}
\centering
\includegraphics[width=0.98\textwidth,height=0.94\textheight,keepaspectratio]{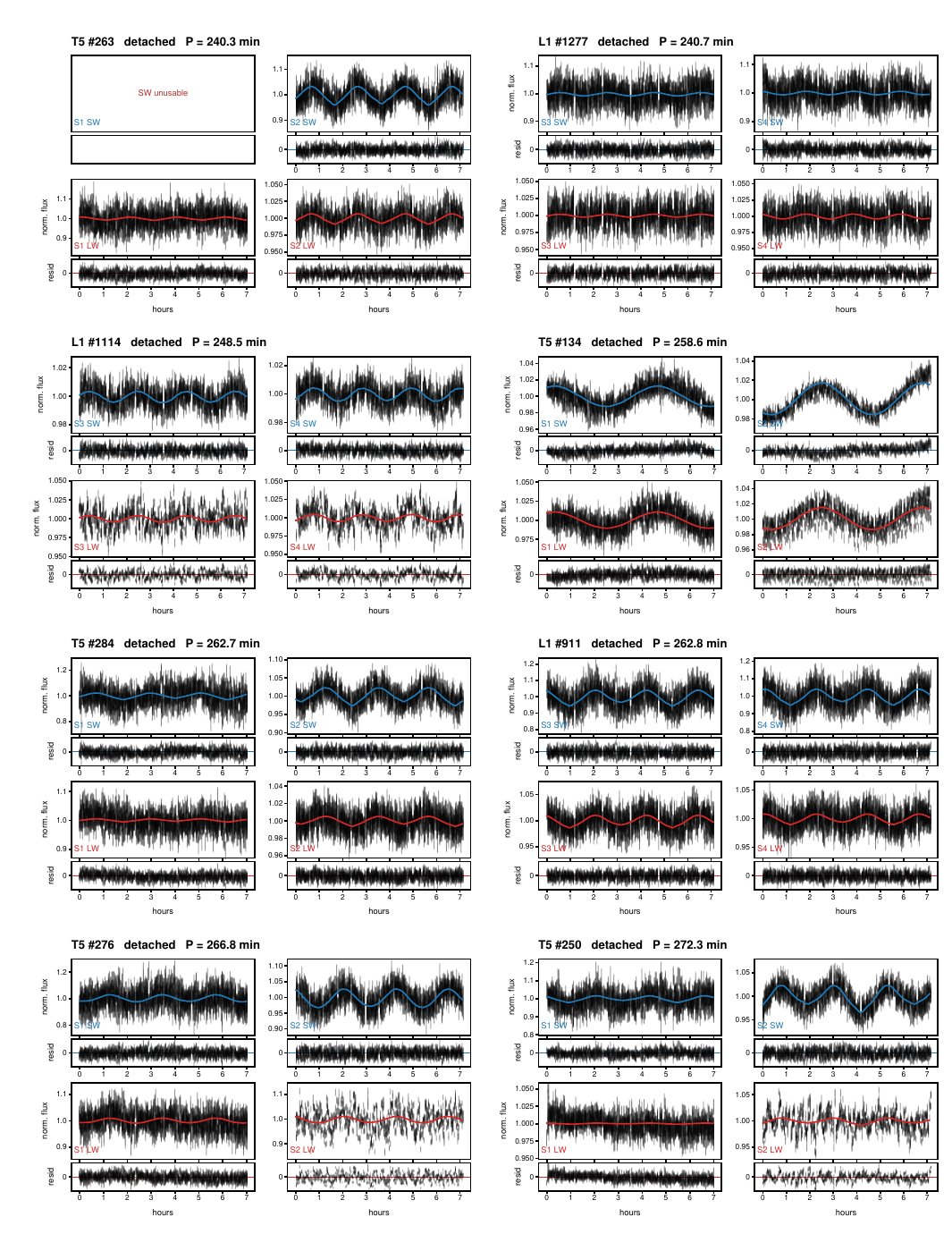}
\caption{Detached eclipsing binaries, continued (page 2 of 36).}
\end{figure*}
\clearpage

\begin{figure*}
\centering
\includegraphics[width=0.98\textwidth,height=0.94\textheight,keepaspectratio]{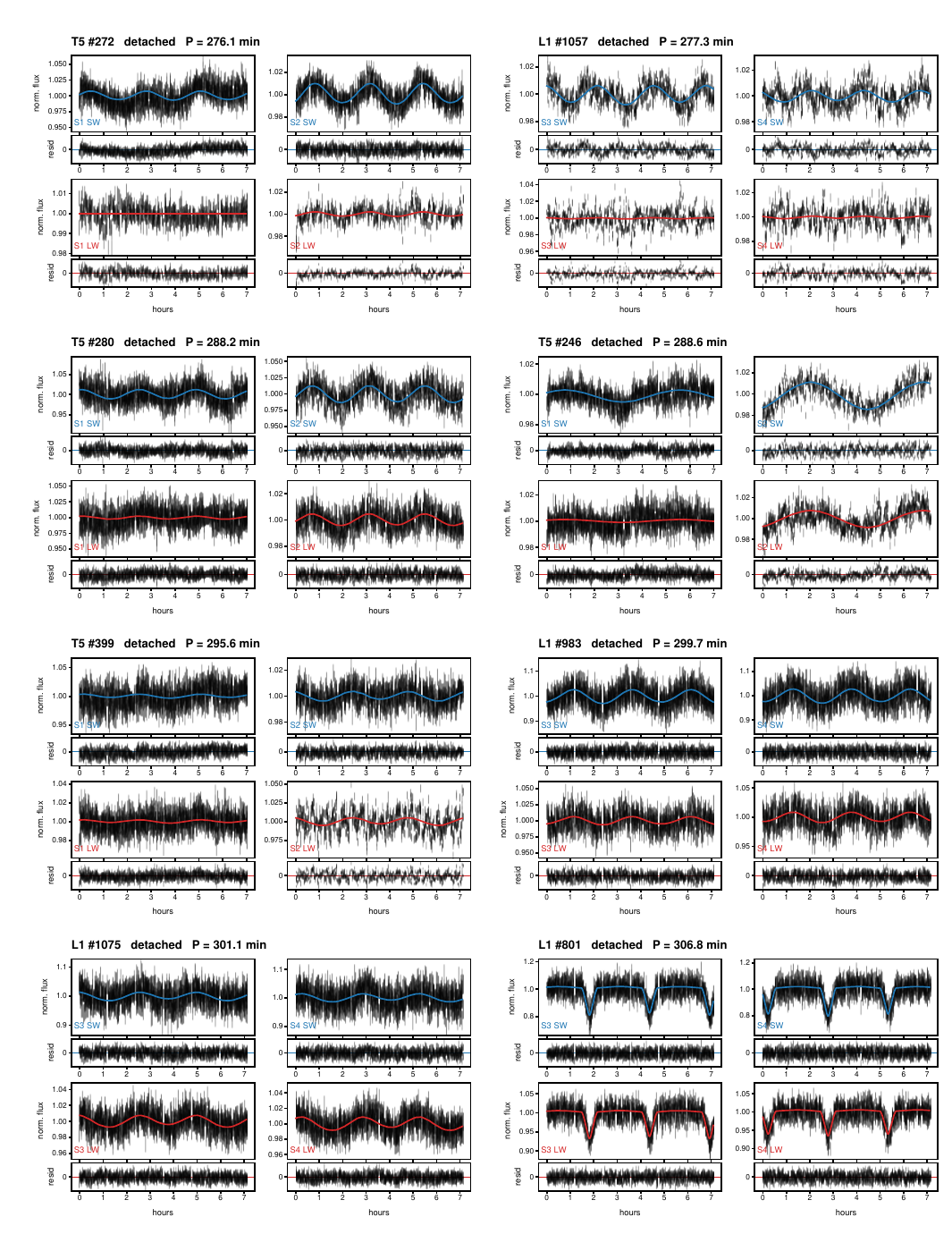}
\caption{Detached eclipsing binaries, continued (page 3 of 36).}
\end{figure*}
\clearpage

\begin{figure*}
\centering
\includegraphics[width=0.98\textwidth,height=0.94\textheight,keepaspectratio]{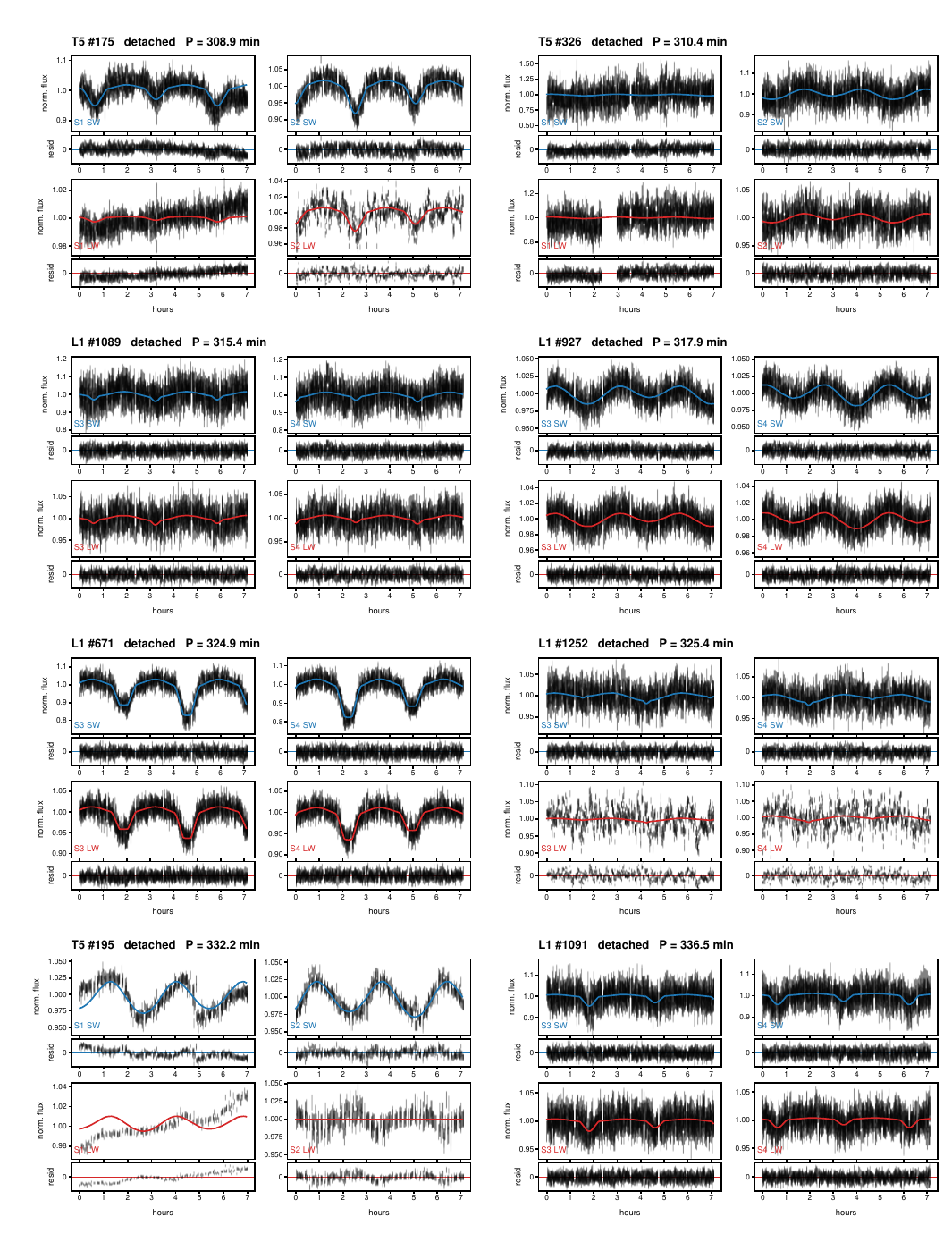}
\caption{Detached eclipsing binaries, continued (page 4 of 36).}
\end{figure*}
\clearpage

\begin{figure*}
\centering
\includegraphics[width=0.98\textwidth,height=0.94\textheight,keepaspectratio]{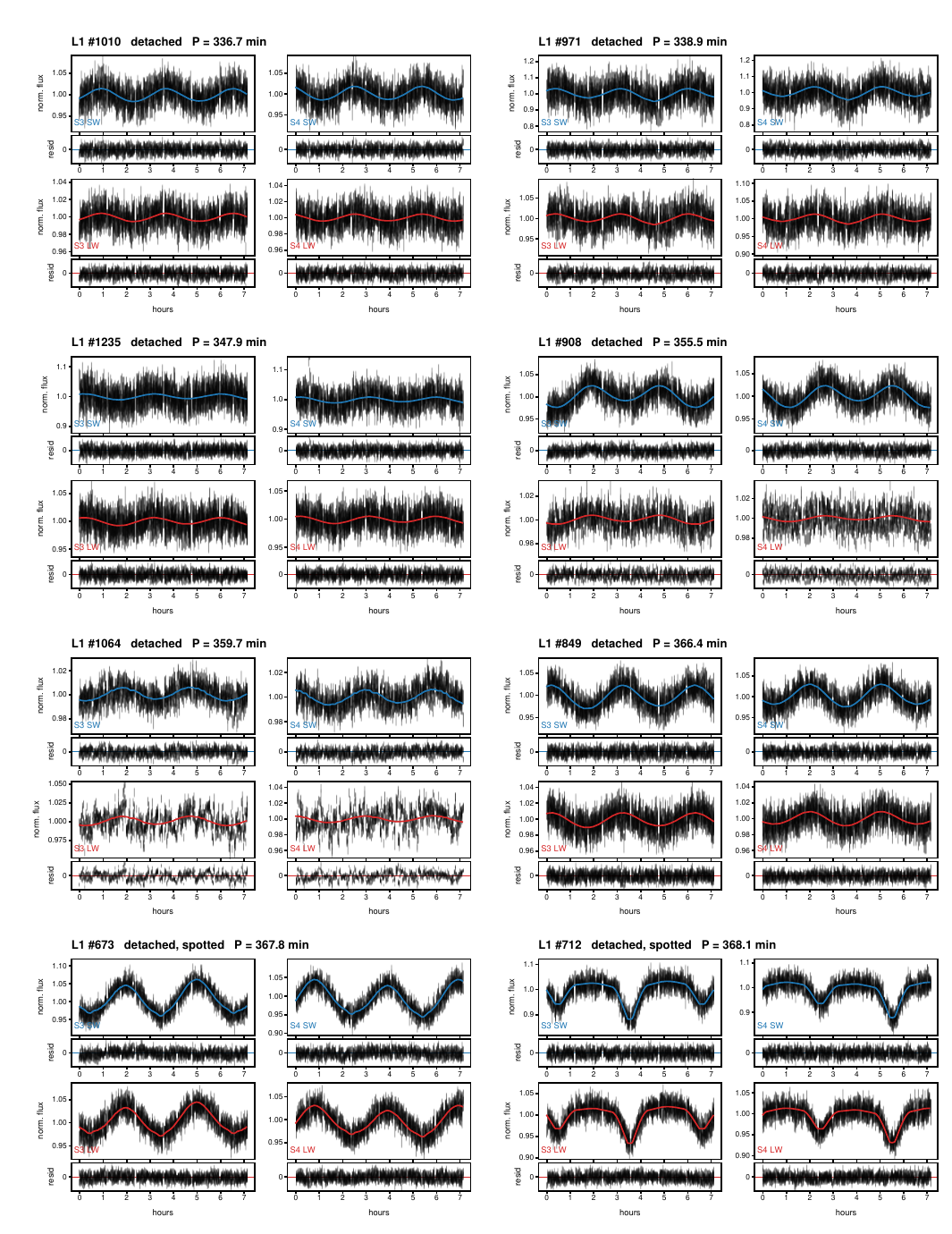}
\caption{Detached eclipsing binaries, continued (page 5 of 36).}
\end{figure*}
\clearpage

\begin{figure*}
\centering
\includegraphics[width=0.98\textwidth,height=0.94\textheight,keepaspectratio]{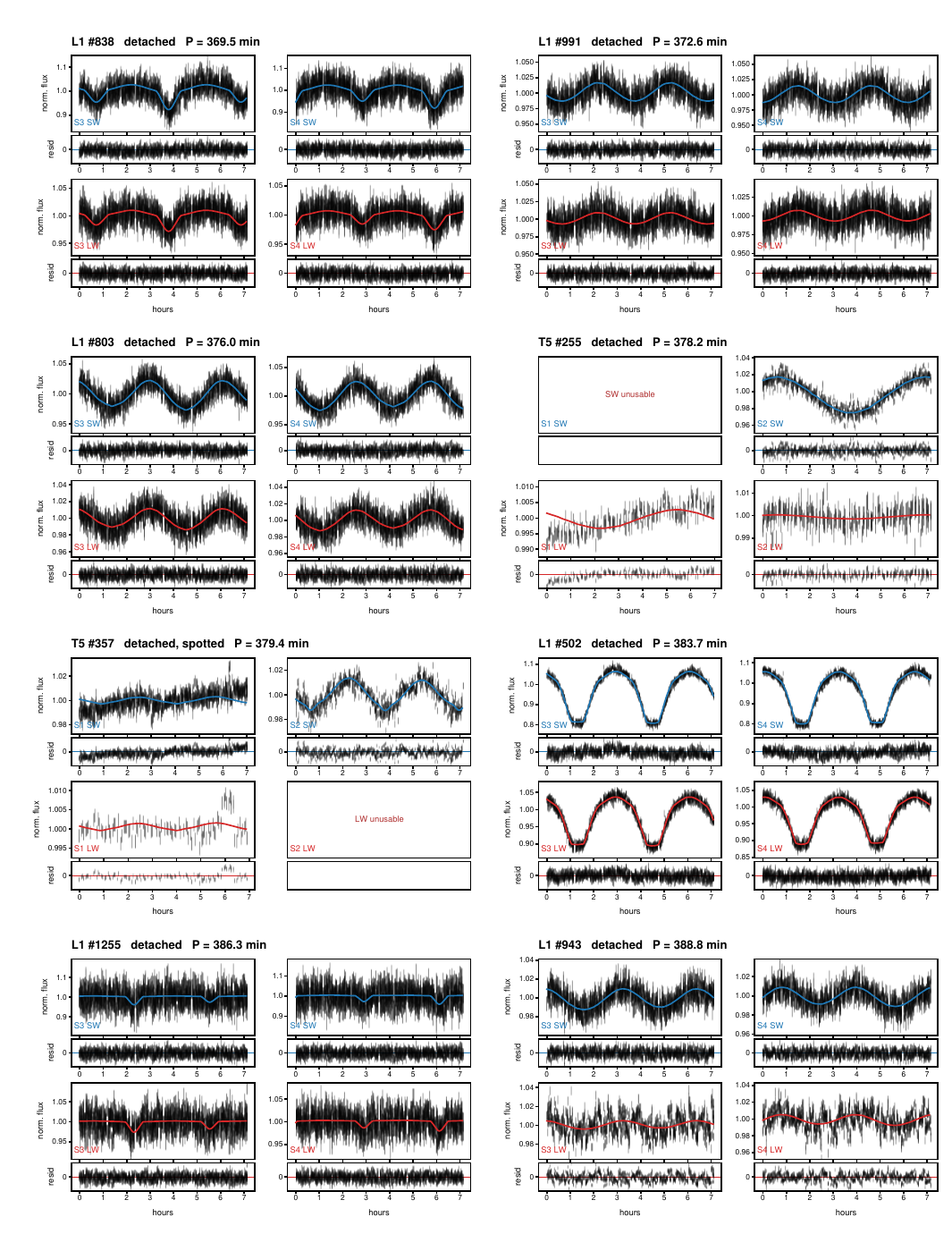}
\caption{Detached eclipsing binaries, continued (page 6 of 36).}
\end{figure*}
\clearpage

\begin{figure*}
\centering
\includegraphics[width=0.98\textwidth,height=0.94\textheight,keepaspectratio]{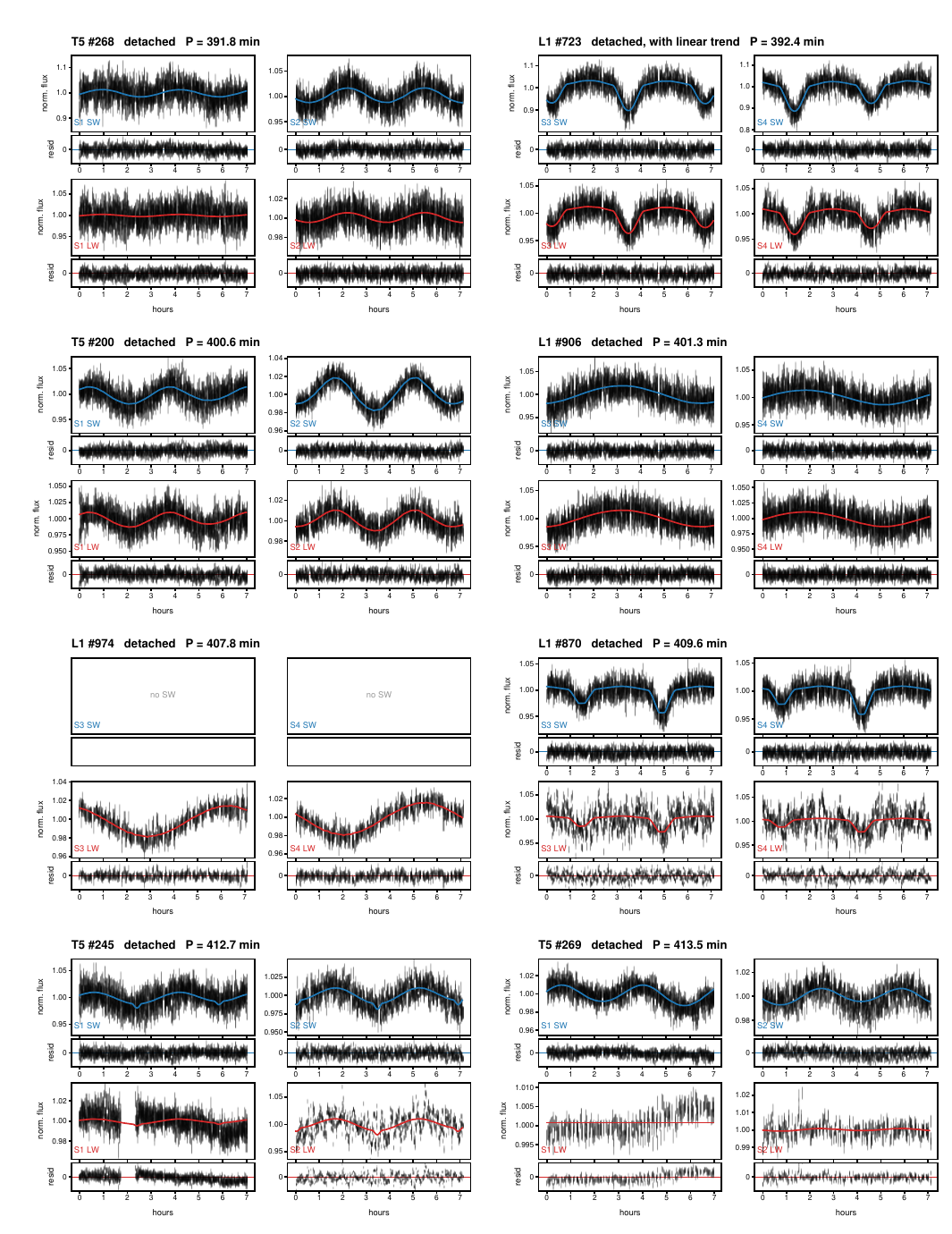}
\caption{Detached eclipsing binaries, continued (page 7 of 36).}
\end{figure*}
\clearpage

\begin{figure*}
\centering
\includegraphics[width=0.98\textwidth,height=0.94\textheight,keepaspectratio]{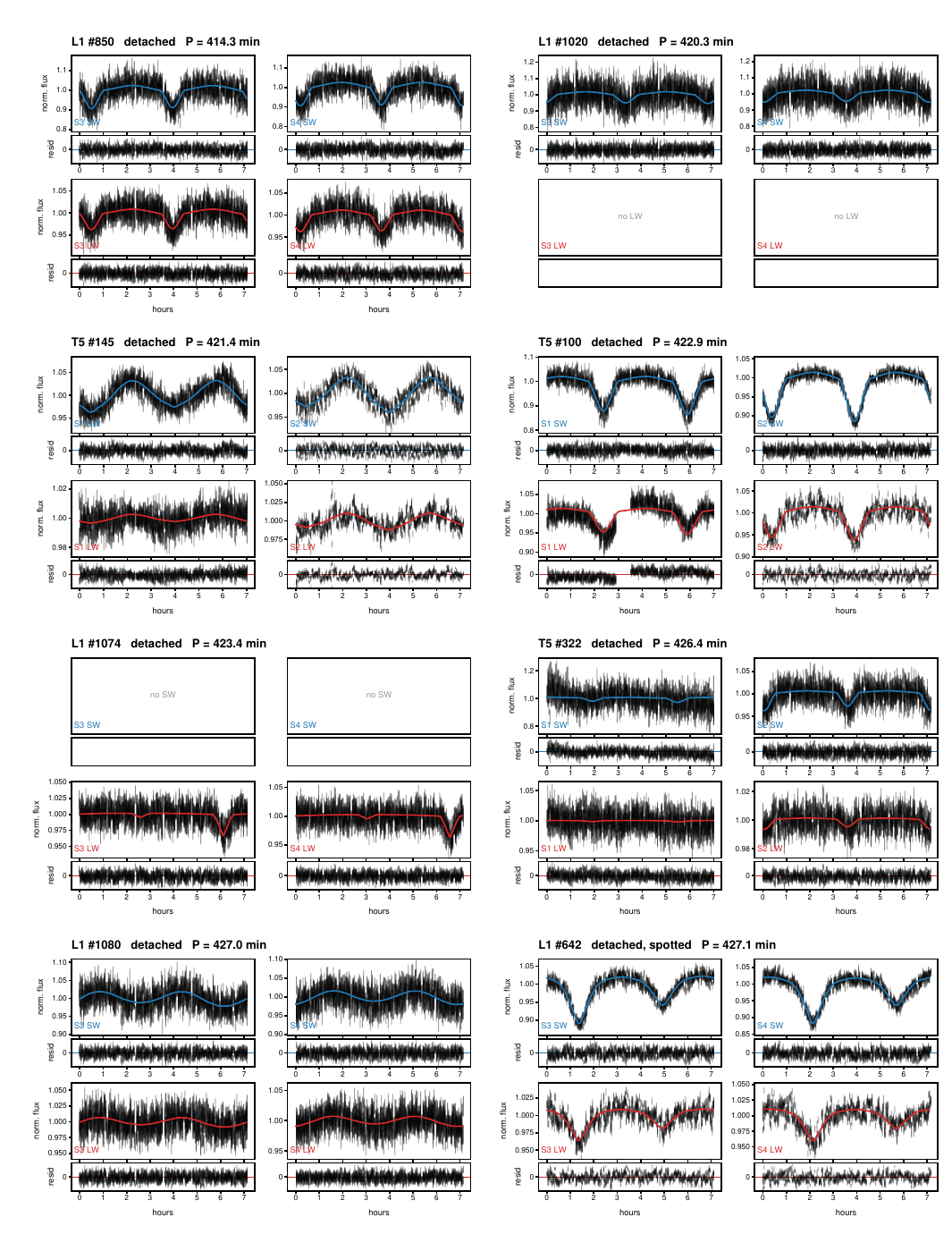}
\caption{Detached eclipsing binaries, continued (page 8 of 36).}
\end{figure*}
\clearpage

\begin{figure*}
\centering
\includegraphics[width=0.98\textwidth,height=0.94\textheight,keepaspectratio]{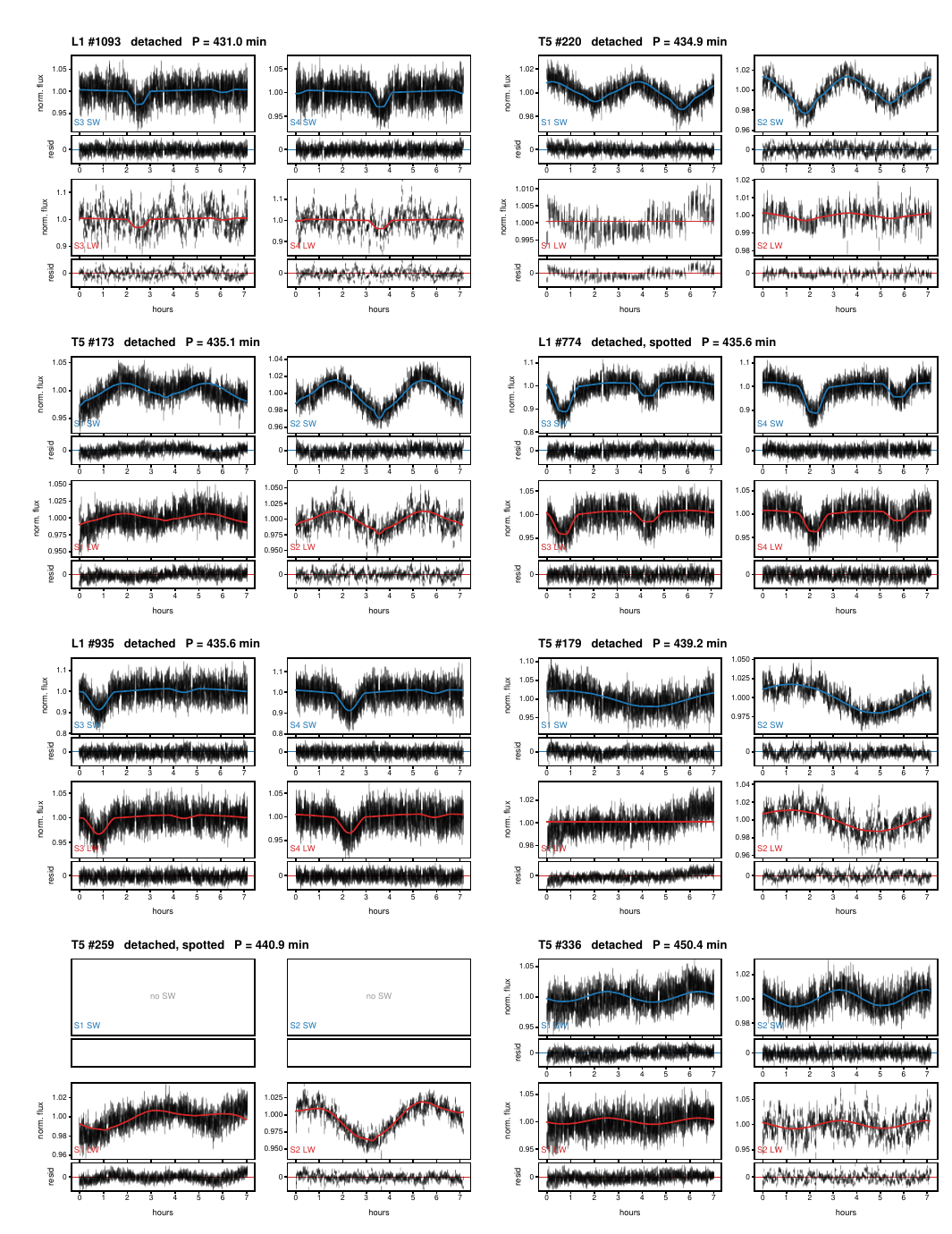}
\caption{Detached eclipsing binaries, continued (page 9 of 36).}
\end{figure*}
\clearpage

\begin{figure*}
\centering
\includegraphics[width=0.98\textwidth,height=0.94\textheight,keepaspectratio]{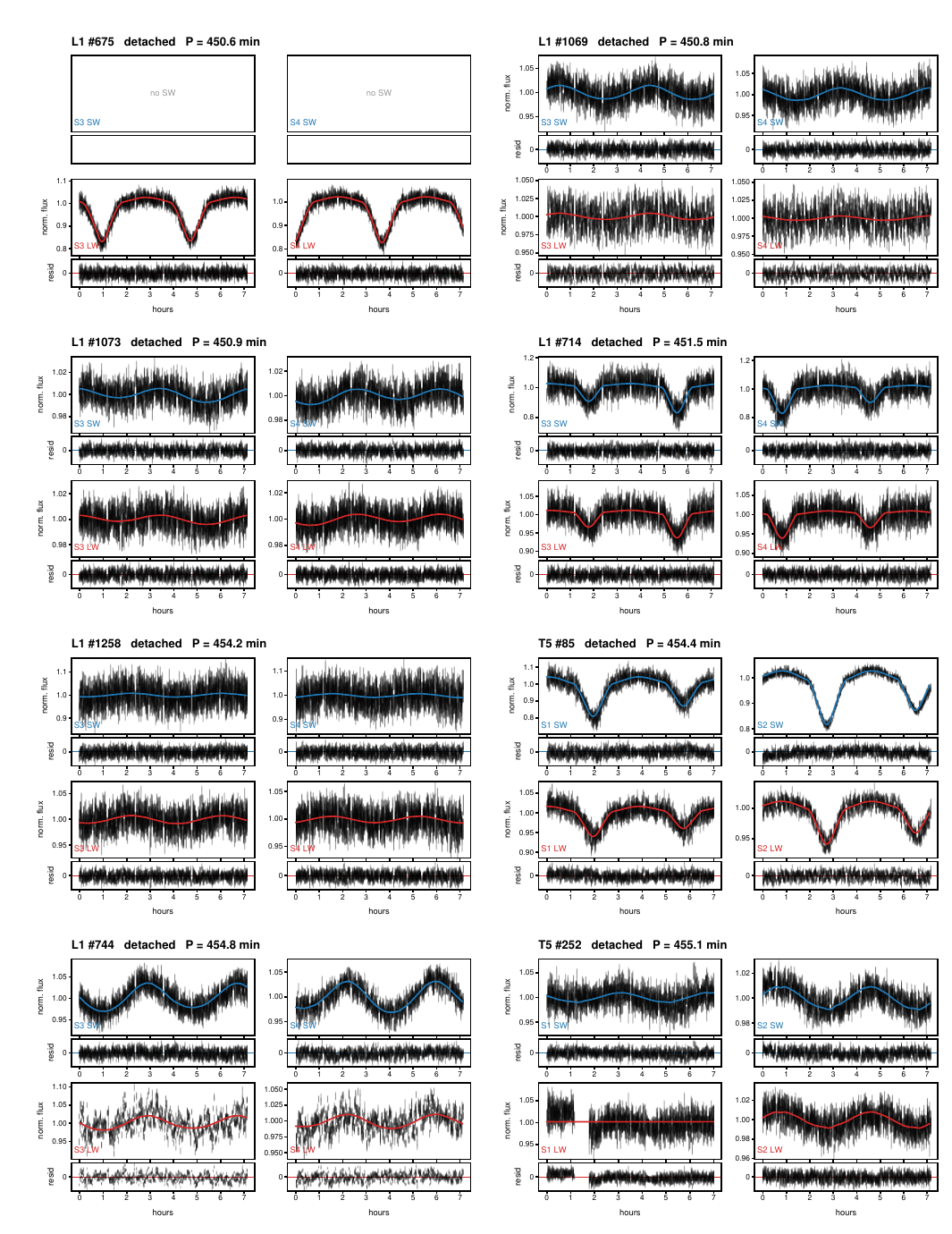}
\caption{Detached eclipsing binaries, continued (page 10 of 36).}
\end{figure*}
\clearpage

\begin{figure*}
\centering
\includegraphics[width=0.98\textwidth,height=0.94\textheight,keepaspectratio]{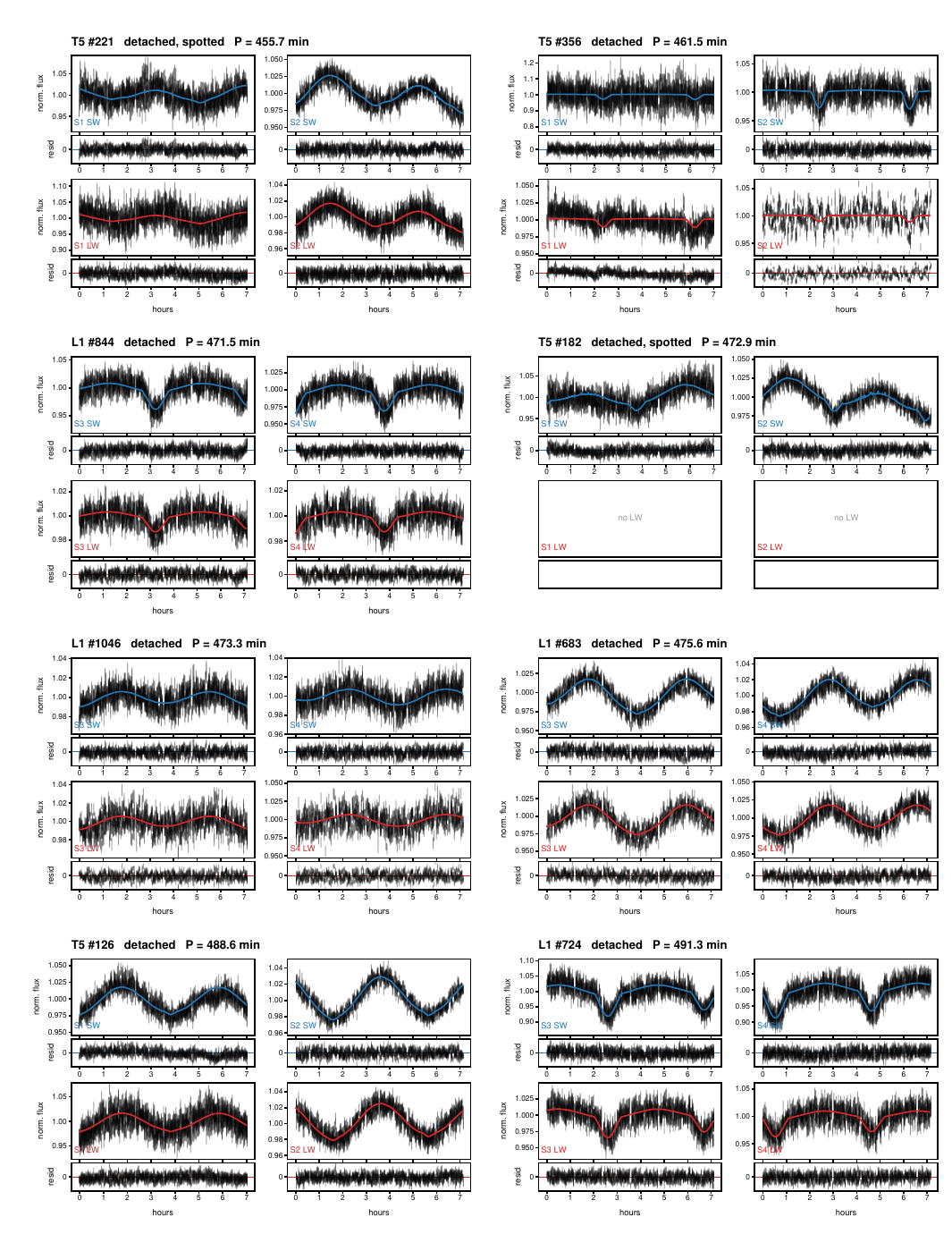}
\caption{Detached eclipsing binaries, continued (page 11 of 36).}
\end{figure*}
\clearpage

\begin{figure*}
\centering
\includegraphics[width=0.98\textwidth,height=0.94\textheight,keepaspectratio]{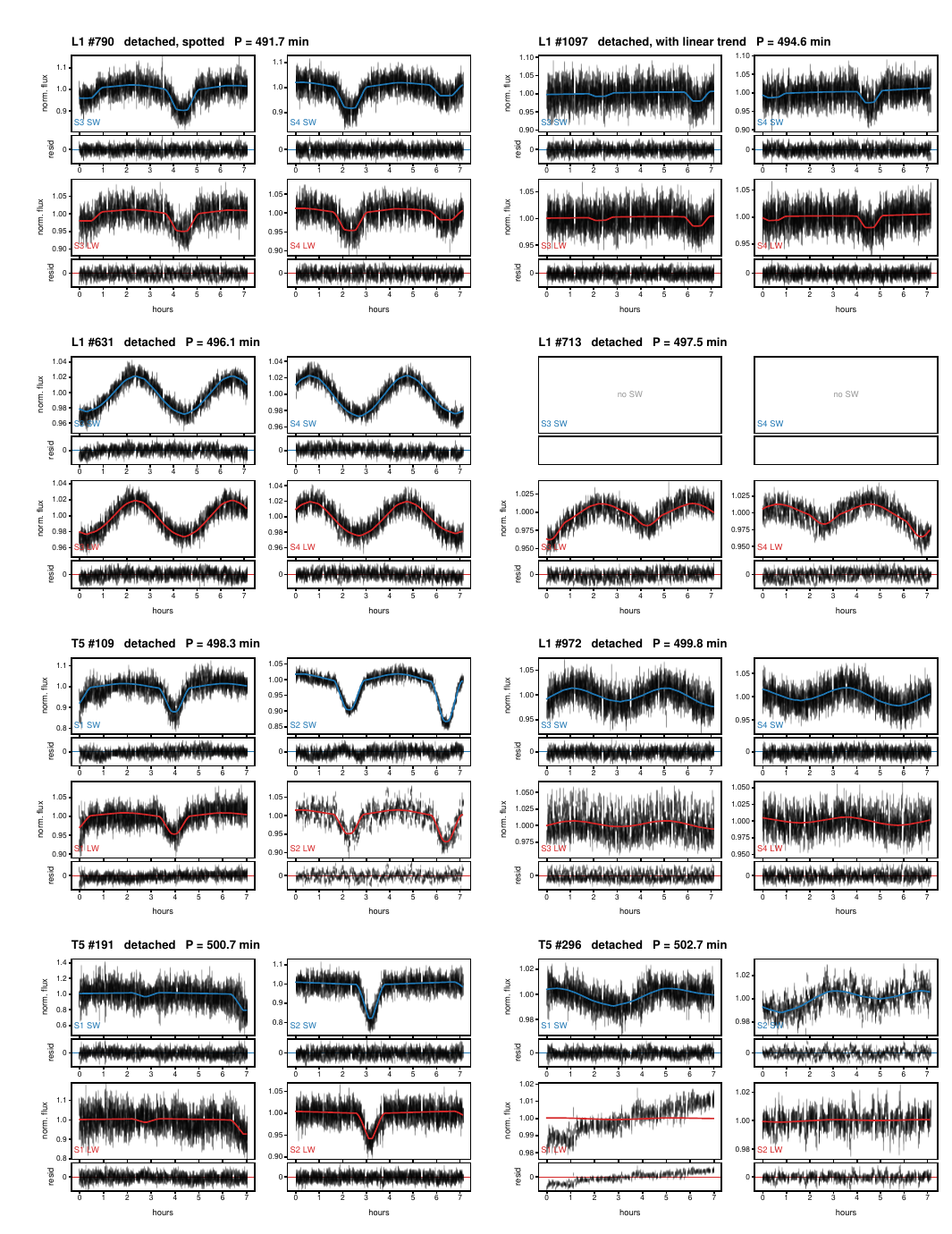}
\caption{Detached eclipsing binaries, continued (page 12 of 36).}
\end{figure*}
\clearpage

\begin{figure*}
\centering
\includegraphics[width=0.98\textwidth,height=0.94\textheight,keepaspectratio]{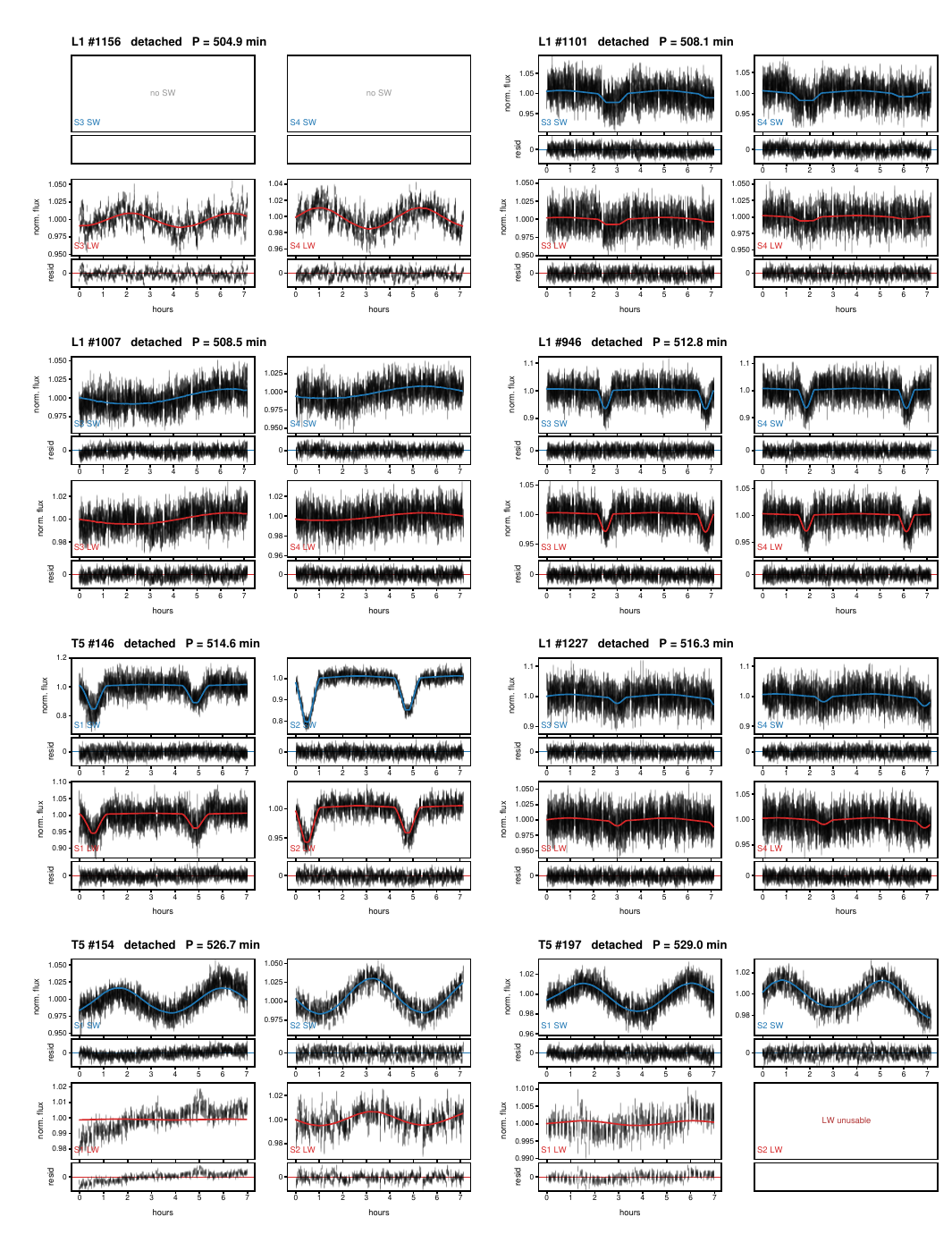}
\caption{Detached eclipsing binaries, continued (page 13 of 36).}
\end{figure*}
\clearpage

\begin{figure*}
\centering
\includegraphics[width=0.98\textwidth,height=0.94\textheight,keepaspectratio]{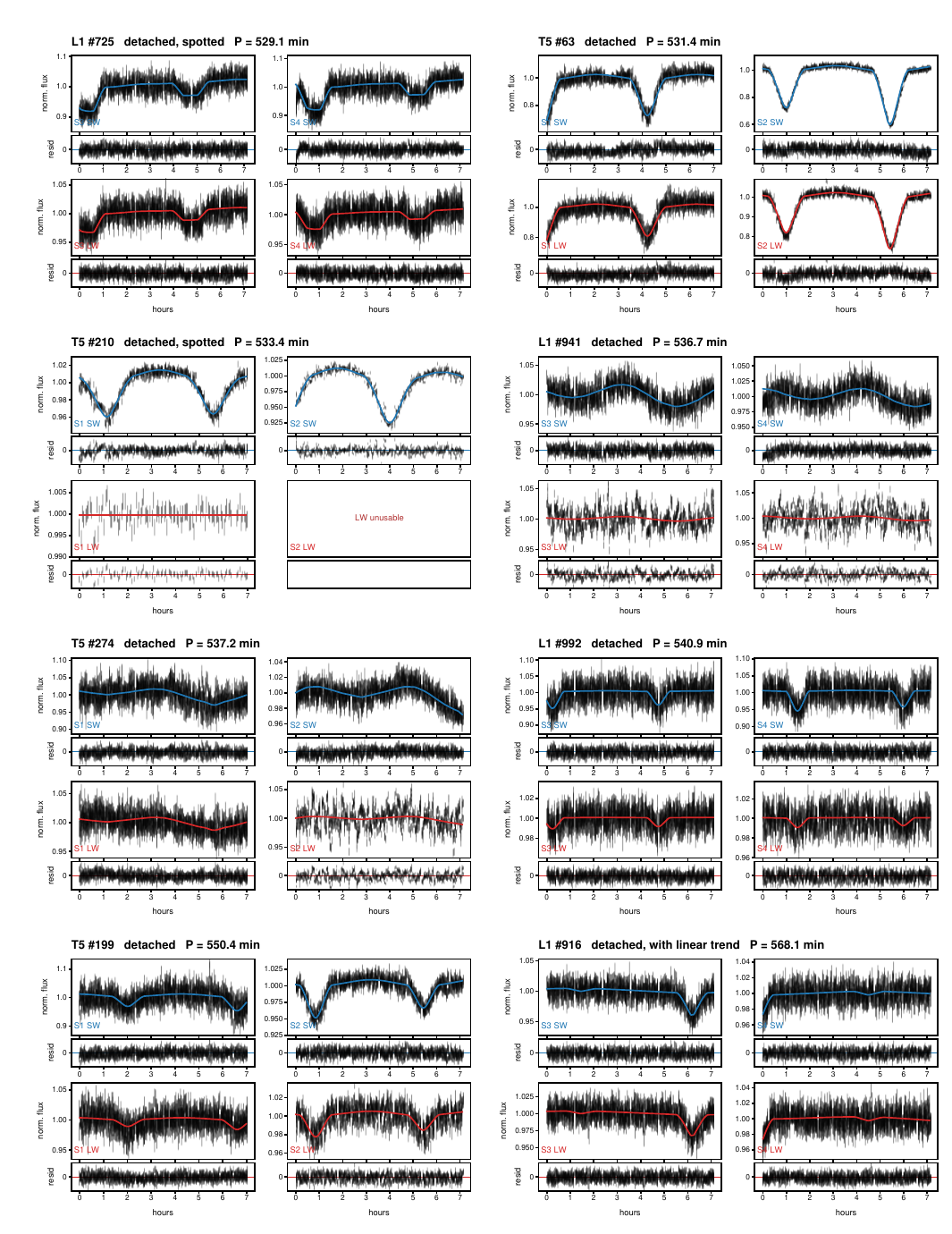}
\caption{Detached eclipsing binaries, continued (page 14 of 36).}
\end{figure*}
\clearpage

\begin{figure*}
\centering
\includegraphics[width=0.98\textwidth,height=0.94\textheight,keepaspectratio]{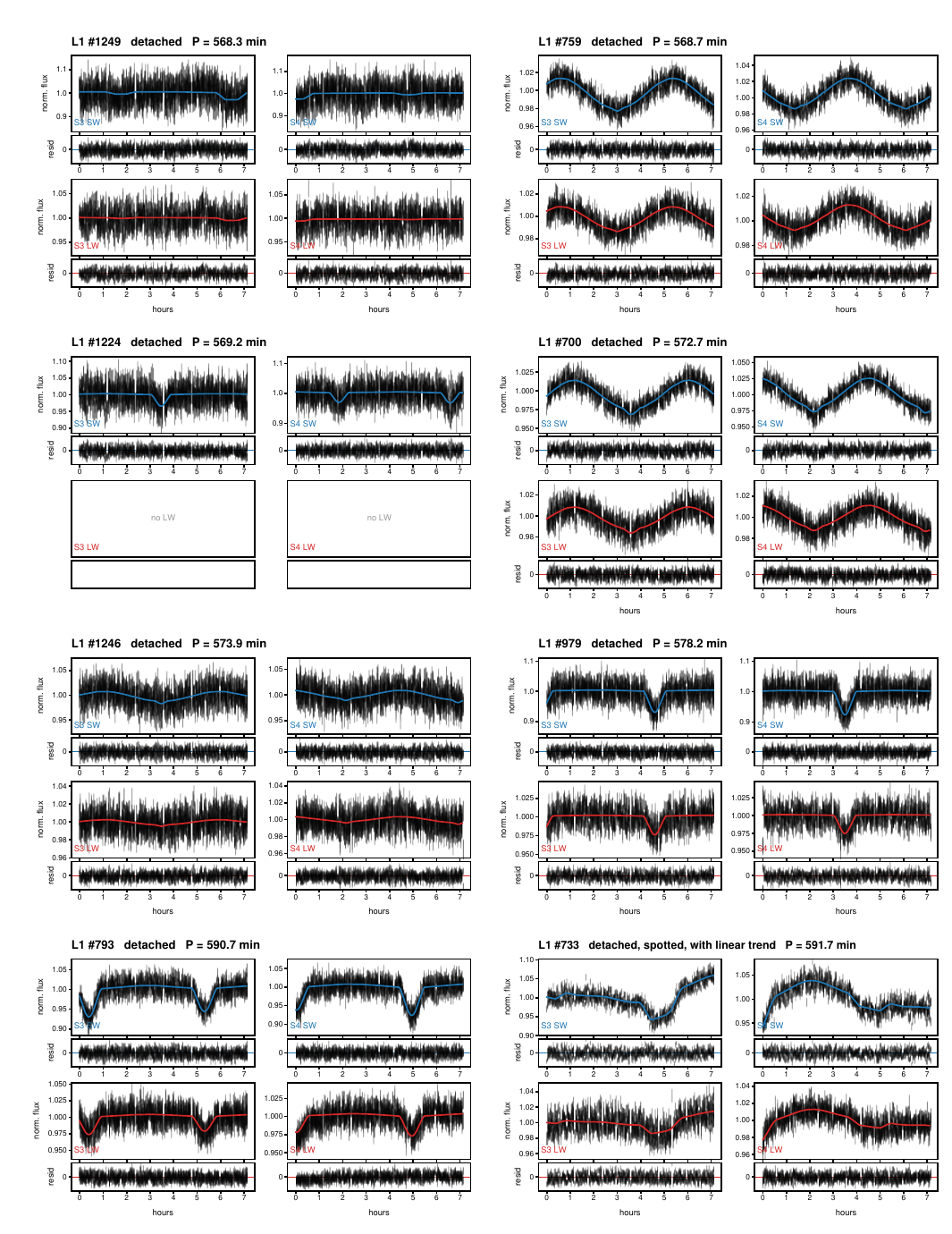}
\caption{Detached eclipsing binaries, continued (page 15 of 36).}
\end{figure*}
\clearpage

\begin{figure*}
\centering
\includegraphics[width=0.98\textwidth,height=0.94\textheight,keepaspectratio]{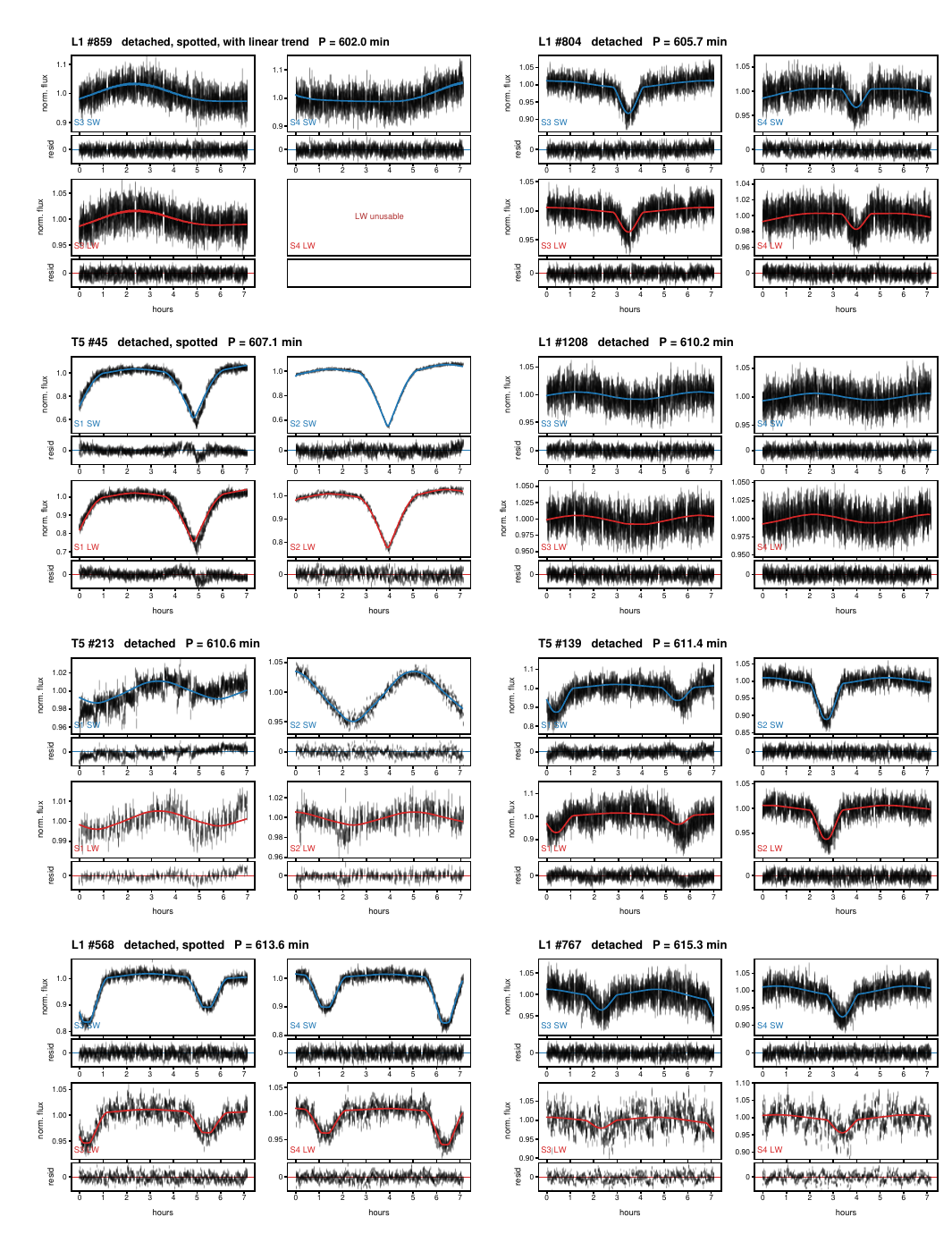}
\caption{Detached eclipsing binaries, continued (page 16 of 36).}
\end{figure*}
\clearpage

\begin{figure*}
\centering
\includegraphics[width=0.98\textwidth,height=0.94\textheight,keepaspectratio]{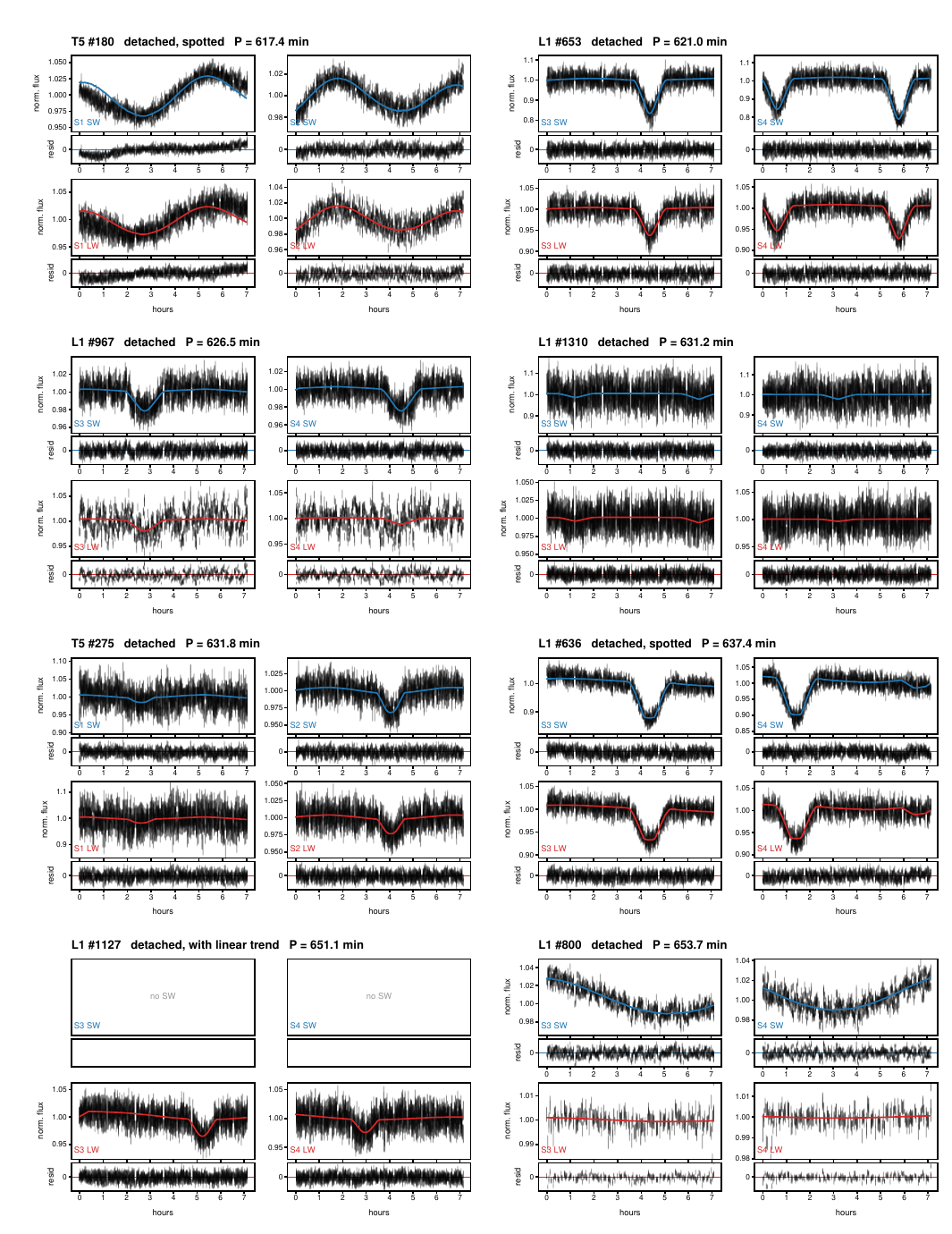}
\caption{Detached eclipsing binaries, continued (page 17 of 36).}
\end{figure*}
\clearpage

\begin{figure*}
\centering
\includegraphics[width=0.98\textwidth,height=0.94\textheight,keepaspectratio]{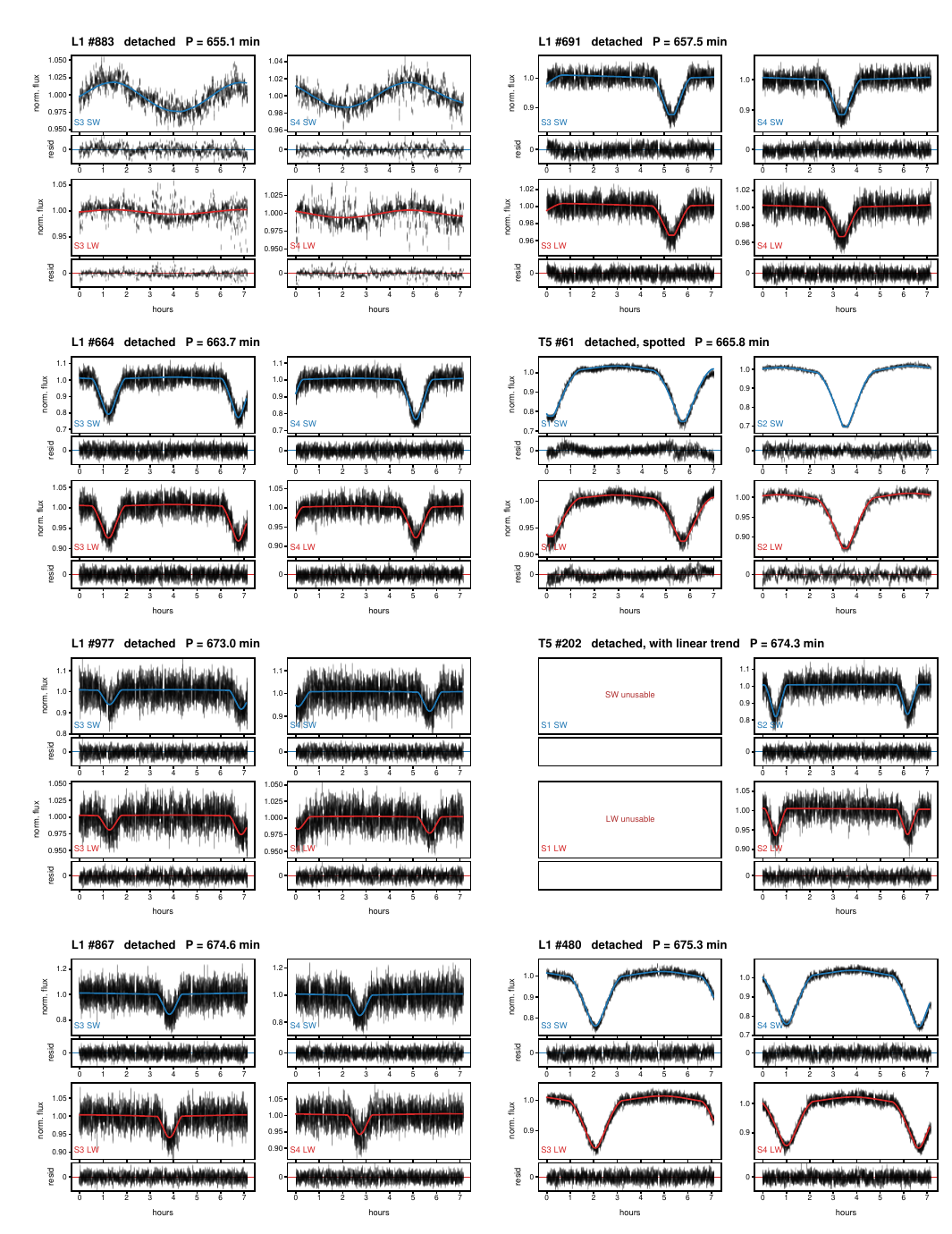}
\caption{Detached eclipsing binaries, continued (page 18 of 36).}
\end{figure*}
\clearpage

\begin{figure*}
\centering
\includegraphics[width=0.98\textwidth,height=0.94\textheight,keepaspectratio]{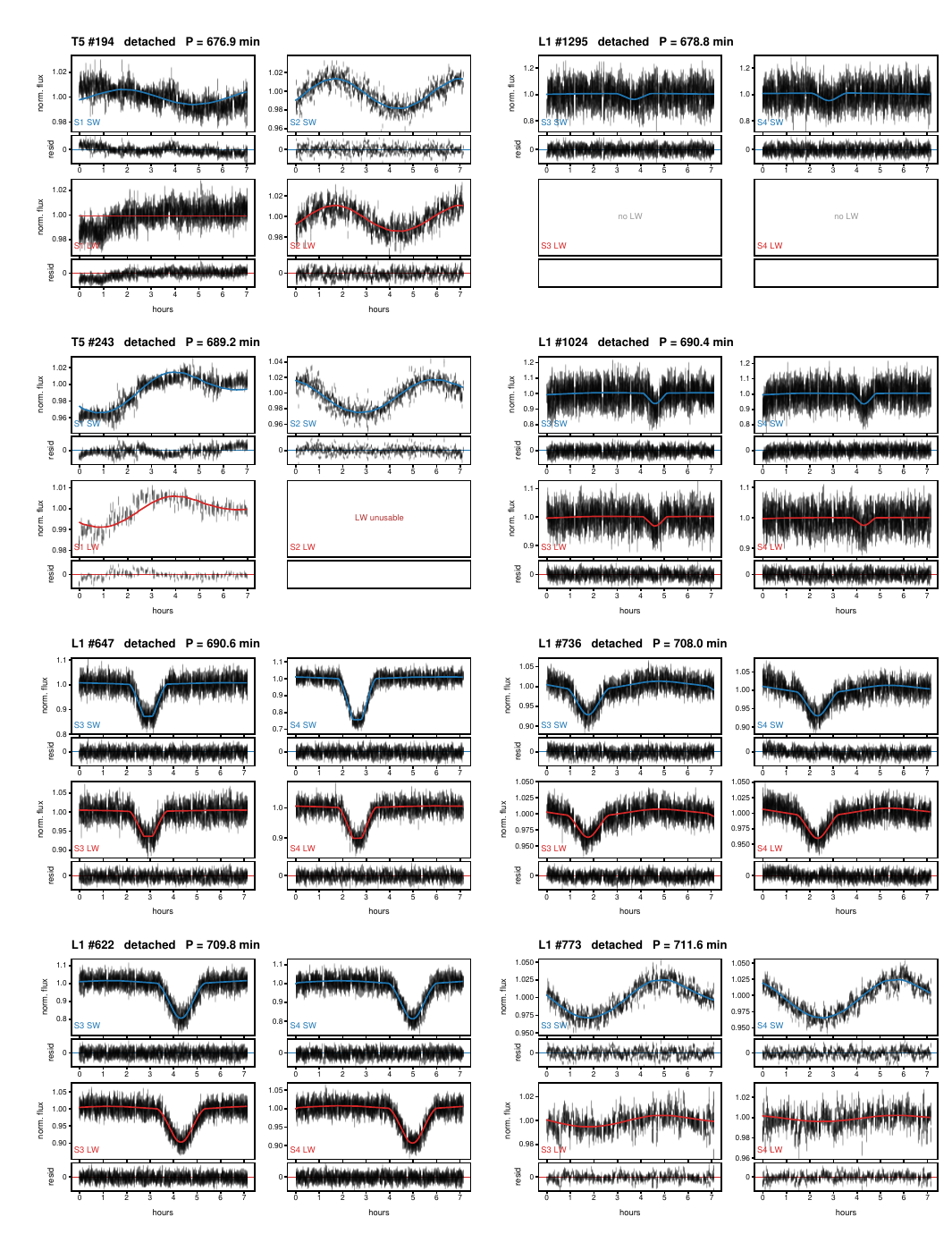}
\caption{Detached eclipsing binaries, continued (page 19 of 36).}
\end{figure*}
\clearpage

\begin{figure*}
\centering
\includegraphics[width=0.98\textwidth,height=0.94\textheight,keepaspectratio]{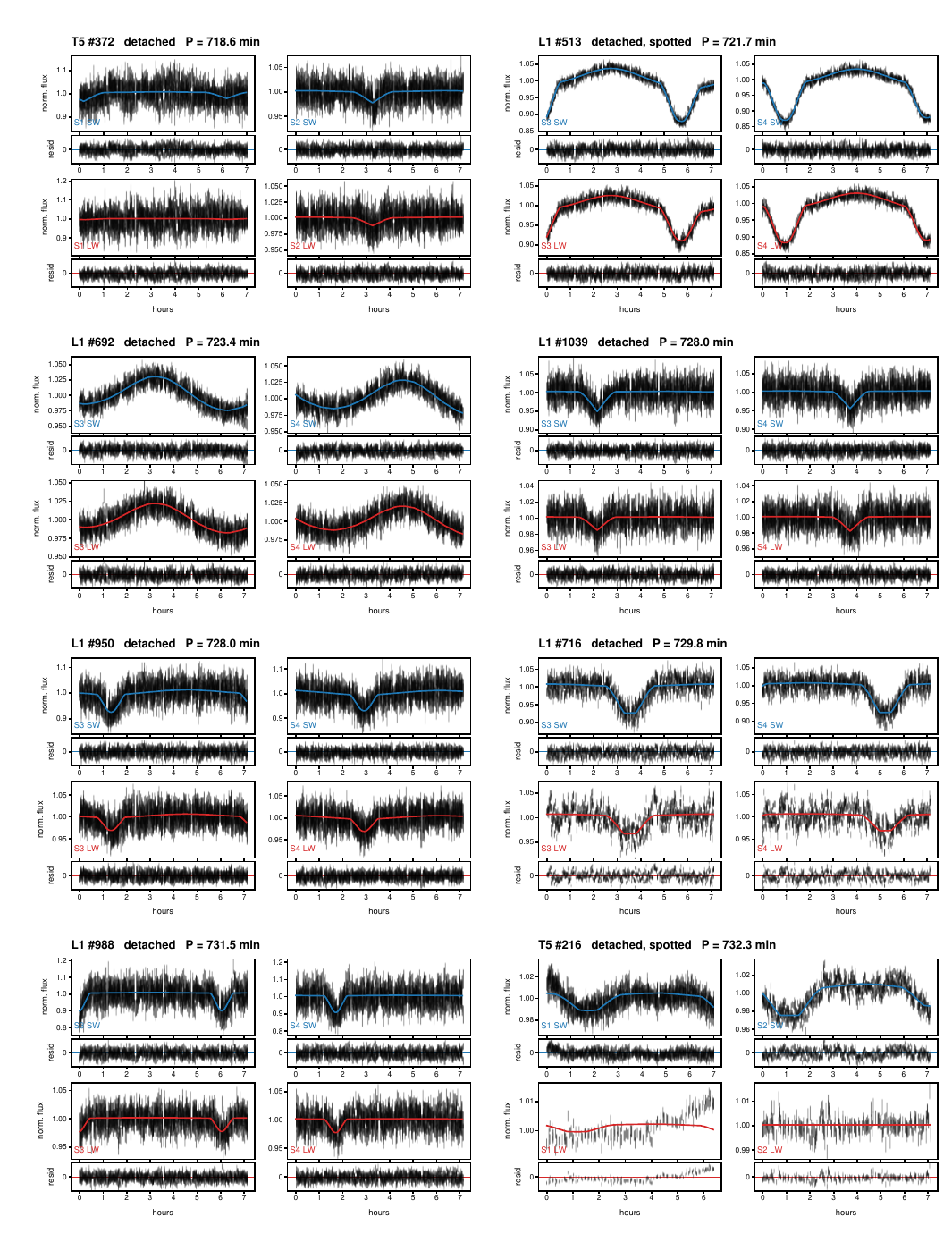}
\caption{Detached eclipsing binaries, continued (page 20 of 36).}
\end{figure*}
\clearpage

\begin{figure*}
\centering
\includegraphics[width=0.98\textwidth,height=0.94\textheight,keepaspectratio]{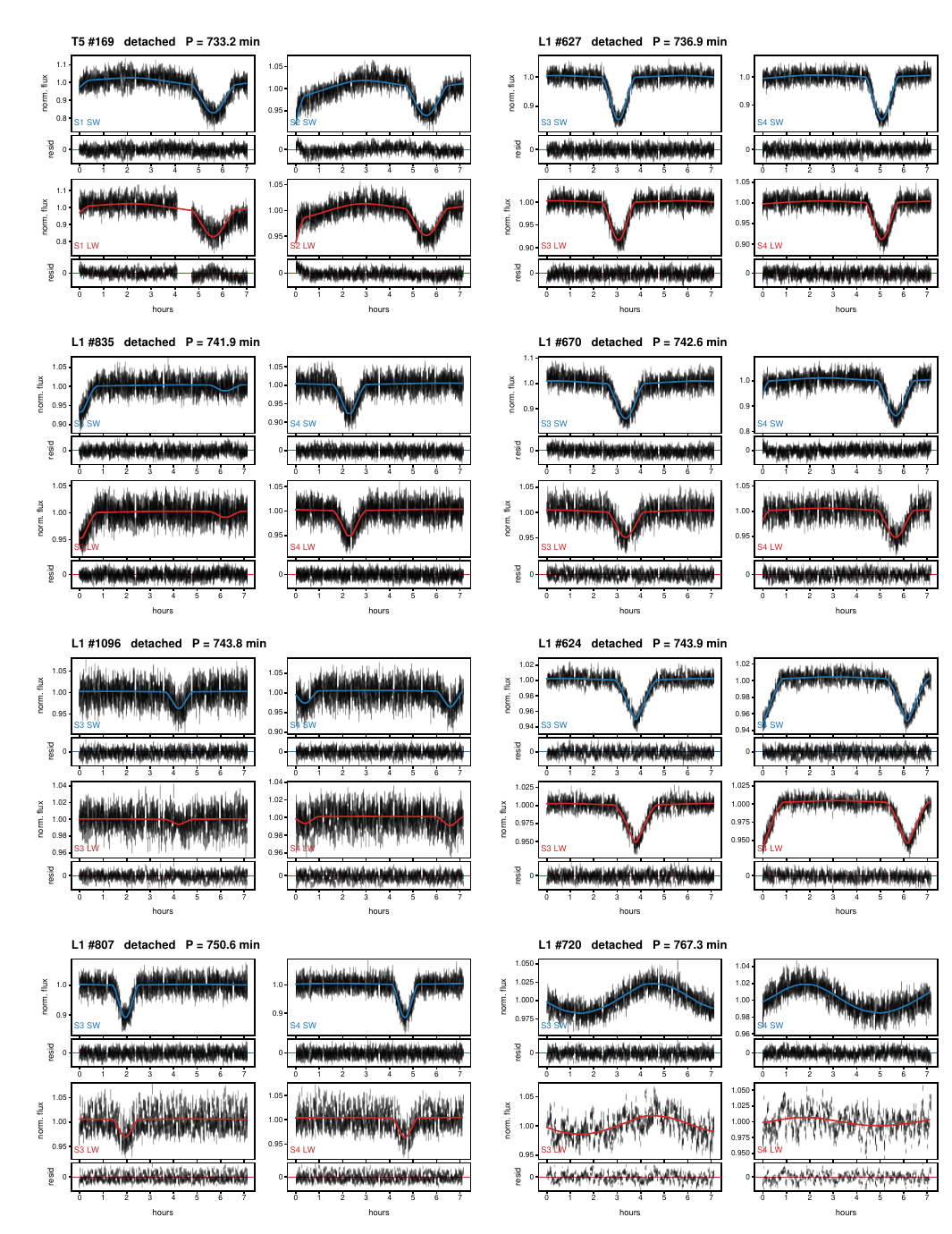}
\caption{Detached eclipsing binaries, continued (page 21 of 36).}
\end{figure*}
\clearpage

\begin{figure*}
\centering
\includegraphics[width=0.98\textwidth,height=0.94\textheight,keepaspectratio]{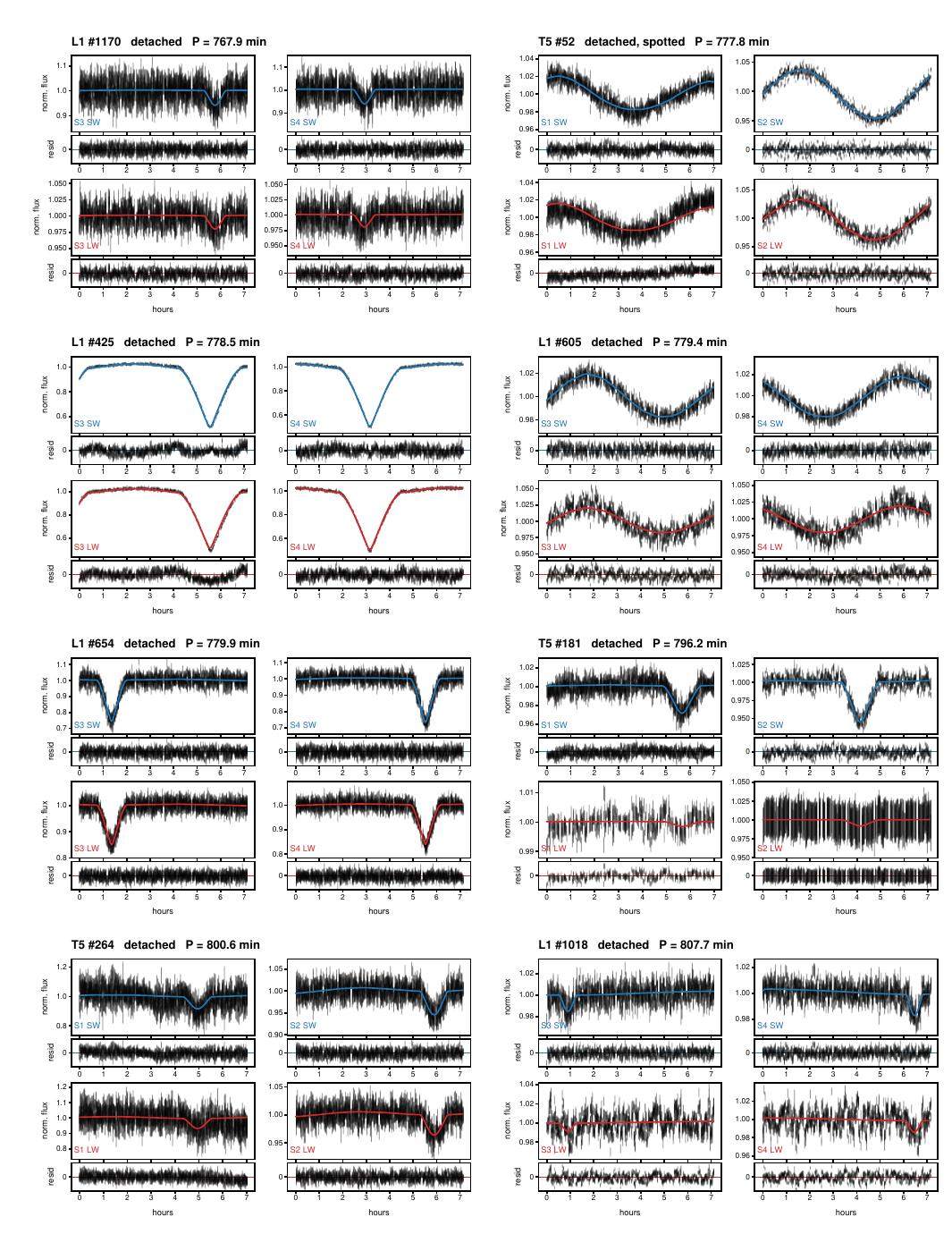}
\caption{Detached eclipsing binaries, continued (page 22 of 36).}
\end{figure*}
\clearpage

\begin{figure*}
\centering
\includegraphics[width=0.98\textwidth,height=0.94\textheight,keepaspectratio]{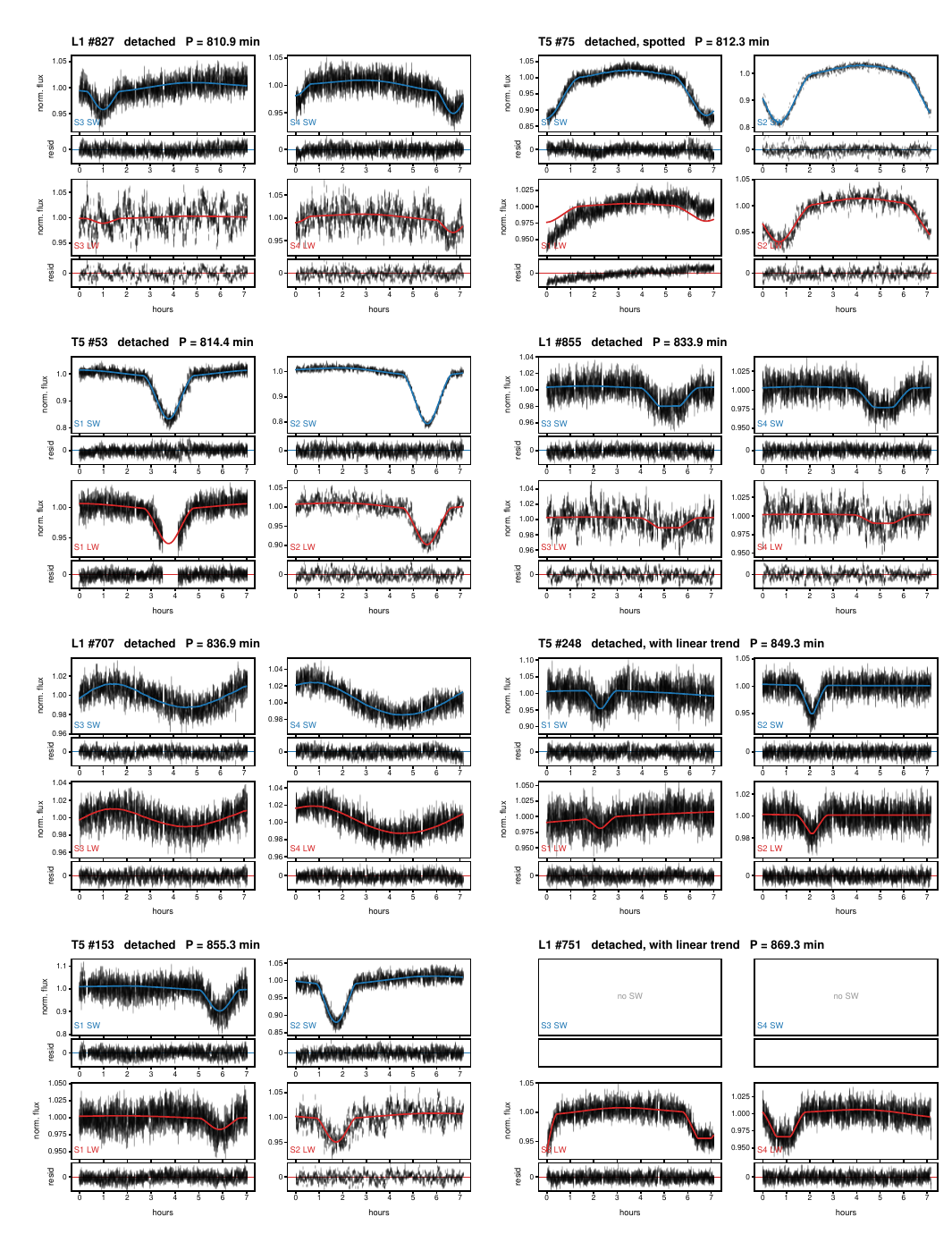}
\caption{Detached eclipsing binaries, continued (page 23 of 36).}
\end{figure*}
\clearpage

\begin{figure*}
\centering
\includegraphics[width=0.98\textwidth,height=0.94\textheight,keepaspectratio]{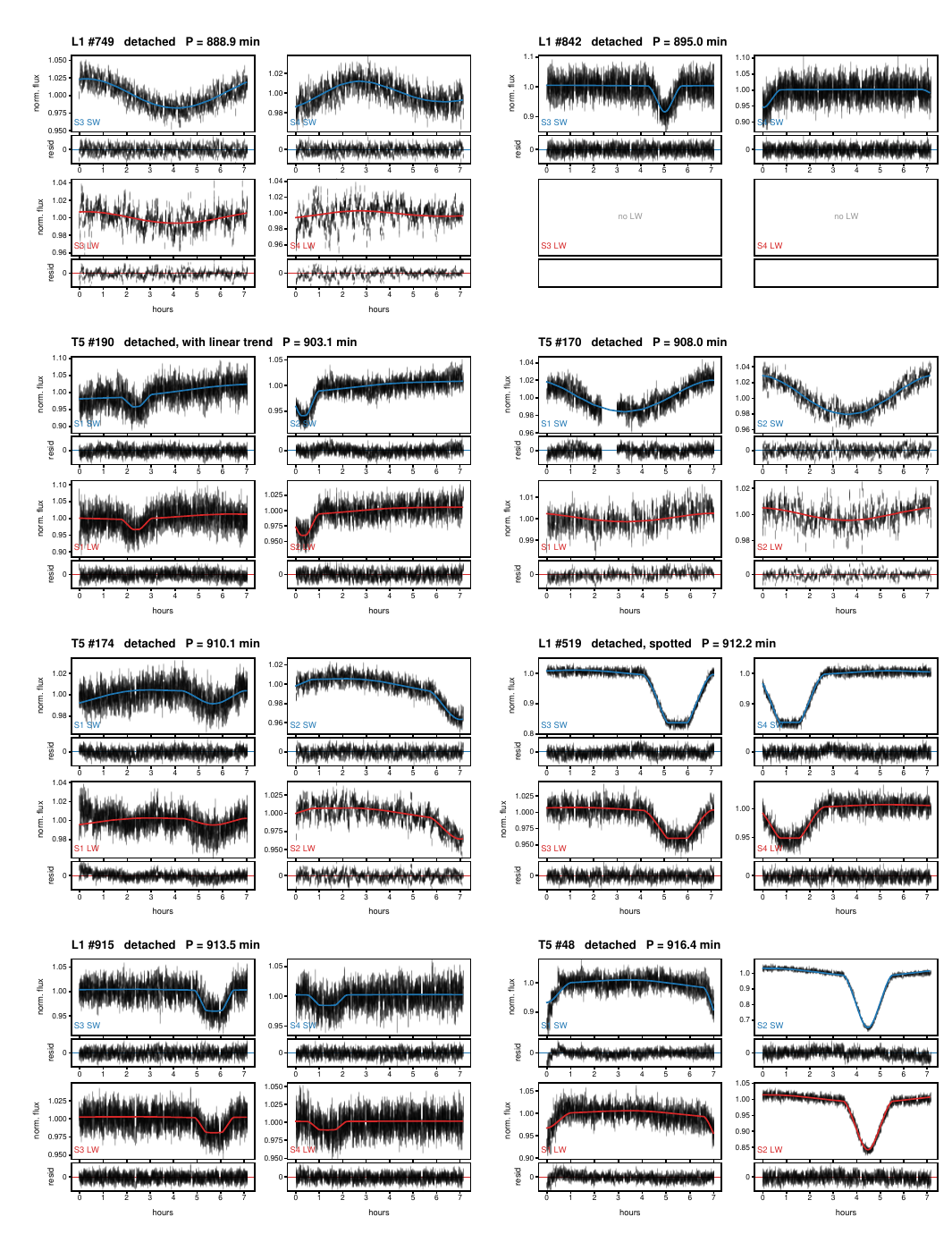}
\caption{Detached eclipsing binaries, continued (page 24 of 36).}
\end{figure*}
\clearpage

\begin{figure*}
\centering
\includegraphics[width=0.98\textwidth,height=0.94\textheight,keepaspectratio]{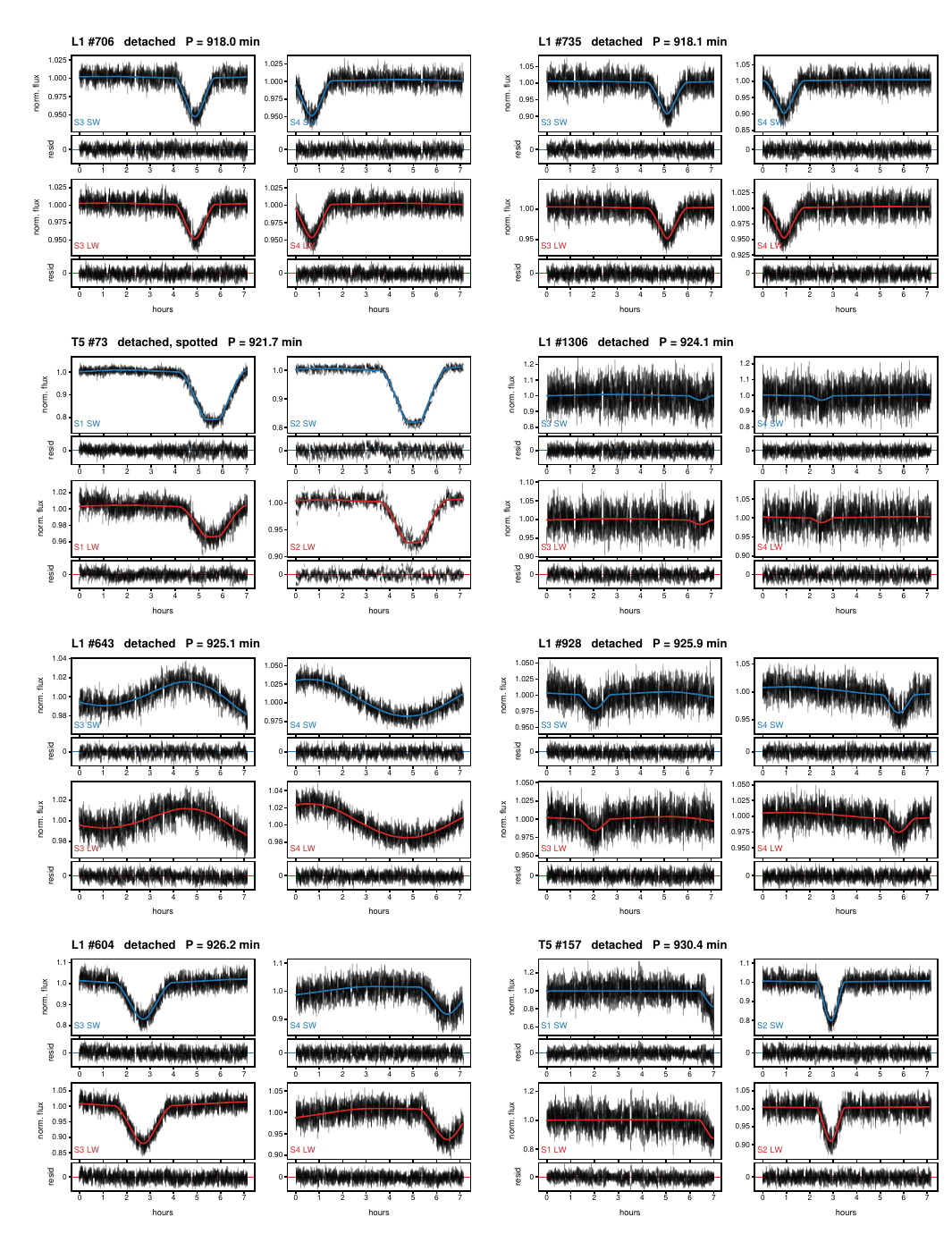}
\caption{Detached eclipsing binaries, continued (page 25 of 36).}
\end{figure*}
\clearpage

\begin{figure*}
\centering
\includegraphics[width=0.98\textwidth,height=0.94\textheight,keepaspectratio]{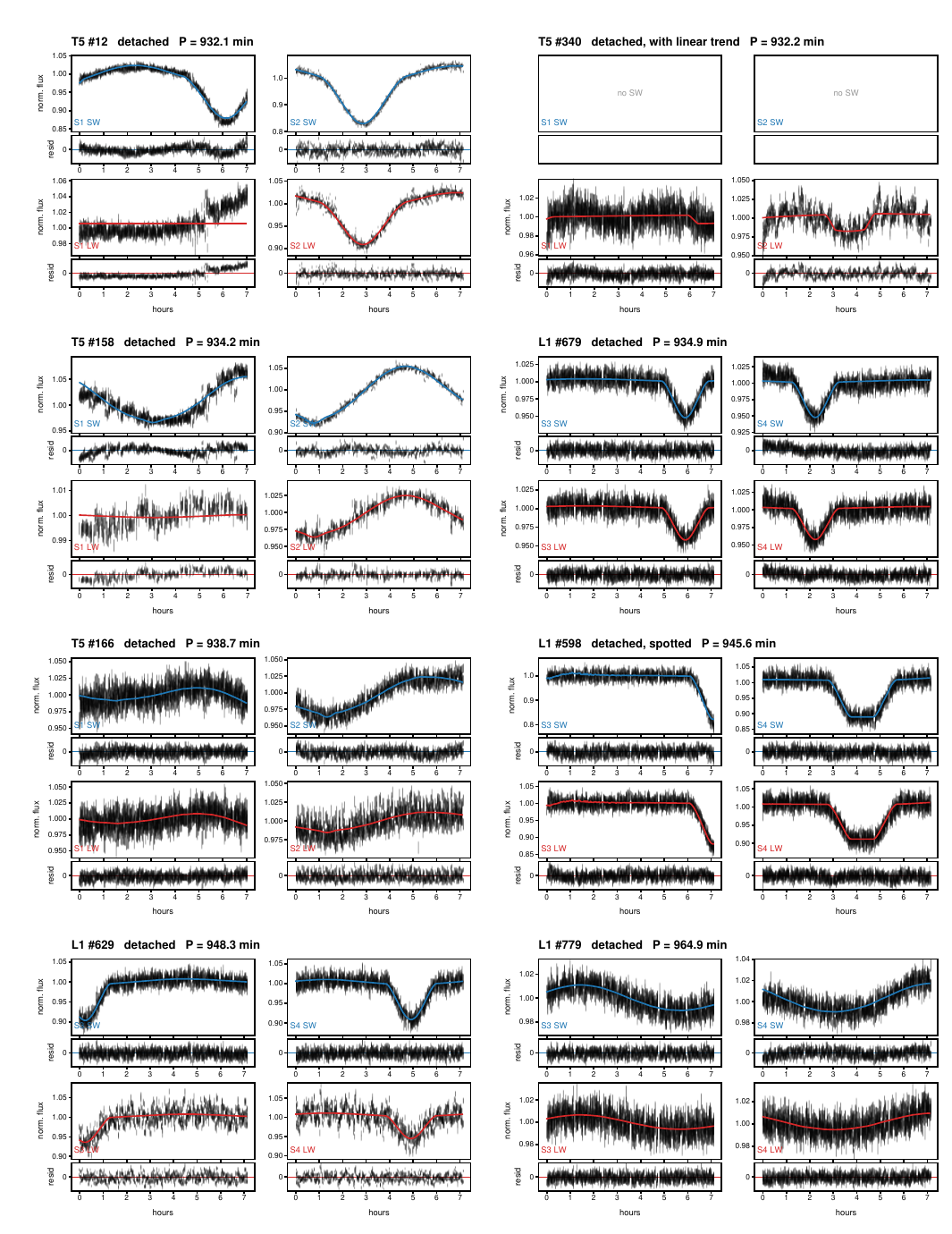}
\caption{Detached eclipsing binaries, continued (page 26 of 36).}
\end{figure*}
\clearpage

\begin{figure*}
\centering
\includegraphics[width=0.98\textwidth,height=0.94\textheight,keepaspectratio]{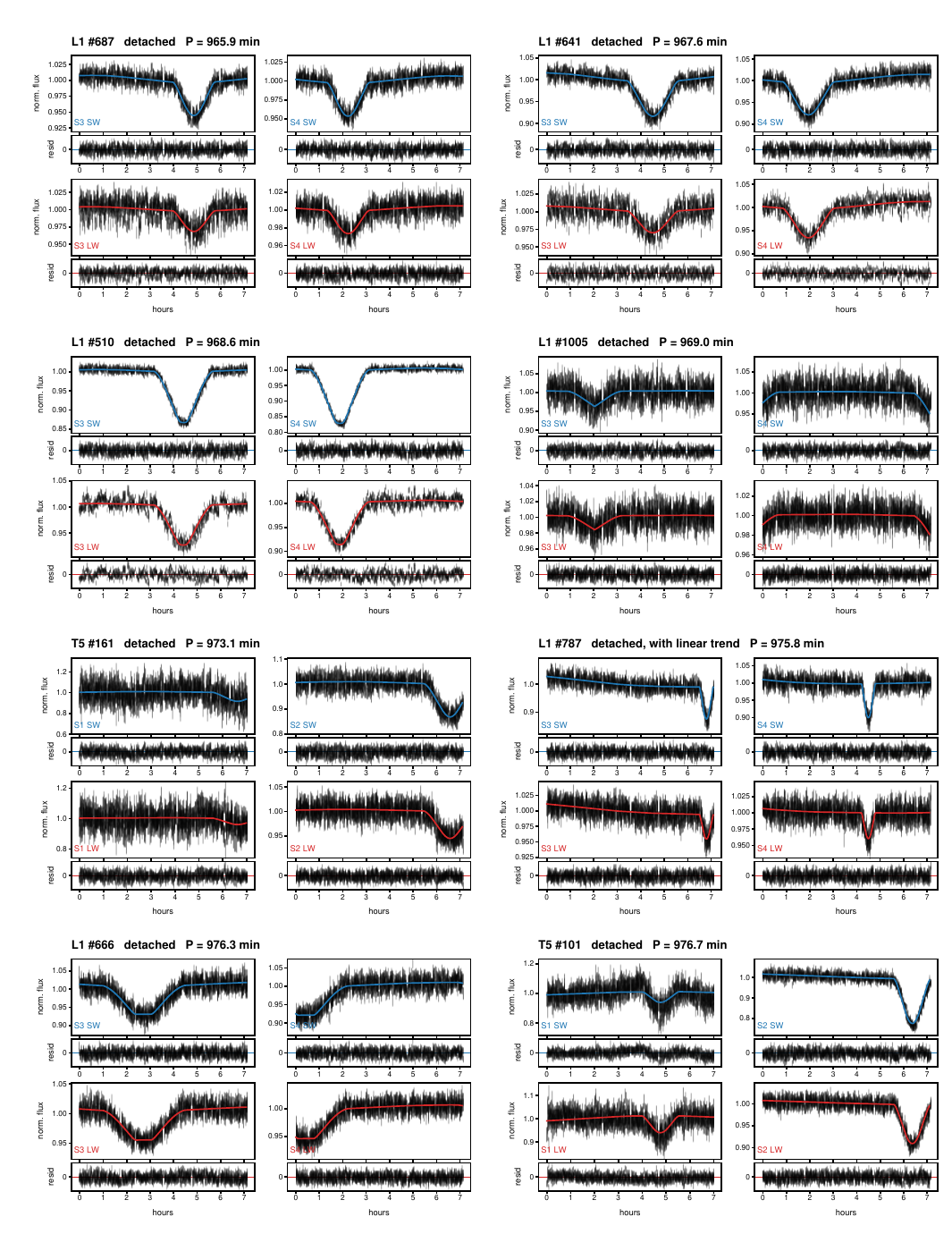}
\caption{Detached eclipsing binaries, continued (page 27 of 36).}
\end{figure*}
\clearpage

\begin{figure*}
\centering
\includegraphics[width=0.98\textwidth,height=0.94\textheight,keepaspectratio]{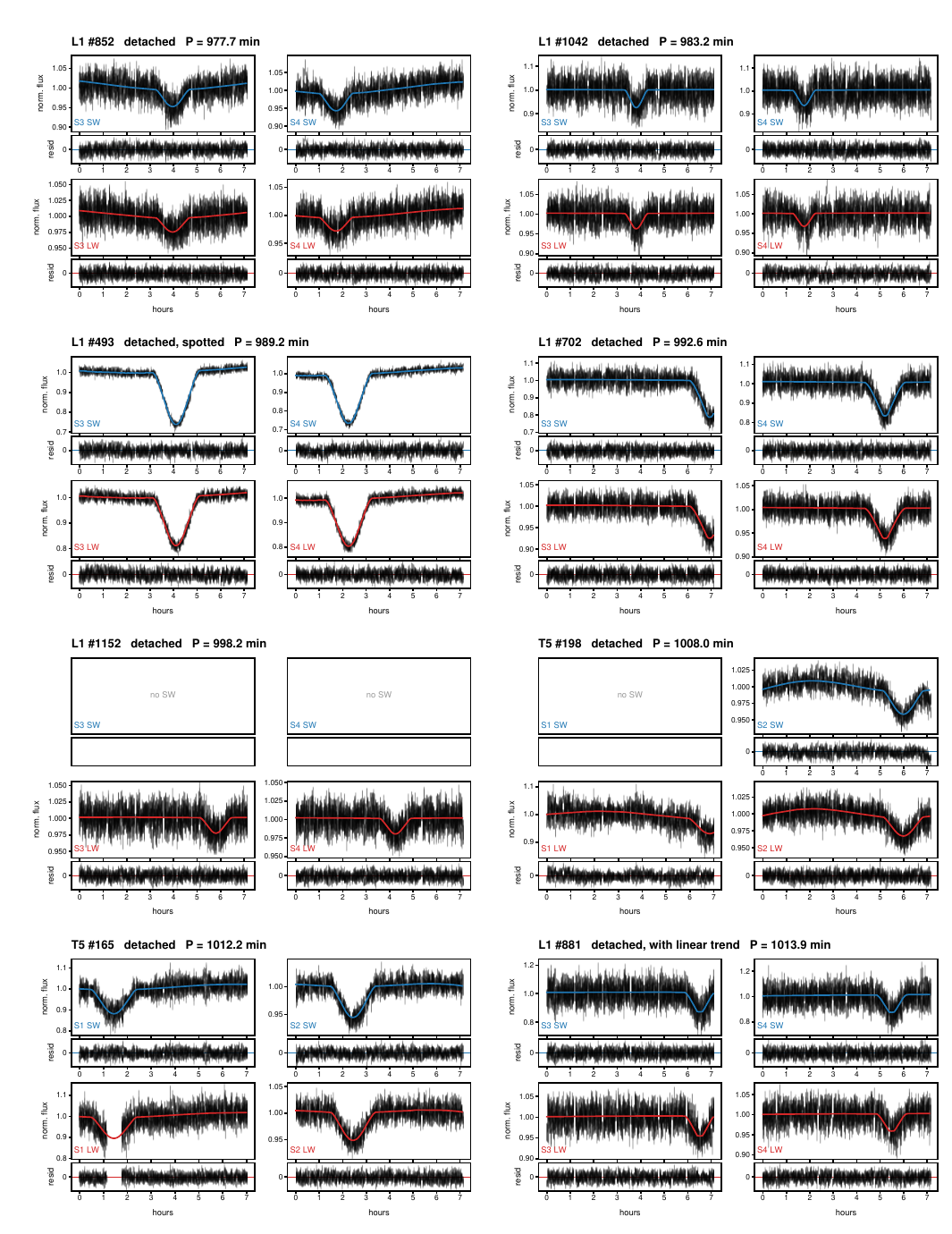}
\caption{Detached eclipsing binaries, continued (page 28 of 36).}
\end{figure*}
\clearpage

\begin{figure*}
\centering
\includegraphics[width=0.98\textwidth,height=0.94\textheight,keepaspectratio]{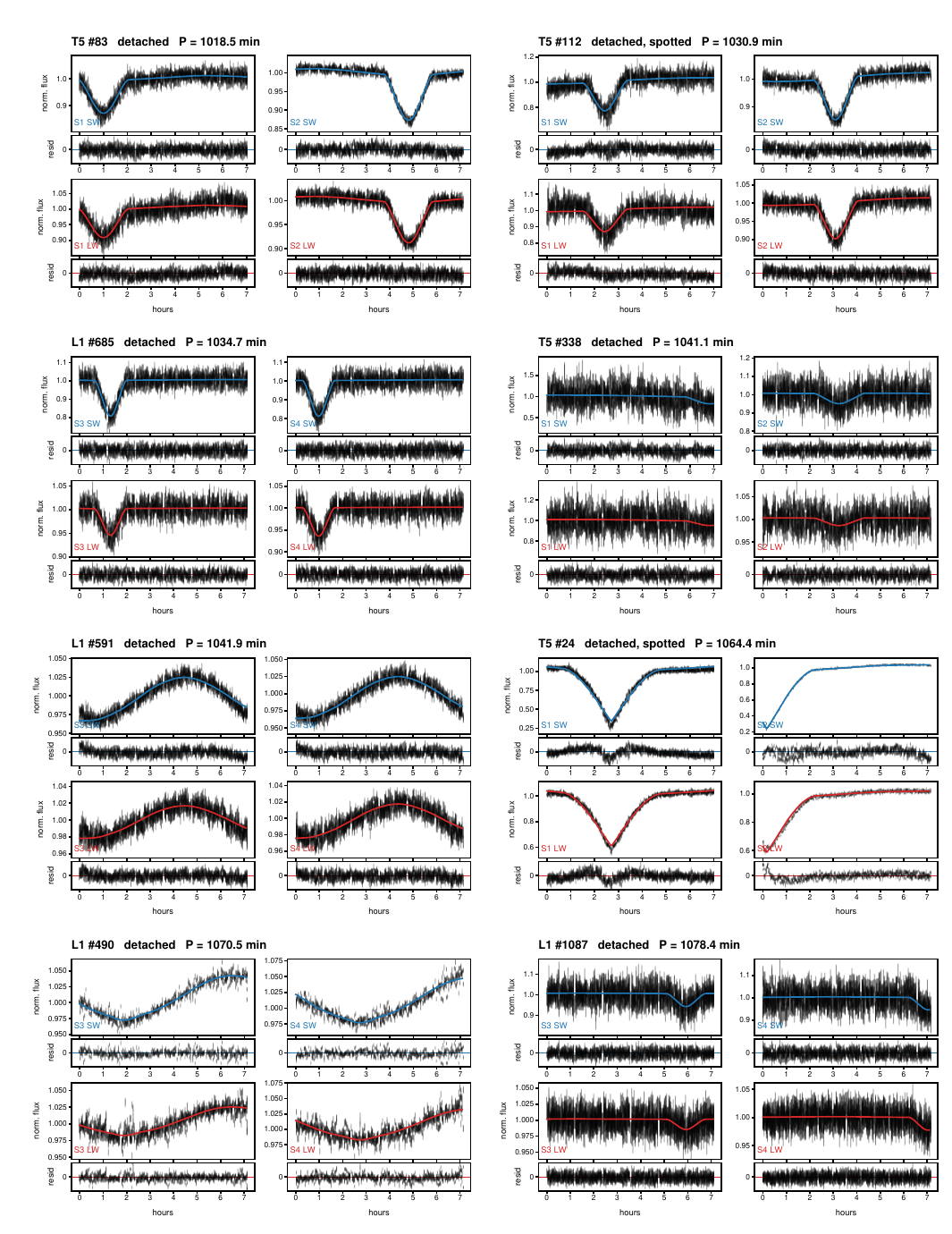}
\caption{Detached eclipsing binaries, continued (page 29 of 36).}
\end{figure*}
\clearpage

\begin{figure*}
\centering
\includegraphics[width=0.98\textwidth,height=0.94\textheight,keepaspectratio]{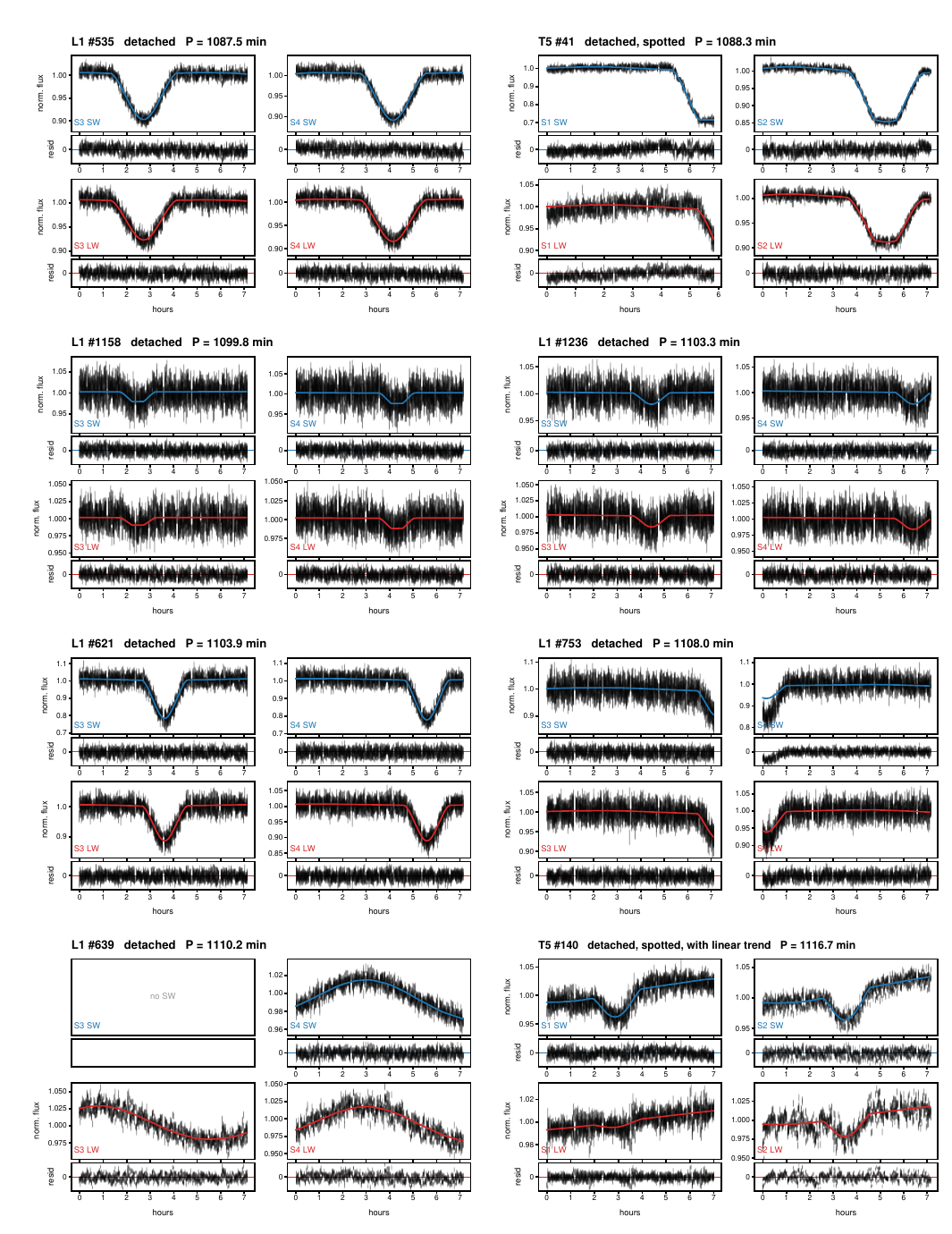}
\caption{Detached eclipsing binaries, continued (page 30 of 36).}
\end{figure*}
\clearpage

\begin{figure*}
\centering
\includegraphics[width=0.98\textwidth,height=0.94\textheight,keepaspectratio]{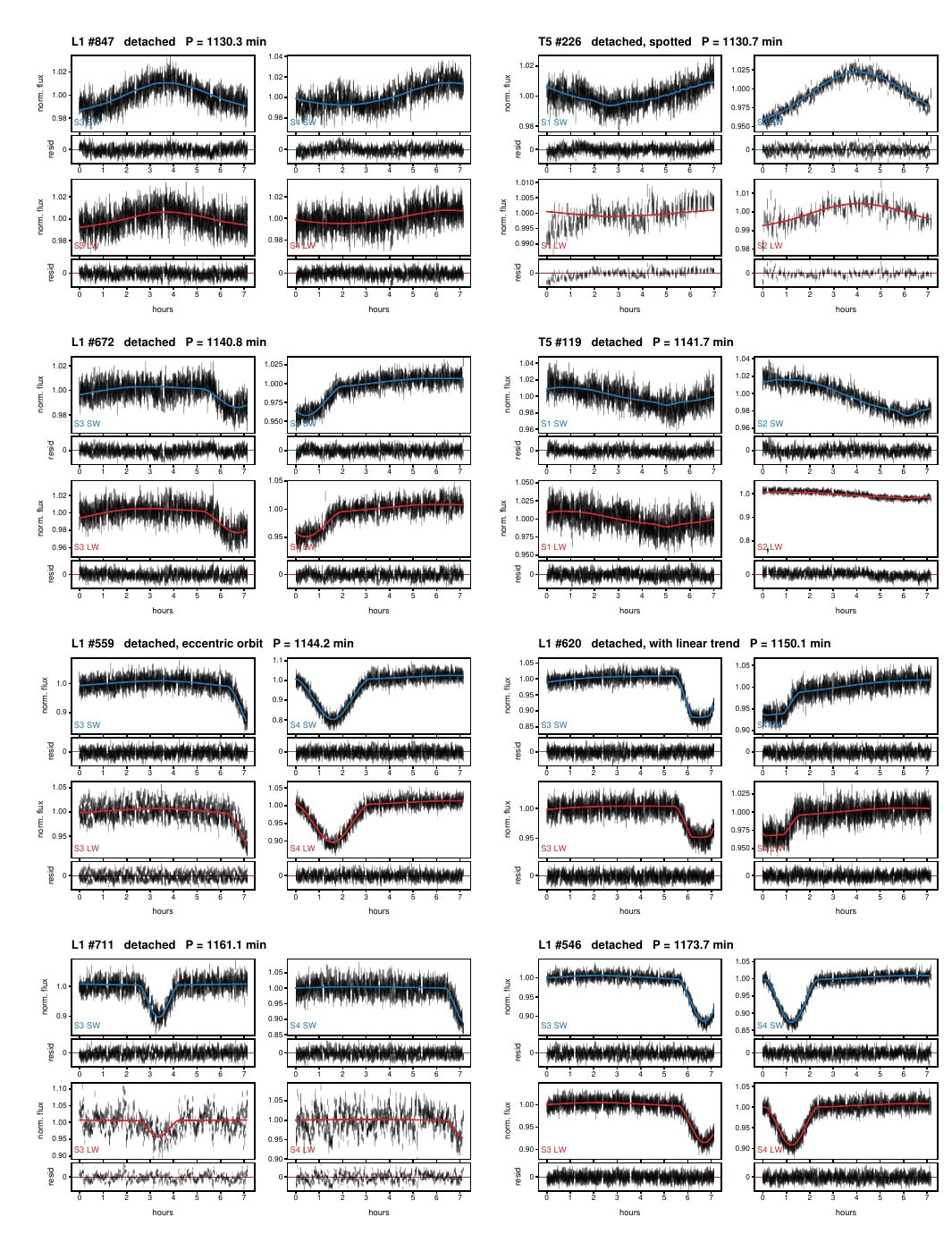}
\caption{Detached eclipsing binaries, continued (page 31 of 36).}
\end{figure*}
\clearpage

\begin{figure*}
\centering
\includegraphics[width=0.98\textwidth,height=0.94\textheight,keepaspectratio]{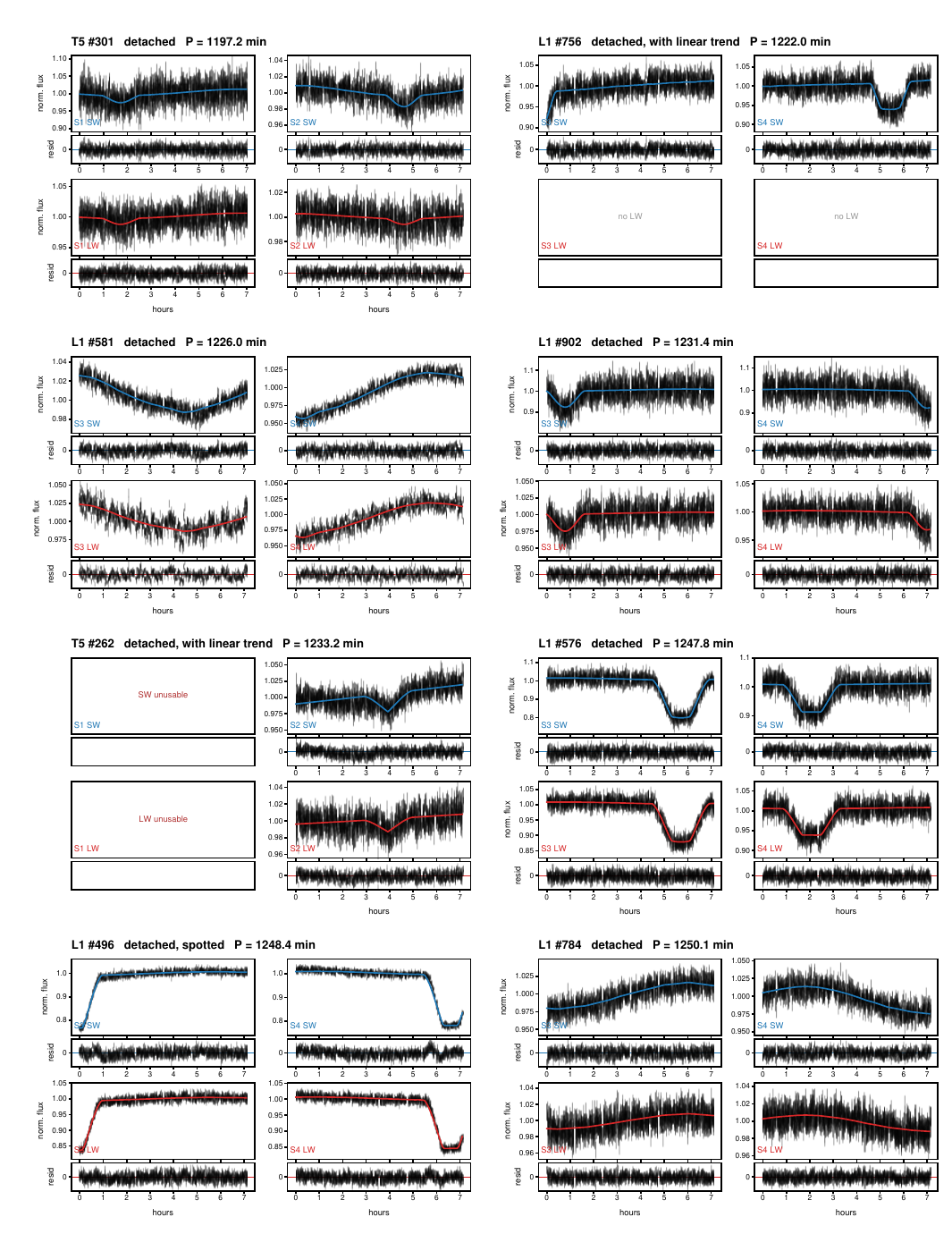}
\caption{Detached eclipsing binaries, continued (page 32 of 36).}
\end{figure*}
\clearpage

\begin{figure*}
\centering
\includegraphics[width=0.98\textwidth,height=0.94\textheight,keepaspectratio]{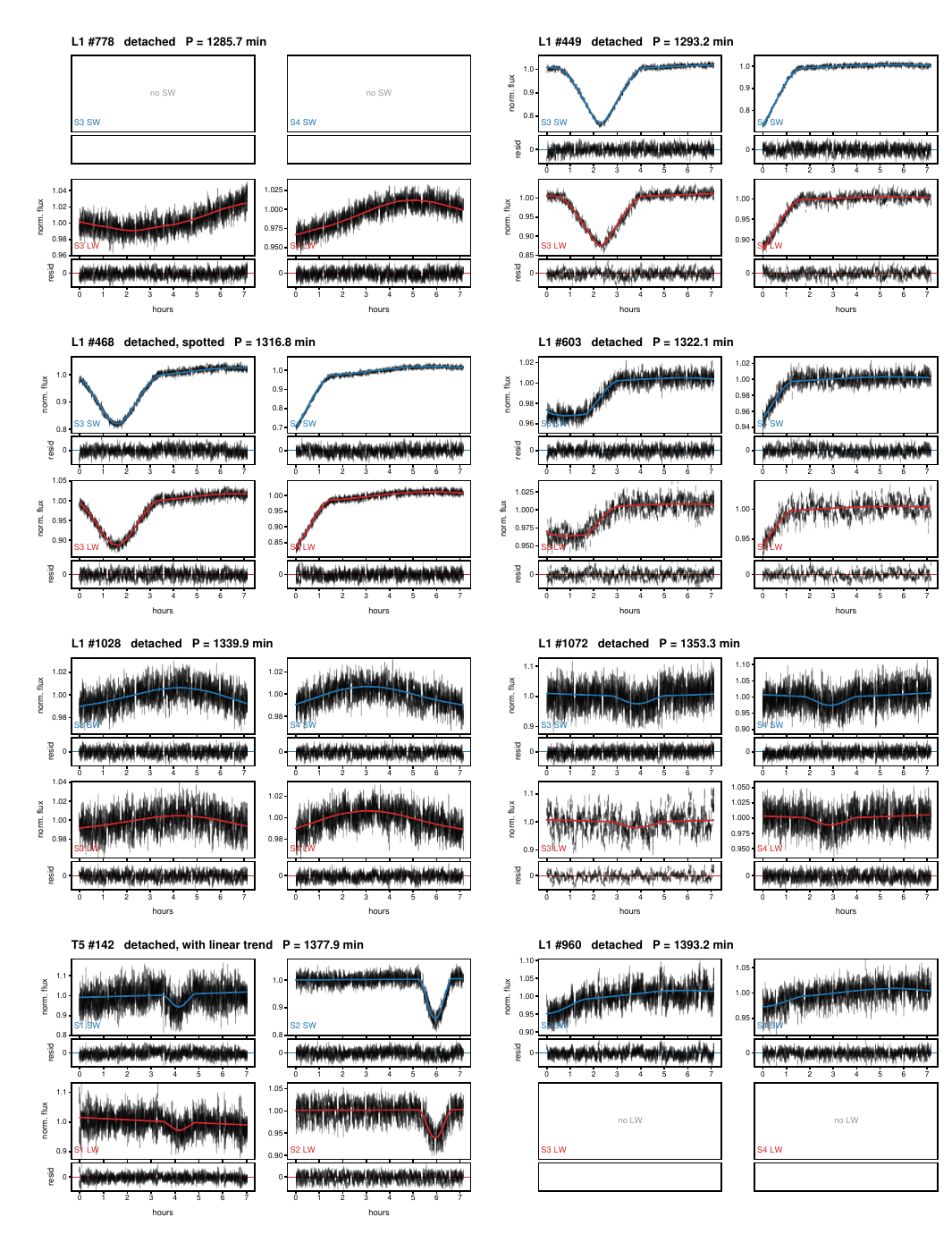}
\caption{Detached eclipsing binaries, continued (page 33 of 36).}
\end{figure*}
\clearpage

\begin{figure*}
\centering
\includegraphics[width=0.98\textwidth,height=0.94\textheight,keepaspectratio]{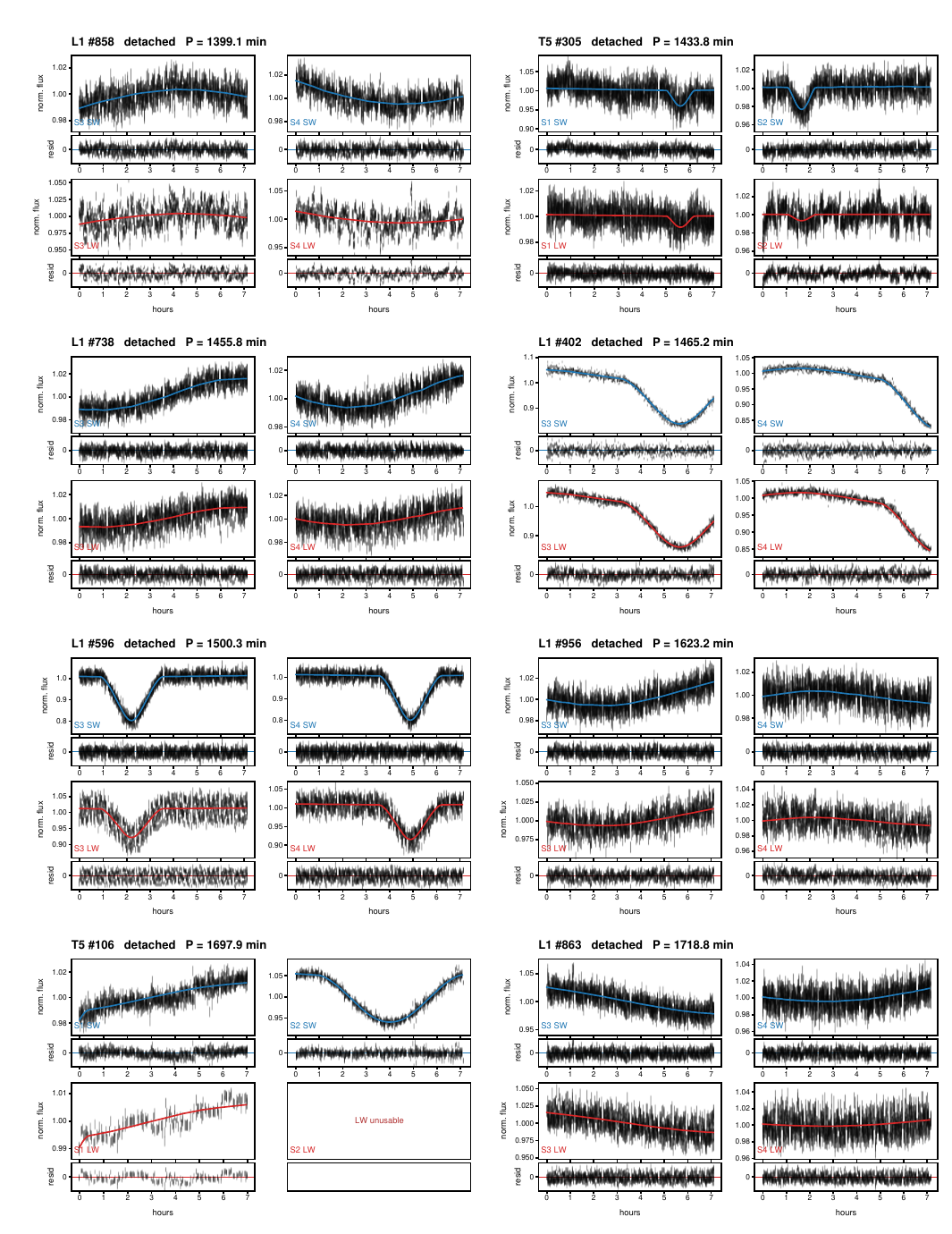}
\caption{Detached eclipsing binaries, continued (page 34 of 36).}
\end{figure*}
\clearpage

\begin{figure*}
\centering
\includegraphics[width=0.98\textwidth,height=0.94\textheight,keepaspectratio]{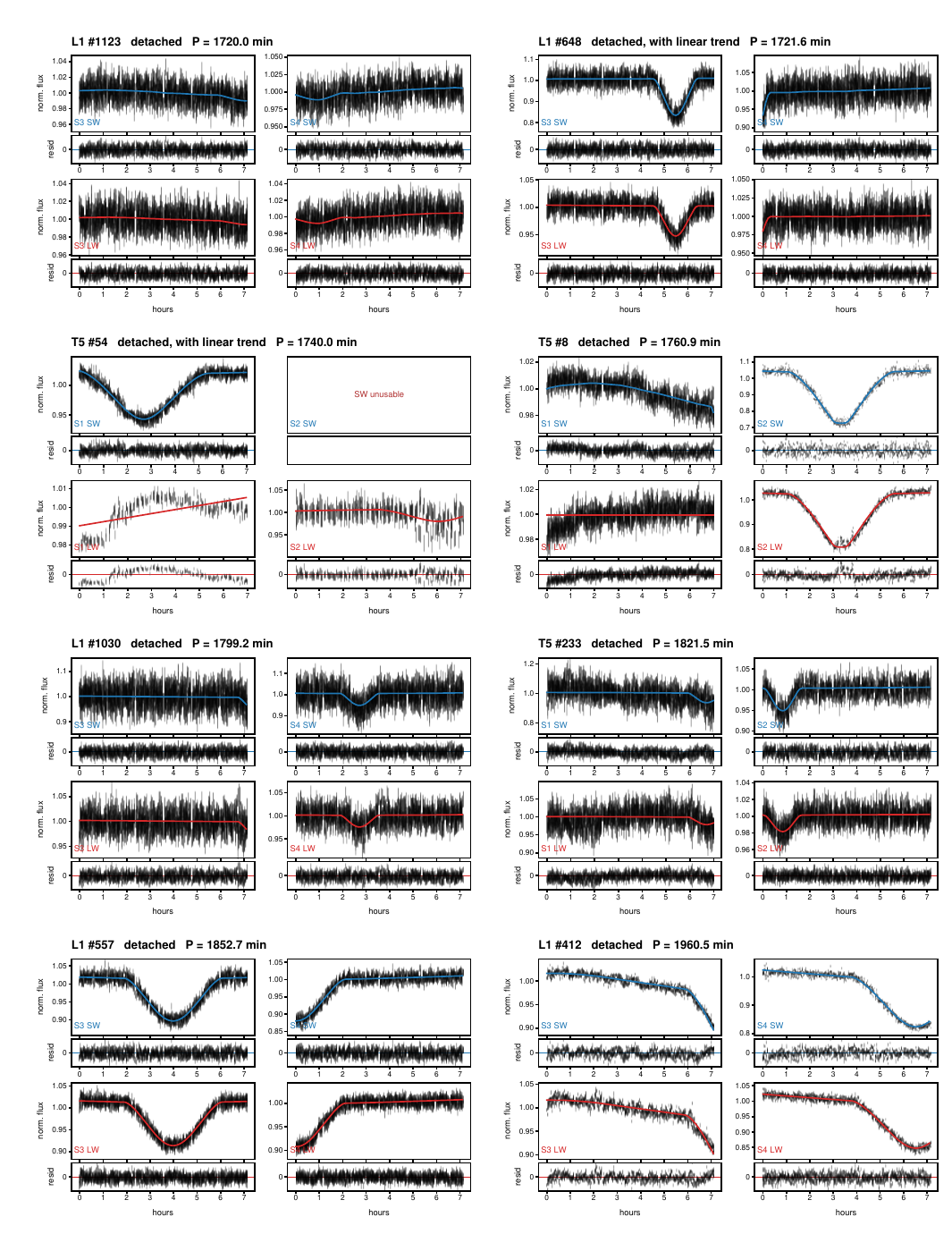}
\caption{Detached eclipsing binaries, continued (page 35 of 36).}
\end{figure*}
\clearpage

\begin{figure*}
\centering
\includegraphics[width=0.98\textwidth,height=0.94\textheight,keepaspectratio]{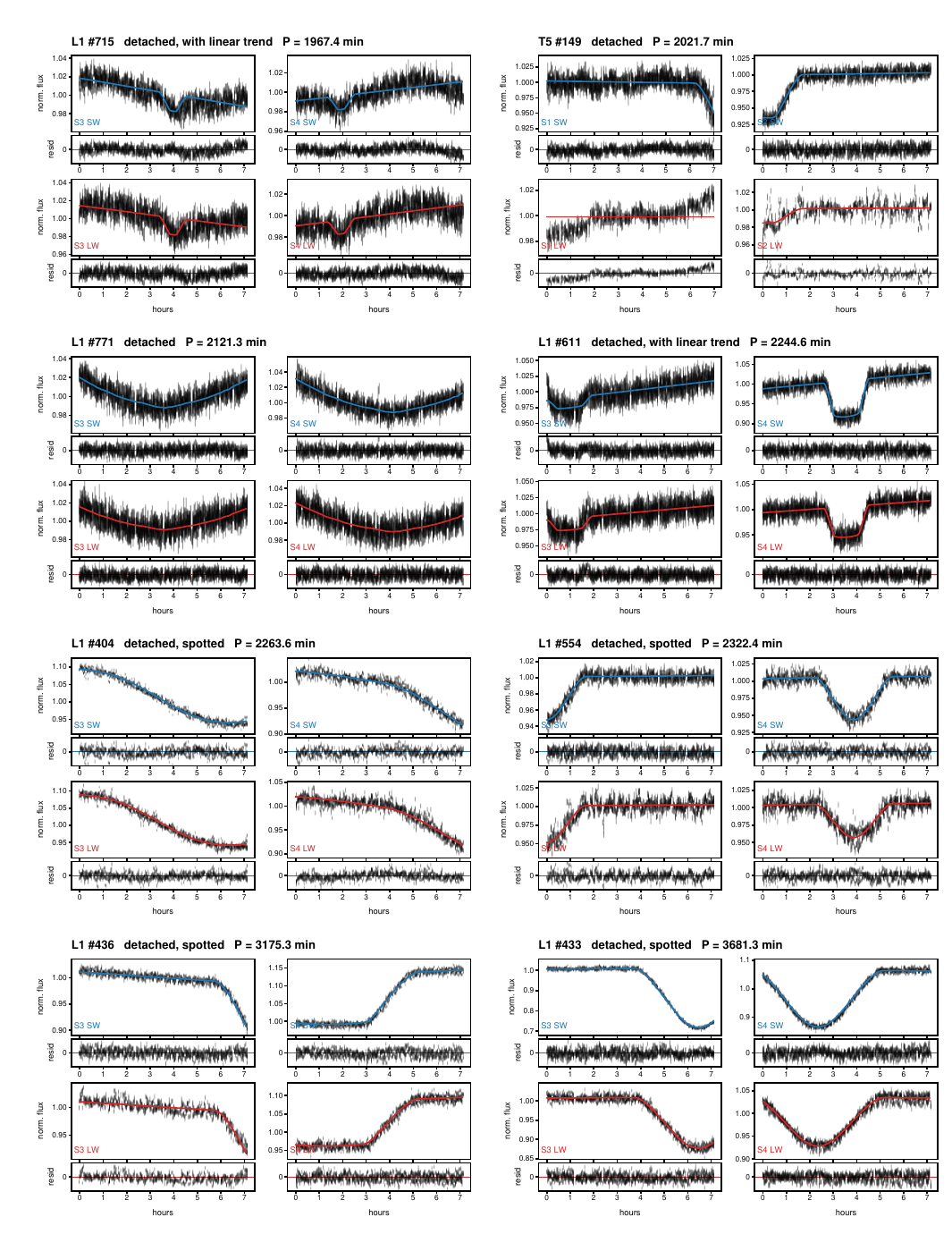}
\caption{Detached eclipsing binaries, continued (page 36 of 36).}
\end{figure*}
\clearpage

\subsection{Single-transit sources}\label{app:atlas:transit}

\begin{figure*}
\centering
\includegraphics[width=0.98\textwidth,height=0.79\textheight,keepaspectratio]{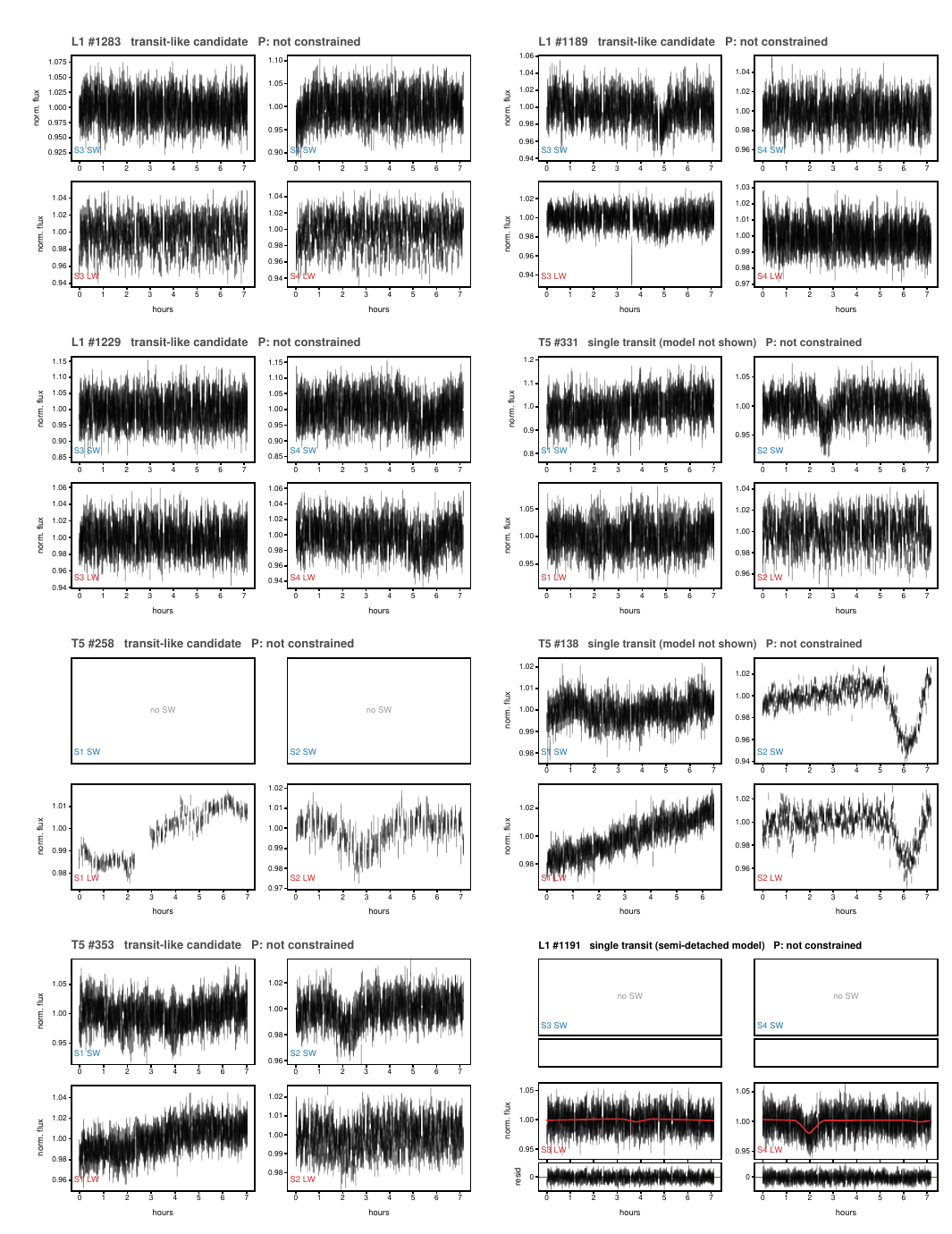}
\caption{Single-transit sources: unfolded JWST NIRCam lightcurves of all 205 sources in this class (page 1 of 26), sorted by the fitted period used only for ordering (sources without a fitted period last). Each source is shown as a four-panel block. The two observing segments run left to right, with the short-wavelength F200W lightcurve (SW, \textbf{blue}) on top and the long-wavelength F356W lightcurve (LW, \textbf{red}) below, and the segment and band are labelled inside every panel. Time is hours from the start of that segment, and lightcurves are never phase-folded. These are sources whose fitted model we accepted but whose period the data do not constrain, in most cases because the visits capture a single eclipse, together with transit-like candidates for which no model was accepted. Black vertical bars are the adopted lightcurve (\S\ref{sec:strategy}), each spanning the $1\sigma$ uncertainty of one plotted sample. Where a model was accepted, it is overplotted as the coloured curve, scaled per panel as in the binary sections, with residuals below. The header then names the Roche geometry of the fitted model, which is not a classification. Two sources whose accepted fit is not drawn are marked ``model not shown''. A single event does not constrain the orbital period, so no period is printed, and the fitted period only orders the figures.}
\end{figure*}
\clearpage

\begin{figure*}
\centering
\includegraphics[width=0.98\textwidth,height=0.94\textheight,keepaspectratio]{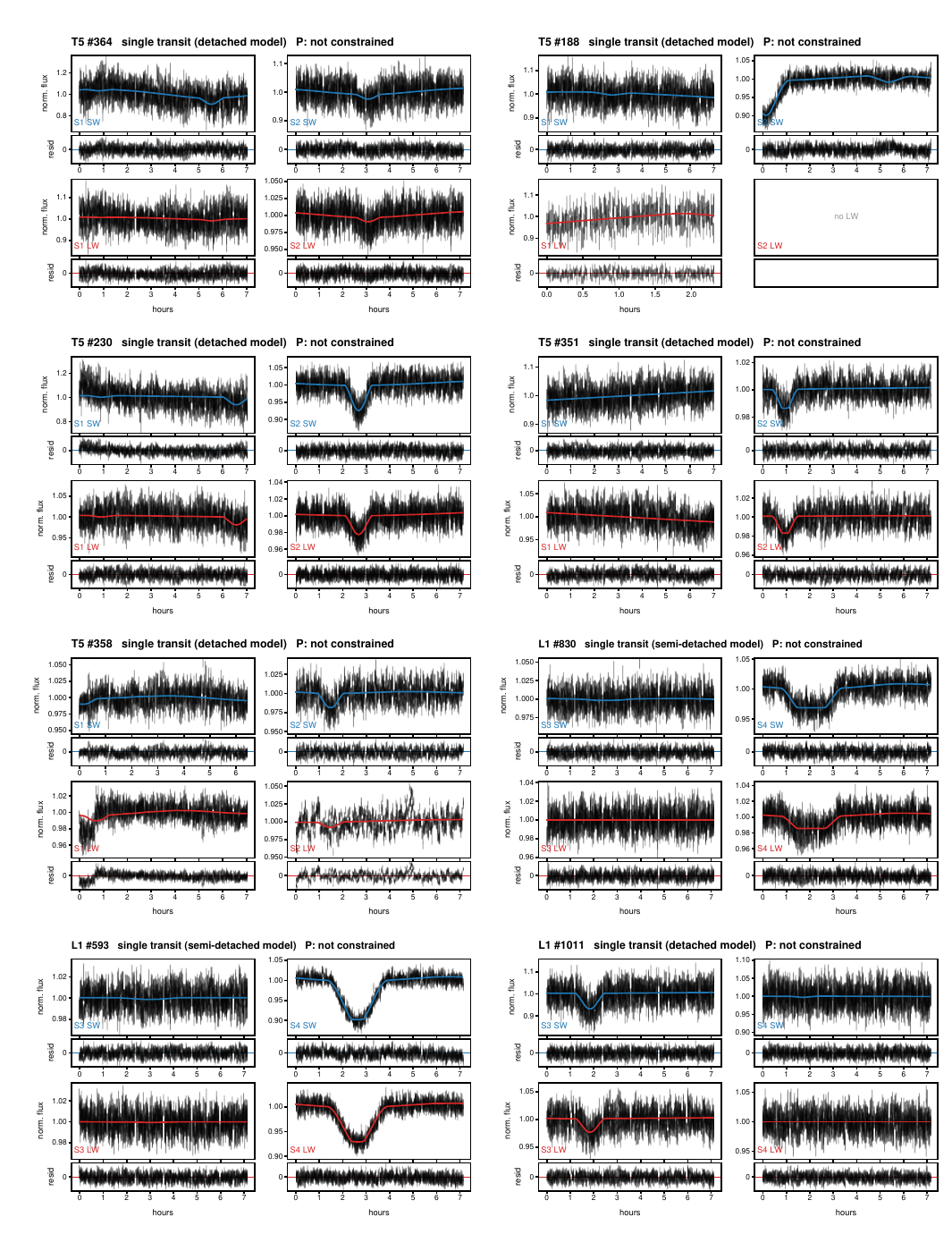}
\caption{Single-transit sources, continued (page 2 of 26).}
\end{figure*}
\clearpage

\begin{figure*}
\centering
\includegraphics[width=0.98\textwidth,height=0.94\textheight,keepaspectratio]{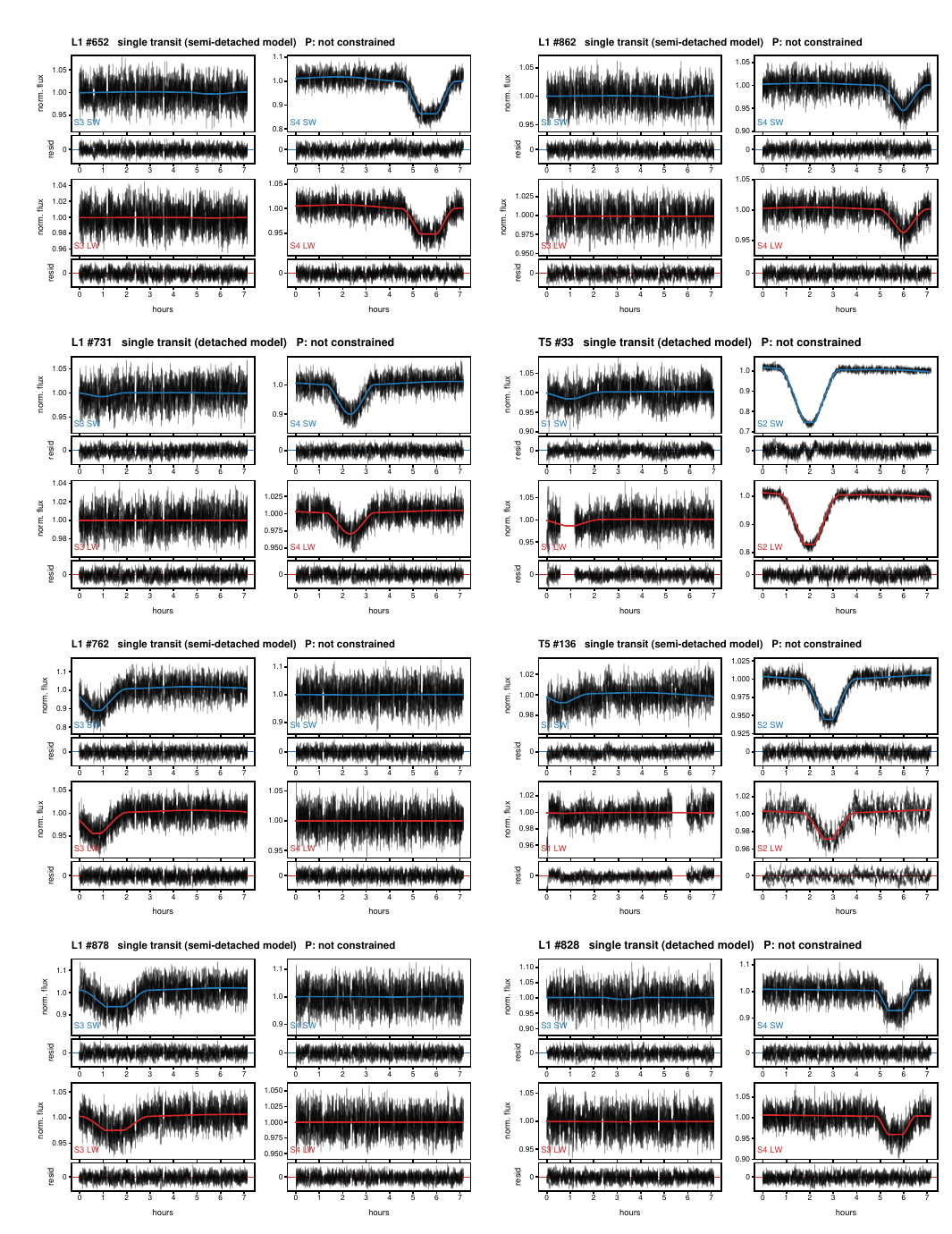}
\caption{Single-transit sources, continued (page 3 of 26).}
\end{figure*}
\clearpage

\begin{figure*}
\centering
\includegraphics[width=0.98\textwidth,height=0.94\textheight,keepaspectratio]{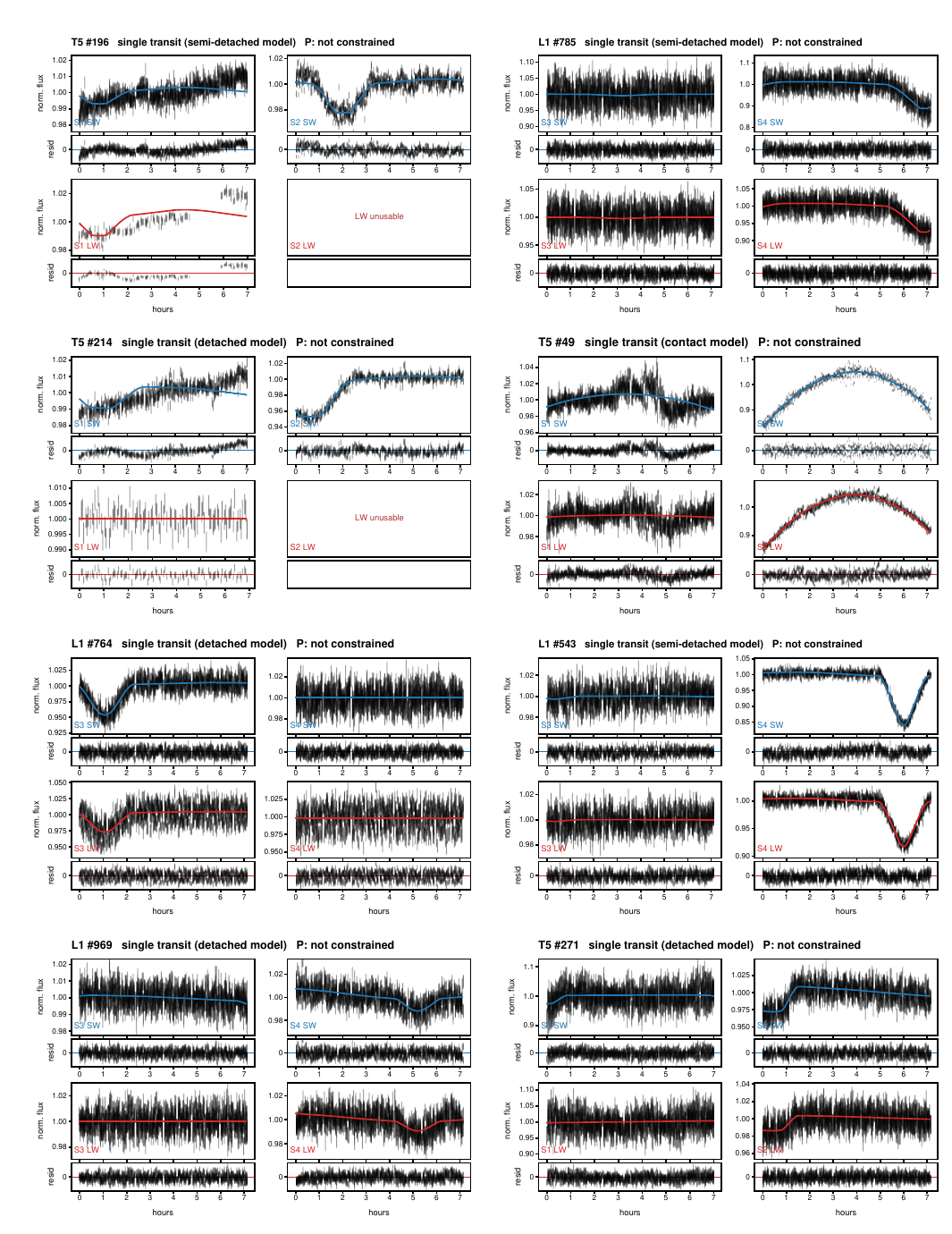}
\caption{Single-transit sources, continued (page 4 of 26).}
\end{figure*}
\clearpage

\begin{figure*}
\centering
\includegraphics[width=0.98\textwidth,height=0.94\textheight,keepaspectratio]{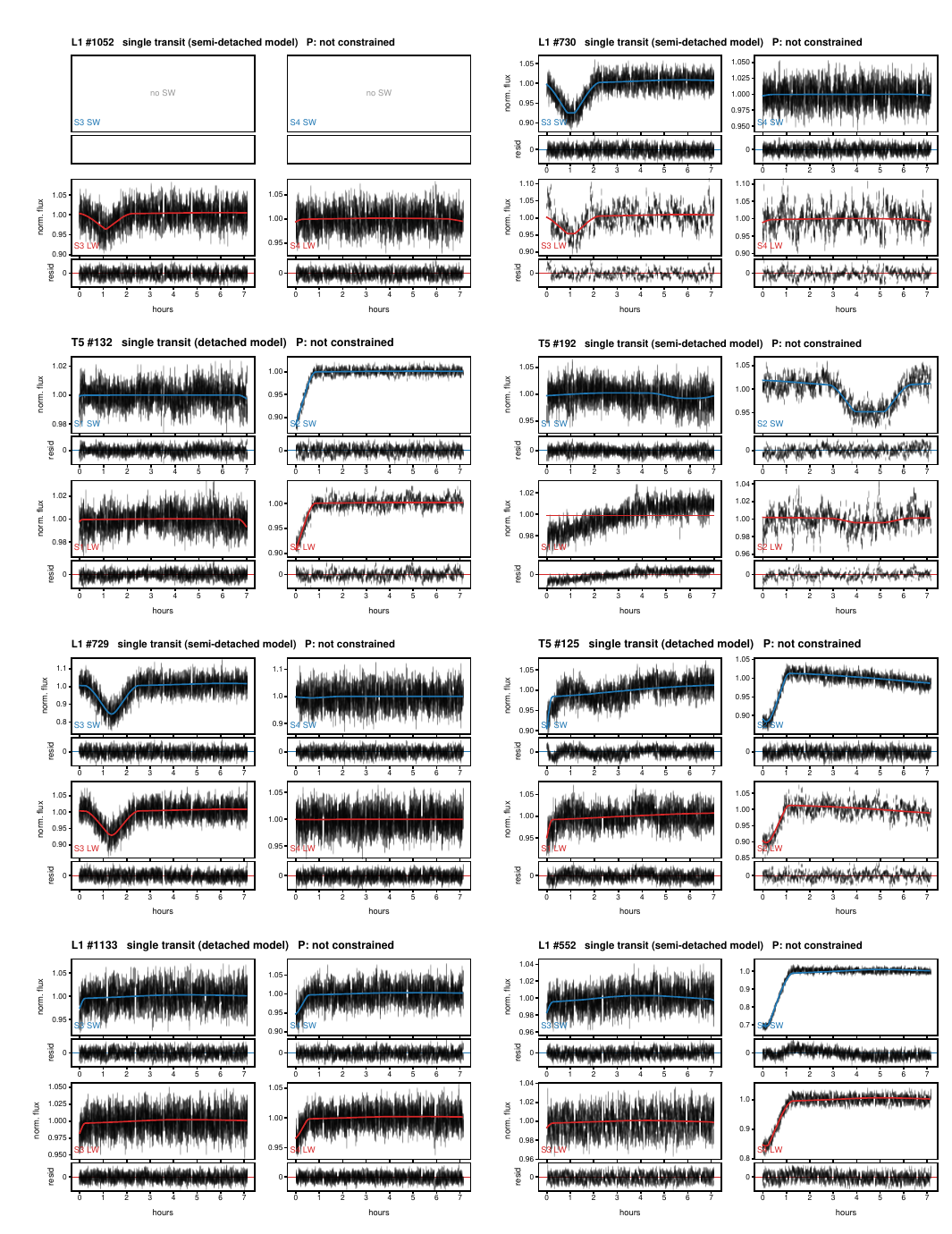}
\caption{Single-transit sources, continued (page 5 of 26).}
\end{figure*}
\clearpage

\begin{figure*}
\centering
\includegraphics[width=0.98\textwidth,height=0.94\textheight,keepaspectratio]{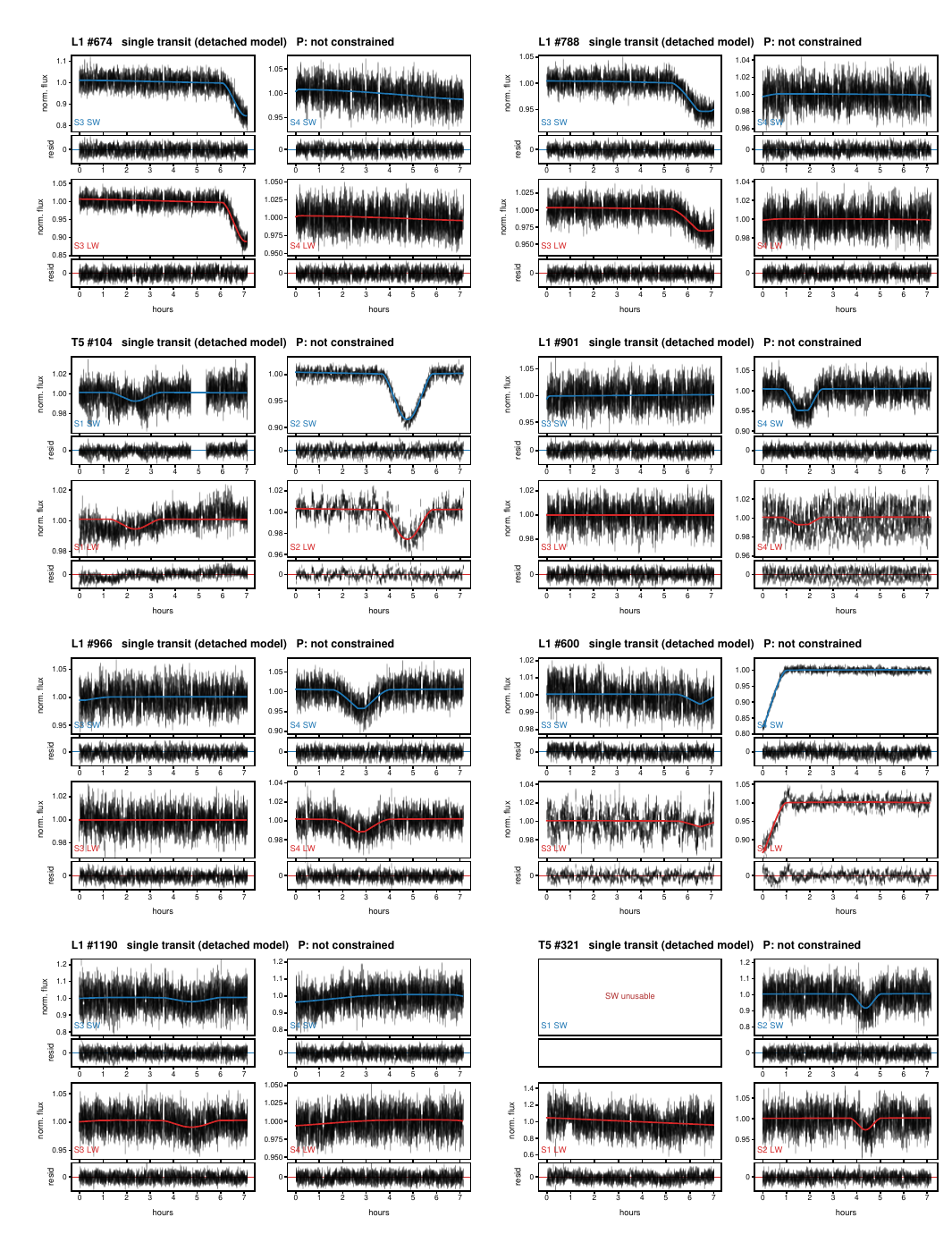}
\caption{Single-transit sources, continued (page 6 of 26).}
\end{figure*}
\clearpage

\begin{figure*}
\centering
\includegraphics[width=0.98\textwidth,height=0.94\textheight,keepaspectratio]{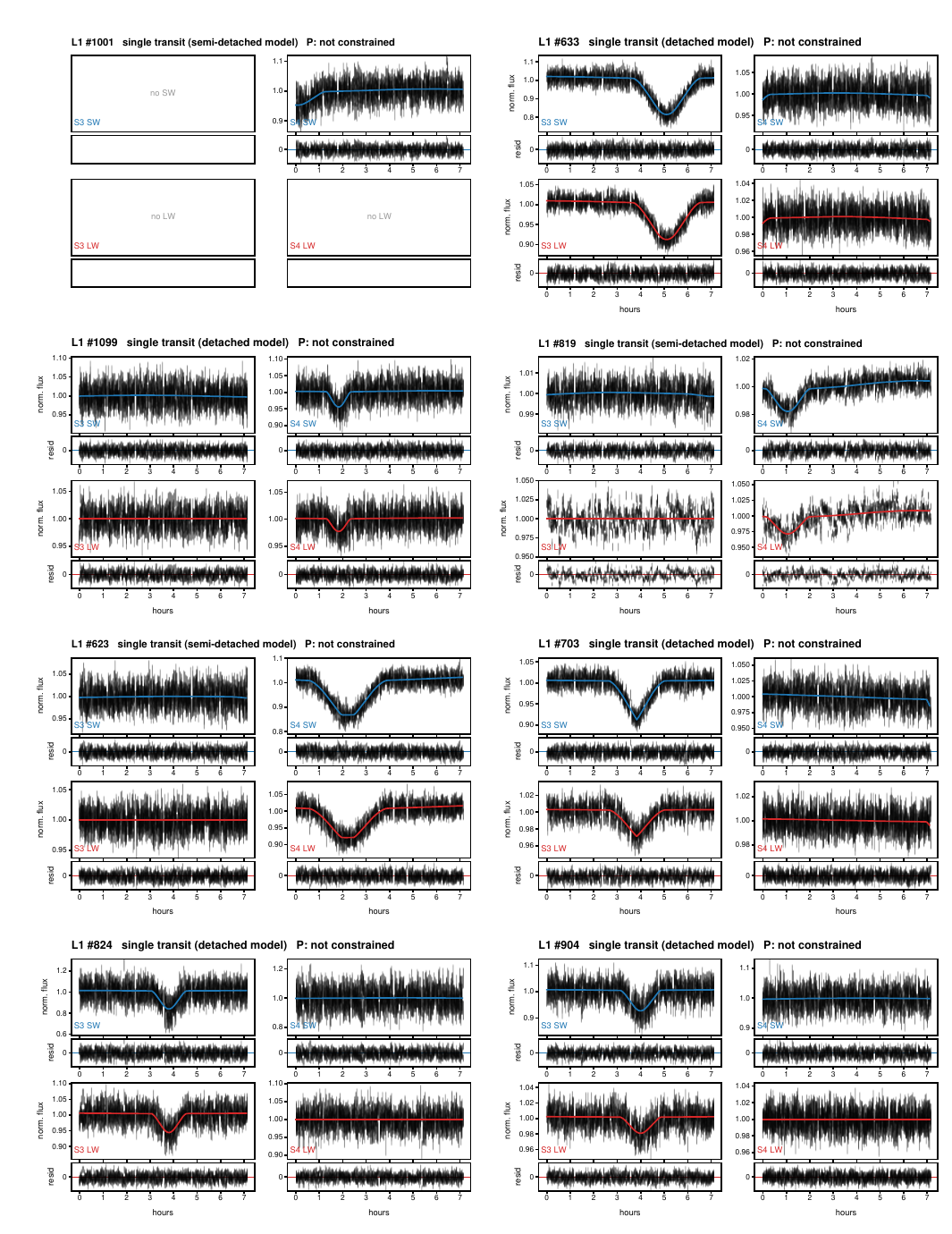}
\caption{Single-transit sources, continued (page 7 of 26).}
\end{figure*}
\clearpage

\begin{figure*}
\centering
\includegraphics[width=0.98\textwidth,height=0.94\textheight,keepaspectratio]{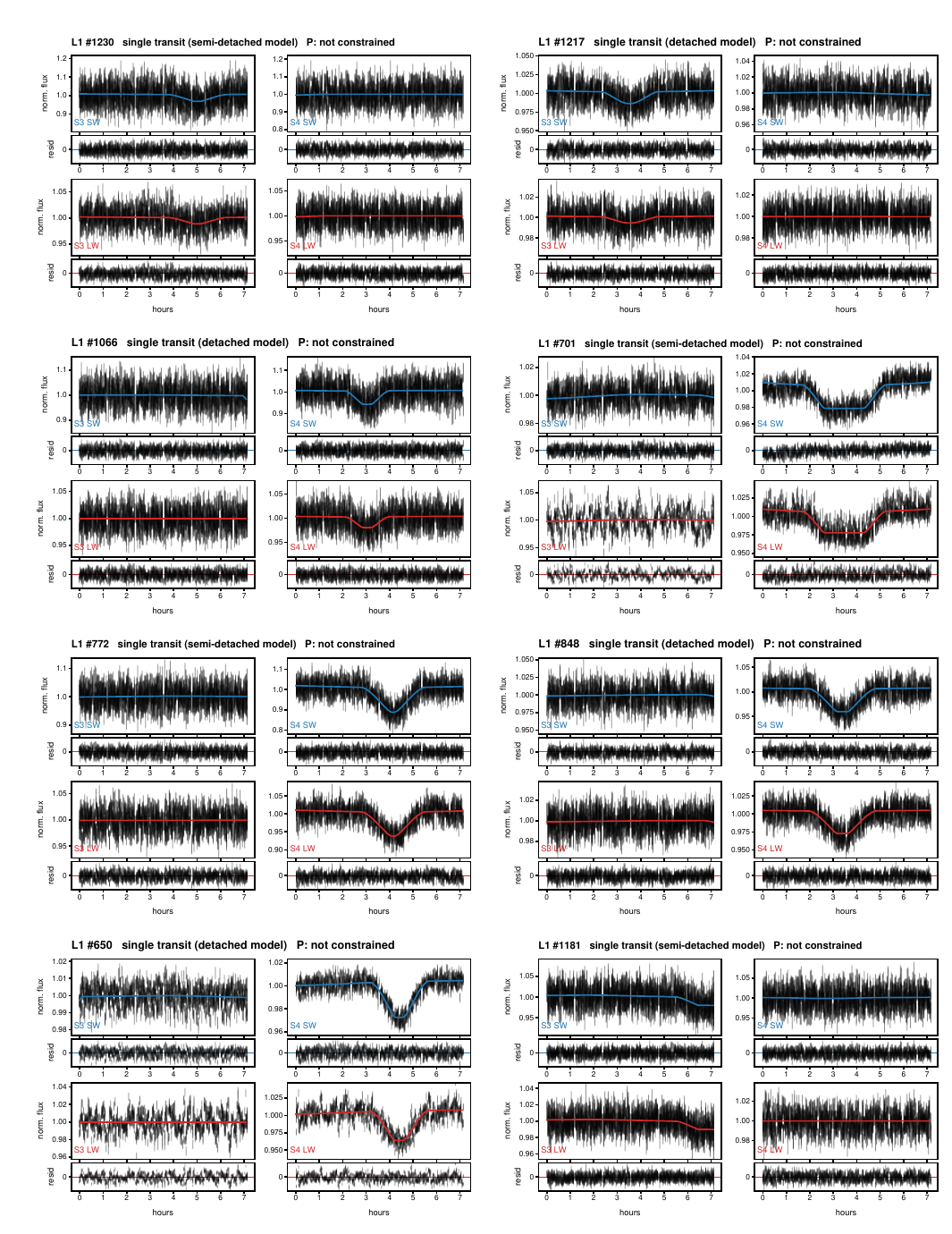}
\caption{Single-transit sources, continued (page 8 of 26).}
\end{figure*}
\clearpage

\begin{figure*}
\centering
\includegraphics[width=0.98\textwidth,height=0.94\textheight,keepaspectratio]{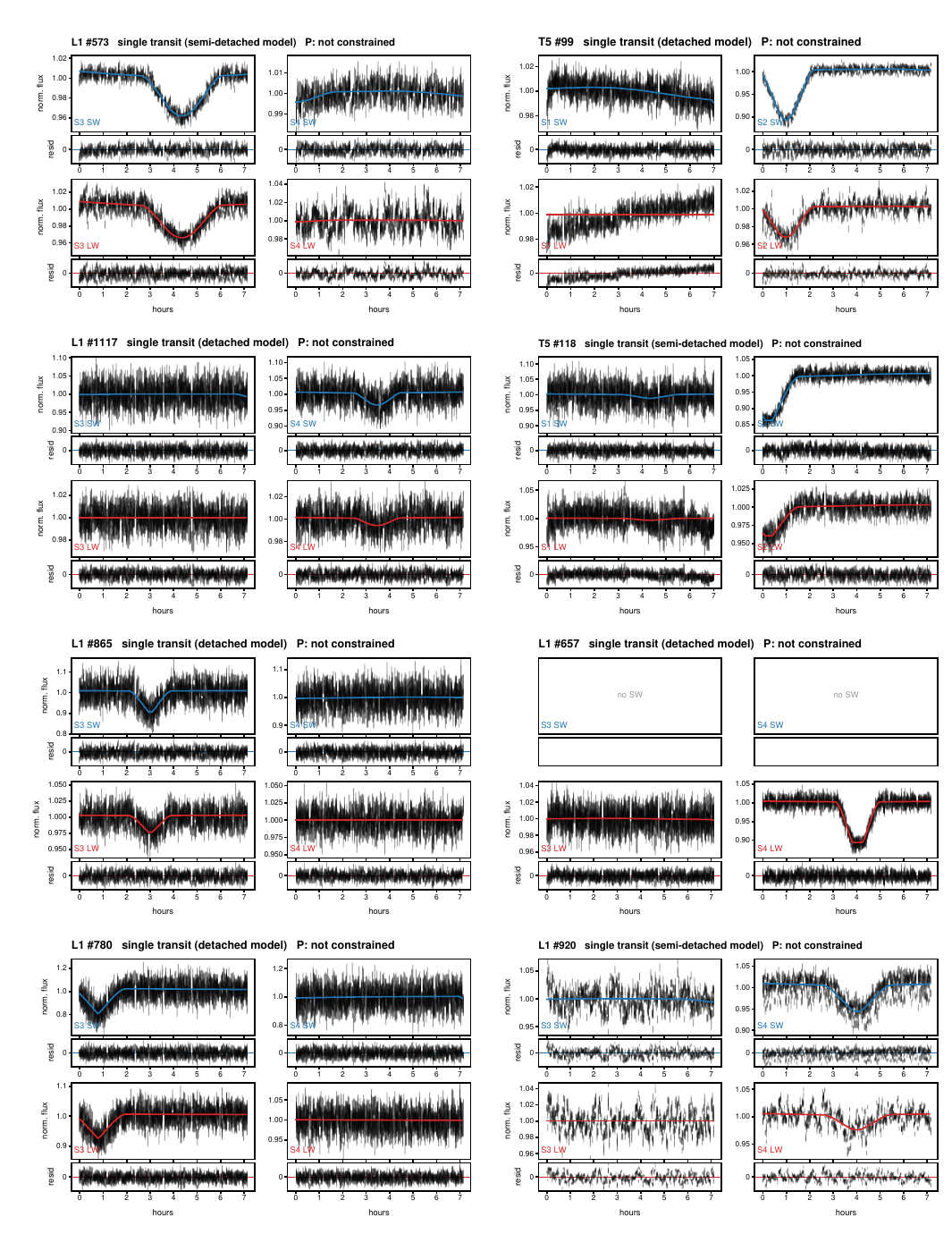}
\caption{Single-transit sources, continued (page 9 of 26).}
\end{figure*}
\clearpage

\begin{figure*}
\centering
\includegraphics[width=0.98\textwidth,height=0.94\textheight,keepaspectratio]{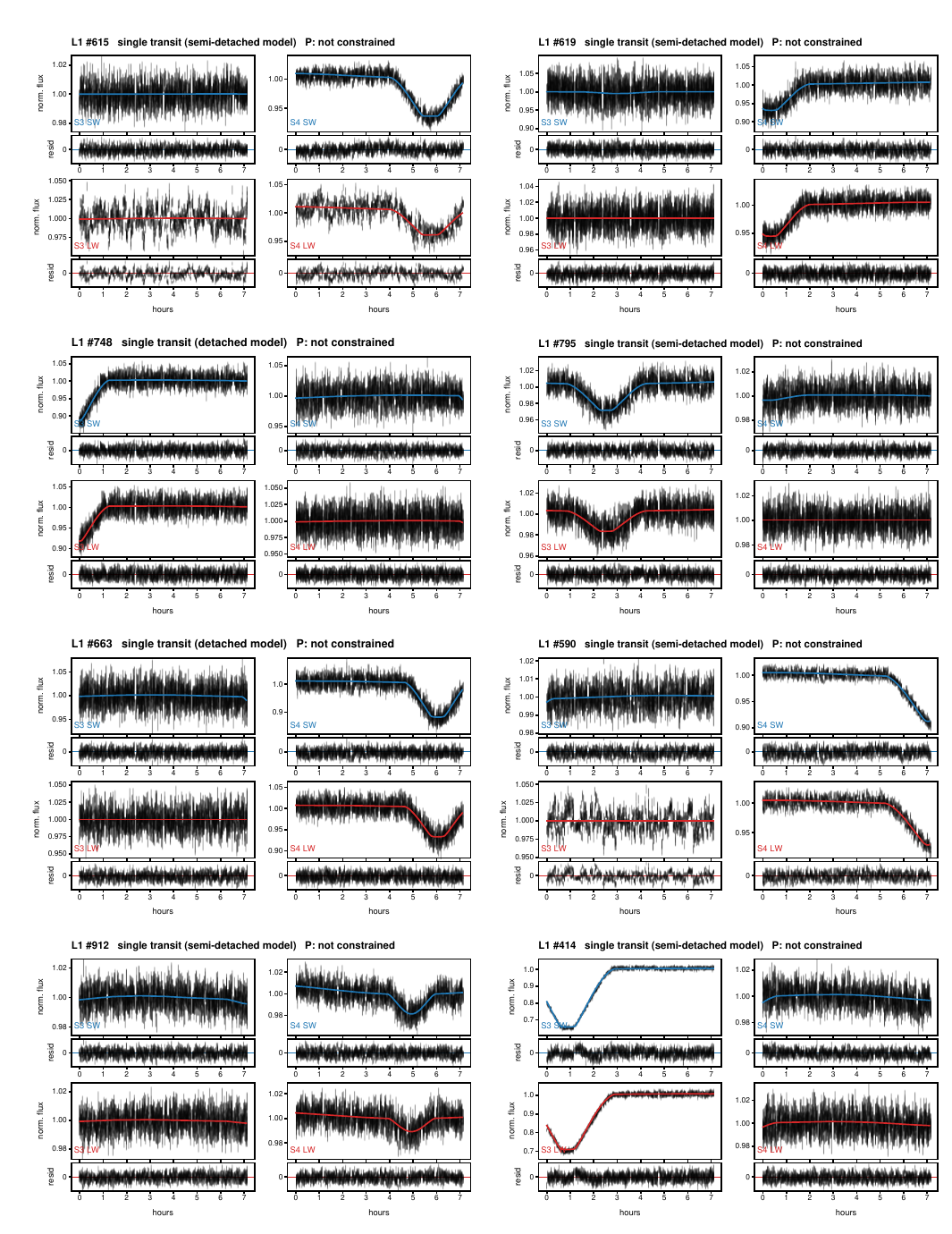}
\caption{Single-transit sources, continued (page 10 of 26).}
\end{figure*}
\clearpage

\begin{figure*}
\centering
\includegraphics[width=0.98\textwidth,height=0.94\textheight,keepaspectratio]{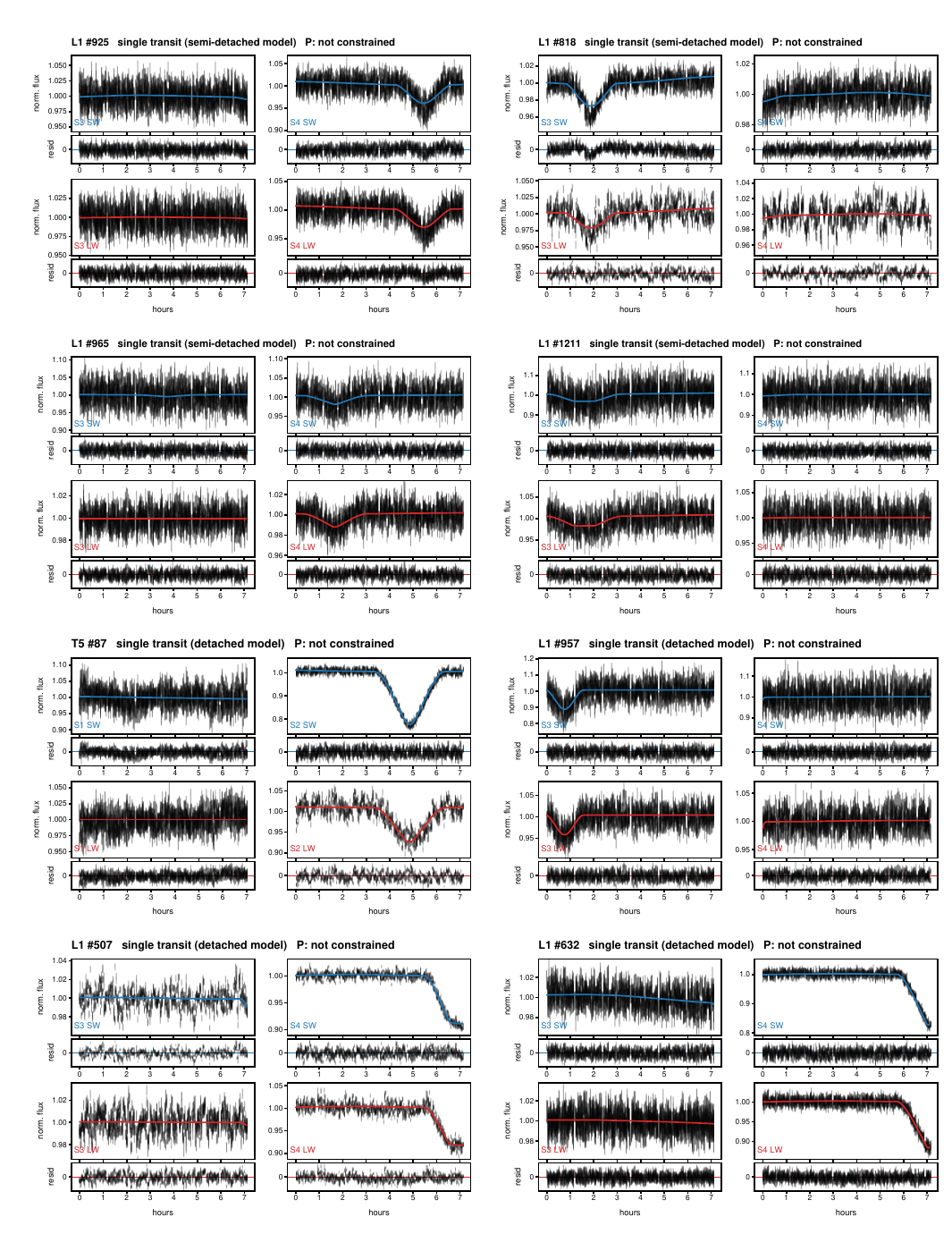}
\caption{Single-transit sources, continued (page 11 of 26).}
\end{figure*}
\clearpage

\begin{figure*}
\centering
\includegraphics[width=0.98\textwidth,height=0.94\textheight,keepaspectratio]{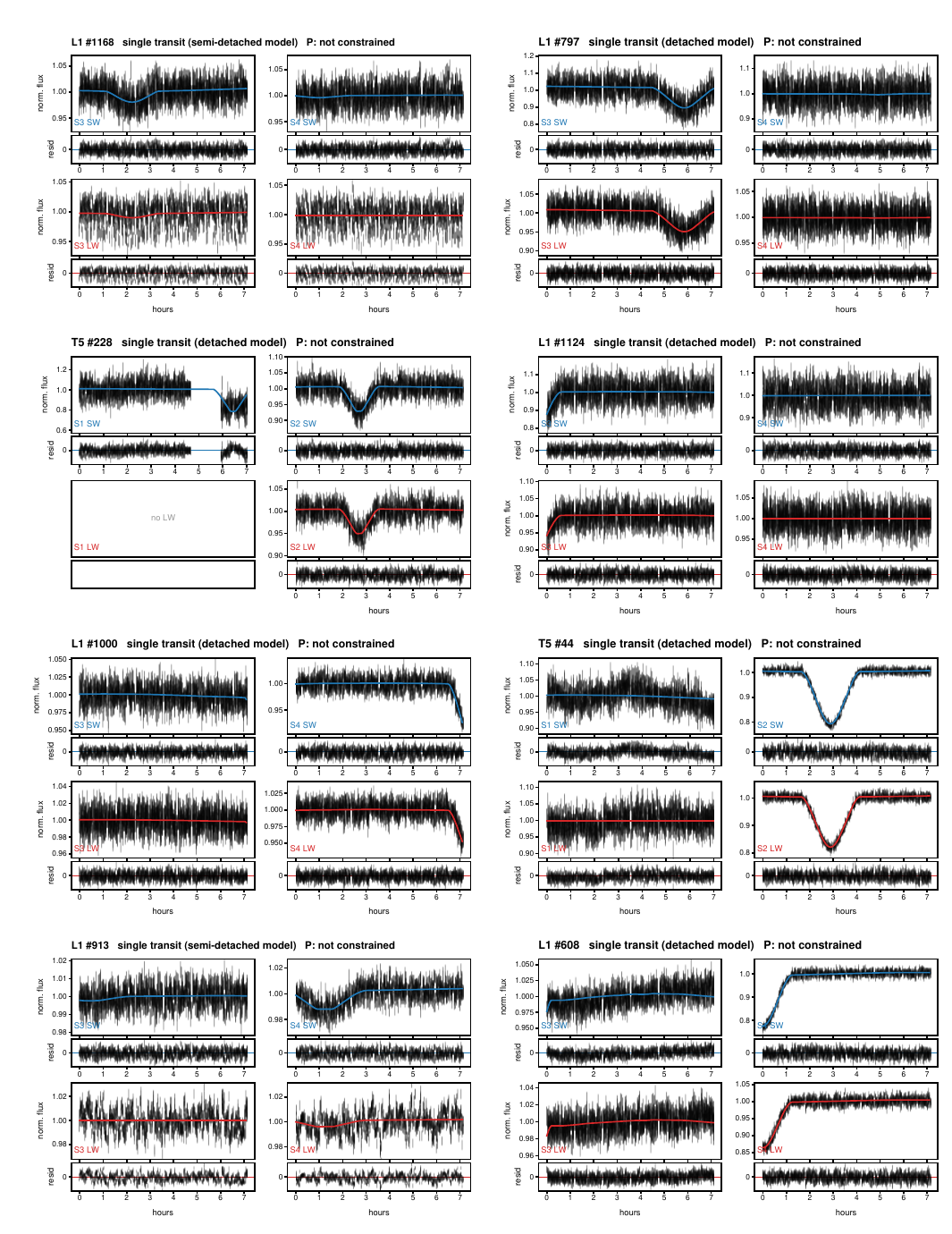}
\caption{Single-transit sources, continued (page 12 of 26).}
\end{figure*}
\clearpage

\begin{figure*}
\centering
\includegraphics[width=0.98\textwidth,height=0.94\textheight,keepaspectratio]{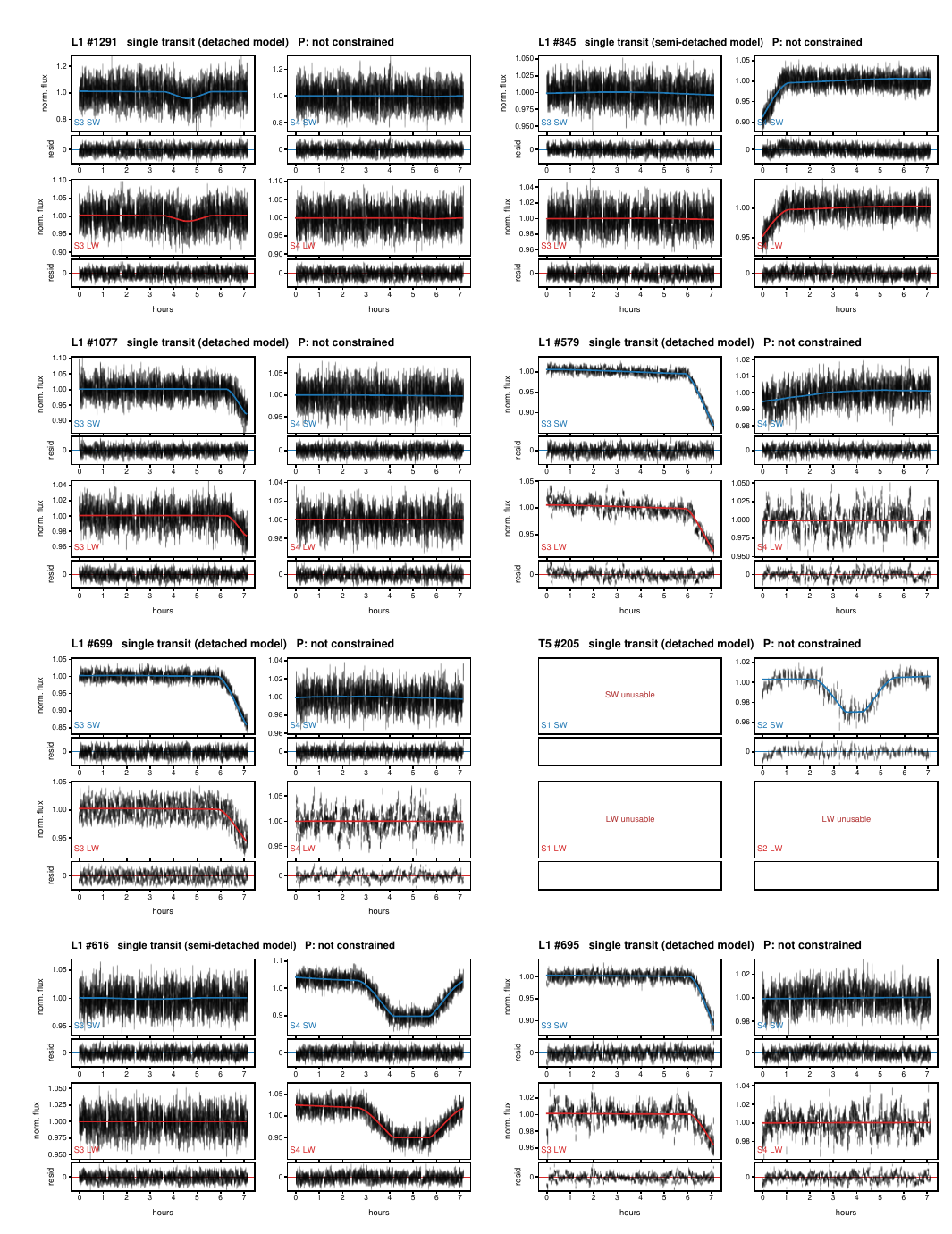}
\caption{Single-transit sources, continued (page 13 of 26).}
\end{figure*}
\clearpage

\begin{figure*}
\centering
\includegraphics[width=0.98\textwidth,height=0.94\textheight,keepaspectratio]{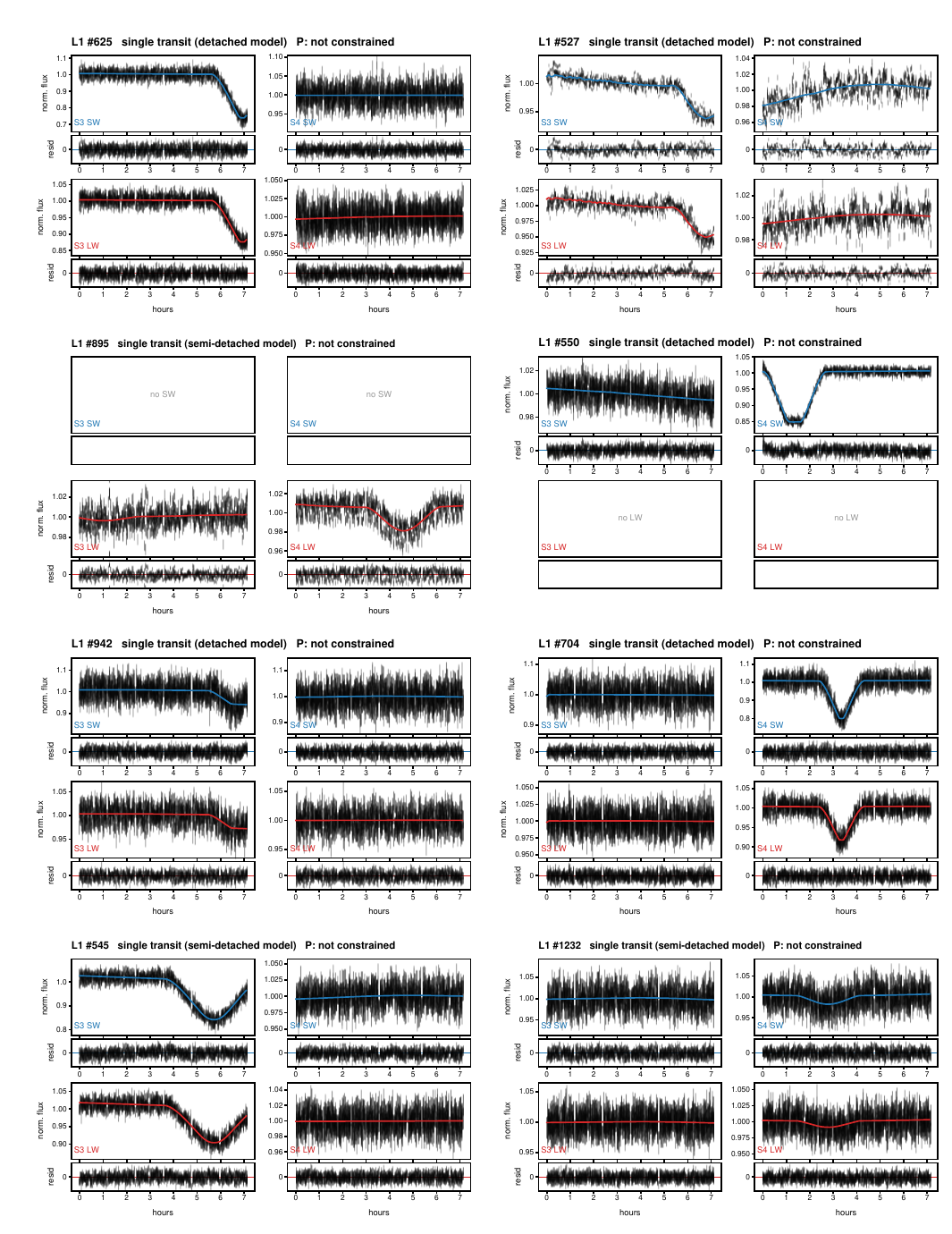}
\caption{Single-transit sources, continued (page 14 of 26).}
\end{figure*}
\clearpage

\begin{figure*}
\centering
\includegraphics[width=0.98\textwidth,height=0.94\textheight,keepaspectratio]{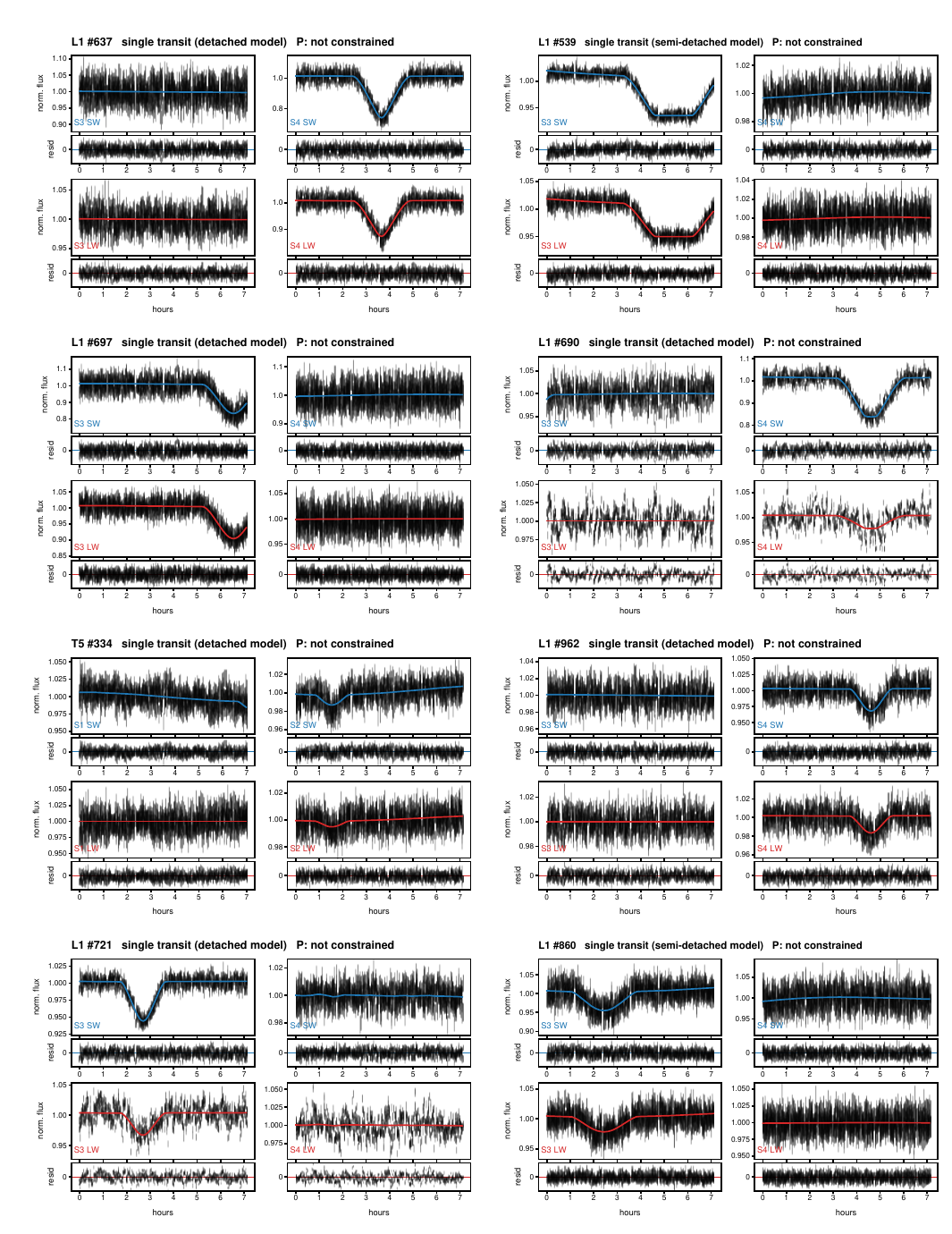}
\caption{Single-transit sources, continued (page 15 of 26).}
\end{figure*}
\clearpage

\begin{figure*}
\centering
\includegraphics[width=0.98\textwidth,height=0.94\textheight,keepaspectratio]{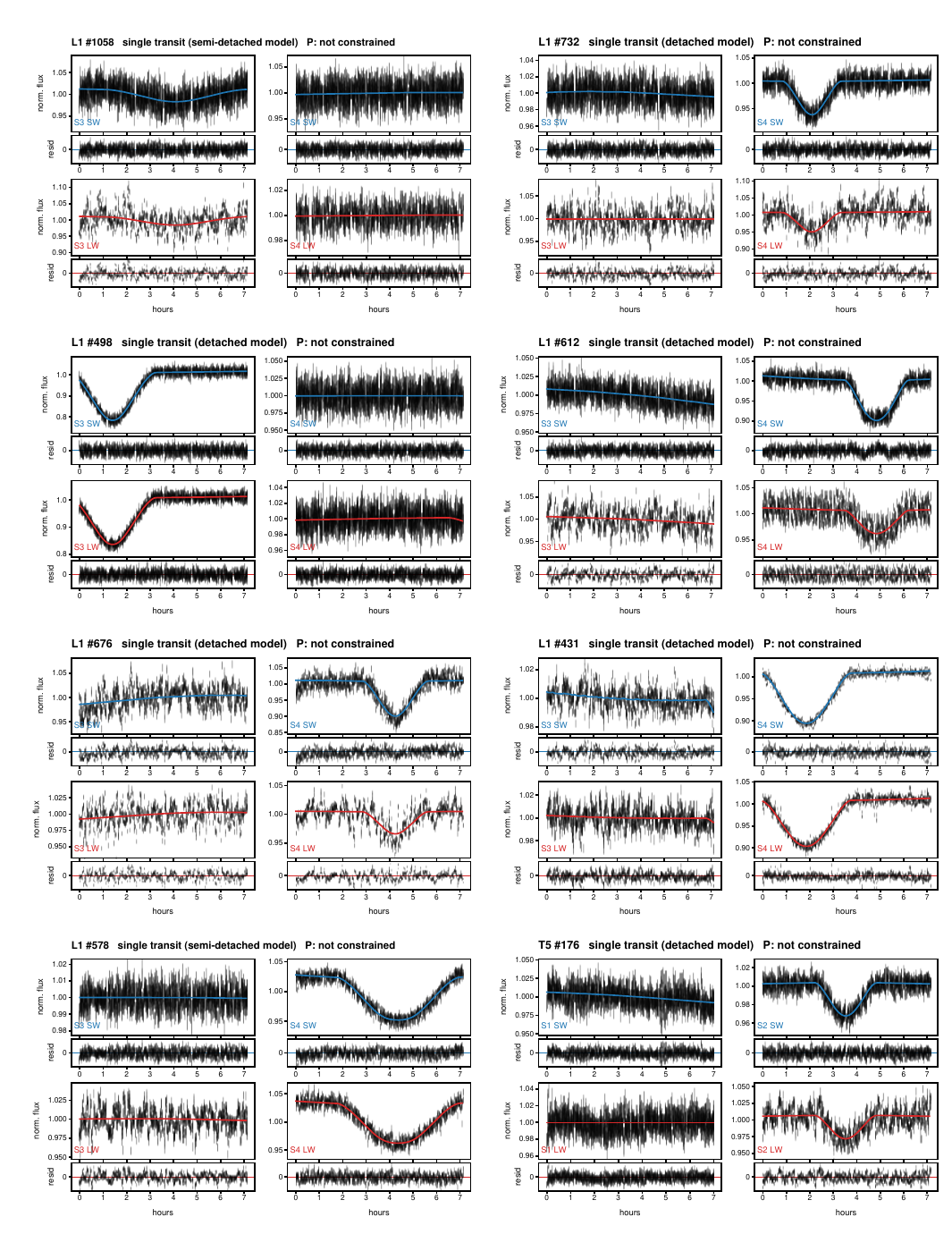}
\caption{Single-transit sources, continued (page 16 of 26).}
\end{figure*}
\clearpage

\begin{figure*}
\centering
\includegraphics[width=0.98\textwidth,height=0.94\textheight,keepaspectratio]{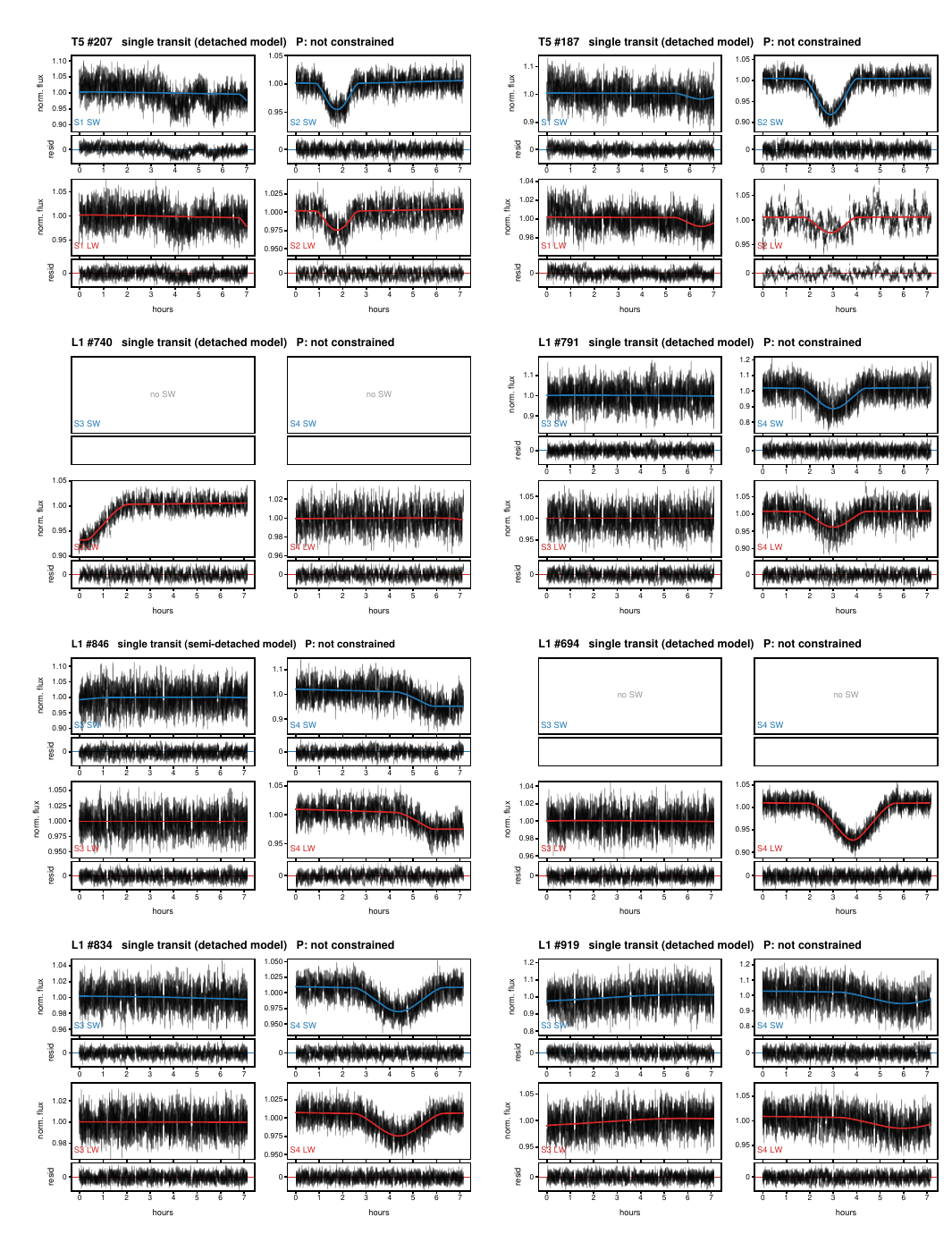}
\caption{Single-transit sources, continued (page 17 of 26).}
\end{figure*}
\clearpage

\begin{figure*}
\centering
\includegraphics[width=0.98\textwidth,height=0.94\textheight,keepaspectratio]{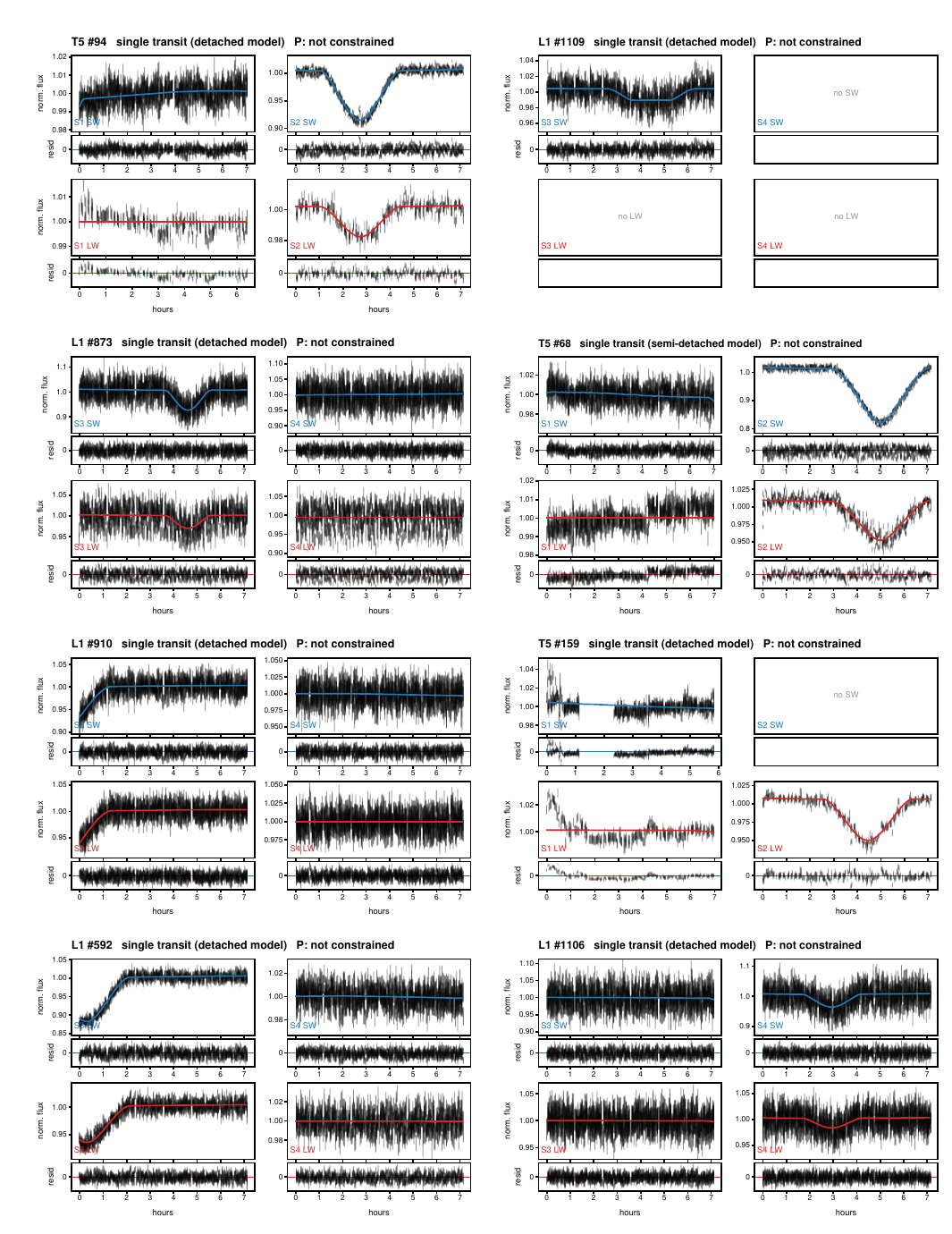}
\caption{Single-transit sources, continued (page 18 of 26).}
\end{figure*}
\clearpage

\begin{figure*}
\centering
\includegraphics[width=0.98\textwidth,height=0.94\textheight,keepaspectratio]{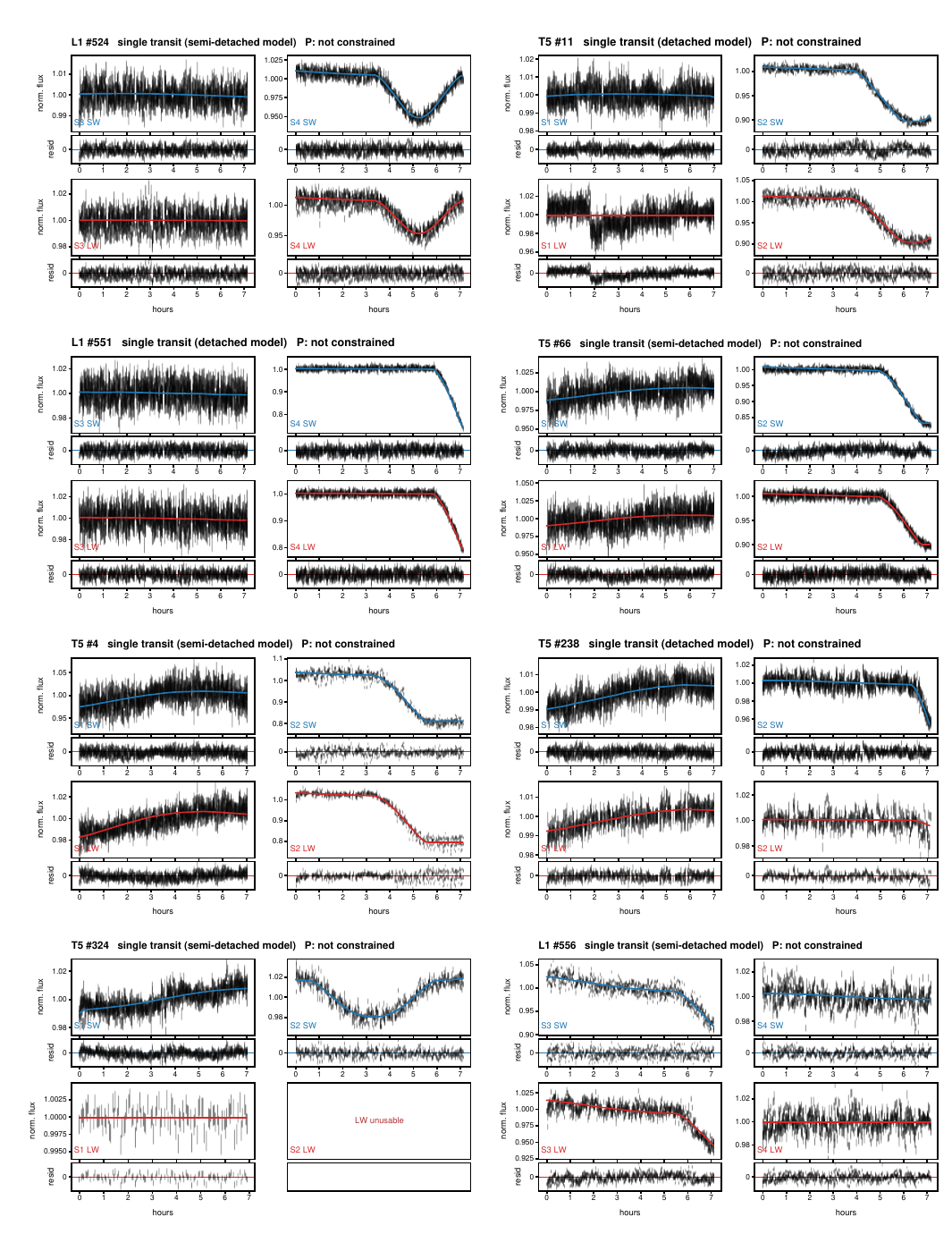}
\caption{Single-transit sources, continued (page 19 of 26).}
\end{figure*}
\clearpage

\begin{figure*}
\centering
\includegraphics[width=0.98\textwidth,height=0.94\textheight,keepaspectratio]{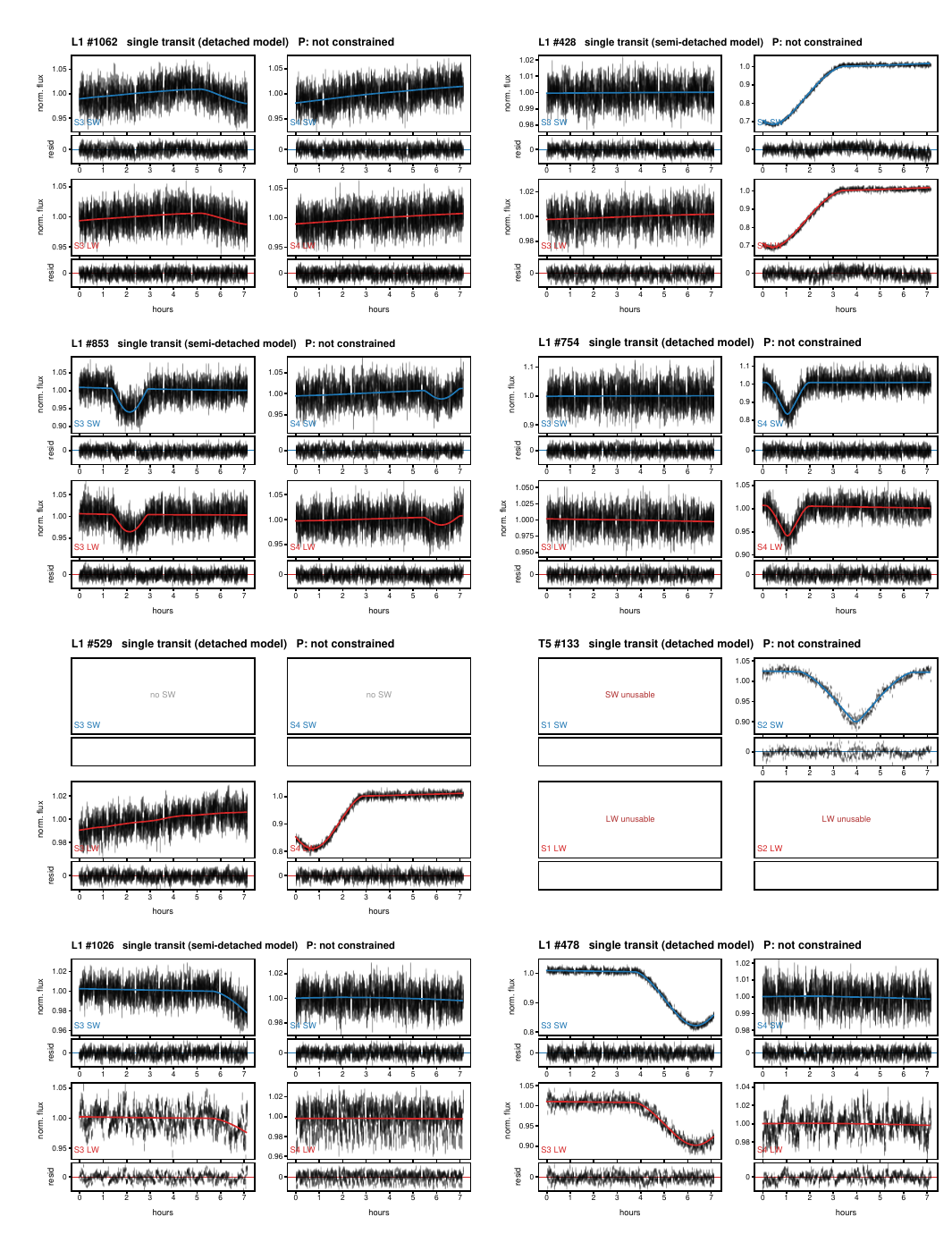}
\caption{Single-transit sources, continued (page 20 of 26).}
\end{figure*}
\clearpage

\begin{figure*}
\centering
\includegraphics[width=0.98\textwidth,height=0.94\textheight,keepaspectratio]{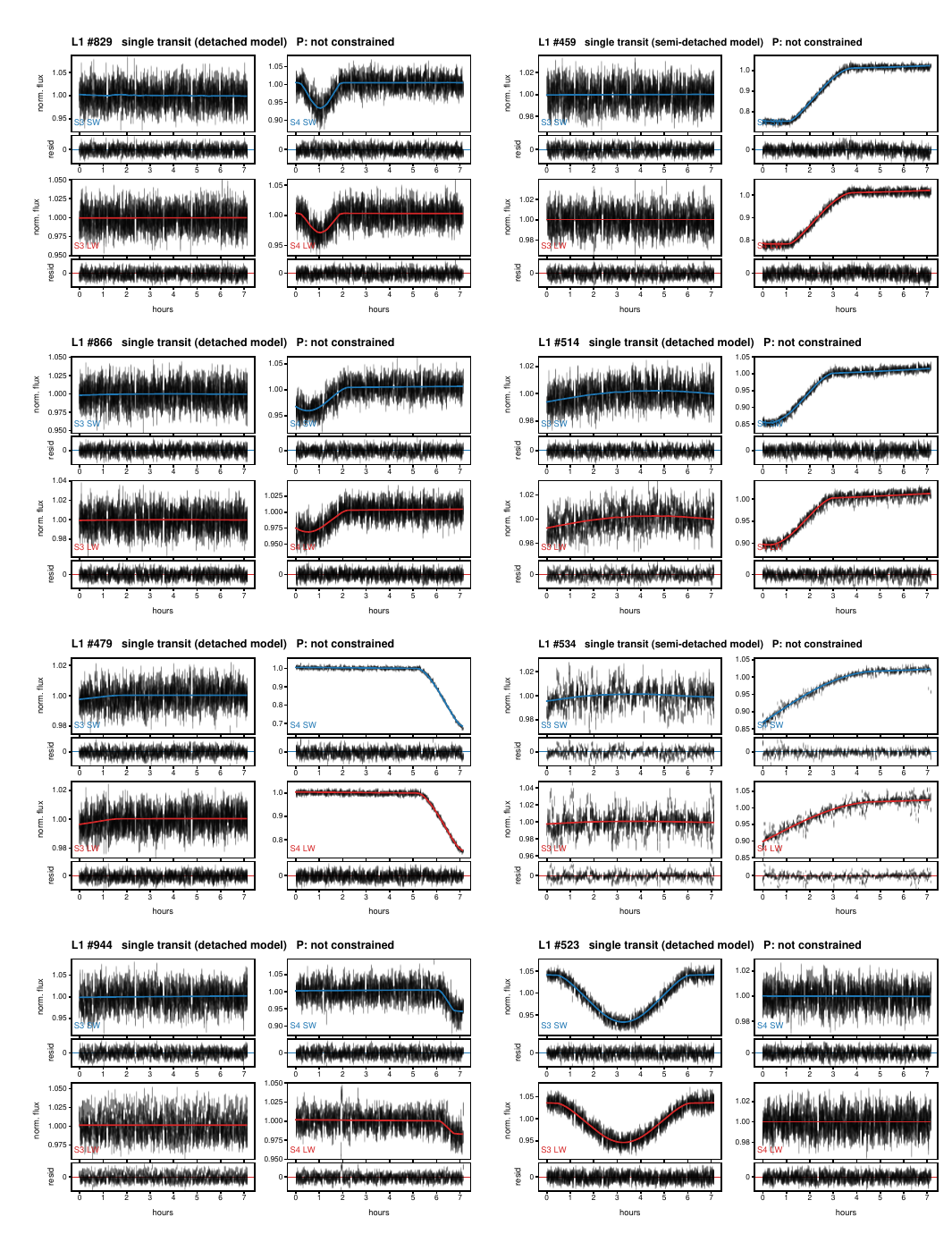}
\caption{Single-transit sources, continued (page 21 of 26).}
\end{figure*}
\clearpage

\begin{figure*}
\centering
\includegraphics[width=0.98\textwidth,height=0.94\textheight,keepaspectratio]{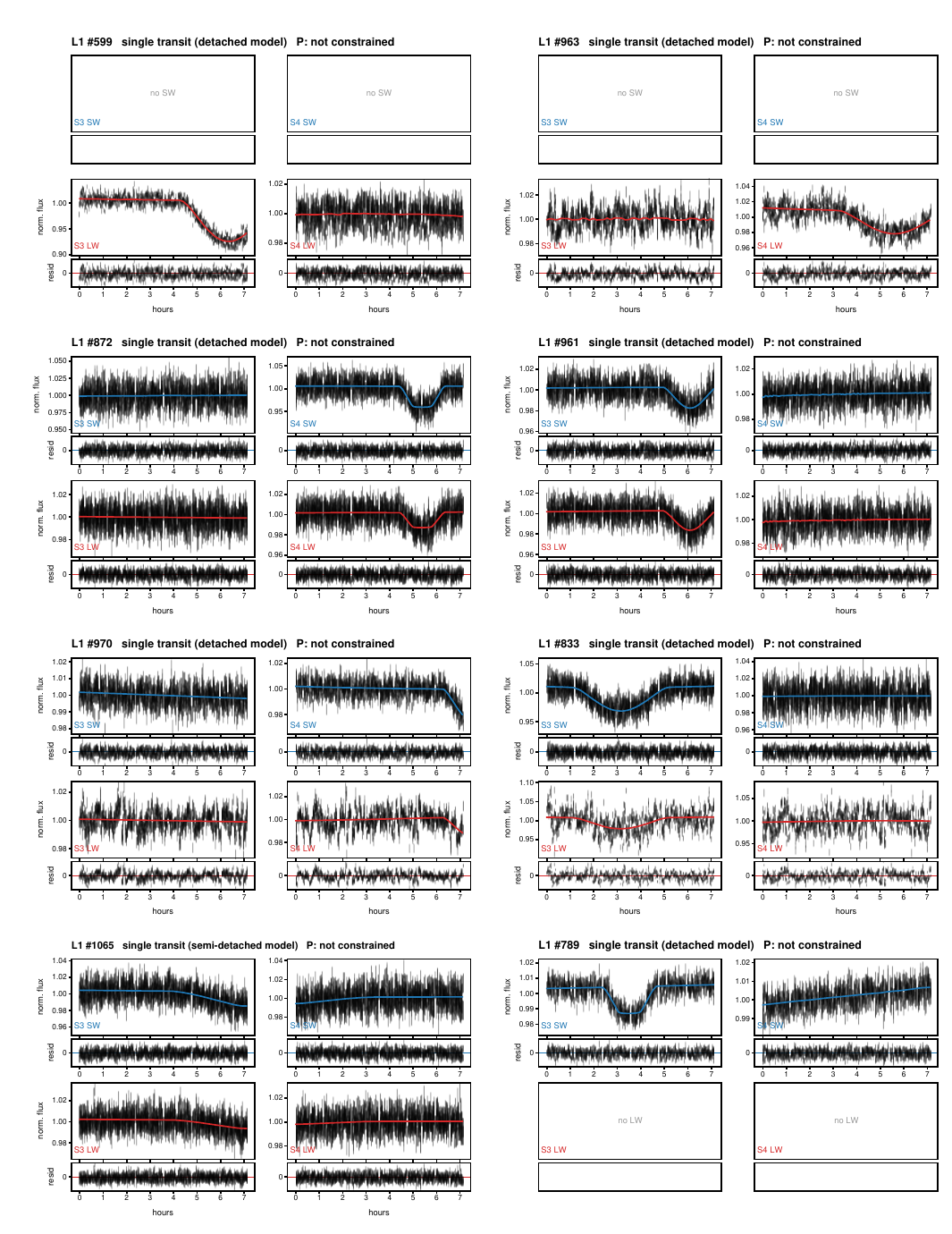}
\caption{Single-transit sources, continued (page 22 of 26).}
\end{figure*}
\clearpage

\begin{figure*}
\centering
\includegraphics[width=0.98\textwidth,height=0.94\textheight,keepaspectratio]{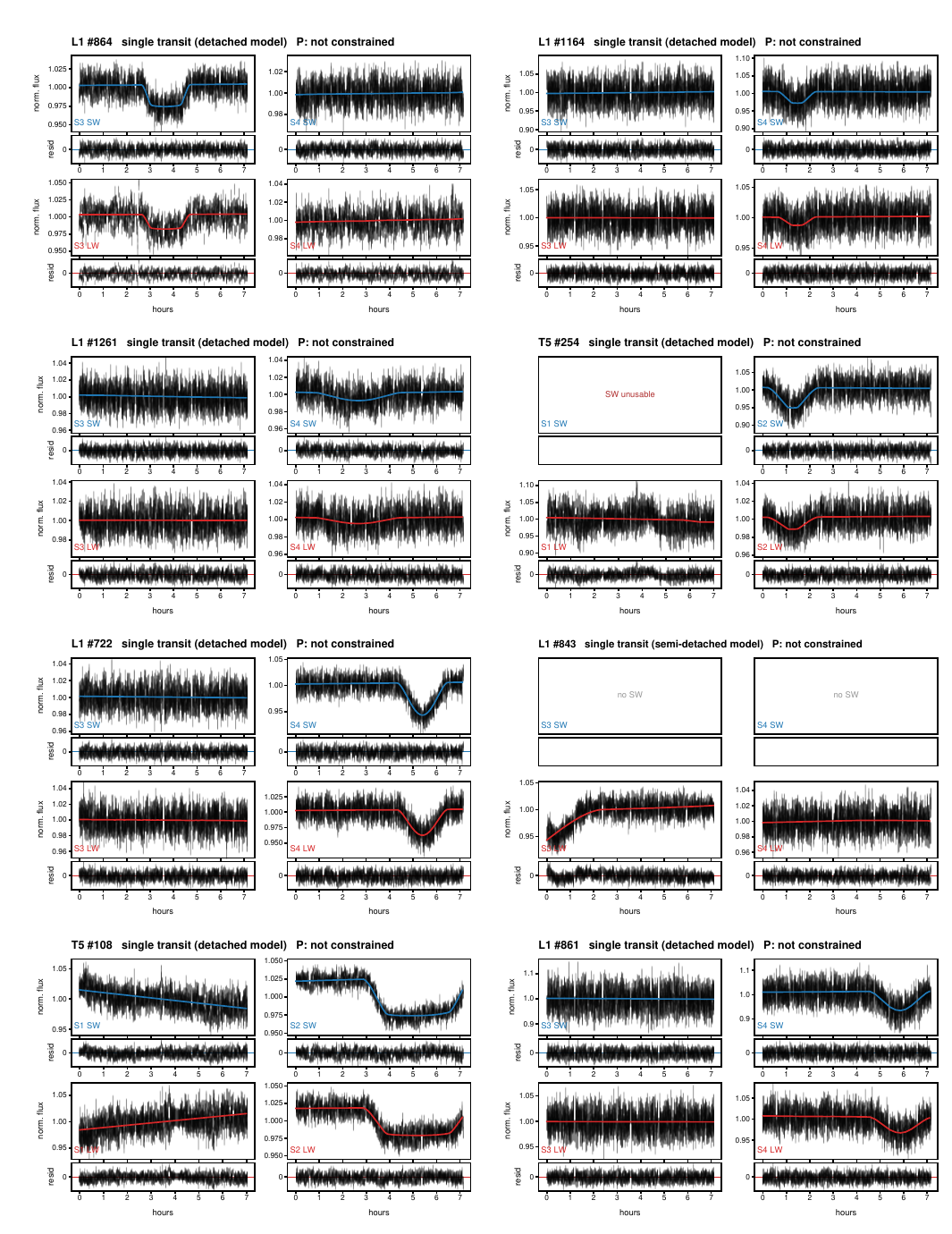}
\caption{Single-transit sources, continued (page 23 of 26).}
\end{figure*}
\clearpage

\begin{figure*}
\centering
\includegraphics[width=0.98\textwidth,height=0.94\textheight,keepaspectratio]{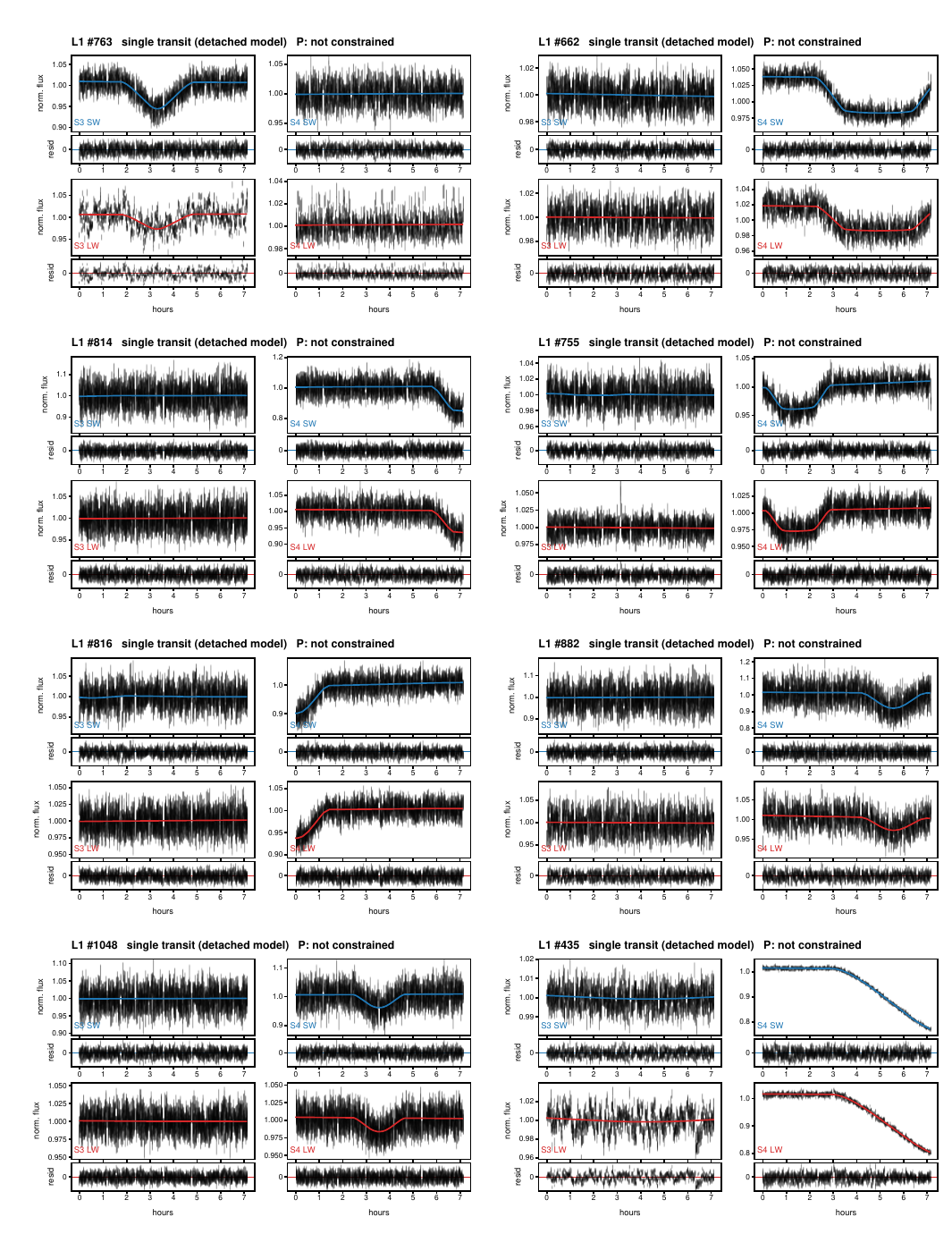}
\caption{Single-transit sources, continued (page 24 of 26).}
\end{figure*}
\clearpage

\begin{figure*}
\centering
\includegraphics[width=0.98\textwidth,height=0.94\textheight,keepaspectratio]{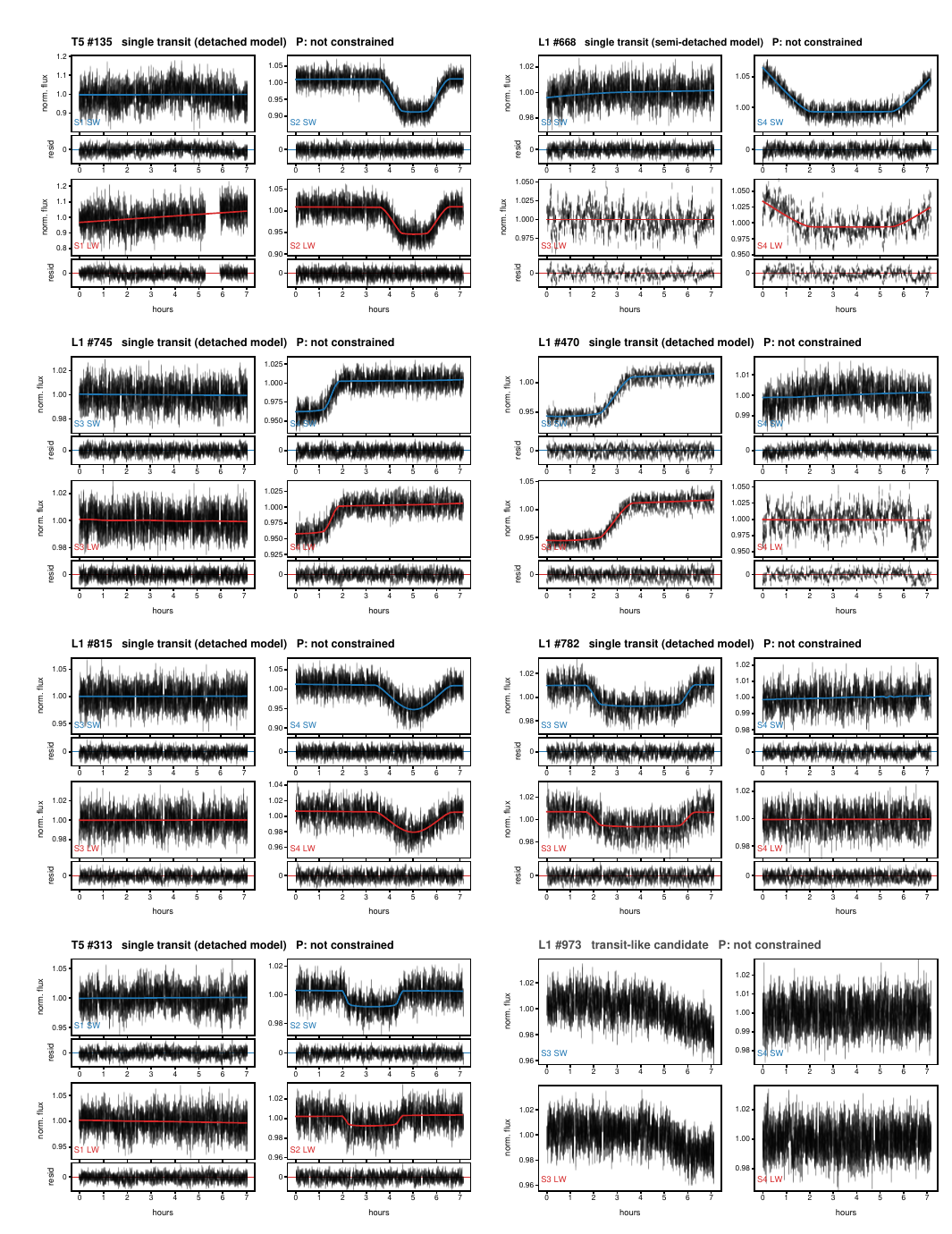}
\caption{Single-transit sources, continued (page 25 of 26).}
\end{figure*}
\clearpage

\begin{figure*}
\centering
\includegraphics[width=0.98\textwidth,height=0.94\textheight,keepaspectratio]{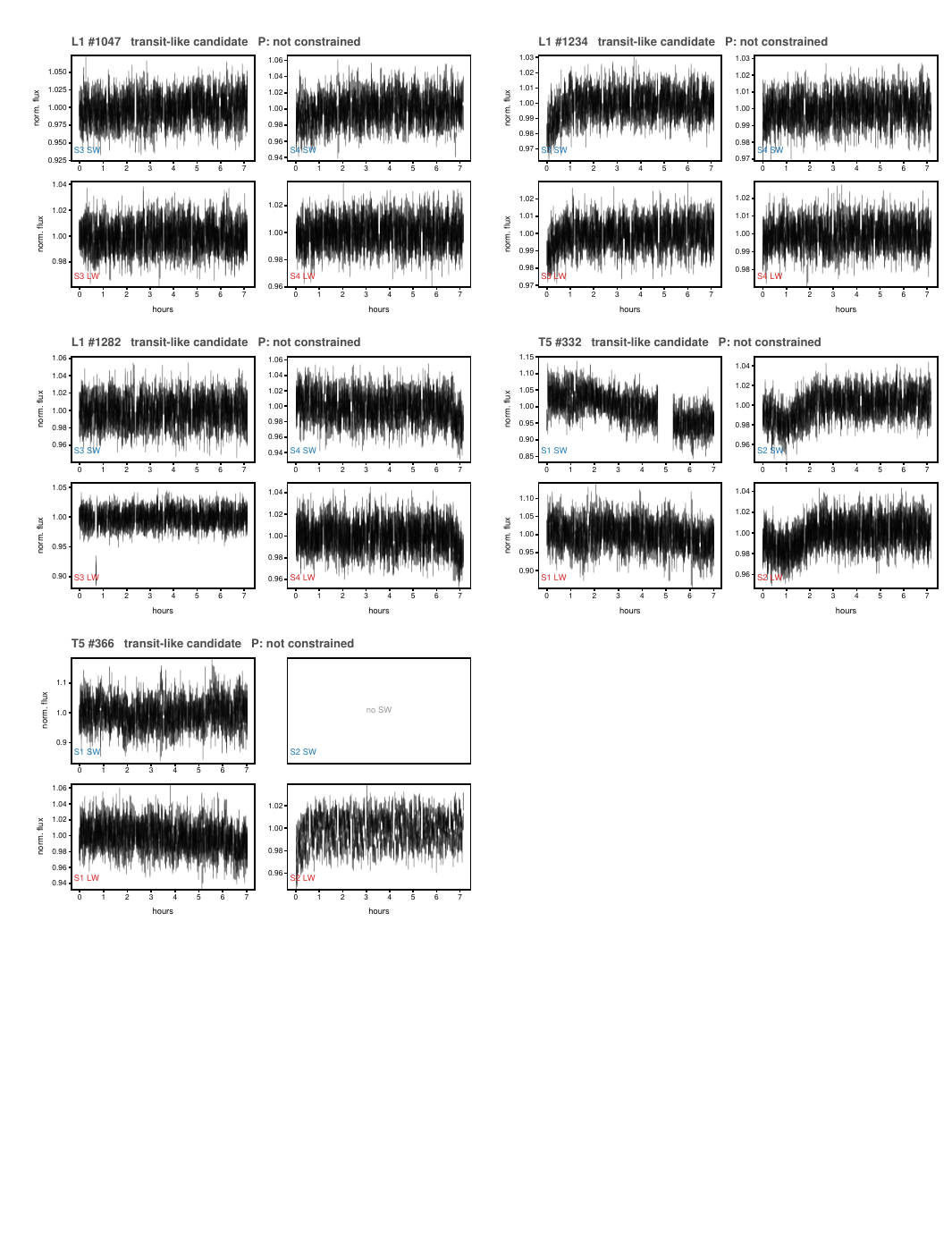}
\caption{Single-transit sources, continued (page 26 of 26).}
\end{figure*}
\clearpage

\subsection{Other classified variables}\label{app:atlas:otherclass}

\begin{figure*}
\centering
\includegraphics[width=0.98\textwidth,height=0.79\textheight,keepaspectratio]{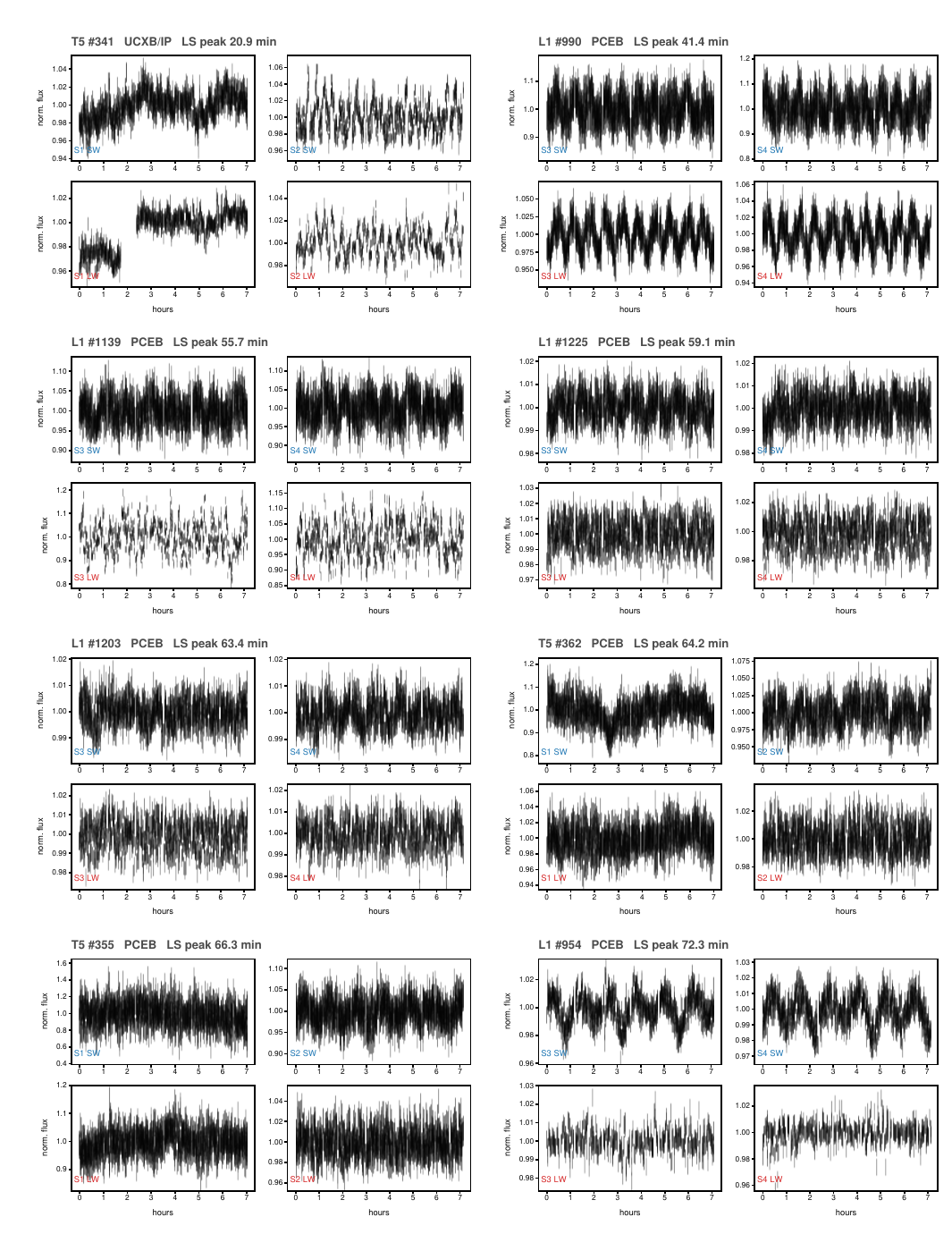}
\caption{Other classified variables: unfolded JWST NIRCam lightcurves of all 41 sources in this class (page 1 of 6), sorted by adopted period (sources without a period last). Each source is shown as a four-panel block. The two observing segments run left to right, with the short-wavelength F200W lightcurve (SW, \textbf{blue}) on top and the long-wavelength F356W lightcurve (LW, \textbf{red}) below, and the segment and band are labelled inside every panel. Time is hours from the start of that segment, and lightcurves are never phase-folded. Black vertical bars are the adopted lightcurve (\S\ref{sec:strategy}), each spanning the $1\sigma$ uncertainty of one plotted sample. No model is overplotted: these sources belong to classes a binary model cannot express (\S\ref{sec:classification}), and all but the redback are candidate designations from lightcurve morphology alone. The header gives the source, its by-eye class or ``no class'', and any Lomb--Scargle peak, marked ``LS peak'', a formal search result, not an adopted orbital period (for the redback it lies within 1\% of the radio-timing orbital period, \S\ref{sec:psrj1748}).}
\end{figure*}
\clearpage

\begin{figure*}
\centering
\includegraphics[width=0.98\textwidth,height=0.94\textheight,keepaspectratio]{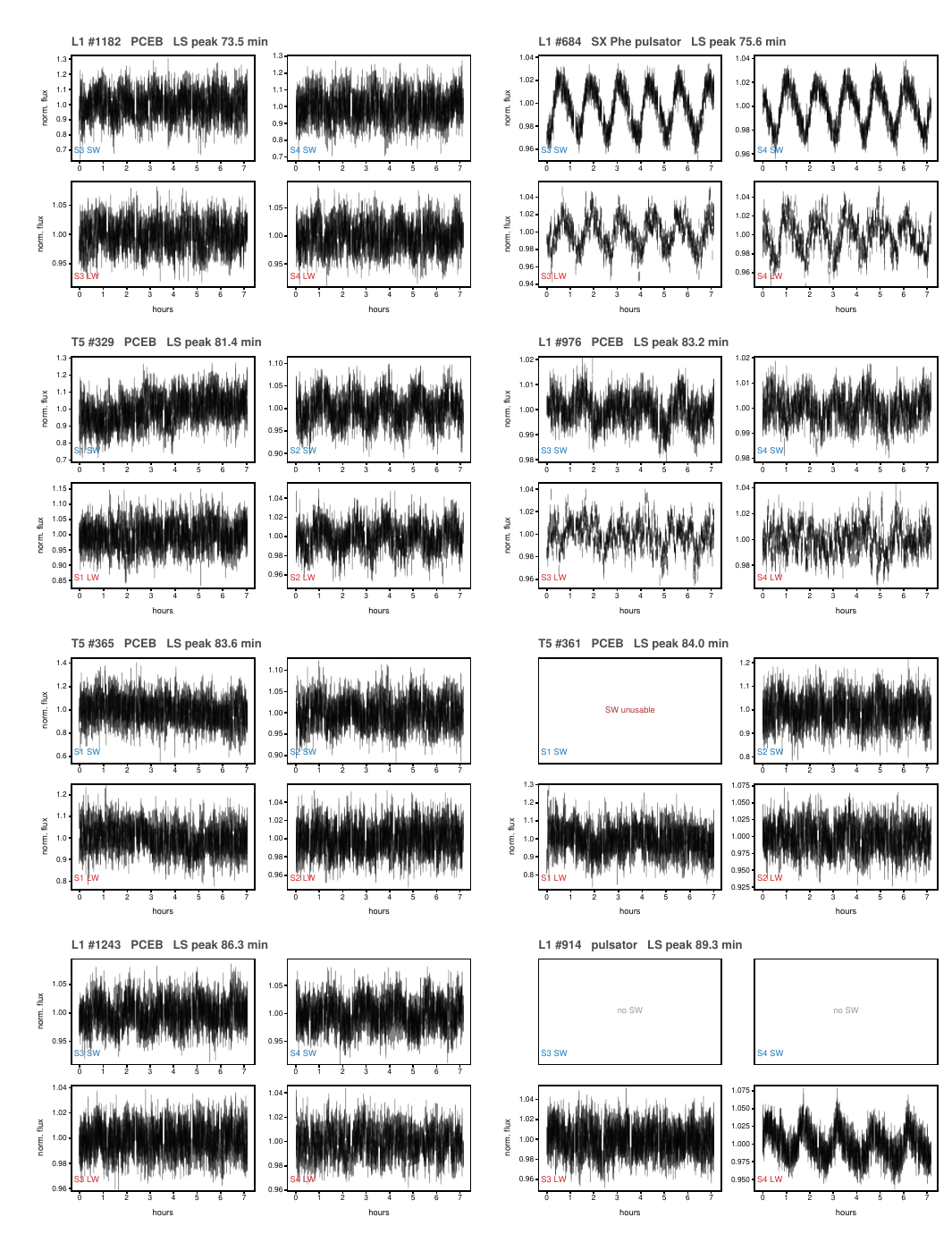}
\caption{Other classified variables, continued (page 2 of 6).}
\end{figure*}
\clearpage

\begin{figure*}
\centering
\includegraphics[width=0.98\textwidth,height=0.94\textheight,keepaspectratio]{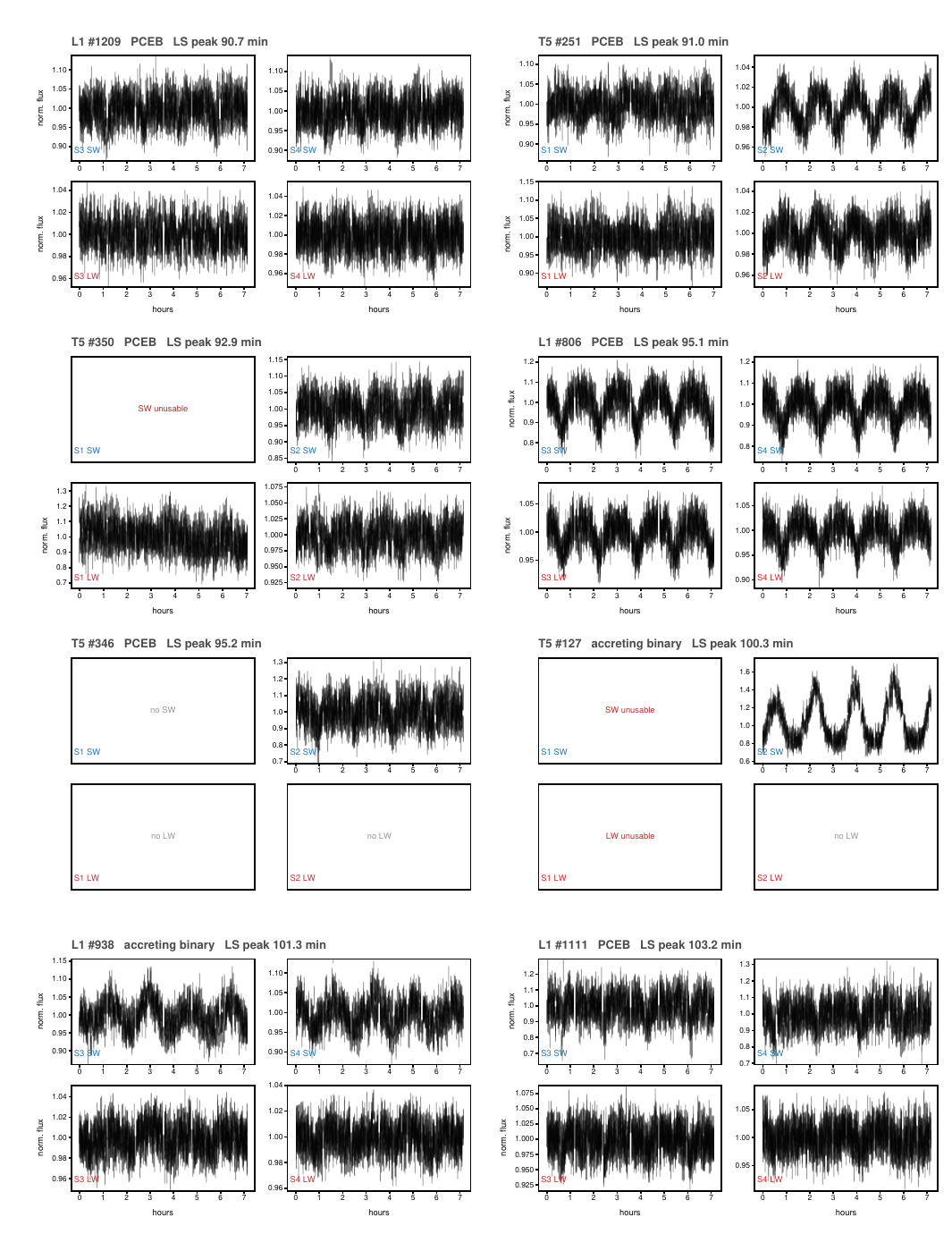}
\caption{Other classified variables, continued (page 3 of 6).}
\end{figure*}
\clearpage

\begin{figure*}
\centering
\includegraphics[width=0.98\textwidth,height=0.94\textheight,keepaspectratio]{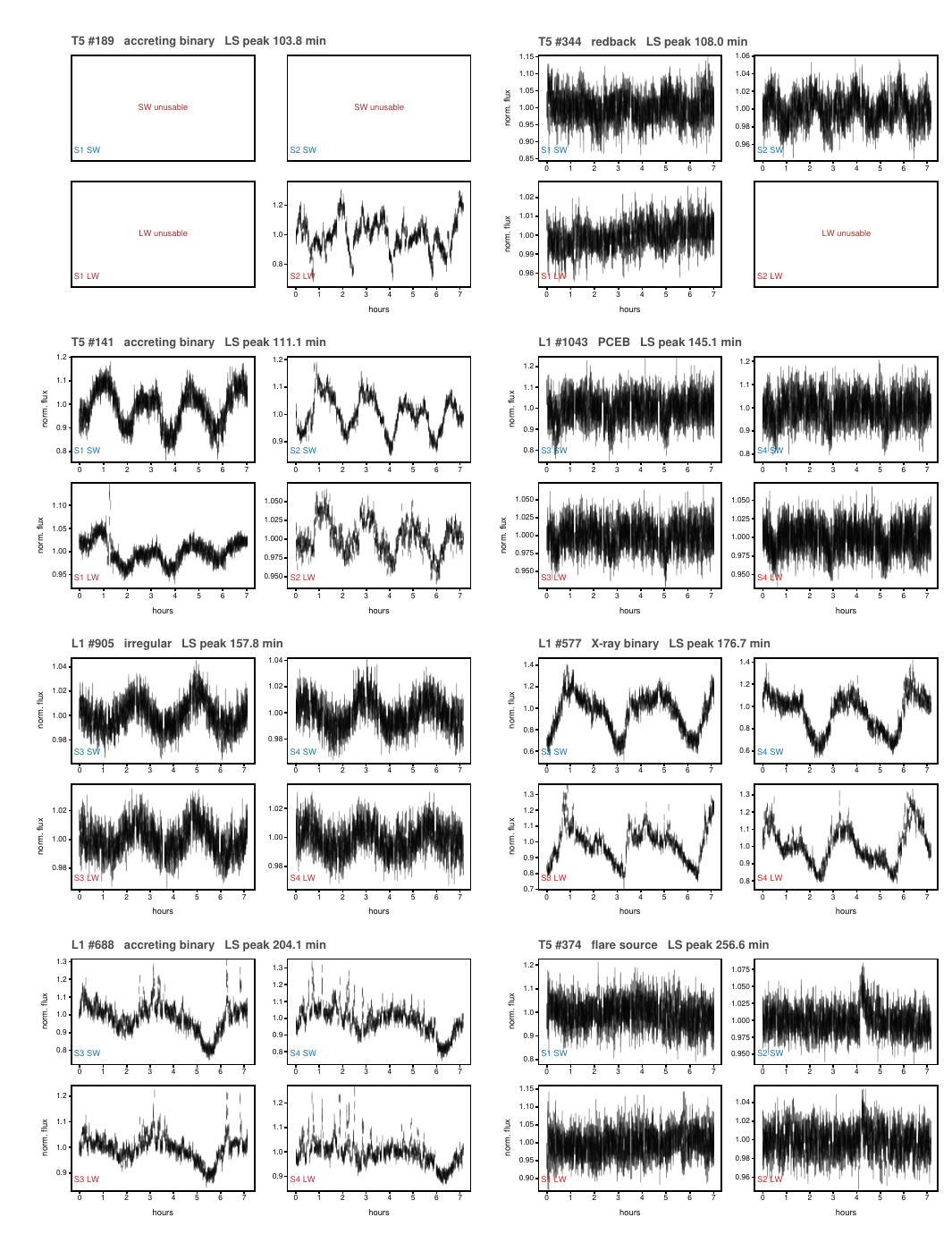}
\caption{Other classified variables, continued (page 4 of 6).}
\end{figure*}
\clearpage

\begin{figure*}
\centering
\includegraphics[width=0.98\textwidth,height=0.94\textheight,keepaspectratio]{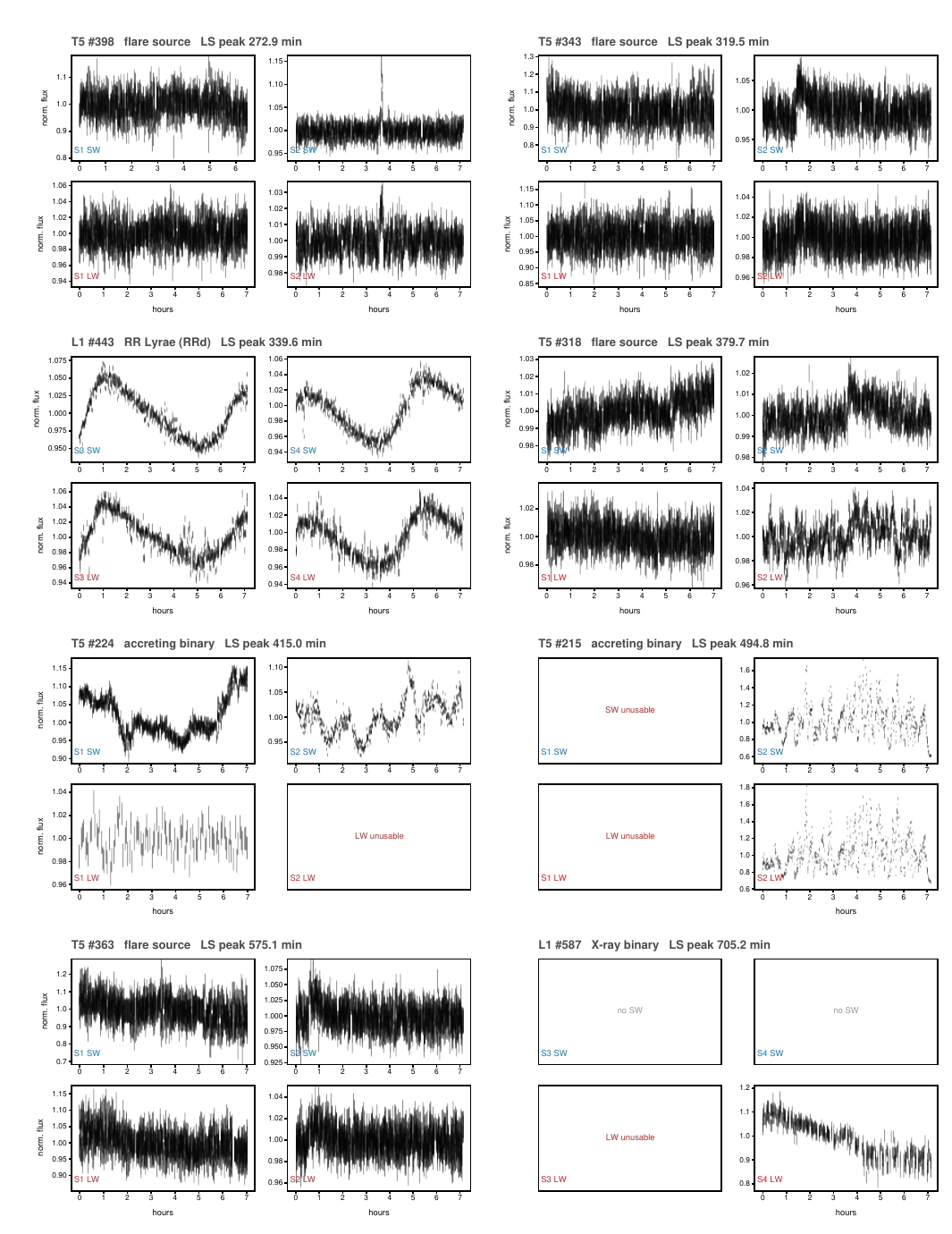}
\caption{Other classified variables, continued (page 5 of 6).}
\end{figure*}
\clearpage

\begin{figure*}
\centering
\includegraphics[width=0.98\textwidth,height=0.94\textheight,keepaspectratio]{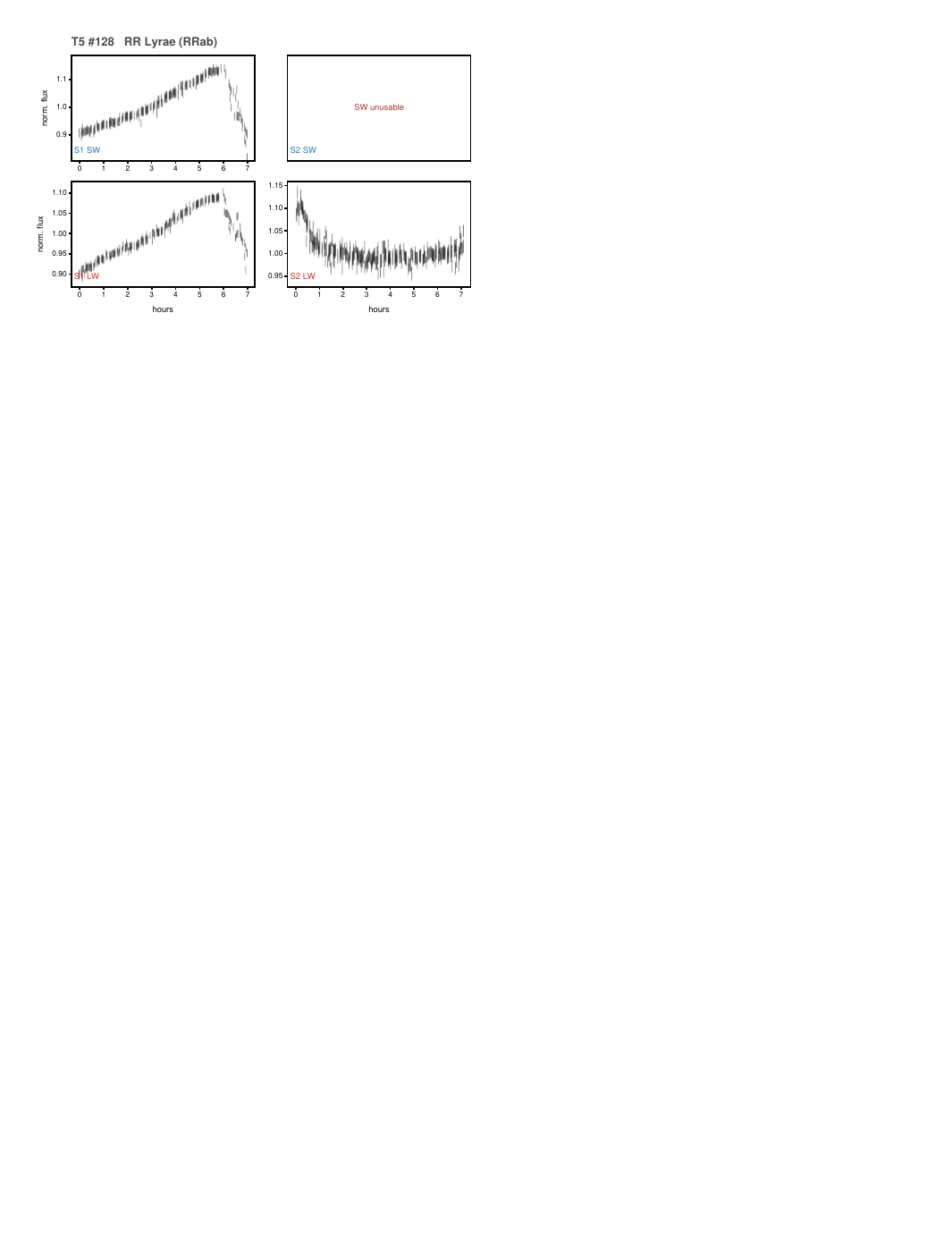}
\caption{Other classified variables, continued (page 6 of 6).}
\end{figure*}
\clearpage

\subsection{Variables with no accepted model}\label{app:atlas:nomodel}

\begin{figure*}
\centering
\includegraphics[width=0.98\textwidth,height=0.79\textheight,keepaspectratio]{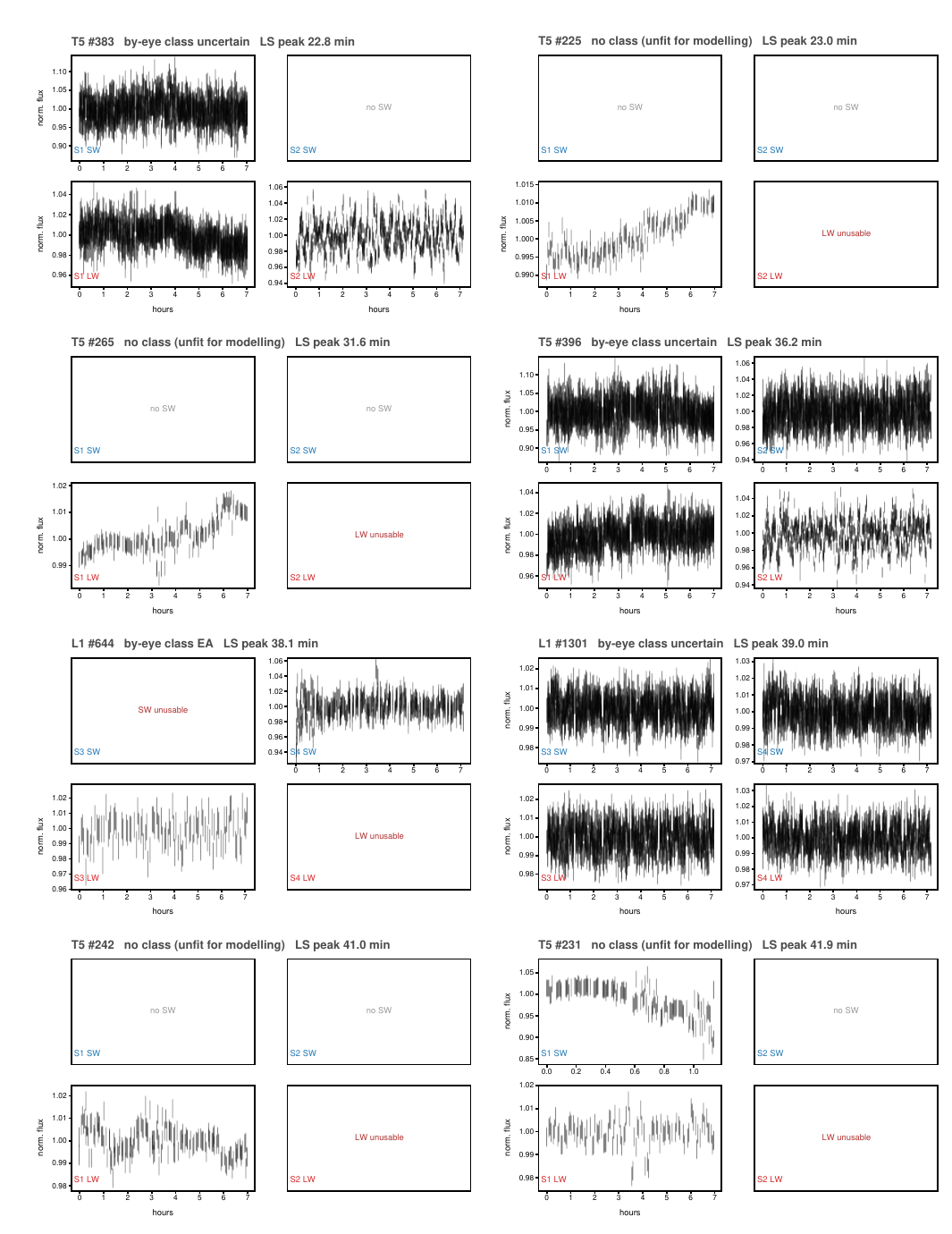}
\caption{Variables with no accepted model: unfolded JWST NIRCam lightcurves of all 348 sources in this class (page 1 of 44), sorted by adopted period (sources without a period last). Each source is shown as a four-panel block. The two observing segments run left to right, with the short-wavelength F200W lightcurve (SW, \textbf{blue}) on top and the long-wavelength F356W lightcurve (LW, \textbf{red}) below, and the segment and band are labelled inside every panel. Time is hours from the start of that segment, and lightcurves are never phase-folded. Black vertical bars are the adopted lightcurve (\S\ref{sec:strategy}), each spanning the $1\sigma$ uncertainty of one plotted sample. No model is overplotted: no \texttt{PHOEBE} model was accepted for these sources, either because we did not accept a fit on visual inspection or because we judged the lightcurve unfit for modelling (\S\ref{sec:classification}). The header gives the source, its by-eye class or ``no class'', and any Lomb--Scargle peak, marked ``LS peak'', a formal search result, not an adopted orbital period.}
\end{figure*}
\clearpage

\begin{figure*}
\centering
\includegraphics[width=0.98\textwidth,height=0.94\textheight,keepaspectratio]{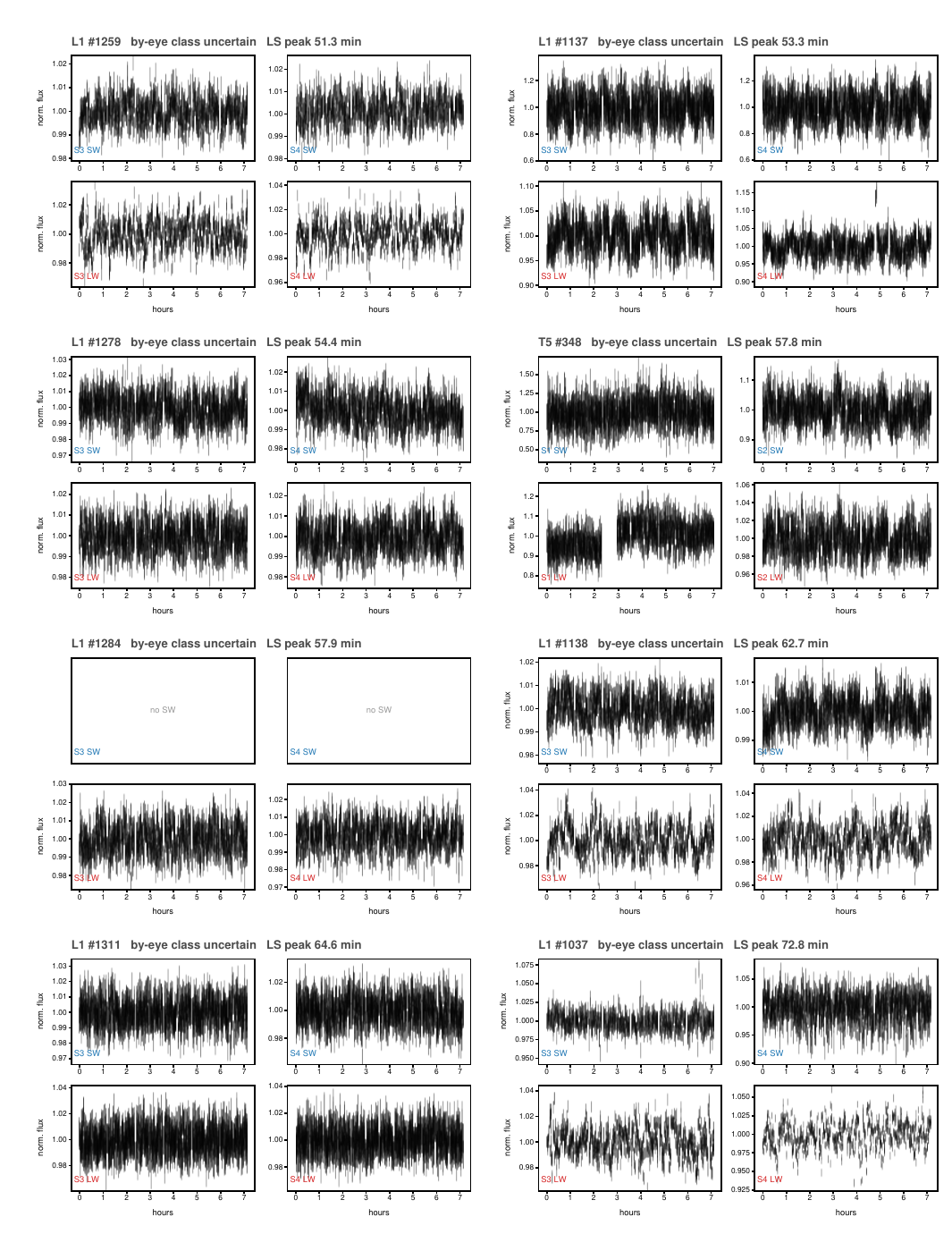}
\caption{Variables with no accepted model, continued (page 2 of 44).}
\end{figure*}
\clearpage

\begin{figure*}
\centering
\includegraphics[width=0.98\textwidth,height=0.94\textheight,keepaspectratio]{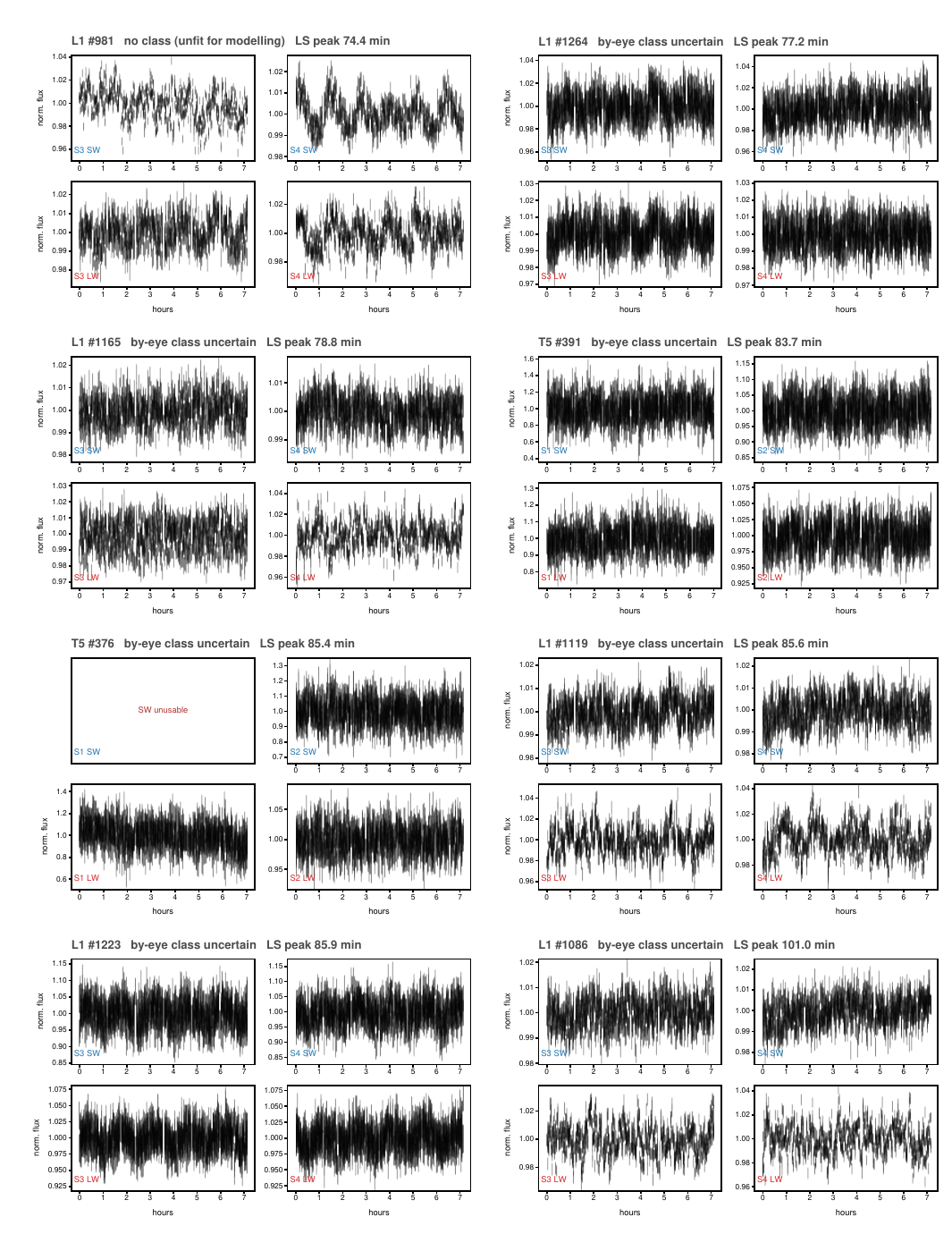}
\caption{Variables with no accepted model, continued (page 3 of 44).}
\end{figure*}
\clearpage

\begin{figure*}
\centering
\includegraphics[width=0.98\textwidth,height=0.94\textheight,keepaspectratio]{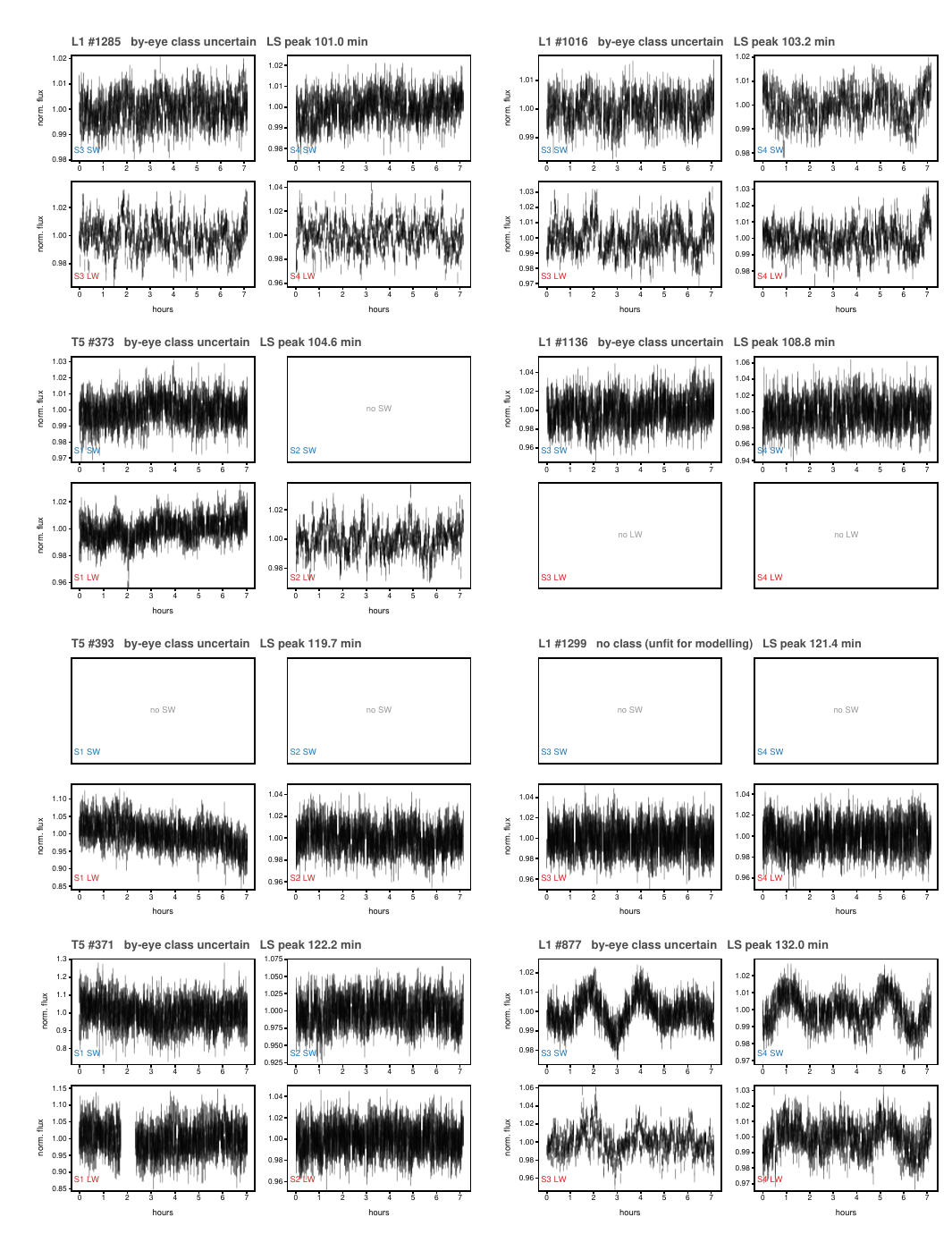}
\caption{Variables with no accepted model, continued (page 4 of 44).}
\end{figure*}
\clearpage

\begin{figure*}
\centering
\includegraphics[width=0.98\textwidth,height=0.94\textheight,keepaspectratio]{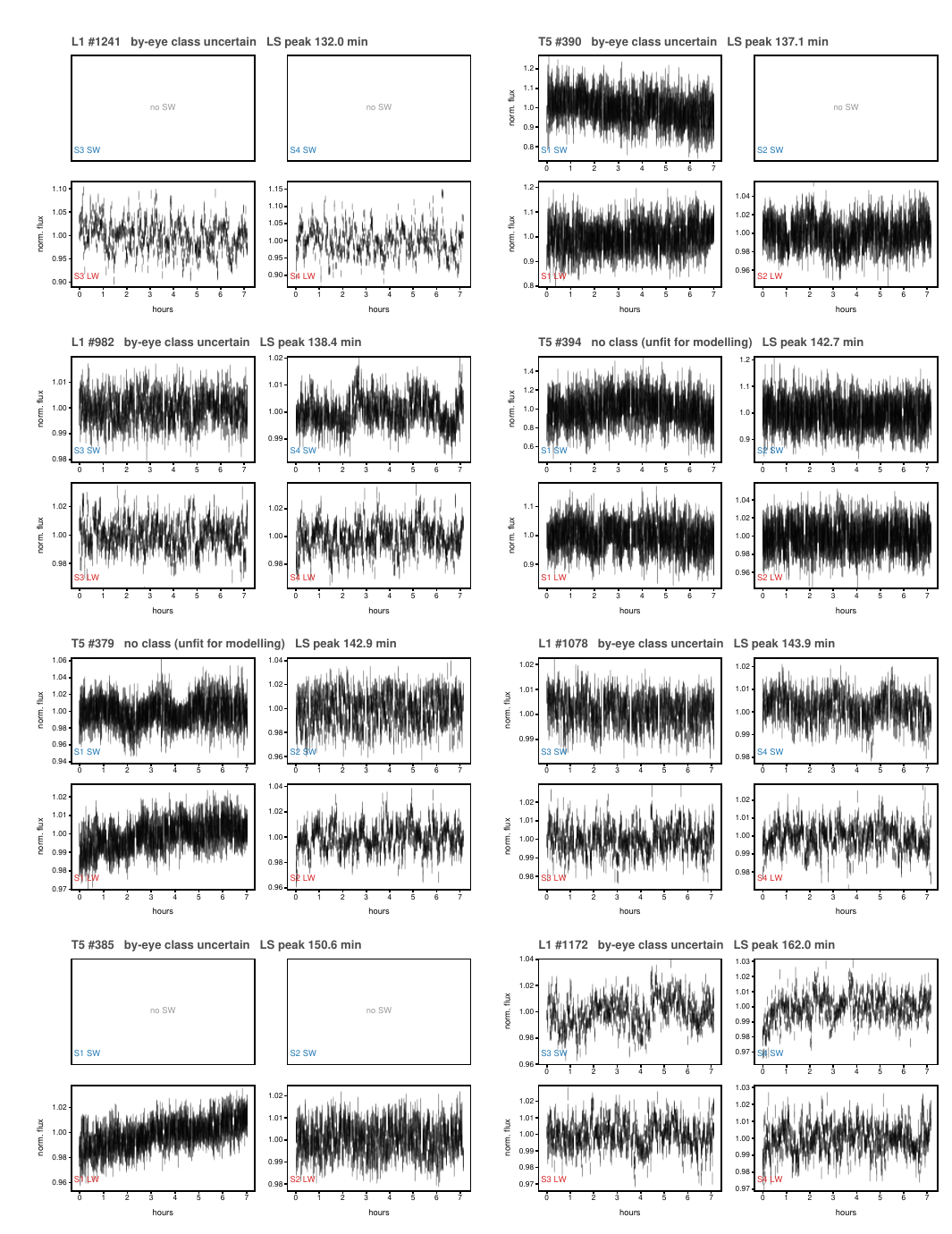}
\caption{Variables with no accepted model, continued (page 5 of 44).}
\end{figure*}
\clearpage

\begin{figure*}
\centering
\includegraphics[width=0.98\textwidth,height=0.94\textheight,keepaspectratio]{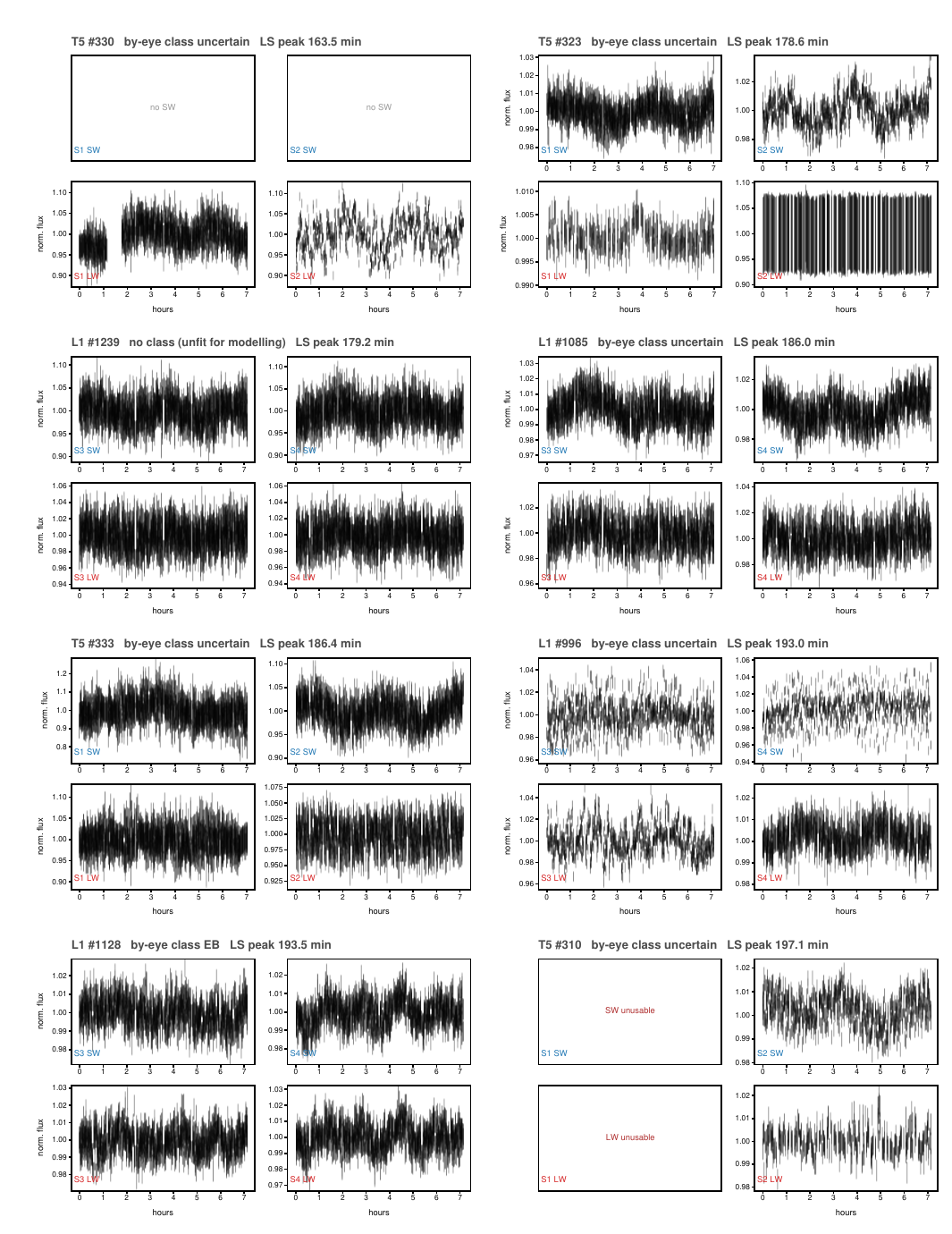}
\caption{Variables with no accepted model, continued (page 6 of 44).}
\end{figure*}
\clearpage

\begin{figure*}
\centering
\includegraphics[width=0.98\textwidth,height=0.94\textheight,keepaspectratio]{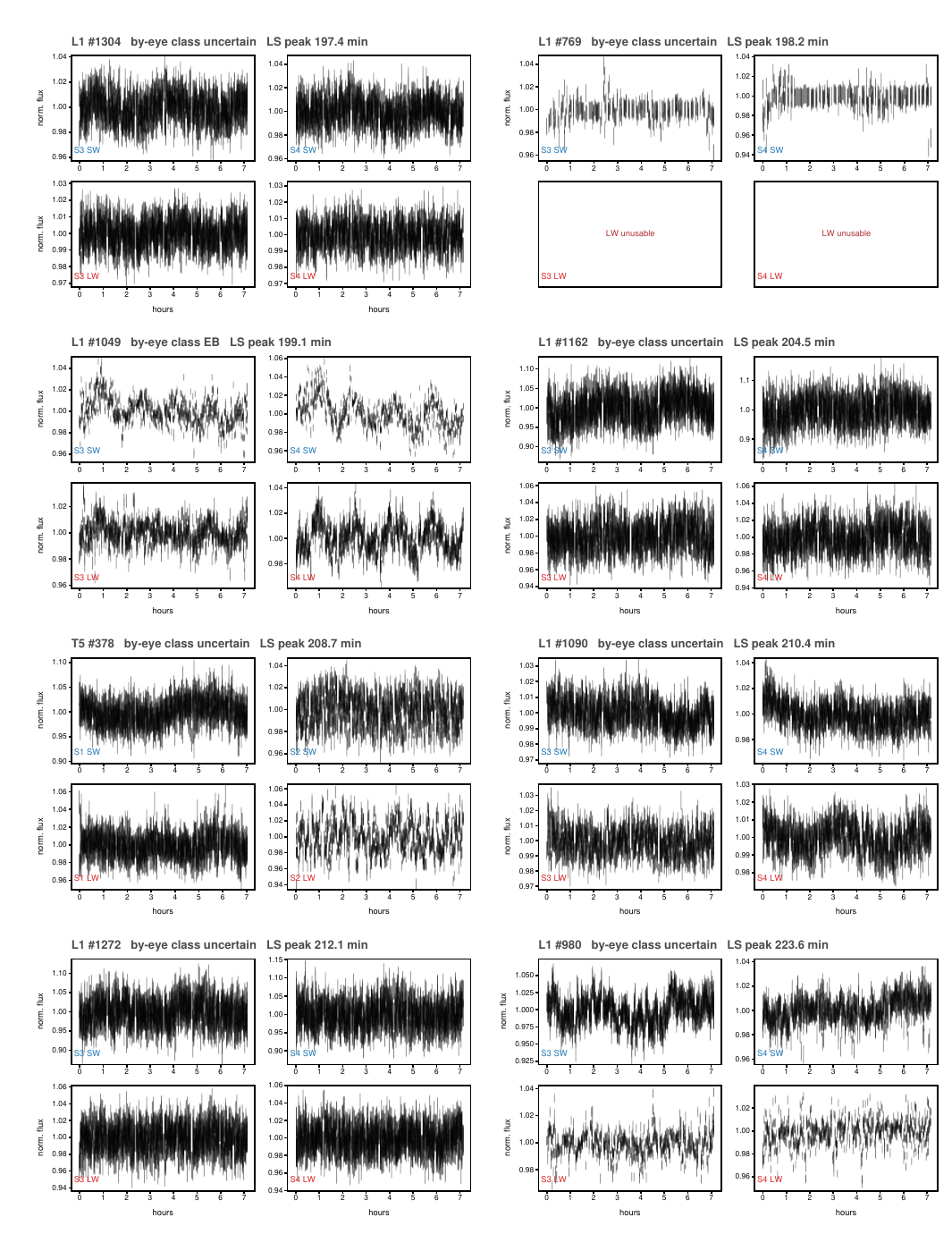}
\caption{Variables with no accepted model, continued (page 7 of 44).}
\end{figure*}
\clearpage

\begin{figure*}
\centering
\includegraphics[width=0.98\textwidth,height=0.94\textheight,keepaspectratio]{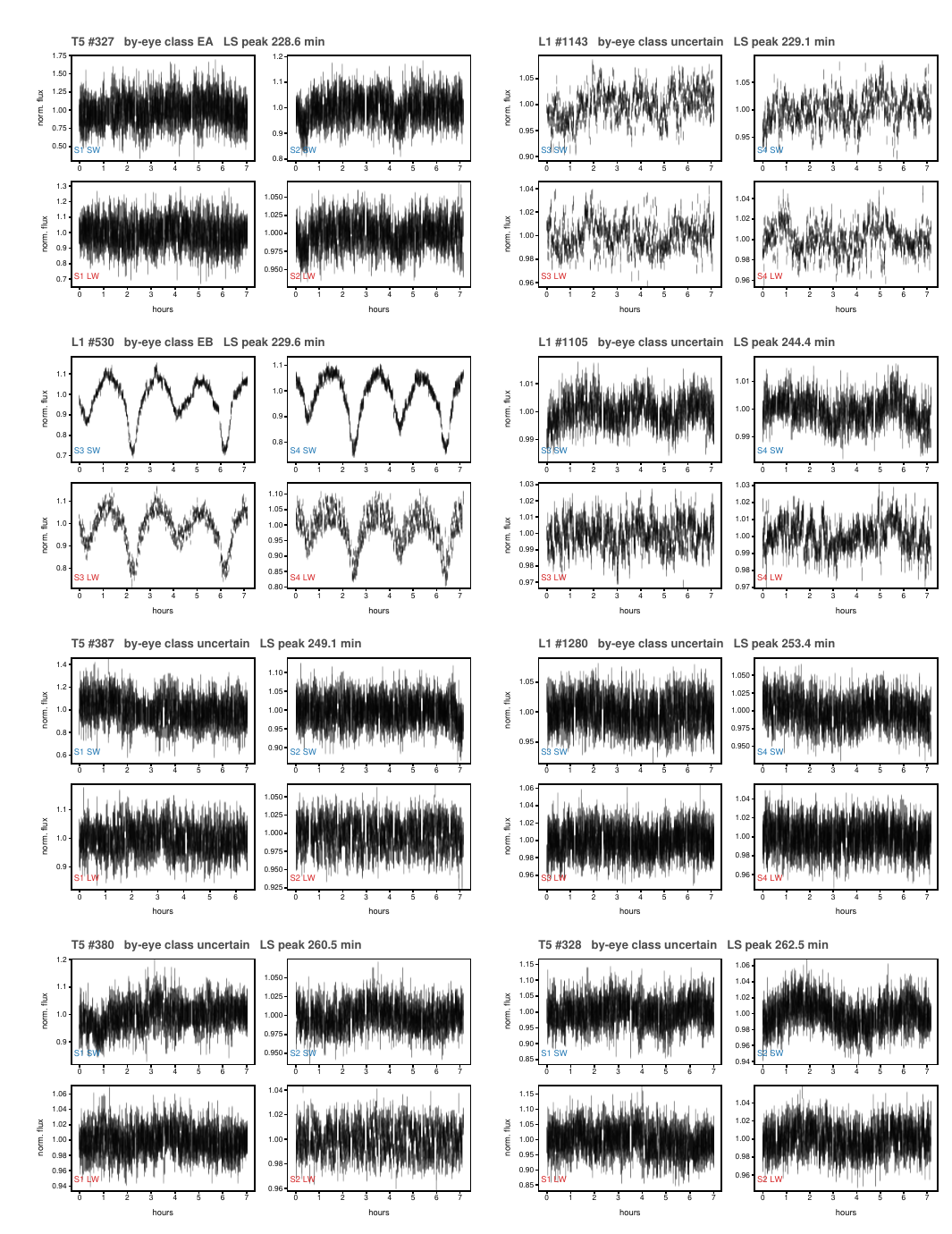}
\caption{Variables with no accepted model, continued (page 8 of 44).}
\end{figure*}
\clearpage

\begin{figure*}
\centering
\includegraphics[width=0.98\textwidth,height=0.94\textheight,keepaspectratio]{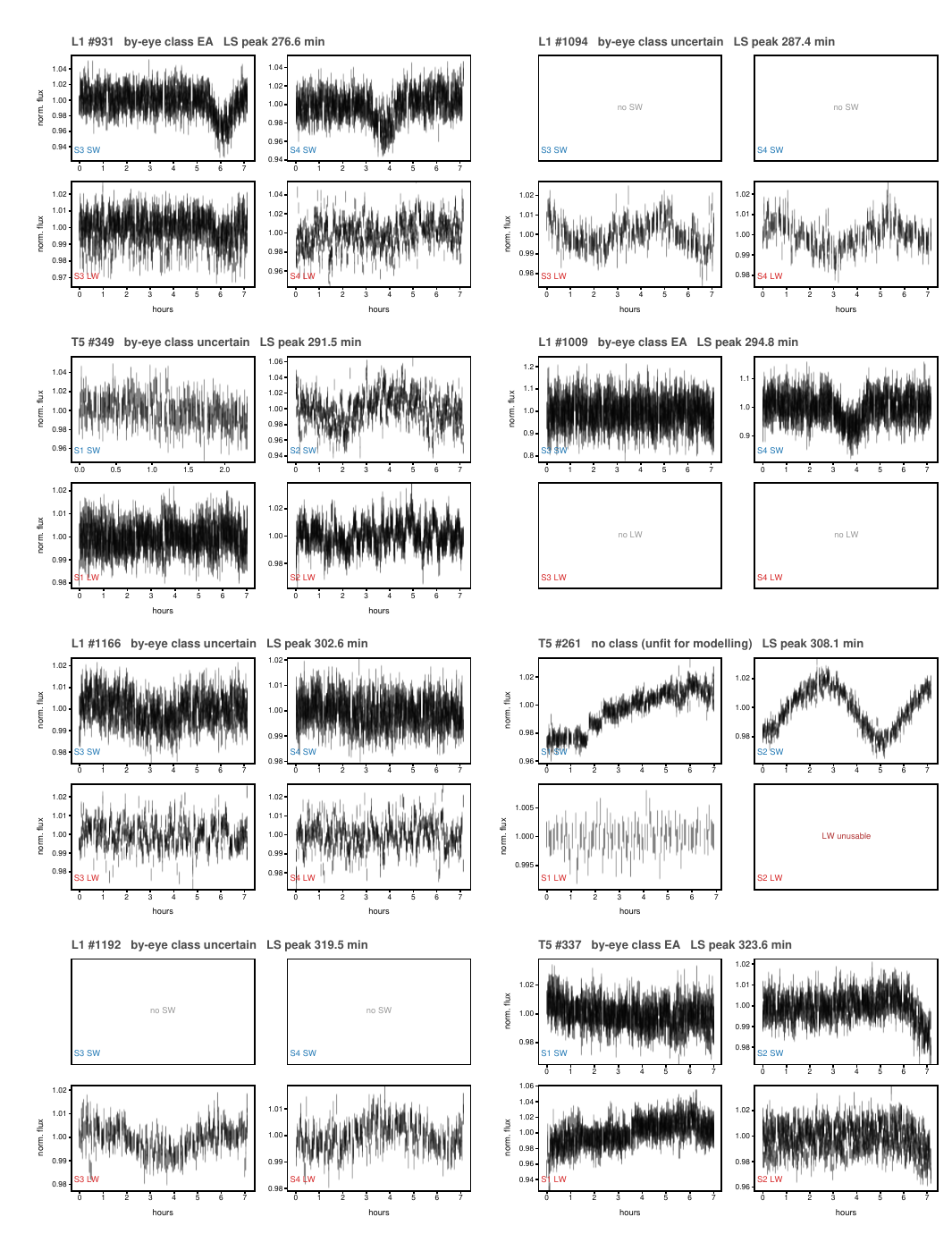}
\caption{Variables with no accepted model, continued (page 9 of 44).}
\end{figure*}
\clearpage

\begin{figure*}
\centering
\includegraphics[width=0.98\textwidth,height=0.94\textheight,keepaspectratio]{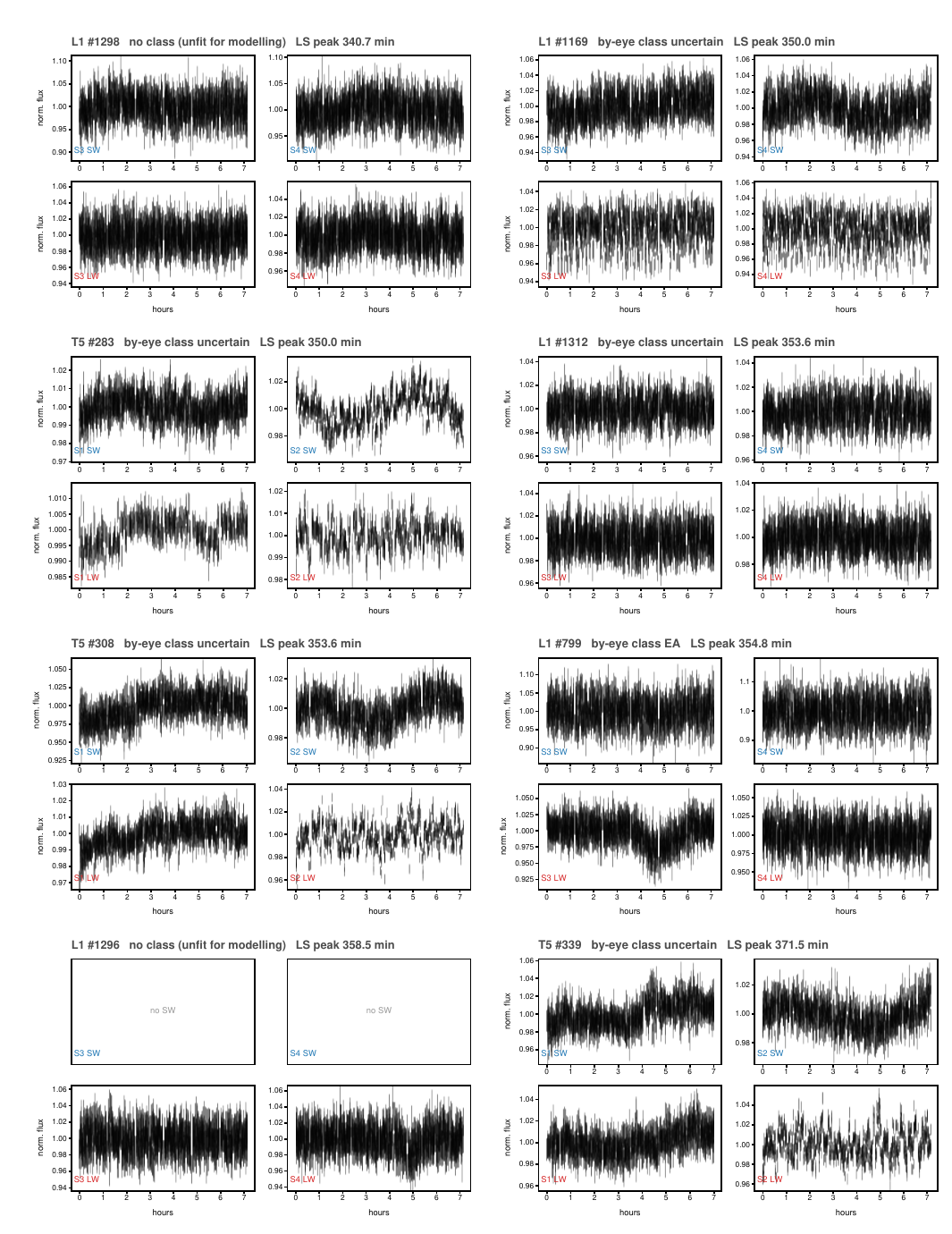}
\caption{Variables with no accepted model, continued (page 10 of 44).}
\end{figure*}
\clearpage

\begin{figure*}
\centering
\includegraphics[width=0.98\textwidth,height=0.94\textheight,keepaspectratio]{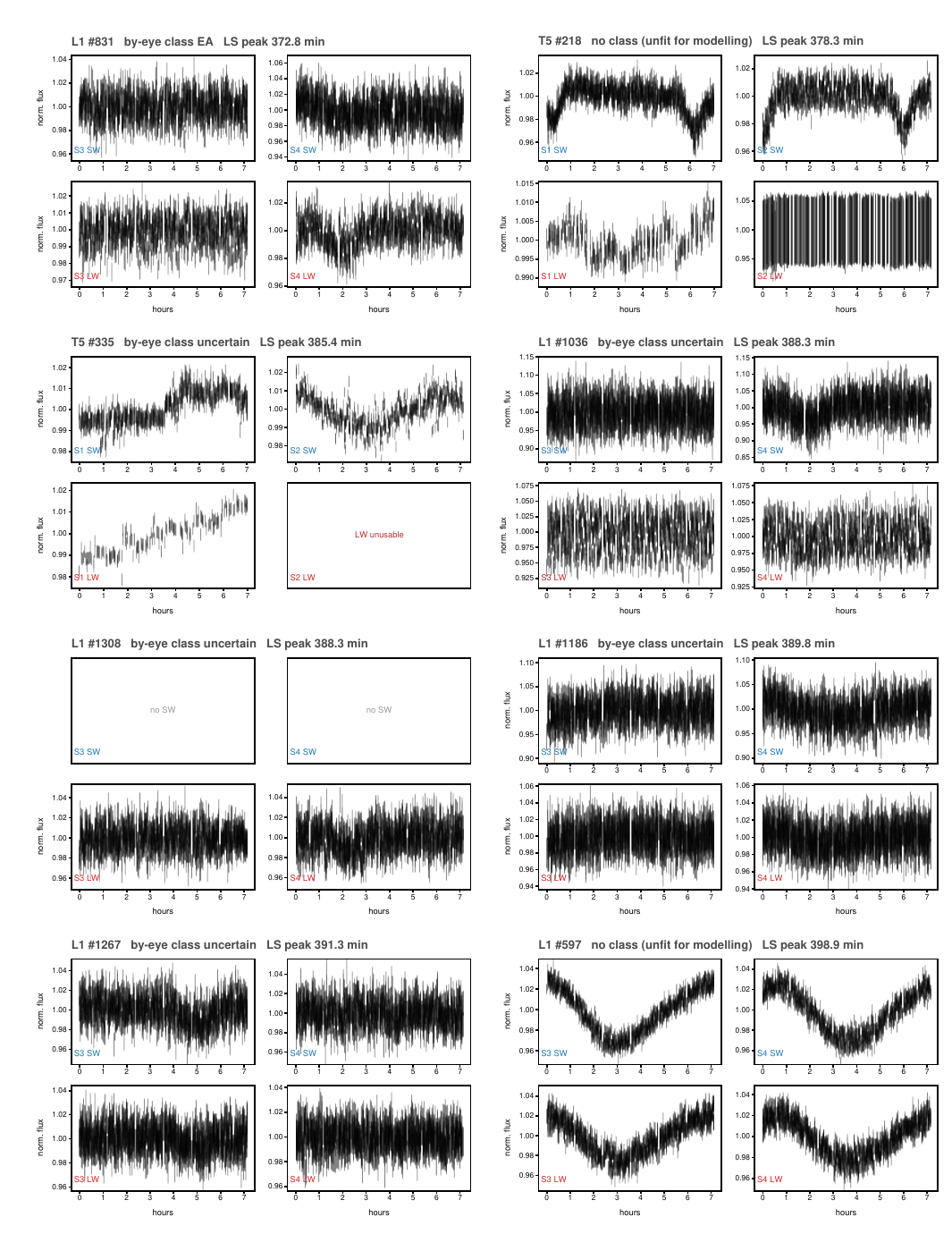}
\caption{Variables with no accepted model, continued (page 11 of 44).}
\end{figure*}
\clearpage

\begin{figure*}
\centering
\includegraphics[width=0.98\textwidth,height=0.94\textheight,keepaspectratio]{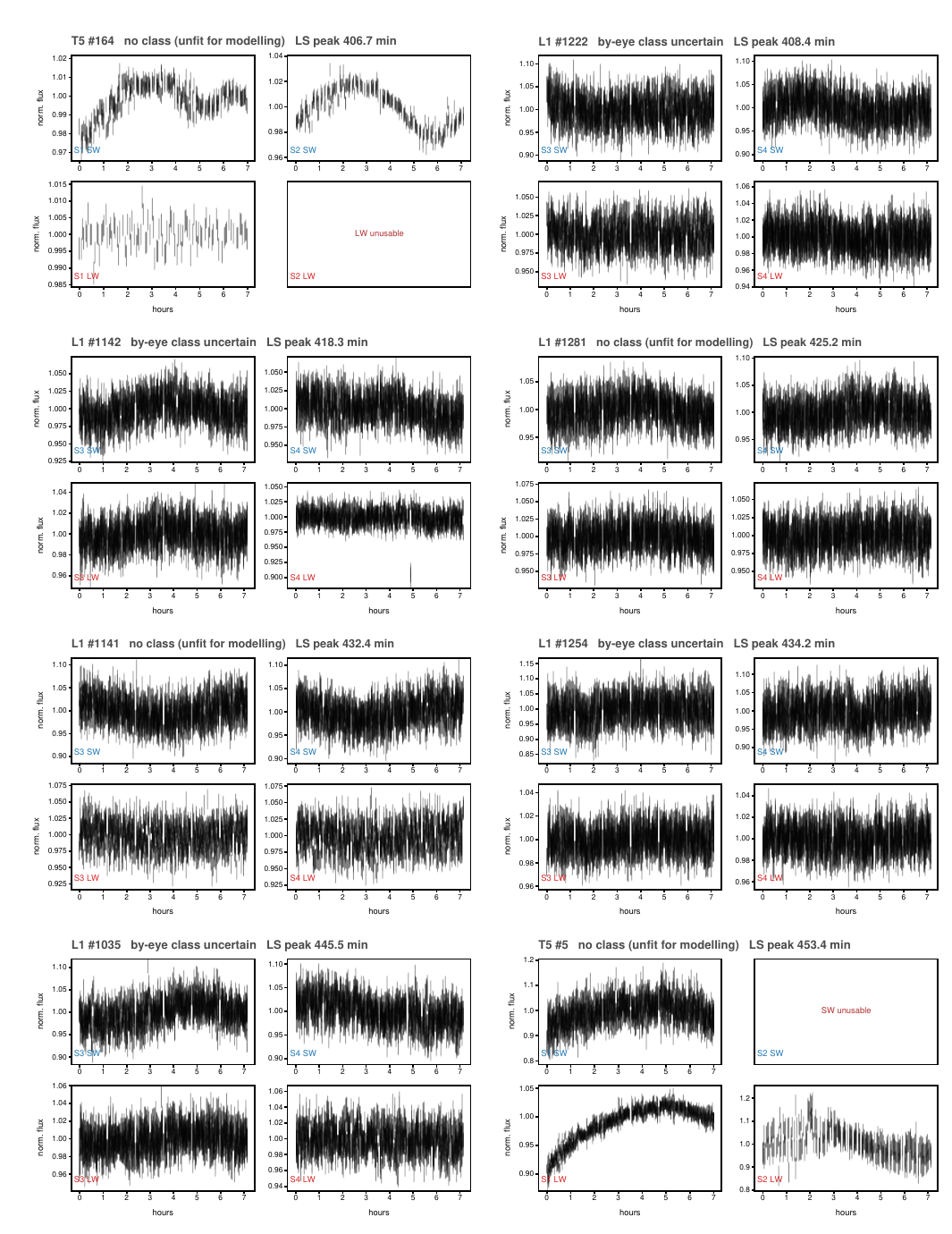}
\caption{Variables with no accepted model, continued (page 12 of 44).}
\end{figure*}
\clearpage

\begin{figure*}
\centering
\includegraphics[width=0.98\textwidth,height=0.94\textheight,keepaspectratio]{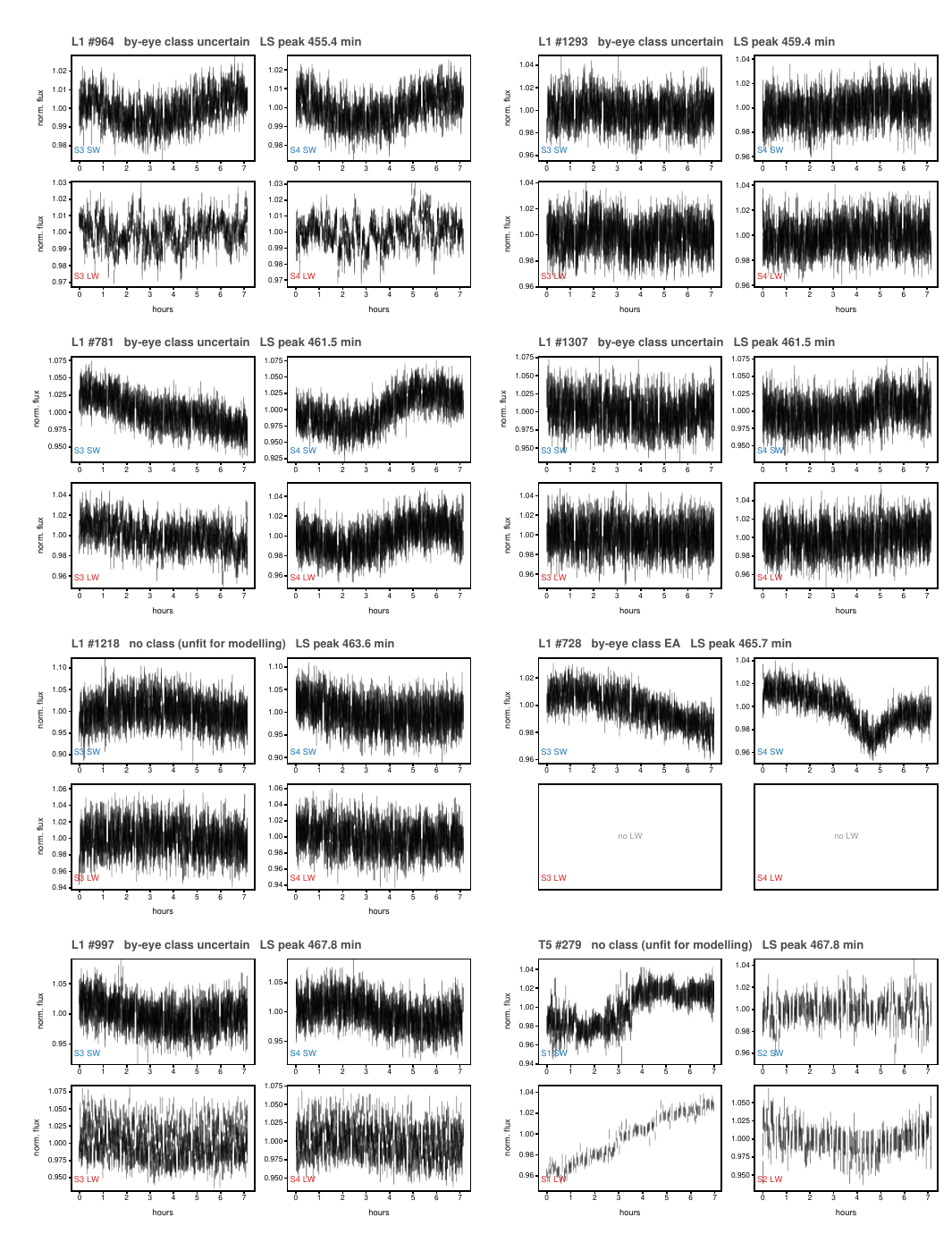}
\caption{Variables with no accepted model, continued (page 13 of 44).}
\end{figure*}
\clearpage

\begin{figure*}
\centering
\includegraphics[width=0.98\textwidth,height=0.94\textheight,keepaspectratio]{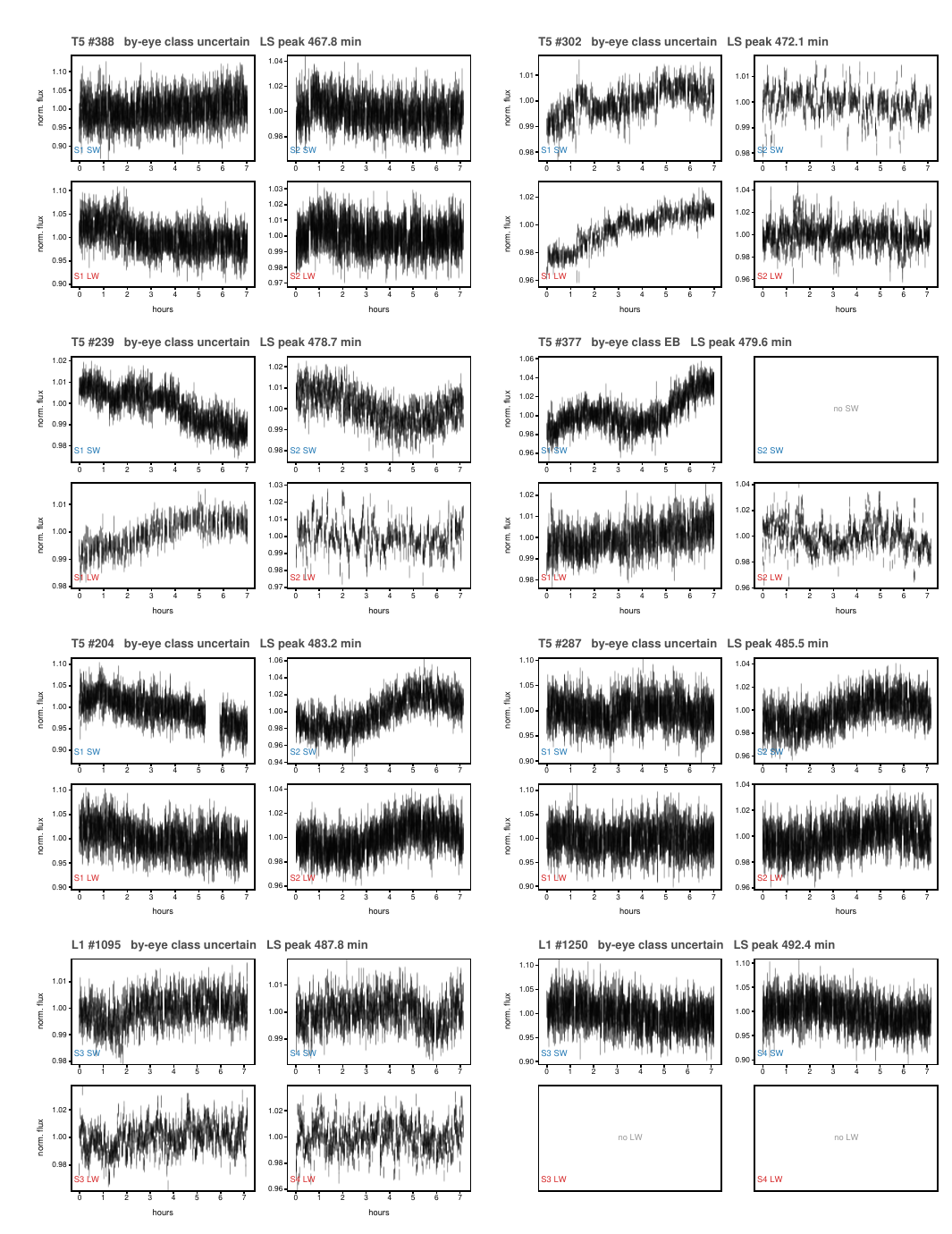}
\caption{Variables with no accepted model, continued (page 14 of 44).}
\end{figure*}
\clearpage

\begin{figure*}
\centering
\includegraphics[width=0.98\textwidth,height=0.94\textheight,keepaspectratio]{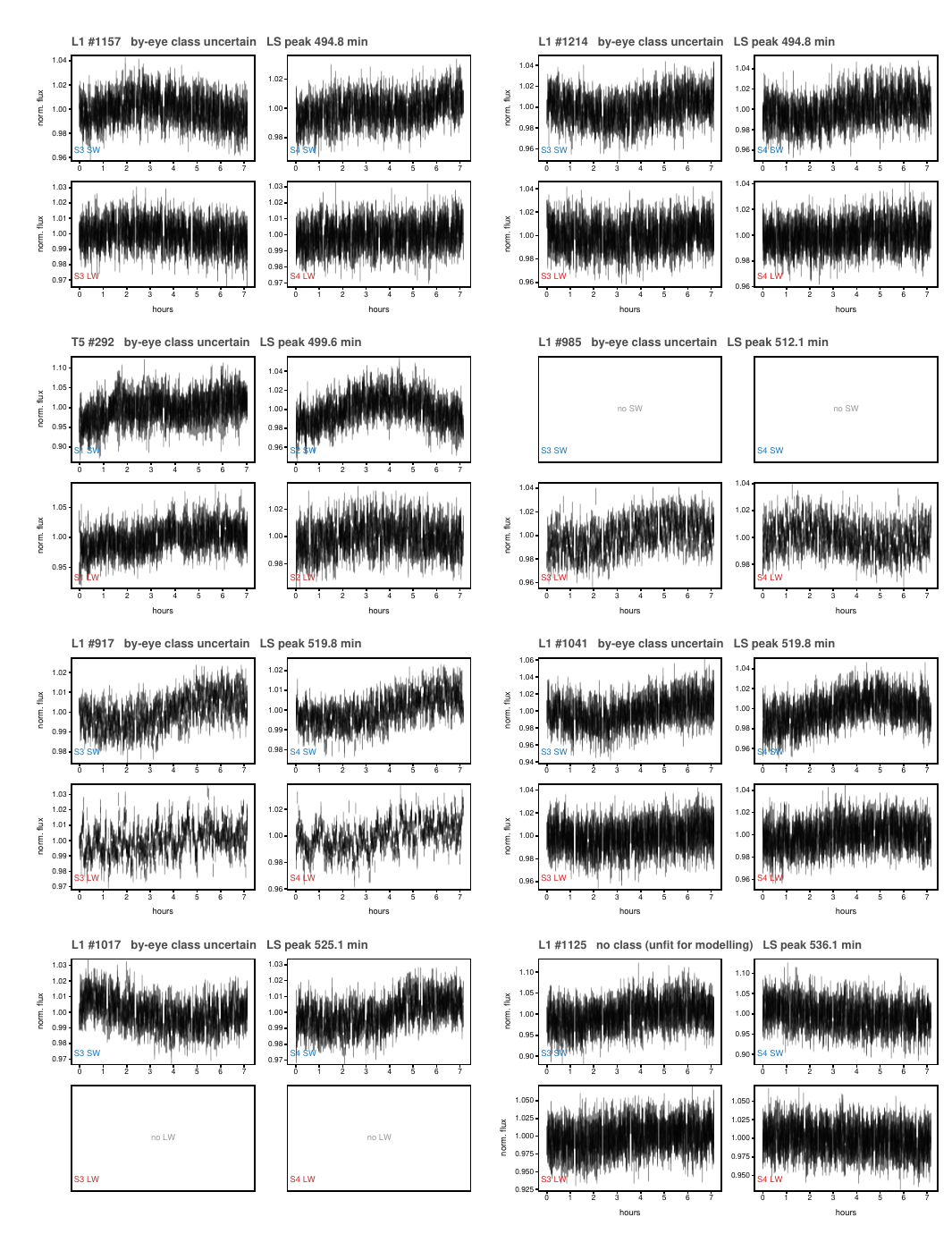}
\caption{Variables with no accepted model, continued (page 15 of 44).}
\end{figure*}
\clearpage

\begin{figure*}
\centering
\includegraphics[width=0.98\textwidth,height=0.94\textheight,keepaspectratio]{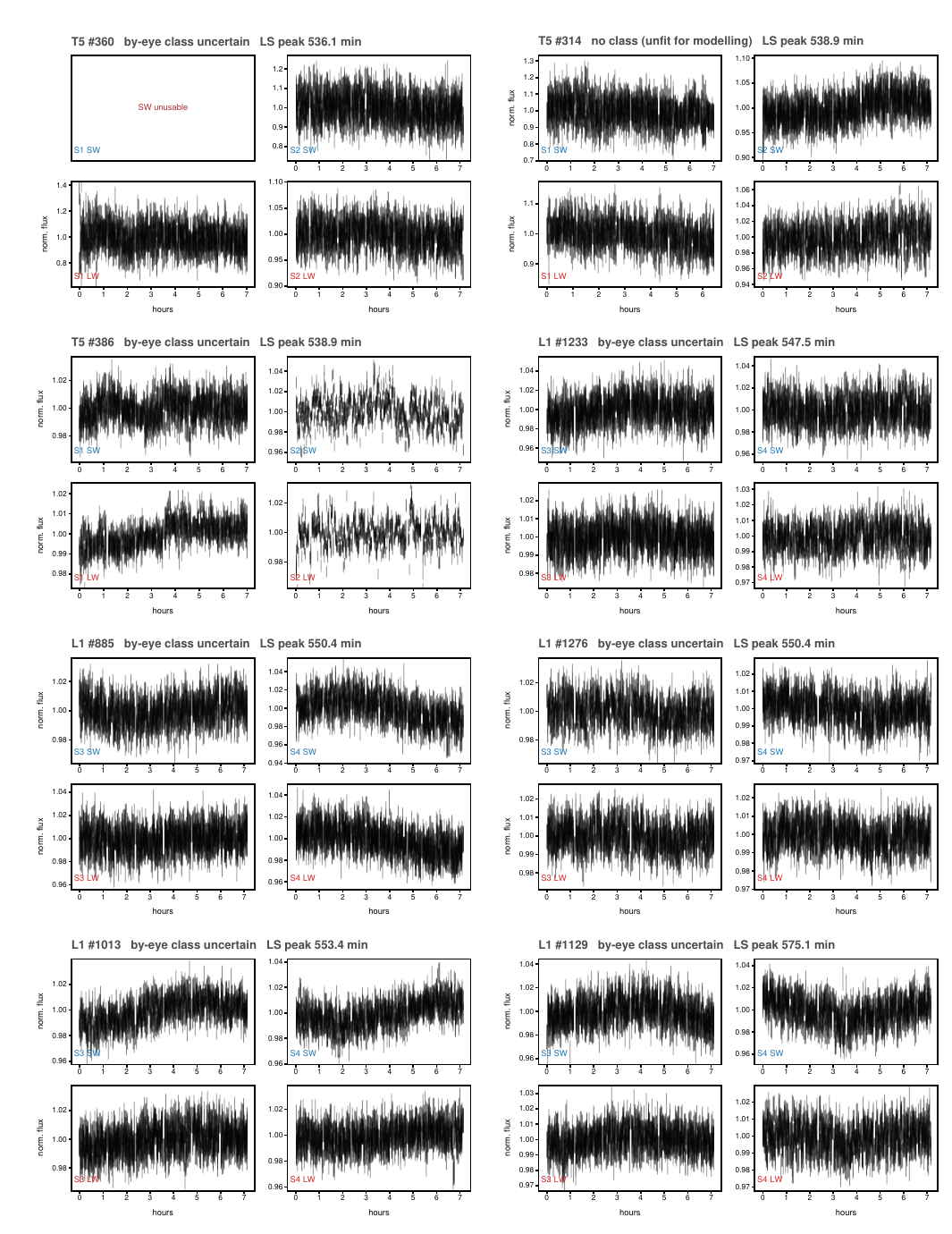}
\caption{Variables with no accepted model, continued (page 16 of 44).}
\end{figure*}
\clearpage

\begin{figure*}
\centering
\includegraphics[width=0.98\textwidth,height=0.94\textheight,keepaspectratio]{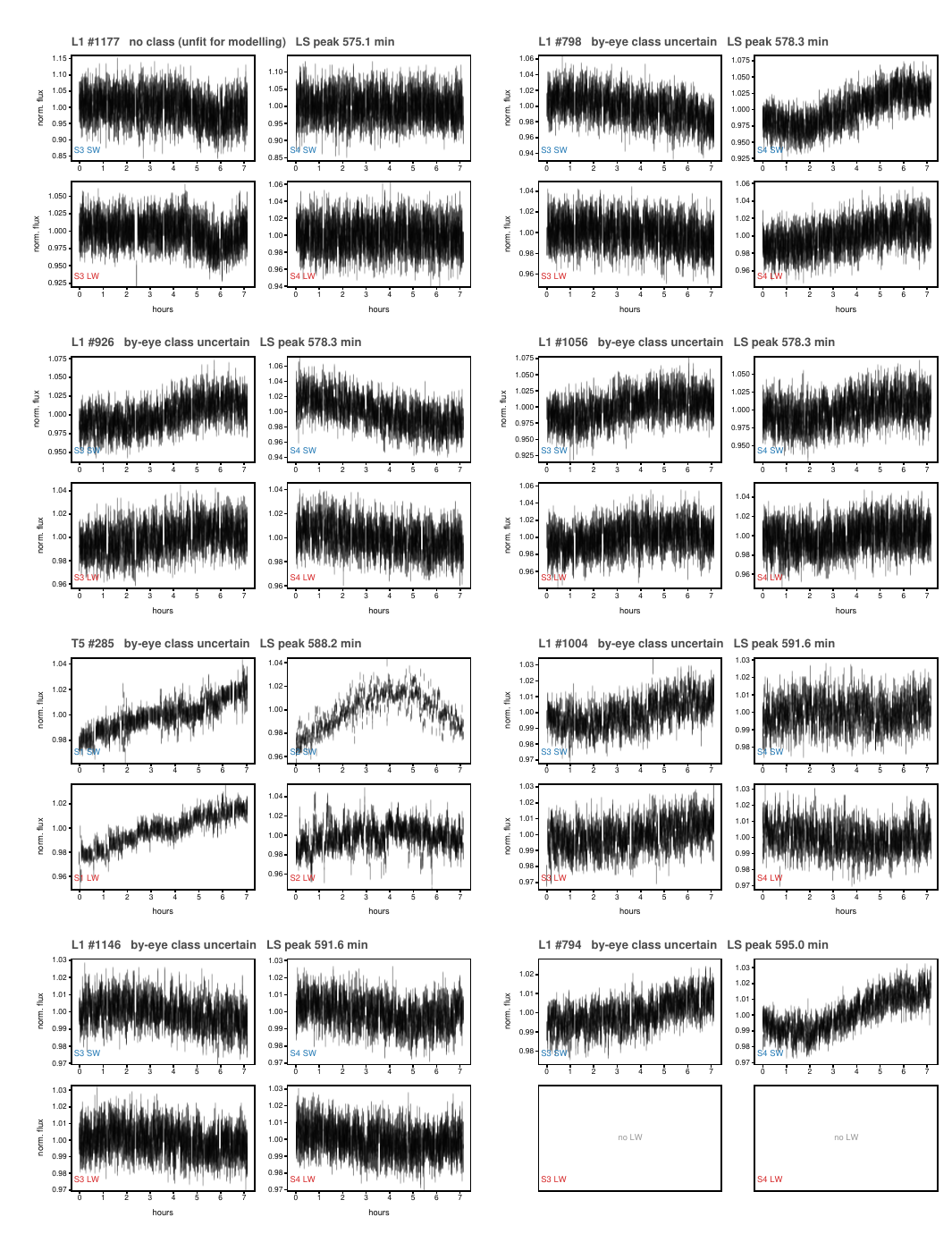}
\caption{Variables with no accepted model, continued (page 17 of 44).}
\end{figure*}
\clearpage

\begin{figure*}
\centering
\includegraphics[width=0.98\textwidth,height=0.94\textheight,keepaspectratio]{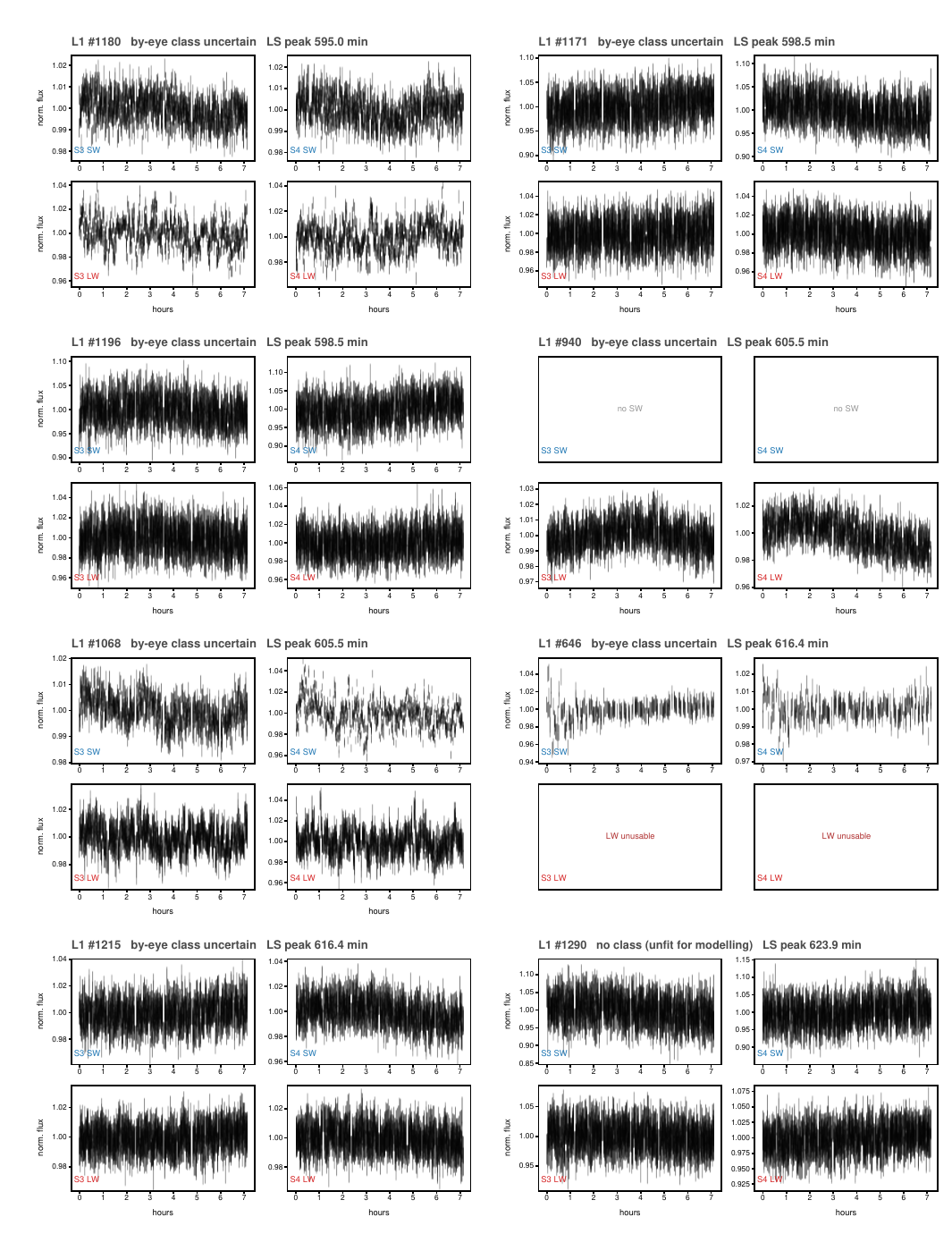}
\caption{Variables with no accepted model, continued (page 18 of 44).}
\end{figure*}
\clearpage

\begin{figure*}
\centering
\includegraphics[width=0.98\textwidth,height=0.94\textheight,keepaspectratio]{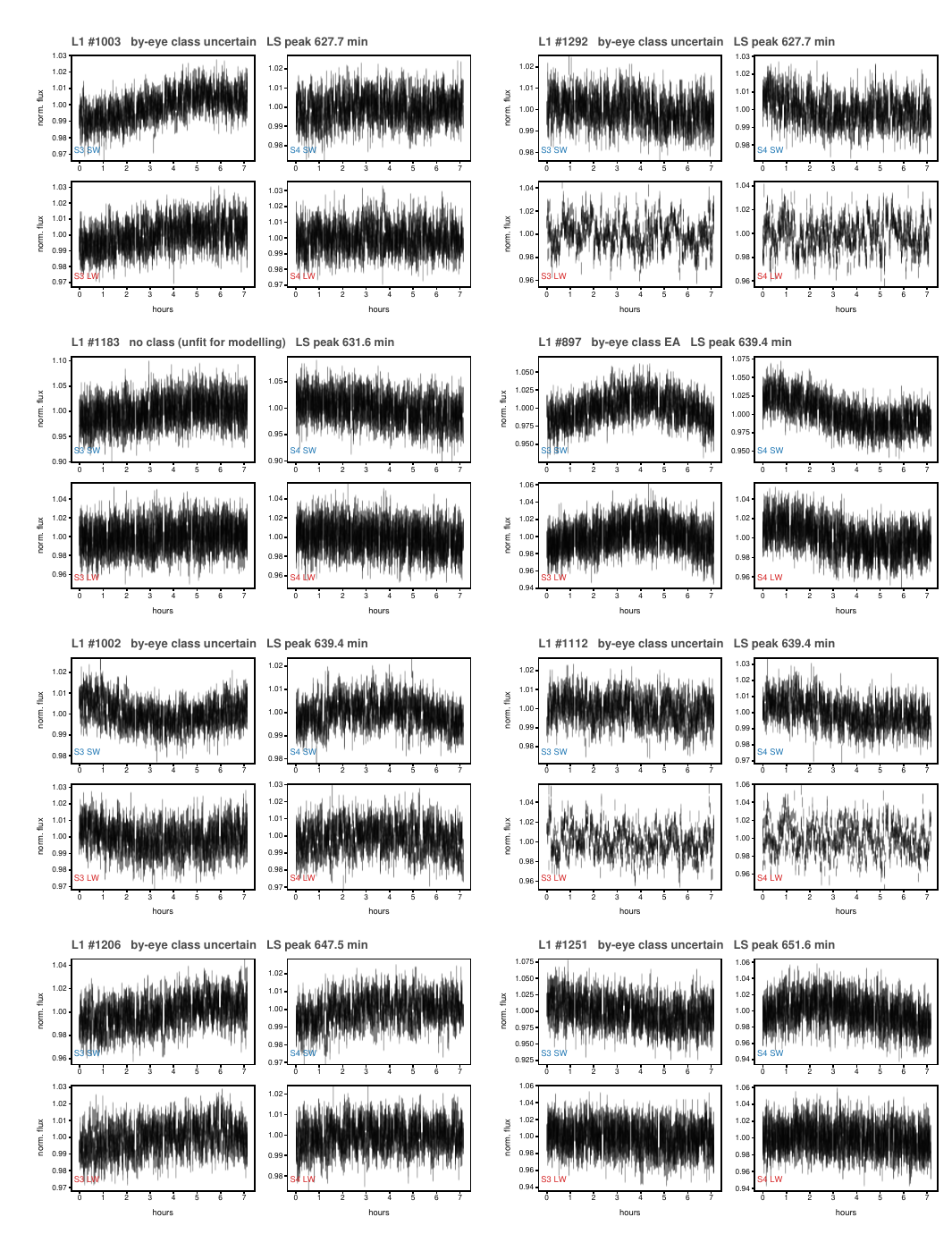}
\caption{Variables with no accepted model, continued (page 19 of 44).}
\end{figure*}
\clearpage

\begin{figure*}
\centering
\includegraphics[width=0.98\textwidth,height=0.94\textheight,keepaspectratio]{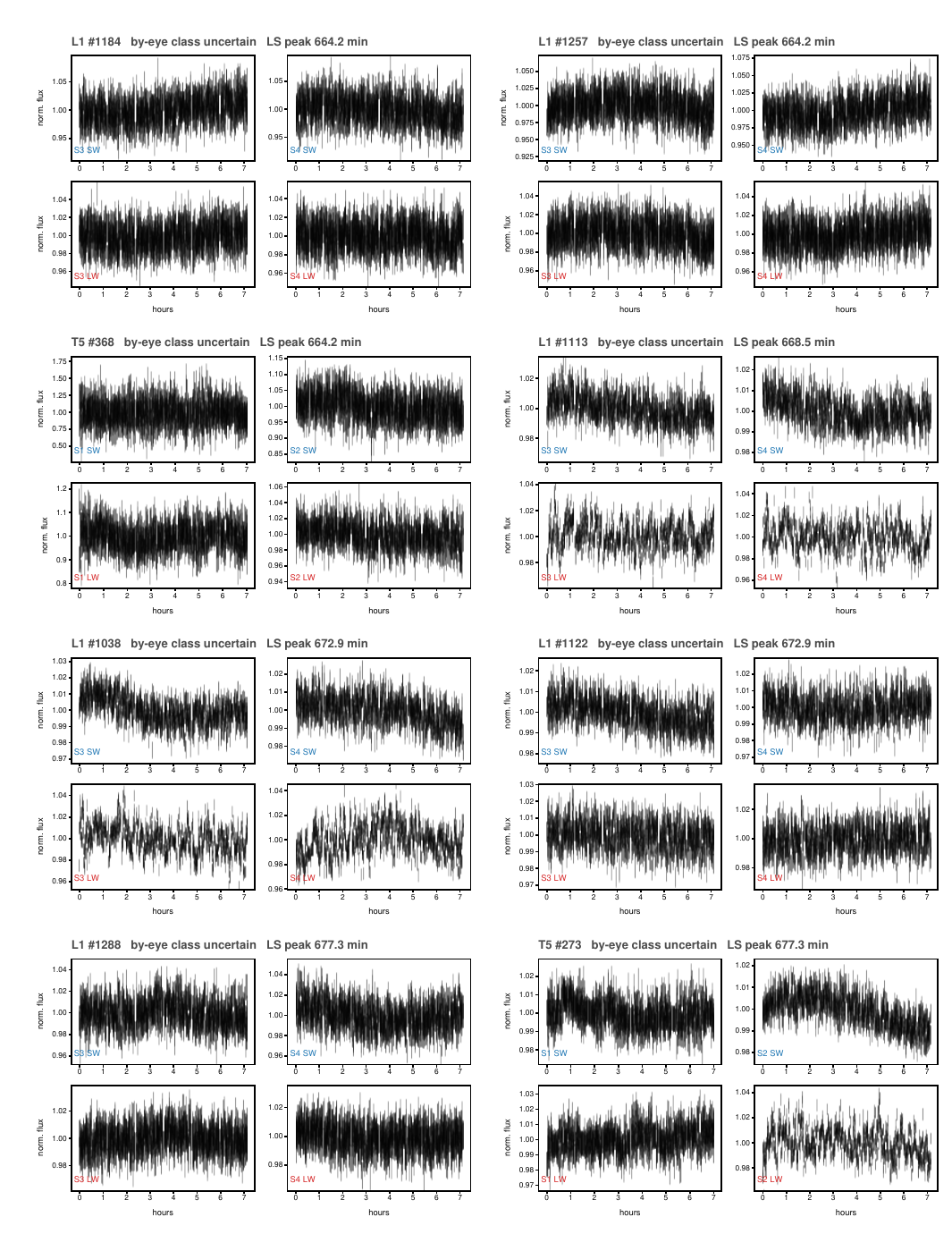}
\caption{Variables with no accepted model, continued (page 20 of 44).}
\end{figure*}
\clearpage

\begin{figure*}
\centering
\includegraphics[width=0.98\textwidth,height=0.94\textheight,keepaspectratio]{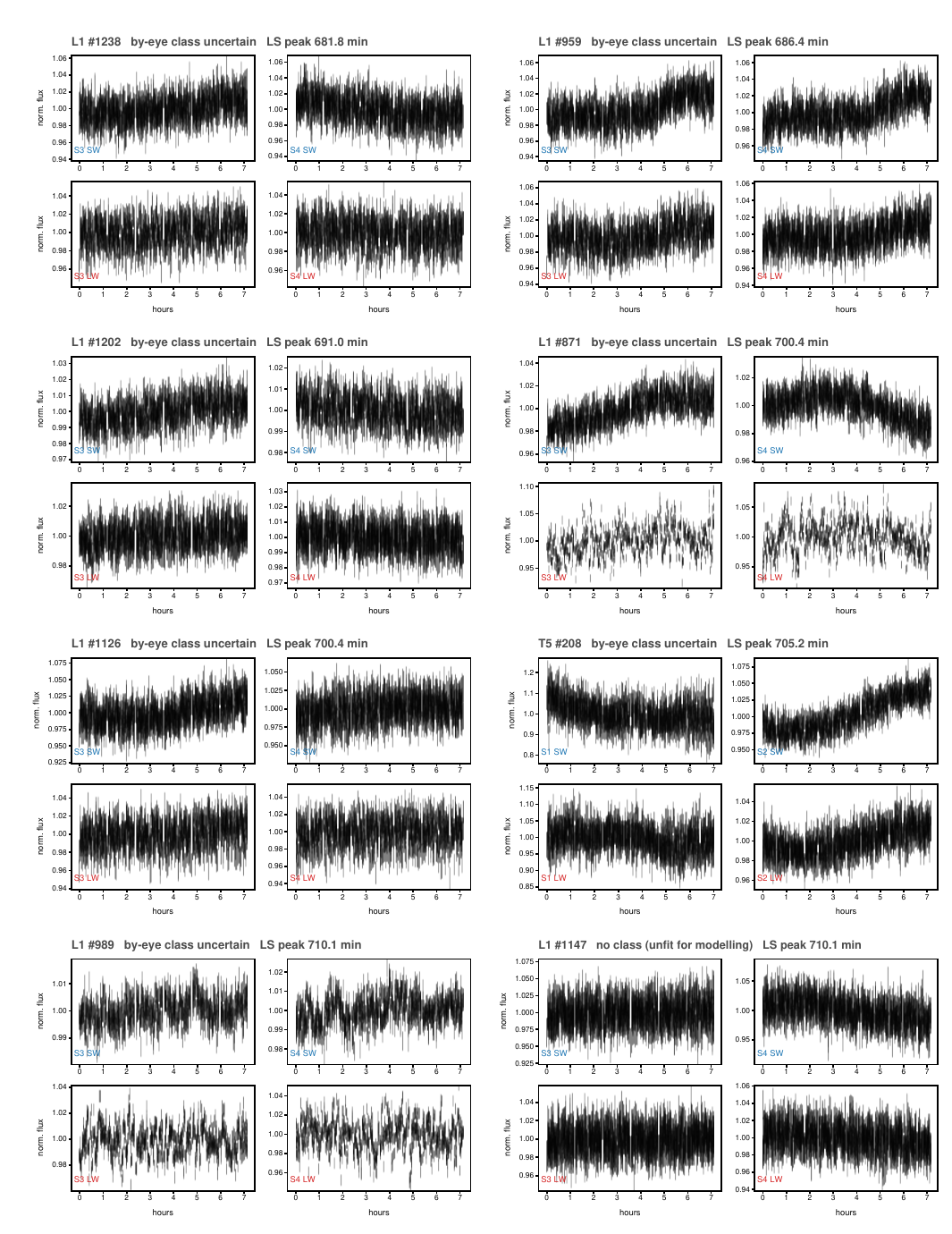}
\caption{Variables with no accepted model, continued (page 21 of 44).}
\end{figure*}
\clearpage

\begin{figure*}
\centering
\includegraphics[width=0.98\textwidth,height=0.94\textheight,keepaspectratio]{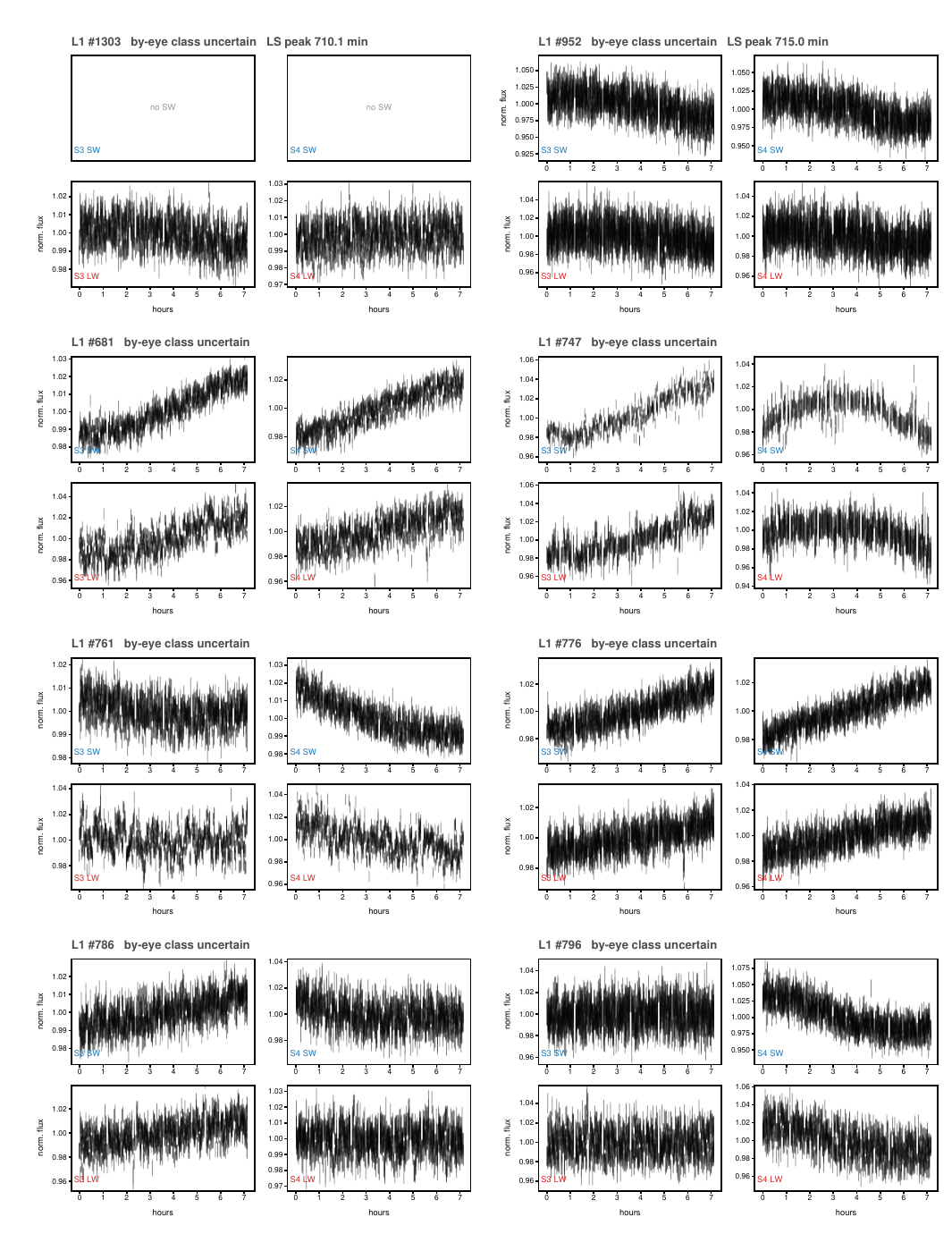}
\caption{Variables with no accepted model, continued (page 22 of 44).}
\end{figure*}
\clearpage

\begin{figure*}
\centering
\includegraphics[width=0.98\textwidth,height=0.94\textheight,keepaspectratio]{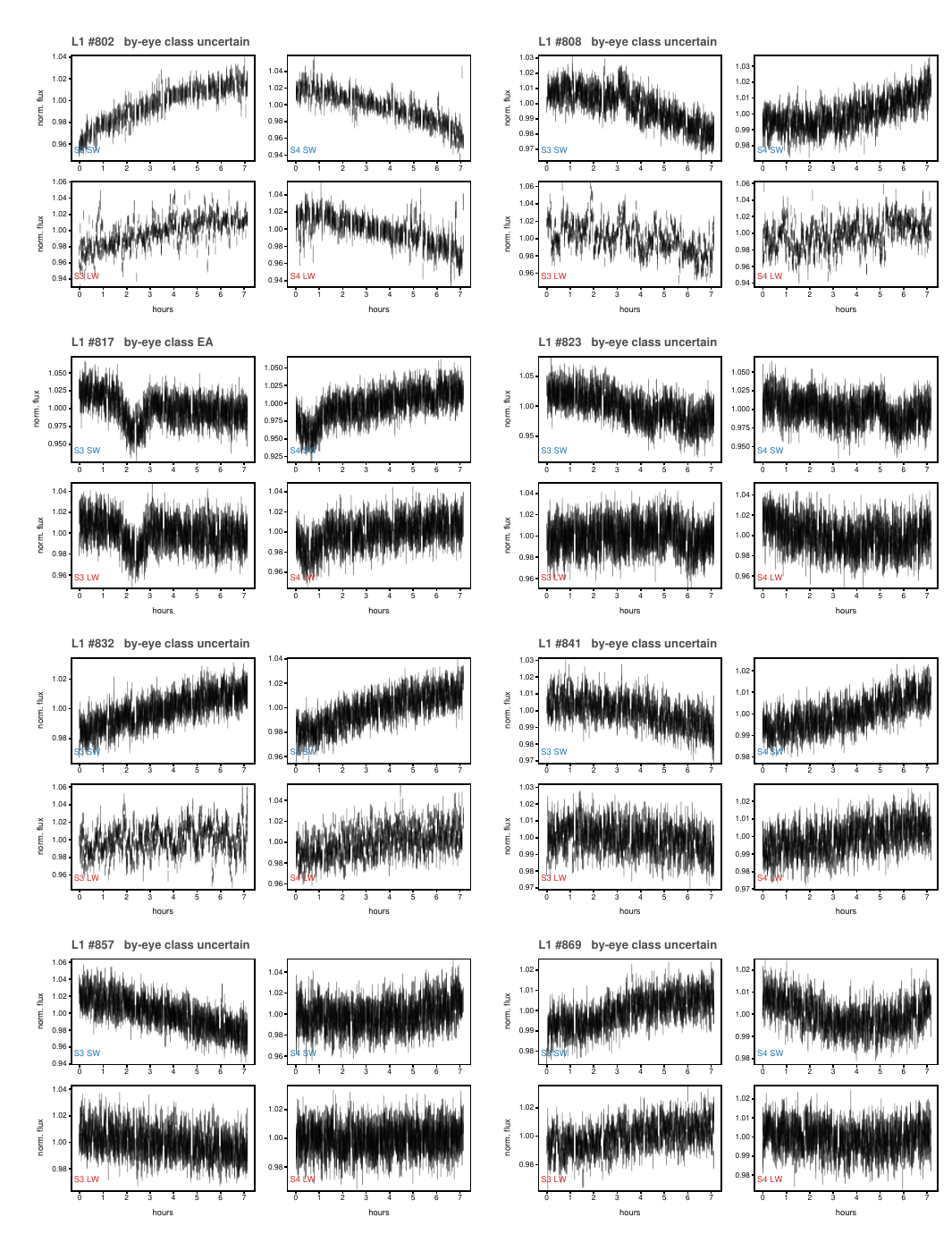}
\caption{Variables with no accepted model, continued (page 23 of 44).}
\end{figure*}
\clearpage

\begin{figure*}
\centering
\includegraphics[width=0.98\textwidth,height=0.94\textheight,keepaspectratio]{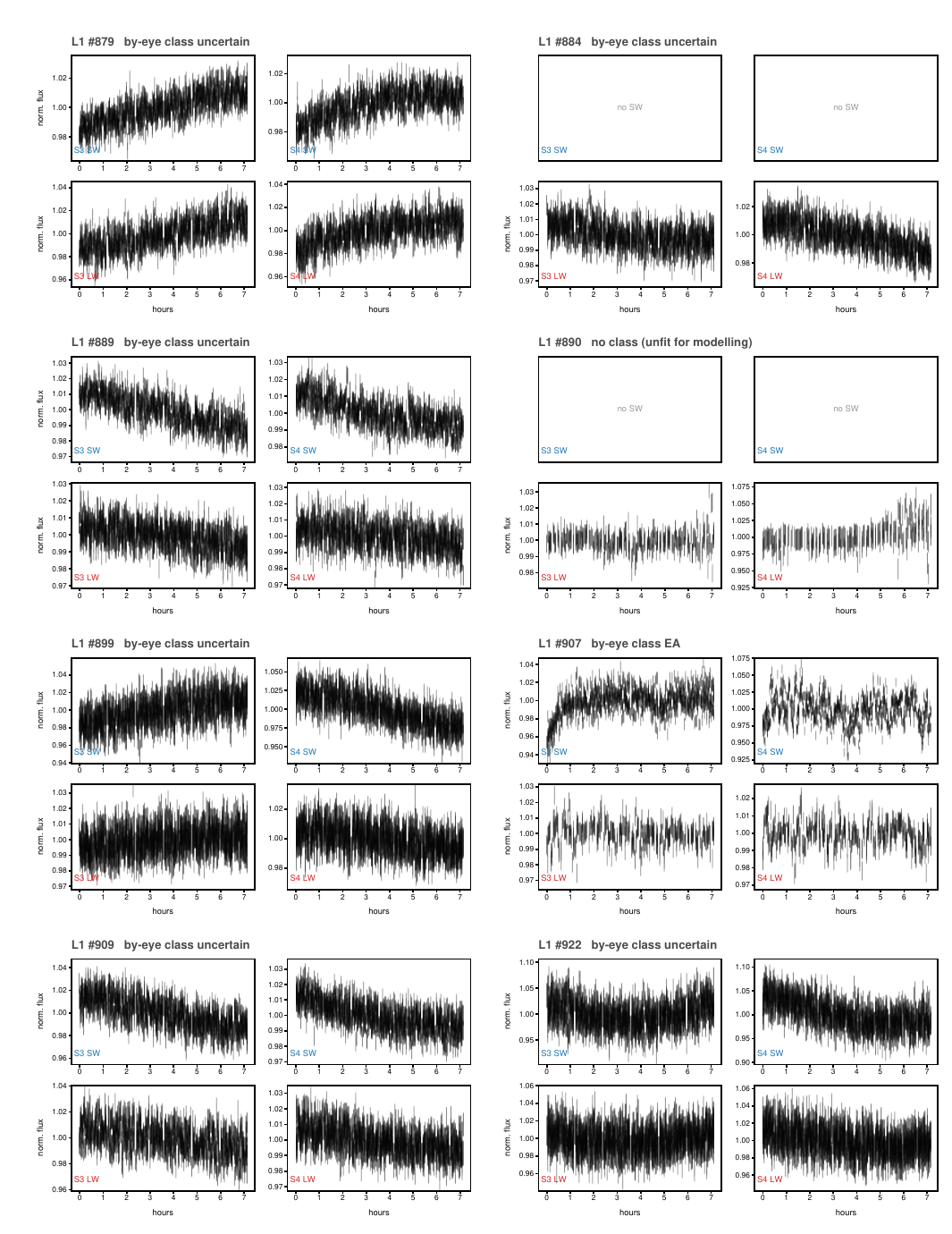}
\caption{Variables with no accepted model, continued (page 24 of 44).}
\end{figure*}
\clearpage

\begin{figure*}
\centering
\includegraphics[width=0.98\textwidth,height=0.94\textheight,keepaspectratio]{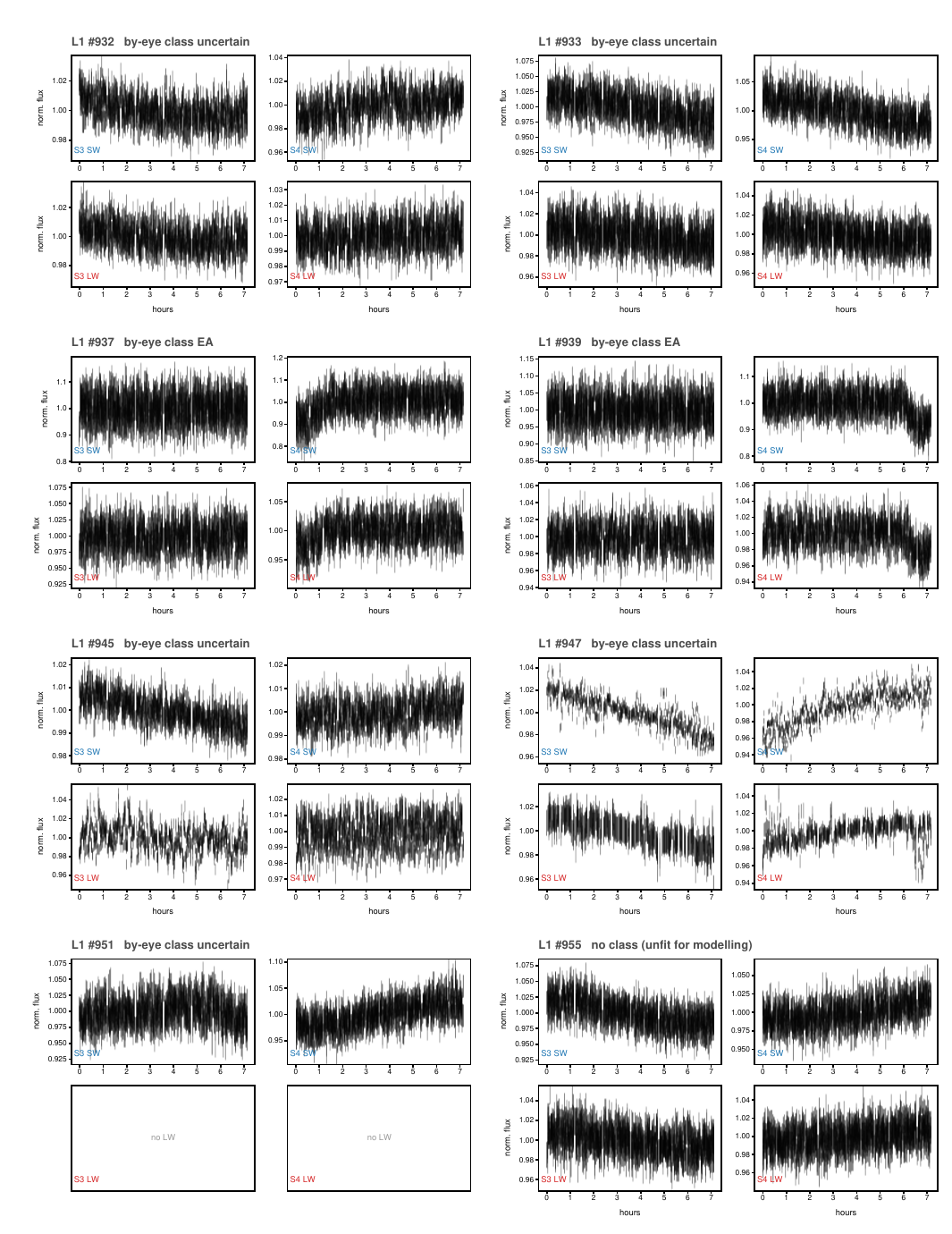}
\caption{Variables with no accepted model, continued (page 25 of 44).}
\end{figure*}
\clearpage

\begin{figure*}
\centering
\includegraphics[width=0.98\textwidth,height=0.94\textheight,keepaspectratio]{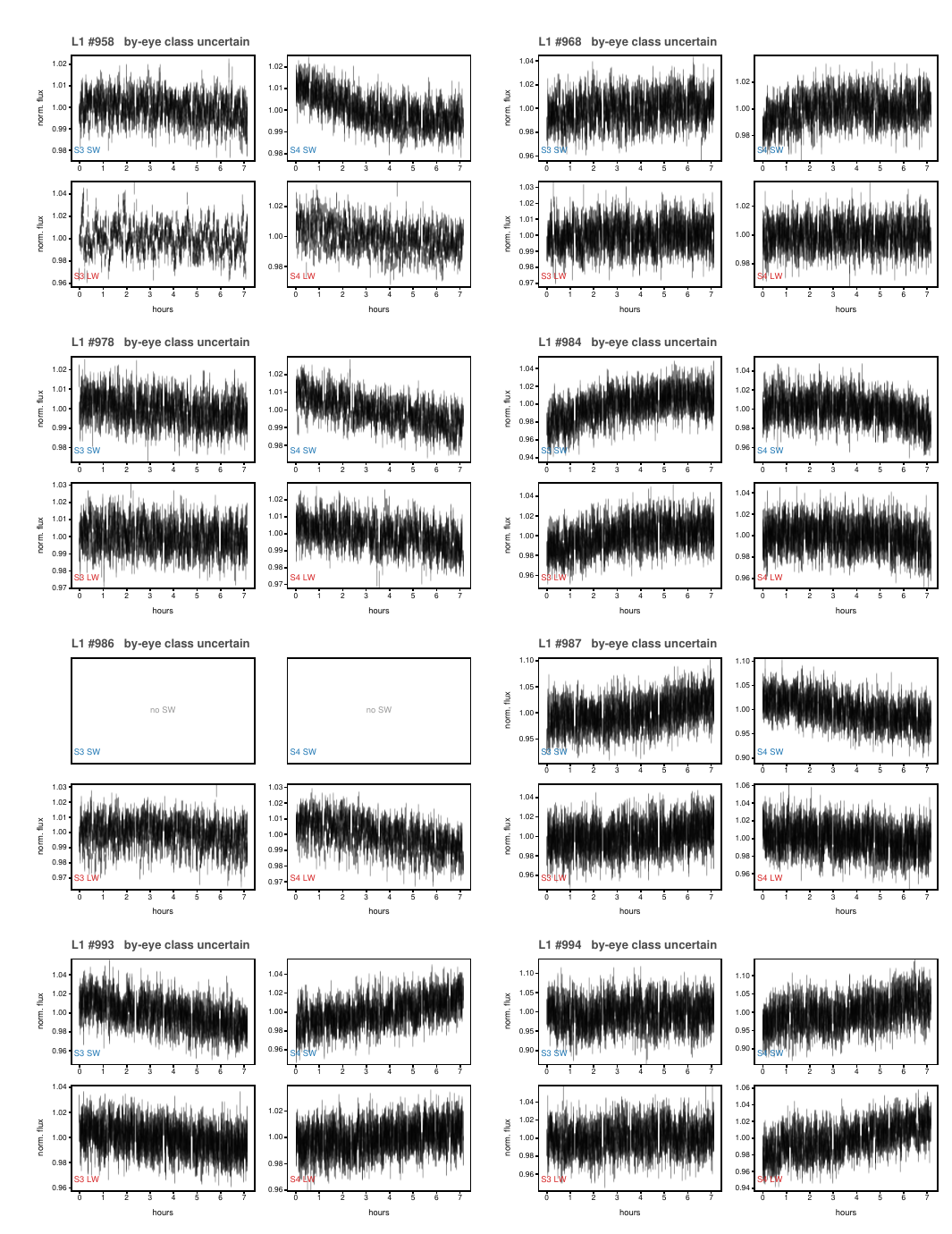}
\caption{Variables with no accepted model, continued (page 26 of 44).}
\end{figure*}
\clearpage

\begin{figure*}
\centering
\includegraphics[width=0.98\textwidth,height=0.94\textheight,keepaspectratio]{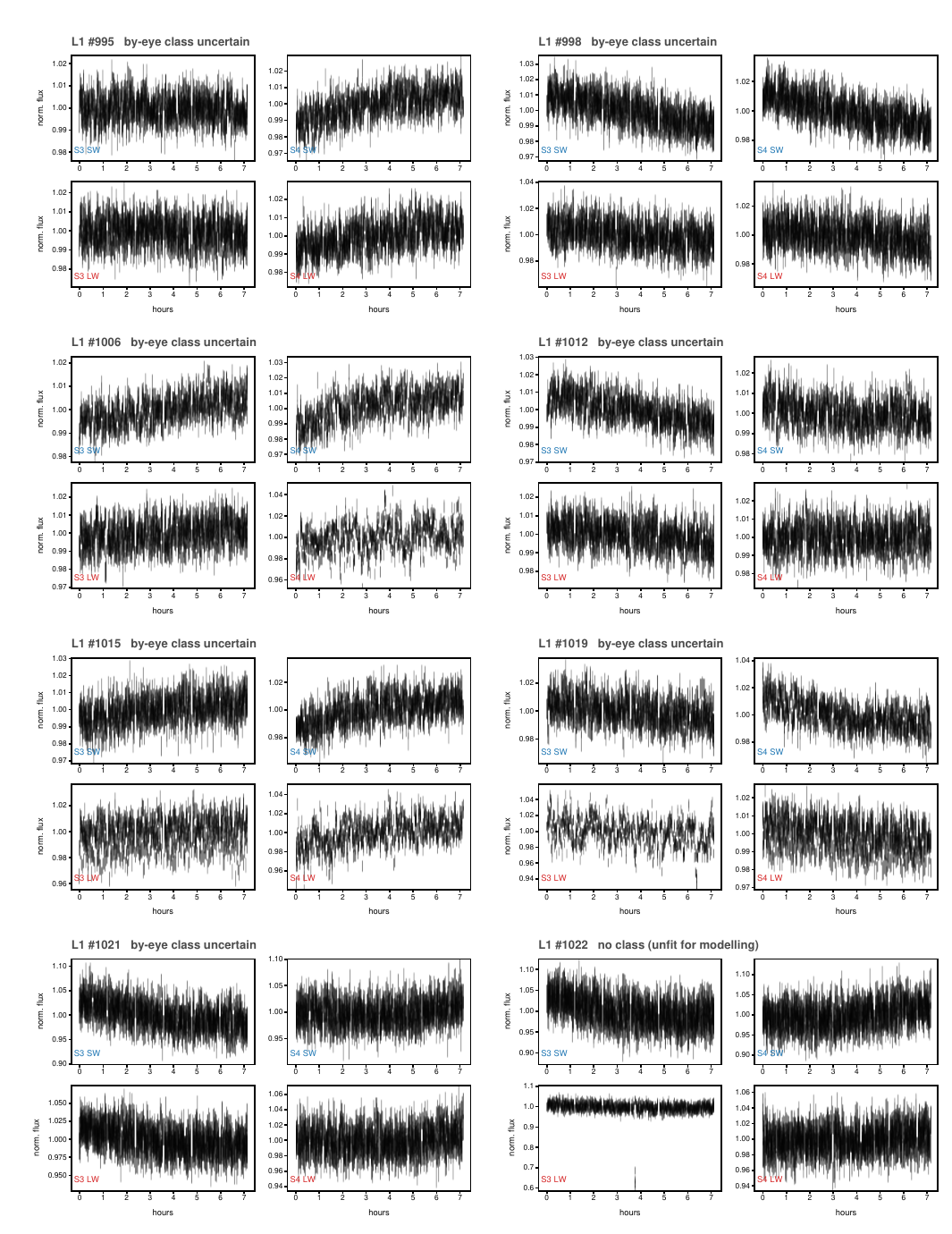}
\caption{Variables with no accepted model, continued (page 27 of 44).}
\end{figure*}
\clearpage

\begin{figure*}
\centering
\includegraphics[width=0.98\textwidth,height=0.94\textheight,keepaspectratio]{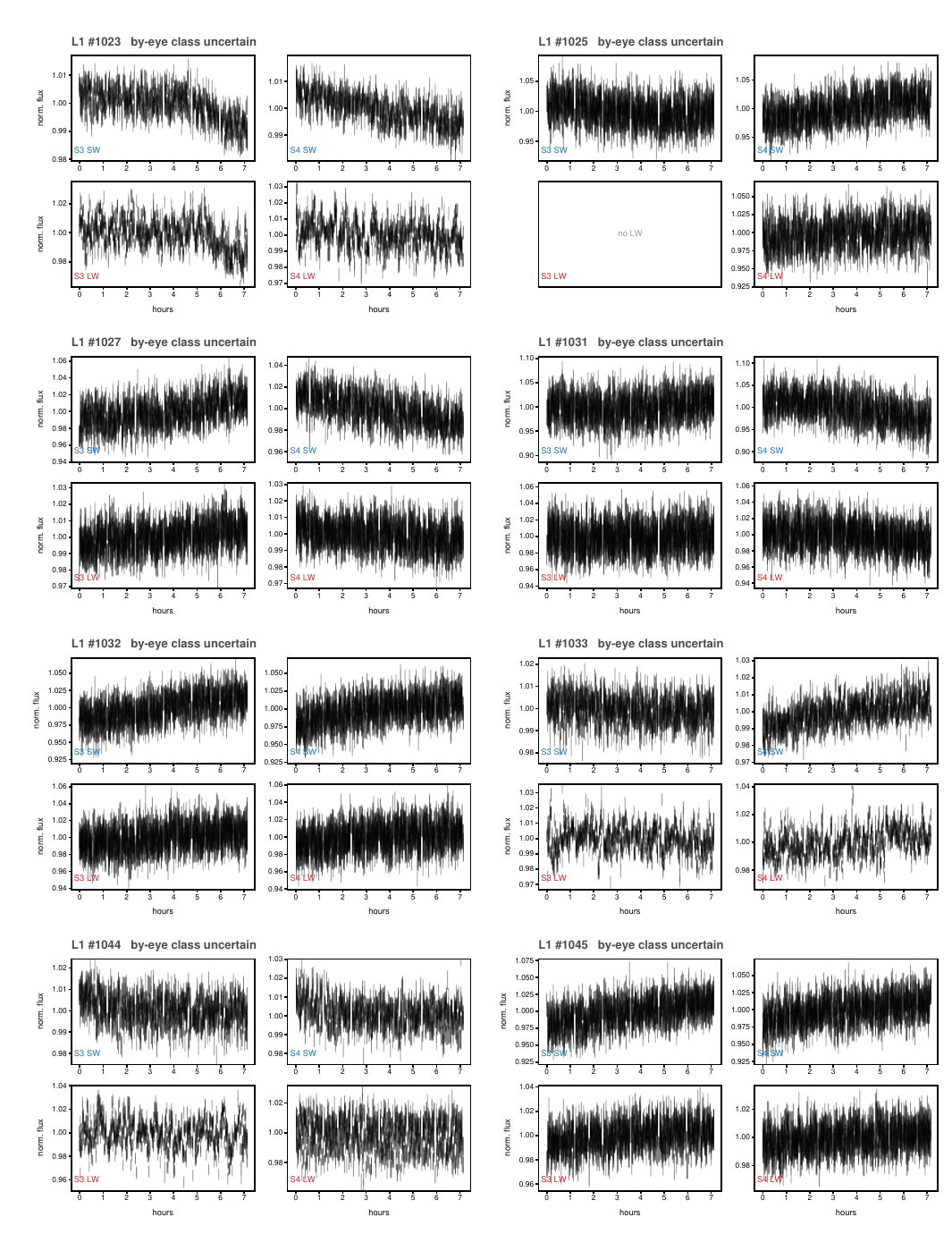}
\caption{Variables with no accepted model, continued (page 28 of 44).}
\end{figure*}
\clearpage

\begin{figure*}
\centering
\includegraphics[width=0.98\textwidth,height=0.94\textheight,keepaspectratio]{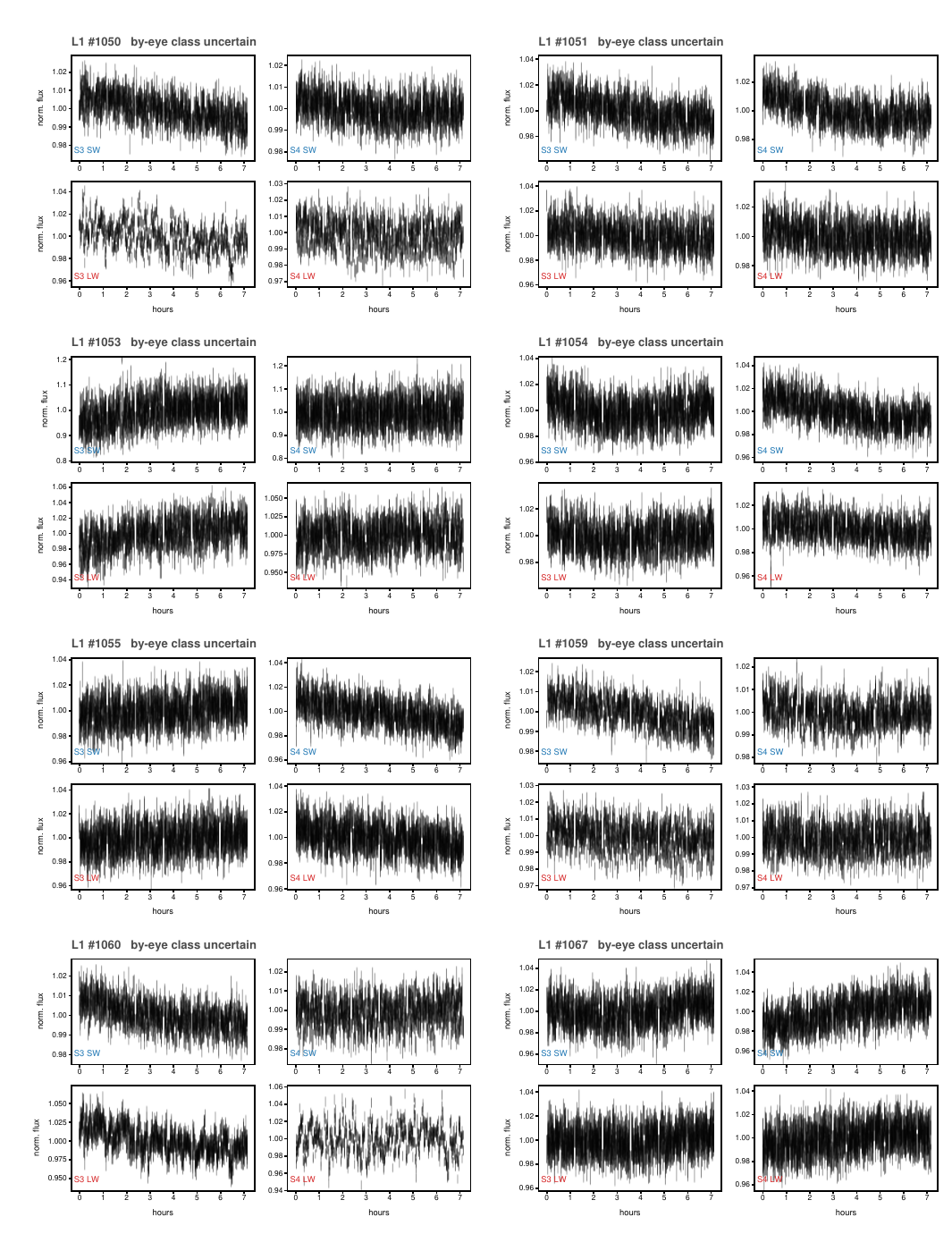}
\caption{Variables with no accepted model, continued (page 29 of 44).}
\end{figure*}
\clearpage

\begin{figure*}
\centering
\includegraphics[width=0.98\textwidth,height=0.94\textheight,keepaspectratio]{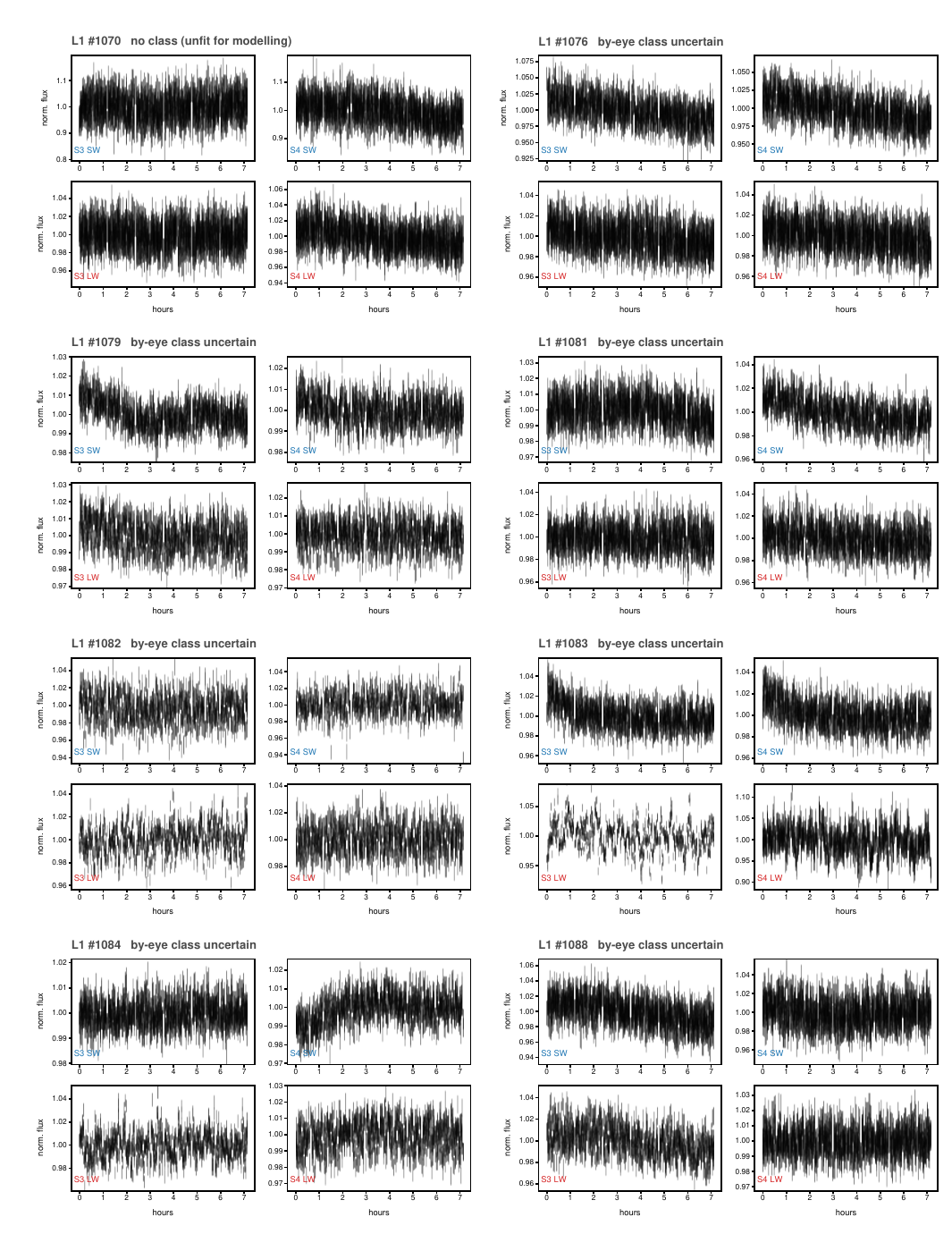}
\caption{Variables with no accepted model, continued (page 30 of 44).}
\end{figure*}
\clearpage

\begin{figure*}
\centering
\includegraphics[width=0.98\textwidth,height=0.94\textheight,keepaspectratio]{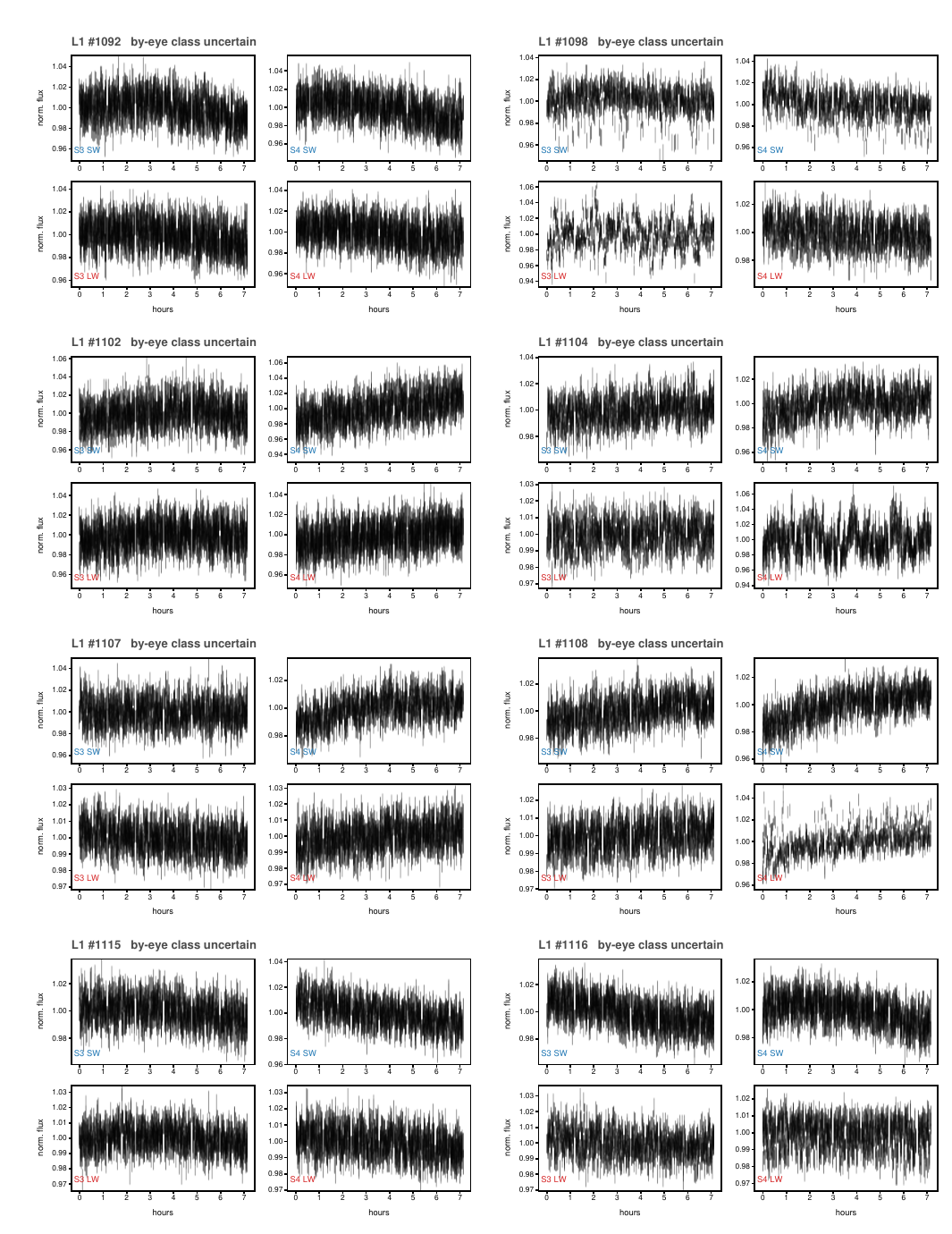}
\caption{Variables with no accepted model, continued (page 31 of 44).}
\end{figure*}
\clearpage

\begin{figure*}
\centering
\includegraphics[width=0.98\textwidth,height=0.94\textheight,keepaspectratio]{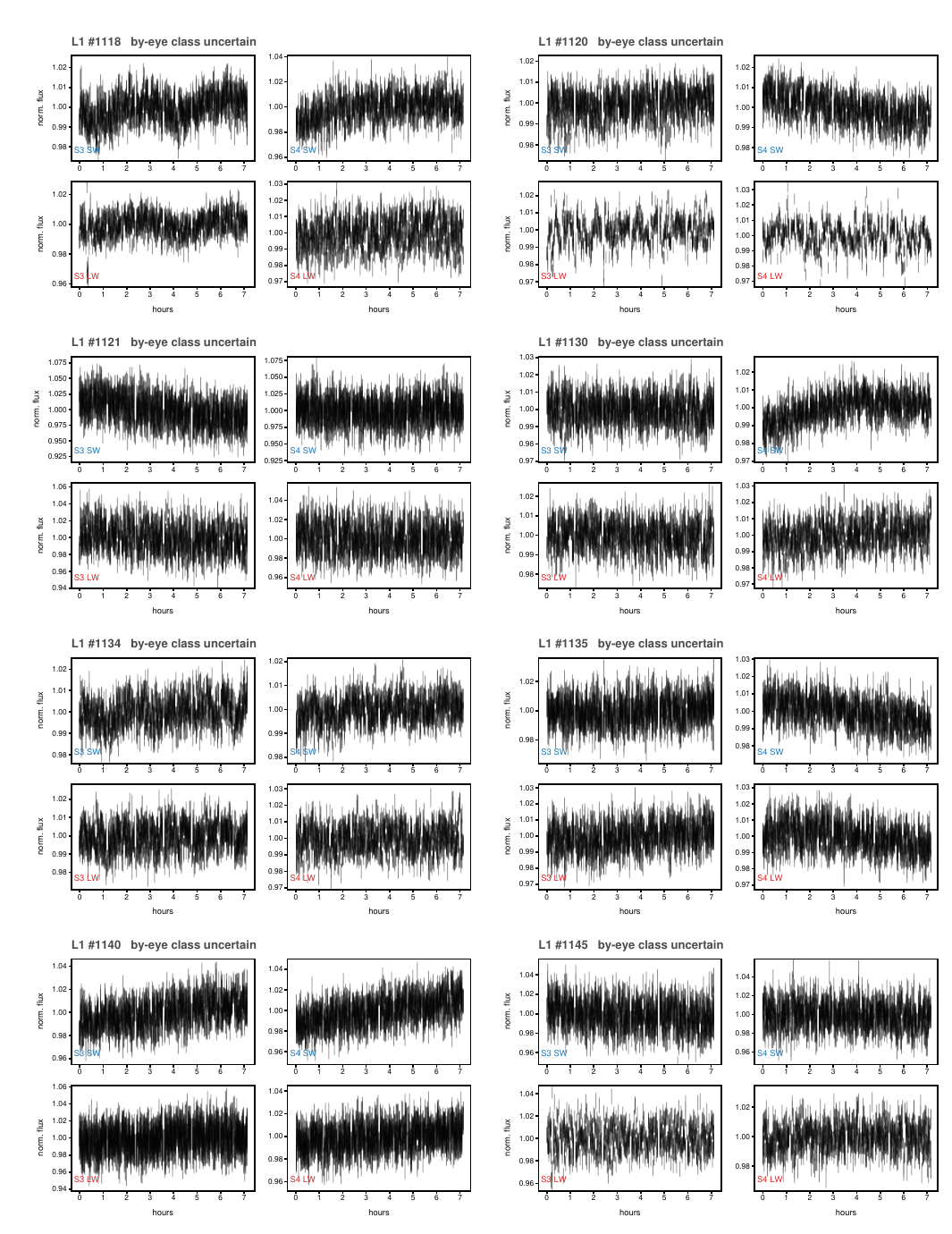}
\caption{Variables with no accepted model, continued (page 32 of 44).}
\end{figure*}
\clearpage

\begin{figure*}
\centering
\includegraphics[width=0.98\textwidth,height=0.94\textheight,keepaspectratio]{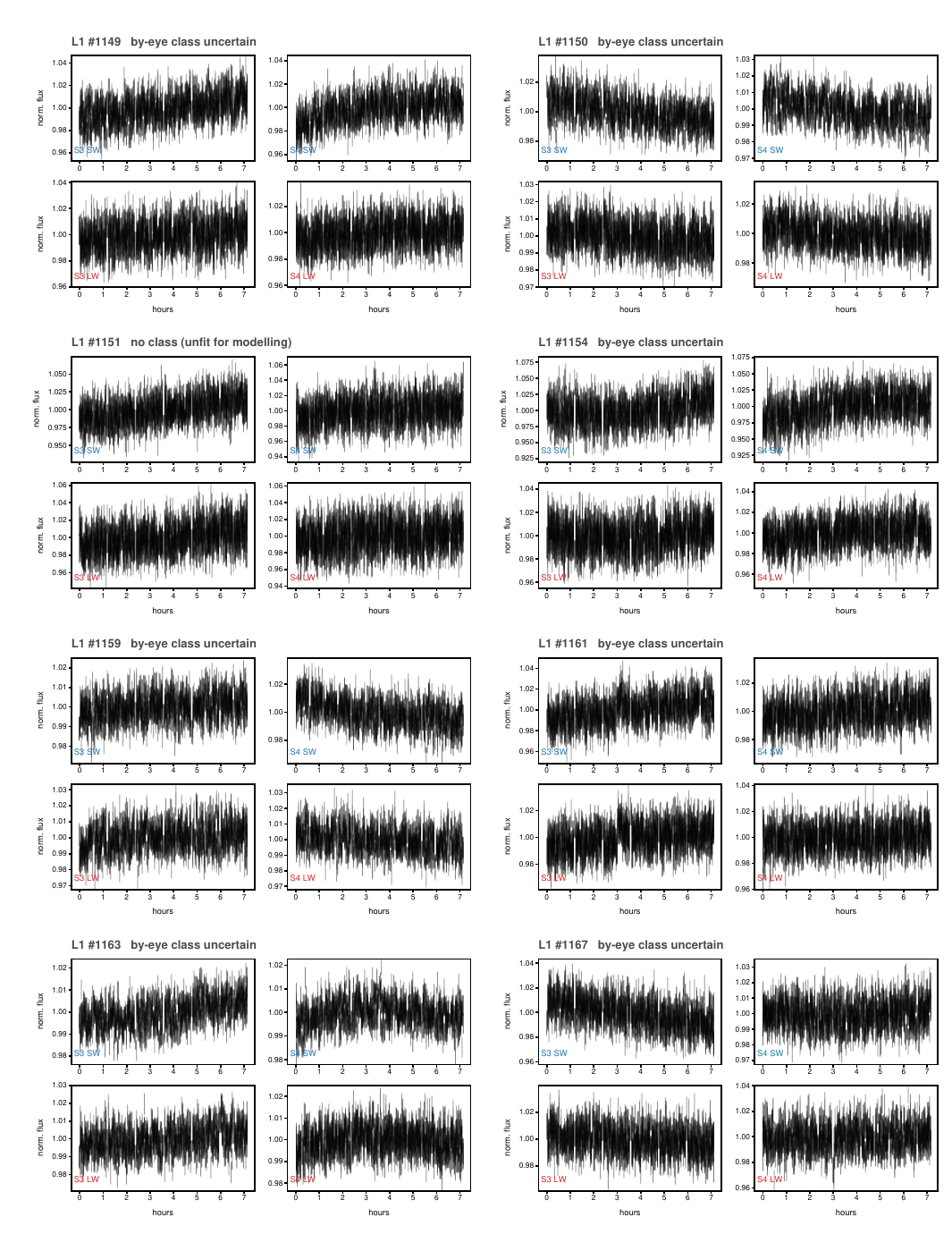}
\caption{Variables with no accepted model, continued (page 33 of 44).}
\end{figure*}
\clearpage

\begin{figure*}
\centering
\includegraphics[width=0.98\textwidth,height=0.94\textheight,keepaspectratio]{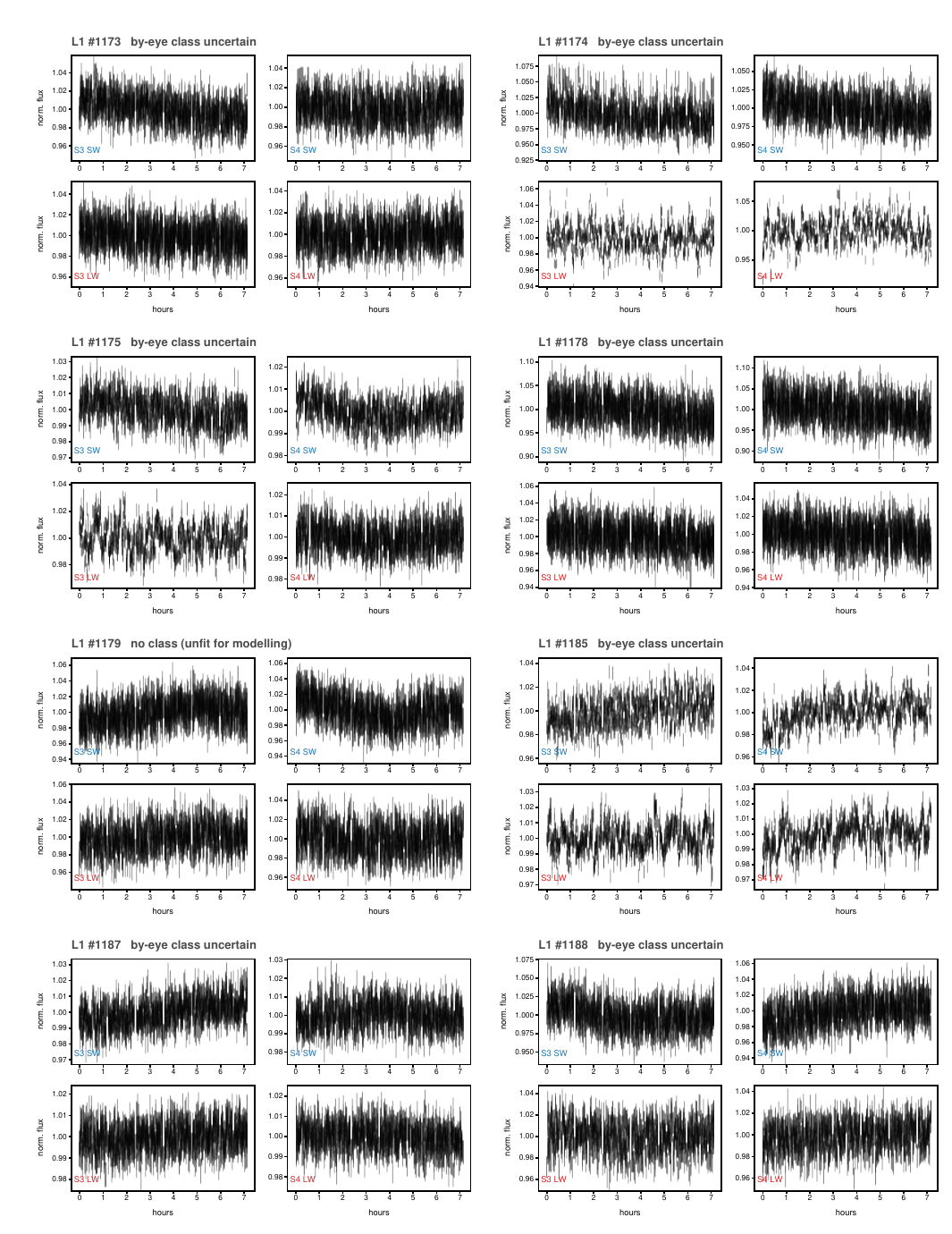}
\caption{Variables with no accepted model, continued (page 34 of 44).}
\end{figure*}
\clearpage

\begin{figure*}
\centering
\includegraphics[width=0.98\textwidth,height=0.94\textheight,keepaspectratio]{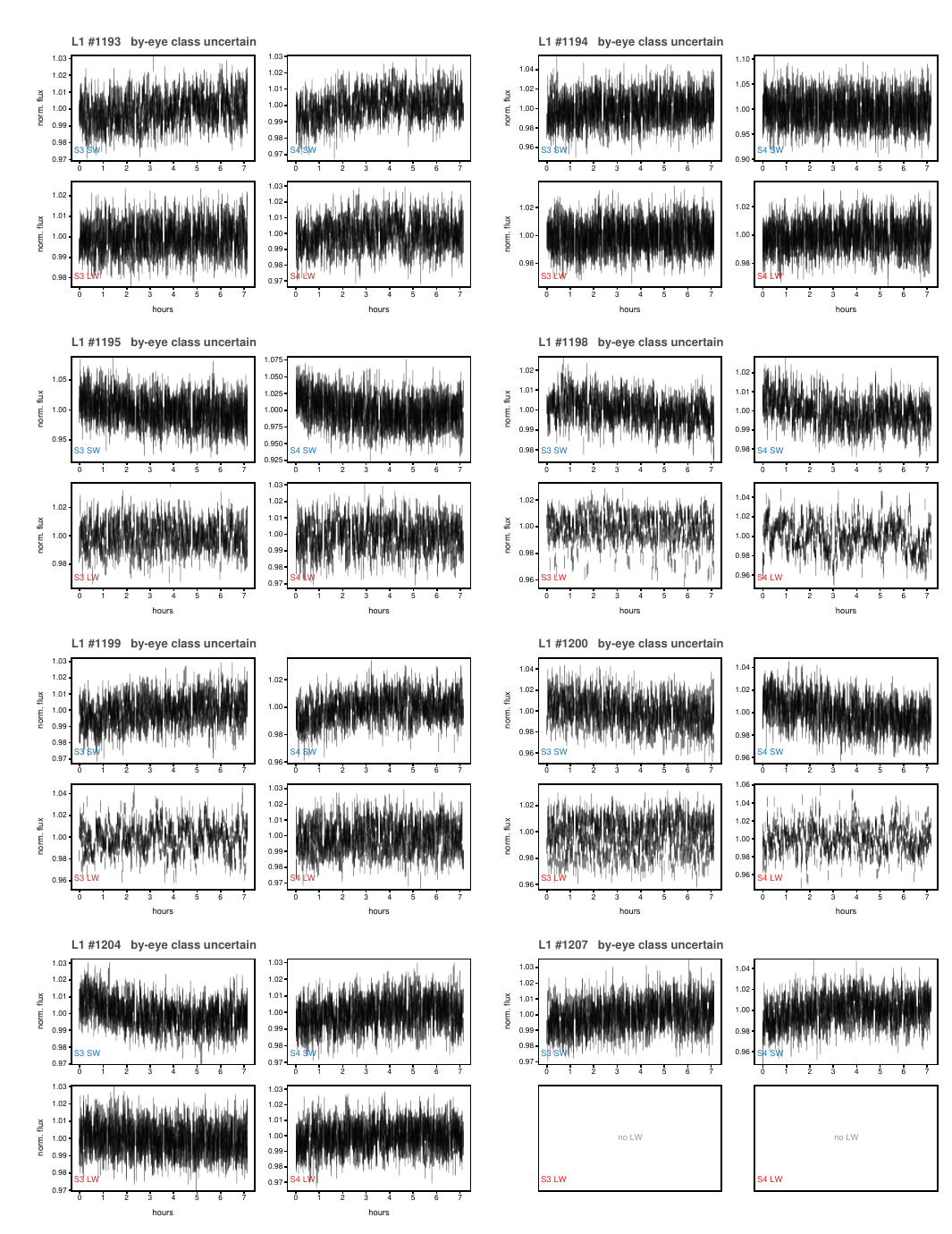}
\caption{Variables with no accepted model, continued (page 35 of 44).}
\end{figure*}
\clearpage

\begin{figure*}
\centering
\includegraphics[width=0.98\textwidth,height=0.94\textheight,keepaspectratio]{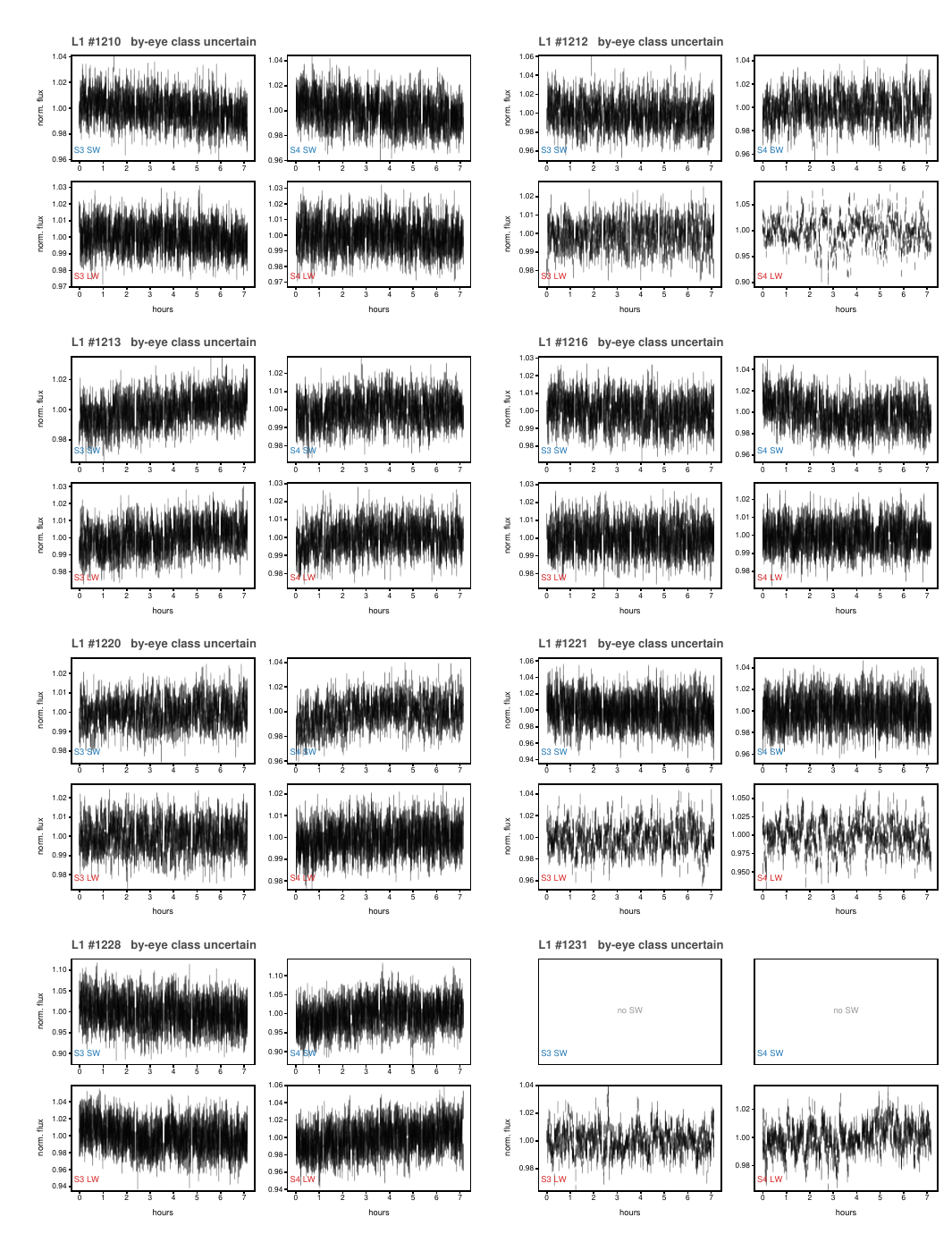}
\caption{Variables with no accepted model, continued (page 36 of 44).}
\end{figure*}
\clearpage

\begin{figure*}
\centering
\includegraphics[width=0.98\textwidth,height=0.94\textheight,keepaspectratio]{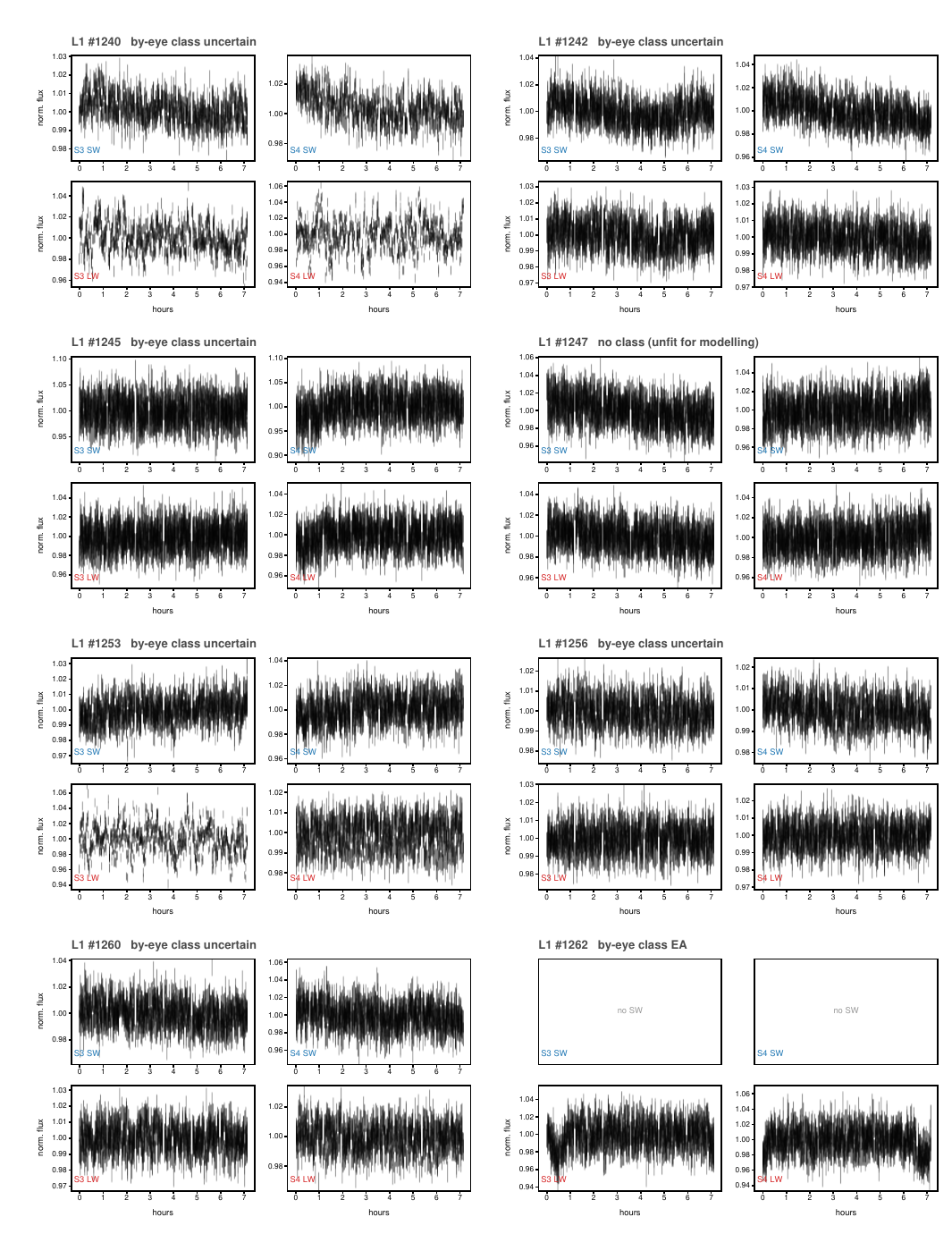}
\caption{Variables with no accepted model, continued (page 37 of 44).}
\end{figure*}
\clearpage

\begin{figure*}
\centering
\includegraphics[width=0.98\textwidth,height=0.94\textheight,keepaspectratio]{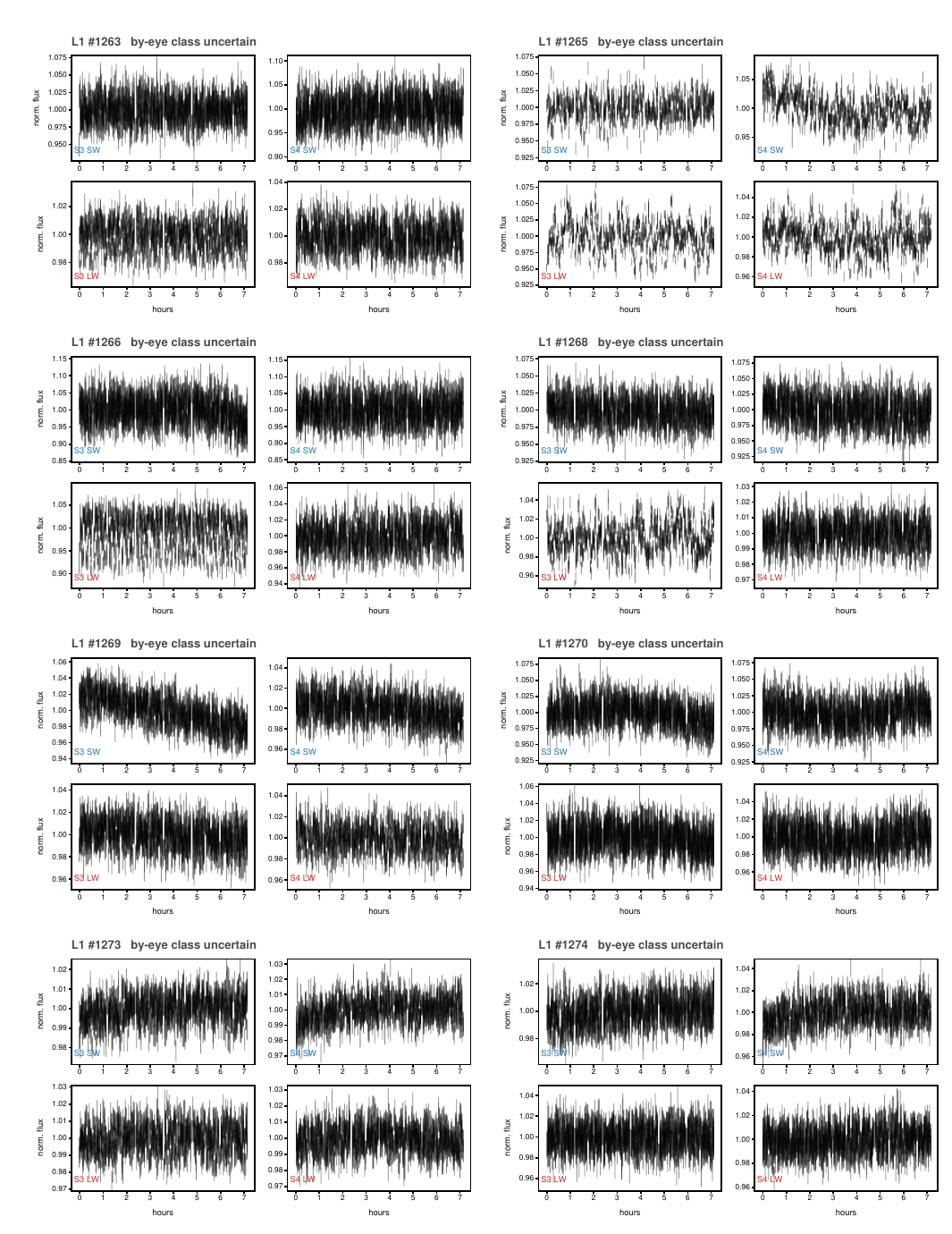}
\caption{Variables with no accepted model, continued (page 38 of 44).}
\end{figure*}
\clearpage

\begin{figure*}
\centering
\includegraphics[width=0.98\textwidth,height=0.94\textheight,keepaspectratio]{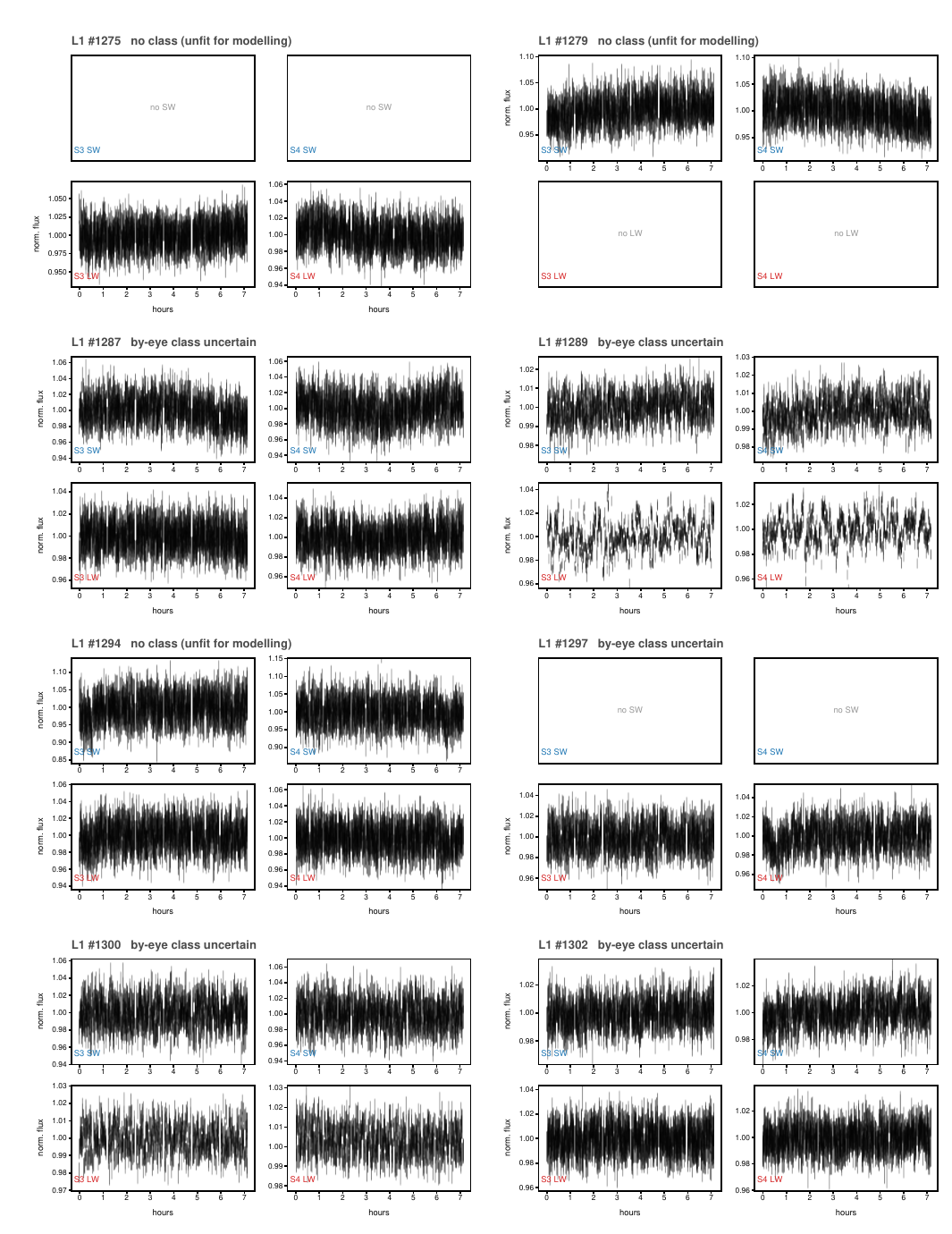}
\caption{Variables with no accepted model, continued (page 39 of 44).}
\end{figure*}
\clearpage

\begin{figure*}
\centering
\includegraphics[width=0.98\textwidth,height=0.94\textheight,keepaspectratio]{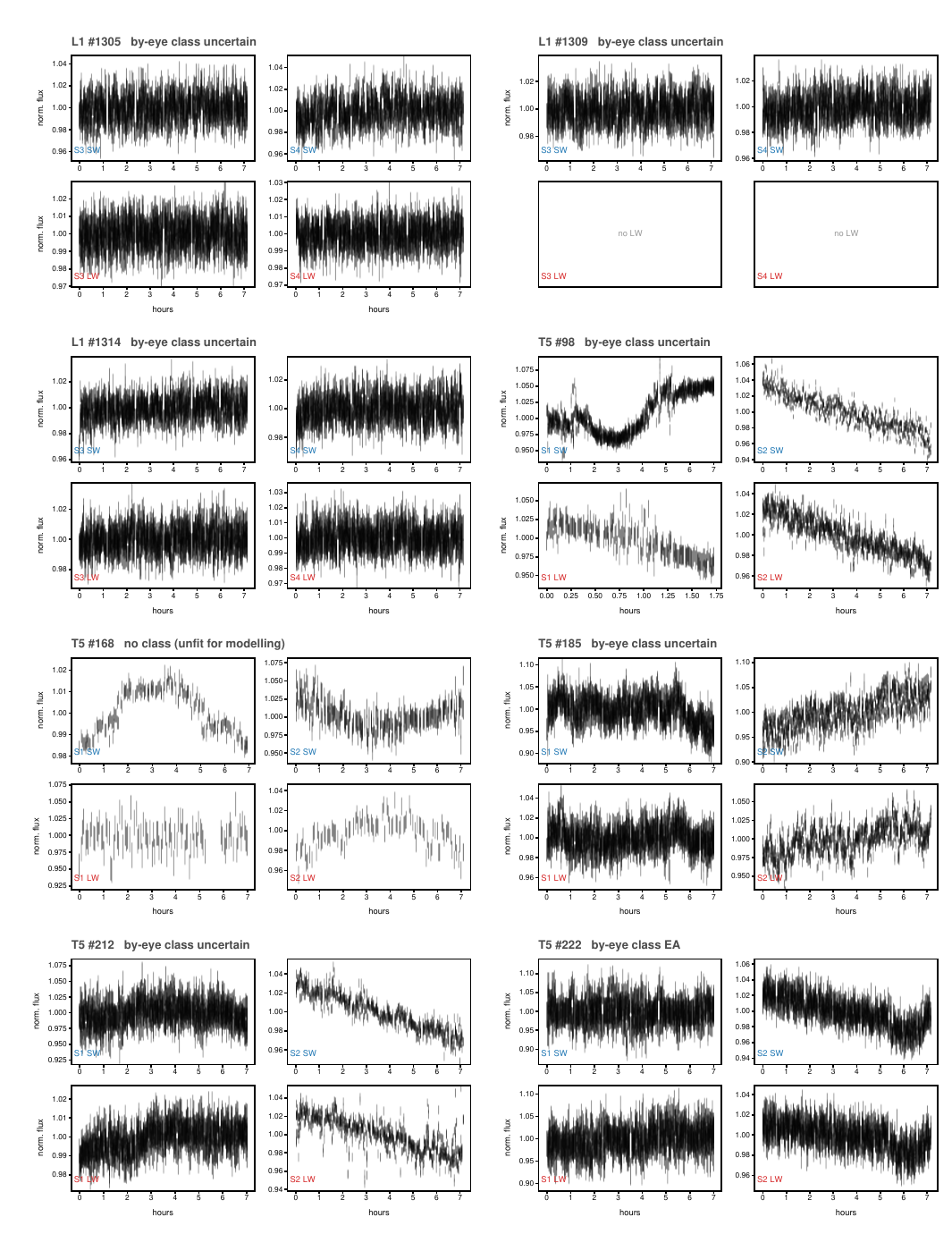}
\caption{Variables with no accepted model, continued (page 40 of 44).}
\end{figure*}
\clearpage

\begin{figure*}
\centering
\includegraphics[width=0.98\textwidth,height=0.94\textheight,keepaspectratio]{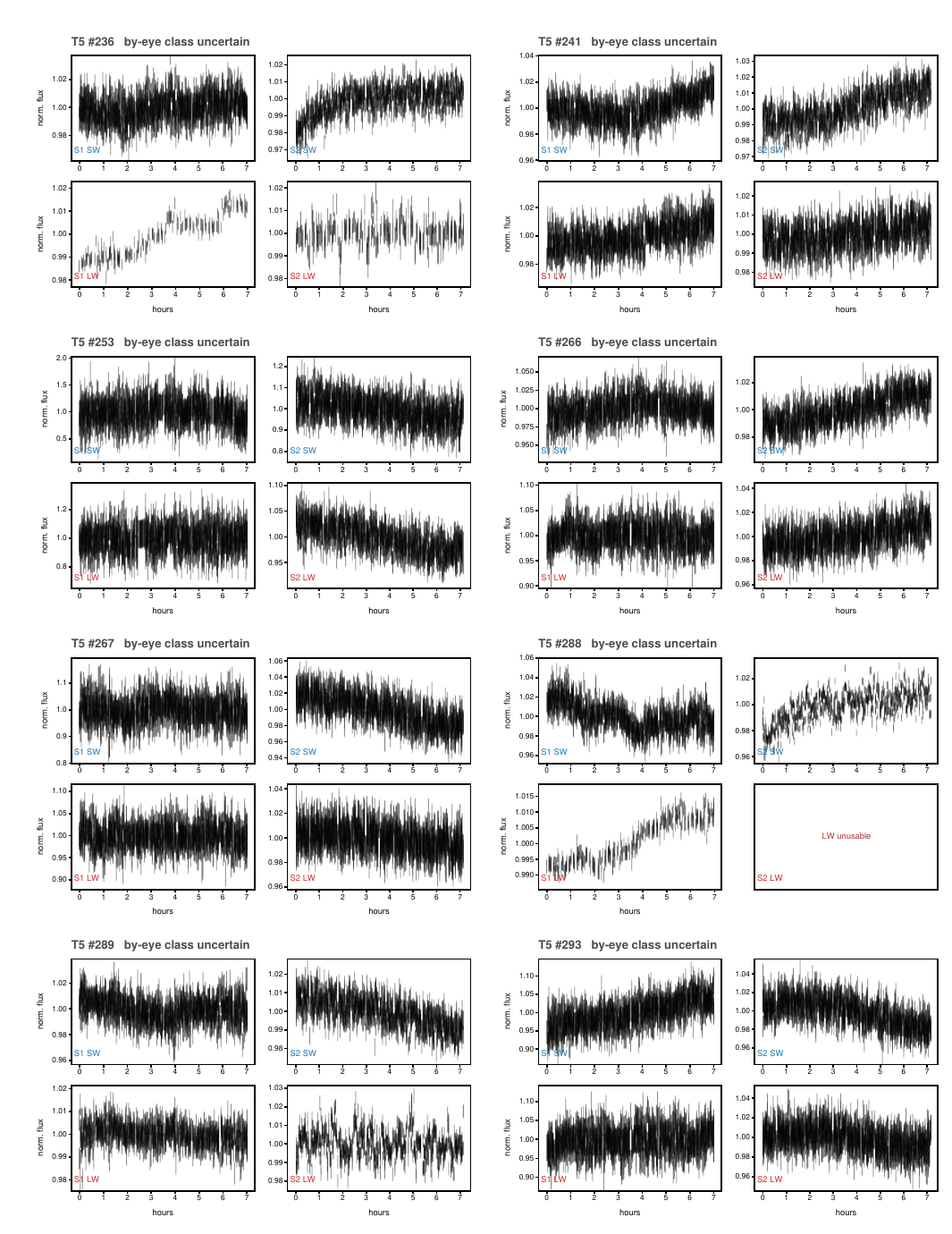}
\caption{Variables with no accepted model, continued (page 41 of 44).}
\end{figure*}
\clearpage

\begin{figure*}
\centering
\includegraphics[width=0.98\textwidth,height=0.94\textheight,keepaspectratio]{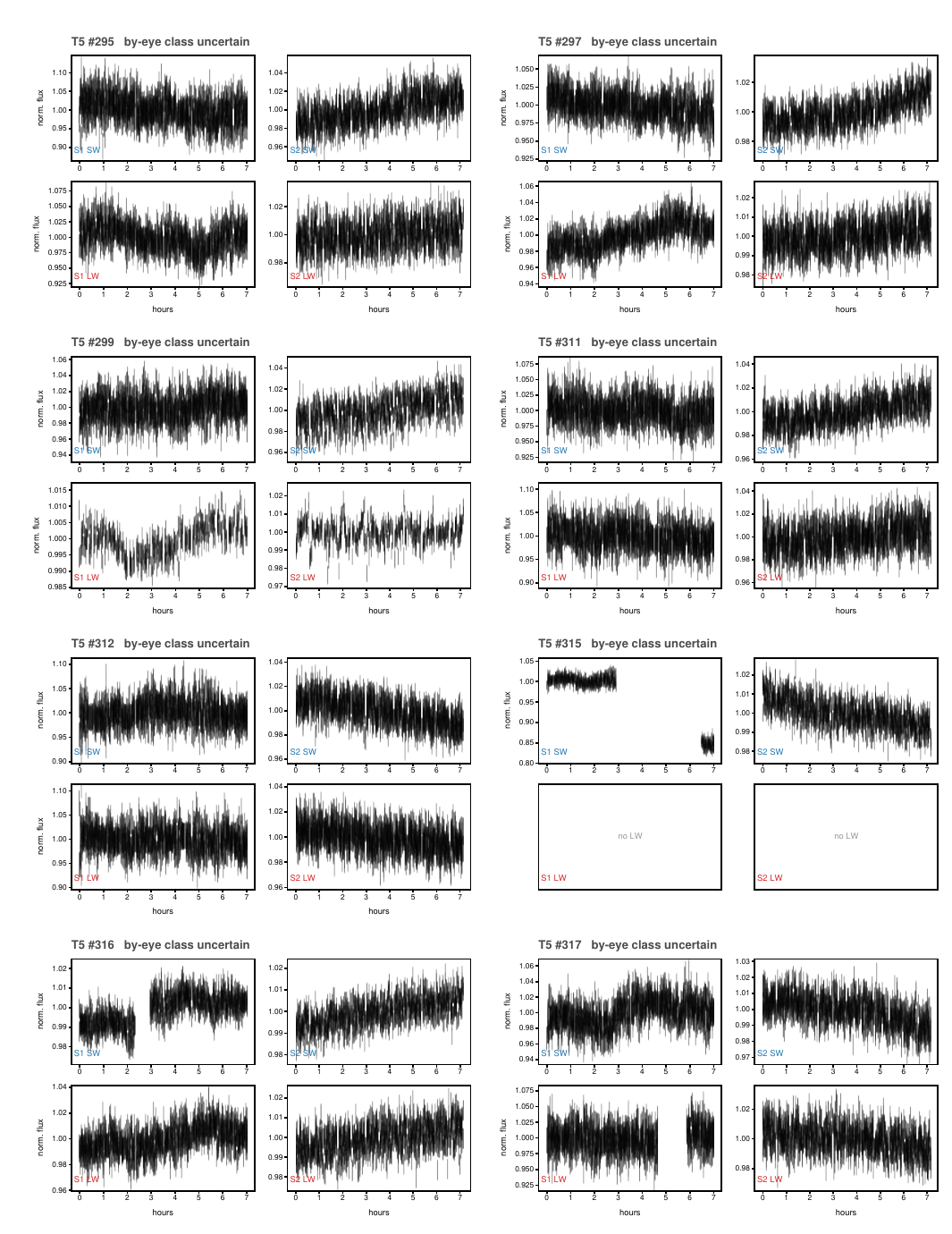}
\caption{Variables with no accepted model, continued (page 42 of 44).}
\end{figure*}
\clearpage

\begin{figure*}
\centering
\includegraphics[width=0.98\textwidth,height=0.94\textheight,keepaspectratio]{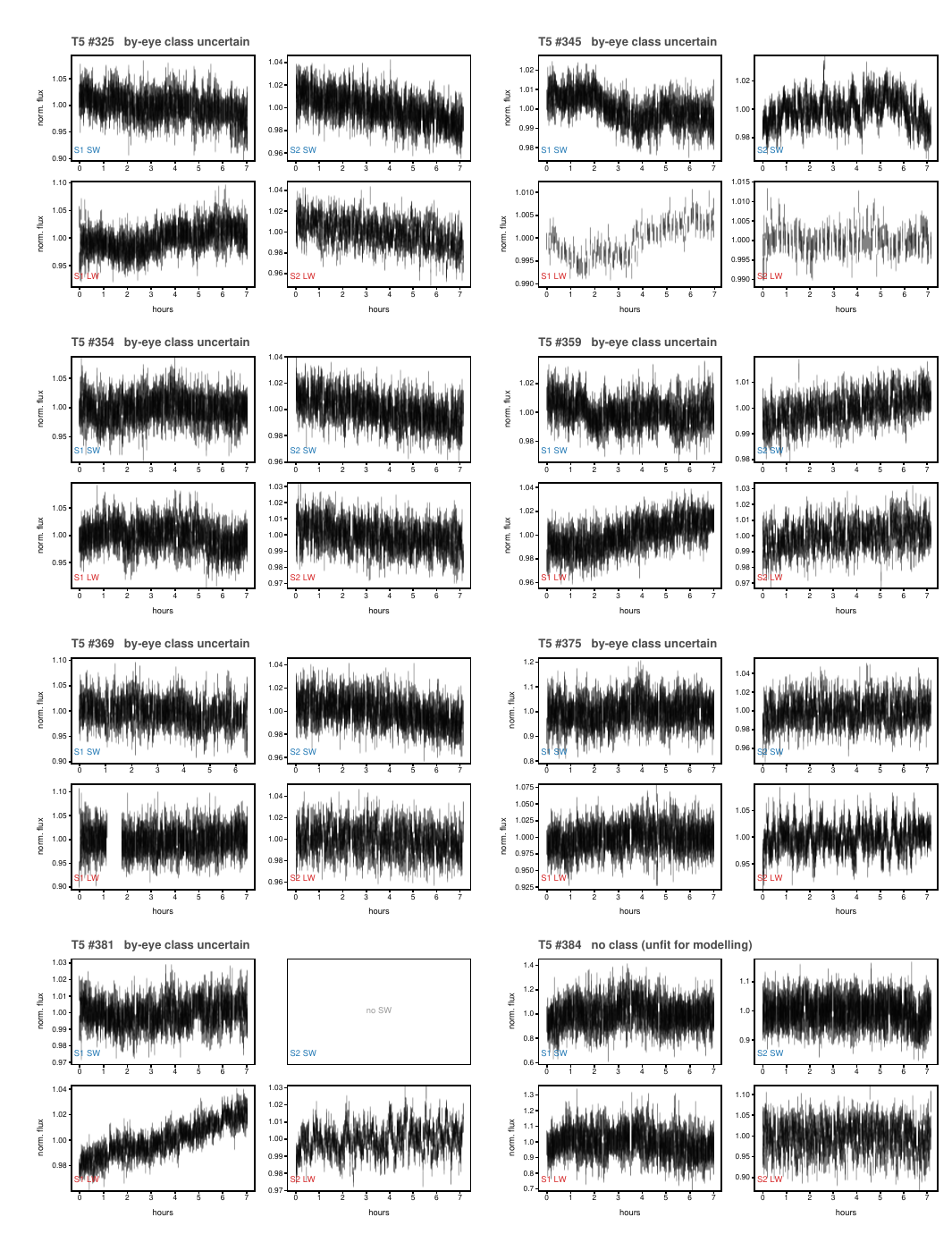}
\caption{Variables with no accepted model, continued (page 43 of 44).}
\end{figure*}
\clearpage

\begin{figure*}
\centering
\includegraphics[width=0.98\textwidth,height=0.94\textheight,keepaspectratio]{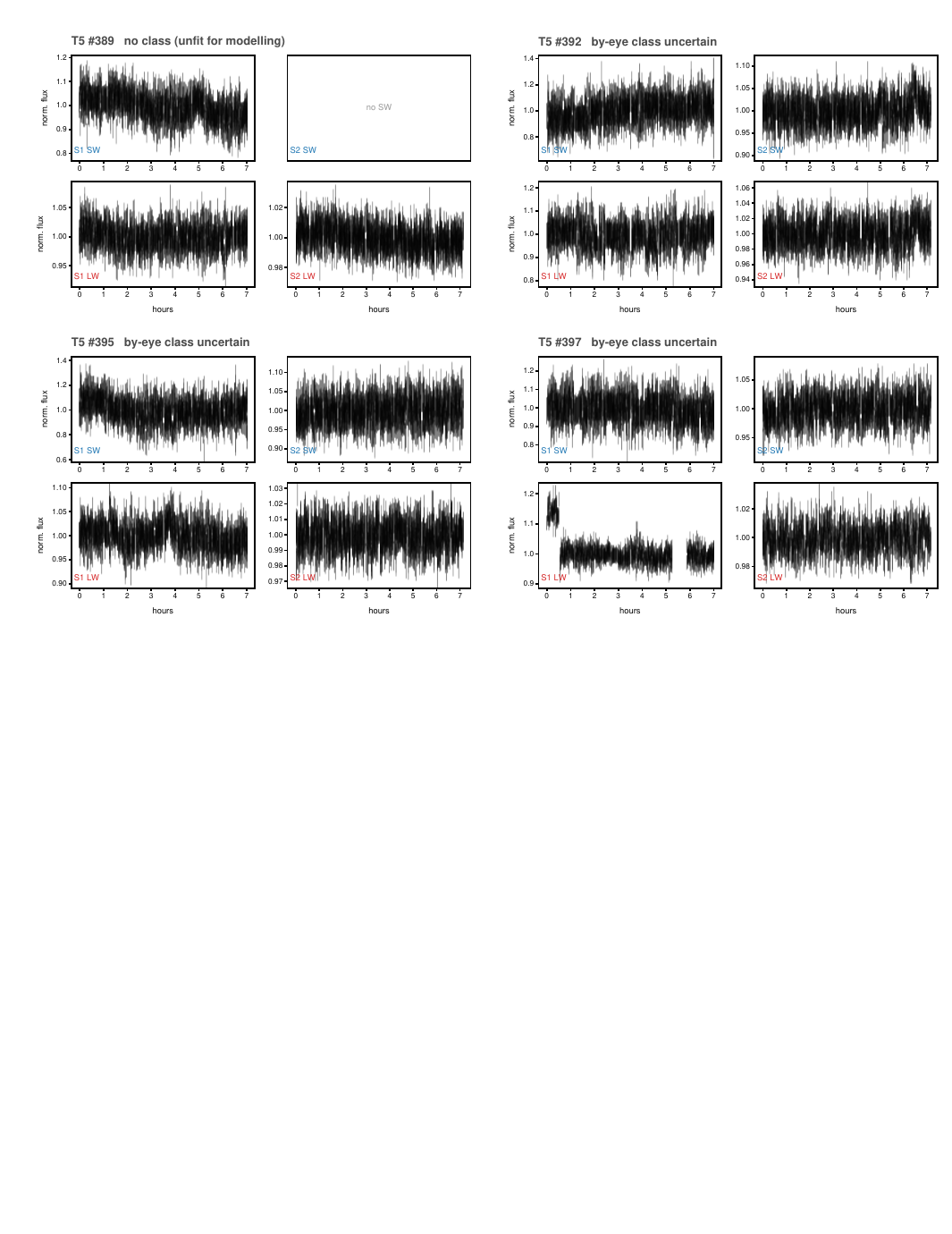}
\caption{Variables with no accepted model, continued (page 44 of 44).}
\end{figure*}
\clearpage

\fi

\bibliography{sample7}{}

@ARTICLE{Lomb1976,
       author = {{Lomb}, N.~R.},
        title = "{Least-Squares Frequency Analysis of Unequally Spaced Data}",
      journal = {\apss},
         year = 1976,
        month = feb,
       volume = {39},
       number = {2},
        pages = {447-462},
          doi = {10.1007/BF00648343},
       adsurl = {https://ui.adsabs.harvard.edu/abs/1976Ap&SS..39..447L}
}

@ARTICLE{Scargle1982,
       author = {{Scargle}, J.~D.},
        title = "{Studies in astronomical time series analysis. II. Statistical aspects of spectral analysis of unevenly spaced data.}",
      journal = {\apj},
         year = 1982,
        month = dec,
       volume = {263},
        pages = {835-853},
          doi = {10.1086/160554},
       adsurl = {https://ui.adsabs.harvard.edu/abs/1982ApJ...263..835S}
}

@ARTICLE{Kovacs2002,
       author = {{Kov{\'a}cs}, G. and {Zucker}, S. and {Mazeh}, T.},
        title = "{A box-fitting algorithm in the search for periodic transits}",
      journal = {\aap},
         year = 2002,
        month = aug,
       volume = {391},
        pages = {369-377},
          doi = {10.1051/0004-6361:20020802},
archivePrefix = {arXiv},
       eprint = {astro-ph/0206099},
 primaryClass = {astro-ph},
       adsurl = {https://ui.adsabs.harvard.edu/abs/2002A&A...391..369K}
}

@misc{photutils,
       author = {{Bradley}, Larry and {Sip{\H o}cz}, Brigitta and {Robitaille}, Thomas and {Tollerud}, Erik and {Vin{\'\i}cius}, Z{\`e} and {Deil}, Christoph and {Barbary}, Kyle and {Wilson}, Tom J. and {G{\"u}nther}, Hans Moritz and {Lim}, Pey Lian and {others}},
        title = "{astropy/photutils: 1.8.0}",
         year = 2023,
       adsurl = {https://doi.org/10.5281/zenodo.596036}
}

@ARTICLE{scipy,
       author = {{Virtanen}, Pauli and {Gommers}, Ralf and {Oliphant}, Travis E. and {Haberland}, Matt and {Reddy}, Tyler and {Cournapeau}, David and {Burovski}, Evgeni and {Peterson}, Pearu and {Weckesser}, Warren and {Bright}, Jonathan and {van der Walt}, St{\'e}fan J. and {Brett}, Matthew and {Wilson}, Joshua and {Millman}, K. Jarrod and {Mayorov}, Nikolay and {Nelson}, Andrew R.~J. and {Jones}, Eric and {Kern}, Robert and {Larson}, Eric and {Carey}, C.~J. and {Polat}, {\.I}lhan and {Feng}, Yu and {Moore}, Eric W. and {VanderPlas}, Jake and {Laxalde}, Denis and {Perktold}, Josef and {Cimrman}, Robert and {Henriksen}, Ian and {Quintero}, E.~A. and {Harris}, Charles R. and {Archibald}, Anne M. and {Ribeiro}, Ant{\^o}nio H. and {Pedregosa}, Fabian and {van Mulbregt}, Paul and {SciPy 1.  0 Contributors}},
        title = "{SciPy 1.0: fundamental algorithms for scientific computing in Python}",
      journal = {Nature Methods},
         year = 2020,
        month = feb,
       volume = {17},
        pages = {261-272},
          doi = {10.1038/s41592-019-0686-2},
archivePrefix = {arXiv},
       eprint = {1907.10121},
 primaryClass = {cs.MS},
       adsurl = {https://ui.adsabs.harvard.edu/abs/2020NaMet..17..261V}
}

@ARTICLE{matplotlib,
       author = {{Hunter}, John D.},
        title = "{Matplotlib: A 2D Graphics Environment}",
      journal = {Computing in Science and Engineering},
         year = 2007,
        month = jan,
       volume = {9},
       number = {3},
        pages = {90-95},
          doi = {10.1109/MCSE.2007.55},
       adsurl = {https://ui.adsabs.harvard.edu/abs/2007CSE.....9...90H}
}

@ARTICLE{Rigby2023,
       author = {{Rigby}, Jane and {Perrin}, Marshall and {McElwain}, Michael and {Kimble}, Randy and {Friedman}, Scott and {Lallo}, Matt and {Doyon}, Ren{\'e} and {Feinberg}, Lee and {Ferruit}, Pierre and {Glasse}, Alistair and {Rieke}, Marcia and {Rieke}, George and {Wright}, Gillian and {Willott}, Chris and {Colon}, Knicole and {Milam}, Stefanie and {Neff}, Susan and {Stark}, Christopher and {Valenti}, Jeff and {Abell}, Jim and {Abney}, Faith and {Abul-Huda}, Yasin and {Acton}, D. Scott and {Adams}, Evan and {Adler}, David and {Aguilar}, Jonathan and {Ahmed}, Nasif and {Albert}, Lo{\"\i}c and {Alberts}, Stacey and {Aldridge}, David and {Allen}, Marsha and {Altenburg}, Martin and {{\'A}lvarez-M{\'a}rquez}, Javier and {Alves de Oliveira}, Catarina and {Andersen}, Greg and {Anderson}, Harry and {Anderson}, Sara and {Argyriou}, Ioannis and {Armstrong}, Amber and {Arribas}, Santiago and {Artigau}, Etienne and {Arvai}, Amanda and {Atkinson}, Charles and {Bacon}, Gregory and {Bair}, Thomas and {Banks}, Kimberly and {Barrientes}, Jaclyn and {Barringer}, Bruce and {Bartosik}, Peter and {Bast}, William and {Baudoz}, Pierre and {Beatty}, Thomas and {Bechtold}, Katie and {Beck}, Tracy and {Bergeron}, Eddie and {Bergkoetter}, Matthew and {Bhatawdekar}, Rachana and {Birkmann}, Stephan and {Blazek}, Ronald and {Blome}, Claire and {Boccaletti}, Anthony and {B{\"o}ker}, Torsten and {Boia}, John and {Bonaventura}, Nina and {Bond}, Nicholas and {Bosley}, Kari and {Boucarut}, Ray and {Bourque}, Matthew and {Bouwman}, Jeroen and {Bower}, Gary and {Bowers}, Charles and {Boyer}, Martha and {Bradley}, Larry and {Brady}, Greg and {Braun}, Hannah and {Breda}, David and {Bresnahan}, Pamela and {Bright}, Stacey and {Britt}, Christopher and {Bromenschenkel}, Asa and {Brooks}, Brian and {Brooks}, Keira and {Brown}, Bob and {Brown}, Matthew and {Brown}, Patricia and {Bunker}, Andy and {Burger}, Matthew and {Bushouse}, Howard and {Cale}, Steven and {Cameron}, Alex and {Cameron}, Peter and {Canipe}, Alicia and {Caplinger}, James and {Caputo}, Francis and {Cara}, Mihai and {Carey}, Larkin and {Carniani}, Stefano and {Carrasquilla}, Maria and {Carruthers}, Margaret and {Case}, Michael and {Catherine}, Riggs and {Chance}, Don and {Chapman}, George and {Charlot}, St{\'e}phane and {Charlow}, Brian and {Chayer}, Pierre and {Chen}, Bin and {Cherinka}, Brian and {Chichester}, Sarah and {Chilton}, Zack and {Chonis}, Taylor and {Clampin}, Mark and {Clark}, Charles and {Clark}, Kerry and {Coe}, Dan and {Coleman}, Benee and {Comber}, Brian and {Comeau}, Tom and {Connolly}, Dennis and {Cooper}, James and {Cooper}, Rachel and {Coppock}, Eric and {Correnti}, Matteo and {Cossou}, Christophe and {Coulais}, Alain and {Coyle}, Laura and {Cracraft}, Misty and {Curti}, Mirko and {Cuturic}, Steven and {Davis}, Katherine and {Davis}, Michael and {Dean}, Bruce and {DeLisa}, Amy and {deMeester}, Wim and {Dencheva}, Nadia and {Dencheva}, Nadezhda and {DePasquale}, Joseph and {Deschenes}, Jeremy and {Hunor Detre}, {\"O}rs and {Diaz}, Rosa and {Dicken}, Dan and {DiFelice}, Audrey and {Dillman}, Matthew and {Dixon}, William and {Doggett}, Jesse and {Donaldson}, Tom and {Douglas}, Rob and {DuPrie}, Kimberly and {Dupuis}, Jean and {Durning}, John and {Easmin}, Nilufar and {Eck}, Weston and {Edeani}, Chinwe and {Egami}, Eiichi and {Ehrenwinkler}, Ralf and {Eisenhamer}, Jonathan and {Eisenhower}, Michael and {Elie}, Michelle and {Elliott}, James and {Elliott}, Kyle and {Ellis}, Tracy and {Engesser}, Michael and {Espinoza}, Nestor and {Etienne}, Odessa and {Etxaluze}, Mireya and {Falini}, Patrick and {Feeney}, Matthew and {Ferry}, Malcolm and {Filippazzo}, Joseph and {Fincham}, Brian and {Fix}, Mees and {Flagey}, Nicolas and {Florian}, Michael and {Flynn}, Jim and {Fontanella}, Erin and {Ford}, Terrance and {Forshay}, Peter and {Fox}, Ori and {Franz}, David and {Fu}, Henry and {Fullerton}, Alexander and {Galkin}, Sergey and {Galyer}, Anthony and {Garc{\'\i}a Mar{\'\i}n}, Macarena and {Gardner}, Jonathan P. and {Gardner}, Lisa and {Garland}, Dennis and {Garrett}, Bruce and {Gasman}, Danny and {Gaspar}, Andras and {Gaudreau}, Daniel and {Gauthier}, Peter and {Geers}, Vincent and {Geithner}, Paul and {Gennaro}, Mario and {Giardino}, Giovanna and {Girard}, Julien and {Giuliano}, Mark and {Glassmire}, Kirk and {Glauser}, Adrian},
        title = "{The Science Performance of JWST as Characterized in Commissioning}",
      journal = {\pasp},
         year = 2023,
        month = apr,
       volume = {135},
       number = {1046},
          eid = {048001},
        pages = {048001},
          doi = {10.1088/1538-3873/acb293},
archivePrefix = {arXiv},
       eprint = {2207.05632},
 primaryClass = {astro-ph.IM},
       adsurl = {https://ui.adsabs.harvard.edu/abs/2023PASP..135d8001R}
}

@ARTICLE{Rieke2023,
       author = {{Rieke}, Marcia J. and {Kelly}, Douglas M. and {Misselt}, Karl and {Stansberry}, John and {Boyer}, Martha and {Beatty}, Thomas and {Egami}, Eiichi and {Florian}, Michael and {Greene}, Thomas P. and {Hainline}, Kevin and {Leisenring}, Jarron and {Roellig}, Thomas and {Schlawin}, Everett and {Sun}, Fengwu and {Tinnin}, Lee and {Williams}, Christina C. and {Willmer}, Christopher N.~A. and {Wilson}, Debra and {Clark}, Charles R. and {Rohrbach}, Scott and {Brooks}, Brian and {Canipe}, Alicia and {Correnti}, Matteo and {DiFelice}, Audrey and {Gennaro}, Mario and {Girard}, Julien H. and {Hartig}, George and {Hilbert}, Bryan and {Koekemoer}, Anton M. and {Nikolov}, Nikolay K. and {Pirzkal}, Norbert and {Rest}, Armin and {Robberto}, Massimo and {Sunnquist}, Ben and {Telfer}, Randal and {Wu}, Chi Rai and {Ferry}, Malcolm and {Lewis}, Dan and {Baum}, Stefi and {Beichman}, Charles and {Doyon}, Ren{\'e} and {Dressler}, Alan and {Eisenstein}, Daniel J. and {Ferrarese}, Laura and {Hodapp}, Klaus and {Horner}, Scott and {Jaffe}, Daniel T. and {Johnstone}, Doug and {Krist}, John and {Martin}, Peter and {McCarthy}, Donald W. and {Meyer}, Michael and {Rieke}, George H. and {Trauger}, John and {Young}, Erick T.},
        title = "{Performance of NIRCam on JWST in Flight}",
      journal = {\pasp},
         year = 2023,
        month = feb,
       volume = {135},
       number = {1044},
          eid = {028001},
        pages = {028001},
          doi = {10.1088/1538-3873/acac53},
archivePrefix = {arXiv},
       eprint = {2212.12069},
 primaryClass = {astro-ph.IM},
       adsurl = {https://ui.adsabs.harvard.edu/abs/2023PASP..135b8001R}
}

@ARTICLE{Ferraro2026,
       author = {{Ferraro}, F.~R. and {Vesperini}, E. and {Lanzoni}, B. and {Romano}, D. and {Origlia}, L. and {Pallanca}, C. and {Fanelli}, C. and {Calura}, F. and {Dalessandro}, E. and {Massari}, D. and {Zullo}, G. and {Cadelano}, M.},
        title = "{Bulge fossil fragments as a new population of factories of gravitational wave sources in the Galaxy}",
      journal = {\aap},
         year = 2026,
        month = may,
       volume = {709},
          eid = {A163},
        pages = {A163},
          doi = {10.1051/0004-6361/202556993},
archivePrefix = {arXiv},
       eprint = {2603.25127},
 primaryClass = {astro-ph.GA},
       adsurl = {https://ui.adsabs.harvard.edu/abs/2026A&A...709A.163F}
}

@ARTICLE{Crociati2023,
       author = {{Crociati}, Chiara and {Valenti}, Elena and {Ferraro}, Francesco R. and {Pallanca}, Cristina and {Lanzoni}, Barbara and {Cadelano}, Mario and {Fanelli}, Cristiano and {Origlia}, Livia and {Leanza}, Silvia and {Dalessandro}, Emanuele and {Mucciarelli}, Alessio and {Rich}, R. Michael},
        title = "{First Evidence of Multi-iron Subpopulations in the Bulge Fossil Fragment Candidate Liller 1}",
      journal = {\apj},
         year = 2023,
        month = jul,
       volume = {951},
       number = {1},
          eid = {17},
        pages = {17},
          doi = {10.3847/1538-4357/acd382},
archivePrefix = {arXiv},
       eprint = {2305.04595},
 primaryClass = {astro-ph.GA},
       adsurl = {https://ui.adsabs.harvard.edu/abs/2023ApJ...951...17C}
}

@ARTICLE{Pallanca2021,
       author = {{Pallanca}, Cristina and {Ferraro}, Francesco R. and {Lanzoni}, Barbara and {Crociati}, Chiara and {Saracino}, Sara and {Dalessandro}, Emanuele and {Origlia}, Livia and {Rich}, Michael R. and {Valenti}, Elena and {Geisler}, Douglas and {Mauro}, Francesco and {Villanova}, Sandro and {Moni Bidin}, Christian and {Beccari}, Giacomo},
        title = "{High-resolution Extinction Map in the Direction of the Strongly Obscured Bulge Fossil Fragment Liller 1}",
      journal = {\apj},
         year = 2021,
        month = aug,
       volume = {917},
       number = {2},
          eid = {92},
        pages = {92},
          doi = {10.3847/1538-4357/ac0889},
archivePrefix = {arXiv},
       eprint = {2106.02448},
 primaryClass = {astro-ph.GA},
       adsurl = {https://ui.adsabs.harvard.edu/abs/2021ApJ...917...92P}
}

@ARTICLE{Ferraro2021,
       author = {{Ferraro}, F.~R. and {Pallanca}, C. and {Lanzoni}, B. and {Crociati}, C. and {Dalessandro}, E. and {Origlia}, L. and {Rich}, R.~M. and {Saracino}, S. and {Mucciarelli}, A. and {Valenti}, E. and {Geisler}, D. and {Mauro}, F. and {Villanova}, S. and {Moni Bidin}, C. and {Beccari}, G.},
        title = "{A new class of fossil fragments from the hierarchical assembly of the Galactic bulge}",
      journal = {Nature Astronomy},
         year = 2021,
        month = jan,
       volume = {5},
        pages = {311-318},
          doi = {10.1038/s41550-020-01267-y},
archivePrefix = {arXiv},
       eprint = {2011.09966},
 primaryClass = {astro-ph.GA},
       adsurl = {https://ui.adsabs.harvard.edu/abs/2021NatAs...5..311F}
}

@ARTICLE{Antonini2019,
       author = {{Antonini}, Fabio and {Gieles}, Mark and {Gualandris}, Alessia},
        title = "{Black hole growth through hierarchical black hole mergers in dense star clusters: implications for gravitational wave detections}",
      journal = {\mnras},
         year = 2019,
        month = jul,
       volume = {486},
       number = {4},
        pages = {5008-5021},
          doi = {10.1093/mnras/stz1149},
archivePrefix = {arXiv},
       eprint = {1811.03640},
 primaryClass = {astro-ph.HE},
       adsurl = {https://ui.adsabs.harvard.edu/abs/2019MNRAS.486.5008A}
}

@ARTICLE{Rodriguez2016,
       author = {{Rodriguez}, Carl L. and {Chatterjee}, Sourav and {Rasio}, Frederic A.},
        title = "{Binary black hole mergers from globular clusters: Masses, merger rates, and the impact of stellar evolution}",
      journal = {\prd},
         year = 2016,
        month = apr,
       volume = {93},
       number = {8},
          eid = {084029},
        pages = {084029},
          doi = {10.1103/PhysRevD.93.084029},
archivePrefix = {arXiv},
       eprint = {1602.02444},
 primaryClass = {astro-ph.HE},
       adsurl = {https://ui.adsabs.harvard.edu/abs/2016PhRvD..93h4029R}
}

@ARTICLE{Saracino2015,
       author = {{Saracino}, S. and {Dalessandro}, E. and {Ferraro}, F.~R. and {Lanzoni}, B. and {Geisler}, D. and {Mauro}, F. and {Villanova}, S. and {Moni Bidin}, C. and {Miocchi}, P. and {Massari}, D.},
        title = "{GEMINI/GeMS Observations Unveil the Structure of the Heavily Obscured Globular Cluster Liller 1.}",
      journal = {\apj},
         year = 2015,
        month = jun,
       volume = {806},
       number = {2},
          eid = {152},
        pages = {152},
          doi = {10.1088/0004-637X/806/2/152},
archivePrefix = {arXiv},
       eprint = {1505.00568},
 primaryClass = {astro-ph.SR},
       adsurl = {https://ui.adsabs.harvard.edu/abs/2015ApJ...806..152S}
}

@INPROCEEDINGS{Perrin2014,
       author = {{Perrin}, Marshall D. and {Sivaramakrishnan}, Anand and {Lajoie}, Charles-Philippe and {Elliott}, Erin and {Pueyo}, Laurent and {Ravindranath}, Swara and {Albert}, Lo{\"\i}c.},
        title = "{Updated point spread function simulations for JWST with WebbPSF}",
    booktitle = {Space Telescopes and Instrumentation 2014: Optical, Infrared, and Millimeter Wave},
         year = 2014,
       editor = {{Oschmann}, Jr., Jacobus M. and {Clampin}, Mark and {Fazio}, Giovanni G. and {MacEwen}, Howard A.},
       series = {Society of Photo-Optical Instrumentation Engineers (SPIE) Conference Series},
       volume = {9143},
        month = aug,
          eid = {91433X},
        pages = {91433X},
          doi = {10.1117/12.2056689},
       adsurl = {https://ui.adsabs.harvard.edu/abs/2014SPIE.9143E..3XP}
}

@ARTICLE{Origlia2013,
       author = {{Origlia}, L. and {Massari}, D. and {Rich}, R.~M. and {Mucciarelli}, A. and {Ferraro}, F.~R. and {Dalessandro}, E. and {Lanzoni}, B.},
        title = "{The Terzan 5 Puzzle: Discovery of a Third, Metal-poor Component}",
      journal = {\apjl},
         year = 2013,
        month = dec,
       volume = {779},
       number = {1},
          eid = {L5},
        pages = {L5},
          doi = {10.1088/2041-8205/779/1/L5},
archivePrefix = {arXiv},
       eprint = {1311.1706},
 primaryClass = {astro-ph.GA},
       adsurl = {https://ui.adsabs.harvard.edu/abs/2013ApJ...779L...5O}
}

@ARTICLE{Origlia2011,
       author = {{Origlia}, L. and {Rich}, R.~M. and {Ferraro}, F.~R. and {Lanzoni}, B. and {Bellazzini}, M. and {Dalessandro}, E. and {Mucciarelli}, A. and {Valenti}, E. and {Beccari}, G.},
        title = "{Spectroscopy Unveils the Complex Nature of Terzan 5}",
      journal = {\apjl},
         year = 2011,
        month = jan,
       volume = {726},
       number = {2},
          eid = {L20},
        pages = {L20},
          doi = {10.1088/2041-8205/726/2/L20},
archivePrefix = {arXiv},
       eprint = {1012.2047},
 primaryClass = {astro-ph.GA},
       adsurl = {https://ui.adsabs.harvard.edu/abs/2011ApJ...726L..20O}
}

@ARTICLE{Lanzoni2010,
       author = {{Lanzoni}, B. and {Ferraro}, F.~R. and {Dalessandro}, E. and {Mucciarelli}, A. and {Beccari}, G. and {Miocchi}, P. and {Bellazzini}, M. and {Rich}, R.~M. and {Origlia}, L. and {Valenti}, E. and {Rood}, R.~T. and {Ransom}, S.~M.},
        title = "{New Density Profile and Structural Parameters of the Complex Stellar System Terzan 5}",
      journal = {\apj},
         year = 2010,
        month = jul,
       volume = {717},
       number = {2},
        pages = {653-657},
          doi = {10.1088/0004-637X/717/2/653},
archivePrefix = {arXiv},
       eprint = {1005.2847},
 primaryClass = {astro-ph.GA},
       adsurl = {https://ui.adsabs.harvard.edu/abs/2010ApJ...717..653L}
}

@ARTICLE{Ferraro2009,
       author = {{Ferraro}, F.~R. and {Dalessandro}, E. and {Mucciarelli}, A. and {Beccari}, G. and {Rich}, R.~M. and {Origlia}, L. and {Lanzoni}, B. and {Rood}, R.~T. and {Valenti}, E. and {Bellazzini}, M. and {Ransom}, S.~M. and {Cocozza}, G.},
        title = "{The cluster Terzan 5 as a remnant of a primordial building block of the Galactic bulge}",
      journal = {\nat},
         year = 2009,
        month = nov,
       volume = {462},
       number = {7272},
        pages = {483-486},
          doi = {10.1038/nature08581},
archivePrefix = {arXiv},
       eprint = {0912.0192},
 primaryClass = {astro-ph.GA},
       adsurl = {https://ui.adsabs.harvard.edu/abs/2009Natur.462..483F}
}

@ARTICLE{Ivanova2008,
       author = {{Ivanova}, N. and {Heinke}, C.~O. and {Rasio}, F.~A. and {Belczynski}, K. and {Fregeau}, J.~M.},
        title = "{Formation and evolution of compact binaries in globular clusters - II. Binaries with neutron stars}",
      journal = {\mnras},
         year = 2008,
        month = may,
       volume = {386},
       number = {1},
        pages = {553-576},
          doi = {10.1111/j.1365-2966.2008.13064.x},
archivePrefix = {arXiv},
       eprint = {0706.4096},
 primaryClass = {astro-ph},
       adsurl = {https://ui.adsabs.harvard.edu/abs/2008MNRAS.386..553I}
}

@ARTICLE{Heinke2006,
       author = {{Heinke}, C.~O. and {Wijnands}, R. and {Cohn}, H.~N. and {Lugger}, P.~M. and {Grindlay}, J.~E. and {Pooley}, D. and {Lewin}, W.~H.~G.},
        title = "{Faint X-Ray Sources in the Globular Cluster Terzan 5}",
      journal = {\apj},
         year = 2006,
        month = nov,
       volume = {651},
       number = {2},
        pages = {1098-1111},
          doi = {10.1086/507884},
archivePrefix = {arXiv},
       eprint = {astro-ph/0606253},
 primaryClass = {astro-ph},
       adsurl = {https://ui.adsabs.harvard.edu/abs/2006ApJ...651.1098H}
}

@ARTICLE{Hessels2006,
       author = {{Hessels}, Jason W.~T. and {Ransom}, Scott M. and {Stairs}, Ingrid H. and {Freire}, Paulo C.~C. and {Kaspi}, Victoria M. and {Camilo}, Fernando},
        title = "{A Radio Pulsar Spinning at 716 Hz}",
      journal = {Science},
         year = 2006,
        month = mar,
       volume = {311},
       number = {5769},
        pages = {1901-1904},
          doi = {10.1126/science.1123430},
archivePrefix = {arXiv},
       eprint = {astro-ph/0601337},
 primaryClass = {astro-ph},
       adsurl = {https://ui.adsabs.harvard.edu/abs/2006Sci...311.1901H}
}

@ARTICLE{Ransom2005,
       author = {{Ransom}, Scott M. and {Hessels}, Jason W.~T. and {Stairs}, Ingrid H. and {Freire}, Paulo C.~C. and {Camilo}, Fernando and {Kaspi}, Victoria M. and {Kaplan}, David L.},
        title = "{Twenty-One Millisecond Pulsars in Terzan 5 Using the Green Bank Telescope}",
      journal = {Science},
         year = 2005,
        month = feb,
       volume = {307},
       number = {5711},
        pages = {892-896},
          doi = {10.1126/science.1108632},
archivePrefix = {arXiv},
       eprint = {astro-ph/0501230},
 primaryClass = {astro-ph},
       adsurl = {https://ui.adsabs.harvard.edu/abs/2005Sci...307..892R}
}

@ARTICLE{Pooley2003,
       author = {{Pooley}, David and {Lewin}, Walter H.~G. and {Anderson}, Scott F. and {Baumgardt}, Holger and {Filippenko}, Alexei V. and {Gaensler}, Bryan M. and {Homer}, Lee and {Hut}, Piet and {Kaspi}, Victoria M. and {Makino}, Junichiro and {Margon}, Bruce and {McMillan}, Steve and {Portegies Zwart}, Simon and {van der Klis}, Michiel and {Verbunt}, Frank},
        title = "{Dynamical Formation of Close Binary Systems in Globular Clusters}",
      journal = {\apjl},
         year = 2003,
        month = jul,
       volume = {591},
       number = {2},
        pages = {L131-L134},
          doi = {10.1086/377074},
archivePrefix = {arXiv},
       eprint = {astro-ph/0305003},
 primaryClass = {astro-ph},
       adsurl = {https://ui.adsabs.harvard.edu/abs/2003ApJ...591L.131P}
}

@ARTICLE{Grindlay2001,
       author = {{Grindlay}, J.~E. and {Heinke}, C.~O. and {Edmonds}, P.~D. and {Murray}, S.~S. and {Cool}, A.~M.},
        title = "{Chandra Exposes the Core-collapsed Globular Cluster NGC 6397}",
      journal = {\apjl},
         year = 2001,
        month = dec,
       volume = {563},
       number = {1},
        pages = {L53-L56},
          doi = {10.1086/338499},
archivePrefix = {arXiv},
       eprint = {astro-ph/0108265},
 primaryClass = {astro-ph},
       adsurl = {https://ui.adsabs.harvard.edu/abs/2001ApJ...563L..53G}
}

@ARTICLE{Homer2001,
       author = {{Homer}, L. and {Deutsch}, Eric W. and {Anderson}, Scott F. and {Margon}, Bruce},
        title = "{The Rapid Burster in Liller 1: The Chandra X-Ray Position and a Search for an Infrared Counterpart}",
      journal = {\aj},
         year = 2001,
        month = nov,
       volume = {122},
       number = {5},
        pages = {2627-2633},
          doi = {10.1086/323545},
archivePrefix = {arXiv},
       eprint = {astro-ph/0106140},
 primaryClass = {astro-ph},
       adsurl = {https://ui.adsabs.harvard.edu/abs/2001AJ....122.2627H}
}

@ARTICLE{Clement2001,
       author = {{Clement}, Christine M. and {Muzzin}, Adam and {Dufton}, Quentin and {Ponnampalam}, Thivya and {Wang}, John and {Burford}, Jay and {Richardson}, Alan and {Rosebery}, Tara and {Rowe}, Jason and {Hogg}, Helen Sawyer},
        title = "{Variable Stars in Galactic Globular Clusters}",
      journal = {\aj},
         year = 2001,
        month = nov,
       volume = {122},
       number = {5},
        pages = {2587-2599},
          doi = {10.1086/323719},
archivePrefix = {arXiv},
       eprint = {astro-ph/0108024},
 primaryClass = {astro-ph},
       adsurl = {https://ui.adsabs.harvard.edu/abs/2001AJ....122.2587C}
}

@ARTICLE{PortegiesZwart2000,
       author = {{Portegies Zwart}, Simon F. and {McMillan}, Stephen L.~W.},
        title = "{Black Hole Mergers in the Universe}",
      journal = {\apjl},
         year = 2000,
        month = jan,
       volume = {528},
       number = {1},
        pages = {L17-L20},
          doi = {10.1086/312422},
archivePrefix = {arXiv},
       eprint = {astro-ph/9910061},
 primaryClass = {astro-ph},
       adsurl = {https://ui.adsabs.harvard.edu/abs/2000ApJ...528L..17P}
}

@ARTICLE{Hut1992,
       author = {{Hut}, Piet and {McMillan}, Steve and {Goodman}, Jeremy and {Mateo}, Mario and {Phinney}, E.~S. and {Pryor}, Carlton and {Richer}, Harvey B. and {Verbunt}, Frank and {Weinberg}, Martin},
        title = "{Binaries in Globular Clusters}",
      journal = {\pasp},
         year = 1992,
        month = nov,
       volume = {104},
        pages = {981},
          doi = {10.1086/133085},
       adsurl = {https://ui.adsabs.harvard.edu/abs/1992PASP..104..981H}
}

@ARTICLE{Clark1975,
       author = {{Clark}, G.~W.},
        title = "{X-ray binaries in globular clusters.}",
      journal = {\apjl},
         year = 1975,
        month = aug,
       volume = {199},
        pages = {L143-L145},
          doi = {10.1086/181869},
       adsurl = {https://ui.adsabs.harvard.edu/abs/1975ApJ...199L.143C}
}

@ARTICLE{2022ApJ...935..167A,
       author = {{Astropy Collaboration} and {Price-Whelan}, Adrian M. and {Lim}, Pey Lian and {Earl}, Nicholas and {Starkman}, Nathaniel and {Bradley}, Larry and {Shupe}, David L. and {Patil}, Aarya A. and {Corrales}, Lia and {Brasseur}, C.~E. and {N{\"o}the}, Maximilian and {Donath}, Axel and {Tollerud}, Erik and {Morris}, Brett M. and {Ginsburg}, Adam and {Vaher}, Eero and {Weaver}, Benjamin A. and {Tocknell}, James and {Jamieson}, William and {van Kerkwijk}, Marten H. and {Robitaille}, Thomas P. and {Merry}, Bruce and {Bachetti}, Matteo and {G{\"u}nther}, H. Moritz and {Aldcroft}, Thomas L. and {Alvarado-Montes}, Jaime A. and {Archibald}, Anne M. and {B{\'o}di}, Attila and {Bapat}, Shreyas and {Barentsen}, Geert and {Baz{\'a}n}, Juanjo and {Biswas}, Manish and {Boquien}, M{\'e}d{\'e}ric and {Burke}, D.~J. and {Cara}, Daria and {Cara}, Mihai and {Conroy}, Kyle E. and {Conseil}, Simon and {Craig}, Matthew W. and {Cross}, Robert M. and {Cruz}, Kelle L. and {D'Eugenio}, Francesco and {Dencheva}, Nadia and {Devillepoix}, Hadrien A.~R. and {Dietrich}, J{\"o}rg P. and {Eigenbrot}, Arthur Davis and {Erben}, Thomas and {Ferreira}, Leonardo and {Foreman-Mackey}, Daniel and {Fox}, Ryan and {Freij}, Nabil and {Garg}, Suyog and {Geda}, Robel and {Glattly}, Lauren and {Gondhalekar}, Yash and {Gordon}, Karl D. and {Grant}, David and {Greenfield}, Perry and {Groener}, Austen M. and {Guest}, Steve and {Gurovich}, Sebastian and {Handberg}, Rasmus and {Hart}, Akeem and {Hatfield-Dodds}, Zac and {Homeier}, Derek and {Hosseinzadeh}, Griffin and {Jenness}, Tim and {Jones}, Craig K. and {Joseph}, Prajwel and {Kalmbach}, J. Bryce and {Karamehmetoglu}, Emir and {Ka{\l}uszy{\'n}ski}, Miko{\l}aj and {Kelley}, Michael S.~P. and {Kern}, Nicholas and {Kerzendorf}, Wolfgang E. and {Koch}, Eric W. and {Kulumani}, Shankar and {Lee}, Antony and {Ly}, Chun and {Ma}, Zhiyuan and {MacBride}, Conor and {Maljaars}, Jakob M. and {Muna}, Demitri and {Murphy}, N.~A. and {Norman}, Henrik and {O'Steen}, Richard and {Oman}, Kyle A. and {Pacifici}, Camilla and {Pascual}, Sergio and {Pascual-Granado}, J. and {Patil}, Rohit R. and {Perren}, Gabriel I. and {Pickering}, Timothy E. and {Rastogi}, Tanuj and {Roulston}, Benjamin R. and {Ryan}, Daniel F. and {Rykoff}, Eli S. and {Sabater}, Jose and {Sakurikar}, Parikshit and {Salgado}, Jes{\'u}s and {Sanghi}, Aniket and {Saunders}, Nicholas and {Savchenko}, Volodymyr and {Schwardt}, Ludwig and {Seifert-Eckert}, Michael and {Shih}, Albert Y. and {Jain}, Anany Shrey and {Shukla}, Gyanendra and {Sick}, Jonathan and {Simpson}, Chris and {Singanamalla}, Sudheesh and {Singer}, Leo P. and {Singhal}, Jaladh and {Sinha}, Manodeep and {Sip{\H{o}}cz}, Brigitta M. and {Spitler}, Lee R. and {Stansby}, David and {Streicher}, Ole and {{\v{S}}umak}, Jani and {Swinbank}, John D. and {Taranu}, Dan S. and {Tewary}, Nikita and {Tremblay}, Grant R. and {de Val-Borro}, Miguel and {Van Kooten}, Samuel J. and {Vasovi{\'c}}, Zlatan and {Verma}, Shresth and {de Miranda Cardoso}, Jos{\'e} Vin{\'\i}cius and {Williams}, Peter K.~G. and {Wilson}, Tom J. and {Winkel}, Benjamin and {Wood-Vasey}, W.~M. and {Xue}, Rui and {Yoachim}, Peter and {Zhang}, Chen and {Zonca}, Andrea and {Astropy Project Contributors}},
        title = "{The Astropy Project: Sustaining and Growing a Community-oriented Open-source Project and the Latest Major Release (v5.0) of the Core Package}",
      journal = {\apj},
         year = 2022,
        month = aug,
       volume = {935},
       number = {2},
          eid = {167},
        pages = {167},
          doi = {10.3847/1538-4357/ac7c74},
archivePrefix = {arXiv},
       eprint = {2206.14220},
 primaryClass = {astro-ph.IM},
       adsurl = {https://ui.adsabs.harvard.edu/abs/2022ApJ...935..167A}
}

@ARTICLE{2018AJ....156..123A,
       author = {{Astropy Collaboration} and {Price-Whelan}, A.~M. and {Sip{\H{o}}cz}, B.~M. and {G{\"u}nther}, H.~M. and {Lim}, P.~L. and {Crawford}, S.~M. and {Conseil}, S. and {Shupe}, D.~L. and {Craig}, M.~W. and {Dencheva}, N. and {Ginsburg}, A. and {VanderPlas}, J.~T. and {Bradley}, L.~D. and {P{\'e}rez-Su{\'a}rez}, D. and {de Val-Borro}, M. and {Aldcroft}, T.~L. and {Cruz}, K.~L. and {Robitaille}, T.~P. and {Tollerud}, E.~J. and {Ardelean}, C. and {Babej}, T. and {Bach}, Y.~P. and {Bachetti}, M. and {Bakanov}, A.~V. and {Bamford}, S.~P. and {Barentsen}, G. and {Barmby}, P. and {Baumbach}, A. and {Berry}, K.~L. and {Biscani}, F. and {Boquien}, M. and {Bostroem}, K.~A. and {Bouma}, L.~G. and {Brammer}, G.~B. and {Bray}, E.~M. and {Breytenbach}, H. and {Buddelmeijer}, H. and {Burke}, D.~J. and {Calderone}, G. and {Cano Rodr{\'\i}guez}, J.~L. and {Cara}, M. and {Cardoso}, J.~V.~M. and {Cheedella}, S. and {Copin}, Y. and {Corrales}, L. and {Crichton}, D. and {D'Avella}, D. and {Deil}, C. and {Depagne}, {\'E}. and {Dietrich}, J.~P. and {Donath}, A. and {Droettboom}, M. and {Earl}, N. and {Erben}, T. and {Fabbro}, S. and {Ferreira}, L.~A. and {Finethy}, T. and {Fox}, R.~T. and {Garrison}, L.~H. and {Gibbons}, S.~L.~J. and {Goldstein}, D.~A. and {Gommers}, R. and {Greco}, J.~P. and {Greenfield}, P. and {Groener}, A.~M. and {Grollier}, F. and {Hagen}, A. and {Hirst}, P. and {Homeier}, D. and {Horton}, A.~J. and {Hosseinzadeh}, G. and {Hu}, L. and {Hunkeler}, J.~S. and {Ivezi{\'c}}, {\v{Z}}. and {Jain}, A. and {Jenness}, T. and {Kanarek}, G. and {Kendrew}, S. and {Kern}, N.~S. and {Kerzendorf}, W.~E. and {Khvalko}, A. and {King}, J. and {Kirkby}, D. and {Kulkarni}, A.~M. and {Kumar}, A. and {Lee}, A. and {Lenz}, D. and {Littlefair}, S.~P. and {Ma}, Z. and {Macleod}, D.~M. and {Mastropietro}, M. and {McCully}, C. and {Montagnac}, S. and {Morris}, B.~M. and {Mueller}, M. and {Mumford}, S.~J. and {Muna}, D. and {Murphy}, N.~A. and {Nelson}, S. and {Nguyen}, G.~H. and {Ninan}, J.~P. and {N{\"o}the}, M. and {Ogaz}, S. and {Oh}, S. and {Parejko}, J.~K. and {Parley}, N. and {Pascual}, S. and {Patil}, R. and {Patil}, A.~A. and {Plunkett}, A.~L. and {Prochaska}, J.~X. and {Rastogi}, T. and {Reddy Janga}, V. and {Sabater}, J. and {Sakurikar}, P. and {Seifert}, M. and {Sherbert}, L.~E. and {Sherwood-Taylor}, H. and {Shih}, A.~Y. and {Sick}, J. and {Silbiger}, M.~T. and {Singanamalla}, S. and {Singer}, L.~P. and {Sladen}, P.~H. and {Sooley}, K.~A. and {Sornarajah}, S. and {Streicher}, O. and {Teuben}, P. and {Thomas}, S.~W. and {Tremblay}, G.~R. and {Turner}, J.~E.~H. and {Terr{\'o}n}, V. and {van Kerkwijk}, M.~H. and {de la Vega}, A. and {Watkins}, L.~L. and {Weaver}, B.~A. and {Whitmore}, J.~B. and {Woillez}, J. and {Zabalza}, V. and {Astropy Contributors}},
        title = "{The Astropy Project: Building an Open-science Project and Status of the v2.0 Core Package}",
      journal = {\aj},
         year = 2018,
        month = sep,
       volume = {156},
       number = {3},
          eid = {123},
        pages = {123},
          doi = {10.3847/1538-3881/aabc4f},
archivePrefix = {arXiv},
       eprint = {1801.02634},
 primaryClass = {astro-ph.IM},
       adsurl = {https://ui.adsabs.harvard.edu/abs/2018AJ....156..123A}
}

@ARTICLE{2013A&A...558A..33A,
       author = {{Astropy Collaboration} and {Robitaille}, Thomas P. and {Tollerud}, Erik J. and {Greenfield}, Perry and {Droettboom}, Michael and {Bray}, Erik and {Aldcroft}, Tom and {Davis}, Matt and {Ginsburg}, Adam and {Price-Whelan}, Adrian M. and {Kerzendorf}, Wolfgang E. and {Conley}, Alexander and {Crighton}, Neil and {Barbary}, Kyle and {Muna}, Demitri and {Ferguson}, Henry and {Grollier}, Fr{\'e}d{\'e}ric and {Parikh}, Madhura M. and {Nair}, Prasanth H. and {Unther}, Hans M. and {Deil}, Christoph and {Woillez}, Julien and {Conseil}, Simon and {Kramer}, Roban and {Turner}, James E.~H. and {Singer}, Leo and {Fox}, Ryan and {Weaver}, Benjamin A. and {Zabalza}, Victor and {Edwards}, Zachary I. and {Azalee Bostroem}, K. and {Burke}, D.~J. and {Casey}, Andrew R. and {Crawford}, Steven M. and {Dencheva}, Nadia and {Ely}, Justin and {Jenness}, Tim and {Labrie}, Kathleen and {Lim}, Pey Lian and {Pierfederici}, Francesco and {Pontzen}, Andrew and {Ptak}, Andy and {Refsdal}, Brian and {Servillat}, Mathieu and {Streicher}, Ole},
        title = "{Astropy: A community Python package for astronomy}",
      journal = {\aap},
         year = 2013,
        month = oct,
       volume = {558},
          eid = {A33},
        pages = {A33},
          doi = {10.1051/0004-6361/201322068},
archivePrefix = {arXiv},
       eprint = {1307.6212},
 primaryClass = {astro-ph.IM},
       adsurl = {https://ui.adsabs.harvard.edu/abs/2013A&A...558A..33A}
}

@ARTICLE{Albrow2001,
       author = {{Albrow}, Michael D. and {Gilliland}, Ronald L. and {Brown}, Timothy M. and {Edmonds}, Peter D. and {Guhathakurta}, Puragra and {Sarajedini}, Ata},
        title = "{The Frequency of Binary Stars in the Core of 47 Tucanae}",
      journal = {\apj},
         year = 2001,
        month = oct,
       volume = {559},
       number = {2},
        pages = {1060-1081},
          doi = {10.1086/322353},
archivePrefix = {arXiv},
       eprint = {astro-ph/0105441},
 primaryClass = {astro-ph},
       adsurl = {https://ui.adsabs.harvard.edu/abs/2001ApJ...559.1060A}
}

@ARTICLE{Lyne1990,
       author = {{Lyne}, A.~G. and {Manchester}, R.~N. and {D'Amico}, N. and {Staveley-Smith}, L. and {Johnston}, S. and {Lim}, J. and {Fruchter}, A.~S. and {Goss}, W.~M. and {Frail}, D.},
        title = "{An eclipsing millisecond pulsar in the globular cluster Terzan 5}",
      journal = {\nat},
         year = 1990,
        month = oct,
       volume = {347},
       number = {6294},
        pages = {650-652},
          doi = {10.1038/347650a0},
       adsurl = {https://ui.adsabs.harvard.edu/abs/1990Natur.347..650L}
}

@ARTICLE{Padmanabh2024,
       author = {{Padmanabh}, P.~V. and {Ransom}, S.~M. and {Freire}, P.~C.~C. and {Ridolfi}, A. and {Taylor}, J.~D. and {Choza}, C. and {Clark}, C.~J. and {Abbate}, F. and {Bailes}, M. and {Barr}, E.~D. and {Buchner}, S. and {Burgay}, M. and {DeCesar}, M.~E. and {Chen}, W. and {Corongiu}, A. and {Champion}, D.~J. and {Dutta}, A. and {Geyer}, M. and {Hessels}, J.~W.~T. and {Kramer}, M. and {Possenti}, A. and {Stairs}, I.~H. and {Stappers}, B.~W. and {Venkatraman Krishnan}, V. and {Vleeschower}, L. and {Zhang}, L.},
        title = "{Discovery and timing of ten new millisecond pulsars in the globular cluster Terzan 5}",
      journal = {\aap},
         year = 2024,
        month = jun,
       volume = {686},
          eid = {A166},
        pages = {A166},
          doi = {10.1051/0004-6361/202449303},
archivePrefix = {arXiv},
       eprint = {2403.17799},
 primaryClass = {astro-ph.HE},
       adsurl = {https://ui.adsabs.harvard.edu/abs/2024A&A...686A.166P}
}

@ARTICLE{Bogdanov2021,
       author = {{Bogdanov}, Slavko and {Bahramian}, Arash and {Heinke}, Craig O. and {Freire}, Paulo C.~C. and {Hessels}, Jason W.~T. and {Ransom}, Scott M. and {Stairs}, Ingrid H.},
        title = "{A Deep Chandra X-Ray Observatory Study of the Millisecond Pulsar Population in the Globular Cluster Terzan 5}",
      journal = {\apj},
         year = 2021,
        month = may,
       volume = {912},
       number = {2},
          eid = {124},
        pages = {124},
          doi = {10.3847/1538-4357/abee78},
archivePrefix = {arXiv},
       eprint = {2012.12944},
 primaryClass = {astro-ph.HE},
       adsurl = {https://ui.adsabs.harvard.edu/abs/2021ApJ...912..124B}
}

@ARTICLE{Cadelano2015,
       author = {{Cadelano}, M. and {Pallanca}, C. and {Ferraro}, F.~R. and {Salaris}, M. and {Dalessandro}, E. and {Lanzoni}, B. and {Freire}, P.~C.~C.},
        title = "{Optical Identification of He White Dwarfs Orbiting Four Millisecond Pulsars in the Globular Cluster 47 Tucanae}",
      journal = {\apj},
         year = 2015,
        month = oct,
       volume = {812},
       number = {1},
          eid = {63},
        pages = {63},
          doi = {10.1088/0004-637X/812/1/63},
archivePrefix = {arXiv},
       eprint = {1509.01397},
 primaryClass = {astro-ph.SR},
       adsurl = {https://ui.adsabs.harvard.edu/abs/2015ApJ...812...63C}
}

@ARTICLE{Petigura2018,
       author = {{Petigura}, Erik A. and {Marcy}, Geoffrey W. and {Winn}, Joshua N. and {Weiss}, Lauren M. and {Fulton}, Benjamin J. and {Howard}, Andrew W. and {Sinukoff}, Evan and {Isaacson}, Howard and {Morton}, Timothy D. and {Johnson}, John Asher},
        title = "{The California-Kepler Survey. IV. Metal-rich Stars Host a Greater Diversity of Planets}",
      journal = {\aj},
         year = 2018,
        month = feb,
       volume = {155},
       number = {2},
          eid = {89},
        pages = {89},
          doi = {10.3847/1538-3881/aaa54c},
archivePrefix = {arXiv},
       eprint = {1712.04042},
 primaryClass = {astro-ph.EP},
       adsurl = {https://ui.adsabs.harvard.edu/abs/2018AJ....155...89P}
}

@ARTICLE{Masuda2017,
       author = {{Masuda}, Kento and {Winn}, Joshua N.},
        title = "{Reassessment of the Null Result of the HST Search for Planets in 47 Tucanae}",
      journal = {\aj},
         year = 2017,
        month = apr,
       volume = {153},
       number = {4},
          eid = {187},
        pages = {187},
          doi = {10.3847/1538-3881/aa647c},
archivePrefix = {arXiv},
       eprint = {1703.06136},
 primaryClass = {astro-ph.EP},
       adsurl = {https://ui.adsabs.harvard.edu/abs/2017AJ....153..187M}
}

@ARTICLE{Weldrake2005,
       author = {{Weldrake}, David T.~F. and {Sackett}, Penny D. and {Bridges}, Terry J. and {Freeman}, Kenneth C.},
        title = "{An Absence of Hot Jupiter Planets in 47 Tucanae: Results of a Wide-Field Transit Search}",
      journal = {\apj},
         year = 2005,
        month = feb,
       volume = {620},
       number = {2},
        pages = {1043-1051},
          doi = {10.1086/427258},
archivePrefix = {arXiv},
       eprint = {astro-ph/0411233},
 primaryClass = {astro-ph},
       adsurl = {https://ui.adsabs.harvard.edu/abs/2005ApJ...620.1043W}
}

@ARTICLE{Sigurdsson2003,
       author = {{Sigurdsson}, Steinn and {Richer}, Harvey B. and {Hansen}, Brad M. and {Stairs}, Ingrid H. and {Thorsett}, Stephen E.},
        title = "{A Young White Dwarf Companion to Pulsar B1620-26: Evidence for Early Planet Formation}",
      journal = {Science},
         year = 2003,
        month = jul,
       volume = {301},
       number = {5630},
        pages = {193-196},
          doi = {10.1126/science.1086326},
archivePrefix = {arXiv},
       eprint = {astro-ph/0307339},
 primaryClass = {astro-ph},
       adsurl = {https://ui.adsabs.harvard.edu/abs/2003Sci...301..193S}
}

@ARTICLE{Gilliland2000,
       author = {{Gilliland}, Ronald L. and {Brown}, T.~M. and {Guhathakurta}, P. and {Sarajedini}, A. and {Milone}, E.~F. and {Albrow}, M.~D. and {Baliber}, N.~R. and {Bruntt}, H. and {Burrows}, A. and {Charbonneau}, D. and {Choi}, P. and {Cochran}, W.~D. and {Edmonds}, P.~D. and {Frandsen}, S. and {Howell}, J.~H. and {Lin}, D.~N.~C. and {Marcy}, G.~W. and {Mayor}, M. and {Naef}, D. and {Sigurdsson}, S. and {Stagg}, C.~R. and {Vandenberg}, D.~A. and {Vogt}, S.~S. and {Williams}, M.~D.},
        title = "{A Lack of Planets in 47 Tucanae from a Hubble Space Telescope Search}",
      journal = {\apjl},
         year = 2000,
        month = dec,
       volume = {545},
       number = {1},
        pages = {L47-L51},
          doi = {10.1086/317334},
       adsurl = {https://ui.adsabs.harvard.edu/abs/2000ApJ...545L..47G}
}

@misc{Bushouse2023,
       author = {{Bushouse}, Howard and {Eisenhamer}, Jonathan and {Dencheva}, Nadia and {Davies}, James and {Greenfield}, Perry and {Morrison}, Jane and {Hodge}, Phil and {Simon}, Bernie and {Grumm}, David and {Droettboom}, Michael and {Slavich}, Edward and {Sosey}, Megan and {Pauly}, Tyler and {Miller}, Todd and {Jedrzejewski}, Robert and {Hack}, Warren and {Davis}, David and {Crawford}, Steven and {Law}, David and {Gordon}, Karl and {Regan}, Michael and {Cara}, Mihai and {MacDonald}, Ken and {Bradley}, Larry and {Shanahan}, Clare and {Jamieson}, William and {Teodoro}, Mairan and {Williams}, Thomas},
        title = "{JWST Calibration Pipeline}",
         year = 2023,
        month = mar,
          eid = {10.5281/zenodo.7714020},
          doi = {10.5281/zenodo.7714020},
      version = {1.9.6},
    publisher = {Zenodo},
       adsurl = {https://ui.adsabs.harvard.edu/abs/2023zndo...7714020B}
}

@ARTICLE{FruchterHook2002,
       author = {{Fruchter}, A.~S. and {Hook}, R.~N.},
        title = "{Drizzle: A Method for the Linear Reconstruction of Undersampled Images}",
      journal = {\pasp},
         year = 2002,
        month = feb,
       volume = {114},
       number = {792},
        pages = {144-152},
          doi = {10.1086/338393},
archivePrefix = {arXiv},
       eprint = {astro-ph/9808087},
 primaryClass = {astro-ph},
       adsurl = {https://ui.adsabs.harvard.edu/abs/2002PASP..114..144F}
}

@TECHREPORT{Anderson2009,
       author = {{Anderson}, Jay},
        title = "{Dither Patterns for NIRCam Imaging}",
  institution = {STScI},
         year = 2009,
       number = {Technical Report JWST-STScI-001738},
 howpublished = {Technical Report JWST-STScI-001738},
       adsurl = {https://ui.adsabs.harvard.edu/abs/2009jwst.rept.1738A}
}

@TECHREPORT{Anderson2011,
       author = {{Anderson}, Jay},
        title = "{NIRCam Dithering Strategies I: Least Square Approach to Image Combination}",
  institution = {STScI},
         year = 2011,
       number = {Technical Report JWST-STScI-002199},
 howpublished = {Technical Report JWST-STScI-002199},
       adsurl = {https://ui.adsabs.harvard.edu/abs/2011jwst.rept.2199A}
}

@ARTICLE{Beichman2014,
       author = {{Beichman}, Charles and {Benneke}, Bjoern and {Knutson}, Heather and {Smith}, Roger and {Dressing}, Courtney and {Latham}, David and {Deming}, Drake and {Lunine}, Jonathan and {Lagage}, Pierre-Olivier and {Sozzetti}, Alessandro and {Beichman}, Charles and {Sing}, David and {Kempton}, Eliza and {Ricker}, George and {Bean}, Jacob and {Kreidberg}, Laura and {Bouwman}, Jeroen and {Crossfield}, Ian and {Christiansen}, Jessie and {Ciardi}, David and {Fortney}, Jonathan and {Albert}, Lo{\"\i}c and {Doyon}, Ren{\'e} and {Rieke}, Marcia and {Rieke}, George and {Clampin}, Mark and {Greenhouse}, Matt and {Goudfrooij}, Paul and {Hines}, Dean and {Keyes}, Tony and {Lee}, Janice and {McCullough}, Peter and {Robberto}, Massimo and {Stansberry}, John and {Valenti}, Jeff and {Deroo}, Pieter D. and {Mandell}, Avi and {Ressler}, Michael E. and {Shporer}, Avi and {Swain}, Mark and {Vasisht}, Gautam and {Carey}, Sean and {Krick}, Jessica and {Birkmann}, Stephan and {Ferruit}, Pierre and {Giardino}, Giovanna and {Greene}, Tom and {Howell}, Steve},
        title = "{Observations of Transiting Exoplanets with the James Webb Space Telescope (JWST), Publications of the Astronomical Society of the Pacific (PASP), December 2014}",
      journal = {arXiv e-prints},
         year = 2014,
        month = nov,
          eid = {arXiv:1411.1754},
        pages = {arXiv:1411.1754},
          doi = {10.48550/arXiv.1411.1754},
archivePrefix = {arXiv},
       eprint = {1411.1754},
 primaryClass = {astro-ph.IM},
       adsurl = {https://ui.adsabs.harvard.edu/abs/2014arXiv1411.1754B}
}

@ARTICLE{Schlawin2023,
       author = {{Schlawin}, Everett and {Beatty}, Thomas and {Brooks}, Brian and {Nikolov}, Nikolay K. and {Greene}, Thomas P. and {Espinoza}, N{\'e}stor and {Glidic}, Kayli and {Baka}, Keith and {Egami}, Eiichi and {Stansberry}, John and {Boyer}, Martha and {Gennaro}, Mario and {Leisenring}, Jarron and {Hilbert}, Bryan and {Misselt}, Karl and {Kelly}, Doug and {Canipe}, Alicia and {Beichman}, Charles and {Correnti}, Matteo and {Knight}, J. Scott and {Jurling}, Alden and {Perrin}, Marshall D. and {Feinberg}, Lee D. and {McElwain}, Michael W. and {Bond}, Nicholas and {Ciardi}, David and {Kendrew}, Sarah and {Rieke}, Marcia},
        title = "{JWST NIRCam Defocused Imaging: Photometric Stability Performance and How It Can Sense Mirror Tilts}",
      journal = {\pasp},
         year = 2023,
        month = jan,
       volume = {135},
       number = {1043},
          eid = {018001},
        pages = {018001},
          doi = {10.1088/1538-3873/aca718},
archivePrefix = {arXiv},
       eprint = {2211.16727},
 primaryClass = {astro-ph.IM},
       adsurl = {https://ui.adsabs.harvard.edu/abs/2023PASP..135a8001S}
}

@ARTICLE{YusefZadeh2025,
       author = {{Yusef-Zadeh}, F. and {Bushouse}, H. and {Arendt}, R.~G. and {Wardle}, M. and {Michail}, J.~M. and {Chandler}, C.~J.},
        title = "{Nonstop Variability of Sgr A* Using JWST at 2.1 and 4.8 {\ensuremath{\mu}}m Wavelengths: Evidence for Distinct Populations of Faint and Bright Variable Emission}",
      journal = {\apjl},
         year = 2025,
        month = feb,
       volume = {980},
       number = {2},
          eid = {L35},
        pages = {L35},
          doi = {10.3847/2041-8213/ada88b},
archivePrefix = {arXiv},
       eprint = {2501.04096},
 primaryClass = {astro-ph.GA},
       adsurl = {https://ui.adsabs.harvard.edu/abs/2025ApJ...980L..35Y}
}

@ARTICLE{Rosenthal2025,
       author = {{Rosenthal}, Alexandra C. and {Ransom}, Scott M. and {Corcoran}, Kyle A. and {DeCesar}, Megan E. and {Freire}, Paulo C.~C. and {Hessels}, Jason W.~T. and {Keith}, Michael J. and {Lynch}, Ryan S. and {Lyne}, Andrew and {Nice}, David J. and {Stairs}, Ingrid H. and {Stappers}, Ben and {Strader}, Jay and {Thorsett}, Stephen E. and {Urquhart}, Ryan},
        title = "{A 34 yr Timing Solution of the Redback Millisecond Pulsar Terzan 5A}",
      journal = {\apj},
         year = 2025,
        month = apr,
       volume = {982},
       number = {2},
          eid = {170},
        pages = {170},
          doi = {10.3847/1538-4357/adb8cd},
archivePrefix = {arXiv},
       eprint = {2410.21648},
 primaryClass = {astro-ph.HE},
       adsurl = {https://ui.adsabs.harvard.edu/abs/2025ApJ...982..170R}
}

@ARTICLE{Dolphin2000,
       author = {{Dolphin}, Andrew E.},
        title = "{WFPC2 Stellar Photometry with HSTPHOT}",
      journal = {\pasp},
         year = 2000,
        month = oct,
       volume = {112},
       number = {776},
        pages = {1383-1396},
          doi = {10.1086/316630},
archivePrefix = {arXiv},
       eprint = {astro-ph/0006217},
 primaryClass = {astro-ph},
       adsurl = {https://ui.adsabs.harvard.edu/abs/2000PASP..112.1383D}
}

@misc{Dolphin2016,
       author = {{Dolphin}, Andrew},
        title = "{DOLPHOT: Stellar photometry}",
 howpublished = {Astrophysics Source Code Library, record ascl:1608.013},
         year = 2016,
        month = aug,
          eid = {ascl:1608.013},
archivePrefix = {ascl},
       eprint = {1608.013},
       adsurl = {https://ui.adsabs.harvard.edu/abs/2016ascl.soft08013D}
}

@ARTICLE{Bahramian2013,
       author = {{Bahramian}, Arash and {Heinke}, Craig O. and {Sivakoff}, Gregory R. and {Gladstone}, Jeanette C.},
        title = "{Stellar Encounter Rate in Galactic Globular Clusters}",
      journal = {\apj},
         year = 2013,
        month = apr,
       volume = {766},
       number = {2},
          eid = {136},
        pages = {136},
          doi = {10.1088/0004-637X/766/2/136},
archivePrefix = {arXiv},
       eprint = {1302.2549},
 primaryClass = {astro-ph.HE},
       adsurl = {https://ui.adsabs.harvard.edu/abs/2013ApJ...766..136B}
}

@ARTICLE{Tam2011,
       author = {{Tam}, P.~H.~T. and {Kong}, A.~K.~H. and {Hui}, C.~Y. and {Cheng}, K.~S. and {Li}, C. and {Lu}, T.-N.},
        title = "{Gamma-ray Emission from the Globular Clusters Liller 1, M80, NGC 6139, NGC 6541, NGC 6624, and NGC 6752}",
      journal = {\apj},
         year = 2011,
        month = mar,
       volume = {729},
       number = {2},
          eid = {90},
        pages = {90},
          doi = {10.1088/0004-637X/729/2/90},
archivePrefix = {arXiv},
       eprint = {1101.4106},
 primaryClass = {astro-ph.HE},
       adsurl = {https://ui.adsabs.harvard.edu/abs/2011ApJ...729...90T}
}

@ARTICLE{Kumawat2025,
       author = {{Kumawat}, Gourav and {Heinke}, Craig O. and {Zhao}, Jiaqi and {Bahramian}, Arash and {Cohn}, Haldan N. and {Lugger}, Phyllis M.},
        title = "{A Comprehensive Analysis of X-Ray Sources in Terzan 5 Using Chandra Observations}",
      journal = {\apj},
         year = 2025,
        month = sep,
       volume = {990},
       number = {2},
          eid = {218},
        pages = {218},
          doi = {10.3847/1538-4357/adf724},
archivePrefix = {arXiv},
       eprint = {2508.02526},
 primaryClass = {astro-ph.HE},
       adsurl = {https://ui.adsabs.harvard.edu/abs/2025ApJ...990..218K}
}

@ARTICLE{Pallanca2025,
       author = {{Pallanca}, Cristina and {Ferraro}, Francesco R. and {Lanzoni}, Barbara and {Cadelano}, Mario and {Heinke}, Craig O. and {van den Berg}, Maureen and {Homan}, Jeroen and {Crociati}, Chiara and {Guillot}, Sebastien},
        title = "{Potential discovery of the long-sought optical counterpart to the Rapid Burster in the bulge fossil fragment Liller 1}",
      journal = {\aap},
         year = 2025,
        month = nov,
       volume = {703},
          eid = {A182},
        pages = {A182},
          doi = {10.1051/0004-6361/202556238},
archivePrefix = {arXiv},
       eprint = {2509.10640},
 primaryClass = {astro-ph.SR},
       adsurl = {https://ui.adsabs.harvard.edu/abs/2025A&A...703A.182P}
}

@ARTICLE{Court2018,
       author = {{Court}, J.~M.~C. and {Altamirano}, D. and {Albayati}, A.~C. and {Sanna}, A. and {Belloni}, T. and {Overton}, T. and {Degenaar}, N. and {Wijnands}, R. and {Yamaoka}, K. and {Hill}, A.~B. and {Knigge}, C.},
        title = "{The evolution of X-ray bursts in the `Bursting Pulsar' GRO J1744-28}",
      journal = {\mnras},
         year = 2018,
        month = dec,
       volume = {481},
       number = {2},
        pages = {2273-2298},
          doi = {10.1093/mnras/sty2312},
archivePrefix = {arXiv},
       eprint = {1808.06916},
 primaryClass = {astro-ph.HE},
       adsurl = {https://ui.adsabs.harvard.edu/abs/2018MNRAS.481.2273C}
}

@ARTICLE{Crisp2025,
       author = {{Crisp}, Alison L. and {Kl{\"u}ter}, Jonas and {Newman}, Marz L. and {Penny}, Matthew T. and {Beatty}, Thomas G. and {Cleeves}, L. Ilsedore and {Collins}, Karen A. and {Johnson}, Jennifer A. and {Johnson}, Marshall C. and {Lund}, Michael B. and {Mart{\'\i}nez-V{\'a}zquez}, Clara E. and {Ness}, Melissa K. and {Rodriguez}, Joseph E. and {Siverd}, Robert and {Stevens}, Daniel J. and {Villanueva}, Steven and {Ziegler}, Carl},
        title = "{The Multiband Imaging Survey for High-alpha PlanetS (MISHAPS). I. Preliminary Constraints on the Occurrence Rate of Hot Jupiters in 47 Tucanae}",
      journal = {\aj},
         year = 2025,
        month = aug,
       volume = {170},
       number = {2},
          eid = {104},
        pages = {104},
          doi = {10.3847/1538-3881/ade054},
archivePrefix = {arXiv},
       eprint = {2412.09705},
 primaryClass = {astro-ph.EP},
       adsurl = {https://ui.adsabs.harvard.edu/abs/2025AJ....170..104C}
}

@ARTICLE{Wirth2025,
       author = {{Wirth}, James A. and {Clarke}, Cathie J. and {Winter}, Andrew J.},
        title = "{Hot Jupiter formation in dense stellar clusters: a Monte Carlo model applied to 47 Tucanae}",
      journal = {\mnras},
         year = 2025,
        month = sep,
       volume = {542},
       number = {3},
        pages = {1761-1775},
          doi = {10.1093/mnras/staf1325},
archivePrefix = {arXiv},
       eprint = {2508.08406},
 primaryClass = {astro-ph.EP},
       adsurl = {https://ui.adsabs.harvard.edu/abs/2025MNRAS.542.1761W}
}

@ARTICLE{FischerValenti2005,
       author = {{Fischer}, Debra A. and {Valenti}, Jeff},
        title = "{The Planet-Metallicity Correlation}",
      journal = {\apj},
         year = 2005,
        month = apr,
       volume = {622},
       number = {2},
        pages = {1102-1117},
          doi = {10.1086/428383},
       adsurl = {https://ui.adsabs.harvard.edu/abs/2005ApJ...622.1102F}
}

@ARTICLE{Kremer2018,
       author = {{Kremer}, Kyle and {Chatterjee}, Sourav and {Breivik}, Katelyn and {Rodriguez}, Carl L. and {Larson}, Shane L. and {Rasio}, Frederic A.},
        title = "{LISA Sources in Milky Way Globular Clusters}",
      journal = {\prl},
         year = 2018,
        month = may,
       volume = {120},
       number = {19},
          eid = {191103},
        pages = {191103},
          doi = {10.1103/PhysRevLett.120.191103},
archivePrefix = {arXiv},
       eprint = {1802.05661},
 primaryClass = {astro-ph.HE},
       adsurl = {https://ui.adsabs.harvard.edu/abs/2018PhRvL.120s1103K}
}

@ARTICLE{Edmonds2001,
       author = {{Edmonds}, Peter D. and {Gilliland}, Ronald L. and {Heinke}, Craig O. and {Grindlay}, Jonathan E. and {Camilo}, Fernando},
        title = "{Optical Detection of a Variable Millisecond Pulsar Companion in 47 Tucanae}",
      journal = {\apjl},
         year = 2001,
        month = aug,
       volume = {557},
       number = {1},
        pages = {L57-L60},
          doi = {10.1086/323122},
archivePrefix = {arXiv},
       eprint = {astro-ph/0107096},
 primaryClass = {astro-ph},
       adsurl = {https://ui.adsabs.harvard.edu/abs/2001ApJ...557L..57E}
}

@ARTICLE{Edmonds2002,
       author = {{Edmonds}, Peter D. and {Gilliland}, Ronald L. and {Camilo}, Fernando and {Heinke}, Craig O. and {Grindlay}, Jonathan E.},
        title = "{A Millisecond Pulsar Optical Counterpart with Large-Amplitude Variability in the Globular Cluster 47 Tucanae}",
      journal = {\apj},
         year = 2002,
        month = nov,
       volume = {579},
       number = {2},
        pages = {741-751},
          doi = {10.1086/342985},
archivePrefix = {arXiv},
       eprint = {astro-ph/0207426},
 primaryClass = {astro-ph},
       adsurl = {https://ui.adsabs.harvard.edu/abs/2002ApJ...579..741E}
}

@ARTICLE{FigueraJaimes2024,
       author = {{Figuera Jaimes}, R. and {Catelan}, M. and {Horne}, K. and {Skottfelt}, J. and {Snodgrass}, C. and {Dominik}, M. and {J{\o}rgensen}, U.~G. and {Southworth}, J. and {Hundertmark}, M. and {Longa-Pe{\~n}a}, P. and {Sajadian}, S. and {Tregolan-Reed}, J. and {Hinse}, T.~C. and {Andersen}, M.~I. and {Bonavita}, M. and {Bozza}, V. and {Burgdorf}, M.~J. and {Haikala}, L. and {Khalouei}, E. and {Korhonen}, H. and {Peixinho}, N. and {Rabus}, M. and {Rahvar}, S.},
        title = "{Digging deeper into the dense Galactic globular cluster Terzan 5 with electron-multiplying CCDs: Variable star detection and new discoveries}",
      journal = {\aap},
         year = 2024,
        month = sep,
       volume = {689},
          eid = {A108},
        pages = {A108},
          doi = {10.1051/0004-6361/202347406},
archivePrefix = {arXiv},
       eprint = {2406.18733},
 primaryClass = {astro-ph.SR},
       adsurl = {https://ui.adsabs.harvard.edu/abs/2024A&A...689A.108F}
}

@ARTICLE{Ye2020,
       author = {{Ye}, Claire S. and {Fong}, Wen-fai and {Kremer}, Kyle and {Rodriguez}, Carl L. and {Chatterjee}, Sourav and {Fragione}, Giacomo and {Rasio}, Frederic A.},
        title = "{On the Rate of Neutron Star Binary Mergers from Globular Clusters}",
      journal = {\apjl},
         year = 2020,
        month = jan,
       volume = {888},
       number = {1},
          eid = {L10},
        pages = {L10},
          doi = {10.3847/2041-8213/ab5dc5},
archivePrefix = {arXiv},
       eprint = {1910.10740},
 primaryClass = {astro-ph.HE},
       adsurl = {https://ui.adsabs.harvard.edu/abs/2020ApJ...888L..10Y}
}

@ARTICLE{Kaluzny2004,
       author = {{Kaluzny}, J. and {Olech}, A. and {Thompson}, I.~B. and {Pych}, W. and {Krzemi{\'n}ski}, W. and {Schwarzenberg-Czerny}, A.},
        title = "{Cluster AgeS ExperimentCatalog of variable stars in the globular cluster {\ensuremath{\omega}} Centauri}",
      journal = {\aap},
         year = 2004,
        month = sep,
       volume = {424},
        pages = {1101-1110},
          doi = {10.1051/0004-6361:20047137},
archivePrefix = {arXiv},
       eprint = {astro-ph/0406456},
 primaryClass = {astro-ph},
       adsurl = {https://ui.adsabs.harvard.edu/abs/2004A&A...424.1101K}
}

@ARTICLE{Origlia2019,
       author = {{Origlia}, L. and {Mucciarelli}, A. and {Fiorentino}, G. and {Ferraro}, F.~R. and {Dalessandro}, E. and {Lanzoni}, B. and {Rich}, R.~M. and {Massari}, D. and {Contreras}, R.~R. and {Matsunaga}, N. and {-}},
        title = "{Variable stars in Terzan 5: additional evidence of multi-age and multi-iron stellar populations}",
      journal = {arXiv e-prints},
         year = 2018,
        month = dec,
          eid = {arXiv:1812.04325},
        pages = {arXiv:1812.04325},
          doi = {10.48550/arXiv.1812.04325},
archivePrefix = {arXiv},
       eprint = {1812.04325},
 primaryClass = {astro-ph.GA},
       adsurl = {https://ui.adsabs.harvard.edu/abs/2018arXiv181204325O}
}

@ARTICLE{Milone2026,
       author = {{Milone}, A.~P. and {Marino}, A.~F. and {Cordoni}, G. and {Dondoglio}, E. and {Legnardi}, M.~V. and {Ziliotto}, T. and {Bortolan}, E. and {Muratore}, F. and {D'Antona}, F. and {Renzini}, A. and {Girardi}, G. and {Gorza}, L. and {Mastrobuono-Battisti}, A. and {Ventura}, C. and {Ventura}, P. and {Altomonte}, V. and {Bisigello}, L. and {Cavecchi}, Y. and {Dell'Agli}, F. and {Dotter}, A. and {Lagioia}, E.~P. and {Li}, C. and {Lionetto}, S. and {Marchuk}, A. and {Qi}, J. and {Rodighiero}, G. and {Tailo}, M. and {Wirth}, H.},
        title = "{Globular Clusters in the Time of the JWST. I. Survey Design and First Results on Multiple Populations and Beyond}",
      journal = {\mnras},
         year = 2026,
        month = aug,
          doi = {10.1093/mnras/stag1597},
archivePrefix = {arXiv},
       eprint = {2606.20979},
 primaryClass = {astro-ph.GA},
       adsurl = {https://ui.adsabs.harvard.edu/abs/2026MNRAS.tmp.1494M}
}

@ARTICLE{Deming2013,
       author = {{Deming}, Drake and {Wilkins}, Ashlee and {McCullough}, Peter and {Burrows}, Adam and {Fortney}, Jonathan J. and {Agol}, Eric and {Dobbs-Dixon}, Ian and {Madhusudhan}, Nikku and {Crouzet}, Nicolas and {Desert}, Jean-Michel and {Gilliland}, Ronald L. and {Haynes}, Korey and {Knutson}, Heather A. and {Line}, Michael and {Magic}, Zazralt and {Mandell}, Avi M. and {Ranjan}, Sukrit and {Charbonneau}, David and {Clampin}, Mark and {Seager}, Sara and {Showman}, Adam P.},
        title = "{Infrared Transmission Spectroscopy of the Exoplanets HD 209458b and XO-1b Using the Wide Field Camera-3 on the Hubble Space Telescope}",
      journal = {\apj},
         year = 2013,
        month = sep,
       volume = {774},
       number = {2},
          eid = {95},
        pages = {95},
          doi = {10.1088/0004-637X/774/2/95},
archivePrefix = {arXiv},
       eprint = {1302.1141},
 primaryClass = {astro-ph.EP},
       adsurl = {https://ui.adsabs.harvard.edu/abs/2013ApJ...774...95D}
}

@ARTICLE{TovarMendoza2023,
       author = {{Tovar Mendoza}, Guadalupe and {Wilson}, Robert F. and {Youngblood}, Allison and {Vega}, Laura D. and {Barclay}, Thomas and {Davenport}, James R.~A. and {Ealy}, Jordan},
        title = "{Enabling Stellar Flare Science in the Roman Galactic Bulge Survey: Cadence, Filters, and the Read-Out Strategy Matter}",
      journal = {arXiv e-prints},
         year = 2023,
        month = jul,
          eid = {arXiv:2307.05806},
        pages = {arXiv:2307.05806},
          doi = {10.48550/arXiv.2307.05806},
archivePrefix = {arXiv},
       eprint = {2307.05806},
 primaryClass = {astro-ph.IM},
       adsurl = {https://ui.adsabs.harvard.edu/abs/2023arXiv230705806T}
}

@ARTICLE{Sokolovsky2017,
       author = {{Sokolovsky}, K.~V. and {Gavras}, P. and {Karampelas}, A. and {Antipin}, S.~V. and {Bellas-Velidis}, I. and {Benni}, P. and {Bonanos}, A.~Z. and {Burdanov}, A.~Y. and {Derlopa}, S. and {Hatzidimitriou}, D. and {Khokhryakova}, A.~D. and {Kolesnikova}, D.~M. and {Korotkiy}, S.~A. and {Lapukhin}, E.~G. and {Moretti}, M.~I. and {Popov}, A.~A. and {Pouliasis}, E. and {Samus}, N.~N. and {Spetsieri}, Z. and {Veselkov}, S.~A. and {Volkov}, K.~V. and {Yang}, M. and {Zubareva}, A.~M.},
        title = "{Comparative performance of selected variability detection techniques in photometric time series data}",
      journal = {\mnras},
         year = 2017,
        month = jan,
       volume = {464},
       number = {1},
        pages = {274-292},
          doi = {10.1093/mnras/stw2262},
archivePrefix = {arXiv},
       eprint = {1609.01716},
 primaryClass = {astro-ph.IM},
       adsurl = {https://ui.adsabs.harvard.edu/abs/2017MNRAS.464..274S}
}

@ARTICLE{Dai2016,
       author = {{Dai}, S. and {Johnston}, S. and {Bell}, M.~E. and {Coles}, W.~A. and {Hobbs}, G. and {Ekers}, R.~D. and {Lenc}, E.},
        title = "{Detecting pulsars with interstellar scintillation in variance images}",
      journal = {\mnras},
         year = 2016,
        month = nov,
       volume = {462},
       number = {3},
        pages = {3115-3122},
          doi = {10.1093/mnras/stw1871},
archivePrefix = {arXiv},
       eprint = {1607.07740},
 primaryClass = {astro-ph.IM},
       adsurl = {https://ui.adsabs.harvard.edu/abs/2016MNRAS.462.3115D}
}

@ARTICLE{Su2022,
       author = {{Su}, Kate Y.~L. and {Rieke}, G.~H. and {Marengo}, M. and {Schlawin}, Everett},
        title = "{Accurate Photometry of Saturated Stars Using the Point-spread-function Wing Technique with Spitzer}",
      journal = {\aj},
         year = 2022,
        month = feb,
       volume = {163},
       number = {2},
          eid = {46},
        pages = {46},
          doi = {10.3847/1538-3881/ac3b5e},
archivePrefix = {arXiv},
       eprint = {2111.10054},
 primaryClass = {astro-ph.SR},
       adsurl = {https://ui.adsabs.harvard.edu/abs/2022AJ....163...46S}
}

@MISC{Gilliland2004,
       author = {{Gilliland}, R.~L.},
        title = "{ACS CCD Gains, Full Well Depths, and Linearity up to and Beyond Saturation}",
 howpublished = {Instrument Science Report ACS 2004-01, 18 pages},
         year = 2004,
        month = jan,
        pages = {1},
       adsurl = {https://ui.adsabs.harvard.edu/abs/2004acs..rept....1G}
}

@ARTICLE{White2017,
       author = {{White}, T.~R. and {Pope}, B.~J.~S. and {Antoci}, V. and {P{\'a}pics}, P.~I. and {Aerts}, C. and {Gies}, D.~R. and {Gordon}, K. and {Huber}, D. and {Schaefer}, G.~H. and {Aigrain}, S. and {Albrecht}, S. and {Barclay}, T. and {Barentsen}, G. and {Beck}, P.~G. and {Bedding}, T.~R. and {Fredslund Andersen}, M. and {Grundahl}, F. and {Howell}, S.~B. and {Ireland}, M.~J. and {Murphy}, S.~J. and {Nielsen}, M.~B. and {Silva Aguirre}, V. and {Tuthill}, P.~G.},
        title = "{Beyond the Kepler/K2 bright limit: variability in the seven brightest members of the Pleiades}",
      journal = {\mnras},
         year = 2017,
        month = nov,
       volume = {471},
       number = {3},
        pages = {2882-2901},
          doi = {10.1093/mnras/stx1050},
archivePrefix = {arXiv},
       eprint = {1708.07462},
 primaryClass = {astro-ph.SR},
       adsurl = {https://ui.adsabs.harvard.edu/abs/2017MNRAS.471.2882W}
}

@ARTICLE{Contreras2010,
       author = {{Contreras}, R. and {Catelan}, M. and {Smith}, H.~A. and {Pritzl}, B.~J. and {Borissova}, J. and {Kuehn}, C.~A.},
        title = "{Time-series Photometry of Globular Clusters: M62 (NGC 6266), the Most RR Lyrae-rich Globular Cluster in the Galaxy?}",
      journal = {\aj},
         year = 2010,
        month = dec,
       volume = {140},
       number = {6},
        pages = {1766-1786},
          doi = {10.1088/0004-6256/140/6/1766},
archivePrefix = {arXiv},
       eprint = {1009.4206},
 primaryClass = {astro-ph.SR},
       adsurl = {https://ui.adsabs.harvard.edu/abs/2010AJ....140.1766C}
}

@ARTICLE{BaumgardtHilker2018,
       author = {{Baumgardt}, H. and {Hilker}, M.},
        title = "{A catalogue of masses, structural parameters, and velocity dispersion profiles of 112 Milky Way globular clusters}",
      journal = {\mnras},
         year = 2018,
        month = aug,
       volume = {478},
       number = {2},
        pages = {1520-1557},
          doi = {10.1093/mnras/sty1057},
archivePrefix = {arXiv},
       eprint = {1804.08359},
 primaryClass = {astro-ph.GA},
       adsurl = {https://ui.adsabs.harvard.edu/abs/2018MNRAS.478.1520B}
}

@ARTICLE{Daylan2016,
       author = {{Daylan}, Tansu and {Finkbeiner}, Douglas P. and {Hooper}, Dan and {Linden}, Tim and {Portillo}, Stephen K.~N. and {Rodd}, Nicholas L. and {Slatyer}, Tracy R.},
        title = "{The characterization of the gamma-ray signal from the central Milky Way: A case for annihilating dark matter}",
      journal = {Physics of the Dark Universe},
         year = 2016,
        month = jun,
       volume = {12},
        pages = {1-23},
          doi = {10.1016/j.dark.2015.12.005},
archivePrefix = {arXiv},
       eprint = {1402.6703},
 primaryClass = {astro-ph.HE},
       adsurl = {https://ui.adsabs.harvard.edu/abs/2016PDU....12....1D}
}

@ARTICLE{vandenEijnden2024,
       author = {{van den Eijnden}, J. and {Robins}, D. and {Sharma}, R. and {S{\'a}nchez-Fern{\'a}ndez}, C. and {Russell}, T.~D. and {Degenaar}, N. and {Miller-Jones}, J.~C.~A. and {Maccarone}, T.},
        title = "{The variable radio jet of the accreting neutron star the Rapid Burster}",
      journal = {\mnras},
         year = 2024,
        month = sep,
       volume = {533},
       number = {1},
        pages = {756-770},
          doi = {10.1093/mnras/stae1826},
archivePrefix = {arXiv},
       eprint = {2405.19827},
 primaryClass = {astro-ph.HE},
       adsurl = {https://ui.adsabs.harvard.edu/abs/2024MNRAS.533..756V}
}

@ARTICLE{Evans2024,
       author = {{Evans}, Ian N. and {Evans}, Janet D. and {Mart{\'\i}nez-Galarza}, J. Rafael and {Miller}, Joseph B. and {Primini}, Francis A. and {Azadi}, Mojegan and {Burke}, Douglas J. and {Civano}, Francesca M. and {D'Abrusco}, Raffaele and {Fabbiano}, Giuseppina and {Graessle}, Dale E. and {Grier}, John D. and {Houck}, John C. and {Lauer}, Jennifer and {McCollough}, Michael L. and {Nowak}, Michael A. and {Plummer}, David A. and {Rots}, Arnold H. and {Siemiginowska}, Aneta and {Tibbetts}, Michael S.},
        title = "{The Chandra Source Catalog Release 2 Series}",
      journal = {\apjs},
         year = 2024,
        month = oct,
       volume = {274},
       number = {2},
          eid = {22},
        pages = {22},
          doi = {10.3847/1538-4365/ad6319},
archivePrefix = {arXiv},
       eprint = {2407.10799},
 primaryClass = {astro-ph.HE},
       adsurl = {https://ui.adsabs.harvard.edu/abs/2024ApJS..274...22E}
}

@INPROCEEDINGS{VerbuntHut1987,
       author = {{Verbunt}, F. and {Hut}, P.},
        title = "{The Globular Cluster Population of X-Ray Binaries}",
    booktitle = {The Origin and Evolution of Neutron Stars},
         year = 1987,
       editor = {{Helfand}, D.~J. and {Huang}, J.-H.},
       series = {IAU Symposium},
       volume = {125},
        month = jan,
        pages = {187},
       adsurl = {https://ui.adsabs.harvard.edu/abs/1987IAUS..125..187V}
}

@INPROCEEDINGS{Roberts2013,
       author = {{Roberts}, Mallory S.~E.},
        title = "{Surrounded by spiders! New black widows and redbacks in the Galactic field}",
    booktitle = {Neutron Stars and Pulsars: Challenges and Opportunities after 80 years},
         year = 2013,
       editor = {{van Leeuwen}, Joeri},
       series = {IAU Symposium},
       volume = {291},
        month = mar,
        pages = {127-132},
          doi = {10.1017/S174392131202337X},
archivePrefix = {arXiv},
       eprint = {1210.6903},
 primaryClass = {astro-ph.HE},
       adsurl = {https://ui.adsabs.harvard.edu/abs/2013IAUS..291..127R}
}

@ARTICLE{Bahramian2020,
       author = {{Bahramian}, Arash and {Strader}, Jay and {Miller-Jones}, James C.~A. and {Chomiuk}, Laura and {Heinke}, Craig O. and {Maccarone}, Thomas J. and {Pooley}, David and {Shishkovsky}, Laura and {Tudor}, Vlad and {Zhao}, Yue and {Li}, Kwan Lok and {Sivakoff}, Gregory R. and {Tremou}, Evangelia and {Buchner}, Johannes},
        title = "{The MAVERIC Survey: Chandra/ACIS Catalog of Faint X-Ray Sources in 38 Galactic Globular Clusters}",
      journal = {\apj},
         year = 2020,
        month = sep,
       volume = {901},
       number = {1},
          eid = {57},
        pages = {57},
          doi = {10.3847/1538-4357/aba51d},
archivePrefix = {arXiv},
       eprint = {2007.04581},
 primaryClass = {astro-ph.HE},
       adsurl = {https://ui.adsabs.harvard.edu/abs/2020ApJ...901...57B}
}

@ARTICLE{Wang2026,
       author = {{Wang}, Bo and {Liu}, Dongdong and {Guo}, Yunlang and {Chen}, Hailiang and {Chen}, Wen-Cong and {Han}, Zhanwen},
        title = "{Ultra-compact X-Ray Binaries: A Review}",
      journal = {Research in Astronomy and Astrophysics},
         year = 2026,
        month = nov,
       volume = {26},
       number = {11},
          eid = {112001},
        pages = {112001},
          doi = {10.1088/1674-4527/ae67a0},
archivePrefix = {arXiv},
       eprint = {2603.13813},
 primaryClass = {astro-ph.HE},
       adsurl = {https://ui.adsabs.harvard.edu/abs/2026RAA....26k2001W}
}

@ARTICLE{Shaw2025,
       author = {{Shaw}, A.~W. and {Kaplan}, D.~L. and {Gandhi}, P. and {Maccarone}, T.~J. and {Borowski}, E.~S. and {Britt}, C.~T. and {Buckley}, D.~A.~H. and {Burdge}, K.~B. and {Charles}, P.~A. and {Dhillon}, V.~S. and {French}, R.~G. and {Heinke}, C.~O. and {Hynes}, R.~I. and {Knigge}, C. and {Littlefair}, S.~P. and {Pawar}, Devraj and {Plotkin}, R.~M. and {Ressler}, M.~E. and {Santos-Sanz}, P. and {Shahbaz}, T. and {Sivakoff}, G.~R. and {Stevens}, A.~L.},
        title = "{Calibrating the Clock of JWST}",
      journal = {\aj},
         year = 2025,
        month = jan,
       volume = {169},
       number = {1},
          eid = {21},
        pages = {21},
          doi = {10.3847/1538-3881/ad8eb1},
archivePrefix = {arXiv},
       eprint = {2411.02238},
 primaryClass = {astro-ph.IM},
       adsurl = {https://ui.adsabs.harvard.edu/abs/2025AJ....169...21S}
}

@ARTICLE{BaumgardtVasiliev2021,
       author = {{Baumgardt}, H. and {Vasiliev}, E.},
        title = "{Accurate distances to Galactic globular clusters through a combination of Gaia EDR3, HST, and literature data}",
      journal = {\mnras},
         year = 2021,
        month = aug,
       volume = {505},
       number = {4},
        pages = {5957-5977},
          doi = {10.1093/mnras/stab1474},
archivePrefix = {arXiv},
       eprint = {2105.09526},
 primaryClass = {astro-ph.GA},
       adsurl = {https://ui.adsabs.harvard.edu/abs/2021MNRAS.505.5957B}
}

@ARTICLE{JohnsonPilachowski2010,
       author = {{Johnson}, Christian I. and {Pilachowski}, Catherine A.},
        title = "{Chemical Abundances for 855 Giants in the Globular Cluster Omega Centauri (NGC 5139)}",
      journal = {\apj},
         year = 2010,
        month = oct,
       volume = {722},
       number = {2},
        pages = {1373-1410},
          doi = {10.1088/0004-637X/722/2/1373},
archivePrefix = {arXiv},
       eprint = {1008.2232},
 primaryClass = {astro-ph.SR},
       adsurl = {https://ui.adsabs.harvard.edu/abs/2010ApJ...722.1373J}
}

@ARTICLE{Clontz2024,
       author = {{Clontz}, C. and {Seth}, A.~C. and {Dotter}, A. and {H{\"a}berle}, M. and {Nitschai}, M.~S. and {Neumayer}, N. and {Feldmeier-Krause}, A. and {Latour}, M. and {Wang}, Z. and {Souza}, S.~O. and {Kacharov}, N. and {Bellini}, A. and {Libralato}, M. and {Pechetti}, R. and {van de Ven}, G. and {Alfaro-Cuello}, M.},
        title = "{oMEGACat. IV. Constraining the Ages of Omega Centauri Subgiant Branch Stars with HST and MUSE}",
      journal = {\apj},
         year = 2024,
        month = dec,
       volume = {977},
       number = {1},
          eid = {14},
        pages = {14},
          doi = {10.3847/1538-4357/ad8621},
archivePrefix = {arXiv},
       eprint = {2409.13855},
 primaryClass = {astro-ph.GA},
       adsurl = {https://ui.adsabs.harvard.edu/abs/2024ApJ...977...14C}
}

@ARTICLE{Siegel2007,
       author = {{Siegel}, Michael H. and {Dotter}, Aaron and {Majewski}, Steven R. and {Sarajedini}, Ata and {Chaboyer}, Brian and {Nidever}, David L. and {Anderson}, Jay and {Mar{\'\i}n-Franch}, Antonio and {Rosenberg}, Alfred and {Bedin}, Luigi R. and {Aparicio}, Antonio and {King}, Ivan and {Piotto}, Giampaolo and {Reid}, I. Neill},
        title = "{The ACS Survey of Galactic Globular Clusters: M54 and Young Populations in the Sagittarius Dwarf Spheroidal Galaxy}",
      journal = {\apjl},
         year = 2007,
        month = sep,
       volume = {667},
       number = {1},
        pages = {L57-L60},
          doi = {10.1086/522003},
archivePrefix = {arXiv},
       eprint = {0708.0027},
 primaryClass = {astro-ph},
       adsurl = {https://ui.adsabs.harvard.edu/abs/2007ApJ...667L..57S}
}

@ARTICLE{Bellazzini2008,
       author = {{Bellazzini}, M. and {Ibata}, R.~A. and {Chapman}, S.~C. and {Mackey}, A.~D. and {Monaco}, L. and {Irwin}, M.~J. and {Martin}, N.~F. and {Lewis}, G.~F. and {Dalessandro}, E.},
        title = "{The Nucleus of the Sagittarius Dsph Galaxy and M54: a Window on the Process of Galaxy Nucleation}",
      journal = {\aj},
         year = 2008,
        month = sep,
       volume = {136},
       number = {3},
        pages = {1147-1170},
          doi = {10.1088/0004-6256/136/3/1147},
archivePrefix = {arXiv},
       eprint = {0807.0105},
 primaryClass = {astro-ph},
       adsurl = {https://ui.adsabs.harvard.edu/abs/2008AJ....136.1147B}
}

@ARTICLE{Rucinski2007,
       author = {{Rucinski}, Slavek M.},
        title = "{The short-period end of the contact binary period distribution based on the All-Sky Automated Survey}",
      journal = {\mnras},
         year = 2007,
        month = nov,
       volume = {382},
       number = {1},
        pages = {393-396},
          doi = {10.1111/j.1365-2966.2007.12377.x},
archivePrefix = {arXiv},
       eprint = {0708.3020},
 primaryClass = {astro-ph},
       adsurl = {https://ui.adsabs.harvard.edu/abs/2007MNRAS.382..393R}
}

@ARTICLE{DexterOLeary2014,
       author = {{Dexter}, Jason and {O'Leary}, Ryan M.},
        title = "{The Peculiar Pulsar Population of the Central Parsec}",
      journal = {\apjl},
         year = 2014,
        month = mar,
       volume = {783},
       number = {1},
          eid = {L7},
        pages = {L7},
          doi = {10.1088/2041-8205/783/1/L7},
archivePrefix = {arXiv},
       eprint = {1310.7022},
 primaryClass = {astro-ph.GA},
       adsurl = {https://ui.adsabs.harvard.edu/abs/2014ApJ...783L...7D}
}

@ARTICLE{Bartels2016,
       author = {{Bartels}, Richard and {Krishnamurthy}, Suraj and {Weniger}, Christoph},
        title = "{Strong Support for the Millisecond Pulsar Origin of the Galactic Center GeV Excess}",
      journal = {\prl},
         year = 2016,
        month = feb,
       volume = {116},
       number = {5},
          eid = {051102},
        pages = {051102},
          doi = {10.1103/PhysRevLett.116.051102},
archivePrefix = {arXiv},
       eprint = {1506.05104},
 primaryClass = {astro-ph.HE},
       adsurl = {https://ui.adsabs.harvard.edu/abs/2016PhRvL.116e1102B}
}

@ARTICLE{Moore2006,
       author = {{Moore}, Andrew C. and {Ninkov}, Zoran and {Forrest}, William J.},
        title = "{Quantum efficiency overestimation and deterministic cross talk resulting from interpixel capacitance}",
      journal = {Optical Engineering},
         year = 2006,
        month = jan,
       volume = {45},
       number = {7},
          eid = {076402},
        pages = {076402},
          doi = {10.1117/1.2219103},
       adsurl = {https://ui.adsabs.harvard.edu/abs/2006OptEn..45g6402M}
}

@ARTICLE{Penny2019,
       author = {{Penny}, Matthew T. and {Gaudi}, B. Scott and {Kerins}, Eamonn and {Rattenbury}, Nicholas J. and {Mao}, Shude and {Robin}, Annie C. and {Calchi Novati}, Sebastiano},
        title = "{Predictions of the WFIRST Microlensing Survey. I. Bound Planet Detection Rates}",
      journal = {\apjs},
         year = 2019,
        month = mar,
       volume = {241},
       number = {1},
          eid = {3},
        pages = {3},
          doi = {10.3847/1538-4365/aafb69},
archivePrefix = {arXiv},
       eprint = {1808.02490},
 primaryClass = {astro-ph.EP},
       adsurl = {https://ui.adsabs.harvard.edu/abs/2019ApJS..241....3P}
}

@ARTICLE{Mosby2020,
       author = {{Mosby}, Gregory and {Rauscher}, Bernard J. and {Bennett}, Chris and {Cheng}, Edward S. and {Cheung}, Stephanie and {Cillis}, Analia and {Content}, David and {Cottingham}, Dave and {Foltz}, Roger and {Gygax}, John and {Hill}, Robert J. and {Kruk}, Jeffrey W. and {Mah}, Jon and {Meier}, Lane and {Merchant}, Chris and {Miko}, Laddawan and {Piquette}, Eric C. and {Waczynski}, Augustyn and {Wen}, Yiting},
        title = "{Properties and characteristics of the Nancy Grace Roman Space Telescope H4RG-10 detectors}",
      journal = {Journal of Astronomical Telescopes, Instruments, and Systems},
         year = 2020,
        month = oct,
       volume = {6},
          eid = {046001},
        pages = {046001},
          doi = {10.1117/1.JATIS.6.4.046001},
archivePrefix = {arXiv},
       eprint = {2005.00505},
 primaryClass = {astro-ph.IM},
       adsurl = {https://ui.adsabs.harvard.edu/abs/2020JATIS...6d6001M}
}

@ARTICLE{Neumayer2020,
       author = {{Neumayer}, Nadine and {Seth}, Anil and {B{\"o}ker}, Torsten},
        title = "{Nuclear star clusters}",
      journal = {\aapr},
         year = 2020,
        month = jul,
       volume = {28},
       number = {1},
          eid = {4},
        pages = {4},
          doi = {10.1007/s00159-020-00125-0},
archivePrefix = {arXiv},
       eprint = {2001.03626},
 primaryClass = {astro-ph.GA},
       adsurl = {https://ui.adsabs.harvard.edu/abs/2020A&ARv..28....4N}
}

@ARTICLE{Braga2018,
       author = {{Braga}, V.~F. and {Stetson}, P.~B. and {Bono}, G. and {Dall'Ora}, M. and {Ferraro}, I. and {Fiorentino}, G. and {Iannicola}, G. and {Marconi}, M. and {Marengo}, M. and {Monson}, A.~J. and {Neeley}, J. and {Persson}, S.~E. and {Beaton}, R.~L. and {Buonanno}, R. and {Calamida}, A. and {Castellani}, M. and {Di Carlo}, E. and {Fabrizio}, M. and {Freedman}, W.~L. and {Inno}, L. and {Madore}, B.~F. and {Magurno}, D. and {Marchetti}, E. and {Marinoni}, S. and {Marrese}, P. and {Matsunaga}, N. and {Minniti}, D. and {Monelli}, M. and {Nonino}, M. and {Piersimoni}, A.~M. and {Pietrinferni}, A. and {Prada-Moroni}, P. and {Pulone}, L. and {Stellingwerf}, R. and {Tognelli}, E. and {Walker}, A.~R. and {Valenti}, E. and {Zoccali}, M.},
        title = "{On the RR Lyrae Stars in Globulars. V. The Complete Near-infrared (JHK $_{ s }$) Census of {\ensuremath{\omega}} Centauri RR Lyrae Variables}",
      journal = {\aj},
         year = 2018,
        month = mar,
       volume = {155},
       number = {3},
          eid = {137},
        pages = {137},
          doi = {10.3847/1538-3881/aaadab},
archivePrefix = {arXiv},
       eprint = {1802.03578},
 primaryClass = {astro-ph.SR},
       adsurl = {https://ui.adsabs.harvard.edu/abs/2018AJ....155..137B}
}

@ARTICLE{Lewin1976,
       author = {{Lewin}, W.~H.~G. and {Doty}, J. and {Clark}, G.~W. and {Rappaport}, S.~A. and {Bradt}, H.~V.~D. and {Doxsey}, R. and {Hearn}, D.~R. and {Hoffman}, J.~A. and {Jernigan}, J.~G. and {Li}, F.~K. and {Mayer}, W. and {McClintock}, J. and {Primini}, F. and {Richardson}, J.},
        title = "{The discovery of rapidly repetitive X-ray bursts from a new source in Scorpius.}",
      journal = {\apjl},
         year = 1976,
        month = jul,
       volume = {207},
        pages = {L95-L99},
          doi = {10.1086/182188},
       adsurl = {https://ui.adsabs.harvard.edu/abs/1976ApJ...207L..95L}
}

@ARTICLE{Desai2026,
       author = {{Desai}, M.~M. and {Burdge}, K.~B. and {Pallanca}, C. and {Rappaport}, S.},
        title = "{JWST Discovers the Infrared Counterpart of the Rapid Burster}",
      journal = {\apjl, submitted},
         year = 2026
}

@ARTICLE{Prsa2016,
       author = {{Pr{\v{s}}a}, A. and {Conroy}, K.~E. and {Horvat}, M. and {Pablo}, H. and {Kochoska}, A. and {Bloemen}, S. and {Giammarco}, J. and {Hambleton}, K.~M. and {Degroote}, P.},
        title = "{Physics Of Eclipsing Binaries. II. Toward the Increased Model Fidelity}",
      journal = {\apjs},
         year = 2016,
        month = dec,
       volume = {227},
       number = {2},
          eid = {29},
        pages = {29},
          doi = {10.3847/1538-4365/227/2/29},
archivePrefix = {arXiv},
       eprint = {1609.08135},
 primaryClass = {astro-ph.SR},
       adsurl = {https://ui.adsabs.harvard.edu/abs/2016ApJS..227...29P}
}

@ARTICLE{Conroy2020,
       author = {{Conroy}, Kyle E. and {Kochoska}, Angela and {Hey}, Daniel and {Pablo}, Herbert and {Hambleton}, Kelly M. and {Jones}, David and {Giammarco}, Joseph and {Abdul-Masih}, Michael and {Pr{\v{s}}a}, Andrej},
        title = "{Physics of Eclipsing Binaries. V. General Framework for Solving the Inverse Problem}",
      journal = {\apjs},
         year = 2020,
        month = oct,
       volume = {250},
       number = {2},
          eid = {34},
        pages = {34},
          doi = {10.3847/1538-4365/abb4e2},
archivePrefix = {arXiv},
       eprint = {2006.16951},
 primaryClass = {astro-ph.SR},
       adsurl = {https://ui.adsabs.harvard.edu/abs/2020ApJS..250...34C}
}

@ARTICLE{Zullo2026,
       author = {{Zullo}, G. and {Pallanca}, C. and {Ferraro}, F.~R. and {Lanzoni}, B. and {Origlia}, L. and {Massari}, D. and {Dalessandro}, E. and {Fanelli}, C. and {Cadelano}, M. and {Vesperini}, E. and {Crociati}, C. and {Rich}, R.~M. and {Valenti}, E.},
        title = "{The multi-age stellar populations of Terzan 5 as revealed by JWST}",
      journal = {\aap},
         year = 2026,
        month = may,
       volume = {709},
          eid = {A212},
        pages = {A212},
          doi = {10.1051/0004-6361/202659349},
archivePrefix = {arXiv},
       eprint = {2604.00098},
 primaryClass = {astro-ph.GA},
       adsurl = {https://ui.adsabs.harvard.edu/abs/2026A&A...709A.212Z}
}

@ARTICLE{ZulloInPrep,
       author = {{Zullo}, G. and {Ferraro}, F.~R. and {Pallanca}, C. and {Lanzoni}, B. and {Burdge}, K.~B. and {Caiazzo}, I. and {Desai}, M.~M. and {Dalessandro}, E. and {Chiappino}, L. and {Origlia}, L. and {Massari}, D. and {Fanelli}, C. and {Cadelano}, M.},
        title = "{The JWST view of the complex stellar population of the Bulge Fossil Fragment Liller 1}",
      journal = {in preparation},
         year = 2026
}

@ARTICLE{Massari2012,
       author = {{Massari}, Davide and {Mucciarelli}, Alessio and {Dalessandro}, Emanuele and {Ferraro}, Francesco R. and {Origlia}, Livia and {Lanzoni}, Barbara and {Beccari}, Giacomo and {Rich}, R. Michael and {Valenti}, Elena and {Ransom}, Scott M.},
        title = "{High-resolution Reddening Map in the Direction of the Stellar System Terzan 5}",
      journal = {\apjl},
         year = 2012,
        month = aug,
       volume = {755},
       number = {2},
          eid = {L32},
        pages = {L32},
          doi = {10.1088/2041-8205/755/2/L32},
archivePrefix = {arXiv},
       eprint = {1209.0990},
 primaryClass = {astro-ph.GA},
       adsurl = {https://ui.adsabs.harvard.edu/abs/2012ApJ...755L..32M}
}

@ARTICLE{ElBadry2022,
       author = {{El-Badry}, Kareem and {Conroy}, Charlie and {Fuller}, Jim and {Kiman}, Rocio and {van Roestel}, Jan and {Rodriguez}, Antonio C. and {Burdge}, Kevin B.},
        title = "{Magnetic braking saturates: evidence from the orbital period distribution of low-mass detached eclipsing binaries from ZTF}",
      journal = {\mnras},
         year = 2022,
        month = dec,
       volume = {517},
       number = {4},
        pages = {4916-4939},
          doi = {10.1093/mnras/stac2945},
archivePrefix = {arXiv},
       eprint = {2208.05488},
 primaryClass = {astro-ph.SR},
       adsurl = {https://ui.adsabs.harvard.edu/abs/2022MNRAS.517.4916E}
}

@ARTICLE{Weisz2023,
       author = {{Weisz}, Daniel R. and {McQuinn}, Kristen B.~W. and {Savino}, Alessandro and {Kallivayalil}, Nitya and {Anderson}, Jay and {Boyer}, Martha L. and {Correnti}, Matteo and {Geha}, Marla C. and {Dolphin}, Andrew E. and {Sandstrom}, Karin M. and {Cole}, Andrew A. and {Williams}, Benjamin F. and {Skillman}, Evan D. and {Cohen}, Roger E. and {Newman}, Max J.~B. and {Beaton}, Rachael and {Bressan}, Alessandro and {Bolatto}, Alberto and {Boylan-Kolchin}, Michael and {Brooks}, Alyson M. and {Bullock}, James S. and {Conroy}, Charlie and {Cooper}, M.~C. and {Dalcanton}, Julianne J. and {Dotter}, Aaron L. and {Fritz}, Tobias K. and {Garling}, Christopher T. and {Gennaro}, Mario and {Gilbert}, Karoline M. and {Girardi}, L{\'e}o and {Johnson}, Benjamin D. and {Johnson}, L. Clifton and {Kalirai}, Jason S. and {Kirby}, Evan N. and {Lang}, Dustin and {Marigo}, Paola and {Richstein}, Hannah and {Schlafly}, Edward F. and {Schmidt}, Judy and {Tollerud}, Erik J. and {Warfield}, Jack T. and {Wetzel}, Andrew},
        title = "{The JWST Resolved Stellar Populations Early Release Science Program. II. Survey Overview}",
      journal = {\apjs},
         year = 2023,
        month = sep,
       volume = {268},
       number = {1},
          eid = {15},
        pages = {15},
          doi = {10.3847/1538-4365/acdcfd},
archivePrefix = {arXiv},
       eprint = {2301.04659},
 primaryClass = {astro-ph.GA},
       adsurl = {https://ui.adsabs.harvard.edu/abs/2023ApJS..268...15W}
}

@ARTICLE{Weisz2024,
       author = {{Weisz}, Daniel R. and {Dolphin}, Andrew E. and {Savino}, Alessandro and {McQuinn}, Kristen B.~W. and {Newman}, Max J.~B. and {Williams}, Benjamin F. and {Kallivayalil}, Nitya and {Anderson}, Jay and {Boyer}, Martha L. and {Correnti}, Matteo and {Geha}, Marla C. and {Sandstrom}, Karin M. and {Cole}, Andrew A. and {Warfield}, Jack T. and {Skillman}, Evan D. and {Cohen}, Roger E. and {Beaton}, Rachael and {Bressan}, Alessandro and {Bolatto}, Alberto and {Boylan-Kolchin}, Michael and {Brooks}, Alyson M. and {Bullock}, James S. and {Conroy}, Charlie and {Cooper}, Michael C. and {Dalcanton}, Julianne J. and {Dotter}, Aaron L. and {Fritz}, Tobias K. and {Garling}, Christopher T. and {Gennaro}, Mario and {Gilbert}, Karoline M. and {Girardi}, Leo and {Johnson}, Benjamin D. and {Johnson}, L. Clifton and {Kalirai}, Jason and {Kirby}, Evan N. and {Lang}, Dustin and {Marigo}, Paola and {Richstein}, Hannah and {Schlafly}, Edward F. and {Tollerud}, Erik J. and {Wetzel}, Andrew},
        title = "{The JWST Resolved Stellar Populations Early Release Science Program. V. DOLPHOT Stellar Photometry for NIRCam and NIRISS}",
      journal = {\apjs},
         year = 2024,
        month = apr,
       volume = {271},
       number = {2},
          eid = {47},
        pages = {47},
          doi = {10.3847/1538-4365/ad2600},
archivePrefix = {arXiv},
       eprint = {2402.03504},
 primaryClass = {astro-ph.GA},
       adsurl = {https://ui.adsabs.harvard.edu/abs/2024ApJS..271...47W}
}

@ARTICLE{Urquhart2026,
       author = {{Urquhart}, Ryan and {Strader}, Jay and {Chomiuk}, Laura and {Ransom}, Scott M. and {Heinke}, Craig O. and {Bahramian}, Arash and {Maccarone}, Thomas J.},
        title = "{New deep radio continuum imaging still indicates a large reservoir of undiscovered millisecond pulsars in Terzan 5}",
      journal = {\mnras},
         year = 2026,
        month = apr,
       volume = {547},
       number = {2},
          eid = {stag276},
        pages = {stag276},
          doi = {10.1093/mnras/stag276},
archivePrefix = {arXiv},
       eprint = {2602.10199},
 primaryClass = {astro-ph.HE},
       adsurl = {https://ui.adsabs.harvard.edu/abs/2026MNRAS.547ag276U}
}

@ARTICLE{Vleeschower2024,
       author = {{Vleeschower}, L. and {Corongiu}, A. and {Stappers}, B.~W. and {Freire}, P.~C.~C. and {Ridolfi}, A. and {Abbate}, F. and {Ransom}, S.~M. and {Possenti}, A. and {Padmanabh}, P.~V. and {Balakrishnan}, V. and {Kramer}, M. and {Venkatraman Krishnan}, V. and {Zhang}, L. and {Bailes}, M. and {Barr}, E.~D. and {Buchner}, S. and {Chen}, W.},
        title = "{Discoveries and timing of pulsars in M62}",
      journal = {\mnras},
         year = 2024,
        month = may,
       volume = {530},
       number = {2},
        pages = {1436-1456},
          doi = {10.1093/mnras/stae816},
archivePrefix = {arXiv},
       eprint = {2403.12137},
 primaryClass = {astro-ph.HE},
       adsurl = {https://ui.adsabs.harvard.edu/abs/2024MNRAS.530.1436V}
}

@ARTICLE{Nieder2020,
       author = {{Nieder}, L. and {Allen}, B. and {Clark}, C.~J. and {Pletsch}, H.~J.},
        title = "{Exploiting Orbital Constraints from Optical Data to Detect Binary Gamma-Ray Pulsars}",
      journal = {\apj},
         year = 2020,
        month = oct,
       volume = {901},
       number = {2},
          eid = {156},
        pages = {156},
          doi = {10.3847/1538-4357/abaf53},
archivePrefix = {arXiv},
       eprint = {2004.11740},
 primaryClass = {astro-ph.HE},
       adsurl = {https://ui.adsabs.harvard.edu/abs/2020ApJ...901..156N}
}

@ARTICLE{Romani2011,
       author = {{Romani}, Roger W. and {Shaw}, Michael S.},
        title = "{The Orbit and Companion of Probable {\ensuremath{\gamma}}-Ray Pulsar J2339-0533}",
      journal = {\apjl},
         year = 2011,
        month = dec,
       volume = {743},
       number = {2},
          eid = {L26},
        pages = {L26},
          doi = {10.1088/2041-8205/743/2/L26},
archivePrefix = {arXiv},
       eprint = {1111.3074},
 primaryClass = {astro-ph.HE},
       adsurl = {https://ui.adsabs.harvard.edu/abs/2011ApJ...743L..26R}
}

@ARTICLE{Romani2012,
       author = {{Romani}, Roger W.},
        title = "{2FGL J1311.7-3429 Joins the Black Widow Club}",
      journal = {\apjl},
         year = 2012,
        month = aug,
       volume = {754},
       number = {2},
          eid = {L25},
        pages = {L25},
          doi = {10.1088/2041-8205/754/2/L25},
archivePrefix = {arXiv},
       eprint = {1207.1736},
 primaryClass = {astro-ph.HE},
       adsurl = {https://ui.adsabs.harvard.edu/abs/2012ApJ...754L..25R}
}

@ARTICLE{Romani2014,
       author = {{Romani}, Roger W. and {Filippenko}, Alexei V. and {Cenko}, S. Bradley},
        title = "{2FGL J1653.6-0159: A New Low in Evaporating Pulsar Binary Periods}",
      journal = {\apjl},
         year = 2014,
        month = sep,
       volume = {793},
       number = {1},
          eid = {L20},
        pages = {L20},
          doi = {10.1088/2041-8205/793/1/L20},
archivePrefix = {arXiv},
       eprint = {1408.2886},
 primaryClass = {astro-ph.HE},
       adsurl = {https://ui.adsabs.harvard.edu/abs/2014ApJ...793L..20R}
}

@ARTICLE{Romani2015,
       author = {{Romani}, Roger W.},
        title = "{A Likely Millisecond Pulsar Binary Counterpart for Fermi Source 2FGL J2039.6-5620}",
      journal = {\apjl},
         year = 2015,
        month = oct,
       volume = {812},
       number = {2},
          eid = {L24},
        pages = {L24},
          doi = {10.1088/2041-8205/812/2/L24},
archivePrefix = {arXiv},
       eprint = {1509.07834},
 primaryClass = {astro-ph.HE},
       adsurl = {https://ui.adsabs.harvard.edu/abs/2015ApJ...812L..24R}
}

@ARTICLE{Cadelano2018,
       author = {{Cadelano}, M. and {Ransom}, S.~M. and {Freire}, P.~C.~C. and {Ferraro}, F.~R. and {Hessels}, J.~W.~T. and {Lanzoni}, B. and {Pallanca}, C. and {Stairs}, I.~H.},
        title = "{Discovery of Three New Millisecond Pulsars in Terzan 5}",
      journal = {\apj},
         year = 2018,
        month = mar,
       volume = {855},
       number = {2},
          eid = {125},
        pages = {125},
          doi = {10.3847/1538-4357/aaac2a},
archivePrefix = {arXiv},
       eprint = {1801.09929},
 primaryClass = {astro-ph.HE},
       adsurl = {https://ui.adsabs.harvard.edu/abs/2018ApJ...855..125C}
}

@ARTICLE{Borowski2025,
       author = {{Borowski}, E.~S. and {Hynes}, R.~I. and {Hunt}, Q. and {Tetarenko}, A.~J. and {Plotkin}, R.~M. and {Shahbaz}, T. and {Gandhi}, P. and {Maccarone}, T.~J. and {Miller-Jones}, J.~C.~A. and {Heinke}, C.~O. and {Shaw}, A.~W. and {Russell}, T.~D. and {Sivakoff}, G.~R. and {Charles}, P.~A. and {Palaiologou}, E.~V. and {Reig}, P.},
        title = "{The Mid-infrared-emitting Jet in the Black Hole V404 Cygni in Quiescence}",
      journal = {\apjl},
         year = 2025,
        month = oct,
       volume = {991},
       number = {2},
          eid = {L31},
        pages = {L31},
          doi = {10.3847/2041-8213/ae0322},
archivePrefix = {arXiv},
       eprint = {2506.20536},
 primaryClass = {astro-ph.HE},
       adsurl = {https://ui.adsabs.harvard.edu/abs/2025ApJ...991L..31B}
}

@ARTICLE{Skrutskie2006,
       author = {{Skrutskie}, M.~F. and {Cutri}, R.~M. and {Stiening}, R. and {Weinberg}, M.~D. and {Schneider}, S. and {Carpenter}, J.~M. and {Beichman}, C. and {Capps}, R. and {Chester}, T. and {Elias}, J. and {Huchra}, J. and {Liebert}, J. and {Lonsdale}, C. and {Monet}, D.~G. and {Price}, S. and {Seitzer}, P. and {Jarrett}, T. and {Kirkpatrick}, J.~D. and {Gizis}, J.~E. and {Howard}, E. and {Evans}, T. and {Fowler}, J. and {Fullmer}, L. and {Hurt}, R. and {Light}, R. and {Kopan}, E.~L. and {Marsh}, K.~A. and {McCallon}, H.~L. and {Tam}, R. and {Van Dyk}, S. and {Wheelock}, S.},
        title = "{The Two Micron All Sky Survey (2MASS)}",
      journal = {\aj},
         year = 2006,
       volume = {131},
       number = {2},
        pages = {1163-1183},
          doi = {10.1086/498708},
       adsurl = {https://ui.adsabs.harvard.edu/abs/2006AJ....131.1163S},
}

@ARTICLE{Samus2017,
       author = {{Samus}, N.~N. and {Kazarovets}, E.~V. and {Durlevich}, O.~V. and {Kireeva}, N.~N. and {Pastukhova}, E.~N.},
        title = "{General catalogue of variable stars: Version GCVS 5.1}",
      journal = {Astronomy Reports},
         year = 2017,
        month = jan,
       volume = {61},
       number = {1},
        pages = {80-88},
          doi = {10.1134/S1063772917010085},
       adsurl = {https://ui.adsabs.harvard.edu/abs/2017ARep...61...80S},
}
\bibliographystyle{aasjournalv7}

\end{document}